\pdfoutput=0
\documentclass[10pt]{report}
\usepackage{axodraw}
\usepackage{epsfig}
\usepackage{amsfonts}
\usepackage{bbm,bm}
\usepackage{makeidx}
\usepackage{cite}

\newcommand{\Eq}{Eq.~}
\newcommand{\Eqs}{Eqs.~}
\newcommand{\gammaE}{\gamma_\rmii{E}}

\input pix.sty
\makeindex

\newcommand{\iB}{\rmi{$B$}}

\newcommand{\iM}{\rmi{$M$}}
\newcommand{\iiM}{\rmii{$M$}}
\newcommand{\iA}{\rmi{$A$}}
\newcommand{\iR}{\rmi{$R$}}
\newcommand{\iT}{\rmi{$T$}}
\newcommand{\ibT}{\rmi{$\overline{T}$}}
\newcommand{\iE}{\rmi{$E$}}
\newcommand{\iiE}{\rmii{$E$}}
\newcommand{\iH}{\rmi{$H$}}
\newcommand{\iI}{\rmi{$I$}}

\newcommand{\E}{\epsilon}
\newcommand{\ses}{secs.~}
\newcommand{\alphas}{\alpha_s}
\newcommand{\CF}{C^{ }_\rmii{F}}
\newcommand{\aL}{a^{ }_\rmii{L}}
\newcommand{\aR}{a^{ }_\rmii{R}}

\newcommand{\Nf}{N_{\rm f}}
\newcommand{\Nc}{N_{\rm c}}

\newcommand{\dA}{d^{ }_\rmii{$\!A$}}
\newcommand{\Tc}{T_{\rm c}}

\newcommand{\mE}{m_\rmii{E}}

\newcommand{\rmO}{{\mathcal{O}}}
\newcommand{\bmu}{\bar\mu}
\newcommand{\CA}{\Nc}

\def\lsi{\raise0.3ex\hbox{$<$\kern-0.75em\raise-1.1ex\hbox{$\sim$}}}
\def\gsi{\raise0.3ex\hbox{$>$\kern-0.75em\raise-1.1ex\hbox{$\sim$}}}
\newcommand{\lsim}{\mathop{\lsi}}
\newcommand{\gsim}{\mathop{\gsi}}

\newcommand{\fe}{\rmi{f}}
\newcommand{\bo}{\rmi{b}}

\newcommand{\disc}{\mathop{\mbox{Disc}}}
\newcommand{\sign}{\mathop{\mbox{sign}}}
\newcommand{\nF}[1]{n_\rmii{F{#1}}} 
\newcommand{\nB}[1]{n_\rmii{B{#1}}} 
\newcommand{\rmii}[1]{{\mbox{\tiny\rm{#1}}}}
\newcommand{\re}{\mathop{\rm {Re}}}
\newcommand{\im}{\mathop{\rm {Im}}}

\newcommand{\Tint}[1]{{\hbox{$\sum$}\!\!\!\!\!\!\!\int\,}_{\!\!\!\!\raise-0.9ex\hbox{$\scriptstyle{#1}$}}}
\newcommand{\Tinti}[1]{{{\Sigma}\!\!\!\!\raise0.3ex\hbox{$\int$}_\rmii{${#1}$}}}

\newcommand{\RR}{{\mathbb{R}}}
\newcommand{\ZZ}{{\mathbb{Z}}}
\newcommand{\CC}{{\mathbb{C}}}
\newcommand{\unit}{{\mathbbm{1}}} 
\newcommand{\bi}{\begin{itemize}}
\newcommand{\ei}{\end{itemize}}
\newcommand{\hide}[1]{ }
\newcommand{\bsl}[1]{\,\slash\!\!\!\!{#1}\,}

\newcommand{\deltabar}{\raise-0.02em\hbox{$\bar{}$}\hspace*{-0.8mm}{\delta}}
\newcommand{\f}{\mbox{\sl f\,}}
\newcommand{\ff}{\rmi{\sl f\,}}
\newcommand{\g}{\mbox{\sl i\,}}

\newcommand{\piB}[1]{\;\parbox[c]{60pt}{\begin{picture}(60,40)(0,0)
\SetWidth{1.0}\SetScale{1.0} #1 \end{picture}}\;}

\def\Daisy{\piB{%
 \SetWidth{1.0} 
 \DashCArc(30,20)(20,0,355){2}%
 \SetWidth{1.0} 
 \CArc(30,50)(10,0,360)%
 \CArc(51.21,41.21)(10,0,360)%
 \CArc(8.79,41.21)(10,0,360)%
 \CArc(0,20)(10,0,360)%
 \CArc(8.79,-1.21)(10,0,360)%
 \CArc(30,-10)(10,0,360)%
 \CArc(51.21,-1.21)(10,0,360)%
 \GCirc(60,20){1}{0}%
 \GCirc(59,25){1}{0}%
 \GCirc(59,15){1}{0}%
}}
\def\ThreeA{\piB{%
 \SetWidth{1.0} 
 \CArc(30,20)(15,0,360)%
 \CArc(30,3.9)(22,47.0,133.0)%
 \CArc(30,36.1)(22,227.0,313.0)%
}}
\def\ThreeB{\piB{%
 \SetWidth{1.0} 
 \CArc(30,20)(15,0,360)%
 \CArc(5,20)(10,0,360)%
 \CArc(55,20)(10,0,360)%
}}
\def\lxi{\piB{%
 \SetWidth{1.0} 
 \Line(0,25)(28,25)%
 \Line(0,15)(28,15)%
 \Lqu(28,25)(100,40)%
 \Laqu(28,15)(100,0)%
 \Lagh(35,20)(100,34)%
 \Lgh(41,18)(90,8)%
 \Lghnoarr(90,8)(100,6)%
 \Laghnoarr(14,3)(35,20)%
 \Lghnoarr(18,0)(41,18)%
 \Photon(5,15)(5,25){1.5}{2}%
 \Photon(15,15)(15,25){1.5}{2}%
 \Photon(25,15)(25,25){1.5}{2}%
 \Photon(43,22)(43,28){1.5}{1.5}%
 \Photon(57,25)(57,31){1.5}{1.5}%
 \Photon(75,29)(75,35){1.5}{1.5}%
 \Photon(89,32)(89,38){1.5}{1.5}%
 \Photon(43,12)(43,18){1.5}{1.5}%
 \Photon(57,9)(57,15){1.5}{1.5}%
 \Photon(75,5)(75,11){1.5}{1.5}%
 \Photon(89,2)(89,8){1.5}{1.5}%
 \Photon(19,7)(22.5,3.5){1.5}{1.5}%
 \Photon(24,11.5)(28,8){1.5}{1.5}%
 \Textl(-6,-8,{\displaystyle e^{-E_\pi/T}})
}}

\makeatletter \@addtoreset{equation}{section} \makeatother
\renewcommand{\theequation}{\arabic{section}.\arabic{equation}}
\makeatletter
\renewcommand\section{\@startsection {section}{1}{\z@}%
                                   {-5.5ex \@plus -1ex \@minus -.2ex}
                                   {2.3ex \@plus.2ex}%
                                   {\normalfont\large\bfseries}}
\renewcommand\subsection{\@startsection{subsection}{2}{\z@}%
                                     {-3.25ex\@plus -1ex \@minus -.2ex}%
                                     {1.5ex \@plus .2ex}%
                                     {\normalfont\normalsize\bfseries}}
\renewcommand\thesection {\@arabic\c@section}
\renewcommand\thesubsection   {\thesection.\@arabic\c@subsection}
\renewcommand{\@seccntformat}[1]{%
\csname the#1\endcsname.\hspace{1.0em}}
\newdimen\rh@wd
\newdimen\rh@hta
\newdimen\rh@htb
\newbox\rh@box
\def\rh@measure#1{\setbox\rh@box=\hbox{$#1$}\rh@wd=\wd\rh@box \rh@hta=\ht\rh@box}
\def\widecheck#1{\rh@measure{#1}%
  \setbox\rh@box=\hbox{$\widehat{\vrule height \rh@hta width\z@ \kern\rh@wd}$}%
  \rh@htb=\ht\rh@box \advance\rh@htb\rh@hta \advance\rh@htb\p@
  \ooalign{$\vrule height \ht\rh@box width\z@ #1$\cr
           \raise\rh@htb\hbox{\scalebox{1}[-1]{\box\rh@box}}\cr}}
\renewcommand\@bibitem[1]{\item\if@filesw \immediate\write\@auxout
  {\string\bibcite{#1}{\theenumiv}}\fi\ignorespaces}
\renewenvironment{thebibliography}[1]
                 {\section*{\refname}
\@mkboth{\MakeUppercase\refname}{\MakeUppercase\refname}%
\list{\@biblabel{\the\c@section.\@arabic\c@enumiv}}%
                        {\settowidth\labelwidth{\@biblabel{#1}}%
                          \leftmargin\labelwidth
                          \advance\leftmargin\labelsep
                          \@openbib@code
                          \usecounter{enumiv}%
                          \let\p@enumiv\@empty
                    \renewcommand\theenumiv{\the\c@section.\@arabic\c@enumiv}}%
                        \sloppy
                        \clubpenalty4000
                        \@clubpenalty \clubpenalty
                        \widowpenalty4000%
                        \sfcode`\.\@m}
                 {\def\@noitemerr
                   {\@latex@warning{Empty `thebibliography' environment}}%
                   \endlist}
\renewcommand\thesection {\@arabic\c@section}
\renewcommand\thesubsection   {\thesection.\@arabic\c@subsection}
\renewcommand{\@seccntformat}[1]{%
\csname the#1\endcsname.\hspace{1.0em}}
\def\refname{{\large Literature}}

\makeatother

\begin{document}

\begin{titlepage} 

\begin{flushright}
January 2022 
\end{flushright}

\vfill

\begin{centering} 

{\Large Basics of Thermal Field Theory}

\vspace*{0.3cm}

{\it\large A Tutorial on Perturbative Computations} \footnote{%
  An earlier version of these notes is available as an ebook   
  (Springer Lecture Notes in Physics 925) 
  at
  {\tt dx.doi.org/10.1007/978-3-319-31933-9};
  an eprint can be found at  
  {\tt arxiv.org/abs/1701.01554};
  the very latest version 
  is kept up to date at
  {\tt www.laine.itp.unibe.ch/basics.pdf}.
}

\vspace*{0.3cm}

Mikko Laine$^\rmi{a}$ and Aleksi Vuorinen$^\rmi{b}$

\vspace*{0.3cm}

$^\rmi{a}$%
{\em
AEC, 
Institute for Theoretical Physics, 
University of Bern, \\ 
Sidlerstrasse 5, CH-3012 Bern, Switzerland \\}

\vspace*{0.3cm}

$^\rmi{b}$%
{\em
Department of Physics, University of Helsinki, \\ 
P.O.\ Box 64, FI-00014 University of Helsinki, Finland \\}

\vspace*{0.8cm}

\mbox{\bf Abstract}

\end{centering}

\vspace*{0.3cm}
 
\noindent
These lecture notes, suitable for a two-semester introductory course or 
self-study, offer an elementary and self-contained exposition of the
basic tools and concepts that are encountered in practical computations in 
perturbative thermal field theory. Selected applications to heavy ion 
collision physics and cosmology are outlined in the last chapter. 

\vfill

\vspace*{4cm}

\end{titlepage}

\thispagestyle{empty}

\mbox{ } 

\newpage

\tableofcontents

\newpage 

\pagenumbering{roman}
\setcounter{page}{1}

\section*{Foreword}

These notes are based on lectures delivered at the 
Universities of Bielefeld and Helsinki, between 2004 and 2015, 
as well as at a number of summer and winter schools, 
between 1996 and 2018. The early 
sections were strongly influenced by lectures by Keijo Kajantie
at the University of Helsinki, in the early 1990s. 
Obviously, the lectures additionally owe an enormous gratitude to 
existing text books and 
literature, particularly the classic monograph by Joseph Kapusta.

There are several good text books on finite-temperature 
field theory, and no attempt is made here to join that group. 
Rather, the goal is to offer an elementary exposition of 
the basics of the field, in an explicit ``hands-on'' 
style which can hopefully more or less directly be transported
to the classroom. The presentation is meant to be self-contained
and display also intermediate steps. 
The idea is, roughly, that each numbered 
section could constitute a single lecture.
Referencing is sparse;
on more advanced topics, as well as 
on historically accurate references, 
the reader is advised to consult the text books and review articles in 
refs.~\cite{jk}--\hspace*{-1.1mm}\cite{rev5}.

These notes could not have been put together without the helpful
influence of many people, varying from students with persistent
requests for clarification; 
colleagues who have used parts of an early
version of these notes in their own lectures and shared their 
experiences with us; colleagues whose interest in specific topics
has inspired us to add corresponding material to these notes; 
alert readers who have informed us about typographic errors 
and suggested improvements; and collaborators from whom
we have learned parts of the material presented here. 
Let us gratefully acknowledge in particular
Gert Aarts, 
Chris Korthals Altes, 
Dietrich B\"odeker, 
Yannis Burnier, 
Stefano Capitani, 
Simon Caron-Huot, 
Jacopo Ghiglieri,  
Ioan Ghisoiu,
Kimmo Kainulainen, 
Keijo Kajantie,
Aleksi Kurkela,
Harvey Meyer,  
Guy Moore, 
Paul Romatschke,
Kari Rummukainen,  
York Schr\"oder,
Mikhail Shaposhnikov, 
Markus Thoma, 
Tanmay Vachaspati, 
 and
Mikko Veps\"al\"ainen. 

\vspace*{5mm}

\hfill Mikko Laine and Aleksi Vuorinen

\addcontentsline{toc}{section}{Foreword}

\newpage 

\section*{Notation}

In thermal field theory, both Euclidean
and Minkowskian spacetimes play a role. 

In the Euclidean case, we write 
\ba
  && X \equiv (\tau,{x}^i) \;, \quad
  x \equiv |\vec{x}| \;, \quad
  S^{ }_\iE = \int_X \, L^{ }_\iE \;,  
\ea
where $i = 1, ..., d$, 
\ba
  && 
  \int_X \equiv \int_0^\beta \! {\rm d}\tau \int_\vec{x} \;, \quad
  \int_\vec{x} \equiv \int \! {\rm d}^d\vec{x} 
  \;, \quad
  \beta \equiv \frac{1}{T} 
  \;, \la{intX}
\ea
and $d$ is the space dimensionality.  Fourier analysis 
is carried out in the Matsubara formalism via 
\ba
  && K \equiv (k^{ }_n,{k}^{ }_i) \;, \quad
  k \equiv |\vec{k}| \;, \quad
  \phi(X) = \Tint{K} \tilde\phi(K)
  \, e^{i K\cdot X}
  \;, 
\ea
where
\ba
  && \Tint{K} \equiv T\sum_{k^{ }_n} \int_\vec{k} \;, \quad
  \int_\vec{k} \equiv \int \! \frac{{\rm d}^d\vec{k}}{(2\pi)^d}
  \;.
\ea
Here, $k^{ }_n$ stands for discrete Matsubara frequencies, which at times are
also denoted by $\omega^{ }_n$. 
In the case of antiperiodic functions, 
the summation is written as $T\sum_{\{ k^{ }_n\} }$. The squares 
of four-vectors read $K^2=k_n^2+k^2$ and $X^2=\tau^2+x^2$, 
but the Euclidean scalar product between $K$ and $X$ is defined as
\be
  K \cdot X = k^{ }_n \tau + \sum_{i=1}^d k^{ }_i x^i
 = k^{ }_n \tau - \vec{k}\cdot\vec{x}
 \;, \la{sprod}
\ee
where the vector notation is reserved for contravariant 
Minkowskian vectors: $\vec{x} = (x^i)$, $\vec{k} = (k^i)$.
If a chemical potential is also present, we denote
$
 \tilde k^{ }_n \equiv k^{ }_n + i \mu
$.

In the Minkowskian case, we have
\ba
  && \mathcal{X} \equiv (t,\vec{x}) \;, \quad
  x \equiv |\vec{x}| \;, \quad
  \mathcal{S}^{ }_\iM = \int_\mathcal{X} \, \mathcal{L}^{ }_\iM 
  \;, 
\ea
where 
$
 \int_\mathcal{X} \equiv \int \! {\rm d}x^0 \int_\vec{x}
$. 
Fourier analysis proceeds via 
\ba
  && \mathcal{K} \equiv (k^0_{ },\vec{k}) \;, \quad
  k \equiv |\vec{k}| \;, \quad
 \phi(\mathcal{X}) = \int_\mathcal{K} \tilde\phi(\mathcal{K}) 
 \, e^{i \mathcal{K}\cdot\mathcal{X}}
 \;,
\ea
where 
$
  \int_\mathcal{K} = \int \frac{{\rm d}k^0_{ }}{2\pi} \int_\vec{k}
$,
and the metric is chosen to be of the ``mostly minus'' form, 
\be
 \mathcal{K}\cdot\mathcal{X} = k^0_{ } x^0 - \vec{k}\cdot\vec{x}
 \;. 
\ee
No special notation is introduced for the case where 
a Minkowskian four-vector is on-shell, i.e.\ when 
$\mathcal{K} = (\E^{ }_k,\vec{k})$; 
this is to be understood from the context.

The argument of a field $\phi$ is taken to indicate 
whether the configuration space is Euclidean or Minkowskian. 
If not specified otherwise, momentum integrations
are regulated by defining the spatial measure   
in $d=3-2\epsilon$ dimensions, whereas the spacetime 
dimensionality is denoted by $D = 4 -2\epsilon$.
A Greek index takes values in the set $\{0,...,d\}$, 
and a Latin one in $\{1,...,d\}$.

Finally, we note that we work consistently in units where the speed of light
$c$ and the Boltzmann constant $k^{ }_\rmii{B}$ have been set to unity. 
The reduced
Planck constant $\hbar$ also equals unity in most places, excluding the
first chapter (on quantum mechanics) as well as some later discussions where
we want to emphasize the distinction between quantum and classical
descriptions.

\addcontentsline{toc}{section}{Notation}

\newpage 

\addcontentsline{toc}{section}{General outline}

\section*{General outline}
\la{literature}

\subsection*{Physics context}

{}From the physics point of view, there are two important contexts in which 
relativistic thermal field theory is being widely applied: cosmology and the 
theoretical description of heavy ion collision experiments. 

In cosmology, the temperatures considered vary hugely, 
ranging from $T \simeq 10^{15}$~GeV
to $T \simeq 10^{-3}$~eV. Contemporary challenges 
in the field include figuring out 
explanations for the existence of dark matter, the observed antisymmetry in 
the amounts of matter and antimatter, and the formation of large-scale
structures from small initial density perturbations. (The origin of 
initial density perturbations itself is generally considered to be  
a non-thermal problem, associated with an early period of inflation.)
An important further issue is that of equilibration, 
i.e.\ details of the processes through which the inflationary
state turned into a thermal plasma, and in particular 
what the highest temperature
reached during this epoch was. It is notable that most of these
topics are assumed to be associated with weak or even superweak interactions, 
whereas strong interactions (QCD) only play a background role. 
A notable exception to this is light element nucleosynthesis, but this
well-studied topic is not in the center of our current focus.

In heavy ion collisions, in contrast, 
strong interactions do play a major role. 
The lifetime of the thermal fireball created 
in such a collision is $\sim 10$~fm/c and the maximal 
temperature reached is in the range of a few hundred MeV. 
Weak interactions are
too slow to take place within the lifetime of the system. 
Prominent observables are
the yields of different particle species, the quenching of energetic jets, 
and the hydrodynamic properties
of the plasma that can be deduced from the observed particle yields. 
An important
issue is again how fast an initial quantum-mechanical state turns into
an essentially incoherent thermal plasma. 

Despite many differences in the physics questions posed and 
in the microscopic forces underlying cosmology 
and heavy ion collision phenomena, there are also
similarities. Most importantly, gauge interactions (whether weak
or strong) are essential in both contexts. Because of asymptotic freedom, 
the strong interactions of QCD also become ``weak'' at sufficiently high
temperatures. It is for this reason that many techniques, such as 
the resummations that are needed for developing a formally consistent
weak-coupling expansion, can be applied in both contexts. The topics 
covered in the present notes have been chosen with both fields of 
application in mind. 

\subsection*{Organization of these notes}

The notes start with the definition and computation of basic ``static''
thermodynamic quantities, such as the partition function 
and free energy density, in various settings. 
Considered are in turn quantum mechanics (\se\ref{se:qm}), 
free and interacting scalar field theories (secs.~\ref{se:free_sft} 
and \ref{se:int}, respectively), fermionic systems (\se\ref{se:fermions}), 
and gauge fields (\se\ref{se:Gauge}). The main points of these 
sections include the introduction
of the so-called imaginary-time formalism; the functioning of 
renormalization at finite temperature; and the issue of infrared
problems that complicates almost every computation 
in relativistic thermal field theory. The last of these
issues leads us to introduce the concept of effective field 
theories (\se\ref{se:EFT}), after which we consider the changes caused by 
the introduction of a finite density 
or chemical potential (\se\ref{se:density}). 
After these topics, we move on to a new set
of observables, so-called real-time quantities, which play an essential
role in many modern phenomenological applications of thermal field 
theory (\se\ref{se:real-time}). In the final chapter of the book, a number
of concrete applications of the techniques introduced 
are discussed (\se\ref{se:app}).

We note that \ses\ref{se:qm}--\ref{se:density} are presented on an elementary
and self-contained level and require no background knowledge beyond 
statistical physics, quantum mechanics, and rudiments of quantum field theory. 
They could constitute the contents of a one-semester basic introduction 
to perturbative thermal field theory. 
In \se\ref{se:real-time}, the level increases
gradually, and parts of 
the discussion in \se\ref{se:app}
are already close to the research level, requiring more background
knowledge. Conceivably the topics of \ses\ref{se:real-time} and 
\ref{se:app} could be covered in an advanced course on perturbative 
thermal field theory, or in a graduate student seminar. 
In addition the whole book is suitable for self-study, 
and is then advised to be 
read in the order in which the material has been presented. 

\subsection*{Recommended literature}

A pedagogical presentation
of thermal field theory, concentrating mostly on Euclidean 
observables and the imaginary-time formalism, can be found in 
ref.~\cite{jk}. The current notes borrow significantly from 
this classic treatise. 

In thermal field theory, the community is somewhat divided between
those who find the imaginary-time formalism more practicable, 
and those who prefer to use the so-called 
real-time formalism from the beginning. 
Particularly for the latter community, 
the standard reference is ref.~\cite{rev2}, 
which also contains an introduction
to particle production rate computations. 

A modern textbook, partly an update of ref.~\cite{jk} but including also 
a full account of real-time observables, as well as reviews on many 
recent developments, is provided by ref.~\cite{rev3}.  

Lecture notes on transport coefficients, infrared resummations,  
and non-equilibrium phenomena such as thermalization, 
can be found in ref.~\cite{rev4}.  
Reviews with varying foci are offered by
refs.~\cite{rev4a}--\hspace*{-1.1mm}\cite{rev4i}.

Finally, an extensive review of efforts 
to approach a non-perturbative understanding of real-time 
thermal field theory has been presented in ref.~\cite{rev5}.  


%

\newpage

\setcounter{enumiv}{1}

\newpage 

\pagenumbering{arabic}
\setcounter{page}{1}

\section{Quantum mechanics}
\la{se:qm}

\paragraph{Abstract:}

After recalling some 
basic concepts of statistical physics and quantum mechanics, 
the partition function of a harmonic oscillator 
is defined and evaluated in the standard
canonical formalism. An imaginary-time path integral 
representation is subsequently developed for the partition function, the path
integral is 
evaluated in momentum space, and the earlier result 
is reproduced upon a careful treatment of the zero-mode contribution.
Finally, the concept of 2-point functions
(propagators) is introduced, and some of their key properties 
are derived in imaginary time. 

\paragraph{Keywords:} 

Partition function, Euclidean path integral, 
imaginary-time formalism, Matsubara modes, 2-point function. 

%
\subsection{Path integral representation of the partition function}
\la{ss:pi_qm}

\subsection*{Basic structure}

\index{Canonical quantization: harmonic oscillator}

The properties of a quantum-mechanical 
system are defined by its {\em Hamiltonian}, 
which for non-relativistic spin-0 particles in one dimension takes the form 
\be 
 \hat H = \frac{\hat p^2}{2 m} + V(\hat x)
 \;, \la{H}
\ee
where $m$ is the particle mass. 
The dynamics of the states $| \psi \rangle$ is governed by the 
{\em Schr\"odinger equation}, 
\be
 i \hbar \frac{\partial}{\partial t} | \psi \rangle = \hat H |\psi \rangle
 \;,
\ee
which can formally be solved in terms of a 
{\em time-evolution operator} $\hat U(t;t^{ }_0)$. 
This operator satisfies the relation
\be
 |\psi(t) \rangle = \hat U(t;t^{ }_0) |\psi (t^{ }_0) \rangle
 \;, 
\ee
and for a time-independent Hamiltonian takes the explicit form 
\be
 \hat U(t;t^{ }_0) = e^{-\frac{i}{\hbar} \hat H(t-t^{ }_0)}
 \;. 
\ee
It is useful to note that in the {\em classical limit}, the system
of \eq\nr{H} can be described by the {\em Lagrangian}
\be
 \mathcal{L} = 
 \mathcal{L}^{ }_\iM = \frac{1}{2} m \dot x^2 - V(x)
 \;, \la{qmLM}
\ee
which is related to the classical
version of the Hamiltonian via a simple Legendre transform: 
\be
 p \equiv \frac{\partial \mathcal{L}^{ }_\iM}{\partial \dot{x}}
 \;, \quad
 {H} = \dot{x} p - \mathcal{L}^{ }_\iM = 
 \frac{p^2}{2 m } + V(x) 
 \;.
\ee

Returning to the quantum-mechanical setting, 
various {\em bases} can be chosen for the state vectors. 
The so-called $|x\rangle$-basis satisfies the relations
\be
 \langle x | \hat x | x' \rangle = x \langle x | x' \rangle  = 
 x\,\delta(x-x')
 \;, \quad
 \langle x | \hat p | x' \rangle = -i \hbar\, \partial^{ }_x  
 \langle x | x' \rangle = -i \hbar\, \partial^{ }_x \,\delta(x-x')
 \;,
\ee
whereas in the energy basis we simply have
\be
 \hat H | n \rangle = \E^{ }_n | n \rangle 
 \;. 
\ee

An important concrete realization of a quantum-mechanical system
is provided by the {\em harmonic oscillator}, defined by the potential
\be
 V (\hat x) \equiv \frac{1}{2} m \omega^2 \hat x^2
 \;. 
\ee 
In this case the energy eigenstates $| n \rangle$ can be found explicitly, 
with the corresponding eigenvalues equalling
\be
 \E^{ }_n = \hbar \omega \Bigl( n + \frac{1}{2} \Bigr) 
 \;, \quad 
 n = 0,1,2, \ldots \;.
\ee
All the states are non-degenerate. 

It turns out to be useful to view (quantum) mechanics formally
as {\em (1+0)-dimensional (quantum) field theory}: the operator
$\hat x$ can be viewed as a field operator $\hat \phi$ at a 
certain point, implying the correspondence 
\be
 \hat x \leftrightarrow \hat \phi (\vec{0})
 \;. \la{qm_qft}
\ee
In quantum field theory operators are usually represented in the 
Heisenberg picture; correspondingly, we then have
\be
 \hat x^{ }_{\! H}(t) \leftrightarrow \hat \phi^{ }_H (t,\vec{0})
 \;.
\ee
In the following we adopt an implicit notation whereby
showing the time coordinate $t$ as an argument of a field automatically
implies the use of the Heisenberg picture, 
and the corresponding subscript is left out. 


\subsection*{Canonical partition function} 

\index{Partition function: harmonic oscillator}

Taking our quantum-mechanical system to a finite temperature $T$, 
the fundamental quantity of interest 
is the partition function, $\mathcal{Z}$. 
We employ the canonical ensemble, whereby $\mathcal{Z}$ is a function
of $T$; introducing units in which $k^{ }_\rmii{B} = 1$ (i.e.,\  
$
 T^{ }_\rmi{here} \equiv k^{ }_\rmii{B} T^{ }_\rmi{SI-units}
$), 
the partition function is defined by
\be
 \mathcal{Z}(T) \;\equiv\; \tr [e^{-\beta \hat H}]
 \;, \quad
 \beta \;\equiv\; \frac{1}{T}
 \;, 
\ee
where the trace is taken over the full Hilbert space. 
From this quantity, other observables, 
such as the free energy $F$, entropy $S$, and
average energy $E$ can be obtained via standard relations: 
\ba
 F & = &  - T \ln\mathcal{Z}
 \;, \\
 S & = & -\frac{\partial F}{\partial T} = 
 \ln\mathcal{Z} + \frac{1}{T \mathcal{Z}} \tr [\hat H e^{-\beta \hat H}]
 = - \frac{F}{T} + \frac{E}{T}
 \;, \\ 
 E & = & \frac{1}{\mathcal{Z}} \tr [\hat H e^{-\beta \hat H}]
 \;.
\ea

Let us now explicitly compute these quantities for the harmonic oscillator. 
This becomes a trivial exercise in the energy basis, 
given that we can immediately write
\ba
 \mathcal{Z} & = & 
 \sum_{n=0}^{\infty} 
 \langle n | e^{-\beta \hat H} | n \rangle 
 = \sum_{n=0}^{\infty} e^{-\beta\hbar \omega (\frac{1}{2} + n)}
 = 
 \frac{e^{-\beta\hbar\omega/2}}{1-e^{-\beta\hbar\omega}}
 = \frac{1}{2\sinh\bigl(\frac{\hbar\omega}{2 T}\bigr)}
 \;. \la{hoZ}
\ea
Consequently, 
\ba
 F & = & T \ln\Bigl( e^{\frac{\hbar\omega}{2 T}} -
  e^{ - \frac{\hbar\omega}{2 T}} \Bigr)
 = \frac{\hbar\omega}{2} + 
 T \ln \Bigl( 1 - e^{-\beta{\hbar\omega}} \Bigr)
 \la{hoF} \\
 & \approx & 
 \left\{ \begin{array}{ll} 
  \displaystyle \frac{\hbar\omega}{2} 
  \;, & T \ll \hbar\omega \\ 
  \displaystyle -T \ln\Bigl( \frac{T}{\hbar \omega} \Bigr)
  \;, & T \gg \hbar\omega 
  \end{array} 
  \right. 
 \;, \\
 S & = & - \ln \Bigl( 1 - e^{-\beta{\hbar\omega}} \Bigr) + 
 \frac{\hbar\omega}{T} \frac{1}{e^{\beta\hbar\omega} - 1}
 \la{hoS} \\
 & \approx & 
 \left\{ \begin{array}{ll} 
 \displaystyle
 \frac{\hbar\omega}{T} e^{-\frac{\hbar\omega}{T}}  
  \;, & T \ll \hbar\omega \\ 
 \displaystyle
 1 + \ln\frac{T}{\hbar\omega} 
  \;, & T \gg \hbar\omega 
  \end{array} 
  \right. 
 \;, \\
 E & = & F + T S = 
 \hbar\omega \, \biggl( \frac{1}{2} + \frac{1}{e^{\beta\hbar\omega} - 1}\biggr)
 \\
 & \approx & 
 \left\{ \begin{array}{ll} 
 \displaystyle
 \frac{\hbar\omega}{2} 
  \;, & T \ll \hbar\omega \\ 
 \displaystyle
 T 
  \;, & T \gg \hbar\omega 
  \end{array} 
  \right. 
 \;. 
\ea 
Note how in most cases one can separate the 
contribution of the ground state, dominating
at low temperatures $T \ll \hbar\omega$, 
from that of the thermally excited states,
characterized by the appearance of the Bose distribution
$\nB{}(\hbar\omega) \equiv 1/[\exp(\beta\hbar\omega) - 1]$. 
Note also 
that $E$ rises linearly with $T$ at high temperatures; 
the coefficient is said to count the number of degrees of freedom
of the system. 


\subsection*{Path integral for the partition function}
\la{se:p3}

\index{Path integral: harmonic oscillator}

In the case of the harmonic oscillator, the energy eigenvalues
are known in an analytic form, and $\mathcal{Z}$ could be easily evaluated. 
In many other cases the $\E^{ }_n$ are, however, difficult to compute. 
A more useful representation of $\mathcal{Z}$ is obtained 
by writing it as a {\em path integral}. 

In order to get started, let us recall some basic relations. 
First of all, it follows from the form of the momentum operator 
in the $|x\rangle$-basis that
\be
 \langle x | \hat p | p \rangle = p \langle x | p \rangle 
 = - i \hbar\, \partial^{ }_x \langle x | p \rangle 
 \Rightarrow \langle x | p \rangle = A\, e^{\frac{i p x}{\hbar}}
 \;, 
\ee
where $A$ is some constant. Second, we need completeness
relations in both $|x\rangle$ 
and $|p\rangle$-bases, which take the respective forms
\be
 \int \! {\rm d} x \, | x \rangle \langle x | = \hat\unit
 \;, \quad
 \int \! \frac{{\rm d}p}{B} \, 
 | p \rangle \langle p | = \hat\unit
 \;, 
\ee
where $B$ is another constant. The choices of $A$ and $B$ are
not independent; indeed, 
\ba
 \hat\unit & = & \int \! {\rm d} x \, 
 \int \! \frac{{\rm d} p }{B} 
 \int \! \frac{{\rm d} p' }{B}
 | p \rangle \langle p | x \rangle \langle x | p' \rangle \langle p' | 
 = \int \! {\rm d} x \, 
 \int \! \frac{{\rm d} p }{B} 
 \int \! \frac{{\rm d} p' }{B}
 | p \rangle |A|^2 e^{\frac{i (p'-p) x}{\hbar}} \langle p' |
 \nn  
 & = &  
 \int \! \frac{{\rm d} p }{B} 
 \int \! \frac{{\rm d} p' }{B}
 | p \rangle |A|^2 2\pi\hbar\, \delta(p'-p) \langle p' |
 = \frac{2\pi\hbar |A|^2}{B}  
 \int \! \frac{{\rm d} p }{B}\, |p\rangle \langle p | 
 = \frac{2\pi\hbar |A|^2}{B}\, \hat\unit
 \;,
\ea
implying that $B = 2 \pi\hbar |A|^2$. We choose $A\equiv 1$ in the following, 
so that $B = 2 \pi \hbar$. 

Next, we move on to evaluate the partition function, 
which we do in the $x$-basis, so that our starting point becomes 
\ba
 \mathcal{Z} & = & \tr[ e^{-\beta \hat H} ]
 = \int \! {\rm d}x \, \langle x | e^{-\beta\hat H} | x \rangle
 = \int \! {\rm d}x \, \langle x | 
 e^{-\frac{\epsilon \hat H}{\hbar}}  \cdots 
 e^{-\frac{\epsilon \hat H}{\hbar}}
 | x \rangle
 \;. \la{Z_x}
\ea
Here we have split $e^{-\beta\hat H}$ into a product of $N\gg 1$ different 
pieces, defining $\epsilon \equiv \beta\hbar/N$. 

A crucial trick at this point is to insert 
\be
 \hat\unit = \int \! \frac{{\rm d} p^{ }_i}{2\pi\hbar} 
 \, |p^{ }_i \rangle \langle p^{ }_i | 
 \;, \quad i=1, \ldots, N\;,
\ee
on the {\em left side} of each exponential, with $i$ increasing
from right to left; and 
\be
 \hat\unit = \int \! {\rm d} x^{ }_i \, 
 |x^{ }_i \rangle \langle x^{ }_i | 
 \;, \quad i=1, \ldots, N\;,
\ee
on the {\em right side} of each exponential, with again $i$ 
increasing from right to left. Thereby we are left 
to consider matrix elements of the type 
\ba
 \langle x^{ }_{i+1} | p^{ }_i \rangle \langle p^{ }_i | 
 e^{-\frac{\epsilon}{\hbar}\hat H(\hat{p},\hat{x})} | x^{ }_i \rangle 
 & = &  e^{ \frac{i p^{ }_i x^{ }_{i+1}}{\hbar} }
 \langle p^{ }_i | 
 e^{-\frac{\epsilon}{\hbar} H({p^{ }_i},{x^{ }_i}) + \rmO(\epsilon^2)}
 | x^{ }_i \rangle
 \nn & = &
 \exp\biggl\{  
 -\frac{\epsilon}{\hbar}
 \biggl[ 
   \frac{p_i^2}{2m} - i p^{ }_i \frac{x^{ }_{i+1}-x^{ }_{i}}{\epsilon}
   + V(x^{ }_i) + \rmO(\epsilon)
 \biggr] 
 \biggr\} 
 \;. \la{ip1i}
\ea
Moreover, we note that at the very right, we have 
\be
 \langle x^{ }_1 | x \rangle = \delta(x^{ }_1 - x)
 \;, 
\ee
which allows us to carry out the integral over $x$. Similarly, at 
the very left, the role of $\langle x^{ }_{i+1} |$ is played
by the state $\langle x | = \langle x^{ }_{1} |$. Finally, 
we remark that the $\rmO(\epsilon)$ correction in 
\eq\nr{ip1i} can be eliminated by sending $N\to\infty$.

In total, we can thus write the partition function in the form
\be
 \mathcal{Z} = 
 \lim_{N\to\infty}
 \int 
 \biggl[ 
  \prod_{i=1}^N \frac{{\rm d} x^{ }_i {\rm d} p^{ }_i}{2\pi\hbar}
 \biggr]
 \left. \exp\biggl\{ 
 -\frac{1}{\hbar} \sum_{j=1}^{N} \epsilon
 \biggl[
  \frac{p_j^2}{2 m} - i p^{ }_j \frac{x^{ }_{j+1} - x^{ }_j}{\epsilon}
  + V(x^{ }_j)
 \biggr]
 \biggr\} 
 \right|_{x^{ }_{N+1} \,\equiv\, x^{ }_1,\, \epsilon \,\equiv\, \beta\hbar/N}
 \;, \la{Z_A}
\ee
which is often symbolically expressed as a ``continuum'' 
path integral
\be
  \mathcal{Z} = 
 \int_{x(\beta\hbar) = x(0)} \!\!\!\!  
 \mathcal{D} x\, \mathcal{D} 
 \biggl( \frac{p}{2\pi\hbar} \biggr) \, 
  \exp\biggl\{ 
 -\frac{1}{\hbar} \int_0^{\beta\hbar} \! {\rm d}\tau \, 
 \biggl[
  \frac{[p(\tau)]^2}{2 m} - i p(\tau) \dot{x}(\tau)
  + V(x(\tau))
 \biggr]
 \biggr\} 
 \;. \la{Z_B}
\ee
The integration measure here is understood
as the limit indicated in \eq\nr{Z_A}; 
the discrete $x^{ }_i$'s have been collected into a function $x(\tau)$; 
and the maximal value of 
the $\tau$-coordinate has been obtained from $\epsilon N = \beta\hbar$.

Returning to the discrete form of the path integral, 
we note that the integral over the momenta $p^{ }_i$ is Gaussian, 
and can thereby be carried out explicitly: 
\be
 \int_{-\infty}^{\infty}
 \! \frac{{\rm d}p^{ }_i}{2\pi\hbar} 
 \exp\biggl\{ 
 -\frac{\epsilon}{\hbar} 
 \biggl[
  \frac{p_i^2}{2 m} - i p^{ }_i \frac{x^{ }_{i+1} - x^{ }_i}{\epsilon}
 \biggr]
 \biggr\}
 = 
 \sqrt{\frac{m}{2\pi\hbar\,\epsilon}}
 \exp\biggl[
   - \frac{m (x^{ }_{i+1}-x^{ }_{i})^2}{2 \hbar\,\epsilon}
 \biggr]
 \;. \la{gaussian}
\ee
Using this, \eq\nr{Z_A} becomes 
\be
 \mathcal{Z} = 
 \lim_{N\to\infty}
 \int 
 \biggl[ 
  \prod_{i=1}^N \frac{{\rm d} x^{ }_i}{\sqrt{2\pi\hbar\,\epsilon/m}}
 \biggr]
 \left. \exp\biggl\{ 
 -\frac{1}{\hbar} \sum_{j=1}^{N} \epsilon
 \biggl[
  \frac{m}{2}\biggl( \frac{x^{ }_{j+1} - x^{ }_{j}}{\epsilon}
  \biggr)^2 + V(x^{ }_j)
 \biggr]
 \biggr\} 
 \right|_{x^{ }_{N+1} \,\equiv\, x^{ }_1,\, \epsilon \,\equiv\, \beta\hbar/N}
 \;, \la{Z_C}
\ee
which may also be written in a continuum form. 
Of course the measure then contains 
a factor which appears quite divergent at large $N$, 
\be
 C \equiv \biggl( \frac{m}{2\pi\hbar\,\epsilon} \biggr)^{N/2}
 = \exp \biggl[
 \frac{N}{2} \ln \biggl( \frac{m N}{2 \pi\hbar^2\beta} \biggr) 
 \biggr] 
 \;. \la{C}
\ee
This factor is, however, 
{\em independent of the properties of the potential $V(x^{ }_j)$} and 
thereby contains {\em no dynamical information}, so that we 
do not need to worry too much about the apparent divergence.
For the moment, then, we can simply write down a continuum 
``functional integral'', 
\be
  \mathcal{Z} = 
 C \; \int_{x(\beta\hbar) = x(0)} \!\!\! \mathcal{D} x 
  \exp\biggl\{ 
 -\frac{1}{\hbar} \int_0^{\beta\hbar} \! {\rm d}\tau \, 
 \biggl[
  \frac{m}{2} \biggl( \frac{{\rm d} x(\tau)}{{\rm d}\tau} \biggr)^2
  + \; V(x(\tau))
 \biggr]
 \biggr\} 
 \;. \la{Z_D}
\ee

Let us end by giving an ``interpretation'' to the result
in \eq\nr{Z_D}. We recall that the usual quantum-mechanical 
path integral at zero temperature contains the exponential 
\be
 \exp\biggl( \frac{i}{\hbar} \int \! {\rm d} t \, 
 \mathcal{L}^{ }_\iM \biggr)
 \;, \quad
 \mathcal{L}^{ }_\iM = \frac{m}{2}
 \biggl( \frac{{\rm d} x}{{\rm d}t} \biggr)^2
  - V(x)
 \;.
\ee
We note that \eq\nr{Z_D} can be obtained from its zero-temperature 
counterpart with the following recipe~\cite{fh}: 
\bi
\item[(i)]
Carry out a Wick rotation, denoting $\tau \equiv i t$. 
\item[(ii)]
Introduce 
\be
 {L}^{ }_\iE \equiv - \mathcal{L}^{ }_\iM(\tau = i t)
  = 
 \frac{m}{2} \biggl( \frac{{\rm d} x }{{\rm d}\tau} \biggr)^2
  + V(x)
 \;. \la{recipe0}
\ee
\item[(iii)]
Restrict $\tau$ to the interval $(0,\beta\hbar)$.
\item[(iv)]
Require periodicity of $x(\tau)$, i.e.\ $x(\beta\hbar) = x(0)$.
\ei
With these steps (and noting that $i{\rm d}t = {\rm d}\tau$), 
the exponential becomes
\be
 \exp\biggl( \frac{i}{\hbar} \int \! {\rm d} t \, 
 \mathcal{L}^{ }_\iM \biggr)
  \;\;
    \stackrel{\rm (i)-(iv)}{\longrightarrow}
  \;\;
 \exp\biggl( -\frac{1}{\hbar} S^{ }_\iE \biggr) \;\equiv\;
 \exp\biggl( -\frac{1}{\hbar}
 \int_0^{\beta\hbar} \! {\rm d}\tau \, {L}^{ }_\iE\biggr) 
 \;, \la{recipe}
\ee
where the subscript $E$ stands for ``Euclidean''.
Because of step (i), the path integral in \eq\nr{recipe}
is also known as the {\em imaginary-time formalism}. 
It turns out that this recipe works, with few modifications, 
also in quantum field theory, and even for spin-1/2 and spin-1 particles, 
although the derivation of the path integral itself looks 
quite different in those cases. 
We return to these issues in later chapters of the book.

\index{Euclidean Lagrangian: harmonic oscillator}

\index{Imaginary-time formalism}

\newpage

\subsection{Evaluation of the path integral for the harmonic oscillator}
\la{se:p5}

As an independent crosscheck of the results of \se\ref{ss:pi_qm}, 
we now explicitly evaluate the path
integral of \eq\nr{Z_D} in the case of a harmonic oscillator,
and compare the result with \eq\nr{hoZ}. To make the exercise
more interesting, we carry out the evaluation in Fourier 
space with respect to the time coordinate $\tau$. 
Moreover we would like to
deduce the information contained in the divergent constant $C$ 
without making use of its actual value, given in \eq\nr{C}. 

Let us start by representing an arbitrary function $x(\tau)$, 
$0 < \tau < \beta\hbar$, with the property $x((\beta\hbar)^-) = x(0^+_{ })$ 
(referred to as ``periodicity'')
as a Fourier sum
\be
 x(\tau) \; \equiv \;  T \sum_{n=-\infty}^{\infty} x^{ }_n\,
  e^{i \omega^{ }_n \tau}
 \;, \la{Fousum}
\ee 
where the factor $T$ is a convention. Imposing periodicity requires that
\be
 e^{i\omega^{ }_n\beta\hbar} = 1 \;, \quad
 \mbox{i.e.} \quad \omega^{ }_n\beta\hbar = 2 \pi n \;, \quad n \in \ZZ
 \;,
\ee
where the values
$
 \omega^{ }_n = 2 \pi T n/\hbar
$
are called {\em Matsubara frequencies}. The corresponding 
amplitudes $x^{ }_n$ are called {\em Matsubara modes}. 

\index{Matsubara frequencies: bosonic}

\index{Fourier representation: harmonic oscillator}

\index{Zero mode: harmonic oscillator}

Apart from periodicity, we also impose reality on $x(\tau)$: 
\be
 x(\tau) \in \RR 
 \; \Rightarrow \; x^*(\tau) = x(\tau)
 \; \Rightarrow \; x_n^* = x^{ }_{-n}
 \;. 
\ee 
If we write $x^{ }_n = a^{ }_n + i b^{ }_n$, it then follows that
\be
 x_n^* = a^{ }_n - i b^{ }_n = 
 x^{ }_{-n} = a^{ }_{-n} + i b^{ }_{-n} 
 \Rightarrow
 \left\{ 
 \begin{array}{c}
   a^{ }_n = a^{ }_{-n} \\
   b^{ }_n = -  b^{ }_{-n}
 \end{array} 
 \right.
 \;, \la{a_n_props}
\ee
and moreover that $b^{ }_0 = 0$ and $x^{ }_{-n} x^{ }_n = a_n^2 + b_n^2$. 
Thereby we now have the representation
\be
 x(\tau) = T 
 \biggl\{
  a^{ }_0 + \sum_{n=1}^{\infty}
  \biggl[
   (a^{ }_n + i b^{ }_n) e^{i \omega^{ }_n \tau} 
 + (a^{ }_n - i b^{ }_n) e^{-i \omega^{ }_n \tau} 
  \biggr]
 \biggr\} 
 \;, \la{Fourier} 
\ee
where $a^{ }_0$ is called (the amplitude of) the Matsubara {\em zero mode}.

With the representation of \eq\nr{Fousum}, general quadratic 
structures can be expressed as 
\ba
 \frac{1}{\hbar} 
 \int_0^{\beta\hbar} \! {\rm d}\tau \, 
 x(\tau) y(\tau) & = &  
 T^2 \sum_{m,n} x^{ }_n y^{ }_m \frac{1}{\hbar}
 \int_0^{\beta\hbar} \! {\rm d}\tau \, 
 e^{i(\omega^{ }_n + \omega^{ }_m)\tau}
 \nn 
 & = &  
  T^2 \sum_{m,n} x^{ }_n y^{ }_m \, \frac{1}{T} \, \delta^{ }_{n,-m}
 = T \sum_n x^{ }_n y^{ }_{-n}
 \;. \la{sum}
\ea
In particular, the argument of the exponential in \eq\nr{Z_D} becomes 
\ba
 & & \hspace*{-3cm} 
  -\frac{1}{\hbar}  
 \int_0^{\beta\hbar} \! {\rm d}\tau \, 
 \frac{m}{2} 
 \biggl[
   \frac{{\rm d}x(\tau)}{{\rm d}\tau} 
   \frac{{\rm d}x(\tau)}{{\rm d}\tau} 
    + \omega^2\, x(\tau) x(\tau)
 \biggr]
 \nn[3mm]
 & \stackrel{\rmi{\nr{sum}}}{=} &
 -\frac{m T}{2}
 \sum_{n=-\infty}^{\infty}
 x^{ }_n \Bigl[ 
   i \omega^{ }_n \, i \omega^{ }_{-n} + \omega^2
 \Bigr] x_{-n}   
 \nn 
 & \stackrel{\omega^{ }_{-n}=-\omega^{ }_n}{=} & 
 -\frac{m T}{2}
 \sum_{n=-\infty}^{\infty} (\omega_n^2 + \omega^2)(a_n^2 + b_n^2)
 \nn 
 & \stackrel{\rmi{\nr{a_n_props}}}{=} &
 -\frac{m T}{2}\, \omega^2 a_0^2 
 - m T \sum_{n=1}^{\infty}
 (\omega_n^2 + \omega^2)(a_n^2 + b_n^2)
 \;. 
\ea

Next, we need to consider the {\em integration measure}. 
To this end, let us make a change of variables from $x(\tau)$, 
$\tau \in (0,\beta\hbar)$, to the Fourier components $a^{ }_n, b^{ }_n$.
As we have seen, the independent variables are  
$a^{ }_0$ and $\{a^{ }_n,b^{ }_n\}$, $n \ge 1$, whereby the measure becomes
\be
 \mathcal{D} x(\tau) = 
 \left| \det \biggl[ \frac{\delta x(\tau)}{\delta x^{ }_n} \biggr]\right|
 \, {\rm d}a^{ }_0 
 \, \Bigl[ \prod_{n\ge 1} {\rm d} a^{ }_n \, {\rm d} b^{ }_n \Bigr]
 \;. 
\ee
The change of bases is purely kinematical and independent
of the potential $V(x)$, implying that we can define
\be
 C' \equiv 
 C \, 
 \left| \det \biggl[ \frac{\delta x(\tau)}{\delta x^{ }_n} \biggr]\right|
 \;, 
\ee
and regard now $C'$ as an unknown coefficient. 

Making use of the Gaussian integral 
$
 \int_{-\infty}^{\infty} {\rm d} x \exp(-c x^2) = \sqrt{\pi/c}
$, 
$c > 0$,
as well as the above integration measure, 
the expression in \eq\nr{Z_D} becomes 
\ba
 \mathcal{Z} & = & 
 C' \, 
 \int_{-\infty}^{\infty} \! {\rm d} a^{ }_0 
 \int_{-\infty}^{\infty}
 \Bigl[ 
   \prod_{n\ge 1} {\rm d} a^{ }_n \, {\rm d} b^{ }_n 
 \Bigr]
 \exp\biggl[ 
   -\frac{1}{2} m T \omega^2 a_0^2 - m T 
 \sum_{n\ge 1} (\omega_n^2 + \omega^2) (a_n^2 + b_n^2)
 \biggr] \hspace*{3mm}
 \la{preZ} \\ & = & 
 C' \sqrt{\frac{2\pi}{m T \omega^2}} \prod_{n=1}^{\infty}
 \frac{\pi}{m T(\omega_n^2 + \omega^2)}
 \;, \quad
 \omega^{ }_n = \frac{2\pi T n}{\hbar}
 \;. \la{preZ2}
\ea
The remaining task is to determine $C'$. 
This can be achieved via the following observations:
\bi
\item
Since $C'$ is independent of $\omega$
(which only appears in $V(x)$), we can determine it 
in the limit $\omega = 0$, whereby the system simplifies. 

\item
The integral over the zero mode $a^{ }_0$ in \eq\nr{preZ} is, 
however, divergent for $\omega\to 0$. We may call such a divergence an 
{\em infrared divergence}: the zero mode is the {\em lowest-energy} mode.

\index{Infrared divergence: harmonic oscillator}

\item
We can still take the $\omega\to 0$ limit, if we momentarily 
{\em regulate} the integration over the zero mode in some way. 
Noting from \eq\nr{Fourier} that 
\be
 \frac{1}{\beta\hbar}\int_0^{\beta\hbar}\! {\rm d}\tau \, x(\tau) = T a^{ }_0 
 \;, 
\ee
we see that $T a^{ }_0$ represents the average value of $x(\tau)$ over
the $\tau$-interval. 
We may thus regulate the system by ``putting it 
in a periodic box'', i.e.\ by restricting the (average) value 
of $x(\tau)$ to 
some (large but finite) interval $\Delta x$.
\ei
With this setup, we can now proceed to find $C'$ via {\em matching}. 

\index{Matching: harmonic oscillator}

\paragraph{``Effective theory computation'':}

In the $\omega \to 0$ limit but in the presence of the regulator, 
\eq\nr{preZ} becomes
\ba
 \lim_{\omega\to 0}  \mathcal{Z}_\rmi{regulated} & = & 
 C' \, 
 \int_{\Delta x/T} \! {\rm d} a^{ }_0 
 \int_{-\infty}^{\infty}
 \Bigl[ 
   \prod_{n\ge 1} {\rm d} a^{ }_n \, {\rm d} b^{ }_n 
 \Bigr]
 \exp\biggl[ - m T 
 \sum_{n\ge 1} \omega_n^2  (a_n^2 + b_n^2)
 \biggr]
 \nn & = & 
 C' \, \frac{\Delta x}{T} \, \prod_{n=1}^{\infty}
 \frac{\pi}{m T\omega_n^2}
 \;, \quad
 \omega^{ }_n = \frac{2\pi T n}{\hbar}
 \;. \la{sideA}
\ea

\paragraph{``Full theory computation'':}

In the presence of the regulator, and in the absence of $V(x)$ 
(implied by the $\omega\to 0$ limit), 
\eq\nr{Z_x} can be computed in a very simple way: 
\ba
 \lim_{\omega\to 0} \mathcal{Z}_\rmi{regulated} & = & 
 \int_{\Delta x} \! {\rm d}x\, 
 \langle x | e^{-\frac{\hat p^2}{2 m T}} | x \rangle 
 \nn & = &
 \int_{\Delta x}  \! {\rm d}x\, \int_{-\infty}^{\infty} \! 
 \frac{{\rm d}p}{2\pi\hbar}
 \langle x | e^{-\frac{\hat p^2}{2 m T}} |p\rangle \langle p | x \rangle 
 \nn & = &
 \int_{\Delta x}  \! {\rm d}x\, \int_{-\infty}^{\infty} \! 
 \frac{{\rm d}p}{2\pi\hbar}
 e^{-\frac{p^2}{2 m T}} 
 \underbrace{\langle x |p\rangle \langle p | x \rangle}_{1} 
 \nn & = & 
 \frac{\Delta x }{2\pi\hbar} \sqrt{2 \pi m T} 
 \;. \la{sideB}
\ea

\paragraph{Matching the two sides:}

Equating \eqs\nr{sideA} and \nr{sideB}, 
we find the formal expression
\be
 C' = \frac{T}{2\pi\hbar} \sqrt{2\pi m T} \prod_{n=1}^{\infty} 
 \frac{m T \omega_n^2}{\pi}
 \;. \la{Cp}
\ee
Since the regulator $\Delta x$ has dropped out, 
we may call $C'$ an ``ultraviolet'' matching coefficient. 

With $C'$ determined, we can now continue with \eq\nr{preZ2}, obtaining
the finite expression
\ba
 \mathcal{Z} & = & 
 \frac{T}{\hbar\omega} \prod_{n=1}^{\infty} 
 \frac{\omega_n^2}{\omega_n^2 + \omega^2}
 \la{hoZ3} \\  & = & 
 \frac{T}{\hbar\omega} 
 \frac{1}{\prod_{n=1}^{\infty} 
 \Bigl[ 1 + \frac{(\hbar\omega/2\pi T)^2}{n^2} \Bigr]}
 \;.  
\ea
Making use of the identity
\be 
 \frac{\sinh\pi x}{\pi x} = \prod_{n=1}^{\infty}
 \biggl( 1 + \frac{x^2}{n^2} \biggr)
\ee
we directly reproduce our earlier result 
for the partition function, \eq\nr{hoZ}.
Thus, we have managed to correctly evaluate
the path integral without ever making recourse to \eq\nr{C} or, 
for that matter, to the discretization that was present in 
\eqs\nr{Z_A} and \nr{Z_C}.

Let us end with a few remarks:
\bi

\item In quantum mechanics, the partition function 
$\mathcal{Z}$ as well as all other observables are
finite functions of the parameters $T$, $m$, 
and $\omega$, if computed properly. We saw that with path 
integrals this is not obvious at every intermediate
step, but at the end it did work out. 
In quantum field theory, on the contrary, ``ultraviolet'' (UV) 
divergences may remain in the results 
even if we compute everything correctly. These are then taken care
of by renormalization. However, 
as our quantum-mechanical example demonstrated, the 
``ambiguity'' of the functional integration measure (through $C'$) 
is not in itself a source of UV divergences.

\item It is appropriate to stress that in many 
physically relevant observables, the coefficient $C'$ drops out
completely, and the above procedure is thereby even simpler. An example of
such a quantity is given in \eq\nr{G_tau_def} below. 

\item Finally, 
some of the concepts and techniques that 
were introduced with this simple example --- zero modes, infrared
divergences, their regularization, matching computations, etc ---  
also play a role in non-trivial quantum field theoretic 
examples that we encounter later on.

\ei


\subsection*{Appendix A: 2-point function}

Defining a Heisenberg-like operator (with $it \to \tau$) 
\be
 \hat x(\tau) \; \equiv \; e^{\frac{\hat H \tau}{\hbar}} 
 \hat x\, e^{-\frac{\hat H \tau}{\hbar}}
 \;, \quad 0 < \tau < \beta\hbar
 \;, 
\ee
we define a ``2-point Green's function'' or a ``propagator'' through 
\be
 G(\tau) \; \equiv \; 
 \frac{1}{\mathcal{Z}}\, \tr \Bigl[ e^{-\beta\hat H}
 \hat x(\tau) \hat x (0) \Bigr]
 \;. \la{G_tau_def}
\ee
The corresponding path integral can be shown to read 
\be
 G(\tau) = 
 \frac{
 \int_{x(\beta\hbar)= x(0)}\! \mathcal{D}x \, 
 x(\tau) x(0) 
 \exp[-S^{ }_\iE/\hbar]}
 {
 \int_{x(\beta\hbar)= x(0)}\! \mathcal{D}x \,
 \exp[-S^{ }_\iE/\hbar]}
 \;, \label{Gtau4}
\ee
whereby the normalization of $\mathcal{D}x$ plays no role. 
In the following, we compute $G(\tau)$  explicitly
for the harmonic oscillator, by making use of 
\bi
\item[(a)]
the canonical formalism, i.e.~expressing $\hat H$ and $\hat x$ 
in terms of the annihilation and creation operators 
$\hat a$ and $\hat a^\dagger$, 

\item[(b)]
the path integral formalism, working in Fourier space.

\ei

Starting with the canonical formalism, we write all quantities 
in terms of $\hat a$ and $\hat a^\dagger$: 
\be
 \hat H = \hbar\omega\, \Bigl(\hat a^\dagger \hat a + \frac{1}{2}\Bigr)
 \;, \quad
 \hat x = \sqrt{\frac{\hbar}{2 m \omega}}(\hat a + \hat a^\dagger)
 \;, \quad
 [\hat a, \hat a^\dagger] = 1
 \;. 
\ee
In order to construct $\hat x(\tau)$, we make use of the expansion
\be
 e^{\hat A} \hat B\, e^{-\hat A}
 = \hat B + [\hat A,\hat B] + 
 \frac{1}{2!} [\hat A,[\hat A,\hat B]] + 
 \frac{1}{3!} [\hat A,[\hat A,[\hat A,\hat B]]] + \ldots \;.
\ee
Noting that 
\ba
 [\hat H, \hat a] & = &  \hbar \omega [\hat a^\dagger \hat a, \hat a]
 = - \hbar \omega \hat a
 \;, \nn
 {[}\hat H, {[} \hat H, \hat a {]}{]} & = & (- \hbar \omega)^2 \hat a
 \;, \nn
 {[}\hat H, \hat a^\dagger] & = &  \hbar \omega [\hat a^\dagger \hat a, 
  \hat a^\dagger]
 =  \hbar \omega \hat a^\dagger
 \;, \nn
 {[}\hat H, {[}\hat H, \hat a^\dagger{]}{]} 
 & = & (\hbar \omega)^2 \hat a^\dagger
 \;, 
\ea
and so forth, we can write
\ba
 e^{\frac{\hat H \tau}{\hbar}} 
 \hat x\, e^{-\frac{\hat H \tau}{\hbar}}
 & = & \sqrt{\frac{\hbar}{2 m \omega}}
 \,\biggl\{ 
   \hat a\biggl[ 1 - \omega\tau + \frac{1}{2!} (-\omega\tau)^2 + \ldots\biggr]
  + \hat a^\dagger
  \biggl[ 1 + \omega\tau + \frac{1}{2!} (\omega\tau)^2 + \ldots\biggr]
 \biggr\} 
 \nn  & = &
 \sqrt{\frac{\hbar}{2 m \omega}}
 \,\Bigl( \hat a\, e^{-\omega \tau} + \hat a^\dagger e^{\omega \tau}\Bigr) 
 \;. 
\ea

Inserting now $\mathcal{Z}$ from \eq\nr{hoZ}, 
\eq\nr{G_tau_def} becomes
\be
 G(\tau) = 2 \sinh \Bigl( \frac{\beta\hbar\omega}{2} \Bigr)
 \sum_{n=0}^{\infty}
 \langle n | e^{-\beta\hbar\omega(n + \frac{1}{2})} \frac{\hbar}{2 m \omega}
 \Bigl( \hat a\, e^{-\omega \tau} + \hat a^\dagger e^{\omega \tau} \Bigr) 
 \bigl( \hat a  + \hat a^\dagger  \bigr) | n \rangle
 \;.
\ee
With the relations
$
 \hat a^\dagger |n\rangle= \sqrt{n+1} |n+1\rangle
$
and 
$
 \hat a |n\rangle = \sqrt{n} |n-1\rangle
$
we can identify the non-zero matrix elements, 
\be
 \langle n |\hat a \hat a^\dagger  | n \rangle = n+1
 \;, \quad
  \langle n |\hat a^\dagger  \hat a  | n \rangle = n
 \;.
\ee
Thereby we obtain
\be
 G(\tau) = \frac{\hbar}{m\omega} 
 \sinh \Bigl( \frac{\beta\hbar\omega}{2} \Bigr) 
 \exp \Bigl( - \frac{\beta\hbar\omega}{2} \Bigr) 
 \sum_{n=0}^{\infty}
 e^{-\beta\hbar\omega n} 
 \Bigl[ e^{-\omega \tau} + n 
 \Bigl(e^{-\omega \tau} +  e^{\omega \tau} \Bigr) \Bigr] 
 \;,
\ee
where the terms are quickly evaluated as geometric sums, 
\ba
 \sum_{n=0}^{\infty}
 e^{-\beta\hbar\omega n} & = & \frac{1}{1-e^{-\beta\hbar\omega}}
 \;, \nn 
 \sum_{n=0}^{\infty}
 n e^{-\beta\hbar\omega n} & = & -\frac{1}{\beta\hbar}
 \frac{{\rm d}}{{\rm d}\omega} \frac{1}{1-e^{-\beta\hbar\omega}}
 =  
 \frac{e^{-\beta\hbar\omega}}{(1-e^{-\beta\hbar\omega})^2}
 \;.
\ea
In total, we then have
\ba
 G(\tau) & = & \frac{\hbar}{2 m\omega} 
 \Bigl( 1 -  e^{- \beta\hbar\omega} \Bigr)
 \biggl[ 
  \frac{e^{-\omega\tau}}{1-e^{-\beta\hbar\omega}} + 
 \Bigl(e^{-\omega \tau} +  e^{\omega \tau} \Bigr)
 \frac{e^{-\beta\hbar\omega}}{(1-e^{-\beta\hbar\omega})^2}
 \biggr] 
 \nn  &= & 
 \frac{\hbar}{2 m\omega}  
 \frac{1}{1-e^{-\beta\hbar\omega}}
 \Bigl[ 
   e^{-\omega\tau} + e^{\omega(\tau - \beta\hbar)}
 \Bigr] 
 \nn[1mm]  &= & 
 \frac{\hbar}{2 m\omega}  
 \frac{ 
   e^{\omega\tau} + e^{\omega(\beta\hbar-\tau)} 
  }{e^{\beta\hbar\omega} - 1 }
 \nn & = & \frac{\hbar}{2 m \omega}
 \frac{\cosh\left[ \left( \frac{\beta\hbar}{2} - \tau\right) \omega\right]}
 {\sinh\left[ \frac{\beta\hbar\omega}{2} \right]} 
 \;. \la{Gtau}
\ea

As far as the path integral treatment goes, 
we employ the same representation as in \eq\nr{preZ}, noting
that $C'$ drops out in the ratio of \eq\nr{Gtau4}.
Recalling the Fourier representation of \eq\nr{Fourier}, 
\ba
 x(\tau) & = &  T 
 \biggl\{
  a^{ }_0 + \sum_{k=1}^{\infty}
  \biggl[
   (a^{ }_k + i b^{ }_k) e^{i \omega^{ }_k \tau} 
 + (a^{ }_k - i b^{ }_k) e^{-i \omega^{ }_k \tau} 
  \biggr]
 \biggr\} 
 \;, \\ 
 x(0) & = &  T 
 \biggl\{
  a^{ }_0 + \sum_{l=1}^{\infty} 2 a^{ }_l 
 \biggr\}
 \;,
\ea
the observable of our interest becomes 
\be
 G(\tau)\; = \;
  \bigl\langle\, x(\tau) x(0) \,\bigr\rangle
 \; \equiv \; \frac{
 \int \! {\rm d} a^{ }_0 \int 
 \! \prod_{n\ge 1} {\rm d}a^{ }_n \, {\rm d}b^{ }_n
 \, x(\tau)\, x(0) \exp[- S^{ }_\iE/\hbar]
 }{
 \int \! {\rm d} a^{ }_0 \int 
 \! \prod_{n\ge 1} {\rm d}a^{ }_n \, {\rm d}b^{ }_n
 \, \exp[-S^{ }_\iE/\hbar] 
 }
 \;. \label{Gtau5}
\ee

At this point, we employ the fact that the exponential is 
quadratic in $a^{ }_0, a^{ }_n, b^{ }_n \in \RR$, which immediately implies 
\be
 \langle a^{ }_0 a^{ }_k \rangle = \langle a^{ }_0 b^{ }_k\rangle = 
 \langle a^{ }_k b^{ }_l \rangle = 0
 \;, \quad
 \langle a^{ }_k a^{ }_l \rangle = \langle b^{ }_k b^{ }_l \rangle 
 \propto \delta^{ }_{kl}
 \;, 
\ee
with the expectation values defined in the 
sense of \eq\nr{Gtau5}. Thereby we obtain
\be
 G(\tau) = T^2 \Bigl\langle 
 a_0^2 + \sum_{k=1}^{\infty} 2 a_k^2 
 \left( 
 e^{i\omega^{ }_k \tau} + e^{- i\omega^{ }_k \tau} 
 \right) \Bigr\rangle
 \;, \la{Gtau1}
\ee
where
\ba
 \langle a_0^2 \rangle 
 & = &  \frac{
 \int \! {\rm d} a^{ }_0 \, a_0^2 \, 
 \exp\left( -\frac{1}{2} m T \omega^2 a_0^2 \right)
 }{
 \int \! {\rm d} a^{ }_0 \, \exp\left( -\frac{1}{2} m T \omega^2 a_0^2 \right)
 }
 \nn 
 & = &
 - \frac{2}{m\omega^2}
 \frac{{\rm d}}{{\rm d}T}
 \biggl[ 
  \ln \int \! {\rm d} a^{ }_0 
 \, \exp\left( -\frac{1}{2} m T \omega^2 a_0^2 \right)
 \biggr] 
  = 
 - \frac{2}{m\omega^2}
 \frac{{\rm d}}{{\rm d}T}
 \biggl[ 
   \ln\sqrt{\frac{2\pi}{m \omega^2 T }} \, 
 \biggr] 
 \nn & = & 
 \frac{1}{m \omega^2 T}
 \;, \\[3mm] 
 \langle a_k^2 \rangle 
 & = &  \frac{
 \int \! {\rm d} a^{ }_k \, a_k^2 \, 
 \exp\left[ - m T ( \omega_k^2 + \omega^2 ) a_k^2 \right]
 }{
 \int \! {\rm d} a^{ }_k \, 
 \exp\left[ - m T ( \omega_k^2 + \omega^2 ) a_k^2 \right]
 }
 \nn & = & 
 \frac{1}{2 m (\omega_k^2 + \omega^2) T}
 \;.   
\ea
Inserting these into \eq\nr{Gtau1} we get
\be
 G(\tau) = \frac{T}{m} 
 \biggl( 
   \frac{1}{\omega^2} +  
 \sum_{k = 1}^{\infty}
 \frac{e^{i \omega^{ }_k\tau} + e^{- i \omega^{ }_k\tau}}
 {\omega_k^2 + \omega^2}
 \biggr)
 = 
 \frac{T}{m} \sum_{k=-\infty}^{\infty} \frac{e^{i\omega^{ }_k\tau}}
 {\omega_k^2 + \omega^2}
 \;, \quad
 \omega^{ }_k = \frac{ 2 \pi T k }{ \hbar }
 \;.  \la{sum2}
\ee

\index{Thermal sums: boson loop}

There are various ways to evaluate the sum in \eq\nr{sum2}. 
We encounter a generic method in \se\ref{se:bsums}, 
so let us present a different approach here. We start by 
noting that 
\be
 \biggl( 
 - \frac{{\rm d}^2}{{\rm d}\tau^2} + \omega^2
 \biggr) G(\tau) 
 = \frac{T}{m} \sum_{k=-\infty}^{\infty} e^{i\omega^{ }_k\tau}
 = \frac{\hbar}{m}\,\delta(\tau \mathop{\mbox{mod}} \beta\hbar)
 \;, \la{sum3}
\ee
where we made use of the standard summation formula
$ 
 \sum_{k=-\infty}^{\infty} e^{i \omega^{ }_k \tau} 
 = \beta\hbar\, \delta(\tau \mathop{\mbox{mod}} \beta\hbar)
$.\footnote{
 ``Proof'': \la{delta_sum}
 $
  \sum_{k=-\infty}^{\infty} e^{i \omega^{ }_k \tau}
  = 1 + \lim_{\epsilon\to 0}
  \sum_{k=1}^{\infty} [(e^{i \frac{2\pi \tau}{\beta\hbar} - \epsilon})^k
 + (e^{- i \frac{2\pi \tau}{\beta\hbar} - \epsilon})^k ]
  = \lim_{\epsilon\to 0} 
  \Bigl[ 
    \frac{1}{1 - e^{i \frac{2\pi \tau}{\beta\hbar} - \epsilon}} - 
    \frac{1}{1 - e^{i \frac{2\pi \tau}{\beta\hbar} + \epsilon}}
  \Bigr] 
 $. 
 If $\tau \neq 0 \, \mbox{mod}\, \beta\hbar$, 
 then the limit $\epsilon\to 0$ can be 
 taken, and the two terms cancel against each other. But if
 $ 
  \frac{2\pi \tau}{\beta\hbar}\approx 0 
 $, 
 we can expand to leading order in a Taylor series, 
 obtaining 
 $ 
  \lim_{\epsilon\to 0} 
  \Bigl[ 
    \frac{i}{ \frac{2\pi \tau}{\beta\hbar} + i \epsilon } - 
    \frac{i}{ \frac{2\pi \tau}{\beta\hbar} - i \epsilon }
  \Bigr] 
 = 2\pi \delta( \frac{2\pi \tau}{\beta\hbar} )
 = \beta\hbar\, \delta (\tau)
 $.  
 }
 
Next, we solve \eq\nr{sum3} for $0 < \tau < \beta\hbar$, obtaining 
\be
 \biggl(  - \frac{{\rm d}^2}{{\rm d}\tau^2} + \omega^2 \biggr)
 G(\tau) = 0 
 \quad\Rightarrow\quad 
 G(\tau) = A\, e^{\omega\tau} + B\, e^{-\omega \tau} 
 \;, 
\ee
where $A, B$ are unknown constants. The solution can be 
further restricted by noting that the definition of $G(\tau)$, \eq\nr{sum2}, 
indicates that $G(\beta\hbar  -\tau) = G(\tau)$. 
Using this condition to obtain $B$, we then get 
\be
 G(\tau) = A 
 \left[ 
   e^{\omega\tau} + e^{\omega(\beta\hbar-\tau)}
 \right]
 \;.  \la{Gtau3}
\ee
The remaining unknown $A$ can be obtained by 
integrating \eq\nr{sum3} over the source at $\tau=0$ and making use of 
the periodicity of $G(\tau)$, 
$G(\tau + \beta\hbar) = G(\tau)$. This finally produces
\be
 G'((\beta\hbar)^{-}) - G'(0^+_{ }) = \frac{\hbar}{ m}
 \quad\Rightarrow\quad 
 2 \omega A \left( e^{\omega\beta\hbar} - 1 \right)
  = \frac{\hbar}{ m}
 \;,
\ee
which together with \eq\nr{Gtau3} yields our earlier result, \eq\nr{Gtau}.  

The agreement of the two different 
computations, \eqs\nr{G_tau_def} and \nr{Gtau4},  
once again demonstrates the
equivalence of the canonical and path integral approaches to solving
thermodynamic quantities in a quantum-mechanical setting. 

%

\newpage


\newpage 

\section{Free scalar fields}
\la{se:free_sft}

\paragraph{Abstract:}

The concepts of \se\ref{se:qm} are generalized to the case of a free
massive scalar field living in a $d+1$ dimensional spacetime.
This can be viewed as a system of infinitely many coupled harmonic
oscillators. The resulting imaginary-time path integral for 
the partition function is expressed 
in Fourier representation. Matsubara sums are evaluated
both in a low-temperature and a high-temperature expansion. 
The numerical convergence of these expansions, as well as 
some of their general properties, are discussed.

\paragraph{Keywords:} 

Field theory, Matsubara sum, 
low-temperature expansion, high-temperature expansion, dimensional 
regularization, chemical potential, Euler gamma function, 
Riemann zeta function. 

%
\subsection{Path integral for the partition function}
\la{ss:pi_sft}

\index{Path integral: scalar field}

\index{Partition function: scalar field}

A path integral representation for the partition function of  
a scalar field theory can be derived from the result obtained for the 
quantum-mechanical harmonic oscillator (HO) in \se\ref{se:p5}. 

In quantum field theory, the form of the theory is 
most economically defined in terms of the corresponding
classical (Minkowskian) Lagrangian $\mathcal{L}^{ }_\iM$, rather 
than the Hamiltonian $\hat H$; for instance,  
Lorentz symmetry is explicit only in $\mathcal{L}^{ }_\iM$. 
Let us therefore start from \eq\nr{qmLM} for the quantum harmonic 
oscillator, and re-interpret
$x$ as an ``internal'' degree of freedom $\phi$, situated
at the origin $\vec{0}$ of $d$-dimensional space,  like 
in \eq\nr{qm_qft}: 
\ba
 \mathcal{S}_\iM^\rmii{HO} & = &  
 \int\!{\rm d}t \, \mathcal{L}_\iM^\rmii{HO}
 \;, \\
 \mathcal{L}_\iM^\rmii{HO} &  = &  
 \frac{m}{2} \biggl( \frac{\partial \phi(t,\vec{0})}{\partial t} \biggr)^2
 - V(\phi(t,\vec{0}))
 \;. \la{SM_QM}
\ea
We may compare this with the usual action of a scalar field theory (SFT)
in $d$-dimensional space, 
\ba
 \mathcal{S}_\iM^\rmii{SFT} & = &  
 \int\!{\rm d}t \int_\vec{x} \, \mathcal{L}_\iM^\rmii{SFT}
 \;, \\
 \mathcal{L}_\iM^\rmii{SFT} & = & 
 \frac{1}{2} \partial^\mu\phi\, \partial^{ }_\mu\phi - V(\phi) 
 =  
 \frac{1}{2} (\partial^{ }_t\phi)^2 - 
 \frac{1}{2} (\partial^{ }_i\phi)(\partial^{ }_i\phi) - V(\phi)
 \;, \la{SM_SFT}
\ea
where we assume that repeated indices are summed over 
(irrespective of whether they are up and down), and 
the metric is 
($+$$-$$-$$-$).

Comparing \eq\nr{SM_QM} with \eq\nr{SM_SFT}, we see that 
scalar field theory is formally nothing but a collection of almost
independent harmonic oscillators with $m=1$, one at every $\vec{x}$.
These oscillators interact via the derivative term 
$(\partial^{ }_i\phi)(\partial^{ }_i\phi)$ which, in the language of 
statistical physics, couples nearest neighbours through 
\be
 \partial^{ }_i \phi \approx \frac{\phi(t,\vec{x} + a\, \vec{e}^{ }_i)
 - \phi(t,\vec{x})}{a}
 \;,
\ee
where $\vec{e}^{ }_i$ is a unit vector in the direction $i$ and 
$a$ is the lattice spacing. 

Next, we note that a coupling of the above type does not change
the derivation of the path integral 
in \se\ref{se:p3} in any essential way: it was only important 
that the Hamiltonian was quadratic in the {\em canonical momenta}, 
$p = m \dot{x} \leftrightarrow \partial^{ }_t\phi$. 
In other words, the derivation of the path integral is only 
concerned with objects having to do with time dependence, 
and these appear
in \eqs\nr{SM_QM} and \nr{SM_SFT} in identical manners. 
Therefore, we can directly 
take over the result of \eqs\nr{Z_D}--\nr{recipe}:  
\ba
  \mathcal{Z}^\rmii{SFT}(T) & = &  
  \int_{\phi(\beta\hbar,\vec{x}) = \phi(0,\vec{x})} 
  \prod_{\vec{x}} [\,C\, \mathcal{D} \phi(\tau,\vec{x})\,]  
  \exp\biggl[ 
  -\frac{1}{\hbar} \int_0^{\beta\hbar} \!\!\! {\rm d}\tau \!  
  \int_\vec{x} \, {L}_\iE^\rmii{SFT}
  \biggr]
 \;, \la{Z_SFT} \\
  {L}_\iE^\rmii{SFT} & = &
  - \mathcal{L}_\iM^\rmii{SFT}(t\to -i \tau)
 \; = \;  
 \frac{1}{2} \biggl(\frac{\partial\phi}{\partial\tau}\biggr)^2 +
 \sum_{i=1}^d 
 \frac{1}{2} \biggl(\frac{\partial\phi}{\partial x^i }\biggr)^2 
 + V(\phi)
 \;. \la{LE_expl}
\ea 
For brevity, we drop out the superscript SFT in the following, and write 
$
 {L}^{ }_\iE = \frac{1}{2} \partial^{ }_\mu \phi\, \partial^{ }_\mu \phi
 + V(\phi)
$.

\index{Euclidean Lagrangian: scalar field}


\subsection*{Fourier representation}
\la{se:p13}

\index{Fourier representation: scalar field}

We now parallel the strategy of \se\ref{se:p5} and rewrite
the path integral in a Fourier representation. In order to simplify
the notation, we measure time in units where
$\hbar = 1.05 \times 10^{-34}$~Js~$=1$. Then the dependence 
of the scalar field on $\tau$ can be expressed as
\be
 \phi(\tau,\vec{x}) = T \sum_{n=-\infty}^{\infty} 
 \tilde \phi(\omega^{ }_n,\vec{x})\, e^{i \omega^{ }_n\tau}
 \;, \quad
 \omega^{ }_n = 2 \pi T n
 \;, \quad n \in \ZZ
 \;. \la{phi_tau}
\ee

With the spatial coordinates, it is useful to make each
direction finite for a moment, 
denoting the corresponding extents by $L^{ }_i$, 
and to impose periodic boundary 
conditions in each direction. 
Then the dependence of the field $\phi$ on a given $x^i$ 
can be represented in the form
\be
 f(x^i) = \frac{1}{L^{ }_i}
 \sum_{n^{ }_i = -\infty}^{\infty}
 \tilde f(n^{ }_i) e^{i k^{ }_i x^i}
 \;, \quad
 k^{ }_i = \frac{2\pi n^{ }_i}{L^{ }_i}
 \;, \quad
 n^{ }_i \in \ZZ
 \;, \la{finV}
\ee
where $1/L^{ }_i$ plays the same role as $T$ in the time direction. 
In the infinite volume limit, the sum in \eq\nr{finV} goes
over to the usual Fourier integral, 
\be
 \frac{1}{L^{ }_i} \sum_{n^{ }_i}
 = \frac{1}{2\pi} \sum_{n^{ }_i} \Delta k^{ }_i
 \; \stackrel{L^{ }_i\to\infty}{\longrightarrow} \;
 \int\! \frac{{\rm d}k^{ }_i}{2\pi}
 \;, \la{Fou_infV}
\ee
where $\Delta k^{ }_i = 2\pi / L^{ }_i$ is the width of the unit shell.
The entire function in \eq\nr{phi_tau} then reads
\be
 \phi(\tau,\vec{x}) 
 =
 T \sum_{\omega^{ }_n} \frac{1}{V} \sum_{\vec{k}}
 \tilde \phi(\omega^{ }_n,\vec{k})
 e^{i\omega^{ }_n\tau - i \vec{k}\cdot\vec{x}}
 \;, \quad
 V \equiv L^{ }_1 L^{ }_2 \ldots L^{ }_d
 \;, \la{finV_final}
\ee
where the sign conventions correspond to those in \eq\nr{sprod}.

Like in \se\ref{se:p5}, the reality of $\phi(\tau,\vec{x})$ implies
that the Fourier modes satisfy
\be
 \left[ \tilde\phi(\omega^{ }_n,\vec{k}) \right]^* 
 = \tilde\phi(-\omega^{ }_n,-\vec{k})
 \;. 
\ee
Thereby only half of the Fourier modes are independent. 
We can choose, for instance, 
\be
 \tilde\phi(\omega^{ }_n,\vec{k})\;, \quad n\ge 1 \;; \quad
 \tilde\phi(0,\vec{k})\;, \quad k^{ }_1 > 0 \;; \quad
 \tilde\phi(0,0,k^{ }_2,...)\;, \quad k^{ }_2 > 0 \;; 
 \; \ldots \;; \quad \mbox{and} \quad
 \tilde\phi(0,\vec{0})
 \la{onlyhalf}
\ee
as the integration variables. Note again the presence of a zero mode. 

\index{Zero mode: scalar field}

With the above conventions, quadratic forms can be written in the form
\be
 \int_0^{\beta} \! {\rm d}\tau \! \int_\vec{x} \, 
 \phi^{ }_1(\tau,\vec{x}) \, 
 \phi^{ }_2(\tau,\vec{x}) = 
 T \sum_{\omega^{ }_n}
 \frac{1}{V} \sum_{\vec{k}}
 \tilde \phi^{ }_1(-\omega^{ }_n,-\vec{k}) \, 
 \tilde \phi^{ }_2(\omega^{ }_n,\vec{k})
 \;, 
\ee
implying that in the {\em free case}, i.e.\ 
for $V(\phi) \equiv \frac{1}{2} m^2\phi^2$, the exponent
in \eq\nr{Z_SFT} becomes
\ba
 \exp(-S^{ }_\iE) & = & 
 \exp\Bigl( - 
  \int_0^{\beta} \! {\rm d}\tau \int_\vec{x} \, {L}^{ }_\iE
 \Bigr) 
 \nn & = & 
 \exp\biggl[ 
   - \frac{1}{2}  T \sum_{\omega^{ }_n}
   \frac{1}{V} \sum_{\vec{k}}
   (\omega_n^2 + \vec{k}^2 + m^2)
 | \tilde \phi(\omega^{ }_n,\vec{k}) |^2
 \biggr]
 \nn & = & 
 \prod_{\vec{k}}\biggl\{ \exp\biggl[ 
   - \frac{T}{2V} \sum_{\omega^{ }_n}
   (\omega_n^2 + \vec{k}^2 + m^2)
 | \tilde \phi(\omega^{ }_n,\vec{k}) |^2
 \biggr]\biggr\} 
 \;. \la{zsft}
\ea
The exponential here is precisely of the same form 
as in \eq\nr{preZ}, with the replacements 
\be
 m^\rmii{(HO)} \to \frac{1}{V}
 \;, \quad
 (\omega^{\rmii{(HO)}})^2 \to \vec{k}^2 + m^2
 \;, \quad
 {|x_n^\rmii{(HO)}|}^{2} \to   
 | \tilde \phi(\omega^{ }_n,\vec{k}) |^2
 \;.  
\ee
Thus, we see that the result for the partition function factorizes 
into a {\em product of 
harmonic oscillator partition functions}, for which 
we know the answer already. 

In order to take advantage of the above observation, we rewrite 
\eqs\nr{preZ}, \nr{hoZ3} and \nr{hoF} for the case $\hbar = 1$. 
This allows us to represent  
the harmonic oscillator partition function in the form
\ba
 \mathcal{Z}^\rmii{HO} & = &             
 C' \int \left[ \prod_{n\ge 0} {\rm d} x^{ }_n \right]
 \exp\biggl[ -\frac{mT}{2}\sum_{n=-\infty}^{\infty} (\omega_n^2 + \omega^2)
 |x^{ }_n|^2 \biggr]
 \la{zho1} \\ & = &
 \frac{T}{\omega} \prod_{n=1}^{\infty} 
 \frac{\omega_n^2}{\omega^2 + \omega_n^2} 
  \\ & = & 
 T \prod_{n=-\infty}^{\infty} (\omega_n^2 + \omega^2)^{-\frac{1}{2}}
   \prod_{n'=-\infty}^{\infty} (\omega_n^2)^{\frac{1}{2}}
 \la{zho2} \\ & = &
 \exp\biggl\{
 -\frac{1}{T} 
 \biggl[ 
  \frac{\omega}{2} + T \ln\Bigl( 1 - e^{-\beta{\omega}}\Bigr)
 \biggr] 
 \biggr\} 
 \;, \la{zho3}
\ea
where $n'$ means that the zero mode $n=0$ is omitted. 

Combining now \eq\nr{zsft} with \eqs\nr{zho1}--\nr{zho3}, 
we obtain two useful representations for $\mathcal{Z}^\rmii{SFT}$. 
First of all, denoting 
\be
 \E^{ }_{k} \equiv \sqrt{\vec{k}^2 + m^2} 
 \;, 
\ee
\eq\nr{zho2} yields
\ba
 \mathcal{Z}^\rmii{SFT} = \exp\biggl( -\frac{F^\rmii{SFT}}{T} \biggr)
 & = & 
 \prod_{\vec{k}} 
 \biggl\{ 
 T \prod_{n} (\omega_n^2 + \E_{k}^2)^{-\frac{1}{2}}
   \prod_{n'} (\omega_n^2)^{\frac{1}{2}}
 \biggr\}
 \\ & = &
 \exp\biggl\{ 
  \sum_\vec{k} 
 \biggl[
  \ln T + \frac{1}{2} \sum_{n'} \ln \omega_n^2  
  -\frac{1}{2} \sum_{n} \ln(\omega_n^2 + \E_{k}^2 ) 
 \biggr]
 \biggr\} 
 \;.
\ea
Taking the infinite-volume limit, the {\rm free-energy density}, 
$F/V$, can thus be written as 
\be
 \lim_{V\to\infty} \frac{F^\rmii{SFT}}{V} 
 = 
 \int \!\! \frac{{\rm d}^d\vec{k}}{(2\pi)^d} \, 
 \biggl[
   T \sum_{\omega^{ }_n} \frac{1}{2} \ln (\omega_n^2 + \E_{k}^2 ) 
 - T \sum_{\omega_n'} \frac{1}{2} \ln (\omega_n^2) - \frac{T}{2} \ln (T^2)
 \biggr]
 \;. \la{f_sft_1}
\ee
Second, making directly use of \eq\nr{zho3}, we get 
the alternative representation
\ba
 \mathcal{Z}^\rmii{SFT} = \exp\biggl( -\frac{F^\rmii{SFT}}{T} \biggr)
 & = & 
 \prod_{\vec{k}} 
 \biggl\{ 
 \exp\biggl[ 
   -\frac{1}{T}
   \biggl( 
     \frac{\E^{ }_{k}}{2} + T \ln 
     \Bigl( 1 - e^{-\beta{\E^{ }_{k}}}\Bigr) 
   \biggr)
 \biggr]
 \biggr\}  
 \;, \\ 
 \lim_{V\to\infty} \frac{F^\rmii{SFT}}{V} 
 & = &  
 \int \! \frac{{\rm d}^d\vec{k}}{(2\pi)^d}
 \biggl[
     \frac{\E^{ }_{k}}{2} + T \ln 
     \Bigl( 1 - e^{-\beta{\E^{ }_{k}}}\Bigr)  
 \biggr]
 \;. \la{f_sft_2}
\ea
We return to the momentum integrations in 
\eqs\nr{f_sft_1} and \nr{f_sft_2} 
in secs.~\ref{se:lowT} and \ref{se:highT}.

\newpage 

\subsection{Evaluation of thermal sums and their low-temperature limit}
\la{se:bsums}

\index{Thermal sums: boson loop}

Thanks to the previously established equality between 
\eqs\nr{zho2} and \nr{zho3}, we have arrived at two different
representations for the free energy density of a free scalar
field theory, namely \eqs\nr{f_sft_1} and \nr{f_sft_2}. The purpose
of this section is to take the step from \eq\nr{f_sft_1}
to \nr{f_sft_2} directly, and learn to carry out 
thermal sums such as those in \eq\nr{f_sft_1} also in more general cases. 

As a first observation, 
we note that the sum in \eq\nr{f_sft_1} contains two physically very
different structures. The first term depends on the energy 
(and thus on the mass of the field), and can be classified as 
a ``physical'' contribution. At the same time, the second and third 
terms represent 
``unphysical'' subtractions, which 
are independent of the energy, but are needed in order to make
the entire sum convergent. It is evident that only 
the contribution of the energy-dependent term
survives in \eq\nr{f_sft_2}. 

In order not to lose the focus of our discussion on the subtraction terms,  
we mostly 
concentrate on another, convergent sum in the following:  
\be
 i(\E^{ }_k) \;\equiv\; 
 \frac{1}{\E^{ }_k} \frac{{\rm d}j(\E^{ }_k)}{{\rm d}\E^{ }_k}
 = 
 T \sum_{\omega^{ }_n} \frac{1}{\omega_n^2 + \E_k^2}
 \;. \la{def_i}
\ee
The term appearing in \eq\nr{f_sft_1}, 
\be
 j(\E^{ }_k) \;\equiv\; T \sum_{\omega^{ }_n}
  \frac{1}{2} \ln ({\omega_n^2 + \E_k^2}) 
  - T \sum_{\omega_n'} \frac{1}{2} \ln (\omega_n^2) - \frac{T}{2} \ln (T^2)
 \;, \quad \omega^{ }_n = 2\pi T n
 \;, \la{def_j}
\ee
can be obtained from $i(\E^{ }_k)$ through integration, 
apart from an $\E^{ }_k$-independent integration constant.

Let now $f(p)$ be a generic function, analytic in the complex plane
(apart from isolated singularities), 
and in particular regular on the real axis. 
We may then consider the sum
\be
 \sigma \; \equiv \; T \sum_{\omega^{ }_n} f(\omega^{ }_n)
 \;, \la{Sfn}
\ee
where the $\omega^{ }_n$ are the Matsubara frequencies defined above
(e.g.\ in \eq\nr{def_j}). 
It turns out to be useful to define the auxiliary function 
\be
 i\, \nB{}(ip) \;\equiv\; \frac{i}{\exp(i\beta p) - 1}
 \;, \la{inB_1}
\ee
where $\nB{}$ is the Bose distribution. 
\Eq\nr{inB_1} can be seen to have poles 
exactly at $\beta p = 2\pi n$, $n\in \ZZ$, 
i.e.~at $p = \omega^{ }_n$. Expanding this function in 
a Laurent series around any of the poles, 
we get 
\be
 i\, \nB{}(i[\omega^{ }_n + z]) \; = \;
 \frac{i}{\exp(i\beta [\omega^{ }_n+z]) - 1}
 \; =  \; \frac{i}{\exp(i\beta z) - 1} \; \approx \; \frac{T}{z} + \rmO(1)
 \;, 
\ee
which implies that the residue at each pole is $T$. 
This means that we can replace the sum in \eq\nr{Sfn} by the complex integral
\ba
 \sigma & = &  \oint \frac{{\rm d}p}{2\pi i} \, f(p)\, i  \nB{}(i p) 
 \; \equiv \;
 \int_{-\infty-i0^+_{ }}^{+\infty-i0^+_{ }}
 \frac{{\rm d}p}{2\pi}\, f(p) \, \nB{}(ip)
 +  
 \int_{+\infty+i0^+_{ }}^{-\infty+i0^+_{ }}
 \frac{{\rm d}p}{2\pi}\, f(p) \, \nB{}(ip)
 \;, \la{Sfn_2} \hspace*{4mm}
\ea
where 
the integration contour runs anti-clockwise around 
the real axis of the complex $p$-plane.

The above result can be further simplified by substituing $p\to -p$ 
in the latter term of \eq\nr{Sfn_2}, and noting that 
\be
 \nB{}(-ip) = \frac{1}{\exp(-i\beta p) - 1}
 = \frac{\exp(i\beta p) - 1 + 1}{1 - \exp(i \beta p)}
 = - 1 - \nB{}(ip) 
 \;.
\ee
This leads to the formula
\ba
 \sigma & = &  
 \int_{-\infty-i0^+_{ }}^{+\infty-i0^+_{ }}
 \frac{{\rm d}p}{2\pi}\,
 \Bigl\{
  f(-p) + [f(p) + f(-p)]\, \nB{}(ip)  
 \Bigr\}
 \nn & = &
 \int_{-\infty}^{+\infty}
 \frac{{\rm d}p}{2\pi}\, f(p) + 
 \int_{-\infty-i0^+_{ }}^{+\infty-i0^+_{ }}
 \frac{{\rm d}p}{2\pi}\,
 [f(p) + f(-p)]\, \nB{}(ip)  
 \;, \la{Sfn_res} 
\ea
where we returned to the real 
axis in the first term, made possible by the lack of singularities 
there. All in all, we have thus  converted the sum of \eq\nr{Sfn} 
into a rather convenient complex integral. 

Inspecting the integral in \eq\nr{Sfn_res}, we note that its first 
term is temperature-independent: 
it gives the zero-temperature, or ``vacuum'', contribution to $\sigma$.
The latter term determines how thermal effects change the result. 
Let us note, furthermore, that in the lower half-plane we have
\be
 |\nB{}(ip)| \stackrel{p=x-iy}{=}
 \left| \frac{1}{e^{i\beta x}e^{\beta y} - 1} \right| 
 \stackrel{y\gg T}{\approx} e^{-\beta y}
 \stackrel{y\gg x}{\approx} e^{-\beta |p|}
 \;. 
\ee
Therefore, it looks likely that 
if the function $f(p)$ grows slower than 
$e^{\beta |p|}$ at large $|p|$ (in particular, polynomially), 
the integration contour for the finite-$T$ term of \eq\nr{Sfn_res}
can be closed in the lower half-plane, whereby the result is 
determined by the poles and residues of the 
function $f(p) + f(-p)$. Physically, we say that the thermal 
contribution to $\sigma$ is related to ``on-shell'' particles.

Let us now apply the general formula in \eq\nr{Sfn_res} to 
the particular example of \eq\nr{def_i}. In fact, without any 
additional cost, we can consider a slight generalization, 
\be
 i(\E^{ }_k;c) \;\equiv\; 
 T \sum_{\omega^{ }_n} \frac{1}{(\omega^{ }_n + c)^2 + \E_k^2}
 \;, \quad c \in \CC
 \;, \la{iEc_def}
\ee
so that in the notation of \eq\nr{Sfn} we have 
\ba
 f(p) & = &  \frac{1}{(p+c)^2 + \E_k^2}
 = \frac{i}{2\E^{ }_k}
 \biggl[
 \frac{1}{p+c+i\E^{ }_k} - \frac{1}{p+c-i\E^{ }_k} 
 \biggr] \; , \la{f_expl2}
 \\ 
 f(p) + f(-p) & = &
 \frac{i}{2\E^{ }_k}
 \biggl[
 \frac{1}{p+c+i\E^{ }_k}  + \frac{1}{p-c+i\E^{ }_k}
 - \frac{1}{p+c-i\E^{ }_k}  - \frac{1}{p-c-i\E^{ }_k} 
 \biggr]
 \;. \la{f_expl}
\ea
For \eq\nr{Sfn_res}, we need the poles of these functions 
in the lower half-plane, which 
for $|\im c\,| < \E^{ }_k$ are located at $p=\pm c-i\E^{ }_k$. According
to \eqs\nr{f_expl2} and \nr{f_expl}, the residue at each 
lower half-plane pole is $i/2\E^{ }_k$. Thus the vacuum
term in \eq\nr{Sfn_res} produces
\be
 \frac{1}{2\pi} (-2\pi i) \frac{i}{2\E^{ }_k} = \frac{1}{2\E^{ }_k}
 \;,  
\ee
whereas the thermal part yields
\be
 \frac{1}{2\pi} (-2\pi i) \frac{i}{2\E^{ }_k}
 \biggl[
   \frac{1}{e^{\beta(\E^{ }_k-ic)} - 1} + 
   \frac{1}{e^{\beta(\E^{ }_k+ic)} - 1}   
 \biggr]
 \;.
\ee
In total, we obtain
\be
 i(\E^{ }_k;c) = \frac{1}{2 \E^{ }_k}
 \Bigl[
  1 + \nB{}(\E^{ }_k-ic) + \nB{}(\E^{ }_k+ic) 
 \Bigr]
 \;, \la{iEc}
\ee
which is clearly periodic in 
$c\to c+2\pi T n$, $n\in \ZZ$,  
as it must be according to \eq\nr{iEc_def}. We also note that the appearance
of $ic$ resembles that of a {\em chemical potential}. Indeed, as shown
around \eqs\nr{FTmu} and \nr{FTmu2}, setting $ic\to -\mu$  
corresponds to a situation where we have averaged over a particle
(chemical potential $\mu$) and an antiparticle 
(chemical potential $-\mu$).\footnote{%
 Apart from a chemical potential, the parameter $c$ can also 
 appear in a system with ``shifted boundary conditions'' over
 a compact direction, cf.\ e.g.\ ref.~\cite{gm}.
 }

To conclude the discussion, we integrate
\eq\nr{iEc} with respect to $\E^{ }_k$ in order to obtain 
the function in \eq\nr{def_j} (generalized to include $c$), 
\be
 j(\E^{ }_k;c) \;\equiv\; T \sum_{\omega^{ }_n} \frac{1}{2} 
  \ln [{(\omega^{ }_n+c)^2 + \E_k^2}]
 - (\mbox{$\E^{ }_k$-independent terms})
 \;.
\ee
Eq.~\nr{def_i} clearly continues to hold in the 
presence of $c$, so noting that 
\be
 \frac{1}{e^x-1} = \frac{e^{-x}}{1-e^{-x}}
 = \frac{{\rm d}\ln \bigl( 1- e^{-x} \bigr)}{{\rm d}x} 
 \;, 
\ee
\eq\nr{iEc} immediately yields
\be
 j(\E^{ }_k;c) = \mbox{const.} 
 + \frac{\E^{ }_k}{2} + 
 \frac{T}{2} 
 \biggl\{ 
   \ln 
     \Bigl[ 1 - e^{-\beta (\E^{ }_k  -i c) }\Bigr]
    + \ln \Bigl[ 1 - e^{-\beta (\E^{ }_k  + i c) }\Bigr]
 \biggr\}
 \;. \la{jEc}
\ee
The constant term in this result can depend 
both on $T$ and $c$, but not on $\E^{ }_k$. 

For $c=0$, a comparison of \eq\nr{jEc} with \eq\nr{f_sft_2} 
shows that the role of the extra terms in \eq\nr{f_sft_1}
is to eliminate the integration constant
in \eq\nr{jEc}. This implies that the full physical result for $j(\E^{ }_k;0)$
can be deduced directly from $i(\E^{ }_k;0)$. The same is true even 
for $\mu\equiv -ic \neq 0$, if we interpret $j(\E^{ }_k;c)$ as a free energy
density averaged over a particle and an antiparticle,  
as we next show.

\subsection*{Extension to a chemical potential}
\la{exe2}

\index{Chemical potential: scalar field}

Considering a harmonic oscillator in the presence of 
a chemical potential, our task becomes to compute the partition function  
\be
 e^{-\beta F(T,\mu)} \;\equiv\; \mathcal{Z}(T,\mu)
 \;\equiv\; \tr\Bigl[ e^{-\beta(\hat H - \mu \hat N)} \Bigr] 
 \;,  \la{FTmu}
\ee
where $\hat N \equiv \hat a^\dagger \hat a$. 
We show that the expression 
\be
 \frac{1}{2} \Bigl[ 
  F(T,ic) + F(T,-ic)
 \Bigr] \la{FTmu2}
\ee
agrees with the $\E^{ }_k$-dependent part of \eq\nr{jEc}.

To start with, we observe that
\be
 \langle n | (\hat H - \mu \hat N) | n \rangle 
 = \hbar\omega \Bigl( n + \frac{1}{2} \Bigr) - \mu n
 = (\hbar\omega - \mu) n + \frac{\hbar\omega}{2}
 \;,
\ee
so that evaluating the partition function in the energy basis yields
\be
 \mathcal{Z}^\rmii{HO} = 
 \sum_{n=0}^{\infty} \exp\biggl( -\frac{\hbar\omega}{2 T}
 - \frac{\hbar\omega-\mu}{T} n \biggr)
 = \frac{\exp\Bigl( -\frac{\hbar\omega}{2 T} \Bigr)}
   {1 - \exp\Bigl( - \frac{\hbar\omega-\mu}{T} \Bigr)}
 \;.
\ee
Setting now $\hbar\to 1$, $\omega\to \E^{ }_k$, $\mu\to -ic$, 
we can rewrite the result as
\be
 \mathcal{Z}^\rmii{HO} =
 \exp\biggl\{ 
   -\frac{1}{T} \biggl[
 \frac{\E^{ }_k}{2} + T \ln
 \biggl( 
   1 - e^{-\frac{\E^{ }_k+ic}{T}}
 \biggr)
 \biggr]
 \biggr\}
 \;. \la{ZHO_mu}
\ee
Reading from here $F(T,\mu)$ according to \eq\nr{FTmu}, 
and computing 
$
 \frac{1}{2} \Bigl[ 
  F(T,ic) + F(T,-ic)
 \Bigr]
$,
clearly yields exactly the $\E^{ }_k$-dependent part of \eq\nr{jEc}.


\subsection*{Low-temperature expansion}
\la{se:lowT}

\index{Thermal sums: low-temperature expansion}

Our next goal is to carry out the momentum integration 
in \eq\nr{f_sft_1} and/or \nr{f_sft_2}. To this end, we denote 
\ba
 J(m,T) & \equiv & 
 \int \! \frac{{\rm d}^d\vec{k}}{(2\pi)^d}
 \biggl[
     \frac{\E^{ }_{k}}{2} + T \ln 
     \Bigl( 1 - e^{-\beta{\E^{ }_{k}}}\Bigr)  
 \biggr]
 \la{JmT_1} \\  & = &  
 T \sum_{\omega^{ }_n} \int \! \frac{{\rm d}^d\vec{k}}{(2\pi)^d}
 \biggl[
    \frac{1}{2} \ln (\omega_n^2 + \E_{k}^2 ) 
 - \mbox{const.}
 \biggr]
 \;, \la{JmT} \\ 
 I(m,T) & \equiv & 
 \frac{1}{m} \frac{{\rm d}}{{\rm d}m} J(m,T) 
 \la{Imt} \\
 & = & 
 \int \! \frac{{\rm d}^d\vec{k}}{(2\pi)^d} 
 \frac{1}{2\E^{ }_{k}}
 \Bigl[
     1 + 2 \nB{} ({\E^{ }_{k}}) 
  \Bigr]
 \la{ImT_1} \\ & = &  
 T \sum_{\omega^{ }_n} \int \! \frac{{\rm d}^d\vec{k}}{(2\pi)^d}
    \frac{1}{\omega_n^2 + \E_{k}^2}
 \;,  \la{ImT_2}
\ea
where $d\equiv 3 - 2\epsilon$ is the space dimensionality, 
$\E^{ }_{k} \equiv \sqrt{{k}^2 + m^2}$, 
and we made use of the fact that inside the integral 
$
 m^{-1} \partial^{ }_m = 
 \E_{k}^{-1} \partial^{ }_{\E^{ }_{k}}
$. 
In order to simplify the notation, we further denote
\be
 \Tint{ K} \;\equiv\; 
 T \sum_{\omega^{ }_n} \int \! \frac{{\rm d}^d\vec{k}}{(2\pi)^d}
 \;, \quad
 \Tint{ K}' \;\equiv\; 
 T \sum_{\omega_n'} \int \! \frac{{\rm d}^d\vec{k}}{(2\pi)^d}
 \;, \quad
 \int_\vec{k} \;\equiv\;
 \int \! \frac{{\rm d}^d\vec{k}}{(2\pi)^d}
 \;, \la{measure}
\ee
where $ K\equiv (\omega^{ }_n,\vec{k})$, and 
a prime denotes that the zero mode ($\omega^{ }_n = 0$) is omitted. 

At low temperatures, $T\ll m$, we may expect the results to
resemble those of the zero-temperature theory. To this end, we write
\be
 J(m,T) \;=\; J_0^{ }(m) + J^{ }_T(m)
 \;, \quad
 I(m,T) \;=\; I_0^{ }(m) + I^{ }_T(m)
 \;, \la{splitup}
\ee
where $J_0^{ }$ is the temperature-independent vacuum energy density, 
\be
 J_0^{ }(m) \;\equiv\; \int_\vec{k} \frac{\E^{ }_{k}}{2}
 \;, \la{J0m_1}
\ee
and $J^{ }_T$ the thermal part of the free energy density, 
\be
 J^{ }_T(m) \;\equiv\; \int_\vec{k} T \ln 
     \Bigl( 1 - e^{-\beta{\E^{ }_{k}}}\Bigr)  
 \;. \la{JTm}
\ee
The sum-integral $I(m,T)$ is divided in a similar way. 
It is clear that $J_0^{ }$ is ultraviolet divergent, and can 
only be evaluated in the presence of a regulator; our choice  
is typically dimensional regularization, as 
indicated in \eq\nr{measure}. In contrast, the integrand
in $J^{ }_T$ is exponentially small for $k \gg T$, and
therefore the integral is convergent.  

Let us start from the evaluation of $J_0^{ }(m)$. Writing 
out the mass dependence
explicitly, the task becomes to compute
\be
 J_0^{ }(m) \;=\; \int_\vec{k} \frac{1}{2} 
 ({k}^2 + m^2)^{\frac{1}{2}}
 \;. \la{J0m}
\ee
For generality 
and future reference, we first consider a somewhat more generic integral,
\be
 \Phi(m,d,A) \;\equiv\; \int\! \frac{{\rm d}^d\vec{k}}{(2\pi)^d}
 \frac{1}{({k}^2 + m^2)^A}
 \;, \la{F_def}
\ee
and obtain then $J_0^{ }$ as  
$J_0^{ }(m) = \frac{1}{2} \Phi(m,d,-\frac{1}{2})$. 

\index{Dimensional regularization}

Owing to the fact that our integrand only depends on ${k}$, all angular 
integrations can be carried out at once, and the integration measure
obtains the well-known form\footnote{%
 A quick derivation: On one hand, 
 $
  \int \! {\rm d}^d\vec{k}\, e^{-t k^2}
 = [\int_{-\infty}^{\infty} {\rm d} k^{ }_1 e^{-t k_1^2}]^d
 = (\pi/t)^{\fr{d}2}
 $. 
 On the other hand, 
 $
  \int \! {\rm d}^d\vec{k}\, e^{-t k^2}
 = c(d) \int_0^{\infty}{\rm d}k\, k^{d-1} e^{-t k^2}
 = c(d) t^{-\fr{d}2} \int_0^{\infty} {\rm d}x \, x^{d-1} e^{-x^2}
 = c(d) \Gamma(\fr{d}2)/2 t^{\fr{d}2} 
 $. 
 Thereby
 $
 c(d) = 2 \pi^{\fr{d}2} / \Gamma(\frac{d}{2})
 $. 
 }
\be
 {\rm d}^d \vec{k} = \frac{\pi^{\fr{d}2}}{\Gamma(\frac{d}{2})} 
 ({k}^2)^\frac{d-2}{2} {\rm d} ({k}^2) 
 \;,  \la{k_measure}
\ee
where $\Gamma(s)$ is the Euler gamma function, 
discussed in further detail in \se\ref{gamma_zeta}.
Substituting now ${k}^2 \to z \to m^2 t$ in \eq\nr{F_def}, we get
\ba
 \Phi(m,d,A) & = &
 \frac{\pi^{\fr{d}2}}{\Gamma(\frac{d}{2})} \frac{1}{(2\pi)^d}
 \int_0^\infty \! {\rm d}z \, 
 z^{\frac{d-2}{2}} (z + m^2)^{-A} 
 \nn 
 & = &
 \frac{m^{d-2 A}}{(4\pi)^{\fr{d}2} \Gamma(\frac{d}{2})}
 \int_0^\infty \! {\rm d}t \, 
 t^{\fr{d}2 - 1} (1+t)^{-A} 
 \;, 
\ea
from which the further substitution $t \to 1/s - 1$, 
${\rm d}t \to - {\rm d}s/s^2$ yields
\be
 \Phi(m,d,A) = 
 \frac{m^{d-2 A}}{(4\pi)^{\fr{d}2} \Gamma(\frac{d}{2})}
 \int_0^1 \! {\rm d}s \, 
 s^{A-\fr{d}2-1} (1-s)^{\fr{d}2 - 1}
 \;. 
\ee
Here we recognize a standard integral that can
be expressed in terms of the Euler $\Gamma$-function, producing finally
\be
 \fbox{$\displaystyle
 \Phi(m,d,A) = 
 \int\! \frac{{\rm d}^d\vec{k}}{(2\pi)^d}
 \frac{1}{(\vec{k}^2 + m^2)^A}
 =  \frac{1}{(4\pi)^{\fr{d}2}} \frac{\Gamma(A-\frac{d}{2})}{\Gamma(A)}
 \frac{1}{(m^2)^{A - \frac{d}{2}}}
 $}
 \;. \la{fmdA}
\ee

Let us now return to $J_0^{ }(m)$ in \eq\nr{J0m}, setting
$A=-\frac{1}{2}$ and $d=3-2\epsilon$
in \eq\nr{fmdA} and multiplying the result by $\frac{1}{2}$. 
The basic property 
$
 \Gamma(s) = s^{-1} \Gamma(s+1)
$
allows us to transport the arguments of 
the $\Gamma$-functions to the vicinity 
of 1/2 or 1, where Taylor expansions are readily 
carried out, yielding (some helpful 
formulae are listed in \eqs\nr{Gamma_def1}--\nr{Gamma_specs3}): 
\ba
 \Gamma(-2+\epsilon) & = & 
 \frac{ \Gamma(1+\epsilon) }{(-2+\epsilon)(-1+\epsilon)\epsilon} 
 \\ & = & 
 \frac{1}{2 \epsilon} \Bigl( 1 + \frac{\epsilon}{2} \Bigr)
 \Bigl( 1 + \epsilon \Bigr) (1 - \gammaE \epsilon) + \rmO(\epsilon)
 \;, \\ 
 \Gamma(-{\textstyle\frac{1}{2}}) & = & 
 -2 \Gamma({\textstyle\frac{1}{2}}) = -2 \sqrt{\pi}
 \;. 
\ea
The other parts of \eq\nr{fmdA} can be written as 
\ba
 (4\pi)^{-\fr32 + \epsilon} & = & 
 \frac{2\sqrt{\pi}}{(4\pi)^2}  \Bigl[ 1 + \epsilon \ln(4\pi) 
 \Bigr] + \rmO(\epsilon^2)
 \;, \\
 (m^2)^{2-\epsilon} & = & m^4 \mu^{-2\epsilon} 
 \left( \frac{\mu^2}{m^2}\right)^\epsilon 
 =  
 m^4 \mu^{-2\epsilon} \biggl(
 1 + \epsilon \ln  \frac{\mu^2}{m^2}
 \biggr) + \rmO(\epsilon^2)
 \;, 
\ea
where $\mu$ is an arbitrary (renormalization) scale parameter, introduced
through $1 = \mu^{-2\epsilon} \mu^{2\epsilon}$.\footnote{%
 When systems with a finite chemical potential are considered, 
 cf.\ \eq\nr{FTmu}, one has to abandon the standard
 convention of denoting the scale parameter by $\mu$; frequently the 
 notation $\Lambda$
 is used instead, 
 cf.\ \eqs\nr{T0measure1} and \nr{T0measure2}.  
 } 

Collecting everything together, we obtain from above
\be
 J_0^{ }(m) 
 = 
 -\frac{m^4\mu^{-2\epsilon}}{64\pi^2}
 \biggl[
   \frac{1}{\epsilon}
 + \ln  \frac{\mu^2}{m^2} + \ln(4\pi) - \gammaE + \fr32 + \rmO(\epsilon)
 \biggr]
 \;, \la{J_0_1}
\ee
which can further be simplified by introducing the ``$\msbar$ scheme''
scale parameter $\bmu$ through
\be
 \ln\bmu^2 \equiv \ln\mu^2 + \ln(4\pi) - \gammaE
 \;. \la{msbar}
\ee
This leads us to 
\be
 J_0^{ }(m) = 
  -\frac{m^4\mu^{-2\epsilon}}{64\pi^2}
 \biggl[
   \frac{1}{\epsilon}
 + \ln  \frac{\bmu^2}{m^2} + \fr32 + \rmO(\epsilon)
 \biggr]
 \;, 
 \la{J0m_res}
\ee
from which a differentiation with respect to the mass parameter produces
\be
 I_0^{ }(m) = \frac{1}{m} \frac{{\rm d}}{{\rm d}m} J_0^{}(m)
 =  \int_\vec{k} \frac{1}{2 \E^{ }_{k}}
 = 
  -\frac{m^2\mu^{-2\epsilon}}{16\pi^2}
 \biggl[
   \frac{1}{\epsilon}
 + \ln  \frac{\bmu^2}{m^2} + 1 + \rmO(\epsilon)
 \biggr]
 \;. 
 \la{I0m_res}
\ee
Interestingly, we note that 
$
 \int_{-\infty}^{\infty} \! \frac{{\rm d}k^{ }_0}{2\pi}
 \frac{1}{k_0^2 + \E_{k}^2} = \frac{1}{2 \E^{ }_{k}}
$, 
so that 
$I_0^{ }(m)$ can also be written as 
\be
 I_0^{ }(m) = \int \! \frac{{\rm d}^{d+1}\vec{k}}{(2\pi)^{d+1}}
 \frac{1}{{k}^2 + m^2}
 \;. 
\ee
This is a very natural result, considering that the quantity 
we are determining is the $T=0$ limit of the sum-integral 
\be
I(m,T) = \Tint{ K} \frac{1}{K^2+m^2} \;,
\ee
with $\lim_{T\to 0} T\sum_{k^{ }_n} = \int \! \frac{{\rm d}k^{ }_0}{2\pi}$, 
cf.\ \eq\nr{Fou_infV}. 

Next, we consider the finite-temperature 
integrals $J^{ }_T(m)$ and $I^{ }_T(m)$ which, 
as already mentioned, are both finite. 
Therefore we can normally set $d=3$ within them, 
even though it is good to recall that 
in multiloop computations these functions sometimes 
get multiplied by a divergent term, in which case 
contributions of $\rmO(\epsilon)$ (or higher) are needed as well.\footnote{%
 The $\rmO(\epsilon)$ terms could be obtained by noting 
 from \eq\nr{k_measure} that  for $d = 3 - 2\epsilon$,
 $ 
 \mu^{2\epsilon} {\rm d}^d \vec{k}/(2\pi)^d 
 = {\rm d}^3 \vec{k}/(2\pi)^3
 \{ 1 + \epsilon [ \ln(\bmu^2 / 4 {k}^2 ) + 2] 
 + \rmO(\epsilon^2) \} 
 $. 
 }
Neglecting this subtlety for now and substituting 
${k}\to T x$ in \eqs\nr{JTm} and \nr{ImT_1}, we find 
\ba
 J^{ }_T(m) 
 & = &  
 \frac{T^4}{2\pi^2} 
 \int_0^{\infty} \! {\rm d}x\, x^2 \, 
 \ln \Bigl(
   1 - e^{-\sqrt{x^2 + y^2}} 
 \Bigr)^{ }_{y \equiv \fr{m}T}
 \;,  \la{JTy} \\
 I^{ }_T(m) 
 & = &  
 \frac{T^2}{2\pi^2} 
 \int_0^{\infty} \! \frac{{\rm d}x\, x^2 }{\sqrt{x^2 + y^2}}
 \frac{1}{e^{\sqrt{x^2 + y^2}} -1} 
 \biggr|^{ }_{y \equiv \fr{m}T}
 \;.  \la{ITy}
\ea
These integrals cannot be expressed in terms
of elementary functions,\footnote{%
 However the following convergent sum representations apply: 
 $J^{ }_T(m) = - \frac{m^2 T^2}{2\pi^2} \sum_{n=1}^{\infty} \frac{1}{n^2}
 K^{ }_2 (\frac{n m}{T})$, 
 $I^{ }_T(m) = \frac{m T}{2\pi^2} \sum_{n=1}^{\infty} \frac{1}{n}
 K^{ }_1 (\frac{n m}{T})$, 
 with $K^{ }_n$ a modified Bessel function. \label{bessel1}
 }
but their numerical evaluation is rather straightforward. 

Even though \eq\nr{JTy} cannot be evaluated exactly, we can 
still find approximate expressions valid in various limits. 
In this section we are interested in low temperatures, 
i.e.\ $y = m/T \gg 1$. We thus evaluate the leading term of 
\eq\nr{JTy} in an expansion in $\exp(-y)$ and $1/y$, which produces
\ba
 \int_0^{\infty} \! {\rm d}x\, x^2 \, 
 \ln \Bigl(
   1 - e^{-\sqrt{x^2 + y^2}} 
 \Bigr)
 & = & \hspace*{-0.5cm}
 -
 \int_0^{\infty} \! {\rm d}x\, x^2 \, 
   e^{-\sqrt{x^2 + y^2}} 
 + \rmO(e^{-2y})
 \nn 
 & \stackrel{w\equiv\sqrt{x^2 + y^2}}{=} & \hspace*{-0.5cm}
 - \int_y^{\infty} \! {\rm d}w \, w \sqrt{w^2 - y^2} e^{-w} 
 + \rmO(e^{-2y})
 \nn 
 & \stackrel{v\equiv w-y}{=} & \hspace*{-0.5cm}
 - e^{-y }\int_0^{\infty}
  \! {\rm d}v\, (v+y) \sqrt{2 v y + v^2}\, e^{-v}
 + \rmO(e^{-2y})
 \nn 
 & = & \hspace*{-0.5cm}
 -\sqrt{2}\, y^{\fr32} e^{-y}
 \int_0^{\infty}
  \! {\rm d}v \, v^{\frac{1}{2}}
 \bigl( 
  1 + {\textstyle \frac{v}{y} } 
 \bigr) 
 \bigl( 
  1 + {\textstyle \frac{v}{2 y} } 
 \bigr)^{\frac{1}{2}}
 e^{-v} 
 + \rmO(e^{-2y})
 \nn 
 & = & \hspace*{-0.5cm}
 -\sqrt{2}\, \Gamma({\textstyle\fr32}) y^{\fr32} e^{-y}
 \Bigl[
 1 + \rmO\bigl({\textstyle\frac{1}{y}}\bigr) + \rmO\bigl(e^{-y}\bigr)
 \Bigr]
 \;,
\ea
where $\Gamma({\textstyle \fr32}) = \sqrt{\pi}/2$.
It may be noted that the power-suppressed terms 
amount to an asymptotic (non-convergent) series, 
but can be accounted for through the 
leading term of a convergent expansion in terms of
modified Bessel functions given in footnote~\ref{bessel1}, 
$
 - y^2 K^{ }_2(y) 
 [
  1 + \rmO\bigl(e^{-y}\bigr)
 ]
$.

Inserting the above expression into \eq\nr{JTy}, we have obtained
\be
 J^{ }_T(m) = - T^4 \Bigl( \frac{m}{2\pi T}\Bigr)^{\fr32} e^{-\frac{m}{T}}
 \biggl[ 1 + \rmO\Bigl( {\textstyle \frac{T}{m} } \Bigr) 
 + \rmO\Bigl(e^{-\fr{m}T }\Bigr)\biggr]
 \;, \la{JTm_final}
\ee
whereas the derivative in \eq\nr{Imt} yields 
\be
 I^{ }_T(m) = \frac{T^3}{m} 
 \Bigl( \frac{m}{2\pi T}\Bigr)^{\fr32} e^{-\frac{m}{T}}
 \biggl[ 1 + \rmO\Bigl( {\textstyle \frac{T}{m} } \Bigr) 
 + \rmO\Bigl(e^{-\fr{m}T }\Bigr) \biggr]
 \;. \la{ITm_final}
\ee
Thereby we have arrived at the main conclusion of this section: 
at low temperatures, $T \ll m$, finite-temperature 
effects in a free theory with a mass gap are 
exponentially suppressed by the Boltzmann factor, $\exp(-m/T)$, 
like in non-relativistic statistical mechanics. 
Consequently, the functions $J(m,T)$ and $I(m,T)$ can be
well approximated by their respective zero-temperature 
limits $J_0^{ }(m)$ and $I_0^{ }(m)$, which are
given in \eqs\nr{J0m_res} and \nr{I0m_res}.

\newpage 

\subsection{High-temperature expansion}
\la{se:highT}

\index{Thermal sums: high-temperature expansion}

Next, we move on to consider a limit opposite to that of the previous section,
i.e.\ $T \gg m$ or, in terms of \eq\nr{JTy}, $y=m/T\ll 1$. It may appear
that the procedure should then be a simple 
Taylor expansion of the integrand in 
\eq\nr{JTy} around $y^2 = 0$. The zeroth order term indeed yields
\be
 J^{ }_T(0) = 
  \frac{T^4}{2\pi^2} 
 \int_0^{\infty} \! {\rm d}x\, x^2 \, 
 \ln \Bigl(
   1 - e^{-\sqrt{x^2}} 
 \Bigr) = -\frac{\pi^2 T^4}{90}
 \;, \la{JT0}
\ee
which is nothing but the free-energy density (minus the pressure)
of black-body radiation with one massless degree of freedom. 
A correction term of order $\rmO(y^2)$ can also be worked out exactly. 

However, $\rmO(y^2)$ is as far as it goes: 
trying to proceed to the next order, $\rmO(y^4)$, 
one finds that the integral for the 
coefficient of $y^4$ is power-divergent at small $x\equiv k/T$.
In other words, the function $J^{ }_T(m)$ is non-analytic
in the variable $m^2$ around the point $m^2=0$. A generalized 
high-temperature expansion nevertheless exists, 
and turns out to take the form
\ba
 J^{ }_T(m)
 \!\!\! & = & \!\!\!  -\frac{\pi^2 T^4}{90}
 + \frac{m^2 T^2}{24}
 - \frac{m^3 T}{12\pi}
 - \frac{m^4}{2(4\pi)^2}
 \biggl[ 
  \ln\biggl( \frac{m e^{\gammaE}}{4\pi T} \biggr) - \fr34
 \biggr]
 + \frac{m^6\zeta(3)}{3 (4\pi)^4 T^2}
 + \rmO\biggl( \frac{m^8}{T^4} \biggr) + \rmO(\epsilon)
 \;, \nn \la{JTm_res}
\ea
where $m\equiv (m^2)^{1/2}$. It is the cubic term 
in \eq\nr{JTm_res} that first indicates that $J^{ }_T(m)$
is non-analytic in $m^2$ --- after all, 
the function $z^{3/2}$ contains a branch cut. 
This term plays a very important role in
certain physics contexts, as will be seen in \se\ref{ss:transition}. 

Our goal in this section is to {\em derive} \eq\nr{JTm_res}. 
A classic derivation, starting directly from the definition in \eq\nr{JTy}, 
was presented by Dolan and Jackiw~\cite{dj}.
It is, however, easier, and ultimately more useful, 
to tackle the task in a slightly different way: 
we start from \eq\nr{JmT} rather 
than \eq\nr{JmT_1}, and carry out {\em first the integration} 
$\int_\vec{k}$, and only then the sum $\sum_{\omega^{ }_n}$
(cf.\ e.g.\ ref.~\cite{az}). A slight drawback in this strategy 
is that \eq\nr{JmT} contains inconvenient constant 
terms. Fortunately, we already know the 
mass-independent value $J(0,T)$: it is given by \eq\nr{JT0}. 
Therefore it is enough to study $I(m,T)$, in which case the starting
point is \eq\nr{ImT_2}, which we may subsequently integrate as 
\be
 J(m,T) = \int_0^m \! {\rm d}m' \, m' \, I(m',T) + J(0,T)
 \;. \la{JI_rel}
\ee

\index{Zero mode: scalar field}

Proceeding now with $I(m,T)$ from \eq\nr{ImT_2}, the essential 
insight is to split the Matsubara sum into the contribution of the zero mode, 
$\omega^{ }_n = 0$, and that of the non-zero modes, $\omega^{ }_n\neq 0$.
Using the notation of \eq\nr{measure}, we thus write
\be
 \Tint{ K} = \Tint{ K}' + T \int_\vec{k}
 \;,
\ee
and first consider the contribution of the last term, 
which is denoted by $I^{(n=0)}$. 

\index{Infrared divergence: scalar field}

To start with, we return to 
the {\em infrared divergences} alluded to above. 
Trying naively a simple Taylor expansion of the integrand 
of  $I^{(n=0)}$ in powers of $m^2$, we would get
\be
 I^{(n=0)} 
 \;=\; T \int_\vec{k} \frac{1}{{k}^2 + m^2}
 \stackrel{\rmi{?}}{=} T \int \! \frac{{\rm d}^d\vec{k}}{(2\pi)^d}
 \biggl[ 
 \frac{1}{{k}^2} 
 - \frac{m^2}{{k}^4} + 
 \frac{m^4}{{k}^6} + \ldots
 \biggr]
 \;.  \la{In0_exp}
\ee
For $d=3-2\epsilon$, the first term is 
``ultraviolet divergent'', i.e.\ grows at large ${k}$, 
whereas the second and subsequent terms are 
``infrared divergent'', i.e. grow at small ${k}$
too fast to be integrable. Of course, in dimensional 
regularization, every expanded term in \eq\nr{In0_exp} appears 
to be zero; the total result is, however, non-zero,  
cf.\ \eq\nr{In0_res} below. The bottom line is that 
the Taylor expansion in \eq\nr{In0_exp}
is not justified.

Next, we compute the integral in \eq\nr{In0_exp} properly. 
The result can be read from \eq\nr{fmdA}, by just
setting $d=3-2\epsilon$, $A=1$: 
\be
 I^{(n=0)} = T \Phi(m,3-2\epsilon,1) = 
 \frac{T}{(4\pi)^{3/2-\epsilon}} 
 \frac{\Gamma(-\frac{1}{2} + \epsilon)}{\Gamma(1)} 
 \frac{1}{(m^2)^{-1/2 + \epsilon}}
 \stackrel{\Gamma(-\frac{1}{2}) = -2 \sqrt{\pi}}{=} 
 - \frac{Tm}{4\pi} + \rmO(\epsilon)
 \;. \la{In0_res}
\ee
We thus see that a linearly divergent integral 
over a manifestly positive function is {\em finite} and {\em negative}
in dimensional regularization! According to \eq\nr{Imt}, the 
corresponding term in $J^{(n=0)}$ reads
\be
 J^{(n=0)}
 = 
 - \frac{Tm^3}{12\pi} + \rmO(\epsilon)
 \;. \la{Jn0_res}
\ee

Given the importance of the result and 
its somewhat counter-intuitive appearance, 
it is worthwhile to demonstrate that 
\eq\nr{In0_res} is {\em not} an artifact 
of dimensional regularization. 
Indeed, let us compute the integral with cutoff regularization, 
by restricting ${k}$ to be smaller than an explicit upper bound $\Lambda$: 
\ba
 I^{(n=0)} & = & 
 T \frac{4\pi}{(2\pi)^3}
 \int_0^{\Lambda} \frac{{\rm d}{k}\, {k}^2}{{k}^2 + m^2}
 =
 \frac{T}{2\pi^2}
 \biggl[ 
  \Lambda - m^2 
  \int_0^{\Lambda} \frac{{\rm d}{k}}{{k}^2 + m^2}
 \biggr]
 \nn & = &
 \frac{T}{2\pi^2}
 \biggl[ 
   \Lambda - m\, \arctan\Bigl( \frac{\Lambda}{m}\Bigr) 
 \biggr]
 \; \stackrel{m\ll \Lambda}{=} \;
 T \biggl[
  \frac{\Lambda}{2\pi^2} - \frac{m}{4\pi} + 
 \rmO\biggl(\frac{m^2}{\Lambda} \biggr) 
 \biggr]
 \;. \la{In0_Lam}
\ea
We observe that, due to the first term, \eq\nr{In0_Lam}
{\em is} positive. This term is unphysical, however: 
it must cancel against similar terms emerging from the non-zero
Matsubara modes, since the temperature-dependent part of \eq\nr{ImT_1}
is manifestly finite. 
Owing to the fact that it represents a power divergence,  
it does not appear in dimensional regularization at all. 
The second term in \eq\nr{In0_Lam} is the physical one, 
and it agrees with \eq\nr{In0_res}. The remaining terms
in \eq\nr{In0_Lam} vanish when the cutoff is taken to infinity, 
and are analogous to the $\rmO(\epsilon)$-terms of \eq\nr{In0_res}. 

Next, we turn to the non-zero Matsubara modes, 
whose contribution to the integral 
is denoted by $I'(m,T)$ (the prime is not to be confused with a derivative). 
It is important to realize that in this case, a Taylor expansion
in $m^2$ {\em can} formally be carried out
(we do not worry about the radius of convergence here): 
the integrals are of the type 
\be
 \int_\vec{k} \frac{(m^2)^n}{(\omega_n^2 + {k}^2)^{n+1}}
 \;, \quad \omega^{ }_n\neq 0
 \;,
\ee
and thus the integrand remains finite for small ${k}$, i.e.,\ 
there are no infrared divergences. For the small-$n$ terms, 
ultraviolet divergences may on the other hand remain, but these 
are taken care of by the regularization.

More explicitly, we obtain 
\ba
 I'(m,T) & = & 
 T \sum_{\omega_n'} \int \! \frac{{\rm d}^d\vec{k}}{(2\pi)^d} 
 \frac{1}{\omega_n^2 + {k}^2 + m^2}
 \nn & \stackrel{\rmi{Taylor}}{=} & 
 2 T \sum_{n=1}^{\infty} 
 \int \! \frac{{\rm d}^d\vec{k}}{(2\pi)^d} 
 \sum_{l=0}^{\infty} (-1)^l \frac{m^{2l}}{[(2\pi n T)^2 + {k}^2]^{l+1}}
 \nn & \stackrel{\rmi{\nr{fmdA}}}{=}&
 2 T \sum_{n=1}^{\infty} \sum_{l=0}^{\infty}
 (-1)^l m^{2l} 
 \frac{1}{(4\pi)^{\fr{d}2}}
 \frac{\Gamma(l+1- \frac{d}{2})}{\Gamma(l+1)} 
 \frac{1}{(2\pi nT)^{2l + 2 - d}} 
 \nn & = & 
 \frac{2 T}{(4\pi)^{\fr{d}2}(2\pi T)^{2-d}}
 \sum_{l=0}^{\infty} 
 \biggl[ \frac{-m^2}{(2\pi T)^2} \biggr]^l
 \frac{\Gamma(l+1- \frac{d}{2})}{\Gamma(l+1)} 
 \, \zeta(2l+2-d)
 \;, \la{IpmT}
\ea
where in the last step
we interchanged the orders of the two summations, 
and identified the sum over $n$ as a Riemann zeta function, 
$\zeta(s) \equiv \sum_{n=1}^{\infty} n^{-s}$.
Some properties of $\zeta(s)$ 
are summarized in appendix A below.

For the sake of illustration, let us work out 
the terms $l=0,1,2$ of the above sum explicitly. 
For $d=3-2\epsilon$, the order
$l=0$ requires evaluating $\Gamma(-\frac{1}{2}+\epsilon)$ 
and $\zeta(-1+2\epsilon)$;
$l=1$ requires evaluating $\Gamma(\frac{1}{2}+\epsilon)$ 
and $\zeta(1+2\epsilon)$; and
$l=2$ requires evaluating $\Gamma(\frac{3}{2}+\epsilon)$ 
and $\zeta(3+2\epsilon)$. 
Applying results listed in appendix A of this section, 
a straightforward computation 
(cf.\ appendix B for intermediate steps) yields
\be
 I'(m,T) = 
 \frac{T^2}{12} - \frac{2 m^2\mu^{-2\epsilon}}{(4\pi)^2} 
 \biggl[
 \frac{1}{2 \epsilon} + \ln\biggl( \frac{\bmu e^{\gammaE}}{4\pi T}\biggr) 
 \biggr] 
 + \frac{2 m^4 \zeta(3)}{(4\pi)^4 T^2} + 
 \rmO\biggl( \frac{m^6}{T^4} \biggr) + \rmO(\epsilon)
 \;. \la{IpmT_res}
\ee
Adding to this the zero-mode contribution from \eq\nr{In0_res}, we get 
\be
 I(m,T) = 
 \frac{T^2}{12} - \frac{mT}{4\pi}
 - \frac{2 m^2 \mu^{-2\epsilon}}{(4\pi)^2} 
 \biggl[
 \frac{1}{2 \epsilon} + \ln\biggl( \frac{\bmu e^{\gammaE}}{4\pi T}\biggr) 
 \biggr] 
 + \frac{2 m^4 \zeta(3)}{(4\pi)^4 T^2} + 
 \rmO\biggl( \frac{m^6}{T^4} \biggr) + \rmO(\epsilon)
 \;. \la{ImT_res}
\ee
Subtracting \eq\nr{I0m_res} to isolate the $T$-dependent 
part finally yields
\be
 I^{ }_T(m) = 
 \frac{T^2}{12} - \frac{mT}{4\pi}
 - \frac{2 m^2}{(4\pi)^2} 
 \biggl[
  \ln\biggl( \frac{m e^{\gammaE}}{4\pi T}\biggr) - \frac{1}{2} 
 \biggr] 
 + \frac{2 m^4 \zeta(3)}{(4\pi)^4 T^2} + 
 \rmO\biggl( \frac{m^6}{T^4} \biggr) + \rmO(\epsilon)
 \;. \la{ITm_res}
\ee
Note how the divergences and $\bmu$ have cancelled in our result 
for $I^{ }_T(m)$, as must be the case.

To transport the above results to various versions
of the function $J$, we make use of \eqs\nr{JT0} and \nr{JI_rel}. 
{}From \eq\nr{IpmT_res}, we first get 
\be
 J'(m,T)
 = -\frac{\pi^2 T^4}{90}
 + \frac{m^2 T^2}{24}
 - \frac{m^4 \mu^{-2\epsilon}}{2(4\pi)^2}
 \biggl[ \frac{1}{2\epsilon} + 
  \ln\biggl( \frac{\bmu e^{\gammaE}}{4\pi T} \biggr)
 \biggr]
 + \frac{m^6\zeta(3)}{3 (4\pi)^4 T^2}
 + \rmO\biggl( \frac{m^8}{T^4} \biggr) + \rmO(\epsilon)
 \;. \la{JpmT_res}
\ee
Adding the zero-mode contribution from \eq\nr{Jn0_res} then leads to 
\be
 J(m,T)
 = -\frac{\pi^2 T^4}{90}
 + \frac{m^2 T^2}{24}
 - \frac{m^3 T}{12\pi}
 - \frac{m^4 \mu^{-2\epsilon}}{2(4\pi)^2}
 \biggl[ \frac{1}{2\epsilon} + 
  \ln\biggl( \frac{\bmu e^{\gammaE}}{4\pi T} \biggr)
 \biggr]
 + \frac{m^6\zeta(3)}{3 (4\pi)^4 T^2}
 + \rmO\biggl( \frac{m^8}{T^4} \biggr) + \rmO(\epsilon)
 \;. \la{JmT_res}
\ee
Subtracting the zero-temperature part, $J_0^{ }(m)$, of \eq\nr{J0m_res} 
leads to the expansion for $J^{ }_T(m)$ that was given in \eq\nr{JTm_res}.
We may again note the cancellation of $1/\epsilon$ and $\bmu$ in $J^{ }_T(m)$.
The numerical convergence of the high-temperature expansion 
is illustrated in \fig\ref{fig:exe3} on p.~\pageref{fig:exe3}. 


\subsection*{Appendix A: Properties of the Euler $\Gamma$ 
and Riemann $\zeta$ functions}
\la{gamma_zeta}

\index{Euler gamma function}

$\fbox{$\displaystyle \Gamma(s)$}$

The function $\Gamma(s)$ is to be viewed as a complex-valued 
function of a complex variable $s$. For $\re(s) > 0$, it can 
be defined as 
\be
 \Gamma(s) \equiv \int_0^{\infty} \! {\rm d}x \, x^{s-1} e^{-x}
 \;, \la{Gamma_def1}
\ee
whereas for $\re(s)\le 0$, the values can be obtained through
the iterative use of the relation 
\be
 \Gamma(s) = \frac{\Gamma(s+1)}{s}
 \;. \la{Gamma_def2}
\ee
On the real axis, $\Gamma(s)$ is regular at $s=1$; 
as a consequence of \eq\nr{Gamma_def2}, it then has first-order
poles at $s=0,-1,-2,...$~.
A useful relation, reflecting the pole structure, reads
$
  \Gamma(s)\Gamma(1-s) = \frac{\pi}{\sin(\pi s)} 
$.

In practical applications, the argument $s$ is typically
close to an integer or a half-integer. In the former
case, we can use \eq\nr{Gamma_def2} to relate the desired value
to the behavior of $\Gamma(s)$ and its derivatives around $s=1$, 
which can in turn be worked out from the convergent integral 
representation in \eq\nr{Gamma_def1}. In particular, 
\be
 \Gamma(1) = 1
 \;, \quad
 \Gamma'(1) = 
 - \gammaE
 \;, 
\ee
where $\gammaE$ is the Euler constant, 
$\gammaE = 0.577215664901...$~.
In the latter case, 
we can similarly use \eq\nr{Gamma_def2} to relate the desired value
to $\Gamma(s)$ and its derivatives around $s=\frac{1}{2}$, 
which can again be worked out from the integral 
representation in \eq\nr{Gamma_def1}, producing 
\be
 \Gamma\bigl(\tfr12\bigr) = \sqrt{\pi}
 \;, \quad
 \Gamma'\bigl(\tfr12\bigr)
 = \sqrt{\pi} (-\gammaE - 2 \ln 2)
 \;. 
\ee
The values required for \eq\nr{IpmT_res} thus become
\ba
 \Gamma\bigl( -\tfr12 + \epsilon \bigr) 
 & = & - 2 \sqrt{\pi} + \rmO(\epsilon)
 \;, \la{Gamma_specs1} \\ 
 \Gamma\bigl( \tfr12 + \epsilon \bigr) 
 & = & 
 \sqrt{\pi} \Bigl[ 1 - \epsilon ( \gammaE + 2 \ln 2) + 
 \rmO(\epsilon^2)
 \Bigr]
 \;, \la{Gamma_specs2} \\ 
 \Gamma\bigl( \tfr32 + \epsilon \bigr) 
 & = & 
 \frac{\sqrt{\pi}}{2} + \rmO(\epsilon)
 \;. \la{Gamma_specs3}
\ea
We have gone one order higher in the middle expansion, because
this function is multiplied by $1/\epsilon$ 
in the result (cf.\ \eq\nr{zeta_specs2}).

$\fbox{$\displaystyle \zeta(s)$}$

\index{Riemann zeta function}

The function $\zeta(s)$ is also to be viewed as a complex-valued 
function of a complex argument $s$. For $\re(s) > 1$, it can 
be defined as 
\be
 \zeta(s) = \sum_{n=1}^{\infty} n^{-s}
 = \frac{1}{\Gamma(s)}
 \int_0^{\infty} \! \frac{ {\rm d}x \, x^{s-1}}{ e^{x} - 1}
 \;, \la{zeta_def1}
\ee
where the equivalence of the two forms can 
be seen by writing 
$
 1/(e^x - 1) = e^{-x}/(1-e^{-x}) = \sum_{n=1}^{\infty} e^{-n x}
$, 
and using the definition of the $\Gamma$-function in \eq\nr{Gamma_def1}. 
Some remarkable properties of $\zeta(s)$ follow from the fact that
by writing 
\be
 \frac{1}{e^x-1} = \frac{1}{(e^{x/2} - 1)(e^{x/2} + 1)}
 = \frac{1}{2}\biggl[ \frac{1}{e^{x/2} - 1}  - \frac{1}{e^{x/2} + 1} 
 \biggr]
 \;, 
\ee
and then substituting integration variables through $x\to 2 x$, 
we can find an alternative integral representation, 
\be
 \zeta(s) 
 = \frac{1}{(1-2^{1-s})\Gamma(s)}
 \int_0^{\infty} \! \frac{ {\rm d}x \, x^{s-1}}{ e^{x} + 1}
 \;, \la{zeta_def2}
\ee
defined for $\re(s) > 0$, $s\neq 1$.  
Even though the integral here clearly diverges at $s\to 0$, 
the function $\Gamma(s)$
also diverges at the same point, making $\zeta(s)$
regular around origin:
\ba
 \zeta(0) & = & -\tfr12 
 \;, \la{zeta0} \\
 \zeta'(0) & = & 
 - \tfr12 \ln(2\pi) 
 \;. \la{zetap0} 
\ea
Finally, for $\re(s) \le 0$, an analytic 
continuation is obtained through the relation
\be
 \zeta(s) = 
 \underbrace{ \frac{(2\pi)^s}{\Gamma(\frac{s}{2})\Gamma(1-\frac{s}{2})} }_
 { 
   2^s \pi^{s-1} \sin ( \frac{\pi s}{2} )
 }
 \Gamma(1-s) \zeta(1-s)
 \;. \la{zeta_def3}
\ee

On the real axis, $\zeta(s)$ has a pole only at $s=1$.
Its values at even arguments are ``easy''; in fact, at even 
negative integers, \eq\nr{zeta_def3} implies that 
\be
 \zeta(-2n) = 0 
 \;, \quad n = 1,2,3, \ldots
 \;,
\ee
whereas at positive even integers the values can be related 
to the Bernoulli numbers, 
\be
 \zeta(2) = \frac{\pi^2}{6}
 \;, \quad
 \zeta(4) = \frac{\pi^4}{90}
 \;, \ldots
 \;. 
\ee
Negative odd integers can be related to positive even ones
through \eq\nr{zeta_def3}, which also allows us to determine
the behaviour of the function around the pole at $s=1$. 
In contrast, odd positive
integers larger than unity, i.e. $s=3,5,...$, yield new transcendental 
numbers.

The values required in \eq\nr{IpmT_res} become
\ba
 \zeta( - 1 + 2 \epsilon ) 
 & = & - \frac{1}{2\pi^2} \Gamma(2) \zeta(2) + \rmO(\epsilon)
 = - \frac{1}{12} + \rmO(\epsilon)
 \;, \la{zeta_specs1} \\ 
 \zeta ( 1 + 2 \epsilon ) 
 & = & 
 2^{1+ 2\epsilon} \pi^{2\epsilon}
 \Bigl[ 
   \sin\Bigl( \frac{\pi}{2} \Bigr) + \pi \epsilon
   \cos\Bigl( \frac{\pi}{2} \Bigr)
 \Bigr]
 \biggl( -\frac{1}{2\epsilon} \biggr)
 \Gamma(1-2\epsilon) \zeta(-2\epsilon)
 \nn & = & 
 2 (1 + 2\epsilon \ln 2)
   (1 + 2 \epsilon \ln \pi)
   \biggl(-\frac{1}{2\epsilon} \biggr)
   ( 1 + 2 \epsilon \gammaE ) 
   \bigl( -\tfr12 \bigr)
   ( 1 - 2 \epsilon \ln 2\pi) + \rmO(\epsilon)
 \nn & = & 
 \frac{1}{2\epsilon} + \gammaE + 
 \rmO(\epsilon)
 \;, \la{zeta_specs2} \\[2mm] 
 \zeta( 3 + 2\epsilon ) 
 & = & 
 \zeta(3) +  \rmO(\epsilon)
 \approx 1.2020569031... + \rmO(\epsilon)
 \;, \la{zeta_specs3}
\ea
where in the first two cases we made use of \eq\nr{zeta_def3}, 
and in the second also of \eqs\nr{zeta0} and \nr{zetap0}.


\subsection*{Appendix B: Numerical convergence}
\la{exe3}

We complete here  the derivation of \eq\nr{IpmT_res}, 
and sketch the regimes
where the low and high-temperature expansions
are numerically accurate by 
inspecting $J^{ }_T(m)$ from \eq\nr{JTm_res}. 

\index{Thermal sums: boson loop}

First of all, 
for the term $l=0$ in \eq\nr{IpmT}, we make use 
of the results of \eqs\nr{Gamma_specs1}, \nr{zeta_specs1}:
\be
 \left. I'(m,T) \right|^{ }_{l = 0}
 = 
 \frac{2 T }{(4\pi)^{3/2}} (2\pi T) \frac{-2\sqrt{\pi}}{1}
 \biggl( -\frac{1}{12} \biggr) + \rmO(\epsilon)
 = \frac{T^2}{12} + \rmO(\epsilon)
 \;. 
\ee
For the term $l=1$, we on the other hand insert 
the values of \eqs\nr{Gamma_specs2} and \nr{zeta_specs2}:
\ba
 \left. I'(m,T) \right|^{ }_{l = 1}
 \hspace*{-0.5cm} & = & \hspace*{-0.5cm}
  2 T \frac{(4\pi)^{\epsilon}}{(4\pi)^{3/2}} 
 (2\pi T)^{1-2\epsilon}
 \biggl[ \frac{-m^2}{(2\pi T)^2}\biggr]
 \sqrt{\pi} 
 \Bigl[ 1 -\epsilon (\gammaE + 2 \ln 2) \Bigr] 
 \frac{1}{2\epsilon} (1 + 2 \epsilon \gammaE ) + \rmO(\epsilon)
 \nn & \stackrel{1=\mu^{-2\epsilon}\mu^{2\epsilon}}{=} & 
 \hspace*{-0.5cm}
 - \frac{m^2\mu^{-2\epsilon}}{(4\pi)^2} 
 \biggl\{ 
    \frac{1}{\epsilon} + \ln\frac{\mu^2}{T^2} + \ln(4\pi) - \gammaE
 + 2 [\gammaE - \ln(4\pi)]
 \biggr\} + \rmO(\epsilon)
 \nn & \stackrel{\rmi{\nr{msbar}}}{=} &  \hspace*{-0.5cm}
 - \frac{m^2\mu^{-2\epsilon}}{(4\pi)^2} 
 \biggl\{ 
    \frac{1}{\epsilon} + \ln\frac{\bmu^2}{T^2} 
 + 2 \ln \Bigl( \frac{e^{\gammaE}}{4\pi} \Bigr)
 \biggr\} 
 + \rmO(\epsilon)
 \;.
\ea
Finally, for the term $l=2$, we make use 
of \eqs\nr{Gamma_specs3} and \nr{zeta_specs3}, giving
\be
 \left. I'(m,T) \right|^{ }_{l = 2}
 = 
 \frac{2 T }{(4\pi)^{3/2}} (2\pi T)
 \frac{m^4}{(2\pi T)^4} \frac{\frac{1}{2}\sqrt{\pi}}{2}
 \zeta(3) + \rmO(\epsilon)
 = \frac{2 m^4 \zeta(3)}{(4\pi)^4 T^2} + \rmO(\epsilon)
 \;. 
\ee

For the numerical evaluation of $J^{ }_T(m)$, 
we again denote $y\equiv m/T$ and inspect the function
\be
 \mathcal{J}(y) \equiv \frac{J^{ }_T(m)}{T^4}
 =  \frac{1}{2\pi^2} 
 \int_0^{\infty} \! {\rm d}x\, x^2 \, 
 \ln \Bigl(
   1 - e^{-\sqrt{x^2 + y^2}} 
 \Bigr)
 \;. \la{calJy}
\ee
We contrast this with the low-temperature results from 
footnote~\ref{bessel1} and from
\eq\nr{JTm_final},
\be
 \mathcal{J}(y) 
 \; \stackrel{y \gsim 1}{\approx} \; 
 - \frac{y^2 K^{ }_2(y) }{2\pi^2}  
 \; \stackrel{y \gg 1}{\approx} \;
 - \biggl( \frac{y}{2\pi} \biggr)^{\fr32} e^{-y} 
 \;, \la{calJlow}
\ee 
as well as with the high-temperature expansion from 
\eq\nr{JTm_res}, 
\be
 \mathcal{J}(y) \stackrel{y \ll 1}{\approx}
  = -\frac{\pi^2}{90}
 + \frac{y^2}{24}
 - \frac{y^3}{12\pi}
 - \frac{y^4}{2(4\pi)^2}
 \biggl[ 
  \ln\biggl( \frac{y e^{\gammaE}}{4\pi} \biggr) - \fr34
 \biggr]
 + \frac{y^6\zeta(3)}{3 (4\pi)^4}  
 \;. \la{calJhigh}
\ee
The result of the comparison is shown in \fig\ref{fig:exe3}. 
We observe that if we keep terms up to $y^6$ in the high-temperature
expansion, its numerical convergence is good for
$y\lsim 3$. On the other hand, the low-temperature 
expansion with power corrections 
converges reasonably well for $y \gsim 6$. 
In between, either 
a numerical evaluation or the 
low-temperature expansion in terms of Bessel functions is necessary. 
It should be stressed that these statements are to be understood 
in a pragmatic sense, rather than as mathematically defined convergence radii.

\begin{figure}[t]

\vspace*{1cm}

\centerline{
  \epsfysize=7.5cm\epsfbox{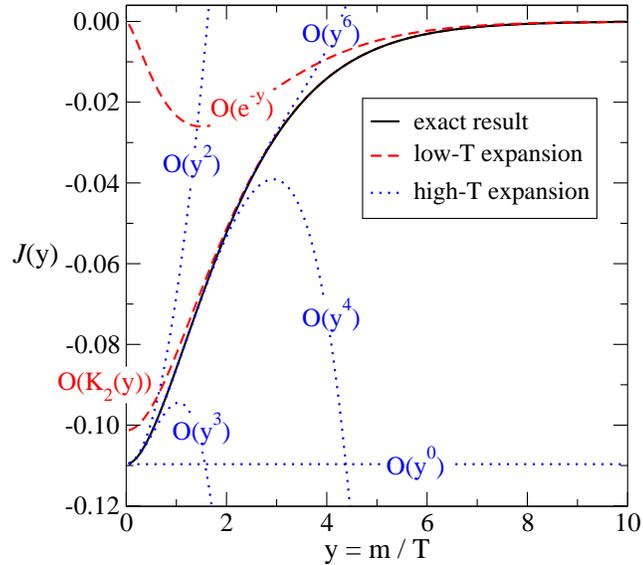} 
}


\caption[a]{\small
The behavior of $\mathcal{J}(y)$ and its various
approximations. Shown are the exact numerical result from \eq\nr{calJy},
the two 
low-temperature approximations from \eq\nr{calJlow} (with exponential
and powerlike corrections, respectively), as well as 
the high-temperature expansion from \eq\nr{calJhigh}.}

\la{fig:exe3}
\end{figure}

%

\newpage


\newpage 

\section{Interacting scalar fields}
\la{se:int}

\paragraph{Abstract:} The key concepts of a perturbative or weak-coupling
expansion are introduced in the context of evaluating the imaginary-time
path integral representation for the partition function of an interacting
scalar field. The issues of ultraviolet
and infrared divergences are brought up. These problems are cured through
renormalization and resummation, respectively. 

\paragraph{Keywords:} 

Weak-coupling expansion, Wick's theorem, propagator, contraction, 
ultraviolet and infrared divergences, renormalization, resummation, 
ring diagrams. 


\subsection{Principles of the weak-coupling expansion}
\la{se:wce}

\index{Weak-coupling expansion: scalar field}

\index{Euclidean Lagrangian: scalar field}

In order to move from a free to an interacting theory, 
we now include a quartic term in the potential in \eq\nr{SM_SFT},
\be
 V(\phi) \;\equiv\; \frac{1}{2} m^2 \phi^2 + \frac{1}{4} \lambda \phi^4 \;,
 \la{Vphi}
\ee
where $\lambda > 0$ is a dimensionless coupling constant. 
Thereby the Minkowskian and Euclidean Lagrangians become
\ba
 \mathcal{L}^{ }_\iM & = & 
 \frac{1}{2} \partial^\mu \phi\, \partial^{ }_\mu \phi
 - \frac{1}{2} m^2 \phi^2 
 - \frac{1}{4} \lambda \phi^4 
 \;, \\
 {L}^{ }_\iE & = & 
 \frac{1}{2} \partial^{ }_\mu \phi\, \partial^{ }_\mu \phi
 + \frac{1}{2} m^2 \phi^2 
 + \frac{1}{4} \lambda \phi^4 
 \;,
\ea
where repeated indices are summed over, 
irrespective of whether they are up and down  or all down. The case with 
all indices down implies the use of Euclidean metric
like in \eq\nr{LE_expl}. 

In the presence of $\lambda > 0$,  it is no longer possible
to determine the partition function of the system exactly, neither
in the canonical formalism nor through a path integral approach. We therefore
need to develop approximation schemes, which could in principle be 
either analytic or numerical. In the following we restrict our attention
to the simplest analytic procedure which, as we will see, already 
teaches us a lot about the nature of the system. 

In a weak-coupling expansion, the theory is solved by formally assuming
that $\lambda \ll 1$, and by expressing the result for the observable 
in question as a (generalized) Taylor series in $\lambda$.
The physical observable that we are interested in is the partition 
function defined according 
to \eq\nr{Z_SFT}. Denoting the free and interacting parts
of the Euclidean action by 
\ba
 S_0^{ } & \equiv &
 \int_0^{\beta} \! {\rm d}\tau \int_\vec{x}
 \biggl[
  \frac{1}{2} \partial^{ }_\mu\phi\,\partial^{ }_\mu\phi
 + \frac{1}{2} m^2 \phi^2 
 \biggr]
 \;, \la{S_0} \\
 S^{ }_\iI & \equiv &
 \lambda 
 \int_0^{\beta} \! {\rm d}\tau \int_\vec{x}
 \biggl[ 
  \frac{1}{4} \phi^4
 \biggr] 
 \;, \la{S_I}
\ea
the partition function can be written in the form 
\ba
 \mathcal{Z}^\rmii{SFT}(T) & = &
 C \, \int \! \mathcal{D}\phi \, \exp\bigl({-S_0^{ } - S^{ }_\iI}\bigr)
 \nn 
 & = &           
 C \, \int \! \mathcal{D}\phi \, e^{-S_0^{ }}
 \biggl[ 1 - S^{ }_\iI + \frac{1}{2} S_\iI^2
 - \frac{1}{6} S_\iI^3 + \ldots \biggr]  
 \nn
 & = & 
 \mathcal{Z}^\rmii{SFT}_{(0)}
 \biggl[ 1 - \langle S^{ }_\iI \rangle_0^{ }+ 
 \frac{1}{2} \langle S_\iI^2 \rangle_0^{ } - \frac{1}{6} 
 \langle S_\iI^3 \rangle_0^{ } + \ldots \biggr]  
 \;. \la{Z_SFT_1}
\ea
Here, 
\be
 \mathcal{Z}^\rmii{SFT}_{(0)}
 \equiv  
 C \, \int \! \mathcal{D}\phi \, e^{-S_0^{ }}
\ee
is the free partition function determined
in \se\ref{se:free_sft}, and the expectation 
value $\langle \cdots \rangle_0^{ }$ 
is defined as
\be \index{Free energy density: scalar field}
 \langle \cdots \rangle_0^{ } \;\equiv\;
 \frac{ \int \! \mathcal{D} \phi \, [\cdots] \exp(-S_0^{ })}
      { \int \! \mathcal{D} \phi \, \exp(-S_0^{ })}
 \;. 
\ee 
With this result, the free energy density reads
\ba
 \frac{F^\rmii{SFT}(T,V)}{V} & = &  -\frac{T}{V}\, \ln \mathcal{Z}^\rmii{SFT} 
 \nn 
 & = &
 \frac{F^\rmii{SFT}_{(0)}}{V}
 - \frac{T}{V}\, \ln 
 \Bigl( 
   1 - \langle S^{ }_\iI \rangle_0^{ }+ 
 \frac{1}{2} \langle S_\iI^2 \rangle_0^{ } - \frac{1}{6} 
 \langle S_\iI^3 \rangle_0^{ } + \ldots
 \Bigr) 
 \la{ante_ln} \\
 & = &
 \frac{F^\rmii{SFT}_{(0)}}{V}
 - \frac{T}{V}\,
 \biggl\{
  - \bigl\langle S^{ }_\iI \bigr\rangle_0^{ } 
 + \frac{1}{2} 
 \biggl[
    \bigl\langle S_\iI^2 \bigr\rangle_0^{ }
  - \bigl\langle S^{ }_\iI \bigr\rangle_0^2 
 \biggr] 
   \nn & & \hspace*{2cm}
 - \frac{1}{6}
 \biggl[
    \bigl\langle S_\iI^3 \bigr\rangle_0^{ }
  - 3 \bigl\langle S^{ }_\iI \bigr\rangle_0^{ } 
  \bigl\langle S_\iI^2 \bigr\rangle_0^{ } 
  + 2 \bigl\langle S^{ }_\iI \bigr\rangle_0^3
 \biggr] 
  + \ldots
 \biggr\}
 \;, \la{f_sft_3}
\ea
where we have Taylor-expanded the logarithm, 
$
 \ln(1-x) = -x - x^2/2 - x^3/3 + ... \,
$. 
The first term, 
${F^\rmii{SFT}_{(0)}}/{V}$, is given in \eq\nr{f_sft_2}, whereas 
the subsequent terms correspond to corrections of orders 
$\rmO(\lambda)$, 
$\rmO(\lambda^2)$, 
and $\rmO(\lambda^3)$, respectively. 
As we will see, the combinations that appear within 
the square brackets in \eq\nr{f_sft_3} have a specific 
significance: \eq\nr{f_sft_3} is {\em simpler} 
than \eq\nr{Z_SFT_1}! 

For future reference, let us denote 
\be
 f(T) \;\equiv\; \lim_{V\to\infty} \frac{F(T,V)}{V}
 \;, 
\ee
where we have dropped the superscript ``SFT'' for simplicity. 
With this definition \eq\nr{f_sft_3} can be compactly represented  
by the formula
$
 f = f^{ }_{(0)} + f^{ }_{(\ge 1)}
$, 
where
\ba
 f^{ }_{(\ge 1)}(T) & = & - \frac{T}{V} 
 \,\Bigl\langle \exp(-S^{ }_\iI) - 1 \Bigr\rangle^{ }_{0,\rmi{c}}
 \nn & = & 
 \Bigl\langle
 S^{ }_\iI - \frac{1}{2} S_\iI^2 + \ldots 
 \Bigr\rangle^{ }_{0,\rmi{c},\,\rmi{drop overall $\int_X$}}
 \;, \la{compact_rule}
\ea
where the subscript $(...)^{ }_\rmi{c}$ refers to ``connected''
contractions, the precise meaning of which is discussed momentarily, 
and an ``overall $\int_X$'' is dropped because it cancels
against the prefactor $T/V$. 

Inserting \eq\nr{S_I} into the various terms of \eq\nr{f_sft_3}, 
we are led to evaluate expectation values of the type 
\be
 \langle \phi(X^{ }_1) \phi(X^{ }_2) \ldots \phi(X^{ }_n) \rangle_0^{ }
 \;. 
\ee
These can be reduced to products of free 2-point correlators, 
$
 \langle \phi(X^{ }_k) \phi(X^{ }_l) \rangle_0^{ }
$, 
through the Wick's theorem, as we now discuss.


\subsection*{Wick's theorem}
\la{se:wick}

\index{Wick's theorem}

Wick's theorem states that free (Gaussian) expectation values
of any number of integration variables can be reduced to products
of 2-point correlators, according to 
\be
 \langle \phi(X^{ }_1) \phi(X^{ }_2) \ldots
 \phi(X^{ }_{n-1}) \phi(X^{ }_n) \rangle_0^{ }
 =
 \sum_\rmi{all combinations} 
 \langle \phi(X^{ }_1) \phi(X^{ }_2) \rangle_0^{ }
   \cdots
 \langle \phi(X^{ }_{n-1}) \phi(X^{ }_n) \rangle_0^{ }
 \;. \la{Wick}
\ee
Before applying this to the terms of \eq\nr{f_sft_3}, we briefly 
recall how the theorem can be derived with 
(path) integration techniques. 

Let us assume that we can discretise spacetime such that the 
coordinates $X$ only take a finite number of values, 
which in particular requires 
the volume to be finite. Then we can collect the 
values $\phi(X), \forall X$, into a single vector $v$, and subsequently
write the free action in the form $S_0^{ } = \frac{1}{2} v^T\! A\, v$, 
where $A$ is a matrix. Here, we assume that $A^{-1}$ exists 
and that $A$ is symmetric, i.e.~$A^T = A$; 
then it also follows that $(A^{-1})^T = A^{-1}$. 

The trick allowing us to evaluate integrals weighted by $\exp(-S_0^{ })$
is to introduce a source vector $b$, and to take derivatives with respect
to its components. Specifically, we define
\ba
 \exp\bigl[ {W(b)} \bigr] & \equiv &
 \int \! {\rm d}v \, \exp\Bigl[ {-\frac{1}{2} 
 v^{ }_i A^{ }_{ij} v^{ }_j + b^{ }_i v^{ }_i} \Bigr]
 \nn & 
 \stackrel{v^{ }_i \to v^{ }_i + A^{-1}_{ij}b^{ }_j}{=} & 
 \exp \Bigl[ {\frac{1}{2} b^{ }_i A^{-1}_{ij} b^{ }_j } \Bigr]  
 \int \! {\rm d}v \, 
 \exp\Bigl[ {-\frac{1}{2} v^{ }_i A^{ }_{ij} v^{ }_j} \Bigr]   
 \;,
\ea
where we made a substitution of integration variables at the second equality. 
We then obtain 
\ba
 \langle 
  v^{ }_k v^{ }_l ... v^{ }_n
 \rangle_0^{ }
 & = & 
 \frac{
 \int \! {\rm d}v \, 
 (v^{ }_k v^{ }_l ... v^{ }_n ) 
 \exp\bigl[ -\frac{1}{2} v^{ }_i A^{ }_{ij} v^{ }_j  \bigr] 
 }{
 \int \! {\rm d}v \, 
 \exp\bigl[ -\frac{1}{2} v^{ }_i A^{ }_{ij} v^{ }_j  \bigr] 
 }
 \nn & = &
 \frac{\Bigl\{  
 \frac{{\rm d}}{{\rm d} b^{ }_k}
 \frac{{\rm d}}{{\rm d} b^{ }_l} ... 
 \frac{{\rm d}}{{\rm d} b^{ }_n} \exp\bigl[ W(b) \bigr] \Bigr\}^{ }_{b=0}
 }{
 \exp\bigl[ W(0) \bigr]
 }
 \nn & = &
 \Bigl\{  
 \frac{{\rm d}}{{\rm d} b^{ }_k}
 \frac{{\rm d}}{{\rm d} b^{ }_l} ... 
 \frac{{\rm d}}{{\rm d} b^{ }_n} 
 \exp\Bigl[ \frac{1}{2} b^{ }_i A^{-1}_{ij} b^{ }_j \Bigr] 
 \Bigr\}^{ }_{b=0}
 \nn & = &
 \biggl\{  
 \frac{{\rm d}}{{\rm d} b^{ }_k}
 \frac{{\rm d}}{{\rm d} b^{ }_l} ... 
 \frac{{\rm d}}{{\rm d} b^{ }_n} 
 \Bigl[ 1 +  \frac{1}{2} b^{ }_i A^{-1}_{ij} b^{ }_j
 + \frac{1}{2} \Bigl( \frac{1}{2} \Bigr)^2 
  b^{ }_i A^{-1}_{ij} b^{ }_j \, b^{ }_r A^{-1}_{rs} b^{ }_s + \ldots
  \Bigr] 
 \biggr\}^{ }_{b=0}
 \;. \hspace*{1cm}
 \la{wick_deriv}
\ea
Taking the derivatives in \eq\nr{wick_deriv}, we observe that: 
\bi
\item
$
 \langle 1 \rangle_0^{ } = 1. 
$

\item
If there is an odd number of components of $v$ in the expectation value, 
the result is zero. 

\item
$
 \langle v^{ }_k v^{ }_l \rangle_0^{ } = A^{-1}_{kl}
$.

\item
$
 \langle 
  v^{ }_k v^{ }_l v^{ }_m v^{ }_n 
 \rangle_0^{ }
 =   
 A^{-1}_{kl} A^{-1}_{mn}
 +  
 A^{-1}_{km} A^{-1}_{ln}
 +  
 A^{-1}_{kn} A^{-1}_{lm}
$ \\ 
\hspace*{1.9cm}$
   =  
  \langle v^{ }_k v^{ }_l \rangle_0^{ }
  \langle v^{ }_m v^{ }_n \rangle_0^{ }
+
  \langle v^{ }_k v^{ }_m \rangle_0^{ }
  \langle v^{ }_l v^{ }_n \rangle_0^{ }
+ 
  \langle v^{ }_k v^{ }_n \rangle_0^{ }
  \langle v^{ }_l v^{ }_m \rangle_0^{ }
$~.

\item
At higher orders, we obtain a discretized 
version of \eq\nr{Wick}.

\item Since all the operations 
were purely combinatorial, removing the discretization
does not modify the result, so that \eq\nr{Wick} holds also in the
infinite volume and continuum limits. 

\ei

Let us now use \eq\nr{Wick} in connection with \eq\nr{f_sft_3}.
{From} \eqs\nr{f_sft_2}, \nr{JmT_1} and \nr{f_sft_3}, 
we read off the familiar leading-order
result, 
\be
 f^{ }_{(0)}(T) = J(m,T)
 \;. \la{f0}
\ee
At the first order, linear in $\lambda$, we on the other hand get
\be
 f^{ }_{(1)}(T) = \lim_{V\to\infty}
 \frac{T}{V}\, \langle S^{ }_\iI \rangle_0^{ }
 = \lim_{V\to\infty} \frac{T}{V} 
 \int_0^\beta \! {\rm d}\tau \int_\vec{x} \,
 \frac{\lambda}{4}\,
 \langle \phi(X) \phi(X) \phi(X) \phi(X) \rangle_0^{ }
 \;, \la{f1_pre} 
\ee
where we can now use Wick's theorem. 
Due to translational invariance, 
$
 \langle \phi(X) \phi(Y) \rangle_0^{ }
$
can only depend on $X-Y$, 
so the spacetime integral becomes trivial, and we obtain 
\be
 f^{ }_{(1)}(T) = 
 \fr34 \lambda \,
 \langle \phi(0) \phi(0) \rangle_0^{ }
 \langle \phi(0) \phi(0) \rangle_0^{ }
 \;. \la{f1}
\ee
Finally, at the second order, we get 
\ba
 f^{ }_{(2)}(T) & = &
    \lim_{V\to\infty}\biggl\{  -\frac{T}{2V}
 \Bigl[ 
   \langle S_\iI^2 \rangle_0^{ } - \langle S^{ }_\iI \rangle_0^2 
 \Bigr] \biggr\}
 \nn & = & 
  \lim_{V\to\infty}\biggl\{  -\frac{T}{2V} \biggl[ 
  \int_{X,Y} 
 \Bigl( \frac{\lambda}{4}\Bigr)^2 
 \langle 
  \phi(X)
  \phi(X)
  \phi(X)
  \phi(X) \, 
  \phi(Y)
  \phi(Y)
  \phi(Y)
  \phi(Y)
 \rangle_0^{ }
 \nn & & \hspace*{1cm} - 
 \int_X \frac{\lambda}{4}\,
 \langle 
  \phi(X)
  \phi(X)
  \phi(X)
  \phi(X) 
 \rangle_0^{ } \, 
 \int_Y \frac{\lambda}{4}\,
 \langle 
  \phi(Y)
  \phi(Y)
  \phi(Y)
  \phi(Y) 
 \rangle_0^{ } 
 \biggr]\biggr\} 
 \;, \hspace*{5mm} \la{pre_f2} 
\ea
where we have again denoted (cf.\ \eq\nr{intX})
\be
 \int_X \equiv 
 \int_0^\beta \! {\rm d}\tau \int_V \! {\rm d}^d\vec{x} 
 \;. \la{def:int_x}
\ee

Upon carrying out the contractions in \eq\nr{pre_f2} 
according to Wick's theorem, 
the role of the ``subtraction term'', i.e.~the second one in \eq\nr{pre_f2}, 
becomes clear: it cancels all {\em disconnected} contractions 
where all fields at point $X$ are 
contracted with other fields at the same point. 
In other words, the combination 
in \eq\nr{pre_f2} amounts to taking into account only the
{\em connected} contractions; this is the meaning of 
the subscript c in \eq\nr{compact_rule}. 
This combinatorial 
effect is caused by the logarithm in \eq\nr{f_sft_3}, i.e., 
by going from the partition function to the free energy.

\index{Wick contractions}

As far as the connected contractions go, 
we obtain through a (repeated) use of Wick's theorem: 
\ba
 && \hspace*{-1cm}
 \langle 
  \phi(X)
  \phi(X)
  \phi(X)
  \phi(X) \, 
  \phi(Y)
  \phi(Y)
  \phi(Y)
  \phi(Y)
 \rangle^{ }_{0,\rmi{c}}
 \nn & = & 
 4 \, \langle \phi(X) \phi(Y) \rangle_0^{ } \;
 \langle 
  \phi(X)
  \phi(X)
  \phi(X) \, 
  \phi(Y)
  \phi(Y)
  \phi(Y)
 \rangle^{ }_{0,\rmi{c}}
  \nn & & \; + \,  
 3 \, \langle \phi(X) \phi(X) \rangle_0^{ } \;
 \langle 
  \phi(X)
  \phi(X) \, 
  \phi(Y)
  \phi(Y)
  \phi(Y)
  \phi(Y)
 \rangle^{ }_{0,\rmi{c}}
 \nn & = &  
 4 \times 3 \,
 \langle \phi(X) \phi(Y) \rangle_0^{ } \;
 \langle \phi(X) \phi(Y) \rangle_0^{ } \;
 \langle 
  \phi(X)
  \phi(X) \, 
  \phi(Y)
  \phi(Y)
 \rangle^{ }_{0,\rmi{c}}
  \nn & & \; + \,  
 4 \times 2 \,
 \langle \phi(X) \phi(Y) \rangle_0^{ } \;
 \langle \phi(X) \phi(X) \rangle_0^{ } \;
 \langle 
  \phi(X) \, 
  \phi(Y)
  \phi(Y)
  \phi(Y)
 \rangle^{ }_{0,\rmi{c}}
 \nn & & \; + \,  
 3 \times 4 \,
 \langle \phi(X) \phi(X) \rangle_0^{ } \;
 \langle \phi(X) \phi(Y) \rangle_0^{ } \;
 \langle 
  \phi(X) \, 
  \phi(Y)
  \phi(Y)
  \phi(Y)
 \rangle^{ }_{0,\rmi{c}}
 \nn & = &  
 4 \times 3 \times 2 \, 
 \langle \phi(X) \phi(Y) \rangle_0^{ } \;
 \langle \phi(X) \phi(Y) \rangle_0^{ } \;
 \langle \phi(X) \phi(Y) \rangle_0^{ } \;
 \langle \phi(X) \phi(Y) \rangle_0^{ } 
  \nn & & \; + \, 
 (4\times 3 + 4\times 2 \times 3 + 3 \times 4 \times 3)
 \langle \phi(X) \phi(X) \rangle_0^{ } \;
 \langle \phi(X) \phi(Y) \rangle_0^{ } \;
 \langle \phi(X) \phi(Y) \rangle_0^{ } \;
 \langle \phi(Y) \phi(Y) \rangle_0^{ } 
 \;. \nn \la{pre2_f2}
\ea
Inspecting the 2-point correlators in this result, we note
that they either depend on $X-Y$, or on neither $X$ nor $Y$, the latter case 
corresponding to the contraction of fields at the same point.
Thereby one of the spacetime integrals is trivial
(just substitute $X\to X+Y$, and note that 
$\langle \phi(X+Y) \phi(Y) \rangle_0^{ }  = 
\langle \phi(X) \phi(0) \rangle_0^{ }$), and 
cancels against the factor $T/V = 1/(\beta V)$ in \eq\nr{pre_f2}. 
In total, we then have 
\be
 f^{ }_{(2)}(T) = 
 - \Bigl( \frac{\lambda}{4} \Bigr)^2
 \biggl[
   12 \int_X \{ \langle \phi(X) \phi(0) \rangle_0^{ } \} ^4 
 + 36\, \{ \langle \phi(0) \phi(0) \rangle_0^{ } \}^2
 \int_X \{ \langle \phi(X) \phi(0) \rangle_0^{ } \}^2
 \biggr]
 \;. \la{f2}
\ee
Graphically this can be represented as 
\ba
 && \hspace*{-1cm}
 \ThreeA \hspace*{-3mm} + \quad \ThreeB \qquad \;,
\ea
where solid lines denote propagators, and the vertices at which they 
cross denote spacetime points, in this case $X$ and $0$. 

We could in principle go on with the third-order terms in \eq\nr{f_sft_3}.
Again, it could be verified that the ``subtraction terms'' cancel
all disconnected contractions, so that only the connected ones
contribute to $f(T)$, and that one spacetime integral cancels
against the explicit factor $T/V$. These features are 
of general nature, and hold at any order in the weak-coupling expansion. 

In summary, Wick's theorem has allowed us to convert the terms in 
\eq\nr{f_sft_3} to various structures made of the 2-point correlator
$\langle \phi(X) \phi(0) \rangle_0^{ }$. 
We now turn to the properties of this function.


\subsection*{Propagator}

\index{Propagator: scalar field}

The 2-point correlator $\langle \phi(X) \phi(Y) \rangle_0^{ }$ 
is usually called the {\em free propagator}. Denoting
\be
 \deltabar ( P+ Q) 
 \;\equiv\; 
 \int_X e^{i ( P +  Q)\cdot X}
 = 
 \beta \delta^{ }_{ p^{ }_n+  q^{ }_n,0} \,
 (2\pi)^d \delta^{(d)} (\vec{p} + \vec{q})
 \;, \la{deltabar}
\ee
where $ P \equiv ( p^{ }_n,\vec{p})$ and 
$ p^{ }_n$ are bosonic Matsubara frequencies, and employing
the representation
\be
 \phi(X) \;\equiv\; \Tint{ P} \tilde \phi( P) \,
 e^{i  P \cdot X}
 \;, \la{tilde_phi}
\ee 
we recall from basic quantum field theory that the (Euclidean) propagator 
can be written as 
\ba
 \langle \tilde \phi( P) \tilde \phi( Q) \rangle_0^{ }
 & = &
 \deltabar( P +  Q)\, \frac{1}{ P^2 + m^2}
 \;, \la{prop_PQ} \\
 \langle \phi(X) \phi(Y) \rangle_0^{ }
 & = &
 \Tint{ P} e^{i  P\cdot (X-Y)}\, \frac{1}{ P^2 + m^2}
 \;. \la{prop_xy}  
\ea
Before inserting these expressions into \eqs\nr{f1} and \nr{f2}, we 
briefly review their derivation, 
working in a finite volume $V$ and proceeding like in \se\ref{se:p13}.

First, we insert \eq\nr{tilde_phi} into the definition
of the propagator, 
\be
 \langle \phi(X) \phi(Y) \rangle_0^{ }
 = \Tint{ P,  Q}
 e^{i  P\cdot X + i  Q\cdot Y}
 \langle \tilde \phi ( P) \tilde \phi ( Q) 
 \rangle_0^{ } 
 \;, \la{raw_prop}
\ee
as well as to the free action, $S_0^{ }$, 
\be
 S_0^{ } = \frac{1}{2} \Tint{ P} \tilde \phi(- P)
 ( P^2 + m^2 ) \tilde \phi( P) 
 = 
 \frac{1}{2} \Tint{ P} ( P^2 + m^2) |\tilde \phi( P)|^2
 \;. \la{S0_a}
\ee
Here, we may further write 
$\tilde\phi( P) = a( P) + i\, b( P)$, 
with $a(- P) = a( P)$, $b(- P) = - b ( P)$, and
subsequently note that only half of the Fourier components are independent.
We may choose these according to \eq\nr{onlyhalf}.  

Restricting the sum to the independent components, and making use 
of the symmetry properties of $a( P)$ and $b( P)$, 
\eq\nr{S0_a} becomes
\be
 S_0^{ } = \frac{T}{V} \sum_{ P^{ }_\rmi{indep.}} ( P^2 + m^2)
 [a^2( P) + b^2( P)]
 \;. \la{S0_b}
\ee
The Gaussian integral, 
\be
 \frac{\int {\rm d} x \, x^2 \exp(-c\, x^2)}
      {\int {\rm d} x \, \exp(-c\, x^2)} = \frac{1}{2c}
 \;, 
\ee
and the symmetries of $a( P)$ and $b( P)$ then imply the results
\ba
 \langle a( P)\, b ( Q) \rangle_0^{ } & = &  0
 \;, \\
 \langle a( P)\, a ( Q) \rangle_0^{ } & = & 
 (\delta^{ }_{ P,  Q} + \delta^{ }_{ P, - Q})
 \,\frac{V}{2T} \frac{1}{ P^2 + m^2}
 \;, \\
 \langle b( P)\, b ( Q) \rangle_0^{ } & = & 
 (\delta^{ }_{ P,  Q} - \delta^{ }_{ P, - Q})
 \,\frac{V}{2T} \frac{1}{ P^2 + m^2}
 \;,
\ea
where the $\delta$-functions are of the Kronecker-type. 
Using these, the momentum-space propagator becomes
\ba
 \langle \tilde \phi( P) \tilde \phi( Q) \rangle_0^{ } 
 & = & 
 \langle 
   a( P) \, a( Q)
   + i\, a( P) \, b( Q)
   + i\, b( P)\, a( Q)
   - b( P)\, b( Q)
 \rangle_0^{ } 
 \nn  & =  & 
 \delta^{ }_{ P, - Q}\, 
 \frac{V}{T}\, 
 \frac{1}{ P^2 + m^2}
 = \beta \delta^{ }_{ p^{ }_n +  q^{ }_n,0}
 V \delta^{ }_{\vec{p} + \vec{q}, \vec{0}} \,
 \frac{1}{ P^2 + m^2}
 \;, 
\ea
which in the infinite-volume limit (cf.\ \eq\nr{Fou_infV}), {\em viz.}\
\be
 \frac{1}{V} \sum_{\vec{p}} \longrightarrow \int \! 
 \frac{{\rm d}^d\vec{p}}{(2\pi)^d}
 \;, \quad
 V \delta^{ }_{\vec{p},\vec{0}} \longrightarrow
 (2\pi)^d \delta^{(d)}(\vec{p})
 \;, 
\ee 
becomes exactly \eq\nr{prop_PQ}. 
Inserting this into \eq\nr{raw_prop} we also recover \eq\nr{prop_xy}.

\index{Thermal sums: boson loop}

It is useful to study the 
behaviour of the propagator $\langle \phi(X)\phi(Y) \rangle_0^{ }$ 
at small and large separations $X-Y$.
For this we may use the result of \eq\nr{Gtau},  
\be
 T \sum_{ p^{ }_n} \frac{e^{i  p^{ }_n \tau} }{ p_n^2 + \E^2}
 = \frac{1}{2 \E}
 \frac{\cosh\left[ \left( \frac{\beta}{2} - \tau\right) \E \right]}
 {\sinh\left[ \frac{\beta \E}{2} \right]} 
 \;, \quad
 \beta=\frac{1}{T}
 \;, \quad
 0 \le \tau \le \beta
 \;. \la{Gtau_2}
\ee
Even though this equation was derived 
for $0 \le \tau \le \beta$, it is clear
from the left-hand side that we can extend its validity to 
$-\beta \le \tau \le \beta$ by replacing $\tau$ by $|\tau|$.
Thereby, the propagator in \eq\nr{prop_xy} becomes
\be
 G_0^{ }(X-Y) \equiv 
 \langle \phi(X) \phi(Y) \rangle_0^{ } 
 = 
 \int \! \frac{{\rm d}^d\vec{p}}{(2\pi)^d}
 e^{i \vec{p}\cdot (\vec{y-x})}
 \left. 
 \frac{1}{2 \E^{ }_{p}}
 \frac{\cosh\left[ \left( \frac{\beta}{2} - |x^{ }_0 - y^{ }_0| \right)
 \E^{ }_{p} \right]}
 {\sinh\left[ \frac{\beta \E^{ }_{p}}{2} \right]} 
 \right|^{ }_{\E^{ }_{p} \equiv \sqrt{{p}^2 + m^2}} \hspace*{-1cm}
 \;, \la{Gxy} \la{Pxy_def}
\ee
where we may set $Y=0$ with no loss of generality. 

Consider first short distances, $|\vec{x}|, |x^{ }_0| \ll \tfr{1}T, \tfr{1}m$. 
We may expect the dominant contribution in the Fourier transform
of \eq\nr{Gxy} to come from the regime $|\vec{p}||\vec{x}|\sim 1$, 
so we assume $|\vec{p}| \gg T,m$. Then 
$\E^{ }_{p} \approx {p}$ 
and $\beta \E^{ }_{p} \approx {p}/T \gg 1$, and consequently, 
\be
 \frac{\cosh\left[ \left( \frac{\beta}{2} - |x^{ }_0| \right)
 \E^{ }_{p} \right]}
 {\sinh\left[ \frac{\beta \E^{ }_{p}}{2} \right]} 
 \;\approx\; 
 \frac{\exp\left[ \left( \frac{\beta}{2} - |x^{ }_0| \right)
 \E^{ }_{p} \right]}
 {\exp\left[ \frac{\beta \E^{ }_{p}}{2} \right]} 
 \;\approx\; 
 e^{-|x^{ }_0|{p}}
 \;. 
\ee
Noting that 
\be
 \frac{1}{2{p}}  e^{-|x^{ }_0|{p}}
 = 
 \int_{-\infty}^{\infty}
 \! \frac{{\rm d} p^{ }_0}{2\pi}
 \frac{e^{i p^{ }_0 x^{ }_0}}{p_0^2 + \vec{p}^2}
 \;,
\ee
this implies
\be
 G_0^{ }(X) \approx 
 \int \! \frac{{\rm d}^{d+1}{P}}{(2\pi)^{d+1}}
 \frac{e^{i {P}\cdot {X}}}{P^2}
 \;, 
\ee
with $P \equiv (p^{ }_0, \vec{p})$. We recognize this as the coordinate space 
propagator of a massless scalar field at zero temperature. 

At this point we make use of the $d+1$-dimensional rotational 
symmetry of Euclidean spacetime, 
and choose $X=(x^{ }_0,\vec{x})$ to point 
in the direction of the component $p^{ }_0$. 
Then, 
\ba
 \int \! \frac{{\rm d}^{d+1}{P}}{(2\pi)^{d+1}}
 \frac{e^{i {P}\cdot {X}}}{P^2}
 & = & 
 \int \! \frac{{\rm d}^{d}\vec{p}}{(2\pi)^{d}}
 \int_{-\infty}^{\infty}
 \! \frac{{\rm d} p^{ }_0}{2\pi}
 \frac{e^{i {p^{ }_0}{|X|}}}{p_0^2 + \vec{p}^2}
 \nn 
 & = & 
 \int \! \frac{{\rm d}^{d}\vec{p}}{(2\pi)^{d}}
 \frac{e^{-{p}|X|}}{2 {p}}
 \nn 
 & \stackrel{\rmi{\nr{k_measure}}}{=} & 
 \frac{1}{(2\pi)^d} \frac{\pi^{\sfr{d}{2}}}{\Gamma(\tfr{d}{2})}
 \int_0^\infty \! {\rm d}{p}\, {p}^{d-2} 
 e^{-{p}|X|}
 \nn 
 & = & 
 \frac{\Gamma(d-1)}{(4\pi)^{\sfr{d}2}\Gamma(\tfr{d}{2})|X|^{d-1}}
 \;, \la{Px_d}
\ea
from which, inserting $d=3$
and $\Gamma(\tfr32) = \sqrt{\pi}/2$, we find 
\be
 G_0^{ }(X) \approx 
 \frac{1}{4\pi^2 |X|^2}
 \;, \quad
 |X| \ll \frac{1}{T}, \frac{1}{m}
 \;.  \la{Px_small}
\ee 
The result is independent of $T$ and $m$, signifying that at short 
distances (in the ``ultraviolet'' regime), 
temperature and masses 
do not play a role. We may further note that 
the propagator rapidly diverges in this regime. 

Next, we consider the opposite limit of large distances, 
$x= |\vec{x}| \gg 1/T$, noting that the periodic  temporal 
coordinate $x^{ }_0$ is always ``small'', 
i.e.\ at most $1/T$. We expect that the Fourier transform of \eq\nr{Gxy} 
is now dominated by small momenta, ${p}\ll T$.
If we simplify the situation further by assuming that we are
also at very high temperatures, $m \ll T$, then 
$\beta \E^{ }_{p} \ll 1$, and we 
can expand the hyperbolic functions in Taylor series, 
approximating $\cosh(\epsilon)\approx 1$, 
$\sinh(\epsilon)\approx \epsilon$. We then obtain from \eq\nr{Gxy}
\be
  G_0^{ }(X) \;\approx\;
  T \int \! \frac{{\rm d}^d\vec{p}}{(2\pi)^d}
 \frac{e^{-i \vec{p}\cdot\vec{x}}}{{p}^2 + m^2}
 \;, \quad {x} \gg \frac{1}{T}  
 \;.
\ee
Note that the integrand here is also the $p^{ }_n = 0$ contribution from 
the left-hand side of \eq\nr{Gtau_2}.
Setting $d=3$,\footnote{
 For a general $d$, 
 $\int \! \frac{{\rm d}^d\vec{p}}{(2\pi)^d}
 \frac{e^{-i \vec{p}\cdot\vec{x}}}{{p}^2 + m^2}
 = 
 {(2\pi)^{-\fr{d}2}} ( \frac{m}{{x}} )^{\fr{d}2-1}
 K^{ }_{\sfr{d}2-1}(m {x})
 $, 
 where $K$ is a modified Bessel function.
 } 
and denoting $z \equiv \vec{p}\cdot\vec{x}/(px)$, the remaining 
integral can be worked out as 
\ba
 G_0^{ }(X) & \approx & 
 \frac{T}{(2\pi)^2} 
 \int_{-1}^{+1} \! {\rm d}z \int_0^\infty \! {\rm d}{p} \, {p}^2
 \frac{e^{-i {p}{x}z}}{{p}^2 + m^2}
 \nn & = & 
 \frac{T}{(2\pi)^2} 
 \int_0^\infty \! \frac{{\rm d}{p} \, {p}^2}{{p}^2 + m^2}
 \frac{e^{i {p}{x}} - e^{-i {p}{x}}}
 {i {p}{x}}
 \nn & = & 
 \frac{T}{(2\pi)^2 i {x}} 
 \int_{-\infty}^\infty \! 
  \frac{{\rm d}{p} \, {p}\, e^{i {p}{x}}}
 {{p}^2 + m^2}
 \nn & = & 
 \frac{T\, e^{-m {x}}}{4\pi {x}}
 \;, \quad {x} \gg \frac{1}{T}  
 \;. \la{Px_large}
\ea
In the last step the integration contour was closed
in the upper half-plane (recalling that $x > 0$). 

We note from \eq\nr{Px_large} 
that at large distances (in the ``infrared'' regime), 
thermal effects modify the behaviour of the propagator in 
an essential way. In particular, if we were to set the mass
to zero, then \eq\nr{Px_small} would be the exact behaviour
at zero temperature, both at small and at large distances, 
whereas \eq\nr{Px_large} shows that a finite temperature would ``slow down''
the long-distance decay to $T/(4\pi |\vec{x}|)$. In other words, 
we can say that at finite temperature the theory is more sensitive 
to infrared physics than at zero temperature. 

The considerations just discussed play a physical role in plasma physics, 
i.e.\ the study of electrodynamics in a statistical environment. We defer
an illustration to appendix~B of \se\ref{ss:expansion}, by which time 
gauge fields will have been introduced.

\newpage 

\subsection{Problems of the naive weak-coupling expansion}

\subsection*{$\rmO(\lambda)$: ultraviolet divergences}
\la{se:naive_lam}

We now proceed with the evaluation of 
the weak-coupling expansion for the free energy density in a scalar field 
theory, the first three orders of which are 
given by \eqs\nr{f0}, \nr{f1} and \nr{f2}. Noting from 
\eqs\nr{ImT_2} and \nr{prop_xy} that $G_0^{ }(0) = I(m,T)$, we obtain
\be
 f(T) = 
 J(m,T) + \fr34 \lambda\, [I(m,T)]^2
 + \rmO(\lambda^2)
 \;. \la{fT_1}
\ee
According to \eqs\nr{J0m_res} and \nr{I0m_res}, we have
\ba
 J(m,T) & = &   -\frac{m^4\mu^{-2\epsilon}}{64\pi^2}
 \biggl[
   \frac{1}{\epsilon}
 + \ln  \frac{\bmu^2}{m^2} + \fr32 + \rmO(\epsilon)
 \biggr]
 + J^{ }_T(m)
 \;, \la{JmT_gen} \\
 I(m,T) & = & 
   -\frac{m^2\mu^{-2\epsilon}}{16\pi^2}
 \biggl[
   \frac{1}{\epsilon}
 + \ln  \frac{\bmu^2}{m^2} + 1 + \rmO(\epsilon)
 \biggr]
 + I^{ }_T(m) \la{ImT_gen}
 \;, 
\ea
where the finite functions $J^{ }_T(m)$ and $I^{ }_T(m)$
were evaluated in various limits 
in \eqs\nr{JTm_final}, \nr{ITm_final}, 
\nr{JTm_res} and \nr{ITm_res}.

Inserting \eqs\nr{JmT_gen} and \nr{ImT_gen} into \eq\nr{fT_1}, 
we note that the result is, in general, {\em ultraviolet divergent}. 
For instance, restricting for simplicity to very high temperatures, 
$T \gg m$, and making use of \eq\nr{ITm_res}, 
\be
 I^{ }_T(m) \approx  \frac{T^2}{12} - \frac{mT}{4\pi} + \rmO(m^2)
\;, \la{IT_appr}
\ee
the dominant term at $\epsilon \to 0$ reads
\be
  f(T) \approx 
 -\frac{\mu^{-2\epsilon}}{64\pi^2\epsilon}
 \biggl\{ 
 m^4 + 
 \lambda \biggl[ \frac{1}{2} T^2 m^2 -\frac{3}{2\pi} T m^3 + \rmO(m^4)
 \biggr]
 + \rmO(\lambda^2) 
 \biggr\} + \rmO(1)
 \;. \la{UV_div}
\ee
This result is clearly non-sensical; 
in particular the divergences  
depend on the temperature, 
i.e.~cannot be removed by subtracting a $T$-independent 
``vacuum'' contribution. To properly handle this issue requires
{\em renormalization}, to which we return in \se\ref{se:UV}. 


\subsection*{$\rmO(\lambda^2)$: infrared divergences}
\la{se:naive_lam2}

\index{Infrared divergence: scalar field}

Let us next consider the $\rmO(\lambda^2)$ correction to \eq\nr{fT_1}, 
given by \eq\nr{f2}. With the notation 
of \eq\nr{Pxy_def}, it can be written as
\be
 f^{ }_{(2)}(T) = 
 -\fr34 \lambda^2\! \int_X [G_0^{ }(X)]^4 
 -\fr94 \lambda^2 [I(m,T)]^2 \int_X [G_0^{ }(X)]^2
 \;. \la{fT_2}
\ee
It is particularly interesting to inspect what happens if we 
take the particle mass $m$ to be very small 
in units of the temperature, $m \ll T$.

As \eqs\nr{JTm_res}, \nr{fT_1} and \nr{IT_appr} show, at $\rmO(\lambda)$
the small-mass limit is perfectly well-defined. 
At the next order, we on the other 
hand must analyze the two terms of \eq\nr{fT_2}. 
Starting with the first one, 
we know from \eq\nr{Px_small} that the behaviour
of $G_0^{ }$ is independent of $m$ at small $x$, and thus nothing particular
happens for ${x} \ll T^{-1}$. On the other hand, for large 
${x}$, $G_0^{ }$ is given by \eq\nr{Px_large}, and we may thus estimate the 
contribution of this region as 
\be
 \int_\rmi{${x} \gsim \beta$} [G_0^{ }(X)]^4
 \sim
 \int_0^\beta \! {\rm d}\tau \int_\rmi{${x} \gsim \beta$}
 \! {\rm d}^3\vec{x} \, 
 \biggl( \frac{T e^{-m {x}}}{4\pi {x}}\biggr)^4
 \;. \la{P4}
\ee
This integral is convergent even for $m\to 0$. 

Consider then the second term of \eq\nr{fT_2}. Repeating the previous 
argument, we see that the long-distance 
contribution to the free energy density 
is proportional to the integral
\be
 \int_\rmi{${x} \gsim \beta$} [G_0^{ }(X)]^2
 \sim
 \int_0^\beta \! {\rm d}\tau \! \int_\rmi{${x} \gsim \beta$}
 \! {\rm d}^3\vec{x}\, \biggl( \frac{T e^{-m {x}}}
 {4\pi {x}}\biggr)^2
 \;. \la{P2}
\ee
If we now attempt to set $m\to 0$, we run into a linearly 
divergent integral. Because this problem emerges from large distances, 
we call this an {\em infrared divergence}. 

In fact, it is easy to be more precise about the form of the 
divergence. We can namely write
\ba
 \int_X [G_0^{ }(X)]^2 & = &  
 \int_X 
 \Tint{ P}  
 \frac{e^{i  P\cdot X}}{ P^2 + m^2}
 \Tint{ Q}  
 \frac{e^{i  Q\cdot X}}{ Q^2 + m^2}
 \nn & = & 
  \Tint{ P  Q} 
 \raise-0.02em\hbox{$\bar{}$}\hspace*{-0.8mm}{\delta}( P +  Q) 
 \frac{1}{( P^2 + m^2)( Q^2 + m^2)}
 \nn & = & 
 \Tint{P} \frac{1}{[ P^2 + m^2]^2}
 \nn & = & 
 - \frac{{\rm d}}{{\rm d}m^2} I(m,T)
 \;. 
\ea
Inserting \eq\nr{IT_appr}, we get 
\be
  \int_X [G_0^{ }(X)]^2 = 
 -\frac{1}{2m} \frac{{\rm d}}{{\rm d}m} I(m,T)
 = \frac{T}{8\pi m} + \rmO(1)
 \;,
\ee
so that for $m\ll T$, \eq\nr{fT_2} evaluates to 
\be
 f^{ }_{(2)}(T) = -\fr94 \lambda^2 \frac{T^4}{144} \frac{T}{8\pi m}
 + \rmO(m^0)
 \;. \la{IR_div}
\ee
This indeed diverges for $m\to 0$. 

It is clear that like the ultraviolet divergence in \eq\nr{UV_div}, 
the infrared divergence in \eq\nr{IR_div} must be an artifact
of some sort: the pressure  and other thermodynamic
properties of a plasma of weakly interacting massless scalar particles
should be finite, as we know to be the case for a plasma 
of massless photons. 
We return to the resolution of this ``paradox'' in \se\ref{se:IR}.

\newpage 

\subsection{Proper free energy density to $\rmO(\lambda)$: 
ultraviolet renormalization}
\la{se:UV}

\index{Renormalization}

In \se\ref{se:naive_lam} we attempted to compute the free energy
density $f(T)$ of a scalar field theory up to $\rmO(\lambda)$, but found 
a result which appeared to be ultraviolet (UV) divergent. 
Let us now show that, as must be the 
case in a renormalizable theory, the divergences disappear order-by-order
in perturbation theory, if we {\em re-express $f(T)$ in terms of 
renormalized parameters}. Furthermore the renormalization 
procedure is identical to that at zero temperature.

In order to proceed, we need to change the notation 
somewhat. The zero-temperature parameters we employed before, 
i.e.\ $m^2,\lambda$, are now re-interpreted to be {\em bare parameters}, 
$m_\rmii{B}^2, \lambda^{ }_\rmii{B}$.\footnote{%
 The temperature, in contrast, is a physical
 property of the system, and is not subject to any modification.}
The expansion in \eq\nr{fT_1} can then be written in the schematic form
\be
 f(T) = \phi^{(0)}(m_\rmii{B}^2,T) + 
 \lambda^{ }_\rmii{B}\, \phi^{(1)}(m_\rmii{B}^2,T)
 + \rmO(\lambda_\rmii{B}^2)
 \;. \la{fT_B}
\ee

As a second step, we introduce the {\em renormalized parameters} 
$m_\rmii{R}^2, \lambda^{ }_\rmii{R}$. 
These could either be directly {\em physical quantities}
(say, the mass of the scalar particle, and the scattering amplitude with
particular kinematics), or quantities which are not directly 
physical, but are related to physical quantities by finite equations
(say, so-called $\msbar$ scheme parameters). In any case, it is natural
to choose the renormalized parameters such that in the limit of
an extremely weak interaction, $\lambda^{ }_\rmii{R} \ll 1$, 
they formally agree
with the bare parameters. In other words, we may write
\ba
 m_\rmii{B}^2 & = & 
 m_\rmii{R}^2 + \lambda^{ }_\rmii{R}\, f^{(1)}(m_\rmii{R}^2)
 + \rmO(\lambda_\rmii{R}^2) 
 \;, \la{m_B} \\ 
 \lambda^{ }_\rmii{B} & = &
 \lambda^{ }_\rmii{R} + \lambda_\rmii{R}^2\, g^{(1)}(m_\rmii{R}^2)
 +  \rmO(\lambda_\rmii{R}^3) 
 \;, \la{l_B}
\ea
where it is important to note that the renormalized parameters are defined at 
zero temperature (no $T$ appears in these relations). 
The functions $f^{(i)}$ and $g^{(i)}$ are in general divergent
in the limit that the regularization is removed; for instance, in 
dimensional regularization, they are expected to contain poles, such as 
$1/\epsilon$ or higher.

The idea now is to convert the expansion in \eq\nr{fT_B}
into an expansion in $\lambda^{ }_\rmii{R}$ by inserting in it the expressions
from \eqs\nr{m_B} 
and \nr{l_B} and Taylor-expanding 
the result in $\lambda^{ }_\rmii{R}$. This produces
\be
 f(T) = \phi^{(0)}(m_\rmii{R}^2,T)
 + \lambda^{ }_\rmii{R}
 \biggl[ 
   \phi^{(1)}(m_\rmii{R}^2,T) + 
   \frac{\partial \phi^{(0)}(m_\rmii{R}^2,T)}{\partial m_\rmii{R}^2}
   f^{(1)}(m_\rmii{R}^2)
 \biggr]
 + \rmO(\lambda_\rmii{R}^2)
 \;, \la{fT_exp}
\ee
where we note that to $\rmO(\lambda_\rmii{R}^2)$ only the mass parameter
needs to be renormalized. 

To carry out renormalization in practice, we need 
to choose a {\em scheme}. We adopt here the so-called
{\em pole mass scheme}, where $m_\rmii{R}^2$ is taken to be the 
physical mass squared of the $\phi$-particle,
denoted by $m_\rmi{phys}^2$. 
In Minkowskian spacetime, this quantity appears 
as an exponential time evolution, 
\be
 e^{-i \E^{ }_0 t}
 \equiv 
 e^{-i m^{ }_\rmii{phys} t}
 \;, \la{mphys_def}
\ee
in the propagator of a particle at rest, $\vec{p} = \vec{0}$. 
In Euclidean spacetime, it on the other hand corresponds to an exponential
fall-off, $\exp(- m^{ }_\rmi{phys} \tau)$, 
in the imaginary-time propagator. 
Therefore, in order to determine $m_\rmi{phys}^2$ to 
$\rmO(\lambda^{ }_\rmii{R})$, we need to compute the {\em full propagator}, 
$G(X)$, to $\rmO(\lambda^{ }_\rmii{R})$ at zero temperature. 

The full propagator can be defined as the generalization 
of \eq\nr{Pxy_def} to the interacting case: 
\ba
 G(X) & \equiv & 
 \frac{\langle \phi(X)\phi(0) \exp(-S^{ }_\iI) \rangle_0^{ }}
 {\langle \exp(-S^{ }_\iI) \rangle_0^{ }}
 \nn  & = & 
\frac{\langle \phi(X)\phi(0) \rangle_0^{ }
 - \langle \phi(X)\phi(0)\, S^{ }_\iI\, \rangle_0^{ }
 + \rmO(\lambda_\rmii{B}^2) }
 {1 - \langle S^{ }_\iI \rangle_0^{ } + \rmO(\lambda_\rmii{B}^2)}
 \nn  & = & 
 \langle \phi(X)\phi(0) \rangle_0^{ }
 - \Bigl[ \langle \phi(X)\phi(0) S^{ }_\iI \rangle_0^{ }
 - \langle \phi(X)\phi(0) \rangle_0^{ }
  \langle S^{ }_\iI \rangle_0^{ } \Bigr]
 + \rmO(\lambda_\rmii{B}^2)
 \;.   \la{P_full}
\ea
We note that just like the subtractions in \eq\nr{f_sft_3}, 
the second term inside the square brackets serves to cancel
disconnected contractions. Therefore, like in \eq\nr{compact_rule},  
we can drop the second term, if we replace
the expectation value in the first one by 
$\langle ... \rangle^{ }_{0,\rmi{c}}$.

Let us now inspect the leading (zeroth order)
term in \eq\nr{P_full}, in order to learn how $m^{ }_\rmi{phys}$ could most 
conveniently be extracted from the propagator. 
Introducing the notation 
\be
 \int_{ P} \equiv \lim_{T\to 0} \Tint{ P} 
 = \int \! \frac{{\rm d}^{d+1} P}{(2\pi)^{d+1}}
 \;, 
\ee
and working in the $T=0$ limit for the time being, 
the free propagator reads (cf.\ \eq\nr{prop_xy})
\be
G_0^{ }(X) = 
\langle \phi(X)\phi(0) \rangle_0^{ }
 = 
  \int_ { P}  \frac{e^{i  P\cdot X}}{ P^2 + m^2}
 \;. \la{P0x}
\ee
For \eq\nr{mphys_def}, we need to project to zero spatial momentum, 
$\vec{p} = \vec{0}$; evidently this can be achieved by taking 
a spatial average of $G_0^{ }(X)$ via
\be
\int_\vec{x} \, 
 \langle \phi(\tau,\vec{x}) \phi(0) \rangle_0^{ } 
 = \int \! \frac{{\rm d}p^{ }_0}{2\pi} \frac{e^{i p^{ }_0 \tau}}{p_0^2 + m^2}
 \;.  \la{spat_av}
\ee
We see that we get an integral which can be evaluated with the 
help of the Cauchy theorem and, in particular, that the exponential
fall-off of the correlation function is determined by the 
pole position of the momentum-space propagator: 
\be
 \int_\vec{x} \,
 \langle \phi(\tau,\vec{x}) \phi(0) \rangle_0^{ } 
  = \frac{1}{2\pi} 2\pi i \frac{e^{-m \tau}}{2 i m}
 \;, \quad \tau \ge 0
 \;. 
\ee
Hence, 
\be
 \left. m_\rmi{phys}^2 \right|^{ }_{\lambda = 0} = m^2
 \;. \la{mphys_res}
\ee
More generally, {\em the physical mass can be extracted by 
determining the pole position of the full propagator in momentum space,
for $\vec{p} = \vec{0}$}.

We then proceed to the second term in \eq\nr{P_full}, keeping still $T=0$:
\ba
 -\langle \phi(X) \phi(0)\, S^{ }_\iI\, \rangle^{ }_{0,\rmi{c}}
 & = & 
 -\frac{\lambda^{ }_\rmii{B}}{4}
 \int_Y \langle
 \phi(X)
 \phi(0) \; 
 \phi(Y)
 \phi(Y)
 \phi(Y)
 \phi(Y)
 \rangle^{ }_{0,\rmi{c}}
 \nn & = & 
 -\frac{\lambda^{ }_\rmii{B}}{4}
 \int_Y 4\times 3 \, 
 \langle
 \phi(X)
 \phi(Y)
 \rangle_0^{ }\;
 \langle
 \phi(Y) 
 \phi(0)
 \rangle_0^{ }\;
 \langle
 \phi(Y)
 \phi(Y)
 \rangle_0^{ }
 \nn & = & 
 -3 \lambda^{ }_\rmii{B} G_0^{ }(0) \int_Y G_0^{ }(Y) G_0^{ }(X-Y)
 \nn & = &
 -3 \lambda^{ }_\rmii{B} \int_{ P} \frac{1}{ P^2 + m_\rmii{B}^2}
 \int_Y \int_{ Q,  R}
 e^{i  Q\cdot Y} e^{i  R\cdot (X-Y)}
 \frac{1}{ Q^2 + m_\rmii{B}^2}
 \frac{1}{ R^2 + m_\rmii{B}^2}
 \nn & = & 
 -3 \lambda^{ }_\rmii{B}\, I_0^{ }(m^{ }_\rmii{B}) 
 \int_{ R} \frac{e^{i  R\cdot X}}{( R^2 + m_\rmii{B}^2)^2}
 \;. 
\ea
Summing this expression together with \eq\nr{P0x}, the full propagator reads
\ba
 G(X) & = & \int_{ P} e^{i  P\cdot X}
 \biggl[ \frac{1}{ P^2 + m_\rmii{B}^2} 
 -  
  \frac{
         3 \lambda^{ }_\rmii{B} I_0^{ }(m^{ }_\rmii{B}) 
       }{( P^2 + m_\rmii{B}^2)^2} + \rmO(\lambda_\rmii{B}^2) 
 \biggr]
 \nn & = & 
 \int_{ P} \frac{e^{i  P\cdot X}}
 { P^2 + m_\rmii{B}^2 
 + 3 \lambda^{ }_\rmii{B} I_0^{ }(m^{ }_\rmii{B})}  + \rmO(\lambda_\rmii{B}^2) 
 \;, \la{resum_Gx}
\ea
where we have resummed a series of higher-order corrections in a way that is 
correct to the indicated order of the weak-coupling expansion.

The same steps that led us from \eq\nr{spat_av} to \nr{mphys_res}
now produce 
\be
 m_\rmi{phys}^2 = m_\rmii{B}^2 + 3 \lambda^{ }_\rmii{B} I_0^{ }(m^{ }_\rmii{B})
  + \rmO(\lambda_\rmii{B}^2)
 \;. \la{pole_mass}
\ee
Recalling from \eq\nr{l_B} that 
$m_\rmii{B}^2 = m_\rmii{R}^2 + \rmO(\lambda^{ }_\rmii{R})$,
$\lambda^{ }_\rmii{B} = \lambda^{ }_\rmii{R} + \rmO(\lambda_\rmii{R}^2)$, 
this relation can be inverted to give 
\be
 m_\rmii{B}^2 = m_\rmi{phys}^2 - 3 \lambda^{ }_\rmii{R}
 I_0^{ }(m^{ }_\rmi{phys})
 + \rmO(\lambda_\rmii{R}^2)
 \;, \la{m_B_2}
\ee
which corresponds to \eq\nr{m_B}. The function 
$I_0^{ }$, given in \eq\nr{I0m_res}, furthermore diverges in the limit
$\epsilon\to 0$, 
\be
  I_0^{ }(m^{ }_\rmi{phys}) 
 = 
  -\frac{m_\rmii{phys}^2 \mu^{-2\epsilon} }{16\pi^2}
 \biggl[
   \frac{1}{\epsilon}
 + \ln  \frac{\bmu^2}{m_\rmii{phys}^2} + 1 + \rmO(\epsilon)
 \biggr]
 \;,
\ee
and we may hope that this divergence cancels those we found in $f(T)$.

Indeed, let us repeat the steps from \eq\nr{fT_B} to \eq\nr{fT_exp}
employing the explicit expression 
for the free energy density from \eq\nr{fT_1}, 
\be
 f(T) = 
 J(m^{ }_\rmii{B},T) + \fr34 \lambda^{ }_\rmii{B} 
 [I(m^{ }_\rmii{B},T)]^2 + \rmO(\lambda_\rmii{B}^2)
 \;. \la{fT_4}
\ee
Recalling from \eq\nr{Imt} that
\be
  I(m,T) = 
 \frac{1}{m} \frac{{\rm d}}{{\rm d}m} J(m,T) 
  = 2 \frac{{\rm d}}{{\rm d}m^2} J(m,T)
 \;, 
\ee 
we can expand the two terms in \eq\nr{fT_4} as a Taylor series
around $m_\rmi{phys}^2$, obtaining
\ba
 J(m^{ }_\rmii{B},T) & = &  
 J(m^{ }_\rmi{phys},T) 
 + (m_\rmii{B}^2 - m_\rmi{phys}^2) 
 \frac{\partial J(m^{ }_\rmii{phys},T)}{\partial m_\rmii{phys}^2} + 
 \rmO(\lambda_\rmii{R}^2)
 \nn & = & 
 J(m^{ }_\rmi{phys},T) - \fr32 \lambda^{ }_\rmii{R} 
 I_0^{ }(m^{ }_\rmi{phys}) I(m^{ }_\rmi{phys},T) + 
 \rmO(\lambda_\rmii{R}^2)
 \;, \la{JmBexp} \\[2mm]
 \lambda^{ }_\rmii{B} [I(m^{ }_\rmii{B},T)]^2 & = & 
 \lambda^{ }_\rmii{R} [I(m^{ }_\rmi{phys},T)]^2 + 
 \rmO(\lambda_\rmii{R}^2)
 \;, 
\ea
where in \eq\nr{JmBexp}
we inserted \eq\nr{m_B_2}. With this input, \eq\nr{fT_4} becomes
\ba
 f(T) & = & 
 J(m^{ }_\rmi{phys},T) + \fr34 \lambda^{ }_\rmii{R} 
 \Bigl[
    I^2(m^{ }_\rmi{phys},T) - 2 
    I_0^{ }(m^{ }_\rmi{phys})\, I(m^{ }_\rmi{phys},T)
 \Bigr] + \rmO(\lambda_\rmii{R}^2)
 \nn 
 & = &
 {\underbrace{\biggl\{
   J_0^{ }(m^{ }_\rmi{phys}) 
   - \fr34 \lambda^{ }_\rmii{R} 
   I_0^2 (m^{ }_\rmi{phys})
 \biggr\}}}
 +  
 \underbrace{\biggl\{ 
  J^{ }_T(m^{ }_\rmi{phys}) 
   + \fr34 \lambda^{ }_\rmii{R} 
   I_T^2 (m^{ }_\rmi{phys})
 \biggr\}} 
 + \rmO(\lambda_\rmii{R}^2)
 \;, \hspace*{0.5cm} \la{fT_5} \\\ 
 & &
 \hspace*{1.5cm} \mbox{$T=0$ part}
 \hspace*{3.5cm} \mbox{$T\neq 0$ part} \nonumber
\ea
where we inserted the definitions 
$J(m,T) = J_0^{ }(m) + J^{ }_T(m)$ and
$I(m,T) = I_0^{ }(m) + I^{ }_T(m)$.

Recalling \eqs\nr{J0m_res} and \nr{I0m_res}, we observe that the first term in 
\eq\nr{fT_5}, the ``$T=0$ part'', is still divergent. However, this 
term is independent of the temperature, and thus 
plays no role in thermodynamics. Rather, it
corresponds to a {\em vacuum energy density} that only
matters in connection with gravity. If we included 
gravity, however, we should also include a bare cosmological constant, 
$\Lambda^{ }_\rmii{B}$, in the bare Lagrangian; this would contribute 
additively to \eq\nr{fT_5}, and we could simply identify the 
physical cosmological constant as
\be
 \Lambda^{ }_\rmi{phys} \equiv 
 \Lambda^{ }_\rmii{B} +    J_0^{ }(m^{ }_\rmi{phys}) 
   - \fr34 \lambda^{ }_\rmii{R} 
   I_0^2 (m^{ }_\rmi{phys}) + \rmO(\lambda_\rmii{R}^2)
 \;. 
\ee
The divergences would now be cancelled by
$\Lambda^{ }_\rmii{B}$, and $\Lambda^{ }_\rmi{phys}$ would be finite. 

In contrast, the second term in 
\eq\nr{fT_5}, the ``$T\neq 0$ part'', is finite: it contains
the functions $J^{ }_T$, $I^{ }_T$ for which we have analytically 
determined various limiting values in 
\eqs\nr{JTm_final}, \nr{ITm_final}, \nr{JTm_res} and \nr{ITm_res}, 
as well as general integral representations in 
\eqs\nr{JTy} and \nr{ITy}.
Therefore all thermodynamic quantities obtained from 
derivatives of $f(T)$, such as the entropy density or 
specific heat, are manifestly finite. 
In other words, the temperature-dependent
ultraviolet divergences that we found in 
\se\ref{se:naive_lam} have disappeared through 
zero-temperature renormalization.

\newpage 

\subsection{Proper free energy density to $\rmO(\lambda^{\fr32})$:
infrared resummation}
\la{se:IR}

\index{Resummation}

We now move on to a topic which is 
in a sense maximally different from the UV issues discussed 
in the previous section, and consider the limit where 
the physical mass of the scalar field, $m^{ }_\rmi{phys}$, 
tends to zero. With a few technical modifications, this would be the case 
(in perturbation theory) for, say,  gluons in QCD. According to 
\eq\nr{m_B_2}, this limit corresponds to $m^{ }_\rmii{B}\to 0$, 
since $I_0^{ }(0)=0$; then we are faced with the infrared
problem discussed in \se\ref{se:naive_lam2}.

In the limit of a small mass, we can employ high-temperature
expansions for the functions $J(m,T)$ and $I(m,T)$, given in 
\eqs\nr{ImT_res} and \nr{JmT_res}. 
Employing \eqs\nr{fT_1} and \nr{IR_div}, 
we write the leading terms in the small-$m^{ }_\rmii{B}$ expansion as
\ba
 \rmO(\lambda_\rmii{B}^0): \quad 
 f^{ }_{(0)}(T) & = & J(m^{ }_\rmii{B},T)
 = -\frac{\pi^2 T^4}{90} + \frac{m_\rmii{B}^2 T^2}{24} 
   -\frac{m_\rmii{B}^3 T}{12 \pi} + \rmO(m_\rmii{B}^4)
 \;, \la{f0_mB} \\ 
 \rmO(\lambda_\rmii{B}^1): \quad 
 f^{ }_{(1)}(T) & = & \fr34 \lambda^{ }_\rmii{B} [I(m^{ }_\rmii{B},T)]^2 
 \nn & = &
 \fr34 \lambda^{ }_\rmii{B} 
 \biggl[ 
   \frac{T^2}{12} - \frac{m^{ }_\rmii{B} T}{4\pi} + \rmO(m_\rmii{B}^2)
 \biggr]^2
 \nn & = & 
 \fr34 \lambda^{ }_\rmii{B} \biggl[ 
  \frac{T^4}{144} - \frac{m^{ }_\rmii{B} T^3}{24\pi} + \rmO(m_\rmii{B}^2 T^2)
 \biggr] 
 \;, \la{f1_mB} \\
 \rmO(\lambda_\rmii{B}^2): \quad 
 f^{ }_{(2)}(T) & = & -\fr94 \lambda_\rmii{B}^2 \frac{T^4}{144} 
 \frac{T}{8\pi m^{ }_\rmii{B}}
 + \rmO(m_\rmii{B}^0)
 \;. \la{f2_mB}
\ea

\index{Zero mode: scalar field}

Let us inspect, in particular, {\em odd powers of $m^{ }_\rmii{B}$}, 
which according to \eqs\nr{f0_mB}--\nr{f2_mB} are becoming
increasingly important as we go further in the expansion. 
We remember from \se\ref{se:highT} that odd powers of $m^{ }_\rmii{B}$ are 
necessarily associated with contributions from the Matsubara zero mode. 
In fact, the odd power in \eq\nr{f0_mB} is directly the 
zero-mode contribution to \eq\nr{Jn0_res}, 
\be
 \delta^{ }_\rmi{odd} f^{ }_{(0)} = J^{(n=0)} = -\frac{m_\rmii{B}^3 T}{12\pi}
 \;. \la{odd0}
\ee
The odd power in \eq\nr{f1_mB} on the other hand originates from 
a cross-term between 
the zero-mode contribution and the leading non-zero mode 
contribution to $I(0,T)$: 
\be
 \delta^{ }_\rmi{odd} f^{ }_{(1)} = \fr32 \lambda^{ }_\rmii{B} \times 
 I'(0,T) \times I^{(n=0)} = 
 -\frac{\lambda^{ }_\rmii{B} m^{ }_\rmii{B} T^3}{32\pi}
 \;. \la{odd1}
\ee 
Finally, the small-$m^{ }_\rmii{B}$ divergence in \eq\nr{f2_mB} 
comes from a product of
two non-zero mode contributions and a particularly infrared
sensitive zero-mode contribution: 
\be
 \delta^{ }_\rmi{odd} f^{ }_{(2)} = \fr94 \lambda_\rmii{B}^2 
 \times [I'(0,T)]^2 \times \frac{{\rm d} I^{(n=0)}}{{\rm d} m_\rmii{B}^2}
 = -\frac{\lambda_\rmii{B}^2 T^5}{8^3\pi m^{ }_\rmii{B}}
 \;. \la{odd2}
\ee
Comparing these structures, we see that the 
``expansion parameter'' related to odd powers is
\be 
 \frac{\delta^{ }_\rmi{odd}
   f^{ }_{(1)}}{\delta^{ }_\rmi{odd} f^{ }_{(0)}} \;\sim\; 
 \frac{\delta^{ }_\rmi{odd}
   f^{ }_{(2)}}{\delta^{ }_\rmi{odd} f^{ }_{(1)}} \;\sim\;
 \frac{\lambda^{ }_\rmii{B} T^2}{8 m_\rmii{B}^2}
 \;.  
\ee
Thus, if we try to set $m_\rmii{B}^2\to 0$ 
(or even just $m_\rmii{B}^2 \ll \lambda^{ }_\rmii{B} T^2/8$), 
the loop expansion shows no convergence. 

In order to cure the problem with the 
infrared (IR) sensitivity of the loop expansion, our goal now becomes to 
{\em identify and sum the divergent terms to all orders}. 
We may then expect that the complete sum obtains a form where we can set
$m_\rmii{B}^2\to 0$ without meeting divergences. 
This procedure is often referred to as {\em resummation}. 

Fortunately, it is indeed possible to identify the problematic terms. 
\Eqs\nr{odd0}--\nr{odd2} already suggest that at order 
$N$ in $\lambda^{ }_\rmii{B}$, they are associated with terms 
containing $N$ non-zero mode contributions $I'(0,T)$, 
and {\em one zero-mode contribution}. Graphically, this corresponds
to a single loop formed by a zero-mode propagator, 
dressed with $N$ non-zero mode
``bubbles''. Such graphs are usually called ``ring'' or 
``daisy'' diagrams, and can be illustrated as follows
(the dashed line is a zero-mode propagator, 
solid lines are non-zero mode propagators):
\ba
 && \nn[5mm]
 && \hspace*{-1cm}
 \Daisy \qquad \;.
 \\[5mm]
 \nonumber
\ea

\index{Ring diagrams}
\index{Daisy resummation}

To be more quantitative, we consider \eq\nr{compact_rule} at order 
$\lambda_\rmii{B}^N$. A straightforward combinatorial analysis then gives
\ba
 f(T) & = & \Bigl\langle
 S^{ }_\iI - \frac{1}{2} S_\iI^2 + \ldots 
 + \frac{(-1)^{N+1}}{N!} S_\iI^N 
 \Bigr\rangle^{ }_{0,\rmi{c},\,\rmi{drop overall $\int_X$}}
 \la{ring} \\[3mm] & \Rightarrow &
 \frac{(-1)^{N+1}}{N!} \biggl( \frac{\lambda^{ }_\rmii{B}}{4}\biggr)^N
  \left\langle
  \begin{minipage}[c]{7.5cm}
   \setlength{\unitlength}{1.0cm}
   \begin{picture}(7.0,1)(0,0)
     \put(0.3,0.5){\makebox(0,0){$\phi$}}
     \put(0.6,0.5){\makebox(0,0){$\phi$}}
     \put(0.9,0.5){\makebox(0,0){$\phi$}}
     \put(1.2,0.5){\makebox(0,0){$\phi$}}
     \put(2.0,0.5){\makebox(0,0){$\phi$}}
     \put(2.3,0.5){\makebox(0,0){$\phi$}}
     \put(2.6,0.5){\makebox(0,0){$\phi$}}
     \put(2.9,0.5){\makebox(0,0){$\phi$}}
     \put(3.7,0.5){\makebox(0,0){$\phi$}}
     \put(4.0,0.5){\makebox(0,0){$\phi$}}
     \put(4.6,0.5){\makebox(0,0){$\phi$}}
     \put(4.3,0.5){\makebox(0,0){$\phi$}}
     \put(5.45,0.5){\makebox(0,0){$\cdots$}}     
     \put(6.3,0.5){\makebox(0,0){$\phi$}}
     \put(6.6,0.5){\makebox(0,0){$\phi$}}
     \put(6.9,0.5){\makebox(0,0){$\phi$}}
     \put(7.2,0.5){\makebox(0,0){$\phi$}}
     \put(0.6,0.7){\line(0,1){0.1}}
     \put(0.9,0.7){\line(0,1){0.1}}
     \put(0.6,0.8){\line(1,0){0.3}}
     \put(0.75,1.0){\makebox(0,0){\small 6}}
     \put(2.3,0.7){\line(0,1){0.1}}
     \put(2.6,0.7){\line(0,1){0.1}}
     \put(2.3,0.8){\line(1,0){0.3}}
     \put(2.45,1.0){\makebox(0,0){\small 6}}
     \put(4.0,0.7){\line(0,1){0.1}}
     \put(4.3,0.7){\line(0,1){0.1}}
     \put(4.0,0.8){\line(1,0){0.3}}
     \put(4.15,1.0){\makebox(0,0){\small 6}}
     \put(6.6,0.7){\line(0,1){0.1}}
     \put(6.9,0.7){\line(0,1){0.1}}
     \put(6.6,0.8){\line(1,0){0.3}}
     \put(6.75,1.0){\makebox(0,0){\small 6}}
     \put(1.2,0.2){\line(0,1){0.1}}
     \put(2.0,0.2){\line(0,1){0.1}}
     \put(1.2,0.2){\line(1,0){0.8}}
     \put(1.6,0.0){\makebox(0,0){\small $2(N-1)$}}
     \put(2.9,0.2){\line(0,1){0.1}}
     \put(3.7,0.2){\line(0,1){0.1}}
     \put(2.9,0.2){\line(1,0){0.8}}
     \put(3.3,0.0){\makebox(0,0){\small $2(N-2)$}}
   \end{picture}
 \end{minipage}
 \right\rangle^{ }_{0,\ldots}
 \nonumber \\[3mm] & = & 
 \frac{(-1)^{N+1}}{N!} \biggl( \frac{\lambda^{ }_\rmii{B}}{4}\biggr)^N
 6^N 
 \underbrace{[2(N-1)][2(N-2)]...[2]}
 \biggl[ \underbrace{\frac{T^2}{12}} \biggr]^N
 \underbrace{T \int\! \frac{{\rm d}^d\vec{p}}{(2\pi)^d}
 \biggl( \frac{1}{{p}^2 + m_\rmii{B}^2} \biggr)^N}
 \;, \nn 
 & & 
 \hspace*{4cm}
 2^{N-1}(N-1)!
 \hspace*{0.8cm}
 I'(0,T)
 \hspace*{1cm}
 \mbox{zero-mode part} 
 \nonumber
\ea
where we have indicated the contractions from which 
the various factors originate. Let us compute the zero-mode part for the first
few orders, omitting for simplicity terms of $\rmO(\epsilon)$: 
\ba
 N=1: & & 
 \int_\vec{p}
 \frac{1}{{p}^2 + m_\rmii{B}^2} = -\frac{m^{ }_\rmii{B}}{4\pi} 
 = \frac{{\rm d}}{{\rm d}m_\rmii{B}^2}
 \biggl( -\frac{m_\rmii{B}^3}{6 \pi}  \biggr)  
 \;, \nn 
 N=2: & & 
 \int_\vec{p}
 \frac{1}{({p}^2 + m_\rmii{B}^2)^2} = 
 - \frac{{\rm d}}{{\rm d}m_\rmii{B}^2}
 \biggl( -\frac{m^{ }_\rmii{B}}{4\pi}  \biggr)
 =  - \frac{{\rm d}}{{\rm d}m_\rmii{B}^2}
 \frac{{\rm d}}{{\rm d}m_\rmii{B}^2}
 \biggl( -\frac{m_\rmii{B}^3}{6 \pi} \biggr)  
 \;, \nonumber \\[2mm] 
 \mbox{generally}: & & 
 \int_\vec{p}
 \frac{1}{({p}^2 + m_\rmii{B}^2)^N} =
 - \frac{1}{N-1}
 \frac{{\rm d}}{{\rm d}m_\rmii{B}^2}
 \int_\vec{p}
 \frac{1}{({p}^2 + m_\rmii{B}^2)^{N-1}}
 \nn & & 
 \hspace*{1cm} = 
 \biggl( \frac{-1}{N-1} \biggr)
 \biggl( \frac{-1}{N-2} \biggr) \cdots
 \biggl( \frac{-1}{1} \biggr)
 \biggl(  \frac{{\rm d}}{{\rm d}m_\rmii{B}^2}
 \biggr)^{N-1} 
 \int_\vec{p}
 \frac{1}{{p}^2 + m_\rmii{B}^2} 
 \nn & & \hspace*{1cm} =  
 \frac{(-1)^N}{(N-1)!}
 \biggl(  \frac{{\rm d}}{{\rm d}m_\rmii{B}^2}
 \biggr)^{N}
 \biggl( \frac{m_\rmii{B}^3}{6\pi} \biggr) 
 \;. \la{ring_expl}
\ea
Combining \eqs\nr{ring} and \nr{ring_expl}, we get
\ba
 \delta^{ }_\rmi{odd} f^{ }_{(N)} & = &
 \frac{(-1)^{N+1}}{N!} \biggl( \frac{3\lambda^{ }_\rmii{B}}{2}\biggr)^N
 2^{N-1} (N-1)! \biggl( \frac{T^2}{12} \biggr)^N
 T  \frac{(-1)^N}{(N-1)!} 
 \biggl(  \frac{{\rm d}}{{\rm d}m_\rmii{B}^2}
 \biggr)^{N}
 \biggl( \frac{m_\rmii{B}^3}{6\pi} \biggr) 
 \nn & = & 
 - \frac{T}{2} 
 \frac{1}{N!} \biggl( \frac{\lambda^{ }_\rmii{B} T^2}{4}\biggr)^N
 \biggl(  \frac{{\rm d}}{{\rm d}m_\rmii{B}^2}
 \biggr)^{N}
 \biggl( \frac{m_\rmii{B}^3}{6\pi} \biggr) 
 \;. \la{geneN}
\ea
As a crosscheck, it can be verified that this 
expression reproduces \eqs\nr{odd0}--\nr{odd2}.

Now, owing to the fact that \eq\nr{geneN} has precisely
the right structure to correspond to a Taylor expansion, 
we can sum the contributions in \eq\nr{geneN} to all orders, obtaining 
\be
 \sum_{N=0}^{\infty}
 \frac{1}{N!} \biggl( \frac{\lambda^{ }_\rmii{B} T^2}{4}\biggr)^N
 \biggl(  \frac{{\rm d}}{{\rm d}m_\rmii{B}^2}
 \biggr)^{N}
 \biggl( - \frac{m_\rmii{B}^3 T}{12 \pi} \biggr) 
 =
 -\frac{T}{12\pi} 
 \biggl( m_\rmii{B}^2 + \frac{\lambda^{ }_\rmii{B} T^2}{4}
 \biggr)^{\fr32}
 \;. \la{ringsum}
\ee
We observe that a ``miracle'' has happened: in \eq\nr{ringsum} the limit 
$m_\rmii{B}^2\to 0$ can be taken without divergences. But there 
is a surprise: setting the mass parameter to zero, 
we arrive at a contribution of $\rmO(\lambda_\rmii{B}^{3/2})$, 
rather than $\rmO(\lambda_\rmii{B}^2)$ as naively expected 
in \se\ref{se:naive_lam2}. In other words, infrared divergences
modify qualitatively the structure of the weak-coupling expansion.

Setting finally $m_\rmii{B}^2\to 0$ everywhere, 
and collecting all finite terms from
\eqs\nr{f0_mB}, \nr{f1_mB} and \nr{ringsum}, 
we find the correct expansion of $f(T)$
in the massless limit, 
\ba
 f(T) & = &
 -\frac{\pi^2 T^4}{90} 
  + \frac{\lambda^{ }_\rmii{B} T^4}{4\times 48} 
  -\frac{T}{12\pi} \biggl( \frac{\lambda^{ }_\rmii{B} T^2}{4} \biggr)^{3/2}
  + \rmO(\lambda_\rmii{B}^2T^4 )
 \la{fT_resum} \\ & = & 
 - \frac{\pi^2 T^4}{90}
 \biggl[
 1 - \frac{15}{32} \frac{\lambda^{ }_\rmii{R}}{\pi^2}
 + \frac{15}{16} \biggl( \frac{\lambda^{ }_\rmii{R}}{\pi^2} \biggr)^{\fr32}
 + \rmO(\lambda_\rmii{R}^2) 
 \biggr]
 \;,  \la{fT_SFT}
\ea
where at the last stage we inserted 
$\lambda^{ }_\rmii{B} = \lambda^{ }_\rmii{R} + \rmO(\lambda_\rmii{R}^2)$.  

\index{Effective field theories: scalar field}

It is appropriate to add that 
despite the complications we have found, higher-order corrections
can be computed to \eq\nr{fT_SFT}. In fact, as of today, 
the coefficients of the seven subsequent terms, of orders
$\rmO(\lambda_\rmii{R}^2)$, 
$\rmO(\lambda_\rmii{R}^{5/2}\ln \lambda^{ }_\rmii{R})$, 
$\rmO(\lambda_\rmii{R}^{5/2})$, 
$\rmO(\lambda_\rmii{R}^3 \ln \lambda^{ }_\rmii{R} )$,  
$\rmO(\lambda_\rmii{R}^3)$, 
$\rmO(\lambda_\rmii{R}^{7/2})$, and 
$\rmO(\lambda_\rmii{R}^8 \ln \lambda^{ }_\rmii{R} )$,  
are known~\cite{phi1,phi2}.
This progress is possible due to the fact that the resummation of
higher-order contributions 
that we carried out explicitly in this section can be 
implemented more elegantly and systematically
with so-called {\em effective field theory 
methods}. We return to this general procedure in \se\ref{se:EFT}, 
but some flavour can be obtained by organizing
the above computation in yet another way, outlined in the appendix below.


\subsection*{Appendix A: An alternative method for resummation}

\index{Thermal mass: scalar}
\index{Self-energy: scalar}

In this appendix we show that the previous resummation can 
also be implemented through the following steps: 
\bi

\item[(i)]
Following the computation 
of $m_\rmi{phys}^2$ in \eq\nr{pole_mass} but 
working now at finite temperature, we determine
a specific $T$-dependent pole mass in the $m^{ }_\rmii{B}\to 0$ limit. 
The result can be 
called an {\em effective thermal mass}, $m_\rmi{eff}^2$.

\item[(ii)]
We argue that in the weak-coupling limit ($\lambda^{ }_\rmii{R} \ll 1$), 
the thermal mass is important only for the Matsubara zero mode~\cite{ae}. 

\item[(iii)]
Writing the Lagrangian (for $m_\rmii{B}^2 = 0$) in the form
\ba
 {L}^{ }_\iE & = &  
 \underbrace{\frac{1}{2} \partial^{ }_\mu \phi\, \partial^{ }_\mu \phi
 + \frac{1}{2} m_\rmi{eff}^2\, \phi_{n=0}^2} + 
 \underbrace{\frac{1}{4} \lambda^{ }_\rmii{B} \phi^4 
 -  \frac{1}{2} m_\rmi{eff}^2\, \phi_{n=0}^2}
 \;, \la{LE_resum} \\ & & 
 \hspace*{1.7cm}
 {L}_0^{ }
 \hspace*{3.4cm}
 {L}_\iI^{ }
 \nonumber
\ea
we treat ${L}_0^{ }$ as the free theory and 
${L}^{ }_\iI$ as an interaction of order $\lambda^{ }_\rmii{R}$. With 
this reorganization of the theory, 
we write down the contributions $f^{ }_{(0)}$ and $f^{ }_{(1)}$ 
to the free energy density, and 
check that we obtain a well-behaved perturbative expansion 
that reproduces \eq\nr{fT_SFT}.

\ei

\index{Effective mass}

Starting with the effective mass parameter, 
the computation proceeds precisely like the one leading 
to \eq\nr{pole_mass}, with just the replacement 
$
 \raise0.3ex\hbox{$\int_{ P}$} \to \Tinti{ P}
$.
Consequently, 
\be
 m_\rmi{eff}^2 
 = \lim_{m_\rmii{B}^2\to 0}
 \Bigl[
   m_\rmii{B}^2 + 3 \lambda^{ }_\rmii{B} I(m^{ }_\rmii{B},T)
 \Bigr]
 =  3 \lambda^{ }_\rmii{B} I(0,T)
 =  \frac{\lambda^{ }_\rmii{R} T^2}{4} + \rmO(\lambda_\rmii{R}^2)
 \;. \la{mmeff}
\ee
We note that for the {\em non-zero} Matsubara modes, 
with $\omega^{ }_n\neq 0$,
we have $ m_\rmi{eff}^2  \ll \omega_n^2$
in the weak-coupling limit $\lambda^{ }_\rmii{R} \ll (4\pi)^2$, 
so that the thermal mass plays
a subdominant role in the propagator. In contrast, 
for the Matsubara zero mode, $m_\rmi{eff}^2$ modifies
the propagator significantly for ${p}^2 \ll m_\rmi{eff}^2$, 
removing any infrared divergences. This observation justifies 
the fact that the thermal mass was only introduced 
for the $n=0$ mode in \eq\nr{LE_resum}.

With our new reorganization, 
the free propagators become different for the Matsubara zero
($\tilde \phi^{ }_{n=0}$)
and non-zero 
($\tilde \phi'$)
modes: 
\ba
 \langle \tilde \phi'( P) \tilde \phi'( Q) \rangle_0^{ } 
 & = &  
 \raise-0.02em\hbox{$\bar{}$}\hspace*{-0.8mm}{\delta}
 ( P +  Q)\, \frac{1}{\omega_n^2 + {p}^2} 
 \;, \\
 \langle \tilde \phi^{ }_{n=0}( P) 
         \tilde \phi^{ }_{n=0}( Q) \rangle^{ }_0 
 & = &  
 \raise-0.02em\hbox{$\bar{}$}\hspace*{-0.8mm}{\delta}
 ( P +  Q)\, \frac{1}{{p}^2 + m_\rmi{eff}^2} 
 \;.
\ea
Consequently, \eq\nr{f0} gets replaced with 
\ba
 f^{ }_{(0)}(T) & = &
 \Tint{ P}' \frac{1}{2} \ln ( P^2)
 + T \int_{\vec{p}} \frac{1}{2} \ln({p}^2 + m_\rmi{eff}^2) - \mbox{const.}
 \nn & = &
 J'(0,T) + J^{(n=0)}(m^{ }_\rmi{eff},T)
 \nn & = & 
 -\frac{\pi^2 T^4}{90} - \frac{m_\rmi{eff}^3 T}{12\pi}
 \;. \la{resu_1}
\ea
In the massless first term, the omission of the zero mode made
no difference. 

With $f^{ }_{(1)}$ now coming from ${L}^{ }_\iI$
in \eq\nr{LE_resum}, \eq\nr{f1} is modified into
\ba
 f^{ }_{(1)}(T) & = &  \fr34 \lambda^{ }_\rmii{B}
 \langle \phi(0) \phi(0) \rangle_0^{ }
 \langle \phi(0) \phi(0) \rangle_0^{ } 
 - \frac{1}{2} m_\rmi{eff}^2 
 \langle \phi^{ }_{n=0}(0) \phi^{ }_{n=0}(0) \rangle_0^{ } 
 \nn & = & 
 \fr34 \lambda^{ }_\rmii{B}
 \Bigl[I'(0,T) + I^{(n=0)}(m^{ }_\rmi{eff},T) \Bigr]^2
 - \fr12 m_\rmi{eff}^2\, I^{(n=0)}(m^{ }_\rmi{eff},T)
 \nn & = & 
 \fr34 \lambda^{ }_\rmii{B}
 \biggl[
   \frac{T^4}{144} - \frac{m^{ }_\rmi{eff} T^3}{24\pi} 
 + \frac{m_\rmi{eff}^2 T^2}{16\pi^2}
 \biggr]
 + \fr12 m_\rmi{eff}^2  \frac{m^{ }_\rmi{eff} T}{4 \pi}
 \;. \la{resu_2}
\ea
Inserting \eq\nr{mmeff} into the last term of \eq\nr{resu_2}, 
we see that this
contribution precisely cancels against the linear term
within the square brackets. As we recall from \eq\nr{odd1}, 
the linear term was part of the problematic series that needed 
to be resummed. Combining \eqs\nr{resu_1} and \nr{resu_2}, we instead get 
\be
 f(T) = -\frac{\pi^2 T^4}{90} + 
 \frac{3 \lambda^{ }_\rmii{R}}{4} \frac{T^4}{144}
 - \frac{m_\rmi{eff}^3 T}{12\pi}
 + \rmO(\lambda_\rmii{R}^2)
 \;,
\ee
which agrees with \eq\nr{fT_SFT}.

The cancellation that took place in \eq\nr{resu_2} can also 
be verified at higher orders. In particular, proceeding
to $\rmO(\lambda_\rmii{R}^2)$, it can be seen that 
the structure in \eq\nr{odd2} gets cancelled as well. Indeed,  
the resummation of infrared divergences that we carried
out explicitly in \eq\nr{ringsum} can be fully
captured by the reorganization in \eq\nr{LE_resum}. 

%

\newpage


\newpage 

\section{Fermions}
\la{se:fermions}

\paragraph{Abstract:}

A fermionic (spin-1/2) field is considered at finite temperature. Starting
with a fermionic analogue of the harmonic oscillator and proceeding to the
case of a field satisfying a Dirac equation, an imaginary-time path
integral representation is derived for the partition function. This leads
to the concept of Grassmann variables satisfying antiperiodic boundary 
conditions. The corresponding Matsubara frequencies are introduced, the 
partition function is evaluated in the low and high-temperature expansions, 
and the structures of these expansions are compared with those of 
a scalar field theory.

\paragraph{Keywords:} 

Fermionic oscillator, Grassmann variables, antiperiodic boundary conditions, 
Dirac field, Dirac matrices, low and high-temperature expansions for fermions. 

%
\subsection{Path integral for the partition function of a
fermionic oscillator}

\index{Path integral: fermionic oscillator}
\index{Partition function: fermionic oscillator}

Just like in the bosonic case, the structure of the path integral 
for the partition function of a 
fermionic field~\cite{fb} can be derived most easily 
by first considering a {\em non-interacting}
field living in a {\em zero-dimensional} space ($d=0$). We refer
to this system as a fermionic oscillator. 

In order to introduce the fermionic oscillator, let us
start by recapitulating the main formulae of the bosonic case. 
In the operator description, the commutation relations, 
the Hamiltonian, the energy eigenstates, as well as the completeness 
relations can be expressed as
\ba
 && [\hat a, \hat a] = 0
 \;, \quad
 [\hat a^\dagger, \hat a^\dagger] = 0
 \;, \quad
 [\hat a, \hat a^\dagger] = 1
 \;; \la{bho_ops} \\ 
 && \hat H = \hbar\omega\, \Bigl( \hat a^\dagger \hat a + \fr12 \Bigr)
 = \frac{\hbar\omega}{2}\, ( \hat a^\dagger \hat a + \hat a \hat a^\dagger )
 \;; \la{bho_H} \\
 && \hat a^\dagger | n \rangle = \sqrt{n+1} | n+1 \rangle
 \;, \quad
 \hat a | n \rangle = \sqrt{n} | n - 1 \rangle
 \;, \quad n = 0,1,2,\ldots 
 \;; \la{bho_states} \\
 && \unit = 
 \sum_{n} |n \rangle \langle n |  
 = \int \! {\rm d}x \, |x\rangle\langle x | = 
 \int\! \frac{{\rm d}p}{2\pi \hbar} \,
 | p \rangle \langle p | 
 \;, \la{bho_unit}  
\ea
where we have momentarily reinstated $\hbar$. 
The observable we are interested in is 
$\mathcal{Z} = \tr[\exp(-\beta \hat H)]$, and the various
path integral representations we obtained for it read
(cf.\ \eqs\nr{Z_B}, \nr{Z_D}, \nr{recipe})
\ba
  \mathcal{Z} & = &  
 \int_{x(\beta\hbar) = x(0)} \!\!\!  
 \mathcal{D} x\, \mathcal{D} \biggl( \frac{ p }{2\pi\hbar} \biggr) \, 
  \exp\biggl\{ 
 -\frac{1}{\hbar} \int_0^{\beta\hbar} \! {\rm d}\tau \, 
 \biggl[
  \frac{p^2(\tau)}{2 m} - i p(\tau) \dot{x}(\tau)
  + V(x(\tau))
 \biggr]
 \biggr\} 
 \la{bho_Zxp} \\
 & = & 
 C \; \int_{x(\beta\hbar) = x(0)} \!\!\! \mathcal{D} x \,
  \exp\biggl\{ 
 -\frac{1}{\hbar} \int_0^{\beta\hbar} \! {\rm d}\tau \, 
 \biggl[
  \frac{m}{2} \biggl( \frac{{\rm d} x(\tau)}{{\rm d}\tau} \biggr)^2
  + V(x(\tau))
 \biggr]
 \biggr\} 
 \la{bho_Zx} \\
 & = & C \; \int_{x(\beta\hbar) = x(0)} \!\!\! \mathcal{D} x \,
 \exp\biggl( -\frac{1}{\hbar}
 \int_0^{\beta\hbar} \! {\rm d}\tau \, {L}^{ }_\iE\biggr) 
 \;, \quad
 {L}^{ }_\iE = - \mathcal{L}^{ }_\iM(t\to -i \tau)
 \;,
\ea
where $C$ is a constant, independent of the potential 
$V(x)=\tfr{1}2 m\omega^2 x^2$. 

\index{Canonical quantization: fermionic oscillator}

In the fermionic case, we replace the algebra of \eq\nr{bho_ops} by
\be 
 \{\hat a, \hat a\} = 0
 \;, \quad
 \{\hat a^\dagger, \hat a^\dagger\} = 0
 \;, \quad
 \{\hat a, \hat a^\dagger\} = 1
 \;. \la{fho_ops}
\ee
Considering the Hilbert space, i.e.~the analogue of \eq\nr{bho_states},
we define the vacuum state $|0\rangle$ by 
\be
 \hat a |0\rangle \equiv 0
 \;,
\ee
and subsequently the one-particle state $|1\rangle$ by
\be
 | 1 \rangle \equiv \hat a^\dagger | 0 \rangle
 \;. 
\ee
It is easy to see that the Hilbert space contains
no other states: operating on $|1\rangle$ with $\hat a$ or 
$\hat a^\dagger$ gives either the already known state $|0\rangle$, 
or nothing: 
\ba
 && 
 \hat a | 1 \rangle = \hat a \hat a^\dagger | 0 \rangle 
 = [1 - \hat a^\dagger \hat a]| 0 \rangle = | 0 \rangle
 \;, \\
 &&
 \hat a^\dagger |1 \rangle = 
 \hat a^\dagger \hat a^\dagger | 0 \rangle = 0 
 \;.   
\ea
Here $0$ stands for a null vector; 
$| 0 \rangle$ is a non-trivial vector representing the vacuum state. 

The Hamiltonian, i.e.~the analogue of \eq\nr{bho_H}, is an
operator acting in this vector space, and can be defined through 
\be
 \hat H 
 \equiv 
 \frac{\hbar\omega}{2}\, ( \hat a^\dagger \hat a - \hat a \hat a^\dagger )
 = 
 \hbar\omega\, \Bigl( \hat a^\dagger \hat a - \fr12 \Bigr)
 \;. \la{fho_H}
\ee
The observable of our interest is the partition function, 
\be
 \mathcal{Z} = \tr \Bigl[ e^{-\beta \hat H}\Bigr]
 = 
 \langle 0 | e^{-\beta\hat H} | 0 \rangle +
 \langle 1 | e^{-\beta\hat H} | 1 \rangle
 \;,  \la{fho_Z_def}
\ee 
which this time can be evaluated almost 
trivially due to the simplicity of the Hilbert space:
\ba
 \mathcal{Z} & = & 
 \biggl[
   \langle 0 | 0 \rangle 
 + \sum_{n=0}^{\infty}
 \frac{(-\beta\hbar\omega)^n}{n!}
 \underbrace{\langle 1 | (\hat a^\dagger \hat a)^n | 1 \rangle}
 \biggr]
 e^{\frac{\beta\hbar\omega}{2}} 
 \nn  
 & & \hspace*{4.3cm} 1 \nonumber \\[3mm] 
 & = & \Bigl[ 1 + e^{-\beta\hbar\omega}\Bigr]  
 e^{\frac{\beta\hbar\omega}{2}} 
 = 2 \cosh\Bigl(\frac{\hbar\omega}{2 T}\Bigr)
 \;. \la{ho_f_Z}
\ea
This can be compared with \eq\nr{hoZ}. 
The corresponding free energy reads
\be
  F  =   - T \ln\mathcal{Z}
  = 
  - T \ln\Bigl( e^{\frac{\hbar\omega}{2 T}} +
  e^{ - \frac{\hbar\omega}{2 T}} \Bigr)
  = - \frac{\hbar\omega}{2} - 
  T \ln \Bigl( 1 + e^{-\beta{\hbar\omega}} \Bigr)
  \;, \la{ho_ferm_F}
\ee
which can be compared with \eq\nr{hoF}.
However, like in the bosonic case, it is ultimately more
useful to write a path integral representation for the partition 
function, and this is indeed our goal.

\index{Grassmann variables}

An essential ingredient in the derivation of the bosonic path integral
was a repeated use of the completeness relations of \eq\nr{bho_unit}
(cf.\ \se\ref{se:p3}). We now need to find some analogues
of these relations for the fermionic system. 
This can be achieved with the help of {\em Grassmann variables}.
In short, the answer is that whereas in the bosonic case
the system of \eqs\nr{bho_ops} leads to commuting  
classical fields, $x(\tau), p(\tau)$, 
in the fermionic case the system of \eqs\nr{fho_ops} leads
to anti-commuting Grassmann fields, 
$c(\tau), c^*(\tau)$.
Furthermore, whereas $x(\tau)$ is periodic, and there is no 
constraint on $p(\tau)$, the fields 
$c(\tau), c^*(\tau)$ are both anti-periodic over the 
compact $\tau$-interval.

We now define the Grassmann variables $c,c^*$ (more generally, the 
Grassmann fields $c(\tau), c^*(\tau)$) through the following axioms: 
\bi

\item
 $c,c^*$ are treated as independent variables, like $x,p$.

\item
 $c^2 = (c^*)^2 \;\equiv\; 0$, 
 $c c^* = - c^* c$. 

\item
 Integration proceeds through
 $
 \int \! {\rm d}c = 
 \int \! {\rm d}c^* \equiv 0 
 $, 
 $
 \int \! {\rm d}c \, c = 
 \int \! {\rm d}c^* \, c^* \equiv 1 
 $.

\item
 The integration is also Grassmann-like, in the sense that
 $
  \{ c,{\rm d}c \} = 
  \{ c,{\rm d}c^* \} = 
  \{ c^*,{\rm d}c \} = 
  \{ c^*,{\rm d}c^* \} = 0 
 $, and similarly
 $
  \{ {\rm d}c,{\rm d}c \} = 
  \{ {\rm d}c,{\rm d}c^* \} = 
  \{ {\rm d}c^*,{\rm d}c^* \} = 0 
 $.

\item
 By convention, we write the integration measure 
 in the order $\int\! {\rm d}c^* {\rm d}c$.

\item
 A field $c(\tau)$ is a collection of independent Grassmann 
 variables, one at each point $\tau \in (0,\beta\hbar)$. 

\item
 $c,c^*$ are defined to anticommute with $\hat a, \hat a^\dagger$
 as well, so that products like $c\, \hat a^\dagger$ act as regular
 bosonic operators, e.g.\ $[c \hat a^\dagger,c^*]=0$.  

\ei

We now define a ket-state, $|c\rangle$, and a bra-state, $\langle c |$, 
which are eigenstates of $\hat a$ (``from the left'')
and $\hat a^\dagger$ (``from the right''), respectively, 
\ba
 | c \rangle & \equiv & 
 e^{- c \hat a^\dagger} | 0 \rangle 
 = (1 - c \hat a^\dagger) | 0 \rangle
 \;; \quad
 \hat a | c \rangle = c | 0 \rangle = c | c \rangle
 \;, \la{eq1} \\
 \langle c | & \equiv & 
 \langle 0 | e^{-\hat a c^*} 
 = \langle 0 | (1 - \hat a c^*)
 \;; \quad
 \langle c | \hat a^\dagger = 
 \langle 0 | c^* 
 = \langle c | c^* 
 \;, \la{eq2}
\ea
where we used $\langle 0 | \hat a^\dagger = 0$, corresponding to 
 $\hat a | 0 \rangle = 0$.
Such states possess the transition amplitude
\be
 \langle c' | c \rangle
 = 
 \langle 0 | (1 - \hat a c'^*)
 (1 - c \hat a^\dagger) | 0 \rangle
 = 
 1 + 
 \langle 0 | \hat a  c'^* c \hat a^\dagger | 0 \rangle
 = 1 + c'^* c 
 = e^{c'^* c}
 \;. \la{eq3}
\ee

With these states, we can define the objects needed,
\ba
 \int \! {\rm d}c^* {\rm d}c \, e^{-c^*c}
 |c\rangle\langle c | 
 & = & 
  \int \! {\rm d}c^* {\rm d}c \,
 (1 - c^* c) 
 (1 - c \hat a^\dagger) |0 \rangle \langle 0 | 
 (1 - \hat a c^*)
 \nn & = & 
 | 0 \rangle \langle 0 | + 
  \int \! {\rm d}c^* {\rm d}c \,
 c \, \hat a^\dagger |0 \rangle \langle 0 | \hat a\, c^*
 \nonumber \\[2mm] & = & 
 | 0 \rangle \langle 0 | + 
 | 1 \rangle \langle 1 | = \unit  
 \;, \la{fho_unit} \\ 
 \int \! {\rm d}c^* {\rm d}c \, e^{-c^*c}
 \langle -c | \hat A |c\rangle 
 & = & 
  \int \! {\rm d}c^* {\rm d}c \,
 (1 - c^* c) 
 \langle 0 | (1 + \hat a c^*) \hat A 
 (1 - c \hat a^\dagger) |0 \rangle 
  \nn & = & 
 \langle 0 | \hat A | 0 \rangle  - 
  \int \! {\rm d}c^* {\rm d}c \,
  \langle 0 | \hat a\, c^* \hat A \, 
  c \, \hat a^\dagger |0 \rangle 
 \nn & = & 
 \langle 0 | \hat A | 0 \rangle  - 
 \langle 1 |
 \int \! {\rm d}c^* {\rm d}c \,
  c^* c  \, \hat A 
  |1 \rangle 
 \nn & = & 
 \langle 0 | \hat A | 0 \rangle  + 
 \langle 1 | \hat A| 1 \rangle  = \tr[\hat A]  
 \;, \la{fho_trace}
\ea
where we assumed $\hat A$ to be a ``bosonic'' operator, 
for instance the Hamiltonian. The minus sign in the bra-state 
on the left-hand side
of \eq\nr{fho_trace} originates essentially from interchanging the
order of the state vectors on the left-hand side of \eq\nr{fho_unit}. 

Representing now the trace 
in \eq\nr{fho_Z_def} as in \eq\nr{fho_trace}, and splitting
the exponential into a product of $N$ small terms like in 
\eq\nr{Z_x}, we can write
\be
  \mathcal{Z}
 = 
 \int \! {\rm d}c^* {\rm d}c \, e^{-c^*c}
 \langle -c | 
 e^{-\frac{\epsilon \hat H}{\hbar}}  \cdots 
 e^{-\frac{\epsilon \hat H}{\hbar}} 
 |c\rangle
 \;, 
 \quad 
 \epsilon \equiv \frac{\beta\hbar}{N}
 \;. \la{Z_c}
\ee
Then we insert \eq\nr{fho_unit} in between the exponentials, as 
$
 \unit = 
 \int \! {\rm d}c_i^* {\rm d}c^{ }_i \, e^{-c_i^* c^{ }_i}
 |c^{ }_i \rangle\langle c^{ }_i | 
$,
whereby we are faced with objects like
\ba
 e^{-c_{i+1}^*c^{ }_{i+1}}
 \langle c^{ }_{i+1} |
  e^{-\frac{\epsilon}{\hbar} \hat H (\hat a^\dagger, \hat a)}
  | c^{ }_i \rangle
 & \stackrel{\rmi{\nr{eq1}, \nr{eq2}}}{=} & 
 \exp\Bigl({-c_{i+1}^*c^{ }_{i+1}}\Bigr) \,
  \langle c^{ }_{i+1} | c^{ }_i \rangle \,
  \exp\Bigl[{-\frac{\epsilon}{\hbar} H (c^*_{i+1},c^{ }_i)}\Bigr]
 \nn 
 & \stackrel{\rmi{\nr{eq3}}}{=} & 
 \exp\Bigl[ {-c_{i+1}^*c^{ }_{i+1} +  c_{i+1}^*c^{ }_{i}
 -\frac{\epsilon}{\hbar} H (c^*_{i+1},c^{ }_i)}\Bigr]
 \nn 
 & = & 
 \exp
 \biggl\{ 
   -\frac{\epsilon}{\hbar}
  \biggl[
    \hbar c^*_{i+1} \frac{c^{ }_{i+1} - c^{ }_i}{\epsilon}
   + H (c^*_{i+1},c^{ }_i)
  \biggr]
 \biggr\}  
 \;.
\ea
Finally, attention needs to be paid to the right-most
and left-most exponentials in \eq\nr{Z_c}.
We may {\em define} $c^{ }_1 \equiv c$, which clarifies the fate of 
the right-most exponential, but the left-most one needs 
to be inspected in detail: 
\ba
 & &
 \int \! {\rm d}c_1^* {\rm d}c^{ }_1 \, e^{-c_1^*c^{ }_1}
 \langle - c^{ }_1 | 
 e^{-\frac{\epsilon}{\hbar} \hat H (\hat a^\dagger, \hat a)} | 
  \int \! {\rm d}c_N^* {\rm d}c^{ }_N | c^{ }_N \rangle
 \nn & = & 
 \int \! {\rm d}c_1^* {\rm d}c^{ }_1 
 \int \! {\rm d}c_N^* {\rm d}c^{ }_N
 \exp\Bigl[
  - c_1^* c^{ }_1 - c_1^* c^{ }_N - 
 \frac{\epsilon}{\hbar} H(-c_1^*,c^{ }_N) 
 \Bigr]
 \nn & = & 
 \int \! {\rm d}c_1^* {\rm d}c^{ }_1 
 \int \! {\rm d}c_N^* {\rm d}c^{ }_N
 \exp\biggl\{ 
  -\frac{\epsilon}{\hbar}
 \biggl[ 
   \hbar c_1^* \frac{c^{ }_1 + c^{ }_N}{\epsilon}
  + H(-c_1^*,c^{ }_N) 
 \biggr]
 \biggr\}
 \nn & = & 
 \int \! {\rm d}c_1^* {\rm d}c^{ }_1 
 \int \! {\rm d}c_N^* {\rm d}c^{ }_N
 \exp\biggl\{ 
  -\frac{\epsilon}{\hbar}
 \biggl[ 
   -\hbar c_1^* \frac{- c^{ }_1 - c^{ }_N}{\epsilon}
  + H(-c_1^*,c^{ }_N) 
 \biggr]
 \biggr\}
 \;.
\ea
Thereby, we obtain in total
\ba
 \mathcal{Z} & = & 
 \int \! {\rm d}c_N^* {\rm d}c^{ }_N
 \cdots
 \int \! {\rm d}c_1^* {\rm d}c^{ }_1 
 \exp\biggl( -\frac{ S^{ }_\iE }{\hbar}  \biggr)
 \;, \\ 
 S^{ }_\iE & = & 
 \epsilon \left. \sum_{i=1}^N
 \biggl[
   \hbar c^*_{i+1} \frac{c^{ }_{i+1}-c^{ }_i}{\epsilon}
 + H (c^*_{i+1},c^{ }_i)
 \biggr]
 \right|^{ }_{c^{ }_{N+1} \equiv - c^{ }_1, c^*_{N+1} \equiv - c_1^*}
 \;.
\ea

Finally, taking the formal limit
$N\to\infty$, $\epsilon\to 0$, with
$\beta\hbar = \epsilon N$ kept fixed, we arrive at 
\be
 \mathcal{Z} = 
 \int_{\begin{array}{c} \scriptstyle
       {c(\beta\hbar) = - c(0)}, \\
       \scriptstyle {c^*(\beta\hbar) = - c^*(0)}
       \end{array}
   } \!\!\! 
 \mathcal{D} c^*(\tau) 
 \mathcal{D} c (\tau) \, 
 \exp\biggl\{
  -\frac{1}{\hbar} \int_0^{\beta\hbar}
 \! {\rm d}\tau \, 
 \biggl[
   \hbar c^*(\tau) 
   \frac{{\rm d}c(\tau)}{{\rm d}\tau} 
  + H\Bigl(c^*(\tau),c(\tau)\Bigr)
 \biggr]
 \biggr\} 
 \;. \la{fho_final}
\ee
In other words, the fermionic path integral resembles the bosonic
one in \eq\nr{bho_Zxp}, but the Grassmann fields obey {\em antiperiodic
boundary conditions} over the Euclidean time interval. 
In fact, the analogy between bosonic and fermionic
path integrals can be pushed even further, but for that we need
to specify precisely the form of the fermionic Hamiltonian.
For this, we turn to Dirac fields. 

\index{Antiperiodic boundary conditions}

\newpage 

\subsection{The Dirac field at finite temperature}
\la{se:Dirac}

In order to make use of the results of the previous section, 
we need to construct the Hamiltonian of the Dirac field and
identify the objects that play the roles of the operators 
$\hat a$ and  $\hat a^\dagger$. Our starting
point is the ``classical'' Minkowskian Lagrangian, 
\be
 \mathcal{L}^{ }_\iM = \bar \psi (i \gamma_{ }^\mu\partial^{ }_\mu - m) \psi
 \;, \la{Dirac_LM}
\ee
where 
$
 \bar\psi \equiv \psi^\dagger \gamma_{ }^0
$
and
$
 m \equiv m \cdot \unit^{ }_{4\times 4} 
$.
The Dirac $\gamma$-matrices obey the relations
\be
 \{ \gamma_{ }^\mu, \gamma_{ }^\nu \} \equiv 2 \eta^{\mu\nu}
 \;, \quad
 (\gamma_{ }^\mu)^\dagger = \gamma_{ }^0 \gamma_{ }^\mu \gamma_{ }^0 
 \;. \la{M_gammas}
\ee
The conjugate momentum is defined by
\be
 \pi = \frac{\partial \mathcal{L}^{ }_\iM }
 {\partial(\partial_0^{ }\psi)} = \bar \psi i \gamma_{ }^0 = i \psi^\dagger
 \;, 
\ee
and the Hamiltonian density subsequently becomes
\be
 \mathcal{H} = \pi \partial_0^{ } \psi - \mathcal{L}^{ }_\iM  
   = \bar\psi[-i \gamma_{ }^k \partial^{ }_k + m ] \psi
 \;.
\ee

If we now switch to operator language and recall the canonical
(anti)commutation relations, 
\ba
 & & 
 \{ \hat\psi^{ }_\alpha(x^0,\vec{x}) , \hat\psi^{ }_\beta(x^0,\vec{y}) \} =
 \{ \hat\psi_\alpha^\dagger(x^0,\vec{x}) , 
    \hat\psi_\beta^\dagger(x^0,\vec{y}) \} = 0
 \;, \\
 & & 
 \{ \hat\psi^{ }_\alpha(x^0,\vec{x}) , \hat\pi^{ }_\beta(x^0,\vec{y}) \}  
 = \{ \hat\psi^{ }_\alpha(x^0,\vec{x}) ,
    i \hat\psi_\beta^\dagger(x^0,\vec{y}) \}  
 = i \delta^{(d)}(\vec{x-y}) \delta^{ }_{\alpha\beta}
 \;, \la{Dirac_can}
\ea
where the subscripts refer to the Dirac indices, $\alpha,\beta\in\{1,...,4\}$,
we note that $\hat\psi^{ }_\alpha, \hat\psi^\dagger_\beta$ play precisely
the same roles as $\hat a, \hat a^\dagger$ in \eq\nr{fho_ops}. 
Furthermore, from
the operator point of view,
the Hamiltonian has indeed the structure of \eq\nr{fho_H}
(apart from the constant term), 
\be
 \hat H = \int_\vec{x} \, \hat\psi^\dagger (x^0,\vec{x})
 [-i \gamma_{ }^0\gamma_{ }^k \partial^{ }_k +
  m \gamma_{ }^0] \hat\psi (x^0,\vec{x})
 \;.
\ee
Rephrasing \eq\nr{fho_final} 
by denoting $c^*\to \psi^\dagger$, $c\to\psi$, and setting again $\hbar = 1$, 
the object within the square brackets, 
which we {\em define} to be 
the Euclidean Lagrangian, then reads 
\be
 {L}^{ }_\iE \;\equiv\;
 \psi^\dagger \partial^{ }_\tau \psi + 
 \psi^\dagger 
 [-i \gamma_{ }^0\gamma_{ }^k \partial^{ }_k + m \gamma_{ }^0] \psi 
 = 
 \bar\psi [\gamma_{ }^0 \partial^{ }_\tau - i \gamma_{ }^k\partial^{ }_k + m]
 \psi
 \;. \la{Dirac_LE}
\ee 
Most remarkably, a comparison of \eqs\nr{Dirac_LM} and \nr{Dirac_LE}
shows that our recipe from \eq\nr{recipe0}, 
$
  {L}^{ }_\iE = - \mathcal{L}^{ }_\iM(\tau = i t)
$, 
apparently again works. 
(Note that $i\partial_0^{ } = i \partial^{ }_t \to -\partial^{ }_\tau$.)

\index{Canonical quantization: Dirac field}
\index{Euclidean Lagrangian: Dirac field}
\index{Dirac matrices}

It is conventional and convenient to simplify 
the appearance of \eq\nr{Dirac_LE} by introducing 
so-called {\em Euclidean Dirac matrices}, through
\be
 \tilde \gamma_0^{ } \equiv \gamma_{ }^0 
 \;, \quad
 \tilde \gamma^{ }_k \equiv -i \gamma_{ }^k
 \;, \quad k = 1, \ldots, d 
 \;, \la{Dirac_E}
\ee
which according to \eq\nr{M_gammas} satisfy the algebra
\be
 \{ \tilde \gamma^{ }_\mu, \tilde \gamma^{ }_\nu \} = 2 \delta^{ }_{\mu\nu}
 \;, \quad
 \tilde \gamma_\mu^\dagger = \tilde \gamma^{ }_\mu
 \;.  \la{eucl_cliff}
\ee
We also denote 
\be
  \partial_0^{ } \equiv \partial^{ }_\tau 
\ee
from now on, understanding that repeated lower indices
imply the use of the Euclidean metric, 
and furthermore drop tildes from $\tilde\gamma^{ }_\mu$'s. 
Thereby \eq\nr{Dirac_LE} can be written in the simple form
\be
 {L}^{ }_\iE = 
 \bar\psi [\gamma^{ }_\mu  \partial^{ }_\mu + m ] \psi
 \;, \la{Dirac_LE_2}
\ee
and the partition function becomes (in continuum)
\be 
 \mathcal{Z} = 
  \int_{\begin{array}{c} \scriptstyle
       {\psi(\beta,\vec{x}) = - \psi(0,\vec{x})}, \\
       \scriptstyle {\bar\psi(\beta,\vec{x}) = - \bar\psi(0,\vec{x})}
       \end{array}
   } \!\!\! 
 \mathcal{D} \bar\psi(\tau,\vec{x}) 
 \mathcal{D} \psi (\tau,\vec{x}) 
 \, 
 \exp\biggl\{
  -\int_0^{\beta}
 \! {\rm d}\tau \int_\vec{x} \,  
 {L}^{ }_\iE
 \biggr\} 
 \;, \la{Dirac_Z}
\ee
where we substituted the integration variables from 
$\psi^\dagger$ to $\bar\psi$. Note that in the path integral formulation, 
$\psi$ and $\bar\psi$ are to be regarded 
as independent integration variables. 

\index{Euclidean Dirac matrices}

In order to evaluate $\mathcal{Z}$, 
it is useful to go to Fourier space; to this end, we write
\be
 \psi(X) \equiv \Tint{\{ P\} } e^{i  P\cdot X} 
 \tilde \psi( P)
 \;, \quad
 \bar\psi(X) \equiv
 \Tint{ \{ P\} } e^{- i  P\cdot X} 
 \,\tilde{\!\bar\psi}( P)
 \;, \la{Fou_fe}
\ee
where the curly brackets remind us of the fermionic nature of thermal sums. 
The anti-periodicity in \eq\nr{Dirac_Z} requires that 
$ P$ be of the form
\be
  P = (\omega_n^\fe,\vec{p})
 \;, \quad
 e^{i \omega_n^\fe\beta} = -1
 \;, 
\ee
whereby the {\em fermionic Matsubara frequencies} become
\be
 \omega_n^\fe = 2\pi T \Bigl(n + \fr12 \Bigr)
 \;, \quad n \in \ZZ
 \;, \la{fMatsu}
\ee
i.e.\ 
$
 \omega_n^\fe = \pm\pi T, \pm 3 \pi T, ...
$~.
Note in particular that anti-periodicity removes the 
Matsubara zero mode from the spectrum, 
implying (recalling the discussion of \se\ref{se:IR}) that there are
{\em no infrared problems associated with fermions}, at least when it comes to
``static'' observables like the partition function. In the following, 
we drop the superscript from $\omega_n^\fe$ and
indicate the fermionic nature of the Matsubara frequency with curly brackets
like in \eq\nr{Fou_fe}.

\index{Fourier representation: fermion}

\index{Matsubara frequencies: fermionic}

In the Fourier representation, the exponent in \eq\nr{Dirac_Z} becomes
\ba
 S^{ }_\iE & \equiv &
 \int_0^{\beta}
 \! {\rm d}\tau \int_\vec{x} \, 
 \bar\psi(X)
 [\gamma^{ }_\mu  \partial^{ }_\mu + m ] 
 \psi(X)
 \nn & = &
 \int_X \Tint{ \{ P \}} \Tint{ \{ Q \}}
 e^{i( P -  Q)\cdot X}
 \,\tilde{\!\bar\psi}( Q)
 [i  \gamma^{ }_\mu  P^{ }_\mu + m ]
 \tilde \psi( P)
 \nn & = & 
 \Tint{ \{ P \}} 
 \,\tilde{\!\bar\psi}( P)
 [i \bsl{ P} + m ]
 \tilde \psi( P)
 \;, \la{Dirac_SE}
\ea
where we made use of \eq\nr{deltabar}, 
and defined
$
 \bsl{ P} \equiv  \gamma^{ }_\mu  P^{ }_\mu
$. 
In contrast to real scalar fields, all Fourier modes 
are independent in the fermionic case. Up to an overall constant, 
we can then change the integration variables
in \eq\nr{Dirac_Z} to be the Fourier modes. 
Another useful result is the generalization of the simple identities  
\ba
 &&
 \int \! {\rm d}c^* {\rm d} c\, e^{-c^* a c}
 = \int \! {\rm d}c^* {\rm d} c\, [-c^* a c] = a
 \;, \\ 
 &&
 \frac{\int \! {\rm d}c^* {\rm d} c \, c \, c^* \, e^{-c^* a c}}
 {\int \! {\rm d}c^* {\rm d} c\, e^{-c^* a c}}
 = \frac{\int \! {\rm d}c^* {\rm d} c \, c \, c^*}
 {\int \! {\rm d}c^* {\rm d} c\, [-c^* a c]} = \frac{1}{a}
 \;,  
\ea
to a multicomponent case:
\ba
 &&
 \int \Bigl\{ \prod_i {\rm d}c_i^* {\rm d} c^{ }_i  \Bigr\}
 \exp\Bigl( - c_i^* M^{ }_{ij} c^{ }_j \Bigr) = 
 \det(M)
 \;, \la{fZ} \\
 &&
 \frac{
 \int \bigl\{ \prod_i {\rm d}c_i^* {\rm d} c^{ }_i  \bigr\} 
 \, c^{ }_k c^*_l \, 
 \exp\Bigl( - c_i^* M^{ }_{ij} c^{ }_j \Bigr)
 }{
 \int \bigl\{ \prod_i {\rm d}c_i^* {\rm d} c_i  \bigr\}
 \exp\Bigl( - c_i^* M^{ }_{ij} c^{ }_j \Bigr)
 }
 = (M^{-1})^{ }_{kl}
 \;. \la{fprop}
\ea

Armoured with this knowledge, we can derive explicit results for the partition
function $\mathcal{Z}$ as well as for the fermion propagator, needed
for computing perturbative corrections to the partition function. 
From \eqs\nr{Dirac_Z}, \nr{Dirac_SE} and \nr{fZ}, we first obtain
\ba
 \mathcal{Z} & = & 
 \tilde C \prod_{ \{ P\} } \det [i \bsl{ P} + m ] 
 \nn & = & 
 \tilde C 
 \Bigl( 
 \prod_{ \{ P\} } \det [i \bsl{ P} + m ] 
 \prod_{ \{ P\} } \det [-i \bsl{ P} + m ] 
 \Bigr)^{\fr12}
 \;, 
\ea
where $\tilde C$ is some constant, and have we ``replicated'' 
the determinant and compensated for that by taking 
the square root of the result.
The reason for the replication is that we may now write
\be
 [i \bsl{ P} + m ][-i \bsl{ P} + m ]
 = 
 \bsl{ P}\bsl{ P} + m^2 
 = ( P^2 + m^2) \unit^{ }_{4\times 4}
 \;, 
\ee
where we applied \eq\nr{eucl_cliff}. Thereby we get
\be \index{Partition function: Dirac field}
 \mathcal{Z} = 
 \tilde C 
 \Bigl( 
 \prod_{ \{ P\} } \det [ ( P^2 + m^2) \unit^{ }_{4\times 4} ] 
 \Bigr)^{\fr12}
 = 
 \tilde C 
 \prod_{ \{ P\} } ( P^2 + m^2)^2
 \;, 
\ee
and the free energy density $f(T)$ becomes 
\ba
 f(T) & = & \lim_{V\to\infty} \frac{F}{V}
 = \lim_{V\to\infty} \biggl( -\frac{T}{V} \ln\mathcal{Z} \biggr)
 \nn & = & 
 - \lim_{V\to\infty}  \frac{T}{V} \times 2 \sum_{ \{ P\} } 
 \ln( P^2 + m^2) + \mbox{const.}
 \nn & = & 
 - 4\; \Tint{ \{ P\} } \fr12 \ln( P^2 + m^2)
 + \mbox{const.}  
 \;, \la{f_f_final}
\ea
where we identified the sum-integration 
measure from \eq\nr{finV_final}.

The following remarks are in order: 
\bi

\item
The sum-integral appearing in \eq\nr{f_f_final} is similar to 
the bosonic one in  \eq\nr{JmT}, but is preceded by a minus sign, 
and contains fermionic Matsubara frequencies.  
These are the characteristic properties of fermions.

\item
The factor 4 in \eq\nr{f_f_final} corresponds to the 
four spin degrees of freedom of a Dirac spinor. 

\item
Like for scalar field theory in \eq\nr{JmT} or \nr{f_sft_1},
there is a constant part in $f(T)$, independent of the particle mass. 
We do not specify this term explicitly here; rather, there will be
an implicit specification below (cf.\ \eq\nr{SfSb_rel}), 
where we relate generic fermionic thermal
sums to the known bosonic ones. 
Alternatively, the result can be extracted from \eq\nr{ho_ferm_F}.

\ei

Finally, from \eqs\nr{Dirac_SE} and \nr{fprop}, we find the propagator
\be
 \langle \tilde \psi^{ }_\alpha( P) \, 
 \,\tilde{\!\bar\psi}^{ }_\beta( Q) \rangle_0^{ } 
 = \deltabar( P -  Q) 
 [i \bsl{ P} + m \unit ]^{-1}_{\alpha\beta}
 = \deltabar( P -  Q) 
 \frac{[- i \bsl{ P} + m \unit ]^{ }_{\alpha\beta}}
 { P^2 + m^2}
 \;, \la{Dirac_prop}
\ee
where the argument of the $\delta$-function is $P-Q$ 
(instead of $P+Q$)
due to the form of \eq\nr{Dirac_SE}. 
Once interactions are added, their effects 
can again be reduced to sum-integrals over products
of propagators through the application of Wick's theorem, 
cf.~\se\ref{se:wick}. However, the Grassmann
nature of the Dirac fields produces a minus sign in every 
commutation. 


\subsection*{Fermionic thermal sums}
\la{se:fsums}

\index{Thermal sums: fermion loop}

Let us now consider the same problem as in \se\ref{se:bsums}, 
but with fermionic Matsubara frequencies. That is, we need to 
perform sums of the type
\be
 \sigma^{ }_\fe \equiv T \sum_{ \{ \omega^{ }_n \} } f(\omega^{ }_n)
 \;. \la{Sfe}
\ee
Denoting for clarity the corresponding sum 
in \eq\nr{Sfn} by $\sigma^{ }_\bo$, we can write: 
\ba
 \sigma^{ }_\fe(T) & = & 
 T[ ... + f(-3\pi T) + f(-\pi T) + f(\pi T) 
  + ...]
 \nn & = & 
 T[ ... 
 + f(-3\pi T) + f(-2\pi T)+ f(-\pi T) 
 + f(0) + f(\pi T) +  f(2\pi T) 
 + ...]
 \nn &  &  - 
 T[ ... 
 + f(-2\pi T)+ f(0) + f(2\pi T) 
 + ...]
 \nn & = & 
 2 \times \frac{T}{2}\Bigl[ ... 
 + f\bigl(-6\pi \tfr{T}{2}\bigr) 
 + f\bigl(-4\pi \tfr{T}{2}\bigr)
 + f\bigl(-2 \pi \tfr{T}{2}\bigr) 
 + f\bigl(0\bigr) 
 + f\bigl(2 \pi \tfr{T}{2}\bigr) 
 + f\bigl(4 \pi \tfr{T}{2}\bigr) 
 + ...\Bigr]
 \nn & &
  - 
 T[ ... 
 + f(-2\pi T)+ f(0) + f(2\pi T) 
 + ...]
 \nn & = & 
 2 \sigma^{ }_\bo\bigl( \tfr{T}{2} \bigr) - \sigma^{ }_\bo(T)
 \;. \la{SfSb_rel}
\ea
Thereby all fermionic sums follow from the known bosonic ones, 
while the converse is {\em not} true.

To give a concrete example, consider \eq\nr{Sfn_res}, 
\be
 \sigma^{ }_\bo(T) = \int_{-\infty}^{+\infty}
 \frac{{\rm d}p}{2\pi}\, f(p) + 
 \int_{-\infty-i0^+_{ }}^{+\infty-i0^+_{ }}
 \frac{{\rm d}p}{2\pi}\,
 [f(p) + f(-p)] \nB{}(ip)
 \;.   
\ee
\Eq\nr{SfSb_rel} implies
\be
 \sigma^{ }_\fe(T) = \int_{-\infty}^{+\infty}
 \frac{{\rm d}p}{2\pi}\, f(p) + 
 \int_{-\infty-i0^+_{ }}^{+\infty-i0^+_{ }}
 \frac{{\rm d}p}{2\pi}\,
 [f(p) + f(-p)] \Bigl[ 2 \nB{}^{(\sfr{T}2)}(ip) - \nB{}^{(T)}(ip) \Bigr]
 \;,
\ee
so that the finite-temperature part has the new weight
\ba
 2 \nB{}^{(\sfr{T}2)}(ip) - \nB{}^{(T)}(ip) & = & 
 \frac{2}{\exp(2 i p \beta) -1} - \frac{1}{\exp(i p \beta) - 1}
 \nn & = & 
 \frac{1}{{\exp(i p \beta) - 1}}
 \biggl[
   \frac{2}{{\exp(i p \beta) + 1}} - 1  
 \biggr]
 = \frac{{1 - \exp(i p \beta)}}
   {[{\exp(i p \beta) - 1}][{\exp(i p \beta) + 1}]}
 \nn & = & 
 - \nF{}^{(T)}(ip)
 \;, \la{nBnF}
\ea
where $\nF{}(p) \equiv 1/[\exp(\beta p) + 1]$ is the Fermi
distribution. In total, then, fermionic sums can 
be converted to integrals according to 
\be
  \sigma^{ }_\fe(T) = \int_{-\infty}^{+\infty}
 \frac{{\rm d}p}{2\pi}\, f(p) - 
 \int_{-\infty-i0^+_{ }}^{+\infty-i0^+_{ }}
 \frac{{\rm d}p}{2\pi}\,
 [f(p) + f(-p)] \nF{}(ip)
 \;. \la{Sfe_res}
\ee


\subsection*{Appendix A: Low and high-temperature expansions for fermions}

\index{Thermal sums: high-temperature fermion}
\index{Thermal sums: low-temperature fermion}
\index{Free energy density: Dirac field}

Defining the fermionic sum-integrals
\ba
 \tilde J(m,T) & \equiv & \fr12\, \Tint{ \{ P\} } \!\!
 \Bigl[ \ln( P^2 + m^2)  - \mbox{const.}
 \Bigr] 
 \;, \la{Jf_def} \\
 \tilde I(m,T) & \equiv & \Tint{ \{ P\} } 
 \frac{1}{ P^2 + m^2}
 \;, 
\ea
we first divide them into zero and finite-temperature parts via
\be
 \tilde J(m,T) = J_0^{ }(m) + \tilde J^{ }_T(m)
 \;, \quad
 \tilde I(m,T) = I_0^{ }(m) + \tilde I^{ }_T(m)
 \;, 
\ee
where we have used the fact that at $T=0$, 
the integrals reduce to their bosonic counterparts, 
i.e.~$J_0^{ }(m)$ and $I_0^{ }(m)$ from 
\eqs\nr{J_0_1} and \nr{I0m_res}, respectively. 
In the following, our goal is to find general expressions for the functions
$\tilde J^{ }_T(m)$ and 
$\tilde I^{ }_T(m)$.
In addition we  work out their 
low and high-temperature expansions, 
noting in particular the absence of odd powers
of $m$ in the high-temperature limits. 
Finally, we also determine
the fermionic version of \eq\nr{sum2}, 
\be
 \tilde G (\tau) \;\equiv\;
 T \sum_{ \{ \omega^{ }_n \} }  
 \frac{e^{i\omega^{ }_n\tau}}
 {\omega_n^2 + \omega^2}
 \;, \quad 0 \le \tau \le \beta
 \;.
\ee

Let us proceed according to \eq\nr{SfSb_rel}. 
From \eq\nr{JmT_1}, i.e.\ 
\be
 J^{ }_T(m) 
 = 
 \int_\vec{k}  
 T \ln \Bigl( 1 - e^{-\beta{\E^{ }_{k}}}\Bigr)
 \;,
\ee
we immediately obtain 
\ba
 \tilde J^{ }_T(m) & = & 
 \int_\vec{k}  
 T \Bigl[
  \ln \Bigl( 1 - e^{-2 \beta{\E^{ }_{k}}}\Bigr)
 - \ln \Bigl( 1 - e^{-\beta{\E^{ }_{k}}}\Bigr)
 \Bigr]
 \nn & = & 
 \int_\vec{k}  
 T 
  \ln \Bigl( 1 + e^{-\beta{\E^{ }_{k}}}\Bigr)
 \;, 
\ea
whereas from \eq\nr{ImT_1}, i.e.\
\be
 I^{ }_T(m) 
 = 
 \int_\vec{k} \frac{  \nB{}(\E^{ }_{k})  }{\E^{ }_{k}}
 \;,
\ee
the same steps as in \eq\nr{nBnF} lead us to 
\be
 \tilde I^{ }_T(m) 
 = 
 - \int_\vec{k} \frac{ \nF{}(\E^{ }_{k})  }{\E^{ }_{k}} 
 \;.
\ee
Unfortunately these integrals cannot be expressed in terms
of known elementary functions.\footnote{%
Rapidly convergent sum representations in terms of 
the modified Bessel function can, however, be obtained: 
 $\tilde J^{ }_T(m) =
  - \frac{m^2 T^2}{2\pi^2} \sum_{n=1}^{\infty} \frac{(-1)^n}{n^2}
 K^{ }_2 (\frac{n m}{T})$, 
 $\tilde I^{ }_T(m) = \frac{m T}{2\pi^2} \sum_{n=1}^{\infty} \frac{(-1)^n}{n}
 K^{ }_1 (\frac{n m}{T})$. 
 }

Concerning the expansions, 
we know from \eq\nr{JTm_final} that the low-temperature limit
of $J^{ }_T$ reads 
\be
  J^{ }_T(m)  
 \approx - T^4 \Bigl( \frac{m}{2\pi T}\bigr)^{\fr32} e^{-\beta m}
 \;. 
\ee
In \eq\nr{SfSb_rel}, the first term is exponentially suppressed, 
and thus we obtain in the fermionic case
\be
  \tilde J^{ }_T(m)  
 \approx  T^4 \Bigl( \frac{m}{2\pi T}\Bigr)^{\fr32} e^{-\beta {m} }
 \;. 
\ee
{}From \eq\nr{ITm_final}, the low-temperature expansion
for $I^{ }_T$ reads
\be
 I^{ }_T(m) 
 \approx \frac{T^3}{m} 
 \Bigl( \frac{m}{2\pi T}\Bigr)^{\fr32} e^{-\beta {m} }
 \;.
\ee
Again, the first term in \eq\nr{SfSb_rel} is exponentially suppressed, 
so that we get
\be
 \tilde I^{ }_T(m) 
 \approx - \frac{T^3}{m} 
 \Bigl( \frac{m}{2\pi T}\Bigr)^{\fr32} e^{-\beta {m} }
 \;.
\ee

Moving on to the high-$T$ limit, the high-temperature expansion of 
$J^{ }_T$ reads  from \eq\nr{JTm_res}
\be
 J^{ }_T(m)
 = -\frac{\pi^2 T^4}{90}
 + \frac{m^2 T^2}{24}
 - \frac{m^3 T}{12\pi}
 - \frac{m^4}{2(4\pi)^2}
 \biggl[ 
  \ln\biggl( \frac{m e^{\gammaE}}{4\pi T} \biggr) - \fr34
 \biggr]
 + \frac{m^6\zeta(3)}{3 (4\pi)^4 T^2}
 + \ldots 
 \;.
\ee
According to \eq\nr{SfSb_rel}, we then get
\ba
 \tilde J^{ }_T(m)
 \!\!\! & = & \!\!\!
 -\fr18 \frac{\pi^2 T^4}{90}
 + \fr12 \frac{m^2 T^2}{24}
 - \frac{m^3 T}{12\pi}
 - 2 \frac{m^4}{2(4\pi)^2}
 \biggl[ 
  \ln\biggl( \frac{m e^{\gammaE}}{4\pi T} \biggr) - \fr34 + \ln2
 \biggr]
 + \frac{8 m^6\zeta(3)}{3 (4\pi)^4 T^2}
  \nn & + & 
  \frac{\pi^2 T^4}{90}
 - \frac{m^2 T^2}{24}
 + \frac{m^3 T}{12\pi}
 + \frac{m^4}{2(4\pi)^2}
 \biggl[ 
  \ln\biggl( \frac{m e^{\gammaE}}{4\pi T} \biggr) - \fr34
 \biggr]
 - \frac{m^6\zeta(3)}{3 (4\pi)^4 T^2}
 - \ldots 
 \nn & = & 
 \fr78 \frac{\pi^2 T^4}{90}
 - \frac{m^2 T^2}{48}
 - \frac{m^4}{2(4\pi)^2}
 \biggl[ 
  \ln\biggl( \frac{m e^{\gammaE}}{\pi T} \biggr) - \fr34
 \biggr]
 +  \frac{7 m^6\zeta(3)}{3 (4\pi)^4 T^2}
 + \ldots  
 \;, \la{Jf_highT}
\ea
where we note in particular the disappearance of the term cubic in $m$. 
Finally, {}from \eq\nr{ITm_res}, we may read off the 
high-temperature expansion 
of $I^{ }_T$,
\be
 I^{ }_T(m) = 
 \frac{T^2}{12} - \frac{mT}{4\pi}
 - \frac{2 m^2}{(4\pi)^2} 
 \biggl[
  \ln\biggl( \frac{m e^{\gammaE}}{4\pi T}\biggr) - \fr12 
 \biggr] 
 + \frac{2 m^4 \zeta(3)}{(4\pi)^4 T^2} + \ldots
 \;,
\ee
which together with \eq\nr{SfSb_rel} yields
\ba
 \tilde I^{ }_T(m) & = & 
 \fr12 \frac{T^2}{12}
 - \frac{mT}{4\pi}
 - \frac{4 m^2}{(4\pi)^2} 
 \biggl[
  \ln\biggl( \frac{m e^{\gammaE}}{4\pi T}\biggr) - \fr12 + \ln2 
 \biggr] 
 + \frac{16 m^4 \zeta(3)}{(4\pi)^4 T^2} 
 \nn & - & 
 \frac{T^2}{12} + \frac{mT}{4\pi}
 + \frac{2 m^2}{(4\pi)^2} 
 \biggl[
  \ln\biggl( \frac{m e^{\gammaE}}{4\pi T}\biggr) - \fr12 
 \biggr] 
 - \frac{2 m^4 \zeta(3)}{(4\pi)^4 T^2} + \ldots
 \nn & = & 
 - \frac{T^2}{24} 
 - \frac{2 m^2}{(4\pi)^2} 
 \biggl[
  \ln\biggl( \frac{m e^{\gammaE}}{\pi T}\biggr) - \fr12 
 \biggr] 
 + \frac{14 m^4 \zeta(3)}{(4\pi)^4 T^2} + \ldots
 \;. \la{If}
\ea
Again, the term odd in $m$ has disappeared. Note also that the term
$-2 m^2 \ln (m) / (4\pi)^2$ is $T$-independent and cancels against
a corresponding logarithm in $I^{ }_0(m)$ if we consider
$\tilde I(m,T)$, cf.\ \eq\nr{I0m_res}, so that 
$\tilde I(m,T)$ is formally a genuine power series in $m^2$.

Finally, we know from \eq\nr{Gtau} that the bosonic 
imaginary-time propagator reads 
\ba
 G(\tau) & = &   
 T \sum_{\omega^{ }_n} \frac{e^{i\omega^{ }_n \tau}}
 {\omega_n^2 + \omega^2}
 \nn  &  =  &
 \frac{1}{2\omega} \frac{e^{(\beta-\tau)\omega} + e^{\tau\omega}}
 {e^{\beta\omega} - 1}
 = \frac{ \nB{}(\omega) }{2 \omega} 
 \Bigl[
   e^{(\beta-\tau)\omega} + e^{\tau\omega}
 \Bigr]
 \;,
 \quad 0 \le \tau \le \beta
 \;. \la{Gebo}
\ea
Employing \eq\nr{SfSb_rel}, we obtain from here
\ba
 \tilde G(\tau) & = &
 \frac{1}{2\omega}
 \biggl\{ 
 \frac{2}{e^{2\beta \omega} - 1}
 \Bigl[
   e^{(2 \beta-\tau)\omega} + e^{\tau\omega}
 \Bigr]
  - 
 \frac{1}{e^{\beta \omega} - 1}
 \Bigl[
   e^{(\beta-\tau)\omega} + e^{\tau\omega}
 \Bigr]
 \biggr\}
 \nn & = & 
 \frac{1}{2\omega} \frac{1}{(e^{\beta \omega} - 1)(e^{\beta \omega} + 1)}
 \underbrace{\Bigl\{
    2 e^{(2 \beta-\tau)\omega} + 2 e^{\tau\omega}
 - (e^{\beta \omega} + 1)
  \Bigl[
  e^{(\beta-\tau)\omega} + e^{\tau\omega}
 \Bigr]
 \Bigr\}}
 \nn & & \hspace{5.5cm}
 (e^{\beta \omega} - 1)
 \Bigl[ 
  e^{(\beta-\tau)\omega} - e^{\tau\omega}  
 \Bigr]  
 \nn & = & 
 \frac{ \nF{}(\omega) }{2 \omega} 
 \Bigl[
   e^{(\beta-\tau)\omega} - e^{\tau\omega}
 \Bigr]
 \;,
 \quad 0 \le \tau \le \beta
 \;. \la{Gefe}
\ea \index{Thermal sums: fermion loop}

For good measure, we end the section by rederiving \eq\nr{Gefe} more 
directly, similarly to the procedure applied around \eq\nr{inB_1}.
Consider the auxiliary function 
\be
 \nF{}(ip) \equiv \frac{1}{e^{i\beta p} + 1}
 \;,
\ee
which has poles at $\exp(i\beta p) = -1$, 
i.e.\ $p = \omega_n^\fe$. The residue around each of these poles is 
\be
 \nF{}(i[\omega_n^\fe + z]) = 
 \frac{1}{-e^{i\beta z} + 1}
 \approx \frac{i T}{z} + \rmO(1)
 \;.
\ee
Therefore, letting $f(p)$ be a generic function regular
along the real axis, we can write 
\ba
 T \sum_{\{ \omega^{ }_n \}} f(\omega^{ }_n) = \frac{-i}{2\pi i}
 \oint \! {\rm d}p \, f(p)\, \nF{}(i p) 
 \;, 
\ea
where the integration contour runs anti-clockwise around 
the real axis of the complex $p$-plane.
Choosing then 
\be
 f(p) \equiv \frac{e^{i p \tau}}{p^2 + \omega^2}
 \;, 
\ee
we note that for $0 < \tau < \beta$ both half-planes are ``safe'', 
i.e.\ 
$
 e^{i p \tau} \nF{}(ip)
$
vanishes fast for $p\to \pm i \infty$. 
(The values for $\tau = 0$ and $\tau = \beta$ can be obtained
in the end from continuity.)
Therefore we can close the two parts of the integration contour in a
clockwise manner in the upper and lower half-planes, respectively. Picking
up the poles of $f(p)$ from $\pm i \omega$, this yields
\be
 T \sum_{\{ \omega^{ }_n \}} 
 \frac{e^{i \omega^{ }_n \tau}}{\omega_n^2 + \omega^2} 
 = \frac{-i}{2\pi i}(-2\pi i)
 \biggl[ 
   \frac{e^{-\omega \tau}}{2 i \omega} \nF{}(-\omega) + 
   \frac{e^{\omega\tau}}{-2 i \omega} \nF{}(\omega)
 \biggr]
 \;,
\ee
and noting that 
\be
 \nF{}(-\omega) = \frac{1}{e^{-\beta\omega}+1} = 
 \frac{e^{\beta\omega}}{e^{\beta\omega} + 1} = 
 e^{\beta\omega} \nF{}(\omega) 
 \;, 
\ee
we directly obtain \eq\nr{Gefe}. 


%

\newpage


\newpage 

\section{Gauge fields}
\la{se:Gauge}

\paragraph{Abstract:}

After introducing the concept of non-Abelian gauge invariance, 
associated with 
the existence of spin-1 gauge fields, the main elements of the canonical
quantization of gauge fields are recalled. Subsequently, an imaginary-time 
path integral expression is motivated for the partition function of such 
fields. The rules for carrying out 
a weak-coupling expansion of this quantity are formulated, 
and the corresponding Feynman rules are derived. 
This machinery is employed for defining and 
computing a thermal gluon mass, also known as a Debye
mass. Finally, the free energy density of non-Abelian black-body radiation is 
determined up to third order in the coupling constant, revealing
a highly non-trivial structure in this asymptotic series.   

\paragraph{Keywords:} 

Yang-Mills theory, gauge invariance, covariant derivative, 
Gauss law, gauge fixing, ghosts, black-body radiation, 
Stefan-Boltzmann law, Debye mass, screening, QED, QCD.

%
\subsection{Path integral for the partition function}
\la{ss:gauge_path}

\index{Path integral: gauge field}
\index{Partition function: gauge field} 
\index{Covariant derivative}
\index{Gauge invariance}

Like with fermions in \se\ref{se:Dirac}, 
our starting point with new fields is their classical Lagrangian 
in Minkowskian spacetime. 
For non-Abelian gauge fields, this has the familiar Yang-Mills form
\be \index{Yang-Mills theory}
 \mathcal{L}^{ }_\iM = 
 - \fr14 F^{a\mu\nu}F^{a}_{\mu\nu}
 \;
 \qquad
 F^a_{\mu\nu} = 
 \partial^{ }_\mu A^a_\nu - \partial^{ }_\nu A^a_\mu
 + g f^{abc} A^b_\mu A^c_\nu
 \;, \la{LM_gauge}
\ee
where $g$ is the (bare) gauge coupling and $f^{abc}$
are the \textit{structure constants} of the gauge group, 
typically taken to be  SU($\Nc$) with some $\Nc$. 
Introducing a {\em covariant derivative in the adjoint representation}, 
\be
 \mathcal{D}^{ac}_\mu 
 \equiv \partial^{ }_\mu \delta^{ac} + g f^{abc} A^b_\mu
 \;, \la{covD_ad}
\ee
we note for later reference that $F^a_{\mu\nu}$ can be expressed 
in the equivalent forms
\be
  F^a_{\mu\nu} = \partial^{ }_\mu A^a_\nu - \mathcal{D}^{ac}_\nu A^c_\mu
   = \mathcal{D}^{ac}_\mu A^c_\nu - \partial^{ }_\nu A^a_\mu  
 \;. \la{F_covD}
\ee
We can also supplement \eq\nr{LM_gauge} with matter fields: 
for instance, letting $\psi$ be a fermion in the fundamental 
representation, $\phi$ a scalar in the fundamental representation,
and $\Phi$ a scalar in the adjoint representation, we could add
the terms 
\be
 \delta \mathcal{L}^{ }_\iM
 \;=\;
 \bar \psi (i \gamma_{ }^\mu D^{ }_\mu - m ) \psi 
 \;+\; (D^\mu\phi)^\dagger D^{ }_\mu \phi 
 \;+\; \frac{1}{2}\mathcal{D}^{ac\mu} \Phi^c\, \mathcal{D}^{ad}_\mu \Phi^d 
 \;-\; V(\phi^\dagger \phi, \Phi^a\Phi^a)
 \;, \la{LM_matter}
\ee
where
$
 D^{ }_\mu = \partial^{ }_\mu  - i g A^a_\mu T^a 
$ is a
{\em covariant derivative in the fundamental representation}. 
The $\Nc\times \Nc$ matrices $T^a$ are the Hermitean generators 
of SU($\Nc$), satisfying the algebra 
$
 [T^a, T^b] = i f^{abc}T^c
$, and conventionally normalized as
$
 \tr[T^a T^b] = \delta^{ab}/2
$.

The construction principle behind \eqs\nr{LM_gauge} and \nr{LM_matter}
is that of {\em local gauge invariance}. With 
$
 U \equiv \exp[{i g \theta^a (x) T^a}] 
$, 
the Lagrangian is invariant in the transformations 
$
 A^{ }_\mu \to A_\mu'
$, 
$
 \psi\to\psi'
$, 
$
 \phi\to\phi'
$, 
$
 \Phi\to\Phi'
$, with
\ba
 A_\mu' \equiv A_\mu'^a T^a & = & 
 U A^{ }_\mu U^{-1} + \frac{i}{g} U \partial^{ }_\mu U^{-1}
 = A^{ }_\mu + i g \theta^a [T^a, A^{ }_\mu] + T^a \partial^\mu \theta^a 
 + \rmO(\theta^2) \la{Atrans0}
 \\
 \Leftrightarrow
 A_\mu'^a & = & 
 A_\mu^a + \mathcal{D}_\mu^{ac} \theta^c + \rmO(\theta^2)
 \;, \la{Atrans} \\[2mm]
 \psi' & = &  U \psi = (\unit + i g \theta^a T^a ) \psi + \rmO(\theta^2)
 \;, \la{psitrans} \\[2mm]
 \phi' & = &  U \phi = (\unit + i g \theta^a T^a ) \phi + \rmO(\theta^2)
 \;, \la{phitrans} \\[2mm]
 \Phi' \equiv \Phi'^a T^a & = & 
 U \Phi U^{-1} = \Phi + i g \theta^a [T^a, \Phi] + \rmO(\theta^2)
 \la{Phitrans0} \\[2mm]
 \Leftrightarrow
 \Phi'^a & = & 
 \Phi^a + g f^{abc}
 \Phi^b \theta^c + \rmO(\theta^2)
 \;. \la{Phitrans}
\ea

We would now like to quantize the theory of \eqs\nr{LM_gauge} and
\nr{LM_matter}, and in particular derive 
a path integral representation for its partition function. 
At this point, the role of gauge invariance becomes
conceptually slightly convoluted. It will namely turn out that: \vspace*{-2mm}
\bi
\item
The classical theory is constructed by insisting on gauge invariance.

\item
Canonical quantization and the derivation of the Euclidean path integral 
necessitate an explicit breaking of gauge invariance. 

\item
The final Euclidean path integral again displays gauge invariance.

\item
Formulating perturbation theory within the Euclidean path integral
necessitates yet again an explicit breaking of gauge invariance.

\item 
Nevertheless, only gauge invariant observables are considered physical.
 
\ei
A proper discussion of these issues goes beyond the scope of this book
but we note in passing that a deeper reason for why the breaking
of gauge invariance by gauge fixing is not considered to be a serious issue
is that the theory nevertheless maintains a certain global symmetry, 
called the BRST symmetry, which is sufficient for guaranteeing many
basic properties of the theory, such as the existence of Slavnov-Taylor
identities, or physical (``gauge-invariant'') states in its 
Hilbert space. 

\index{BRST symmetry}

As far as canonical quantization and the derivation of 
the Euclidean path integral are concerned, there are (at least) two procedures 
followed in the literature. The idea of the perhaps 
most common one is to carry 
out a complete gauge fixing (going to the axial gauge $A_3^a = 0$), 
identifying physical degrees of freedom,\footnote{%
 These are $A_1^a, A_2^a$ and the 
 corresponding canonical momenta; $A_0^a$ is expressed in terms of
 these by imposing a further constraint, the Gauss law,
 which reads $\mathcal{D}^{ab}_i F^{b}_{0 i} = 0$ 
 if no matter fields are present.} 
and then following the quantization procedure of scalar field theory. 

We take here a different approach, where the idea is to do
{\em as little gauge fixing as possible}; the price to pay is 
that one then has to be careful about the {\em states} over 
which the physical Hilbert space is constructed.\footnote{%
 This approach dates back to ref.~\cite{cwb}, and 
 can be given a precise meaning within  
 lattice gauge theory~\cite{kogut,ml}. \la{fund}
 } 
The advantage
of this approach is that the role of gauge invariance remains less 
compromised during quantization. If the evaluation of the resulting
Euclidean path integral were also to be carried out non-perturbatively
(within lattice regularization, for instance), then it would 
become rather transparent why only gauge invariant observables are physical.


\subsection*{Canonical quantization}

For simplicity, let us restrict to \eq\nr{LM_gauge} in the following, 
omitting the matter fields for the time being. 
For canonical quantization, the first step is to construct the Hamiltonian. 
We do this after setting 
\be 
 A^a_0 \equiv 0
 \;, 
\ee
which, however, fixes the gauge 
only partially; according to \eq\nr{Atrans}, 
time-independent gauge transformations are still allowed, 
given that $A^a_0$ remains zero in them.
In some sense, our philosophy is to break gauge invariance only 
to the same ``soft'' degree that Lorentz invariance is necessarily broken
in the canonical formulation through the special role that is given
to the time coordinate. 

The spatial components $A^a_i$ are now treated as 
the canonical fields (coordinates). 
According to \eq\nr{F_covD}, $F^a_{0i} = \partial_0^{ } A^a_i$, and 
\eq\nr{LM_gauge} thus becomes
\be
  \mathcal{L}^{ }_\iM = \fr12 \partial_0^{ } A^a_i \partial_0^{ } A^a_i 
 - \fr14 F^{a}_{ij}F^{a}_{ij}
 \;. 
\ee
The canonical momenta corresponding to $A^a_i$, 
denoted by $E^a_i$, take the form
\be
 E^a_i \equiv \frac{\partial \mathcal{L}^{ }_\iM}
                   {\partial (\partial_0^{ } A^a_i)}
 = \partial_0^{ } A^a_i
 \;, \la{Eai}
\ee
and the Hamiltonian density subsequently reads
\be
 \mathcal{H} = E^a_i \partial_0^{ } A^a_i - \mathcal{L}^{ }_\iM
 = \fr12 E^a_i E^a_i + \fr14 F^{a}_{ij}F^{a}_{ij}
 \;. \la{H_gauge_cl}
\ee
We also note that the ``multiplier'' of $A^a_0$ 
in the action (before gauge fixing) 
reads, according to \eq\nr{F_covD}, 
\be
 \frac{\delta \mathcal{S}^{ }_\iM}{\delta A^a_0} = 
 \frac{\delta}{\delta A^a_0}
 \int_\mathcal{X} 
 \biggl[ \fr12(\partial_0^{ } A^b_i - \mathcal{D}^{bc}_i A^c_0) 
 F^{b}_{0 i} \biggr]
 = \mathcal{D}^{ab}_i F^{b}_{0 i}
 \;, \la{G_gauge_cl}
\ee
where we made use of the identity
\be
   \int_\mathcal{X} f^a(\mathcal{X}) \mathcal{D}^{ab}_\mu g^b(\mathcal{X}) = 
 - \int_\mathcal{X} g^a(\mathcal{X}) \mathcal{D}^{ab}_\mu f^b(\mathcal{X})
 \;. \la{ibp}
\ee
The object in \eq\nr{G_gauge_cl} is identified as 
the left-hand side of the non-Abelian Gauss law. 

\index{Canonical quantization: gauge field}
\index{Gauss law}

The theory can now be canonically quantized by promoting $A^a_i$ and 
$E^a_i$ into operators, and by imposing standard bosonic 
equal-time commutation relations between them, 
\be
 [\hat A^a_i(t,\vec{x}),\hat E^b_j(t,\vec{y})] 
 = i \delta^{ab} \delta^{ }_{ij} \delta(\vec{x - y})
 \;. \la{AcE}
\ee
According to \eq\nr{H_gauge_cl}, the Hamiltonian then becomes
\be
 \hat H = \int_\vec{x} 
 \biggl( \fr12 \hat E^a_i \hat E^a_i + \fr14 \hat F^{a}_{ij} \hat F^{a}_{ij}
 \biggr)
 \;. \la{H_gauge_qm}
\ee

A very important role in the quantization is played 
by the so-called Gauss law operators, 
cf.~\eq\nr{G_gauge_cl}. Combining this expression 
with $F^b_{0i} = \partial_0^{ } A^b_i = E^b_i$, we write them in the form 
\be
 \hat G^a = \hat \mathcal{D}_i^{ab} \hat E^b_i 
 \;, \quad a = 1, \ldots, \Nc^2 - 1
 \;, \la{G_gauge_qm}
\ee
and furthermore define an operator parametrized by time-independent
gauge transformations, 
\be
 \hat U \;\equiv\; \exp\Bigl\{
  - i \int_\vec{x}\, \theta^a(\vec{x}) \hat G^a(\vec{x}) 
 \Bigr\}
 \;. \la{hatU}
\ee

We now claim that $\hat U$ 
{\em generates time-independent gauge transformations}.
Let us prove this to the leading non-trivial order in $\theta^a$.
First of all, 
\ba
 \hat U \hat A^b_j(\vec{y}) \hat U^{-1} 
 \!\! & = & \!\!  \hat A^b_j(\vec{y}) 
 - i \int_\vec{x}\,
 \theta^a(\vec{x}) [\hat G^a(\vec{x}), \hat A^b_j(\vec{y})] + 
 \rmO(\theta^2)
 \nn 
 \!\! & = & \!\!  \hat A^b_j(\vec{y}) 
 - i \int_\vec{x}\,
 \theta^a(\vec{x}) 
 \Bigl\{ 
 \partial^\vec{x}_i [\hat E^a_i(\vec{x}), \hat A^b_j(\vec{y})]
 + g f^{acd} \hat A^c_i(\vec{x}) [\hat E^d_i(\vec{x}), \hat A^b_j(\vec{y})]
 \Bigr\} 
 + \rmO(\theta^2)
 \nn  
 \!\! & = & \!\!  \hat A^b_j(\vec{y}) 
 - \int_\vec{x}\,
 \theta^a(\vec{x}) 
 \Bigl\{ 
 \partial^\vec{x}_i \delta^{ab}\delta^{ }_{ij}\delta(\vec{x-y})
 + g f^{acd} \hat A^c_i(\vec{x})
   \delta^{db} \delta^{ }_{ij} \delta(\vec{x-y})
 \Bigr\} 
 + \rmO(\theta^2)
 \nn  
 \!\! & = & \!\!  \hat A^b_j(\vec{y}) 
 + \partial^{ }_j \theta^b(\vec{y}) 
 + g f^{bca} \hat A^c_i(\vec{y}) \theta^a(\vec{y})
 + \rmO(\theta^2)
 \nn[2mm] 
 \!\! & = & \!\!  \hat A^b_j(\vec{y}) 
 + \hat \mathcal{D}_j^{ba} \theta^a(\vec{y})  + \rmO(\theta^2)
 \nn[2mm] 
 \!\! & = & \!\!
 \hat A'^b_j(\vec{y}) + \rmO(\theta^2) 
 \;, \la{hatAp}
\ea
where we used the antisymmetry of the structure constants 
as well as \eq\nr{Atrans}. Similarly, 
\ba
 \hat U \hat E^b_j(\vec{y}) \hat U^{-1} 
 \!\! & = & \!\!  \hat E^b_j(\vec{y}) 
 - i \int_\vec{x}\,
 \theta^a(\vec{x}) [\hat G^a(\vec{x}), \hat E^b_j(\vec{y})] + 
 \rmO(\theta^2)
 \nn 
 \!\! & = & \!\!  \hat E^b_j(\vec{y}) 
 - i \int_\vec{x}\,
 \theta^a(\vec{x}) 
 \Bigl\{ 
 + g f^{acd}  [\hat A^c_i(\vec{x}), \hat E^b_j(\vec{y})] \hat E^d_i(\vec{x})
 \Bigr\} 
 + \rmO(\theta^2)
 \nn  
 \!\! & = & \!\!  \hat E^b_j(\vec{y}) 
 + \int_\vec{x}\,
 \theta^a(\vec{x}) 
 g f^{acd}  \delta^{cb}\delta^{ }_{ij} \delta(\vec{x-y}) \hat E^d_i(\vec{x})
 + \rmO(\theta^2)
 \nn  
 \!\! & = & \!\!  \hat E^b_j(\vec{y}) 
 + g f^{bda}  \hat E^d_j(\vec{y}) \theta^a(\vec{y}) 
 + \rmO(\theta^2)
 \nn[2mm]
 \!\! & = & \!\!
  \hat E'^b_j(\vec{y})
 + \rmO(\theta^2)
 \;, \la{hatEp}
\ea
where the result corresponds to the transformation law 
of an adjoint scalar, cf.~\eq\nr{Phitrans}.

One important consequence of \eqs\nr{hatAp} and \nr{hatEp} is that 
{\em the operators $\hat G^a$ commute with the Hamiltonian $\hat H$}.
This follows from the fact that the Hamiltonian 
of \eq\nr{H_gauge_qm} is gauge-invariant in time-independent
gauge transformations, as long as $\hat E^a_i$ transforms 
as an adjoint scalar. This leads to  
\be
 \hat U \hat H \hat U^{-1} = \hat H 
 \quad\Rightarrow \quad
 [\hat G^a(\vec{x}), \hat H] = 0 \quad \forall\, \vec{x} 
 \;. \la{GcH}
\ee
Another implication of these results is that $\hat U$ transforms
eigenstates as well: if 
$
 \hat A^a_i | A^a_i \rangle = A^a_i | A^a_i \rangle
$, 
then
\ba
 \hat A^a_i \hat U^{-1} | A^a_i \rangle
 & = &  \hat U^{-1} \hat A'^a_i | A^a_i \rangle
 = \hat U^{-1} \bigl[     
   \hat A^a_i
 + \hat \mathcal{D}_j^{ab} \theta^b  + \rmO(\theta^2)
  \bigr] | A^a_i \rangle
 \nn 
 & = & 
 \hat U^{-1} \bigl[     
   A^a_i
 + \mathcal{D}_j^{ab} \theta^b  + \rmO(\theta^2)
  \bigr] | A^a_i \rangle
 = \hat U^{-1} A'^a_i | A^a_i \rangle
 \nn 
 & = & A'^a_i\hat U^{-1}  | A^a_i \rangle
 \;, 
\ea
where we made use of \eq\nr{hatAp}.
Consequently, we can identify
\be
 \hat U^{-1} | A^a_i \rangle = 
 | A'^a_i \rangle
 \;.
\ee

Let us now define a physical state, ``$|\mbox{phys}\rangle$'', 
to be one which is gauge-invariant: 
$
 \hat U^{-1} |\mbox{phys}\rangle 
 = |\mbox{phys}\rangle
$.
Expanding to first order in $\theta^a$, we see that 
these states must satisfy
\be
 \hat G^a(\vec{x}) |\mbox{phys}\rangle = 0 \quad \forall\, \vec{x}
 \;,
\ee
which is an operator manifestation of the statement
that {\em physical states must obey the Gauss law}. 
Moreover, given that the Hamiltonian commutes with $\hat G^a$, 
we can choose the basis vectors of the Hilbert space to be 
simultaneous eigenstates of $\hat H$ and $\hat G^a$. Among all of 
these states, only the ones with zero eigenvalue of $\hat G^a$
are physical; it is then only these states which are to be used in 
the evaluation of $\mathcal{Z} = \tr[\exp(-\beta \hat H)]$.

After these preparations, we are finally in a position 
to derive a path integral expression for $\mathcal{Z}$. In terms
of quantum mechanics, we have a system with a Hamiltonian $\hat H$
and a commuting operator, $\hat Q$, whose role is played by $\hat G^a$.
One could in principle consider the grand canonical partition function, 
$
 \mathcal{Z}(T,\mu) = \tr\{ \exp[-\beta (\hat H - \mu \hat Q) ] \}
$, but 
according to the discussion above we are only interested in the  
contribution to $\mathcal{Z}$ from the states with zero ``charge'',
$\hat Q|\mbox{phys}\rangle = 0$. To this end, 
it is more natural to remain in the canonical picture, 
and we thus label the states with 
the eigenvalues $\E^{ }_q,q$, so that  
$
 \hat H |\E^{ }_q,q \rangle = \E^{ }_q |\E^{ }_q,q \rangle
$, 
$
 \hat Q |\E^{ }_q,q \rangle = q |\E^{ }_q,q \rangle
$. 
Assuming for concreteness that 
the eigenvalues $q$ of $\hat Q$ are integers,
we can write the relevant partition function 
by taking a trace over {\em all states}, but inserting a 
Kronecker-$\delta$ inside the trace,
\ba
 \mathcal{Z}^{ }_\rmi{phys} & \equiv &
 \sum_{\E_0^{ }} \langle \E_0^{ },0| e^{-\beta \E_0^{ }} | \E_0^{ },0 \rangle 
 = 
 \sum_{\E^{ }_q,q} \langle \E^{ }_q,q|
  \delta^{ }_{q,0} e^{-\beta \E^{ }_q}  | \E^{ }_q,q \rangle 
 =
 \tr\Bigl[ \delta^{ }_{\hat Q,\hat 0} e^{-\beta \hat H}  \Bigr]
 \;, \la{Z_constrained}
\ea
where 
$
 \delta^{ }_{\hat Q,\hat 0} | \E^{ }_q,q \rangle \equiv 
 \delta^{ }_{q,0}  | \E^{ }_q,q \rangle
$.

Given that 
$
 \delta^{ }_{\hat Q,\hat 0} =
 \delta^{ }_{\hat Q,\hat 0} \delta^{ }_{\hat Q,\hat 0}
$ and 
$
 [\hat H, \hat Q] = 0
$, 
we can write 
\be
 \mathcal{Z}^{ }_\rmi{phys} = 
 \tr 
 \Bigl[
 \underbrace{
 \delta^{ }_{\hat Q,\hat 0} e^{-{\epsilon} \hat H} 
 \delta^{ }_{\hat Q,\hat 0} e^{-{\epsilon} \hat H} 
 \ldots
 \delta^{ }_{\hat Q,\hat 0} e^{-{\epsilon} \hat H}
 }_{N\;\rmi{parts}}  
 \Bigr]
 \;,
\ee
where $\epsilon = \beta/N$ and $N\to\infty$ as before. 
Here, we may further represent 
\be
 \delta^{ }_{\hat Q,\hat 0} = \int_{-\pi}^{\pi}
 \! \frac{{\rm d}\theta^{ }_i}{2\pi} \, e^{i \theta^{ }_i \hat Q}
 = 
 \int_{-\pi/\epsilon}^{\pi/\epsilon}
 \! \frac{{\rm d}y^{ }_i}{2\pi \epsilon^{-1}} \, e^{i \epsilon y^{ }_i \hat Q}
 \;, 
\ee
and insert unit operators as in \eq\nr{ip1i}, but placing now the momentum 
state representation between 
$
 \delta^{ }_{\hat Q,\hat 0}
$
and
$
 \exp({-{\epsilon} \hat H})
$. The typical building block of the discretised path integral then reads
\ba
 & &  \hspace*{-1cm} \langle x^{ }_{i+1} | 
 e^{i \epsilon y^{ }_i \hat Q(\hat x,\hat p)} 
 | p^{ }_i \rangle \langle p^{ }_i | 
 e^{-{\epsilon} \hat H(\hat{p},\hat{x})} | x^{ }_i \rangle 
 \nn 
 & & =   
 \exp\biggl\{  
 -{\epsilon}
 \biggl[ - i y^{ }_i Q(x^{ }_{i+1},p^{ }_i)
   + \frac{p_i^2}{2m} - i p^{ }_i \frac{x^{ }_{i+1}-x^{ }_{i}}{\epsilon}
   + V(x^{ }_i) + \rmO(\epsilon)
 \biggr] 
 \biggr\} 
 \;. \la{Zn0}
\ea

It remains to
take the limit $\epsilon\to 0$, whereby $x^{ }_i,p^{ }_i,y^{ }_i$
become functions, $x(\tau), p(\tau), y(\tau)$, 
and  to replace 
$x(\tau) \to A^a_i$, 
$p(\tau) \to E^a_i$, 
$y(\tau) \to \tilde A_0^a$, 
$Q \to \mathcal{D}^{ab}_i E^b_i$, 
$m\to 1$. 
Then the integral over the square brackets in \eq\nr{Zn0} becomes
\ba
 & & \hspace*{-2cm} \int_X \biggl[ 
 -i \tilde A^a_0 \mathcal{D}^{ab}_i E^b_i + \fr12 E^a_i E^a_i 
 -i E^a_i \partial^{ }_\tau A^a_i + \fr14 F^a_{ij} F^a_{ij}
 \biggr] \nn
 & = &  \int_X \biggl[
 \fr12 E^a_i E^a_i - i E^a_i 
 \Bigl( \partial^{ }_\tau A^a_i - \mathcal{D}^{ab}_i \tilde A^b_0 \Bigr) 
  + \fr14 F^a_{ij} F^a_{ij}
 \biggr]
 \;,  \la{Zn0p}
\ea
where we made use of \eq\nr{ibp}. 

\index{Euclidean Lagrangian: gauge field}

At this point, we make a curious observation: 
inside the round brackets in \eq\nr{Zn0p} there is
an expression of the form we encountered in \eq\nr{F_covD}. 
Of course, the field 
$\tilde A^a_0$ is {\em not} the original $A^a_0$-field, which was
set to zero, but rather a new field, which we are however free
to {\em rename} as $A^a_0$. Indeed, in the following we leave out
the tilde from $\tilde A^a_0$, and redefine a {\em Euclidean} field 
strength tensor according to 
\be
 F^a_{0i} \equiv \partial^{ }_\tau A^a_i - \mathcal{D}^{ab}_i  A^b_0
 \;.  \la{F0i}
\ee 
Noting furthermore that 
\be
 \fr12 E^a_i E^a_i - i E^a_i F^a_{0i} 
 = \fr12 (E^a_i - i F^a_{0i})^2 + \fr12 F^a_{0i} F^a_{0i}
 \;, 
\ee
we can carry out the Gaussian integral over $E^a_i$, and end
up with the desired path integral expression for 
the partition function of the theory: 
\be \index{Yang-Mills theory}
 \mathcal{Z}^{ }_\rmi{phys} 
 = C \int \! \mathcal{D} A^a_0 
 \int_{A^a_i(\beta,\vec{x}) = A^a_i(0,\vec{x})}
 \!\!\! \mathcal{D} A^a_i \, 
 \exp\biggl\{ 
  - \int_0^\beta \! {\rm d}\tau \! \int_\vec{x} \,
 {L}^{ }_\iE  
 \biggr\} 
 \;, \quad
 {L}^{ }_\iE = \fr14 F^a_{\mu\nu} F^a_{\mu\nu}
 \;. \la{Z_gauge} \la{LE_gauge}
\ee
In the next section, we address the evaluation of this quantity using 
a weak-coupling expansion, which requires us to return 
to the question of gauge fixing.

Two final remarks are in order: 
\bi
\item
The field $A^a_0$ was introduced in order to 
impose the Gauss law at every $\tau$, and therefore 
the integrations at each $\tau$ are independent of each other. In other words, 
it is not obvious from the derivation of the path integral 
whether the field $A^a_0$ should satisfy periodic boundary 
conditions like the spatial components $A^a_i$ do. 

It may be noted, however, that the fields to which the $A^a_0$
couple in \eq\nr{LE_gauge} do obey periodic boundary conditions. 
This suggests that we can consider them to live on a circle, 
and therefore make the same choice for $A^a_0$ itself. 
Further evidence comes from a perturbative computation around
\eq\nr{f_QCD_0}, showing that only a periodic $A^a_0(\tau)$ 
leads to physical results in a simple way. We make
this choice in the following. It is perhaps also appropriate to 
remark that in lattice gauge theory, the fields $A^a_0$ live on 
timelike ``links'', rather than ``sites'', and the question of
periodicity does not directly concern them. 

\item \la{AMAE_rel}
For scalar field theory and fermions, 
\eqs\nr{LE_expl} and \nr{Dirac_LE}, 
we found after a careful derivation of the Euclidean path integral
that the result could be interpreted in terms of a simple recipe: 
${L}^{ }_\iE = - \mathcal{L}^{ }_\iM(t\to - i \tau)$. We may now ask
whether the same is true for gauge fields.

A comparison
of \eqs\nr{LM_gauge} and \nr{LE_gauge} shows that, indeed, the recipe
again works. The only complication is that the Minkowskian 
$A_0^a$ needs to be replaced with $i \tilde A_0^a$ (of which we 
have normally left out the tilde), just like 
$\partial^{ }_t$ gets replaced with $i \partial^{ }_\tau$. This reflects
the structure of gauge invariance, implying that 
covariant derivatives change as $D^{ }_t \to i D^{ }_\tau$. 

\ei

\newpage 

\subsection{Weak-coupling expansion}
\la{ss:expansion}

\index{Weak-coupling expansion: gauge field}
\index{Gauge fixing and ghosts}

\subsection*{Gauge fixing and ghosts}

The path integral representation in \eq\nr{LE_gauge} 
is manifestly gauge invariant and could 
in principle (after a suitable regularization) be evaluated as such. 
As before we restrict our treatment here 
to perturbation theory; in this case, it turns out that
gauge invariance needs to be broken once again, because 
the quadratic part of ${L}^{ }_\iE$ otherwise contains 
a non-invertible matrix, so that no propagators can be defined.  
For completeness, let us recall the main steps of this procedure. 

\index{Gribov ambiguity}

Let $G^a$ now be some function 
of the path integration variables in \eq\nr{LE_gauge}, 
for instance $G^a(X) = A^a_3(X)$ or 
$G^a(X) = -\partial^{ }_\mu A^a_\mu(X)$ (note that our notation has changed
here, and this function has no relation to the Gauss law). Ideally
the function should be so chosen that the equation
$G^a = 0$ has a unique solution for $A^a_\mu$; 
otherwise we are faced with the so-called Gribov ambiguity.
The idea is then to insert the object 
\be
 \prod_{X,Y,a,b}\delta(G^a) 
 \det\biggl[ \frac{\delta G^a(X)}{\delta \theta^b(Y)} \biggr]
 \la{gauge_fix}
\ee
as a multiplier in front of the exponential in \eq\nr{LE_gauge}, 
in order to remove the (infinite) redundancy related to integrating 
over physically equivalent gauge configurations, the ``gauge orbits''. 
Indeed, it appears that this insertion
does not change the value of gauge invariant expectation values, 
but merely induces an overall constant in $\mathcal{Z}$, 
analogous to $C$. First of all, since ${L}^{ }_\iE$ is 
gauge invariant, its value within each gauge orbit 
does not depend on the particular form 
of the constraint $G^a = 0$. Second, inspecting the integration measure, 
we can imagine dividing the integration into one over 
gauge non-equivalent fields, $\bar A_\mu$, and another over 
gauge transformations thereof, parametrized by~$\theta$. Then
\ba
 & & \hspace*{-1cm} \int \! \mathcal{D} A^{ }_\mu \, 
 \delta(G^a) \det\Bigl[ \frac{\delta G^a}{\delta \theta^b}\Bigr]
 \exp\Bigl\{  - \int_X {L}^{ }_\iE( A^{ }_\mu) \Bigr\}
 \nn 
 & = & 
 \int \! \mathcal{D} \bar A^{ }_\mu \int \! \mathcal{D}\theta^b \, 
 \delta(G^a) \det\Bigl[ \frac{\delta G^a}{\delta \theta^b}\Bigr]
 \exp\Bigl\{  - \int_X {L}^{ }_\iE( \bar A^{ }_\mu) \Bigr\}
 \nn & = & 
 \int \! \mathcal{D} \bar A^{ }_\mu \int \! \mathcal{D}G^a \, 
 \delta(G^a) 
 \exp\Bigl\{  - \int_X {L}^{ }_\iE( \bar A^{ }_\mu) \Bigr\}
 \nn & = & 
 \int \! \mathcal{D} \bar A^{ }_\mu 
 \exp\Bigl\{  - \int_X {L}^{ }_\iE( \bar A^{ }_\mu) \Bigr\}
 \;. \la{G_indep}
\ea
In other words, the result seems to exhibit no dependence on 
the particular choice of $G^a$.\footnote{%
  The arguments presented are heuristic in nature. 
  The manipulations can be given a precise
  meaning in lattice regularization, where the integration measure
  is well defined as the gauge invariant Haar measure on SU($\Nc$).}

Given that the outcome is independent of $G^a$, it is conventional
and convenient to replace $\delta(G^a)$ by
$\delta(G^a - f^a)$, where $f^a$ is some $A_\mu^a$-independent function, 
and then to average over the $f^a$'s with a Gaussian weight. 
This implies writing
\ba
 \delta(G^a) & \rightarrow &
 \int \! \mathcal{D} f^a \,  \delta(G^a - f^a) 
 \exp\biggl( -\frac{1}{2\xi} 
 \int_X f^a f^a \biggr)
 \nn & = &  
 \exp\biggl( -\frac{1}{2\xi} 
 \int_X G^a G^a \biggr)
 \;,
\ea
where an arbitrary parameter, $\xi$, 
has been introduced. Its presence at intermediate stages of a
perturbative calculation permits for a very efficient 
and non-trivial crosscheck, since all dependence on it 
must vanish in the final results for physical (gauge invariant) quantities.

\index{Faddeev-Popov ghosts}

Finally, the other structure in \eq\nr{gauge_fix}, namely the determinant, 
can be written in terms of Faddeev-Popov ghosts~\cite{fp}, making
use of \eq\nr{fZ}, 
\be
 \det (M) = \int \! \mathcal{D} \bar c\, \mathcal{D} c \, 
 \exp\Bigl( - \bar c\, M c \Bigr)
 \;.  
\ee 
Given that the ``matrix'' 
${\delta G^a}/{\delta \theta^b}$ is purely bosonic, 
ghost fields should obey {\em the same boundary conditions
as gauge fields}, i.e.\ be periodic in spite of their 
Grassmann nature. 

\index{Euclidean Lagrangian: QCD}

In total, then, we can write the gauge-fixed version of \eq\nr{LE_gauge}, 
adding now also Dirac fermions to complete the theory into QCD. 
The result reads 
\ba \index{BRST symmetry}
 \mathcal{Z}^{ }_\rmi{phys} & = &  
 C 
 \int_\rmi{periodic} \!\! \mathcal{D} A_0^a \, \mathcal{D} A_k^a
 \int_\rmi{periodic} \!\! \mathcal{D} \bar c^{\,a} \, \mathcal{D} c^a 
 \int_\rmi{anti-periodic} \!\! \mathcal{D} \bar\psi \, \mathcal{D} \psi
 \nn & \times &  
 \exp\biggl\{
  - \int_0^\beta \! {\rm d}\tau \! \int_\vec{x} \,  
  \biggl[
   \fr14 F^a_{\mu\nu}F^a_{\mu\nu}
   + \frac{1}{2\xi} G^a G^a
   + \bar c^{\,a} \Bigl( \frac{\delta G^a}{\delta \theta^b}\Bigr) c^b 
  + \bar \psi ( \gamma^{ }_\mu D^{ }_\mu + m ) \psi   
  \biggr] 
 \biggr\} 
 \;, \nn \la{Z_full}
\ea
where we have on purpose simplified the quark mass term by assuming the
existence of one flavour-degenerate mass $m$.\footnote{%
 A more general Euclidean Lagrangian, incorporating
 all fields of the Standard Model, is given on p.~\pageref{SM}.
 }  
We remark again, although do not prove here, 
that the argument of the exponent in \eq\nr{Z_full}
is invariant under BRST symmetry. 
It will turn out to be 
convenient to make a particular choice for the functions $G^a$ by 
selecting {\em covariant gauges}, defined by
\ba
 G^a & \equiv & -\partial^{ }_\mu A^a_\mu
 \;, \la{Ga_cov} \\
 \frac{1}{2\xi} G^aG^a & =  &
 \frac{1}{2\xi} \partial^{ }_\mu A^a_\mu \, \partial^{ }_\nu A^a_\nu
 \;, \\
 \frac{\delta G^a}{\delta \theta^b} & =  & 
 + \overleftarrow{\!\partial}^{ }_{\!\!\mu}
   \frac{\delta A^a_\mu}{\delta \theta^b}
 = \overleftarrow{\!\partial}^{ }_{\!\!\mu}
 \Bigl[ 
  \overrightarrow{\!\partial}^{ }_{\!\!\mu} \delta^{ab} + g f^{acb} A^c_\mu
 \Bigr]
 \;, \\
 \bar c^{\,a} \Bigl( \frac{\delta G^a}{\delta \theta^b}\Bigr) c^b & = & 
 \partial^{ }_\mu \bar c^{\,a} \partial^{ }_\mu c^a + 
 g f^{abc} \partial^{ }_\mu \bar c^{\,a} A^b_\mu c^c
 \;.
\ea
Here we made use of \eqs\nr{covD_ad} and \nr{Atrans}.


\subsection*{Feynman rules for Euclidean continuum QCD}

\index{Feynman rules: Euclidean QCD}
\index{QCD}

For completeness, we now collect together the Feynman rules 
that apply to computations within the theory defined by \eq\nr{Z_full}, 
when the gauge is fixed according to \eq\nr{Ga_cov}.

Consider first the free (quadratic) part of the Euclidean action. 
Expressing everything in the Fourier representation, this becomes
\ba
 S^{ }_\rmi{$E$,$0$} & = &
 \Tint{ P  Q}
 \! \deltabar( P +  Q)
 \biggl\{
  \fr12 i  P^{ }_\mu \tilde A^a_\nu( P)
 \Bigl[ i  Q^{ }_\mu \tilde A^a_\nu( Q) - 
        i  Q^{ }_\nu \tilde A^a_\mu( Q) \Bigr] 
  + \frac{1}{2\xi} i  P^{ }_\mu \tilde A^a_\mu( P)
 \, i  Q^{ }_\nu \tilde A^a_\nu( Q)
 \biggr\}
  \nn & + & 
 \Tint{ P  Q}
 \! \deltabar(-  P +  Q)
 \Bigl[ - i  P^{ }_\mu \tilde{\bar c}^{\,a}( P)
 \, i  Q^{ }_\mu \tilde c^a( Q) \Bigr]
 + 
 \Tint{\{  P  Q \} }
 \! \deltabar(-  P +  Q)
 \,\,\tilde{\!\bar\psi}^{ }_A( P)
 [i  \gamma^{ }_\mu  Q^{ }_\mu + m ] \tilde \psi^{ }_A( Q) 
 \nn & = & 
 \Tint{ P  Q}
 \! \deltabar( P +  Q)
 \biggl\{
  \fr12 \tilde A^a_\mu( P) \tilde A^a_\nu( Q)
 \Bigl[  P^2 \delta^{ }_{\mu\nu} - 
 \Bigl( 1 - \frac{1}{\xi} \Bigr)   P^{ }_\mu  P^{ }_\nu
 \Bigr]
 \biggr\}
 \nn & + & 
 \Tint{ P  Q}
 \! \deltabar(-  P +  Q)
 \Bigl[ \tilde{\bar c}^{\,a}( P) \tilde c^a( Q)  P^2  \Bigr]
 + 
 \Tint{\{  P  Q \} }
 \! \deltabar(-  P +  Q)
 \,\,\tilde{\!\bar\psi}^{ }_A( P)
 [i \bsl{ P} + m ] \tilde \psi^{ }_A( Q) 
 \;, \la{SE_2}
\ea
where the index $A$ for the quarks is assumed to comprise both 
colour and flavour indices, whereas in the Dirac space $\bar\psi$ 
and $\psi$ are treated as vectors. 
The propagators are obtained by inverting the matrices in this expression: 
\ba
 \Bigl\langle \tilde A^a_\mu( P)
 \tilde A^b_\nu( Q) \Bigr\rangle_0^{ } 
 & = & 
 \delta^{ab} \;\,
 \deltabar( P +  Q)
 \biggl[  
     \frac{\delta^{ }_{\mu\nu} - 
     \frac{ P^{ }_\mu  P^{ }_\nu }{ P^2} }{ P^2}
 + \frac{ \frac{\xi\,  P^{ }_\mu  P^{ }_\nu }{ P^2} }{ P^2}
 \biggr]
 \;, \la{Aprop} \\
 \Bigl\langle \tilde c^a ( P)
 \tilde{\bar c}^{\,b} ( Q) \Bigr\rangle_0^{ } 
 & = & 
 \delta^{ab} \;\,
 \deltabar( P -  Q) \,
 \frac{1}{ P^2}
 \;, \la{cprop} \\ 
 \Bigl\langle \tilde \psi^{ }_A ( P)
 \,\tilde{\!\bar\psi}^{ }_B ( Q) \Bigr\rangle_0^{ } 
 & = & 
 \delta^{ }_{AB} \;\,
 \deltabar( P -  Q) \,
 \frac{- i \bsl{ P} + m }{ P^2 + m^2}
 \;.   \la{psiprop}
\ea

\index{Propagator: gauge field}
\index{Propagator: Dirac fermion}

Finally, we list the interactions, which are most conveniently 
written in a maximally symmetric form, obtained through changes 
of integration and summation variables. Thereby the three-gluon vertex becomes
\ba
 S_\iI^{(AAA)} & = &
 \int_X \fr12 (\partial^{ }_\mu A^a_\nu - \partial^{ }_\nu A^a_\mu)
 g f^{abc} A^b_\mu A^c_\nu 
 \nn & = & 
 \Tint{ P  Q  R}
 \! \frac{1}{3!}\,
 \tilde A^a_\mu ( P)
 \tilde A^b_\nu ( Q)
 \tilde A^c_\rho ( R)
 \; \deltabar( P +  Q +  R)
 \; 
 \nn & & \hspace*{2cm} \times \, i g f^{abc}
 \Bigl[
   \delta^{ }_{\mu\rho} ( P^{ }_\nu -  R^{ }_\nu ) 
   + \delta^{ }_{\rho\nu} ( R^{ }_\mu -  Q^{ }_\mu ) 
   + \delta^{ }_{\nu\mu} ( Q^{ }_\rho -  P^{ }_\rho ) 
 \Bigr]
 \;, \la{AAA}
\ea
the four-gluon vertex  
\ba
 & & \hspace*{-1cm} S_\iI^{(AAAA)} = 
 \int_X \fr14 g^2 f^{abc}f^{ade} A^b_\mu A^c_\nu A^d_\mu A^e_\nu 
 \nn & = & 
 \Tint{ P  Q  R  S} 
 \frac{1}{4!} \,
 \tilde A^a_\mu( P) 
 \tilde A^b_\nu( Q) 
 \tilde A^c_\rho( R) 
 \tilde A^d_\sigma( S) 
 \; \deltabar( P +  Q +  R +  S) 
 \nn & \times & 
 g^2 \,  \Bigl[
   f^{eab}f^{ecd} (\delta^{ }_{\mu\rho}\delta^{ }_{\nu\sigma} - 
  \delta^{ }_{\mu\sigma} \delta^{ }_{\nu\rho}) 
   + f^{eac}f^{ebd} (\delta^{ }_{\mu\nu}\delta^{ }_{\rho\sigma} - 
   \delta^{ }_{\mu\sigma} \delta^{ }_{\nu\rho}) 
   + f^{ead}f^{ebc} (\delta^{ }_{\mu\nu}\delta^{ }_{\rho\sigma} - 
  \delta^{ }_{\mu\rho} \delta^{ }_{\nu\sigma}) 
 \Bigr]
 \;, \nn \la{AAAA}
\ea
the ghost interaction 
\ba
 S_\iI^{(\bar c A c)}
 & = &  \int_X \partial^{ }_\mu \bar{c}^{\,a} g f^{abc} A^b_\mu c^c 
 \nn 
 & = &  \Tint{ P  Q  R}
 \tilde{\bar c}^{\,a} ( P)
  \tilde A^b_\mu( Q) 
 \tilde c^c ( R) 
 \; \deltabar(- P +  Q +  R)   
  \Bigl( - i g f^{abc}  P^{ }_\mu \Bigr) 
 \;, \la{ccA}
\ea
and finally the fermion interaction
\ba
 S_\iI^{(\bar\psi A \psi)}
 & = &  \int_X \bar\psi^{ }_A  \gamma^{ }_\mu \Bigl( - i g T^a_{AB} \Bigr)
  A^a_\mu \psi^{ }_B 
 \nn 
 & = &  \Tint{ Q \{  P  R \} }
 \,\tilde{\!\bar\psi}^{ }_A( P)  \gamma^{ }_\mu 
  \tilde A^a_\mu( Q) 
 \tilde \psi^{ }_B( R) 
 \; \deltabar(- P +  Q +  R)   
  \Bigl( - i g T^a_{AB} \Bigr) 
 \;. \la{qqA}
\ea


\subsection*{Appendix A: Non-Abelian black-body radiation in the free limit}

\index{Blackbody radiation}

In this appendix, we
compute the free energy density $f(T)$ for 
$\Nc$ colours of free gluons and $\Nf$ flavours of massless quarks, 
starting from \eq\nr{SE_2}, and use the outcome 
to deduce a result for the usual 
electromagnetic blackbody radiation. 
This is an interesting exercise because, inspite of 
us being in the free limit, ghosts turn out to play a role
at finite temperature.

To start with, we recall from 
\eqs\nr{JmT}, \nr{JT0}, \nr{Jf_def} and \nr{Jf_highT} that 
\ba
 J(0,T) & = & \fr12\, \Tint{ P}
 \bigl[
 \ln( P^2)  - \mbox{const.}
 \bigr] 
 =  - \frac{\pi^2 T^4}{90}
 \;, \la{J_b0} \\ 
  \tilde J(0,T) & = & \fr12\, \Tint{ \{ P \} } \!\!\!\!
 \bigl[ \ln( P^2)  - \mbox{const.}
 \bigr] 
 =  \fr78 \frac{\pi^2 T^4}{90}
 \;. \la{J_f0}
\ea
Our task is to figure out the prefactors of these terms, 
corresponding to the contributions of gluons, ghosts and quarks. 

In the {\em gluonic} case, we are faced with the matrix 
\be
 M^{ }_{\mu\nu} = 
  P^2 \delta^{ }_{\mu\nu} - 
 \Bigl( 1 - \frac{1}{\xi} \Bigr)   P^{ }_\mu  P^{ }_\nu
 \;, \la{Mmunu} 
\ee
which is conveniently handled by introducing two further matrices, 
\be
 \mathbbm{P}^\rmii{T}_{\mu\nu} \;\equiv\;
 \delta^{ }_{\mu\nu} - \frac{  P^{ }_\mu  P^{ }_\nu }{ P^2}
 \;, \quad
 \mathbbm{P}^\rmii{L}_{\mu\nu} \;\equiv\; 
  \frac{  P^{ }_\mu  P^{ }_\nu }{ P^2}
 \;. 
\ee
As matrices, these satisfy 
$
 \mathbbm{P}^\rmii{T} \mathbbm{P}^\rmii{T} = \mathbbm{P}^\rmii{T}
$, 
$
 \mathbbm{P}^\rmii{L} \mathbbm{P}^\rmii{L} = \mathbbm{P}^\rmii{L}
$, 
$
 \mathbbm{P}^\rmii{T} \mathbbm{P}^\rmii{L} = 0 
$, 
$
 \mathbbm{P}^\rmii{T} +  \mathbbm{P}^\rmii{L} = \unit 
$, making them projection operators and implying that 
their eigenvalues are either zero or unity. The numbers
of the unit eigenvalues can furthermore be found by 
taking the appropriate traces: 
$
 \tr[\mathbbm{P}^\rmii{T}] = \delta^{ }_{\mu\mu} - 1 = d
$, 
$
 \tr[\mathbbm{P}^\rmii{L}] = 1
$.

We can clearly write 
\be
 M^{ }_{\mu\nu} =  P^2 \, \mathbbm{P}^\rmii{T}_{\mu\nu} 
 + \frac{1}{\xi}  P^2 \, \mathbbm{P}^\rmii{L}_{\mu\nu}
 \;,
\ee
from which we see that $M$ has $d$ eigenvalues of $ P^2$ 
and one $ P^2/\xi$.
Also, there are $a = 1, \ldots, \Nc^2-1$ copies of this structure, 
so that in total 
\ba
 \left. f(T) \right|^{ }_\rmi{gluons}
 & = & 
 (\Nc^2 - 1) \biggl\{  
 d \times
 \fr12 \Tint{ P}
 \Bigl[
 \ln( P^2)  - \mbox{const.}
 \Bigr] 
 + 
 \fr12 \Tint{ P}
 \Bigl[
 \ln(\frac{1}{\xi} P^2)  - \mbox{const.}
 \Bigr]
 \biggr\}
 \nn & = & 
 (\Nc^2 - 1) \biggl\{ 
  -\fr12  \Tint{ P} \ln(\xi) + 
 (d+1) J(0,T)
 \biggr\} 
 \;. \la{f_gluons}
\ea
The first term vanishes in dimensional regularization, 
because it contains no scales.

For the {\em ghosts}, the Gaussian integral yields (cf.\ \eq\nr{fZ})
\be
 \int \! \prod_{a} {\rm d} \tilde{\bar c}^{\,a} {\rm d}\tilde c^a
 \exp(-\tilde{\bar c}^{\,a}  P^2 \tilde c^a)
 = \prod_{a}  P^2 = 
 \exp\Bigl\{ - \Bigl[ -2(\Nc^2 - 1) \fr12 \ln( P^2) \Bigr]\Bigr\}
 \;.
\ee
Recalling that ghosts obey periodic boundary conditions, we obtain from here
\be
 \left. f(T) \right|^{ }_\rmi{ghosts}
 = -2 (\Nc^2 - 1) J(0,T)
 \;. \la{f_ghosts} 
\ee
Finally, {\em quarks} function as in \eq\nr{f_f_final}, except that they 
now come in $\Nc$ colours and $\Nf$ flavours, giving 
\be
 \left. f(T) \right|^{ }_\rmi{quarks}
 = -4 \Nc \Nf\, \tilde J(0,T)
 \;. \la{f_quarks} 
\ee

\index{Stefan-Boltzmann law}

Summing together \eqs\nr{f_gluons}, \nr{f_ghosts} and \nr{f_quarks}, 
inserting the values of $J$ and $\tilde J$
from \eqs\nr{J_b0} and \nr{J_f0}, and setting $d=3$, 
we get 
\be
 \left. f(T) \right|^{ }_\rmi{QCD} = 
 - \frac{\pi^2 T^4}{90}
 \Bigl[
 2 (\Nc^2 - 1) + \fr72 \Nf \Nc 
 \Bigr]
 \;. 
 \la{f_QCD_0}
\ee
This result is often referred to as (the QCD-version of) the 
Stefan-Boltzmann law.

It is important to realize
that the contribution from the ghosts was essential above: 
according to \eq\nr{f_ghosts}, it cancels half of 
the result in \eq\nr{f_gluons}, thereby yielding the 
correct number of physical degrees of freedom 
in a massless gauge field as the multiplier
in \eq\nr{f_QCD_0}. In addition, 
the assumption that $A^a_0$ is periodic 
has played a role: had it also had an antiperiodic part, 
\eq\nr{f_QCD_0} would have received a further unphysical 
term. 

To finish the section, we finally note that 
the case of QED can be obtained by setting $\Nc\to 1$ and 
$\Nc^2 - 1 \to 1$, recalling that the gauge group of QED is U(1). 
This produces
\be
 \left. f(T) \right|^{ }_\rmi{QED} = 
 - \frac{\pi^2 T^4}{90}
 \biggl(
 2  + \fr72 \Nf  
 \biggr)
 \;, 
 \la{f_QED_0}
\ee
where the factor 2 inside the square brackets corresponds 
to the two photon polarizations, and  the factor 4 multiplying 
$\fr78\Nf$ to the degrees of freedom of a spin-$\fr12$ particle 
and a spin-$\fr12$ antiparticle.  If left-handed neutrinos were to 
be included here, 
they would contribute an additional term of $2 \times \fr78\Nf = \fr74 \Nf$.
\Eq\nr{f_QED_0} together with the contribution of the neutrinos gives
the free energy density determining the expansion rate of the universe
for temperatures in the MeV range. 

\index{QED}


\subsection*{Appendix B: 2-point correlator of magnetic field fluctuations}

\index{Magnetic fields}

As a second illustration of perturbative computations with thermal gauge
fields, we consider the correlation function of magnetic fields in QED. 
This discussion bears a resemblance to that for scalar fields
at the end of \se\ref{se:wce}, with the difference that the
magnetic field involves derivatives. 
There is 
a clear physics motivation now, in that correlators
of magnetic fields play an important role 
in astrophysics, where they are measureable to an extent
(cf.,\ e.g.,\ ref.~\cite{tv}). 

In analogy with \eq\nr{Gxy}, let us consider the ``equal-time'' 
($x^{ }_0 = y^{ }_0$) correlator
\be
 G^{ }_{ij}(\vec{x-y}) \; \equiv \;  
   \langle\,
      \mathcal{B}^{ }_i(\vec{x}) \mathcal{B}^{ }_j(\vec{y})
 \,\rangle^{ }_0
 \;, \la{Bxy}
\ee
where the magnetic field is defined as 
$
 \mathcal{B}^{ }_i \equiv \fr12 \epsilon^{ }_{ijk} F^{ }_{jk} 
 = 
 \epsilon^{ }_{ijk} \partial^{ }_j A^{ }_k
$, 
with $A^{ }_k$ denoting an Abelian gauge field. Going to momentum
space and inserting the propagator from \eq\nr{Aprop}, 
\eq\nr{Bxy} becomes
\ba
 G^{ }_{ij}(\vec{x-y}) & = & 
 \Tint{P,Q} \epsilon^{ }_{ikl} \epsilon^{ }_{jmn} 
 (-i p^{ }_k) (-i q^{ }_m) e^{-i (\vec{p}\cdot\vec{x} + \vec{q}\cdot\vec{y})}
 \Bigl\langle \tilde A^{ }_l( P)
 \tilde A^{ }_n( Q) \Bigr\rangle_0^{ } 
 \nn 
 & = & 
 \Tint{P} \epsilon^{ }_{ikl} \epsilon^{ }_{jmn}\,
 p^{ }_k p^{ }_m \, e^{i \vec{p}\cdot(\vec{y-x})}
 \,
 \biggl[
     \frac{ \delta^{ }_{ln} }{ P^2}
 +   \frac{(\xi-1)\, p^{ }_l p^{ }_n }{ ( P^2)^2}
  \biggr] 
 \;. 
\ea
The antisymmetry of the Levi-Civita symbol implies that 
the longitudinal part, proportional 
to $p^{ }_l p^{ }_n$, drops out, and simultaneously with it 
any dependence on the gauge parameter~$\xi$.
Employing 
$
 \epsilon^{ }_{ikl} \epsilon^{ }_{jmn}\, \delta^{ }_{ln}
 = 
  \delta^{ }_{ij} \delta^{ }_{km} - \delta^{ }_{im}\delta^{ }_{kj}
$, 
and pulling spatial momenta in front of the sum-integration, we get
\be
 G^{ }_{ij}(\vec{x-y})
 = 
 \bigl( \partial^{ }_i \partial^{ }_j - \delta^{ }_{ij} \nabla^2 \bigr)
 \Tint{P} \frac{e^{i\vec{p}\cdot(\vec{y-x})}}{p_n^2 + p^2}
 \;, 
\ee
where the derivatives operate on $\vec{x}$.
The Matsubara sum can be carried out, with the help of \eq\nr{iEc}, 
however it is actually helpful to first carry out the integral,  
according to \eq\nr{Px_large}, and the Matsubara sum only afterwards
(it is typical of thermal computations that, depending on 
the problem, one or the other ordering may offer for a nice
simplification). 
Showing the first steps in both directions, we get
\ba
 G^{ }_{ij}(\vec{x-y})
 & \stackrel{\rmii{\nr{iEc}}}{=} &  
 \bigl( \partial^{ }_i \partial^{ }_j - \delta^{ }_{ij} \nabla^2 \bigr)
 \int_{\vec{p}} 
 e^{i\vec{p}\cdot(\vec{y-x})}
 \frac{1 + 2 \nB{}(p)}{2p} 
 \nn 
 & \stackrel{\rmii{\nr{Px_large}}}{=} & 
 \bigl( \partial^{ }_i \partial^{ }_j - \delta^{ }_{ij} \nabla^2 \bigr)
 T\sum_{p^{ }_n} \frac{e^{-|p^{ }_n||\vec{x-y}|}}{4\pi |\vec{x-y}|}
 \nn 
 & = & 
 \bigl( \partial^{ }_i \partial^{ }_j - \delta^{ }_{ij} \nabla^2 \bigr)
 \frac{T} {4\pi |\vec{x-y}|} 
 \biggl[ -1 + \frac{2}{ 1 - e^{-2\pi T |\vec{x-y}|}} \biggr]
 \nn[2mm] 
 & = & 
 \bigl( \partial^{ }_i \partial^{ }_j - \delta^{ }_{ij} \nabla^2 \bigr)
 \frac{T \coth(\pi T |\vec{x-y}| )} {4\pi |\vec{x-y}|} 
 \;.
 \la{Bij_result}
\ea
The function here only depends on 
$
 r \equiv |\vec{x-y}|
$, 
so the partial derivatives can be expressed as 
\ba
 &&
 \partial^{ }_i \partial^{ }_j \; = \; 
 \frac{r^{ }_i r^{ }_j \partial_r^2}{r^2}
 + 
 \biggl( \delta^{ }_{ij} - \frac{r^{ }_i r^{ }_j}{r^2} \biggr)
 \frac{\partial^{ }_r}{r}
 \;, \quad
 \nabla^2 = \partial_r^2 + \frac{2 \partial^{ }_r}{r}  
 \nn
 & \Longrightarrow & 
 \partial^{ }_i \partial^{ }_j - \delta^{ }_{ij} \nabla^2
 = 
 \frac{r^{ }_i r^{ }_j}{r^2}
 \biggl( \partial_r^2 - \frac{\partial^{ }_r}{r} \biggr) 
 - 
 \delta^{ }_{ij}
 \biggl( \partial_r^2 + \frac{\partial^{ }_r}{r} \biggr) 
 \;. 
 \la{Bij_derivs}
\ea
The radial derivatives could be taken, 
however that does not make the physics more transparent.

There are a few lessons to be learned from \eq\nr{Bij_result}. 
First, if we consider the limit of short distances, 
then $\coth(\pi T |\vec{x-y}|) \to 1/(\pi T |\vec{x-y}|)$, and 
$
 \frac{T \coth(\pi T |\vec{x-y}| )} {4\pi |\vec{x-y}|} 
 \to 
 1/(4\pi^2 |\vec{x-y}|^2) 
$.
This is just the 
vacuum behaviour from \eq\nr{Px_small}. So, once again, 
thermal effects are not important at short distances. 
At large distances, in turn, $\coth(\pi T |\vec{x-y}|) \to 1$, 
and 
$
 \frac{T \coth(\pi T |\vec{x-y}| )} {4\pi |\vec{x-y}|} 
 \to 
 T/(4\pi |\vec{x-y}|) 
$.
This represents the same physics as discussed below \eq\nr{Px_large}.
Incidentally, in terms of the Matsubara sum in \eq\nr{Bij_result},
the long-distance behaviour originates from the Matsubara zero mode, 
$p^{ }_n = 0$. 
This anticipates its important role, 
as will be discussed in \se\ref{se:Linde}.

\newpage 

\subsection{Thermal gluon mass}
\la{se:mmeff_gluon}

\index{Thermal mass: gluon}

We consider next the gauge field propagator, 
in particular the Matsubara zero-mode sector thereof. 
We wish to see whether an effective thermal mass $m^{ }_\rmi{eff}$
is generated for this field mode, 
as was the case for a scalar field (cf.\ \eq\nr{mmeff}).
The observable to consider is the full propagator, 
i.e.~the analogue of \eq\nr{P_full}. 
Note that we do not consider non-zero Matsubara modes since, 
like in \eq\nr{LE_resum}, the thermal mass corrections 
are parametrically subdominant 
if we assume the coupling to be weak, 
$g^2 T^2 \ll (2\pi T)^2$. 
For the same reason, we do not need to consider thermal mass 
corrections for fermions at the present order.

\index{Feynman gauge}

In order to simplify the task somewhat, 
we choose to carry out the computation in the so-called 
Feynman gauge, $\xi \equiv 1$, whereby the free propagator 
of \eq\nr{Aprop} becomes
\be
 \Bigl\langle \tilde A^a_\mu( K)
 \tilde A^b_\nu( Q) \Bigr\rangle_0^{ } 
  \quad = \quad  
 \delta^{ab} \;
 \deltabar( K +  Q)\,
 \Biggl[  
     \frac{\delta^{ }_{\mu\nu} - 
     \frac{ K^{ }_\mu  K^{ }_\nu }{ K^2} }{ K^2}
 + \frac{ \frac{\xi \,  K^{ }_\mu  K^{ }_\nu }{ K^2} }{ K^2}
 \Biggr]
 \quad \stackrel{\xi = 1}{=} \quad
 \delta^{ab} \;
 \deltabar( K +  Q)
     \frac{\delta^{ }_{\mu\nu}}{ K^2}
 \;. \la{Fprop}
\ee
Specifically, our goal is to compute 
the 1-loop gluon self-energy $\Pi^{ }_{\mu\nu}$, defined via
\be
 \frac{
    \bigl\langle \tilde A^a_\mu( K)
    \tilde A^b_\nu( Q) e^{ - S^{ }_\rmii{I} }\bigr\rangle_0^{ } 
 }{
 \bigl\langle e^{ - S^{ }_\rmii{I} }\bigr\rangle_0^{ } 
 }
 \; = \;  
 \delta^{ab} \;
 \deltabar( K +  Q)
 \, \biggl[
     \frac{\delta^{ }_{\mu\nu}}{ K^2} - \frac{\Pi^{ }_{\mu\nu}(K)}{K^4}
 + \rmO(g^4) \biggr]
 \;, \la{Fprop_Pi}
\ee
where the role of the denominator of the left-hand side 
is to cancel the disconnected contributions.

At 1-loop level, there are several distinct contributions 
to $\Pi^{ }_{\mu\nu}$: two of these involve gauge loops 
(via a quartic vertex from $-S^{ }_\iI$ and two cubic vertices
from $+S_\iI^2/2$, respectively), one involves a ghost loop
(via two cubic vertices from $+S_\iI^2/2$), and one a fermion loop
(via two cubic vertices from $+S_\iI^2/2$). 
If the theory were to contain additional 
scalar fields, then two additional graphs similar to 
the gauge loops would be generated. 
For future purposes we  treat the external momentum 
$ K$ of the graphs as a general Euclidean four-momentum, even though 
for the Matsubara zero modes (that we are ultimately interested in) 
only the spatial part is non-zero.

Let us begin by considering the gauge loop originating from a quartic vertex. 
Denoting the structure in \eq\nr{AAAA} by 
\be
  C^{abcd}_{\mu\nu\rho\sigma} \equiv
   f^{eab}f^{ecd} (\delta^{ }_{\mu\rho}\delta^{ }_{\nu\sigma} - 
  \delta^{ }_{\mu\sigma} \delta^{ }_{\nu\rho}) 
   + f^{eac}f^{ebd} (\delta^{ }_{\mu\nu}\delta^{ }_{\rho\sigma} - 
   \delta^{ }_{\mu\sigma} \delta^{ }_{\nu\rho}) 
   + f^{ead}f^{ebc} (\delta^{ }_{\mu\nu}\delta^{ }_{\rho\sigma} - 
  \delta^{ }_{\mu\rho} \delta^{ }_{\nu\sigma}) 
%
 \;, 
\ee
we get 
\ba
 & & \hspace*{-1cm} \Bigl\langle \tilde A^a_\mu( K)
 \tilde A^b_\nu( Q) (-S^{ }_\iI) \Bigr\rangle^{ }_{0,\rmi{c}} 
 \quad = \quad \TopoST(\Lgl,\Agl)
 \nn 
 & = & 
 -\frac{g^2}{24} 
 \Bigl\langle 
 \tilde A^a_\mu( K)
 \tilde A^b_\nu( Q)
 \Tint{ R  P  T  U} 
 \hspace*{-0.5cm}
 \tilde A^c_\alpha( R) 
 \tilde A^d_\beta( P) 
 \tilde A^e_\rho( T) 
 \tilde A^f_\sigma( U) 
 \; \deltabar( R +  P +  T +  U) 
 \; C^{cdef}_{\alpha\beta\rho\sigma}
 \Bigr\rangle^{ }_{0,\rmi{c}} 
 \nn 
 & = & 
 -\frac{g^2}{2} 
 \Tint{ R  P  T  U} 
 \hspace*{-0.5cm}
 \deltabar( R +  P +  T +  U) 
 \langle
 \tilde A^a_\mu( K)
 \tilde A^c_\alpha( R) 
 \rangle_0^{ } \;
 \langle
 \tilde A^b_\nu( Q)
 \tilde A^d_\beta( P) 
 \rangle_0^{ } \;
 \langle
 \tilde A^e_\rho( T) 
 \tilde A^f_\sigma( U) 
 \rangle_0^{ } \;
 C^{cdef}_{\alpha\beta\rho\sigma}
 \;,   \nn 
\ea
where we made use of the complete symmetry of 
$C^{cdef}_{\alpha\beta\rho\sigma}$.
Inserting here \eq\nr{Fprop}, this becomes
\ba
 \Bigl\langle \tilde A^a_\mu( K)
 \tilde A^b_\nu( Q) (-S^{ }_\iI) \Bigr\rangle^{ }_{0,\rmi{c}}
 & = & -\frac{g^2}{2} 
 \Tint{ R  P  T  U} 
 \hspace*{-0.5cm}
 \deltabar( R +  P +  T +  U) 
 \; \deltabar( K +  R) 
 \; \deltabar( Q +  P) 
 \; \deltabar( T +  U) 
 \nn & & \times \, 
 \frac{1}{ K^2  Q^2  T^2}\, 
 \delta^{ac}
 \delta^{bd}
 \delta^{ef}
 \delta^{ }_{\mu\alpha}
 \delta^{ }_{\nu\beta}
 \delta^{ }_{\rho\sigma}
 C^{cdef}_{\alpha\beta\rho\sigma}
 \;.    
\ea

The sum-integrals over 
$ R,  P,  T$ in the above expression are trivially
carried out, and amount to 
$R\to -K$, $P\to -Q$, $T\to -U$.
The remaining constraint becomes 
$\,\deltabar(R+P+T+U) \to \,\deltabar(K+Q)$, 
whereby $Q\to -K$. 
Moreover, we note that 
\ba
 \delta^{ac}
 \delta^{bd}
 \delta^{ef}
 \delta^{ }_{\mu\alpha}
 \delta^{ }_{\nu\beta}
 \delta^{ }_{\rho\sigma}
 C^{cdef}_{\alpha\beta\rho\sigma}
 & = & 
 \delta^{ef}
 \delta^{ }_{\rho\sigma}\Bigl[ 
 f^{gae}f^{gbf} (\delta^{ }_{\mu\nu}\delta^{ }_{\rho\sigma} - 
   \delta^{ }_{\mu\sigma} \delta^{ }_{\nu\rho}) 
   + f^{gaf}f^{gbe} (\delta^{ }_{\mu\nu}\delta^{ }_{\rho\sigma} - 
  \delta^{ }_{\mu\rho} \delta^{ }_{\nu\sigma}) 
 \Bigr]
 \nn & = & 2 \, d\,  f^{age}f^{bge} \delta^{ }_{\mu\nu}
 \;, 
\ea
where we made use of the antisymmetry of the structure constants, 
as well as of the fact that 
$\delta^{ }_{\sigma\sigma} = d + 1 = 4 - 2\epsilon$. 
Noting that the structure constants furthermore satisfy
$
 f^{age}f^{bge} = \Nc\, \delta^{ab}
$, 
we get in total
\be
 \Bigl\langle \tilde A^a_\mu( K)
 \tilde A^b_\nu( Q) (-S^{ }_\iI) \Bigr\rangle^{ }_{0,\rmi{c}}
 = 
 - g^2 \Nc \, d \, I^{ }_T(0)\, \delta^{ab}
  \; \deltabar( K +  Q) 
  \frac{\delta^{ }_{\mu\nu}}{( K^2)^2}
 \;, \la{Gtadpole}
\ee
where $I^{ }_T(0) = \Tinti{U} \tfr{1}{U^2}$, 
cf.\ \eqs\nr{ImT_2} and \nr{splitup}.
Note that the $\delta$-functions as well as the colour and spacetime 
indices appear here just like in \eq\nr{Fprop_Pi}, 
allowing us to straightforwardly read off the contribution 
of this graph to $\Pi^{ }_{\mu\nu}(K)$.

Next, we move on to the gluon loop originating from 
two cubic interaction vertices. Denoting
the combination of $\delta$-functions and momenta in \eq\nr{AAA} by 
\be
 D^{ }_{\mu\nu\rho}( P, Q, R)
 \;\equiv\;
   \delta^{ }_{\mu\rho} ( P^{ }_\nu -  R^{ }_\nu ) 
   + \delta^{ }_{\rho\nu} ( R^{ }_\mu -  Q^{ }_\mu ) 
   + \delta^{ }_{\nu\mu} ( Q^{ }_\rho -  P^{ }_\rho ) 
%
 \;, \la{Dabc}
\ee
we get for this contribution
\ba
 & & \hspace*{-2cm} 
 \Bigl\langle \tilde A^a_\mu( K)
 \tilde A^b_\nu( Q) \Bigl(\fr12 S_\iI^2 \Bigr) 
 \Bigr\rangle_{0,\rmi{c}}^{(1)} 
 \quad = \quad  \TopoSB(\Lgl,\Agl,\Agl)
 \nn 
 \!\!\!\!\! & = &  \!\!\!
 -\frac{g^2}{72} 
 \Bigl\langle 
 \tilde A^a_\mu( K)
 \tilde A^b_\nu( Q)
 \,\Tint{ R  P  T} 
 \hspace*{-0.5cm}
 \tilde A^c_\alpha( R) 
 \tilde A^d_\beta( P) 
 \tilde A^e_\gamma( T) 
 \,\Tint{ U  V\!  X} 
 \hspace*{-0.5cm}
 \tilde A^g_\zeta( U) 
 \tilde A^h_\eta( V) 
 \tilde A^i_\rho( X) 
 \Bigr\rangle^{ }_{0,\rmi{c}}
 \nn & & \hspace*{0.5cm} \times  
 \; 
 f^{cde} f^{ghi}
 \; \deltabar( R +  P +  T) 
 \; \deltabar( U +  V +  X) 
 \; D^{ }_{\alpha\beta\gamma}( R ,  P ,  T)
 \; D^{ }_{\zeta\eta\rho}( U ,  V ,  X)
 \nn 
 & = &  
 -\frac{g^2}{2} 
 \Tint{ R  P  T  U  V\!  X} 
 \hspace*{-1cm}
 \langle 
 \tilde A^a_\mu( K)
 \tilde A^c_\alpha( R) 
 \rangle_0^{ } \;
 \langle 
 \tilde A^b_\nu( Q)
 \tilde A^g_\zeta( U) 
 \rangle_0^{ } \;
 \langle 
 \tilde A^d_\beta( P) 
 \tilde A^h_\eta( V) 
 \rangle_0^{ } \;
 \langle 
 \tilde A^e_\gamma( T) 
 \tilde A^i_\rho( X) 
 \rangle_0^{ }
 \nn & & \hspace*{0.5cm} \times  
 \; 
 f^{cde} f^{ghi}
 \; \deltabar( R +  P +  T) 
 \; \deltabar( U +  V +  X) 
 \; D^{ }_{\alpha\beta\gamma}( R ,  P ,  T)
 \; D^{ }_{\zeta\eta\rho}( U ,  V ,  X)
 \;, \la{Gloop0} 
\ea
where we made use of the complete symmetry of 
$
 f^{cde}D^{ }_{\alpha\beta\gamma}( R, P, T)
$
in simultaneous interchanges of all indices labelling 
a particular gauge field
(for instance $c,\alpha, R \leftrightarrow d,\beta, P$).

Inserting \eq\nr{Fprop} into the above expression, 
let us inspect in turn the colour indices, 
spacetime indices, and momenta. 
The colour contractions are easily carried out, 
and result in the overall factor
\be
 \delta^{ac}\delta^{bg}\delta^{dh}\delta^{ei} f^{cde} f^{ghi}
 = f^{ade}f^{bde} = \Nc\, \delta^{ab}
 \;. 
\ee
The spacetime contractions can all be 
transported to the $D$-functions, noting that the effect 
can be summarized with the substitution rules
$
 \alpha \to \mu
$, 
$
 \zeta \to \nu
$, 
$
 \eta \to \beta
$, 
$ 
 \rho \to \gamma
$.
The momenta get fixed as 
$
 R\to -K 
$, 
$
 U\to -Q
$, 
$
 V\to -P
$, 
$
 X\to -T
$, 
whereby
\ba
 && \hspace*{-1cm}
  \; \deltabar( R +  P +  T) 
  \;\, \deltabar( U +  V +  X) 
 \to 
 \; \deltabar(-  K +  P +  T) 
 \;\, \deltabar(-  Q -  P -  T) 
 = 
 \; \deltabar(-  K +  P +  T) 
 \;\, \deltabar( K +  Q) 
 \;. \nonumber
\ea
Then $T\to K - P$, and  
\ba
 && \hspace*{-1cm} 
 \; D^{ }_{\mu\beta\gamma}( R ,  P ,  T)
 \; D^{ }_{\nu\beta\gamma}( U ,  V ,  X)
 \to 
 \; D^{ }_{\mu\beta\gamma}(-  K ,  P ,  K -P)
 \; D^{ }_{\nu\beta\gamma}( K , - P , P - K) 
 \;. 
\ea
Only the sum-integral over $P$ is left over, giving us 
\ba
 & & \hspace*{-1cm} 
 \Bigl\langle \tilde A^a_\mu( K)
 \tilde A^b_\nu( Q) \Bigl(\fr12 S_\iI^2 \Bigr) 
 \Bigr\rangle_{0,\rmi{c}}^{(1)} 
 \nn 
 \!\!\!\!\! & = &  \!\!\!
 -\frac{g^2 \Nc}{2} 
 \frac{\delta^{ab} \; \deltabar( K +  Q)}{( K^2)^2}
 \Tint{ P} 
 \frac{1}{ P^2 ( K -  P)^2}
 \; D^{ }_{\mu\beta\gamma}(-  K ,  P ,  K -  P)
 \; D^{ }_{\nu\beta\gamma}( K , - P , -  K +  P)
 \;. \hspace*{1cm} \la{Gloop}
\ea

Finally, we are faced with the tedious task of inserting 
\eq\nr{Dabc} into the above expression 
and carrying out all contractions --- a task 
most conveniently handled using programming languages intended 
for carrying out symbolic manipulations, such as FORM~\cite{form}. 
Here we perform the contractions by hand, obtaining first
\ba
 & & \hspace*{-1.0cm}
 D^{ }_{\mu\beta\gamma}(-  K ,  P ,  K -  P)
 \; D^{ }_{\nu\beta\gamma}( K , - P , -  K +  P)
 \nn 
 & = & - 
 [ 
     \delta^{ }_{\mu\gamma} (- 2  K^{ }_\beta +  P^{ }_\beta ) 
   + \delta^{ }_{\gamma\beta} ( K^{ }_\mu - 2  P^{ }_\mu ) 
   + \delta^{ }_{\beta\mu} ( K^{ }_\gamma +  P^{ }_\gamma ) 
 ] 
 \nn & & \times \, 
  [ 
     \delta^{ }_{\nu\gamma} (- 2  K^{ }_\beta +  P^{ }_\beta ) 
   + \delta^{ }_{\gamma\beta} ( K^{ }_\nu - 2  P^{ }_\nu ) 
   + \delta^{ }_{\beta\nu} ( K^{ }_\gamma +  P^{ }_\gamma ) 
 ]
 \nn & = &
 - \delta^{ }_{\mu\nu}
  (4  K^2 - 4  K\cdot  P +  P^2
   +  K^2 + 2  K \cdot  P +  P^2)
 \nn & & - \,
 (d+1) ( K^{ }_\mu  K^{ }_\nu - 2  K^{ }_\mu  P^{ }_\nu 
             - 2  K^{ }_\nu  P^{ }_\mu + 4  P^{ }_\mu  P^{ }_\nu)
 \nn & & -\, [ 
 (- 2  K^{ }_\mu +  P^{ }_\mu)( K^{ }_\nu - 2  P^{ }_\nu)
 + 
  (- 2  K^{ }_\mu +  P^{ }_\mu)( K^{ }_\nu +  P^{ }_\nu)
 + 
  ( K^{ }_\mu - 2  P^{ }_\mu)( K^{ }_\nu +  P^{ }_\nu) 
 + 
  (\mu\leftrightarrow\nu) ] 
 \nn & = & 
 - \delta^{ }_{\mu\nu}[4  K^2 + ( K -  P)^2 +  P^2]
 - (d-5) K^{ }_\mu  K^{ }_\nu 
 + (2d - 1) ( K^{ }_\mu  P^{ }_\nu +  K^{ }_\nu  P^{ }_\mu)
 - (4d - 2) P^{ }_\mu  P^{ }_\nu
 \;. \nn  \la{DD}
\ea
Because the propagators in \eq\nr{Gloop} are identical, 
we can furthermore simplify the structure 
$
  K^{ }_\mu  P^{ }_\nu +  K^{ }_\nu  P^{ }_\mu
$
by renaming one of the integration variables 
as $P\to K-P$ in ``one half of this term'', i.e.\ by writing
\ba 
  K^{ }_\mu  P^{ }_\nu +  K^{ }_\nu  P^{ }_\mu
 & \rightarrow & 
 \fr12 \Bigl[
   K^{ }_\mu  P^{ }_\nu +  K^{ }_\nu  P^{ }_\mu
 +
   K^{ }_\mu (K^{ }_\nu -  P^{ }_\nu) +  K^{ }_\nu (K^{ }_\mu -  P^{ }_\mu)   
 \Bigr]
 \nn & = & K^{ }_\mu K^{ }_\nu
 \;. \la{P_shift}
\ea
Therefore a representation equivalent to \eq\nr{DD} is 
\ba
  & & \hspace*{-1.0cm}
 D^{ }_{\mu\beta\gamma}(-  K ,  P ,  K -  P)
 \; D^{ }_{\nu\beta\gamma}( K , - P , -  K +  P)
  \nn 
 & \rightarrow & 
 - \delta^{ }_{\mu\nu}[4  K^2 + ( K -  P)^2 +  P^2]
 + (d+4) K^{ }_\mu  K^{ }_\nu 
 - (4d - 2) P^{ }_\mu  P^{ }_\nu
 \;.  \la{DD2}
\ea

Inserting now  \eq\nr{DD2}
into \eq\nr{Gloop}, we observe that the result 
depends in a non-trivial way on the ``external''  
momentum $ K$. This is an important fact 
that plays a role later on. For the moment, 
we however note that since the tree-level gluon propagator of 
\eq\nr{Fprop} is massless, the leading order 
pole position lies at $ K^2 = 0$.
This may get shifted by the loop corrections
that we are currently investigating, 
like in the case of a scalar field theory (cf.\ \eq\nr{mmeff}). 
Since this correction 
is suppressed by a factor of $\rmO(g^2)$, 
in our perturbative calculation we may insert 
$ K = 0$ in \eq\nr{DD2}, making only an error of $\rmO(g^4)$. 
Proceeding this way, we get
\ba
 & & \hspace*{-1cm} 
 \Bigl\langle \tilde A^a_\mu( K)
 \tilde A^b_\nu( Q) \Bigl(\fr12 S_\iI^2 \Bigr) 
  \Bigr\rangle_{0,\rmi{c}}^{(1)} 
 \approx
 \frac{g^2 \Nc}{2} 
 \,\frac{\delta^{ab} \; \deltabar( K +  Q)}{( K^2)^2}
 \,\Tint{ P} 
 \frac{1}{( P^2)^2}
 \Bigl[ 2  P^2 \delta^{ }_{\mu\nu} + (4d - 2) P^{ }_\mu  P^{ }_\nu\Bigr]
 \;. \hspace*{1cm} \la{Gloop2}
\ea

Now, symmetries tell us that the integral in \eq\nr{Gloop2} 
can only depend on two second rank tensors, 
$\delta^{ }_{\mu\nu}$ and $\delta^{ }_{\mu 0}\delta^{ }_{\nu 0}$,
of which the latter originates from the breaking of Lorentz symmetry by 
the rest frame of the heat bath. Denoting $P = (p^{ }_n,\vec{p})$, 
this allows us to split the latter term into two parts according to
(note that $\delta^{ }_{\mu i}\delta^{ }_{\nu i} =
\delta^{ }_{\mu\nu} -  \delta^{ }_{\mu 0}\delta^{ }_{\nu 0}$)
\ba
 \Tint{ P} 
 \frac{ P^{ }_\mu  P^{ }_\nu}
      {( P^2)^2} & = & 
 \delta^{ }_{\mu 0}\delta^{ }_{\nu 0}\,
 \Tint{ P} 
 \frac{ P_0^2}
      {( P^2)^2} + 
 \delta^{ }_{\mu i}\delta^{ }_{\nu i}\,
 \Tint{ P} 
 \frac{ P_i^2}
      {( P^2)^2}
 \nn  & =  &
 \delta^{ }_{\mu 0}\delta^{ }_{\nu 0}\,
 \Tint{ P} 
 \frac{ p_n^2}
      {( P^2)^2} + 
 \delta^{ }_{\mu i}\delta^{ }_{\nu i}\, \frac{1}{d}\;
 \Tint{ P} 
 \frac{{p}^2}
      {( P^2)^2}
 \nn  & =  &
 \delta^{ }_{\mu 0}\delta^{ }_{\nu 0}\,
 \Tint{ P} 
 \frac{ p_n^2}
      {( P^2)^2} + 
 \delta^{ }_{\mu i}\delta^{ }_{\nu i}\, \frac{1}{d}\;
 \Tint{ P} 
 \frac{ P^2 -  p_n^2}
      {( P^2)^2}
 \;. \la{SmuSnu}
\ea
At this point, let us inspect the familiar 
sum-integral (cf.\ \eq\nr{ImT_res})
\be
 I^{ }_T(0) = \Tint{ P} \frac{1}{ P^2}
 = T \sum_{n=-\infty}^{\infty} \int_\vec{p} \frac{1}{(2\pi n T)^2 + {p}^2}
 = \frac{T^2}{12} + \rmO(\epsilon)
 \;. \la{ImT_again}
\ee
Taking the derivative $T^2 \fr{\rm d}{{\rm d} T^2} = 
\fr{T}{2} \, \fr{\rm d}{{\rm d}T}$
on both sides, we find
\be
 \frac{T}{2} 
  \sum_{n=-\infty}^{\infty} \int_\vec{p} \frac{1}{(2\pi n T)^2 + {p}^2}
 - T \sum_{n=-\infty}^{\infty} \int_\vec{p} 
 \frac{(2\pi n T)^2}{[(2\pi n T)^2 + \vec{p}^2]^2}
 = \frac{T^2}{12} + \rmO(\epsilon)
 \;, 
\ee
which can be used in order 
to solve for the only unknown sum-integral in \eq\nr{SmuSnu},  
\be
 \Tint{ P} 
 \frac{ p_n^2}
      {( P^2)^2} = -\frac{ I^{ }_T(0) }{2} + \rmO(\epsilon)
 \;. \la{new_sum_int}
\ee
Inserting this result into \eq\nr{SmuSnu}, 
we thereby obtain in $d=3-2\epsilon$ dimensions
\be \index{Thermal sums: bosonic tensor}
 \Tint{ P} 
 \frac{ P^{ }_\mu  P^{ }_\nu}
      {( P^2)^2}
 = \fr12 ( -  \delta^{ }_{\mu 0}\delta^{ }_{\nu 0}
 +  \delta^{ }_{\mu i}\delta^{ }_{\nu i} )\, \frac{T^2}{12} + \rmO(\epsilon)
 \;, \la{tensor_IT}
\ee 
which turns \eq\nr{Gloop2} finally into
\ba
 \Bigl\langle \tilde A^a_\mu( K)
 \tilde A^b_\nu( Q) 
 \Bigl(\fr12 S_\iI^2 \Bigr) \Bigr\rangle_{0,\rmi{c}}^{(1)} 
  \!\!\! & \approx &  \!\!\! 
 \frac{g^2 \Nc}{2} \,
 \frac{\delta^{ab} \; \deltabar( K +  Q)}{( K^2)^2} \,
 \biggl\{ 
   \delta^{ }_{\mu0}\delta^{ }_{\nu0}
 \biggl[ 2 -\fr12(4d - 2)
 \biggr] 
  \nn & & \hspace*{2.8cm} + \,
  \delta^{ }_{\mu i}\delta^{ }_{\nu i}
 \biggl[ 2 + \fr12(4d - 2)
 \biggr]
 \biggr\} I^{ }_T(0) + \rmO(\epsilon)
 \hspace*{5mm} \nn 
  \!\!\! & \stackrel{ d=3-2\epsilon
                    }{=} & \!\!\!
 \frac{g^2 \Nc}{2} \, 
 \frac{\delta^{ab} \; \deltabar( K +  Q)}{( K^2)^2}
 \, \biggl\{ 
  - 3 \, \delta^{ }_{\mu0}\delta^{ }_{\nu0}
  + 7 \, \delta^{ }_{\mu i}\delta^{ }_{\nu i}
 \biggr\} \frac{T^2}{12} + \rmO(\epsilon)
 \;. \la{Gloop3}
\ea

Moving on to the ghost loop, 
we apply the vertex of \eq\nr{ccA} but otherwise 
proceed as in \eq\nr{Gloop0}. This produces
\ba
 & & \hspace*{-2cm} 
 \Bigl\langle \tilde A^a_\mu( K)
 \tilde A^b_\nu( Q) \Bigl(\fr12 S_\iI^2 \Bigr)
 \Bigr\rangle_{0,\rmi{c}}^{(2)} 
 \quad = \quad \TopoSB(\Lgl,\Agh,\Agh)
 \nn 
 \!\!\!\!\! & = &  \!\!\!
 -\frac{g^2}{2} 
 \Bigl\langle 
 \tilde A^a_\mu( K)
 \tilde A^b_\nu( Q)
 \,\Tint{ R  P  T} 
 \hspace*{-0.5cm}
 \,\tilde{\!\bar c}^{\,c}( R) 
 \tilde A^d_\alpha( P) 
 \tilde c^e( T) 
 \,\Tint{ U  V\!  X} 
 \hspace*{-0.5cm}
 \,\tilde{\!\bar c}^{\,g}( U) 
 \tilde A^h_\beta( V) 
 \tilde c^i( X) 
 \Bigr\rangle^{ }_{0,\rmi{c}}
 \nn & & \hspace*{0.5cm} \times  
 \; 
 f^{cde} f^{ghi}
 \; \deltabar(-  R +  P +  T) 
 \; \deltabar(-  U +  V +  X) 
 \;  R^{ }_\alpha  U^{ }_\beta
 \nn 
 & = &  
 {g^2} 
 \Tint{ R  P  T  U  V\!  X} 
 \hspace*{-1cm}
 \langle 
 \tilde A^a_\mu( K)
 \tilde A^d_\alpha( P) 
 \rangle_0^{ } \;
 \langle 
 \tilde A^b_\nu( Q)
 \tilde A^h_\beta( V) 
 \rangle_0^{ } \;
 \langle 
 \tilde c^e( T) 
 \,\tilde{\!\bar c}^{\,g}( U) 
 \rangle_0^{ } \;
 \langle 
 \tilde c^i( X) 
 \,\tilde{\!\bar c}^{\,c}( R) 
 \rangle_0^{ }
 \nn & & \hspace*{0.5cm} \times  
 \; 
 f^{cde} f^{ghi}
 \; \deltabar(-  R +  P +  T) 
 \; \deltabar(-  U +  V +  X) 
 \;  R^{ }_\alpha  U^{ }_\beta
 \;, 
\ea
where the Grassmann nature of the ghosts induced a minus sign at the second 
equality sign.

Inserting now the gluon propagator from \eq\nr{Fprop}
and the ghost propagator from \eq\nr{cprop}, 
we inspect in turn the colour indices, 
spacetime indices, and momenta. 
The colour contractions result in the familiar factor
\be
 \delta^{ad}\delta^{bh}\delta^{eg}\delta^{ic} f^{cde} f^{ghi}
 = f^{cae}f^{ebc} = - \Nc\, \delta^{ab}
 \;,
\ee
whereas the spacetime indices can be directly transported to the momenta:
$
 \delta^{ }_{\mu\alpha}\delta^{ }_{\nu\beta}
  R^{ }_\alpha  U^{ }_\beta =
  R^{ }_\mu  U^{ }_\nu
$. 
The momenta get fixed as 
$ 
 P\to -K
$, 
$
 V \to - Q
$, 
$
 U \to T
$, 
$
 R\to X
$, 
whereby
\be
 \; \deltabar(-  R +  P +  T) 
 \;\, \deltabar(-  U +  V +  X) 
 \to 
 \; \deltabar(-  X -  K +  T) 
 \;\, \deltabar(-  T -  Q +  X) 
 = 
 \; \deltabar(-  X -  K +  T) 
 \;\, \deltabar( K +  Q)  
 \;. 
\ee
This implies that
$
  R^{ }_\mu  U^{ }_\nu
  \to 
  X^{ }_\mu  T^{ }_\nu
 \to 
  ( T^{ }_\mu -  K^{ }_\mu ) T^{ }_\nu
$. 
Renaming finally $T\to P$, we obtain
\ba
 & & \hspace*{-1cm} 
 \Bigl\langle \tilde A^a_\mu( K)
 \tilde A^b_\nu( Q) \Bigl(\fr12 S_\iI^2 \Bigr) 
 \Bigr\rangle_{0,\rmi{c}}^{(2)} 
 =
 -{g^2 \Nc}\, 
 \frac{\delta^{ab} \; \deltabar( K +  Q)}{( K^2)^2}
 \,\Tint{ P} 
 \frac{1}{ P^2 ( K -  P)^2}
 ( P^{ }_\mu -  K^{ }_\mu ) P^{ }_\nu
 \;. \hspace*{1cm} \la{cloop}
\ea
Repeating the trick of \eq\nr{P_shift}, this can be turned into
\ba
 & & \hspace*{-1cm} 
 \Bigl\langle \tilde A^a_\mu( K)
 \tilde A^b_\nu( Q) \Bigl(\fr12 S_\iI^2 \Bigr) 
 \Bigr\rangle_{0,\rmi{c}}^{(2)} 
 =
 -\frac{g^2 \Nc}{2} 
 \,\frac{\delta^{ab} \; \deltabar( K +  Q)}{( K^2)^2}
 \,\Tint{ P} 
 \frac{1}
 { P^2 ( K -  P)^2}
 (2  P^{ }_\mu  P^{ }_\nu -  K^{ }_\mu K^{ }_\nu)
 \;, \hspace*{1cm} \la{cloop2}
\ea
which in the $K\to 0$ limit produces, 
upon setting $d\to 3$ and using \eq\nr{tensor_IT},  
\ba
 & & \hspace*{-1cm} 
 \Bigl\langle \tilde A^a_\mu( K)
 \tilde A^b_\nu( Q) 
 \Bigl(\fr12 S_\iI^2 \Bigr) \Bigr\rangle_{0,\rmi{c}}^{(2)} 
 \approx
 -\frac{g^2 \Nc}{2} 
 \,\frac{\delta^{ab} \; \deltabar( K +  Q)}{( K^2)^2}
 \,\bigl( 
  -  \delta^{ }_{\mu0}\delta^{ }_{\nu0}
  +  \delta^{ }_{\mu i}\delta^{ }_{\nu i}
 \bigr) \, \frac{T^2}{12} + \rmO(\epsilon)
 \;. \hspace*{1cm} \la{cloop3}
\ea

Finally, we consider the fermion loop, originating from 
the vertex of \eq\nr{qqA}.
Proceeding as above, we obtain
\ba
 & & \hspace*{-1cm} 
 \Bigl\langle \tilde A^a_\mu( K)
 \tilde A^b_\nu( Q) 
 \Bigl(\fr12 S_\iI^2 \Bigr) \Bigr\rangle_{0,\rmi{c}}^{(3)} 
 \quad = \quad \TopoSB(\Lgl,\Aqu,\Aqu)
 \nn 
 \!\!\!\!\! & = &  \!\!\!
 -\frac{g^2}{2} 
 \Bigl\langle 
 \tilde A^a_\mu( K)
 \tilde A^b_\nu( Q)
 \Tint{ P \{  R  T \} } 
 \hspace*{-0.5cm}
 \,\tilde{\!\bar \psi}^{ }_A( R) 
  \gamma^{ }_\alpha \tilde A^c_\alpha( P) 
 \tilde \psi^{ }_B( T) 
 \; \Tint{ V \{  U\!  X \} } 
 \hspace*{-0.5cm}
 \,\tilde{\!\bar \psi}^{ }_C( U) 
  \gamma^{ }_\beta \tilde A^d_\beta( V) 
 \tilde \psi^{ }_D ( X) 
 \Bigr\rangle^{ }_{0,\rmi{c}}
 \nn & & \hspace*{0.5cm} \times  
 \; \deltabar(-  R +  P +  T) 
 \; \deltabar(-  U +  V +  X) T^c_{AB} T^d_{CD}
 \nn[3mm] 
 & = &  
 {g^2} 
 \Tint{ P  V \{  R  T  U\!  X \} } 
 \hspace*{-1cm}
 \langle 
 \tilde A^a_\mu( K)
 \tilde A^c_\alpha( P) 
 \rangle_0^{ } \;
 \langle 
 \tilde A^b_\nu( Q)
 \tilde A^d_\beta( V) 
 \rangle_0^{ } \;
 \tr\Bigl[ 
 \langle 
 \tilde \psi^{ }_D( X) 
 \,\tilde{\!\bar \psi}^{ }_A( R) 
 \rangle_0^{ } \;  \gamma^{ }_\alpha \; 
 \langle 
 \tilde \psi^{ }_B( T) 
 \,\tilde{\!\bar \psi}^{ }_C( U) 
 \rangle_0^{ }  \gamma^{ }_\beta 
 \Bigr]
 \nn & & \hspace*{0.5cm} \times  
 \; 
 \; \deltabar(-  R +  P +  T) 
 \; \deltabar(-  U +  V +  X) T^c_{AB} T^d_{CD}
 \;,  \la{fe_loop}
\ea
where the Grassmann nature of the fermions induced a minus sign.
As noted earlier, the capital indices originating from the quark spinors 
stand both for colour and flavour quantum numbers.

Inserting next the gluon propagator from \eq\nr{Fprop}
and the fermion propagator from \eq\nr{psiprop}, 
let us once more inspect in turn the colour and flavour indices, 
Lorentz indices, and momenta. 
The colour and flavour contractions result this time in the factor
\be
 \delta^{ac}\delta^{bd}\delta^{ }_{DA}\delta^{ }_{BC} T^c_{AB} T^d_{CD}
 = \tr [T^a T^b] = \frac{\Nf}{2}
 \;,
\ee
where we assumed the flavours to be degenerate 
in mass and in addition took advantage of the assumed normalization 
of the fundamental representation generators $T^a$. 
The spacetime indices yield on the other hand 
\ba
 \delta^{ }_{\mu\alpha}\delta^{ }_{\nu\beta}
 \tr[(-i \bsl{ R}\! + m )  \gamma^{ }_\alpha 
     (-i \bsl{ U}\! + m )  \gamma^{ }_\beta]
 & = & 
 4 [-  R^{ }_\sigma  U^{ }_\rho (
   \delta^{ }_{\sigma\mu}\delta^{ }_{\rho\nu}
  - \delta^{ }_{\sigma\rho}\delta^{ }_{\mu\nu}
  + \delta^{ }_{\sigma\nu}\delta^{ }_{\rho\mu}
 ) + m^2 \delta^{ }_{\mu\nu}]
 \nn & = & 
 4 [\delta^{ }_{\mu\nu}( R\cdot  U + m^2) 
 -  R^{ }_\mu  U^{ }_\nu -  R^{ }_\nu  U^{ }_\mu]
 \;, \la{ferm_nume}
\ea
where we used standard results for the traces 
of Euclidean $\gamma$-matrices. Momenta get fixed as 
$
 P\to -K
$, 
$
 V\to -Q
$, 
$
 R \to X
$, 
$
 U \to T
$.
Thereby 
\be
 \; \deltabar(-  R +  P +  T) 
 \;\, \deltabar(-  U +  V +  X) 
 \to 
 \; \deltabar(-  X -  K +  T) 
 \;\, \deltabar(-  T -  Q +  X) 
 = 
 \; \deltabar(-  X -  K +  T) 
 \;\, \deltabar( K +  Q)   
 \;,  
\ee
which implies that we can substitute
$
 f( R,  U)
 \to 
 f(  X ,  T )
 \to 
 f( T -  K,  T)
$
in \eq\nr{ferm_nume}.
Renaming finally $T\to P$, 
the contribution of the quark loop diagram to the self-energy becomes
\ba
 & & \hspace*{-1cm} 
 \Bigl\langle \tilde A^a_\mu( K)
 \tilde A^b_\nu( Q) 
 \Bigl(\fr12 S_\iI^2 \Bigr) \Bigr\rangle_{0,\rmi{c}}^{(3)} 
 \nn 
 \!\!\!\!\! & = &  \!\!\!
 2 {g^2 \Nf} 
 \, \frac{\delta^{ab} \; \deltabar( K +  Q)}{( K^2)^2}
 \, \Tint{ \{  P \} } 
 \frac{\delta^{ }_{\mu\nu}( P^2 -  K\cdot  P + m^2) 
 - 2  P^{ }_\mu  P^{ }_\nu 
 +  K^{ }_\mu  P^{ }_\nu +  K^{ }_\nu  P^{ }_\mu}
 {[ P^2 +m^2][ ( K -  P)^2 + m^2 ]}
 \;. \hspace*{1cm} \la{psiloop}
\ea
For {\em vanishing chemical potential}, a shift like 
in \eq\nr{P_shift} works also with fermionic four-momenta, 
so that this expression further simplifies to
\ba
 & & \hspace*{-1cm} 
 \Bigl\langle \tilde A^a_\mu( K)
 \tilde A^b_\nu( Q) 
 \Bigl(\fr12 S_\iI^2 \Bigr) \Bigr\rangle_{0,\rmi{c}}^{(3)} 
 \nn 
 \!\!\!\!\! & = &  \!\!\!
 {g^2 \Nf} 
 \, \frac{\delta^{ab} \; \deltabar( K +  Q)}{( K^2)^2}
 \, \Tint{ \{  P \} } 
 \frac{\delta^{ }_{\mu\nu}(2  P^2 -  K^2 + 2 m^2) 
 - 4  P^{ }_\mu  P^{ }_\nu 
 + 2  K^{ }_\mu  K^{ }_\nu }
 {[ P^2 +m^2][ ( K -  P)^2 + m^2 ]}
 \;. \hspace*{1cm} \la{psiloop_2}
\ea

\index{Thermal sums: fermionic tensor}

The structure in the numerator of \eq\nr{psiloop_2} is similar to that 
in \eq\nr{DD2}, except that the Matsubara frequencies are fermionic. 
In particular, if we again set the external momentum to zero, 
and for simplicity also consider the limit $T\gg m$, so that quark masses
can be ignored, the entire term becomes proportional to
\ba
 \Tint{\{  P \} } 
 \frac{\delta^{ }_{\mu\nu} P^2 - 2  P^{ }_\mu  P^{ }_\nu}
      {( P^2)^2} & = & 
 \delta^{ }_{\mu 0}\delta^{ }_{\nu 0}\,
 \Tint{ \{ P \}  } 
 \frac{ P^2 - 2  P_0^2}
      {( P^2)^2} + 
 \delta^{ }_{\mu i}\delta^{ }_{\nu i}\,
 \Tint{ \{ P \}  } 
 \frac{ P^2 - 2  P_i^2}
      {( P^2)^2}
 \nn  & =  &
 \delta^{ }_{\mu 0}\delta^{ }_{\nu 0}\,
 \Tint{ \{ P \}  } 
 \frac{ P^2 - 2  p_n^2}
      {( P^2)^2} + 
 \delta^{ }_{\mu i}\delta^{ }_{\nu i}\,
 \Tint{ \{ P \}  } 
 \frac{ P^2 - (\tfr{2}{d}) {p}^2}
      {( P^2)^2}
 \nn  & =  &
 \delta^{ }_{\mu 0}\delta^{ }_{\nu 0}\,
 \Tint{ \{ P \}  } 
 \frac{ P^2 - 2  p_n^2}
      {( P^2)^2} + 
 \delta^{ }_{\mu i}\delta^{ }_{\nu i}\,
 \Tint{ \{ P \}  } 
 \frac{(1-\tfr{2}d) P^2 +(\tfr{2}d)  p_n^2}
      {( P^2)^2}
 \;. \hspace*{8mm} \la{qSmuSnu}
\ea
The relation in \eq\nr{new_sum_int} continues to hold in the fermionic
case, so setting $d\to 3$, we get 
\be
 \Tint{ \{ P \}  } 
 \frac{\delta^{ }_{\mu\nu} P^2 - 2  P^{ }_\mu  P^{ }_\nu}
      {( P^2)^2} = 
 \delta^{ }_{\mu 0}\delta^{ }_{\nu 0} \times 2 \tilde I^{ }_T(0) 
 + \delta^{ }_{\mu i}\delta^{ }_{\nu i} \times \Bigl(\fr13 - \fr13 \Bigr)
  \tilde I^{ }_T(0) + \rmO(\epsilon)
 \;. 
\ee
Inserting here finally  $\tilde I^{ }_T(0) = - T^2/24$ 
from \eq\nr{If}, we arrive at the final result for the fermionic contribution,
\ba
 & & \hspace*{-1cm} 
 \Bigl\langle \tilde A^a_\mu( K)
 \tilde A^b_\nu( Q) \Bigl(\fr12 S_\iI^2 \Bigr) 
 \Bigr\rangle_{0,\rmi{c}}^{(3)} 
 \approx
 -{g^2 \Nf} 
 \,\frac{\delta^{ab} \; \deltabar( K +  Q)}{( K^2)^2}
 \,\biggl\{ 
  \delta^{ }_{\mu0}\delta^{ }_{\nu0}
  + 0 \times \delta^{ }_{\mu i}\delta^{ }_{\nu i}
 \biggr\} \frac{T^2}{6} + \rmO(\epsilon)
 \;. \hspace*{1cm} \la{psiloop2}
\ea

\index{Ward-Takahashi identities}
\index{Slavnov-Taylor identities}

Summing together \eqs\nr{Gtadpole}, \nr{Gloop3}, \nr{cloop3} 
and \nr{psiloop2} and omitting the terms of $\rmO(\epsilon)$, 
we find the surprisingly compact expression
\ba
 & & \hspace*{-1cm} 
 \Bigl\langle \tilde A^a_\mu( K)
 \tilde A^b_\nu( Q) 
 \Bigl(-S^{ }_\iI + \fr12 S_\iI^2 \Bigr) \Bigr\rangle^{ }_{0,\rmi{c}} 
 \nn 
 \!\!\!\!\! &  \approx  &  \!\!\!
 -{g^2} 
 \,\frac{\delta^{ab} \; \deltabar( K +  Q)}{( K^2)^2}
 \,\biggl\{ 
  \biggl[\biggl(3 + \fr32 - \fr12 \biggr)\Nc + 2 \Nf \biggr]
  \delta^{ }_{\mu0}\delta^{ }_{\nu0}
  + \biggl[\biggl(3 - \fr72 + \fr12  \biggr)\Nc \biggr]
   \delta^{ }_{\mu i}\delta^{ }_{\nu i}
 \biggr\} \frac{T^2}{12}
 \hspace*{1cm} \nn 
 & = & 
- \frac{\delta^{ab}\,\delta^{ }_{\mu0}\delta^{ }_{\nu0} 
  \;\, \deltabar( K +  Q)}{( K^2)^2}
  \, \times 
 g^2 T^2 \biggl( \frac{\Nc}{3} + \frac{\Nf}{6} \biggr)
 \;. \la{GselfE}
\ea
It is important to note that 
all corrections have cancelled from the spatial part.\footnote{%
 We checked this for $d=3$ but with some more effort it is 
 possible to verify that the same is true for general $d$. 
 The generalization of the coefficient to 
 general $d$ is given in \eq\nr{mmE_D}.
 } 
Due to Ward-Takahashi identities (or more properly their
non-Abelian generalizations, Slavnov-Taylor identities), 
the gauge field self-energy 
must be transverse with respect to the external four-momentum, 
which in the case of the Matsubara
zero mode takes the form $K = (0,\vec{k})$. Since we computed the 
self-energy with $\vec{k}=\vec{0}$, the transverse structure 
$\delta^{ }_{ij} k^2 - k^{ }_i k^{ }_j$ cannot appear, and the spatial
part must vanish altogether. 

The result obtained above has a direct physical meaning. 
Indeed, we recall from the discussion of scalar field theory, 
\eq\nr{resum_Gx}, that \eqs\nr{Fprop_Pi} and \nr{GselfE} can 
be interpreted as a (resummed) full propagator of the form
\be
 \Bigl\langle \tilde A^a_\mu( K)
 \tilde A^b_\nu( Q) 
 \Bigr\rangle 
 \quad \stackrel{K\approx 0}{\approx} \quad
 \frac{\delta^{ab} \delta^{ }_{\mu\nu}\; 
 \deltabar( K +  Q)}
 { K^2 + \delta^{ }_{\mu0}\delta^{ }_{\nu0}\,\mE^2}
 \;, \la{A0_effprop}
\ee
where
\be \index{Effective mass}
 \mE^2 \;\equiv\; g^2 T^2 \biggl( \frac{\Nc}{3}  + \frac{\Nf}{6} \biggr)
 \la{mmE}
\ee
is called the {\em Debye mass parameter}. 
Its existence corresponds to the fact the colour-electric
field $ A_0^{ }$ 
gets exponentially {\em screened} in a thermal plasma, like
the scalar field propagator in \eq\nr{Px_large}. In contrast, the 
colour-magnetic field $ A_i$ {\em does not get screened}, 
at least at this order. 

We conclude with two remarks: 
\bi
\item
If we consider the full Standard Model rather than QCD
(the corresponding Euclidean Lagrangian is given on p.~\pageref{SM}), 
then there is 
a separate thermal mass for the temporal components of 
all three gauge fields, and for the Matsubara
zero mode of the Higgs field. 
These can be found in ref.~\cite{meg}.

\item
The definition of a Debye mass becomes
ambiguous at higher orders. One possibility is to define it as
a ``matching coefficient'' in a certain ``effective theory''; this
is discussed in more detail in \se\ref{se:DR_QCD}, cf.\ \eq\nr{barm2}. 
In that case higher-order corrections to the expression in \eq\nr{mmE}
can be computed~\cite{Ghi-Sch}. On the other hand, if we want
to define the Debye mass as a {\em physical quantity}, the result becomes
non-perturbative already at the next-to-leading order~\cite{rebhan}, 
and a proper definition and extraction requires a lattice approach~\cite{ay}.

\ei

\index{Debye mass}
\index{Screening}

\newpage 

\subsection{Free energy density to $\rmO(g^3)$}
\la{se:fTg3}

\index{Free energy density: QCD}

As an application of the results of the previous 
section, we now compute the free energy density of QCD up to $\rmO(g^3)$, 
parallelling the method introduced for scalar field theory 
around \eq\nr{LE_resum}.
We recall that the essential insight in this treatment was to supplement
the quadratic part of the Lagrangian for the Matsubara zero modes
by an effective thermal mass computed from the full propagator, and 
to treat minus the same term as part of the interaction 
Lagrangian.
The ``non-interacting'' free energy density computed
with the corrected propagator 
then yields the result for the ring sum, 
whereas the bilinear interaction 
term cancels the corresponding, infrared
(IR) divergent contributions 
order by order in a loop expansion.\footnote{%
 This cancellation is the only role that the subtraction plays
 at the order
 that we are considering, cf.\ the discussion below \eq\nr{resu_2}. In the 
 following we simplify the procedure by computing the contribution
 of $\rmO(g^2)$ with massless propagators, whereby no odd powers of 
 thermal masses are generated and the subtraction can be
 omitted as well. 
 }  
In the present case, given the result 
of \eq\nr{A0_effprop}, we see that only 
the temporal components of the gauge fields need 
to be corrected with a mass term. 
This is in accordance with the gauge transformation properties 
of static colour-electric and colour-magnetic fields, which forbid
the spatial components from having a mass; we return to this 
in \se\ref{se:DR_QCD}.

With the above considerations in mind, 
the correction of $\rmO(g^3)$~\cite{jk2} to the tree-level result
in \eq\nr{f_QCD_0} can immediately be written down, 
if we employ \eq\nr{Jn0_res} and take into account that 
there are $\Nc^2 - 1$ copies of the gauge field. This produces 
\ba
 \left. f^{ }_{(\sfr32)}(T) \right|^{ }_\rmi{QCD}
 & = & 
 (\Nc^2 - 1) \biggl( - \frac{T \mE^3}{12\pi} \biggr)
 \nn & = & 
 (\Nc^2 - 1) T^4 g^3 \biggl( - \frac{1}{12\pi} \biggr)
 \biggl(  \frac{\Nc}{3}  + \frac{\Nf}{6} \biggr)^{\fr32}
 \nn & = & 
  - \frac{\pi^2 T^4}{3} 2 (\Nc^2 - 1)
 \biggl( \frac{g^2}{4\pi^2} \biggr)^{\fr32}
\biggl(  \frac{\Nc}{3}  + \frac{\Nf}{6} \biggr)^{\fr32}
 \;, \la{f_QCD_3} 
\ea
where the effective mass $\mE$ was taken from \eq\nr{mmE}.

Next, we consider the contributions of $\rmO(g^2)$. 
In analogy with \eq\nr{resu_2}, these terms~\cite{es,ch}
come from the non-zero mode contributions 
to the 2-loop ``vacuum''-type graphs in (cf.\ \eq\nr{compact_rule})
\be
 \left. f^{ }_{(1)}(T) \right|^{ }_\rmi{QCD}
 = 
 \Bigl\langle
 S^{ }_\iI - \fr12 S_\iI^2 + \ldots 
 \Bigr\rangle^{ }_{0,\rmi{c},\,\rmi{drop overall $\int_X$}}
 \;. \la{compact_QCD}
\ee
It is useful to compare this expression
with the computation of the full propagator in the previous
section, \eq\nr{GselfE}. We note that, apart from an overall
minus sign, the two computations are quite similar at the present order. 
In fact, we claim that we only need to ``close'' the gluon
line in the results of the previous section and simultaneously
divide the graphs by $-1/2n$, where $n$ is the number of gluon lines 
in the vacuum graph in question. Let us prove this by direct inspection. 

Consider first the $\rmO(g^2)$ contribution from the 4-gluon vertex. 
In vacuum graphs, this leads to the combinatorial factor 
\be
 \langle \tilde A\,\tilde A\,\tilde A\,\tilde A \rangle^{ }_{0,\rmi{c}}
 = 3 \, \langle \tilde A\,\tilde A \rangle_0^{ } \; 
 \langle \tilde A\,\tilde A \rangle_0^{ } \;, 
 \la{ex1}
\ee
whereas in the propagator calculation we arrived at
\be
 - \langle \tilde A\,\tilde A\;\; 
 \tilde A\,\tilde A\,\tilde A\,\tilde A \rangle^{ }_{0,\rmi{c}}
 = - 4\times 3 \, \langle \tilde A\,\tilde A \rangle_0^{ } \; 
             \langle \tilde A\,\tilde A \rangle_0^{ } \; 
            \langle \tilde A\,\tilde A \rangle_0^{ } \;. 
\ee
The difference is $- 4 = - 2 \times n$, 
with $n=2$ being the number of contractions in \eq\nr{ex1}.
Similarly, with the contribution from two 3-gluon vertices,
the vacuum graphs lead to the combinatorial factor 
\be
 - \langle \tilde A\,\tilde A\,\tilde A \;
 \; \tilde A\, \tilde A\, \tilde A \rangle^{ }_{0,\rmi{c}}
 = - 3\times 2 \, \langle \tilde A\,\tilde A \rangle_0^{ } \;
 \langle \tilde A\,\tilde A \rangle_0^{ } 
 \; \langle \tilde A\,\tilde A \rangle_0^{ } \;, 
 \la{ex2}
\ee
whereas when considering the propagator we got
\be
 \langle \tilde A\,\tilde A\;
 \;  \tilde A\,\tilde A\,\tilde A \;
 \; \tilde A\, \tilde A\, \tilde A \rangle^{ }_{0,\rmi{c}}
 =  6\times 3 \times 2 \, 
             \langle \tilde A\,\tilde A \rangle_0^{ } \; 
             \langle \tilde A\,\tilde A \rangle_0^{ } \; 
             \langle \tilde A\,\tilde A \rangle_0^{ } \; 
             \langle \tilde A\,\tilde A \rangle_0^{ } \;.
\ee
There is evidently a difference of $- 6 = - 2 \times n$, 
with $n=3$ the number of contractions in \eq\nr{ex2}.
Finally, the ghost and fermion contributions to  
the vacuum graphs lead to the combinatorial factor 
\be
 - \langle \; \tilde{\!\bar c}\,\tilde A\, \tilde c \;\;
           \tilde{\!\bar c}\, \tilde A\, \tilde c \rangle^{ }_{0,\rmi{c}}
 = \langle \tilde A\,\tilde A \rangle_0^{ } \;
 \langle \tilde c\;\tilde{\!\bar c} \rangle_0^{ } 
\; \langle \tilde c\;\tilde{\!\bar c} \rangle_0^{ } \;, 
 \la{ex3}
\ee
whereas in the propagator computation we obtained
\be
 \langle \tilde A \,\tilde A\;\; 
 \tilde{\!\bar c}\,\tilde A\, \tilde c \;\;
 \tilde{\!\bar c}\, \tilde A\, \tilde c \rangle^{ }_{0,\rmi{c}}
 = - 2 \, \langle \tilde A\,\tilde A \rangle_0^{ } \; 
             \langle \tilde A\,\tilde A \rangle_0^{ } \; 
             \langle \tilde c\;\tilde{\!\bar c} \rangle_0^{ } \; 
             \langle \tilde c\;\tilde{\!\bar c} \rangle_0^{ } \;.
\ee
So once more a difference of $- 2 \times n$, 
with $n=1$ the number of gluon contractions in \eq\nr{ex3}.
 
With the insights gained, 
the contribution of the 4-gluon vertex to the free energy density of QCD
can be extracted directly from \eq\nr{Gtadpole}:
\ba
 \ToptVE(\Agl,\Agl) 
 & = & -\fr14 
 \biggl\{  
 - g^2 \Nc \, d \, I^{ }_T(0)\, 
 \Tint{ K}
 \frac{\delta^{aa}\delta^{ }_{\mu\mu}}{ K^2}
 \biggr\} 
 \nn 
 & = & \frac{g^2}{4} \Nc (\Nc^2 -1 ) d (d+1) [ I^{ }_T(0) ]^2
 \;. \la{ggAAAA}
\ea
The contribution of the 3-gluon vertices is similarly 
obtained from \eq\nr{Gloop}: noting from \eq\nr{DD} that 
\ba
 && \hspace*{-1cm} \delta^{ }_{\mu\nu}
 \; D^{ }_{\mu\beta\gamma}(-  K ,  P ,  K -  P)
 \; D^{ }_{\nu\beta\gamma}( K , - P , -  K +  P)
 \nn & = & 
 - (d+1) [4  K^2 + ( K -  P)^2 +  P^2]
 - (d-5) K^2 
 + 2 (2d - 1)  K \cdot  P
 - (4d - 2) P^2
 \hspace*{1cm} 
 \nn & = &
 -\{
  K^2 [5d + 5 + d - 5]
 +  K\cdot  P [-2d - 2-4d + 2]
 + P^2 [2d + 2 + 4d - 2]
 \}
 \nn & = & 
 - 3d \{  K^2 + ( K -  P)^2 +  P^2 \}
 \;, 
\ea
we get from \eq\nr{Gloop}
\ba
 \ToptVS(\Agl,\Agl,\Lgl) 
 & = & 
 -\fr16\biggl\{ 
 \frac{3 g^2 \Nc}{2} d \;
 \Tint{ K} \frac{\delta^{aa}}{ K^2}
 \Tint{ P} 
 \frac{ K^2 + ( K -  P)^2 +  P^2}
 { P^2 ( K -  P)^2}
 \biggr\}
 \nn & = & 
 - \frac{g^2}{4} \Nc (\Nc^2 - 1)\; d \times 3 \;[I^{ }_T(0)]^2
 \;. \la{ggAAA}
\ea
Note that unlike in \eq\nr{Gloop2}, 
for the present calculation it was crucial to keep the full 
$ K$-dependence in the two-point function, 
because {\em all values of} $ K$ are now integrated over. 

Similarly, the contribution of the ghost loop 
can be extracted from \eq\nr{cloop}, producing 
\ba
 \ToptVS(\Agh,\Agh,\Lgl) 
 & = & 
 -\fr12\biggl\{ 
 -{g^2 \Nc} 
 \Tint{ K}
 \frac{\delta^{aa}}{ K^2}
 \Tint{ P} 
 \frac{  P^2 -  K \cdot  P }
 { P^2 ( K -  P)^2}
 \biggr\}
 \nn 
 & = & 
 \frac{g^2}{4} \Nc (\Nc^2 - 1)
 \Tint{ K  P}
 \frac{  P^2 +  ( K -  P)^2 -  K^2 }
 { K^2  P^2 ( K -  P)^2}
 \nn 
 & = & 
 \frac{g^2}{4} \Nc (\Nc^2 - 1)
 [I^{ }_T(0)]^2
 \;, \la{ggcc}
\ea
whereas the contribution of the fermion loop 
is obtained from \eq\nr{psiloop}: 
\ba
 \ToptVS(\Aqu,\Aqu,\Lgl) 
 & = & -\fr12 \biggl\{ 
 2 {g^2 \Nf} 
 \Tint{ K}
 \frac{\delta^{aa}}{ K^2}
 \Tint{ \{ P \}  } 
 \frac{(d+1)( P^2 -  K\cdot  P + m^2) 
 - 2  P^2 + 2  K \cdot  P}
 {[ P^2 +m^2][ ( K -  P)^2 + m^2 ]}
 \biggr\} 
 \;.
\ea
Simplifying the 
last expression by setting $m/T \to 0$, we get
\ba
 \ToptVS(\Aqu,\Aqu,\Lgl) 
 & = & 
 - {g^2 \Nf}(\Nc^2 - 1) 
 \Tint{ K  \{ P \}  } 
 \frac{(d-1)( P^2 -  K\cdot  P)}
 { K^2  P^2 ( K -  P)^2 }
 \nn 
  & = & 
 - {g^2 \Nf}(\Nc^2 - 1) \frac{d-1}{2}
 \Tint{ K  \{ P \}  } 
 \frac{ P^2 + ( K -  P)^2 -  K^2}
 { K^2  P^2 ( K -  P)^2 }
 \nn 
 & = & 
 -\frac{g^2}{2} \Nf (\Nc^2 - 1) (d-1)
 \bigl\{ 2 I^{ }_T(0) \tilde I^{ }_T(0) - [\tilde I^{ }_T(0)]^2 \bigr\}
 \;. \la{ggpsipsi}
\ea
Here careful attention needed to be paid to the 
nature of the Matsubara frequencies appearing in the propagators. 

Adding together the terms from \eqs\nr{ggAAAA}, \nr{ggAAA}, 
\nr{ggcc} and \nr{ggpsipsi}, 
setting $d=3$ (note the absence of divergences), 
and using 
$
 I^{ }_T(0) = T^2/12
$, 
$ 
 \tilde I^{ }_T(0) = - T^2/24
$, 
we get as the full $\rmO(g^2)$ contribution to the free energy density
\ba
 \left. f^{ }_{(1)}(T) \right|^{ }_\rmi{QCD}
 & = &
 g^2 (\Nc^2 - 1) \frac{T^4}{144}
 \biggl[ 
 \biggl( 3 - \fr94 +\fr14 \biggr)\Nc 
 - \biggl( -2\times\fr12 -\fr14 \biggr)\Nf
 \biggr]
 \nn 
 & = & 
 g^2 (\Nc^2 - 1) \frac{T^4}{144} \biggl( \Nc + \fr54 \Nf\biggr)
 \nn 
 & = &  
 -\frac{\pi^2 T^4}{90} (\Nc^2 - 1)
 \biggl( -\fr52 \frac{g^2}{4\pi^2}\biggr) \biggl( \Nc + \fr54 \Nf\biggr)
 \;.  \la{f_QCD_2}
\ea
Adding to this the effects of \eqs\nr{f_QCD_0} 
and \nr{f_QCD_3}, the final result reads
\ba
 \left. f(T) \right|^{ }_\rmi{QCD}
 & = &
 -\frac{\pi^2 T^4}{45} (\Nc^2 - 1)
 \biggl\{
   1  + \fr74 \frac{\Nf \Nc}{\Nc^2 - 1} 
 - \fr54 \biggl( \Nc + \fr54 \Nf\biggr) \frac{\alphas}{\pi}
  \nn & & \hspace*{3cm}
 +\, 30 \biggl(  \frac{\Nc}{3}  + \frac{\Nf}{6} \biggr)^{\fr32}
 \biggl( \frac{\alphas}{\pi} \biggr)^\fr32
 + \rmO(\alphas^2)
 \biggr\}
 \;, \la{fT_QCD}
\ea
were we have denoted $\alphas \equiv g^2/4\pi$. 

A few remarks are in order:
\bi

\item 
The result in \eq\nr{fT_QCD} can be compared with that for a
scalar field theory in \eq\nr{fT_SFT}. The general structure
is identical, and in particular the first relative correction 
is {\em negative} in both cases. 
This means that the interactions between the particles in a plasma 
tend to {\em decrease} the pressure that the plasma exerts.

\item
The second correction to the pressure turns out to be positive. 
Such an alternating
structure indicates that it may be difficult to quantitatively
estimate the magnitude of radiative corrections to 
the non-interacting result. We may recall, however,  
that $1-\tfr12+\tfr13-\tfr14 ...= \ln 2 = 0.693...$, whereas
$1-\tfr12-\tfr13-\tfr14 ...= -\infty$; in principle an alternating 
structure is beneficial as far as (asymptotic) convergence goes.

\item
The coefficients of the four subsequent terms, 
of orders
$\rmO(\alphas^2\ln \alphas)$, 
$\rmO(\alphas^2)$, 
$\rmO(\alphas^{5/2})$, and
$\rmO(\alphas^3 \ln \alphas )$, 
are also known~\cite{tt}--\hspace*{-1.1mm}\cite{klrs}.
Like for scalar field theory, this progress is possible 
thanks to the use of effective field theory methods that we 
discuss in the next chapter.

\ei


\subsection*{Appendix A: Do ghosts develop a thermal mass?}

\index{Thermal mass: ghost?}
\index{Ghost self-energy}

In the computation of the present section, we have assumed that 
only the Matsubara zero modes of the fields $A^a_0$ need
to be resummed, i.e.\ get an effective thermal mass. 
The fact that fermions do not need to be resummed is clear, 
but the case of ghosts is less obvious. To this end, 
let us finish the section by demonstrating that 
ghosts do {\em not} get any
thermal mass, and thus behave like the {\em spatial} components
of the gauge fields. 

The tree-level ghost propagator is given in \eq\nr{cprop}, and we now
consider corrections to this expression. The relevant vertex is
the one in \eq\nr{ccA}, yielding for the only correction of $\rmO(g^2)$  
\ba
 & & \hspace*{-1cm} 
 \Bigl\langle \tilde c^a( K)
 \,\tilde{\!\bar c}^{\,b}( Q) 
 \Bigl(\fr12 S_\iI^2 \Bigr) \Bigr\rangle^{ }_{0,\rmi{c}} 
 \quad = \quad \TopoSB(\Lhh,\Agh,\Agl)
 \nn 
 \!\!\!\!\! & = &  \!\!\!
 -\frac{g^2}{2} 
 \Bigl\langle 
 \tilde c^a( K)
 \,\tilde{\!\bar c}^{\,b}( Q)
 \,\Tint{ R  P  T} 
 \hspace*{-0.5cm}
 \,\tilde{\!\bar c}^{\,c}( R) 
 \tilde A^d_\alpha( P) 
 \tilde c^e( T) 
 \,\Tint{ U  V \! X} 
 \hspace*{-0.5cm}
 \,\tilde{\!\bar c}^{\,g}( U) 
 \tilde A^h_\beta( V) 
 \tilde c^i( X) 
 \Bigr\rangle^{ }_{0,\rmi{c}}
 \nn & & \hspace*{0.5cm} \times  
 \; 
 f^{cde} f^{ghi}
 \; \deltabar(-  R +  P +  T) 
 \; \deltabar(-  U +  V +  X) 
 \;  R^{ }_\alpha  U^{ }_\beta
 \nn 
 & = &  
 -{g^2} 
 \Tint{ R  P  T  U  V\!  X} 
 \hspace*{-1cm}
 \langle 
 \tilde c^a( K)
 \,\tilde{\!\bar c}^{\,c}( R) 
 \rangle_0^{ } \;
 \langle 
 \tilde c^e( T) 
 \,\tilde{\!\bar c}^{\,g}( U) 
 \rangle_0^{ } \;
 \langle 
 \tilde c^i( X) 
 \,\tilde{\!\bar c}^{\,b}( Q)
 \rangle_0^{ } \;
 \langle 
 \tilde A^d_\alpha( P) 
 \tilde A^h_\beta( V) 
 \rangle_0^{ } \;
 \nn & & \hspace*{0.5cm} \times  
 \; 
 f^{cde} f^{ghi}
 \; \deltabar(-  R +  P +  T) 
 \; \deltabar(-  U +  V +  X) 
 \;  R^{ }_\alpha  U^{ }_\beta
 \;, \la{gh_loop}
\ea
where an even number of minus signs originated from 
the commutations of Grassmann fields. 
Inserting here the gluon propagator from \eq\nr{Fprop}
as well as the free ghost propagator from \eq\nr{cprop}, 
we again end up inspecting colour indices, 
Lorentz indices, and momenta in the resulting expression. 
The colour contractions are seen to result in the factor
\be
 \delta^{ac}\delta^{eg}\delta^{ib}\delta^{dh}  f^{cde} f^{ghi}
 = f^{ade}f^{edb} = -\Nc \, \delta^{ab}
 \;, 
\ee
whereas the spacetime indices yield simply $\delta^{ }_{\alpha\beta}$.
Momenta get fixed as 
$
 R\to K
$, 
$
 U\to T
$, 
$
 X\to Q
$, 
$
 V\to -P
$.
Thereby
\be
 \;\, \deltabar(-  R +  P +  T) 
 \;\, \deltabar(-  U +  V +  X) 
 \to 
 \;\, \deltabar(-  K +  P +  T) 
 \;\, \deltabar(-  T -  P +  Q) 
 =
 \;\, \deltabar(-  K +  P +  T) 
 \;\, \deltabar(-  K +  Q)  
 \;,
\ee
implying that 
$
 T = K - P
$
and 
$
 R\cdot U 
 \to
 K\cdot T
 \to 
 K\cdot (K-P)
$.
Inserting all this, \eq\nr{gh_loop} turns into
\ba
 & & \hspace*{-1cm} 
 \Bigl\langle \tilde c^a( K)
 \,\tilde{\!\bar c}^{\,b}( Q) 
 \Bigl(\fr12 S_\iI^2 \Bigr) \Bigr\rangle^{ }_{0,\rmi{c}} 
 = 
 {g^2 \Nc} 
 \frac{\delta^{ab} \; \deltabar( K -  Q)}{( K^2)^2}
 \Tint{ P} 
 \frac{ K \cdot  (K - P)}
 { P^2  ( K -  P)^2}
 \;. \hspace*{1cm} \la{ghsE}
\ea

The expression in \eq\nr{ghsE} is proportional to the external 
momentum $ K$. Therefore, it does
{\em not} represent an effective mass correction, but is rather 
a ``wave function (re)normalization'' contribution, as can be made 
explicit through a shift like in \eq\nr{P_shift}.

%

\newpage


\newpage 

\section{Low-energy effective field theories}
\la{se:EFT}

\paragraph{Abstract:}

The existence of a so-called infrared (IR) problem in relativistic 
thermal field theory is pointed out, both
from a physical and a formal (imaginary-time) point of view. The notion
of effective field theories is introduced, 
and the main issues related to their
construction and use are illustrated with the help of a simple example. 
Subsequently this methodology is applied to the imaginary-time path 
integral represention for the partition function of non-Abelian gauge
field theory. This leads to the construction of a dimensionally
reduced effective field theory for capturing certain 
(so-called ``static'', i.e.\ time-independent) 
properties of QCD (or more generally Standard Model) 
thermodynamics in the high-temperature limit. 

\paragraph{Keywords:} 

Infrared divergences, power counting, Matsubara zero mode, 
Bose enhancement, Linde problem, 
hard and soft modes, effective theories, Electrostatic QCD, 
Magnetostatic QCD, symmetries, matching, truncation. 

\index{Effective field theories: general}
\index{Infrared divergence: general}
\index{Zero mode: Matsubara formalism}

%
\subsection{The infrared problem of thermal field theory}
\la{se:Linde}

Let us start by considering the types of integrals that appear in 
thermal perturbation theory. According to \eqs\nr{Sfn_res} and \nr{Sfe_res}, 
each new loop order (corresponding to an additional loop 
momentum) produces one of
\ba
 \Tint{ P} 
 \, f(\omega^{ }_n,\vec{p})
 & = & 
 \int_{\vec p}
 \biggl\{  
 \fr12 
 \int_{-\infty-i0^+_{ }}^{+\infty-i0^+_{ }}
 \frac{{\rm d}\omega}{2\pi}\,
 [f(\omega,\vec{p}) + f(-\omega,\vec{p})] [ 1 + 2 \nB{}(i\omega) ]  
 \biggr\}  
 \;, \la{Tint_bo} \\
 \Tint{ \{ P \} } 
 \!\! f(\omega^{ }_n,\vec{p})
 & = & 
 \int_{\vec p}
 \biggl\{  
 \fr12 
 \int_{-\infty-i0^+_{ }}^{+\infty-i0^+_{ }}
 \frac{{\rm d}\omega}{2\pi}\,
 [f(\omega,\vec{p}) + f(-\omega,\vec{p})] 
 [ 1 - 2\nF{}(i\omega) ] 
 \biggr\} 
 \;, \la{Tint_fe}
\ea
depending on whether the new line is bosonic or fermionic. 
The functions $f$ here contain propagators and additional structures
emerging from vertices; in the simplest case, 
$
 f(\omega,\vec{p}) \sim 1 / (\omega^2 + \E_p^2)
$, 
where we denote 
$
 \E^{ }_p \equiv \sqrt{{p}^2 + m^2}
$. 

Now, the structures which are the most important, or yield 
the largest contributions, are those where the functions $f$
are largest. Let us inspect this question in terms of the left
and right-hand sides of \eqs\nr{Tint_bo} and \nr{Tint_fe}.

For bosons, the largest contribution on the left-hand side of 
\eq\nr{Tint_bo} is clearly associated with the {\em Matsubara zero mode}, 
$\omega^{ }_n = 0$; in the case
$
 f(\omega,\vec{p}) \sim 1 / (\omega^2 + \E_p^2)
$, 
this gives simply
\be
 \left. \Bigl\{ T \sum_{\omega^{ }_n} f \Bigr\}
 \right|_{\omega^{ }_n = 0} 
 \sim \frac{T}{\E_p^2}
 \;. \la{ir_prob}
\ee
On the right-hand side, we on the other hand close the contour
in the lower half-plane, whereby the largest contribution is associated with 
{\em Bose enhancement} around the pole $\omega = - i \E^{ }_p$: 
\ba \index{Bose enhancement}
 \{ \ldots \} & \sim & 
 \fr12 \frac{-2\pi i}{2\pi} \frac{2}{-2 i \E^{ }_p}
 \Bigl[ 1 + 2 \nB{} (\E^{ }_p) \Bigr]
 = \frac{1}{\E^{ }_p}
 \biggl( \fr12 + \frac{1}{e^{\E^{ }_p/T} - 1} \biggr)
 \nn & \approx &
 \frac{1}{\E^{ }_p} 
 \biggl( \fr12 + \frac{1}{{\E^{ }_p/T} + \E_p^2/2 T^2 } + \ldots \biggr)
 = \frac{T}{\E_p^2} + \rmO\Bigl(\frac{1}{T} \Bigr)
 \;. 
\ea
On the second row, we performed 
an expansion in powers of $\E^{ }_p/T$, which is valid 
for small energies at high temperatures.

For fermions, there is no Matsubara zero mode on the left-hand side
of \eq\nr{Tint_fe}, 
so that the largest terms have at most (i.e.\ for $\E^{ }_p \ll \pi T$) 
the magnitude
\be
 \left. \Bigl\{ T \sum_{ \{ \omega^{ }_n \} } f \Bigr\}
 \right|^{ }_{\omega^{ }_n = \pm \pi T} 
 \sim \frac{T}{(\pi T)^2} \sim \frac{1}{\pi^2 T}
 \;. \la{no_ir_prob}
\ee
Similarly, in terms of the right-hand side of \eq\nr{Tint_fe}, 
we can estimate
\ba
 \{ \ldots \} & \sim & 
 \fr12 \frac{-2\pi i}{2\pi} \frac{2}{-2 i \E^{ }_p}
 \Bigl[ 1 - 2 \nF{} (\E^{ }_p) \Bigr]
 = \frac{1}{\E^{ }_p}
 \biggl( \fr12 - \frac{1}{e^{\E^{ }_p/T} + 1} \biggr)
 \nn & \approx &
 \frac{1}{\E^{ }_p} 
 \biggl( \fr12 - \frac{1}{2 + {\E^{ }_p/T}} + \ldots \biggr)
 = \rmO\Bigl(\frac{1}{T} \Bigr)
 \;. 
\ea

Given the estimates above, let us construct a dimensionless 
expansion parameter associated with the loop expansion. Apart from 
an additional propagator, each loop order also brings in an additional
vertex or vertices; we denote the corresponding coupling by $g^2$, 
as would be the case in gauge theory. Moreover, the Matsubara summation
involves a factor $T$, so we can assume that the expansion parameter
contains the combination $g^2 T$. We now have to use the other scales
in the problem to transform this into a dimensionless number. For the
Matsubara zero modes, \eq\nr{ir_prob} tells us that we are allowed to 
use inverse powers of $\E^{ }_p$ or, after integration over the spatial 
momenta, inverse powers of $m$. Therefore, we can assume that for
large temperatures, $\pi T \gg m$, 
the largest possible expansion parameter is 
\be
 \epsilon^{ }_\bo \sim \frac{g^2 T}{\pi m}
 \;. \la{eps_bo}
\ee
For fermions, in contrast, \eq\nr{no_ir_prob} suggests that inverse
powers of $\E^{ }_p$ or, after integration over spatial momenta, $m$, 
{\em cannot appear} in the denominator, even if $m \ll \pi T$; 
we are thus led to the estimate
\be
 \epsilon^{ }_\fe \sim \frac{g^2 T}{\pi^2 T} \sim \frac{g^2}{\pi^2}
 \;. \la{eps_fe}
\ee
In these estimates most numerical factors have been omitted 
for simplicity.

Assuming that we work in the weak-coupling limit, $g^2 \ll \pi^2$, 
we can thus conclude the following: 
\bi
\item
{\em Fermions} appear to be purely {\em perturbative} in  
computations concerning ``static'' observables, 
with the corresponding weak-coupling expansion 
proceeding in powers of $g^2/\pi^2$. 

\item
Bosonic {\em Matsubara zero modes} appear to suffer from 
bad convergence in the limit $m\to 0$.

\index{Resummation}
\index{Linde problem}

\item
The {\em resummations} that we saw around \eq\nr{LE_resum} 
for scalar field theory 
and in \se\ref{se:mmeff_gluon} for QCD produce an effective thermal
mass, $m_\rmi{eff}^2 \sim g^2 T^2$. Thus, we may expect the expansion
parameter in \eq\nr{eps_bo} to become $\sim g^2 T/(\pi g T)  = g/\pi$. 
In other words, 
a small expansion parameter exists in principle
if $g \ll \pi$, but the structure of the 
weak-coupling series is peculiar, with odd powers of $g$ appearing. 

\item
As we found in \eq\nr{A0_effprop}, colour-magnetic fields do not develop
a thermal mass squared 
at $\rmO(g^2T^2)$. This might still happen at higher orders, 
so we can state that $m^{ }_\rmi{eff} \lsim g^2 T/\pi$ for these modes. 
Thereby the expansion 
parameter in \eq\nr{eps_bo} reads $\epsilon^{ }_\bo\gsim g^2 T / g^2 T = 1$. 
In other words, {\em colour-magnetic fields cannot 
be treated perturbatively}; 
this is known as the {\em infrared problem} (or ``Linde problem'') 
of thermal gauge theory~\cite{linde}.

\ei

The situation that we have encountered, namely
that infrared problems exist but that they
are related to particular degrees of freedom, 
is common in (quantum) field theory. Correspondingly, 
there is also a generic tool, called 
the {\em effective field theory} approach, 
which allows us to isolate the infrared problems 
into a simple Lagrangian, and treat them in this setting.  
The concept of effective field theories is not restricted
to finite-temperature physics, but applies also at zero temperature, 
if the system possesses a {\em scale hierarchy}. In fact,  
the high-temperature case can be considered a special case of this, 
with the corresponding hierarchy often 
expressed as $g^2 T/\pi \ll gT \ll \pi T$, where the first scale
refers to the non-perturbative one associated with 
colour-magnetic fields. 
Given the generic nature of effective field theories, we first discuss 
the basic idea in a zero-temperature setting, before 
moving on to finite-temperature physics. 

\subsection*{A simple example of an effective field theory}
\la{se:eff_ex}

Let us consider a Lagrangian containing
two different scalar fields, $\phi$ and $H$, 
with masses $m$ and $M$, respectively:\footnote{
 The discussion follows closely that in ref.~\cite{jc}.
 }
\be
 {L}^{ }_\rmi{full} \; \equiv \;
 \fr12 \partial^{ }_\mu\phi\, \partial^{ }_\mu\phi +
 \fr12 m^2 \phi^2 + 
 \fr12 \partial^{ }_\mu H \partial^{ }_\mu H + 
 \fr12 M^2 H^2 + g^2 \phi^2 H^2 + 
 \fr14 \lambda \phi^4 + \fr14 \kappa H^4
 \;. \la{L_full}  
\ee
We assume that there exists a hierarchy $m \ll M$ or, 
to be more precise, $m^{ }_\rmii{R} \ll M^{ }_\rmii{R}$, though 
we leave out the subscripts
in the following. 
Our goal is to study to what extent the physics described
by this theory can be captured by a simpler effective theory of the form
\be
 {L}^{ }_\rmi{eff} = 
 \fr12 \partial^{ }_\mu\bar\phi\, \partial^{ }_\mu\bar\phi +
 \fr12 \bar m^2 \bar\phi^2 + 
 \fr14 \bar\lambda \bar\phi^4 + \ldots
 \;, \la{L_eff} 
\ee
where infinitely many higher-dimensional operators have been 
dropped.\footnote{If we also wanted to describe
gravity with these theories, we could add a ``fundamental'' cosmological
constant $\Lambda$ in ${L}^{ }_\rmii{full}$, and an ``effective'' 
cosmological constant $\bar \Lambda$ in ${L}^{ }_\rmii{eff}$.}

The main statement concerning the effective description 
goes as follows.
Let us assume that $m\lsim g M$ and 
that all couplings are parametrically of 
similar magnitude, $\lambda\sim\kappa\sim g^2$, 
and proceed to  consider external momenta $P \lsim gM$.
Then the one-particle-irreducible Green's functions $\bar \Gamma^{ }_n$, 
computed within the effective theory, reproduce those of the full theory, 
$\Gamma^{ }_n$, with a relative error
\be
 \frac{\delta \bar \Gamma^{ }_n}{\bar \Gamma^{ }_n} 
 \;\equiv\; 
  \frac{|\bar \Gamma^{ }_n - \Gamma^{ }_n|}{\bar \Gamma^{ }_n}
 \; \lsim \; \rmO(g^k)
 \;, \quad k > 0 
 \;, \la{eff_acc}
\ee
if the parameters $\bar m^2$ and $\bar \lambda$ 
of \eq\nr{L_eff} are tuned suitably. 
The number $k$ may depend on the dimensionality of spacetime 
as well as on $n$, although a universal lower bound
should exist. This lower bound can furthermore 
be increased by adding 
suitable higher-dimensional operators to ${L}^{ }_\rmi{eff}$; 
in the limit of infinitely many such operators the effective description
should become exact.

A weaker form of the effective theory statement, although already 
sufficiently strong for practical purposes, 
is that Green's functions are matched only ``on-shell'', 
rather than for arbitrary external momenta. 
This form of the statement is implemented, for instance, in the 
so-called non-perturbative Symanzik 
improvement program of lattice QCD~\cite{kj}
(for a nice review, see ref.~\cite{pw}).

It has been fittingly said that 
the effective theory assertion is 
almost trivial yet very difficult to prove. 
We will not attempt a formal proof here, 
but rather try to get an impression on how it arises, 
by inspecting with some care the 2-point Green's function of the 
light field $\phi$. In the full theory, at 1-loop level, 
the inverse of this (``amputated'') quantity reads
\ba
 G^{-1} & = & \TopoStree(\Ldsc) 
 \quad + \quad \TopoST(\Ldsc,\Adsc)
 \quad + \quad \TopoST(\Ldsc,\Asc)
 \nn[2mm] & = &
 P^2 + m^2 +\; \Pi_l^{(1)}(0;m^2)\; + \; \Pi_h^{(1)}(0;M^2)
 \;,  \la{G2_full}
\ea 
where the dashed line represents the light field and
the solid one the heavy field, while the subscripts
$l,h$ stand for light and heavy, respectively. 
The first argument of the functions $\Pi_l^{(1)}$, 
$\Pi_h^{(1)}$ is the external momentum; 
as the notation indicates, closed bubbles 
contain no dependence on it.  

Within the effective theory, the same computation yields
\ba
 \bar G^{-1} & = & \TopoStree(\Ldsc) 
 \quad + \quad \TopoST(\Ldsc,\Adsc)
 \nn[2mm] & = &
 P^2 + \bar m^2 +\; \bar \Pi_l^{(1)}(0;\bar m^2)
 \;.  \la{G2_eff}
\ea
The equivalence of all Green's functions at the on-shell point 
should imply the equivalence of pole masses, i.e.\ the locations
of the on-shell points. By {\em matching} \eqs\nr{G2_full} and 
\nr{G2_eff}, we see that this can indeed be achieved provided that 
\be
 \bar m^2 = m^2 + \Pi_h^{(1)}(0;M^2) + \rmO(g^4)
 \;. \la{1loop_match}
\ee
Note that within perturbation theory the matching is 
carried out ``order-by-order'': $\bar \Pi_l^{(1)}(0;\bar m^2)$ is 
already of 1-loop order, so inside it $\bar\lambda$ and $\bar m^2$ can 
be replaced by $\lambda$ and $m^2$, respectively, given that the difference
between $\bar \lambda$ and $\lambda$ as well as 
$\bar m^2$ and $m^2$ is itself of 1-loop order. 

The situation becomes considerably more complicated once we go to 
the 2-loop level. To this end, 
let us analyze various types of graphs that exist
in the full theory, and try to understand how they could be matched
onto the simpler contributions within the effective theory. 

First of all, there are graphs involving only light fields, 
\be
  \ToptSS(\Ldsc,\Adsc,\Adsc,\Ldsc)
  \qquad 
  \ToptSDB(\Ldsc,\Ldsc,\Adsc,\Adsc,\Adsc) 
  \qquad
  \;. \la{2loop_ll}
\ee
These can directly be matched with the corresponding graphs 
within the effective theory; as above, the fact that
different parameters appear in the propagators (and vertices) is 
a higher-order effect. 

Second, there are graphs which account for the 
``insignificant higher-order effects'' that we omitted in the 1-loop 
matching, but that would play a role once we go to the 2-loop level: 
\ba
 \ToptSDB(\Ldsc,\Ldsc,\Adsc,\Adsc,\Asc) & \Leftrightarrow &
 (\bar m^2 - m^2)\, \frac{\partial \Pi_l^{(1)}(0;m^2)}{\partial m^2}
 \;, \la{2loop_lh}
 \\[2mm] 
 \ToptSDB(\Ldsc,\Ldsc,\Asc,\Asc,\Adsc) & \Leftrightarrow &
 (\bar \lambda - \lambda)\, \frac{\partial \Pi_l^{(1)}(0;m^2)}{\partial\lambda}
 \;. \la{2loop_hl} 
\ea
As indicated here, these two combine to reproduce (a part of) 
the 1-loop effective theory expression
$\bar \Pi_l^{(1)}(0;\bar m^2)$ with 2-loop full theory accuracy.  

Third, there are graphs only involving heavy fields in the loops: 
\be
  \ToptSDB(\Ldsc,\Ldsc,\Asc,\Asc,\Asc)
 \;. \la{2loop_hh}
\ee
Obviously we can account
for their effects by a 2-loop correction to $\bar m^2$.

Finally, there remain the most complicated graphs: 
structures involving both heavy and light fields, in a way that the momenta 
flowing through the two sets of lines do {\em not} get factorized: 
\be
  \ToptSS(\Ldsc,\Asc,\Asc,\Ldsc) \quad = \quad
  \ToptSSp(\Ldsc,\Asc,\Asc,\Adsc) 
  \;. \la{2loop_mixed} 
\ee
Naively, the representation on the right-hand side might suggest
that this graph is simply part of the correction 
$
  (\bar \lambda - \lambda) 
  {\partial \Pi_l^{(1)}(0;m^2)} / {\partial\lambda}
$,
just like the graph in \eq\nr{2loop_hl}. This, however, is {\em not} 
the case, because the substructure appearing, 
\be
 \TopoFff(\Ldsc,\Ldsc,\Ldsc,\Ldsc,\Asc,\Asc)
 \;, \la{4pt}
\ee
is momentum-dependent, unlike the effective vertex $\bar \lambda$.

\index{Soft and hard modes: vacuum example}
\index{Hard and soft modes: vacuum example}

Nevertheless, it should be possible to split \eq\nr{2loop_mixed} 
into two parts, pictorially represented by
\ba
  \ToptSSp(\Ldsc,\Asc,\Asc,\Adsc) & = & 
  \TopoSTTxt(\Ldsc,\Adsc,\widehat\Pi)
  \qquad + \qquad 
  \TopoSTxt(\Ldsc,\widecheck\Pi) 
   \la{mixed_splitup} \\[7mm] 
  \Leftrightarrow \quad
  \Pi^{(2)}_\rmi{\it mixed}(P^2;m^2,M^2)
  & = & 
  \widehat \Pi^{(2)}_\rmi{\it mixed}(P^2;m^2,M^2)
 + 
  \widecheck \Pi^{(2)}_\rmi{\it mixed}(P^2;m^2,M^2)
 \;.
\ea
The first part $\widehat \Pi^{(2)}$ is, {\em by definition}, 
characterized by the fact that 
it depends {\em non-analytically} on the mass parameter $m^2$ of 
the light field; therefore the internal $\phi$ field is {\em soft} in 
this part, i.e.\ gets a contribution from momenta $Q\sim m$.
In this situation, the momentum dependence  of \eq\nr{4pt}
is of subleading importance.
In other words, this part of the graph {\em does} 
contribute simply to  
$
  (\bar \lambda - \lambda) 
  {\partial \Pi_l^{(1)}(0;m^2)} / {\partial\lambda}
$, as we naively expected. 

The second part $\widecheck \Pi^{(2)}$ is, by definition,
analytic in the mass parameter $m^2$. We associate this 
with a situation where the internal $\phi$ is {\em hard}:
even though its mass is small, it can have a large internal momentum
$Q\sim M$, transmitted to it through interactions with the 
heavy modes. In this situation, the momentum dependence of 
\eq\nr{4pt} plays an essential role. At the same time, 
the fact that all internal momenta are hard, permits for 
a Taylor expansion in the small external momentum: 
\ba
  \widecheck \Pi^{(2)}_\rmi{\it mixed}(P^2;m^2,M^2)
 & = &  
  \widecheck \Pi^{(2)}_\rmi{\it mixed}(0;m^2,M^2)
  + 
  P^2 \frac{\partial}{\partial P^2} 
 \widecheck \Pi^{(2)}_\rmi{\it mixed}(0;m^2,M^2)
 \nn & & 
  + \fr12
  (P^2)^2 \frac{\partial^2}{\partial (P^2)^2} 
 \widecheck \Pi^{(2)}_\rmi{\it mixed}(0;m^2,M^2)
 + \ldots
 \;.  \la{mixed_expansion}
\ea
The first term here represents a 2-loop correction 
to $\bar m^2$, just like the graph in \eq\nr{2loop_hh}, whereas 
the second term can be compensated for by a change of the
normalization of the field $\bar \phi$. Finally, the 
further terms have the appearance of higher-order (derivative)
operators, truncated from the structure shown explicitly 
in \eq\nr{L_eff}. Comparing with the leading kinetic term, the magnitude
of the third term is very small,
\be
 \frac{\textstyle g^4 \frac{(P^2)^2}{M^2}}{P^2} \; \lsim \; g^6
 \;, 
\ee
for $P\lsim g M$, justifying the truncation 
of the effective action up to a certain relative accuracy. The structures
in \eq\nr{mixed_expansion} are collectively denoted by the 2-point
``blob'' in \eq\nr{mixed_splitup}.

To summarize, we see that the explicit construction of 
an effective field theory becomes subtle at higher loop orders. 
Another illuminating example of the difficulties met with ``mixed graphs''
is given around \eq\nr{L_subtle} below. 
Nevertheless, we may formulate the following practical recipe 
for the effective field theory description of a Euclidean theory 
with a scale hierarchy: 

\bi

\item[(1)]
Identify the ``light'' or ``soft'' degrees of freedom, 
i.e.\ the ones that are {\em IR-sensitive}.

\index{Symmetries: general effective theory}
\index{Matching: general}

\item[(2)]
Write down the most general Lagrangian for them, respecting
all the {\em symmetries} of the system, and including local
operators of arbitrary order. 

\item[(3)]
The parameters of this Lagrangian can be determined by {\em matching}: \\[3mm]
$\bullet$ Compute the same observable in the full and effective
theories, applying the same UV-regularization and IR-cutoff. \\[3mm]
$\bullet$ Subtract the results. \\[3mm]
$\bullet$ The IR-cutoff should now disappear, and the result of 
the subtraction be analytic in $P^2$. This allows for a matching 
of the parameters and field normalizations of the effective theory. \\[3mm]
$\bullet$ If the IR-cutoff does not disappear, 
the degrees of freedom, or the form of the effective
theory, have not been correctly identified.

\item[(4)]
{\em Truncate} the effective theory by dropping higher-dimensional
operators suppressed by $1/M^k$, which can only give a relative 
contribution of order
\be
 \sim \Bigl( \frac{m}{M} \Bigr)^k \sim g^k
 \;, \la{param_error}
\ee
where the dimensionless coefficient 
$g$ parametrizes the scale hierarchy.

\ei

\newpage 

\subsection{Dimensionally reduced effective field theory for hot QCD}
\la{se:DR_QCD}

\index{Dimensional reduction}
\index{Soft and hard modes: thermal QCD}
\index{Hard and soft modes: thermal QCD}
\index{Symmetries: thermal QCD}

We now apply the effective theory recipe 
to the problem outlined at the beginning of \se\ref{se:Linde}, 
i.e.~accounting for the soft contributions to the free energy 
density of thermal QCD. In this process, we follow the numbering 
introduced at the end of \se\ref{se:Linde}.

\paragraph{(1) Identification of the soft degrees of freedom.}

As discussed earlier, the soft degrees of freedom in perturbative Euclidean 
thermal field 
theory are the bosonic Matsubara zero modes. Since they do not depend
on the coordinate $\tau$, they live in $d=3-2\epsilon$ spatial 
dimensions; for this reason, the construction of the effective theory
is in this context called 
{\em high-temperature dimensional reduction}~\cite{dr1,dr2}.
For simplicity, we concentrate on the dimensional reduction of QCD 
in the present section, but within perturbation theory the same 
procedure can also be (and indeed has been) 
applied to the full Standard Model~\cite{generic}, as well as
many extensions thereof.

\paragraph{(2) Symmetries.}
Since the heat bath breaks Lorentz invariance, 
the time direction and the space directions are not interchangeable. 
Therefore, the spacetime symmetries of the effective theory 
are merely invariances in 
{\em spatial rotations and translations}.

In addition, the full theory possesses a number of {\em discrete symmetries}: 
QCD is invariant in C, P and T separately. The effective theory
inherits these symmetries, and it turns out 
that ${L}^{ }_\rmi{eff}$ is symmetric 
in $\bar A^{ }_0 \to -\bar A^{ }_0$, where the
low-energy fields are denoted by $\bar A^{ }_\mu$ 
(the symmetry $\bar A^{ }_0 \to -\bar A^{ }_0$ 
is absent if the C symmetry of QCD 
is broken by coupling the quarks to a chemical potential).

Finally, consider the {\em gauge symmetry} from \eq\nr{Atrans0}: 
\ba
 A_\mu'  & = & 
 U  A^{ }_\mu U^{-1} + \frac{i}{g} U \partial^{ }_\mu U^{-1}
 \;.
\ea 
Since we now restrict to {\em static} 
(i.e.\ $\tau$-independent) fields, $U$ 
should not depend on $\tau$, either, and the effective
theory should be invariant under
\ba
 \bar A_i'  & = & 
 U \bar A^{ }_i U^{-1} + \frac{i}{g} U \partial^{ }_i U^{-1}
 \;,
 \\
 \bar A_0'  & = & 
 U \bar A_0^{ } U^{-1}
 \;.
\ea
In other words, the spatial components $\bar A^{ }_i$ remain gauge fields, 
whereas the temporal component $\bar A^{ }_0$ has 
turned into a {\em scalar field
in the adjoint representation} (cf.\ \eq\nr{Phitrans0}).

With these ingredients, we can postulate the general form of 
the effective Lagrangian. It is illuminating to start by simply writing down 
the contribution of the soft degrees of freedom to the 
full Yang-Mills 
Lagrangian, \eq\nr{LE_gauge}. Noting from \eq\nr{F0i}, {\em viz.}  
\be
 F^a_{0i} \equiv \partial^{ }_\tau A^a_i - \mathcal{D}^{ab}_i A^b_0
 \;,  
\ee 
that in the static case
$
 F^a_{i0} = \mathcal{D}^{ab}_i  A^b_0
$, 
we end up with
\be
 {L}^{ }_\iE = \fr14 F^a_{ij}F^a_{ij} 
 + \fr12 
 (\mathcal{D}^{ab}_i  A_0^b)
 (\mathcal{D}^{ac}_i  A^c_0)
 \;. 
\ee
At this point, it is convenient to note that 
\be
 T^a \mathcal{D}^{ab}_i  A_0^b
 = \partial^{ }_i A_0^{ } + g f^{acb} T^a A^c_i  A_0^b
 = \partial^{ }_i A_0^{ } - i g [A^{ }_i, A_0^{ }] 
 = [D^{ }_i,  A_0^{ }]
 \;, 
\ee
where $D^{ }_i = \partial^{ }_i - i g A^{ }_i$ is the covariant
derivative in the fundamental representation. Thereby we obtain 
as the ``tree-level'' terms of our effective theory 
\be
 {L}_\rmi{eff}^{(0)} = \fr14 \bar F^a_{ij} \bar F^a_{ij} 
 + \tr \{ [\bar D^{ }_i, \bar A_0^{ }] [\bar D^{ }_i, \bar A_0^{ }] \}
 \;, \la{Leff_0}
\ee
where we have now replaced $A^{ }_\mu \to \bar A^{ }_\mu$.

Next, we complete the tree-level structure by adding
all mass and interaction terms allowed by symmetries. 
In this process, it is useful to proceed in order of increasing
dimensionality, whereby we obtain at the three lowest orders: 
\ba
 \mbox{dim = 2}: & & 
 \tr[\bar A_0^2]
 \;; \la{dim2} \\  
 \mbox{dim = 4}: & & 
 \tr[\bar A_0^4]
 \;, \quad
 (\tr[\bar A_0^2])^2
 \;; \la{dim4} \\  
 \mbox{dim = 6}: & & 
 \tr\{[\bar D^{ }_i,\bar F_{ij}] [\bar D^{ }_k,\bar F_{kj}]\}
 \;, \ldots
 \;. 
 \la{dim6}
\ea
In the last case, we have only shown one example operator, while many others
are listed~in ref.~\cite{sc}. Note also that for $\Nc=2$ and 3, 
there exists a linear relation between the two operators of 
dimensionality 4, 
but from $\Nc=4$ onwards they are fully independent.

Combining \eqs\nr{Leff_0}--\nr{dim4}, 
we can write the effective action in the form
\be \index{Classical limit}
 S^{ }_\rmi{eff} = \frac{1}{T} \int_\vec{x} \, 
 \biggl\{ 
  \fr14 \bar F^a_{ij} \bar F^a_{ij}
 +\tr ( [\bar D^{ }_i, \bar A_0^{ }] [\bar D^{ }_i, \bar A_0^{ }] )
 + \bar m^2 \tr[\bar A_0^2]
 + \bar \lambda^{(1)}  (\tr[\bar A_0^2])^2
 + \bar \lambda^{(2)} \tr[\bar A_0^4]
 + \ldots
 \biggr\}
 \;. \la{Leff_EQCD}
\ee
The prefactor $1/T$, appearing like in {\em classical} 
statistical physics, comes from the integration 
$\int_0^\beta \! {\rm d}\tau$, since none of the soft fields
depend on $\tau$.
This theory is referred to as EQCD,
for ``Electrostatic QCD''.
Note that in the presence of a finite chemical potential, 
cf.\ \se\ref{se:density}, charge conjugation symmetry is broken 
and the additional 
operator $i \bar{\gamma}\, \tr[\bar A_0^3]$ appears in 
the effective action~\cite{mu}.

\index{Matching: thermal QCD}

\paragraph{(3) Matching.}

If we restrict to 1-loop order, then the matching of the parameters
in \eq\nr{Leff_EQCD} is rather simple, as explained around \eq\nr{1loop_match}:
we just need to compute Green's functions for the soft fields with 
vanishing external momenta, with the heavy modes appearing 
in the internal propagators.
For the parameter $\bar m^2$, this is furthermore precisely the 
computation that we carried out in \se\ref{se:mmeff_gluon}, so 
the result can be directly read off from \eq\nr{mmE}:
\be
  \bar m^2 = g^2 T^2 \biggl( \frac{\Nc}{3}  + \frac{\Nf}{6} \biggr)
  + \rmO(g^4 T^2)
 \;. \la{barm2}
\ee

The parameters $\bar \lambda^{(1)}$, $\bar \lambda^{(2)}$ can, 
in turn, be obtained by considering 4-point functions with soft
modes of $A_0^{ }$ on the external legs, and non-zero Matsubara modes
in the loop:
\be
  \TopoFff(\Ldsc,\Ldsc,\Ldsc,\Ldsc,\Agl,\Agl) +
  \TopoFft(\Ldsc,\Ldsc,\Ldsc,\Ldsc,\Agl,\Agl) +
  \TopoFtt(\Ldsc,\Ldsc,\Ldsc,\Ldsc,\Agl,\Agl) +
  \TopoFtt(\Ldsc,\Ldsc,\Ldsc,\Ldsc,\Agh,\Agh) +
  \TopoFtt(\Ldsc,\Ldsc,\Ldsc,\Ldsc,\Aqu,\Aqu) 
 \;.  
\ee 
These graphs are clearly of $\rmO(g^4)$, 
and the actual values of 
the two parameters read~\cite{nadkarni,npl}
\be
 \bar \lambda^{(1)} = \frac{g^4}{4\pi^2} + \rmO(g^6)
 \;, \quad
 \bar \lambda^{(2)} = \frac{g^4}{12\pi^2}(\Nc - \Nf) + \rmO(g^6)
 \;.  
\ee
The gauge coupling $\bar g$ appearing in $\bar D^{ }_i$ and 
$\bar F^a_{ij}$ is of the form 
$
 \bar g^2 = g^2 + \rmO(g^4)
$
and needs to be matched as well~\cite{hl,gE2}.
If there are non-zero chemical potentials $\mu^{ }_i$ in the problem, 
the same is true for 
$\bar\gamma = \sum_{i=1}^{\Nf} \mu^{ }_i\, g^3 /(3\pi^2)
 + \rmO(g^5)$~\cite{mu}.

\index{Truncation of effective theory}

\paragraph{(4) Truncation of higher-dimensional operators.}

The most non-trivial part of any 
effective theory construction is the quantitative analysis 
of the error made, when operators beyond a given dimensionality 
are dropped. In other words, the challenge is to determine
the constant $k$ in \eq\nr{eff_acc}. We illustrate this
by considering the error made when dropping the operator
in \eq\nr{dim6}. 

First of all, we need to know the parametric magnitude of
the coefficient with which the neglected operator 
would enter ${L}^{ }_\rmi{eff}$, 
if it were kept. The operator of \eq\nr{dim6} 
could be generated through the momentum
dependence of graphs like 
\be
 \TopoFffTxt(\Ldsc,\Ldsc,\Ldsc,\Ldsc,\Agl,\Agl,n\neq 0)
 \quad \sim \quad
 \frac{g^2}{T^2} (\partial^{ }_i \bar F^a_{ij})^2
 \;, \la{ho_full}
\ee
where the dashed lines now stand for the spatial components of the 
gauge field, $\bar A^{ }_i$.
If we drop this term, the corresponding Green's function 
will not be computed correctly; however, it still has {\em some} 
value, namely that which would be obtained within the effective 
theory via the graph
\be
 \TopoFffTxt(\Ldsc,\Ldsc,\Ldsc,\Ldsc,\Agl,\Agl,\bar A_0^{ })
 \quad \sim \quad
 {g^2} (\partial^{ }_i \bar{F}^a_{ij})^2 T \int_\vec{p} 
 \frac{1}{({p}^2 + \bar m^2)^3}
 \; \sim \; \frac{g^2 T}{\bar m^3} (\partial^{ }_i \bar{F}^a_{ij})^2
 \;. \la{ho_eff}
\ee
Here, we have noted that to account 
for the momentum dependence of the graph, represented by 
the derivative $\partial^{ }_i$ in front of $\bar F^a_{ij}$, one needs
to Taylor-expand the integral to the first non-trivial order in 
external momentum, 
explaining why the propagator is raised to power three in \eq\nr{ho_eff}. 
An explicit computation further shows  that the coefficient in 
\eq\nr{ho_eff} comes with a negative sign, but this has
no significance for our general discussion.

Next, we note that
the value of the Green's function within the (truncated) effective theory,
\eq\nr{ho_eff}, is in fact {\em larger} than 
what the contribution of the omitted
operator would have been, cf.~\eq\nr{ho_full}.
Therefore, the error made through the omission of \eq\nr{ho_full}
is {\em small}: 
\be
 \frac{\delta \bar \Gamma}{\bar \Gamma} \sim \frac{g^2}{T^2} 
 \frac{\bar m^3}{g^2 T} \sim 
 \Bigl( \frac{\bar m}{T} \Bigr)^3 \sim g^3
 \;. \la{error}
\ee
In other words, for the Green's function considered and the  
dimensionally reduced effective theory of 
hot QCD truncated beyond dimension 4, 
we can expect the relative accuracy 
exponent of \eq\nr{eff_acc} to take the value $k=3$ \cite{parity}.

\vspace*{5mm}

Having now completed the construction of the effective 
theory of \eq\nr{Leff_EQCD}, we can 
take a further step: the field $\bar A_0^{ }$ is massive, and can thus
be integrated out, should we wish to study distance scales 
longer than $1/\bar{m}$. 
Thereby we arrive at an even simpler effective theory,  
\be
 S_\rmi{eff}' = \frac{1}{T} \int_\vec{x} \, 
 \biggl\{ 
  \fr14\, \bar{\!\bar{F}}^a_{ij}\, \bar{\!\bar F}^a_{ij}
 + \ldots
 \biggr\}
 \;, \la{Leff_MQCD}
\ee
referred to as MQCD,
for ``Magnetostatic QCD''. It is important to realize that 
this theory, i.e.~three-dimensional Yang-Mills theory
(up to higher-order operators such as the one in \eq\nr{dim6}), 
only has one parameter, the gauge coupling.  
Furthermore, if the fields $\,\,\bar{\!\!\bar A}^a_i$ are rescaled by 
an appropriate power of $T^{1/2}$, 
$\,\,\bar{\!\!\bar A}^a_i \to 
 \,\,\bar{\!\!\bar A}^a_i T^{1/2}$, 
then the coefficient $1/T$ in 
\eq\nr{Leff_MQCD} disappears. The coupling constant squared
that appears afterwards is $\,\bar{\!\bar g}^2 T$, and this
is the only scale in the system. Therefore all dimensionful
quantities (correlation lengths, string tension, free energy density, ...)
must be proportional to an appropriate power of $\,\bar{\!\bar g}^2 T$,
with a non-perturbative coefficient. This is the essence
of the non-perturbative physics pointed out by Linde~\cite{linde}.\footnote{%
 In contrast, topological configurations such as instantons,  
 which play an important role for certain 
 non-perturbative phenomena in vacuum, 
 only play a minor
 role at finite temperatures~\cite{gpy,dn}, save for special observables 
 where the anomalous U$^{ }_\rmii{A}$(1)
 breaking dominates the signal (cf.\ ref.~\cite{ua1} and references therein).  
 The reason is that the Euclidean
 topological susceptibility (measuring topological ``activity'') 
 vanishes to all orders in perturbation theory, 
 and is numerically small. 
 }

The implication of the above setup for the
properties of the weak-coupling expansion
is the following. Consider a generic observable $\mathcal{O}$, with an 
expectation value of the form
\be 
 \langle \mathcal{O} \rangle \sim g^m T^n [1 + \alpha\, g^r + ... ] \;.
\ee
There are now four distinct possibilities: 
\bi

\item[(i)] 
$r$ is even, and $\alpha$ is determined 
by the heavy scale $\sim \pi T$ 
and is purely perturbative. This is the case for instance 
for the leading correction to the free energy density
$f(T)$, cf.\ \eq\nr{fT_QCD}.

\item[(ii)] $r$ is odd, and $\alpha$ 
is determined by the intermediate scale $\sim gT$, 
being still purely perturbative. 
This is the case for the next-to-leading order corrections to 
many real-time quantities in thermal QCD, 
for instance to the heavy quark diffusion coefficient~\cite{sch}.

\item[(iii)] $m+r$ is even, and $\alpha$ is non-perturbatively 
determined by the soft scale $\sim g^2 T/\pi$. 
This is the case e.g.\ for the next-to-leading order correction
to the physical Debye screening length~\cite{rebhan,ay} and 
for one of the subleading corrections to $f(T)$
in a non-Abelian plasma~\cite{linde,nspt}. 

\item[(iv)] $r > k$, and $\alpha$ can only be determined correctly 
by adding higher-dimensional operators to the effective theory. 

\ei

A few final remarks are in order:
\bi 
\item
We have seen that the omission of higher-order 
operators in the construction of an effective theory usually 
leads to a small error, since the same Green's function
is produced with a larger coefficient within it. 
It could happen, however, that there is some approximate symmetry
in the full theory, which {\em becomes exact} within the effective
theory, if we truncate its derivation to a given order. 
For instance, many Grand Unified
Theories violate baryon minus lepton number ($B-L$), whereas
in the classic Standard Model this is an exact symmetry, to be broken 
only by some higher-dimensional operator~\cite{sw,wz}.
Therefore, if such a Grand Unified Theory represented a true description
of Nature and we considered $B-L$ violation within the classic Standard Model, 
we would make an {\em infinitely large} relative error. 

\item
There are several reasons why effective theories 
constitute a useful framework. 
First of all, 
they allow us to justify and extend resummations such as those discussed
in \se\ref{se:IR} systematically to higher orders in the 
weak-coupling expansion. As mentioned
below \eqs\nr{fT_SFT} and \nr{fT_QCD}, this has led to the determination
of many subsequent terms in the weak-coupling series. 
Second, effective theories
permit for a simple non-perturbative study of the infrared sector
affected by the Linde problem; examples are provided by 
refs.~\cite{ay_2,mu,nspt,srate}, 
and further ones will be encountered below. 

\index{Background field gauge}

\item
When proceeding to higher orders in the matching computations, 
they are often most conveniently formulated in the so-called {\em background
field gauge}~\cite{abbott}, rather than in the covariant gauge
of \eq\nr{Ga_cov}, cf.\ e.g.\ ref.~\cite{Ghi-Sch-2}.

\ei


\subsection*{Appendix A: Subtleties related to the low-energy expansion}

Let us consider the full theory
\be
 {L}^{ }_\rmi{full} \equiv
 \fr12 \partial^{ }_\mu\phi\, \partial^{ }_\mu\phi +
 \fr12 m^2 \phi^2 + 
 \fr12 \partial^{ }_\mu H \partial^{ }_\mu H + 
 \fr12 M^2 H^2 + \fr16 \gamma H \phi^3 
 \;.  \la{L_subtle}
\ee
For simplicity (more precisely, in order to avoid
ultraviolet divergences), we assume that the dimensionality of spacetime 
is 3, i.e. $d=2-2\epsilon$ in our standard notation, and moreover we 
work at zero temperature, like in \se\ref{se:eff_ex}. We then take 
the following steps: 

\bi

\item[(i)]
Integrating out $H$ in order to construct an effective theory, 
we compute the graph 
\be
 \TopoSix(\Ldsc,\Ldsc,\Ldsc,\Ldsc,\Ldsc,\Ldsc,\Lsc) \la{graph_9a}
 \;. 
\ee
After Taylor-expanding the result in external momenta,   
we write down all the corresponding operators. 

\item[(ii)]
We focus on the 4-point function of the $\bar \phi$ field at vanishing
external momenta, and determine the contributions of the operators 
computed in step (i) to this Green's function.

\item[(iii)]
Finally we consider directly the full theory graph
\be
  \TopoFff(\Ldsc,\Ldsc,\Ldsc,\Ldsc,\Adsc,\Asc) \la{graph_9c}
\ee
at vanishing external momenta. Comparing
with the Taylor-expanded result obtained from step (ii), 
we demonstrate how a ``careless'' Taylor expansion can lead to wrong results.

\ei

The construction of the effective theory 
proceeds essentially 
as in \eq\nr{compact_rule}, except that only the $H$-field is now
integrated out. We get from here
\ba
 S^{ }_\rmi{eff} & \approx &
 \Bigl\langle -\fr12 S_\iI^2 \Bigr\rangle^{ }_\rmi{$H$,c}
 \nn & = & 
 -\frac{\gamma_{ }^2}{72}
 \int_{X,Y} \phi^3(X)\phi^3(Y) \langle H(X) H(Y) \rangle_0^{ } 
 \nn & = &
 -\frac{\gamma_{ }^2}{72}
 \int_{X,Y} \phi^3(X)\phi^3(Y) 
 \int_P \frac{e^{i P\cdot (X-Y)}}{P^2 + M^2}
 \nn & = &
 -\frac{\gamma_{ }^2}{72}
 \int_{X,Y} \phi^3(X)\phi^3(Y) \int_P e^{i P\cdot (X-Y)} 
 \biggl[\sum_{n=0}^\infty \frac{(-1)^n (P^2)^n}{(M^2)^{n+1}} \biggr]
 \nn & = &
 -\frac{\gamma_{ }^2}{72}
 \int_{X,Y} \phi^3(X)\phi^3(Y)  
 \biggl[\sum_{n=0}^\infty \frac{(\nabla_X^2)^n}{(M^2)^{n+1}} \biggr]
 \;\deltabar(X-Y)
 \nn & = &
 -\frac{\gamma_{ }^2}{72}
 \int_{X} 
  \sum_{n=0}^\infty
 \phi^3(X) \,  \frac{(\nabla_X^2)^n}{(M^2)^{n+1}} \,
 \phi^3(X)
 \;, \label{Seffexp}
\ea
where an expansion was carried out assuming $P^2 \ll M^2$, 
and partial integrations were performed at the last step.

Using \eq\nr{Seffexp}, we can extract the corresponding contribution 
to the 4-point function at vanishing momenta: 
\ba
 && \hspace*{-1cm}
 \Bigl\langle 
 \tilde \phi(0)
 \tilde \phi(0)
 \tilde \phi(0)
 \tilde \phi(0)
 e^{-S^{ }_\rmi{eff}}
 \Bigr\rangle 
 \nn & \Rightarrow &
 \frac{\gamma_{ }^2}{72}
 \Bigl\langle 
 \tilde \phi(0)
 \tilde \phi(0)
 \tilde \phi(0)
 \tilde \phi(0)
 \int_{P^{ }_1,...,P^{ }_6} \hspace*{-1cm}
 \; \deltabar(\Sigma^{ }_i P^{ }_i)
 \tilde \phi(P^{ }_1)
 \ldots
 \tilde \phi(P^{ }_6) \Bigr\rangle
 \sum_{n=0}^{\infty} \frac{[- (P^{ }_4+P^{ }_5+P^{ }_6)^2]^n}{(M^2)^{n+1}}
 \nn & = & 
 \frac{\gamma_{ }^2}{72}
 \times 6 \times (2\times 3 \times 2 + 3\times 4 \times 2)
 \int_{P^{ }_1,...,P^{ }_6}\hspace*{-1cm}
 \; \deltabar(\Sigma^{ }_i P^{ }_i)
 \sum_{n=0}^{\infty} \frac{[- (P^{ }_4+P^{ }_5+P^{ }_6)^2]^n}{(M^2)^{n+1}}
 \nn & & 
 \times 
 \langle \tilde \phi(0) \tilde \phi(P^{ }_1) \rangle^{ }_0\,
 \langle \tilde \phi(0) \tilde \phi(P^{ }_2) \rangle^{ }_0\,
 \langle \tilde \phi(0) \tilde \phi(P^{ }_5) \rangle^{ }_0\,
 \langle \tilde \phi(0) \tilde \phi(P^{ }_6) \rangle^{ }_0\,
 \langle \tilde \phi(P^{ }_3) \tilde \phi(P^{ }_4) \rangle^{ }_0
 \nn & = &
 3 \gamma_{ }^2 \frac{\;\deltabar(0)}{(\bar m^2)^4}
 \int_{P^{ }_3} \frac{1}{P_3^2 + \bar m^2} 
 \sum_{n=0}^{\infty} \frac{(- P_3^2)^n}{(M^2)^{n+1}}
 \;, \la{try_expansion}
\ea
where we denoted by $\bar{m}$ the mass of the effective-theory 
field $\bar{\phi}$ and by $\tilde\phi$ its Fourier representation. 
Furthermore we noted that the result vanishes unless the fields
$\tilde\phi(P^{ }_i)$ are contracted so that   
one of the momenta $P^{ }_4$, $P^{ }_5$ and $P^{ }_6$ remains an integration 
variable.
The integrals appearing in the result can be carried out in 
dimensional regularization; for instance, the two leading terms read
\ba
 n=0: && 
 \frac{1}{M^2} \int_{P^{ }_3} \frac{1}{P_3^2 + \bar m^2}
 = \frac{1}{M^2} \biggl( - \frac{\bar m}{4\pi} \biggr)
 \;, \la{try_n0} \\
 n=1: && 
 - \frac{1}{M^4} \int_{P^{ }_3} \frac{P_3^2}{P_3^2 + \bar m^2}
 =  \frac{\bar m^2}{M^4} \int_{P^{ }_3} \frac{1}{P_3^2 + \bar m^2}
 = - \frac{1}{M^4} \frac{\bar m^3}{4\pi}
 \;, \la{try_n1} 
\ea
where we made use of \eq\nr{In0_res} and of the vanishing of 
scale-free integrals in dimensional regularization. We note that
the terms get smaller with increasing $n$, apparently 
justifying {\em a posteriori} the Taylor expansion we carried
out above. 

Let us finally carry out the integral corresponding
to \eq\nr{graph_9c} exactly. The contractions
remain as above, and we simply need to replace the integral
in \eq\nr{try_expansion} by
\ba
 \int_{P^{ }_3} \frac{1}{P_3^2 + \bar m^2} \frac{1}{P_3^2 + M^2}
 & = &
 \int_{P^{ }_3} \frac{1}{M^2-\bar m^2}
 \biggl[\frac{1}{P_3^2 + \bar m^2} -  \frac{1}{P_3^2 + M^2} \biggr] 
 \nn & = & 
 \frac{1}{M^2-\bar m^2} 
 \biggl( \frac{-1}{4\pi} \biggr) (\bar m - M)
 \nn & = & 
 \frac{1}{4\pi(M + \bar m)}
 \nn & = & 
 \frac{1}{4\pi M} \biggl( 1 - \frac{\bar m}{M} 
 + \frac{\bar m^2}{M^2}  - \frac{\bar m^3}{M^3} + \ldots \biggr)
 \;. \la{correct_expansion}
\ea
Comparing \eqs\nr{try_n0} and \nr{try_n1} with \eq\nr{correct_expansion}, 
we note that by carrying out the Taylor expansion, i.e.\ the naive matching
of the effective theory parameters, we missed the leading 
contribution in \eq\nr{correct_expansion}.
The largest term we found, \eq\nr{try_n0}, is only next-to-leading
in \eq\nr{correct_expansion}. 
It furthermore appears that we missed all 
even powers of $\bar{m}$ in the sum of \eq\nr{correct_expansion}.

The reason for the problem encountered is the same as in \eq\nr{mixed_splitup}:
it again has to be taken into account that the light fields $\phi$ can also
carry large momenta $P^{ }_3\sim M$, in which case a Taylor expansion
of $1/(P_3^2 + M^2)$ is not justified. Rather, we have to 
view \eq\nr{graph_9c} in analogy with \eq\nr{mixed_splitup},
\be
  \TopoFff(\Ldsc,\Ldsc,\Ldsc,\Ldsc,\Adsc,\Asc)
  \quad = \quad
 \TopoFbB(\Ldsc,\Ldsc,\Ldsc,\Ldsc,\Adsc) \quad + \quad 
 \TopoFb(\Ldsc,\Ldsc,\Ldsc,\Ldsc) 
 \;,
\ee
where the first term corresponds to a naive replacement of 
\eq\nr{graph_9a} by a momentum-independent 6-point vertex, and the second term
to a contribution from hard $\phi$-modes to an effective 4-point vertex. 
In accordance with our discussion around \eq\nr{mixed_splitup}, 
we see that the result of 
\eq\nr{try_n0} (and more generally \eq\nr{try_expansion}) is indeed 
non-analytic in the parameter $\bar m^2$, whereas 
the supplementary terms in \eq\nr{correct_expansion} 
that the naive Taylor expansion missed are analytic in it.

\hide{
\begin{eqnarray*}
 && \TopoStree(\Lsc) \\
 && \TopoS(\Lsc) \\
 && \TopoSTxt(\Lsc,1) \\
 && \TopoSB(\Lsc,\Asc,\Asc) \\
 && \TopoST(\Lsc,\Asc) \\
 && \TopoSTTxt(\Lsc,\Asc,1) \\
 && \ToptSi(\Lsc) \\
 && \ToptSiTxt(\Lsc,1) \\
 && \ToptSM(\Lsc,\Asc,\Asc,\Asc,\Asc,\Lsc) \\
 && \ToptSAl(\Lsc,\Asc,\Asc,\Asc,\Asc) \\
 && \ToptSAr(\Lsc,\Asc,\Asc,\Asc,\Asc) \\
 && \ToptSE(\Lsc,\Asc,\Asc,\Asc,\Asc) \\
 && \ToptSDB(\Lsc,\Lqu,\Agl,\Agh,\Asc) \\
 && \ToptSS(\Lsc,\Asc,\Asc,\Lsc) \\
 && \ToptSSp(\Lsc,\Asc,\Asc,\Asc) \\
 && \ToptSr(\Lsc) \\
 && \ToptSrTxt(\Lsc,1) \\
 && \ToptSBB(\Lsc,\Asc,\Asc) \\
 && \ToptSBBTxt(\Lsc,\Asc,\Asc,1) \\
 && \ToptSTB(\Lsc,\Asc) \\
 && \ToptSTBTxt(\Lsc,\Asc,1) \\
 && \TopoT(\Lsc,\Lsc,\Lsc) \\
 && \TopoTTxt(\Lsc,\Lsc,\Lsc,1) \\
 && \TopoTS(\Lsc,\Lsc,\Lsc,\Asc,\Asc,\Asc) \\
 && \TopoTAo(\Lsc,\Lsc,\Lsc,\Asc,\Asc) \\
 && \TopoTAr(\Lsc,\Lsc,\Lsc,\Asc,\Asc) \\
 && \TopoTAl(\Lsc,\Lsc,\Lsc,\Asc,\Asc) \\
 && \TopoF(\Lsc,\Lsc,\Lsc,\Lsc) \\
 && \TopoFb(\Lsc,\Lsc,\Lsc,\Lsc) \\
 && \TopoFbB(\Lsc,\Lsc,\Lsc,\Lsc,\Adsc) \\
 && \TopoFff(\Lsc,\Lsc,\Lsc,\Lsc,\Adsc,\Adsc) \\
 && \TopoFffTxt(\Lsc,\Lsc,\Lsc,\Lsc,\Adsc,\Adsc,1) \\
 && \TopoFft(\Lsc,\Lsc,\Lsc,\Lsc,\Adsc,\Adsc) \\
 && \TopoFtt(\Lsc,\Lsc,\Lsc,\Lsc,\Adsc,\Adsc) \\
 && \TopoSix(\Lsc,\Lsc,\Lsc,\Lsc,\Lsc,\Ldsc,\Ldsc) \\
\end{eqnarray*}
}

%

\newpage


\newpage 

\section{Finite density}
\la{se:density}

\paragraph{Abstract:}

The concept of a system at a finite density or, equivalently, 
at a finite chemical potential, 
is introduced. Considering first a complex scalar field, an imaginary-time
path integral representation is derived for the partition function. 
The evaluation of the partition function reveals infrared
problems, which are this time related to the phenomenon of Bose-Einstein
condensation. A generic tool applicable to any scalar field theory, called
the effective potential, is introduced in order to handle this situation. 
Subsequently the case of a Dirac fermion at a finite chemical
potential is discussed. 
The concept of a susceptibility is introduced. The quark number
susceptibility in QCD is evaluated up to second order in the gauge coupling. 

\paragraph{Keywords:} 

Noether's theorem, global symmetry, 
Bose-Einstein condensation, condensate, constrained
effective potential, susceptibility.

\index{Chemical potential: scalar field}
\index{Finite density}

%
\subsection{Complex scalar field and effective potential}
\la{ss:scalar_mu}

Let us consider a system which possesses 
some conserved global charge, $Q$.
We assume the conserved charge to be {\em additive}, i.e.\ the charge
can in principle have any (integer) value. 
Physical examples of possible $Q$'s include:
\bi
\item
The baryon number $B$ and the lepton number $L$.
(In fact, within the classic Standard Model, the combination $B+L$ is 
not conserved because of an anomaly~\cite{anomaly},
so that strictly speaking
only the linear combination $B-L$ is conserved; however, 
the rate of $B+L$ violation is exponentially small 
at $T < 160$~GeV~\cite{srate2}, 
so in this regime we can 
treat both $B$ and $L$ as separate conserved quantities.) 

\item
If weak interactions are switched off (i.e., if we inspect phenomena
at temperatures well below 50 GeV, time scales well shorter 
than $10^{-10}$~s, or distances well below $1$~cm within 
the collision region of a particle experiment), then flavour
quantum numbers 
such as the strangeness $S$ are conserved. One prominent example 
of this is QCD thermodynamics, where one typically considers 
the chemical potentials of all quark flavors to be independent parameters.

\item
In non-relativistic field theories, the particle number $N$ is conserved.

\item
In some supersymmetric theories, there is 
a quantity called the $R$-charge which is conserved. 
(However this is normally a {\em multiplicative} rather 
than an {\em additive} charge. As discussed below, this 
leads to a qualitatively different behaviour.)

\ei

The case of a conserved $Q$ turns out to be  
analogous to the case of gauge fields, treated
in \se\ref{se:Gauge}; indeed the introduction
of a chemical potential, $\mu$, as a conjugate variable to $Q$, 
is closely related to the introduction of the gauge field $ A_0^{ }$
that was needed for imposing 
the Gauss law, ``$Q=0$''.
However, in contrast to that situation, we work in 
a grand canonical ensemble in the following, so that the quantum 
mechanical partition function is of the type
\be
 \mathcal{Z}(T,\mu) \equiv \tr\Bigl[
  e^{-\beta(\hat H - \mu \hat Q)} 
 \Bigr]
 \;. \la{Zmu}
\ee 
In \se\ref{se:Gauge}
the projection operator $\delta^{ }_{\hat Q,\hat 0}$ was effectively
imposed as 
\be
 \delta^{ }_{\hat Q,\hat 0} = 
 \int_{-\pi}^{\pi} \! \frac{{\rm d}\theta }{2\pi}
 e^{i \theta \hat Q}
 \;, \la{delta_Q}
\ee
where $\theta \propto  A_0^{ }$ and we assumed
the eigenvalues of $\hat Q$ to be integers. 
Comparing \eqs\nr{Zmu} and \nr{delta_Q}, 
a chemical potential is seen to 
correspond, roughly speaking, to a 
{\em constant purely imaginary Euclidean gauge field $ A_0^{ }$.}

\index{Noether's theorem}

Now, let us go back to classical field theory for a moment, 
and recall that if the system possesses a {global} U(1)
symmetry, then there exists, according to Noether's theorem, 
a conserved current, $\mathcal{J}^{ }_\mu$. The integral of the 
zeroth component of the current, i.e.\ the charge density, over
the spatial volume, defines the conserved charge, 
\be
 Q \; \equiv \; \int_\vec{x} \, \mathcal{J}_0^{ }(t,\vec{x})
 \;. \la{Qdef}
\ee
Conversely, one can expect that a system
which does have a conserved global charge should also 
display a global U(1) symmetry in its field-theoretic
description. Usually this is indeed the case, 
and we restrict to these situations in the following. (One notable
exception is free field theory where, due to a lack of interactions, 
particle number is conserved even without a global symmetry. Another
is that a discrete symmetry, $\phi\to-\phi$, may also lead to the 
concept of a generalized ``parity'', which acts as a {\em multiplicative} 
quantum number, with possible values $\pm 1$; however, in this case 
no non-trivial {\em charge density} $\rho = \langle\hat Q \rangle/V$
can be defined in the thermodynamic limit.) 

As the simplest example of a system with an 
additive conserved charge and a global U(1) symmetry, 
consider a {\em complex scalar field}. 
The classical Lagrangian of a complex scalar field reads
\be
 \mathcal{L}^{ }_\iM = \partial^\mu \phi^* \partial^{ }_\mu \phi - V(\phi)
 \;, 
\ee
where the potential has the form
\be
 V(\phi) \equiv m^2 \phi^*\phi + \lambda(\phi^* \phi)^2
 \;. 
\ee
The system is invariant in the (position-independent)
phase transformation
\be
 \phi \to e^{-i \alpha} \phi \;, \quad
 \phi^* \to e^{i \alpha} \phi^*
 \;, \la{phi_sym}
\ee
where $\alpha \in \RR$. The corresponding Noether current 
can be defined as
\ba
 \mathcal{J}^{ }_\mu & \equiv & 
 \frac{\partial \mathcal{L}^{ }_\iM}{\partial(\partial^\mu \phi)}
 \frac{\delta \phi}{\delta\alpha} + 
 \frac{\partial \mathcal{L}^{ }_\iM}{\partial(\partial^\mu \phi^*)}
 \frac{\delta \phi^*}{\delta\alpha}
 \nn & = & 
 - \partial^{ }_\mu \phi^* \,i \phi +
 \partial^{ }_\mu \phi \, i \phi^* 
 \nn & = & 
 - i [(\partial^{ }_\mu \phi^*) \phi - \phi^* \partial^{ }_\mu  \phi]
 = - 2 \im [ \phi^* \partial^{ }_\mu  \phi ]
 \;. \la{sft_charge}  
\ea
The overall sign (i.e., what we call particles
and antiparticles) is a matter of convention; we could equally
well have defined the global symmetry through
$
 \phi \to e^{i \alpha} \phi$, 
$
 \phi^* \to e^{-i \alpha} \phi^*
$, 
and then $\mathcal{J}^{ }_\mu$ would have the opposite sign. 

The first task, as always, is to write down a path integral 
expression for the partition function, \eq\nr{Zmu}. Subsequently, 
we may try to evaluate the partition function, in order to see what
kind of phenomena take place in this system. 

In order to write down the path integral, we start from the known expression
of $\mathcal{Z}$ of a real scalar field $\phi^{ }_1$ 
{\em without a chemical potential}, 
i.e.\ the generalization to field theory of \eq\nr{Z_B}: 
\be
 \mathcal{Z} \propto 
 \int_\rmi{periodic} \!\!\!  
 \mathcal{D} \phi^{ }_1 \int \mathcal{D} \pi^{ }_1  \, 
  \exp\biggl\{ 
 - \int_0^{\beta} \! {\rm d}\tau \int_\vec{x} \, 
 \biggl[
  \fr12 \pi_1^2  - i \pi^{ }_1 \partial^{ }_\tau \phi^{ }_1 
  + \fr12 (\partial^{ }_i\phi^{ }_1)^2 + V(\phi^{ }_1)
 \biggr]
 \biggr\} 
 \;, \la{Z_sft_0}
\ee
where $\pi^{ }_1 = \partial \phi^{ }_1/\partial t$ 
(cf.~the discussion in \se\ref{ss:pi_sft}). Here the combination
$ 
  \fr12 \pi_1^2   + \fr12 (\partial^{ }_i\phi^{ }_1)^2 + V(\phi^{ }_1)
$
is nothing but the classical Hamiltonian density, 
$\mathcal{H}(\pi^{ }_1,\phi^{ }_1)$.

In order to make use of \eq\nr{Z_sft_0}, let us 
rewrite the complex scalar 
field $\phi$ as $\phi = (\phi^{ }_1 + i \phi^{ }_2)/\sqrt{2}$, 
$\phi^{ }_i \in \RR$. Then 
\be
 \partial^\mu\phi^* \partial^{ }_\mu\phi = 
 \fr12 \partial^\mu\phi^{ }_1 \partial^{ }_\mu\phi^{ }_1 + 
 \fr12 \partial^\mu\phi^{ }_2 \partial^{ }_\mu\phi^{ }_2
 \;, \quad
 \phi^*\phi = \fr12 (\phi_1^2 + \phi_2^2)
 \;, 
\ee
and the classical Hamiltonian density reads
\be
 \mathcal{H} = 
 \fr12 \Bigl[ \pi_1^2  + \pi_2^2 + (\partial^{ }_i\phi^{ }_1)^2 
 + (\partial^{ }_i\phi^{ }_2)^2 + m^2 \phi_1^2 + m^2 \phi_2^2
 \Bigr]  + \fr14 \lambda (\phi_1^2 + \phi_2^2)^2
 \;. \la{sft_charged_H}
\ee
For the grand canonical ensemble, we need to 
add from \eqs\nr{Qdef} and \nr{sft_charge}
the classical version of $-\mu\hat Q$ to the Hamiltonian, 
cf.\ \eq\nr{Zmu}: 
\ba
 -\mu Q & = &  \mu\int_\vec{x} 
 \im\Bigl[ (\phi^{ }_1 - i \phi^{ }_2)
 (\partial^{ }_t \phi^{ }_1 + i \partial^{ }_t \phi^{ }_2) 
 \Bigr] 
 \nn & = & 
 \int_\vec{x} \mu(\pi^{ }_2 \phi^{ }_1 - \pi^{ }_1 \phi^{ }_2)
 \;. \la{sft_charged_Q}
\ea
Since the charge can be expressed in terms of the canonical variables, 
nothing changes in the derivation of the path integral, and we can 
simply replace the Hamiltonian of \eq\nr{Z_sft_0} by  
the sum of \eqs\nr{sft_charged_H} and \nr{sft_charged_Q}. 

\index{Partition function: complex scalar}
\index{Path integral: complex scalar}

Finally, we carry out the Gaussian integrals over $\pi^{ }_1$, $\pi^{ }_2$: 
\ba
 \int \! {\rm d}\pi^{ }_1 
 \exp\biggl\{
 - {\rm d}^D\! X \, \biggl[
 \fr12 \pi_1^2 + \pi^{ }_1 \biggl( - i \frac{\partial \phi^{ }_1}{\partial\tau}
 - \mu\phi^{ }_2 \biggr)  
 \biggr] 
 \biggr\} \!\! & = & \!\! 
 C \,
 \exp\biggl\{
 -\fr12 {\rm d}^D\! X \, \biggl( \frac{\partial \phi^{ }_1}{\partial\tau}
 - i \mu\phi^{ }_2 \biggr)^2  
 \biggr\}
 \;, \hspace*{0.5cm} \nn \\
 \int \! {\rm d}\pi^{ }_2  
 \exp\biggl\{
 - {\rm d}^D\! X \, \biggl[
 \fr12 \pi_2^2 + \pi^{ }_2 \biggl( - i \frac{\partial \phi^{ }_2}{\partial\tau}
 + \mu\phi^{ }_1 \biggr)  
 \biggr] 
 \biggr\} \!\! & = & \!\!  
 C \, 
 \exp\biggl\{
 -\fr12 {\rm d}^D\! X \, \biggl( \frac{\partial \phi^{ }_2}{\partial\tau}
 + i \mu\phi^{ }_1 \biggr)^2  
 \biggr\}
 \;. \hspace*{0.5cm} \nn 
\ea
Afterwards we go back to the complex notation, writing 
\ba
 & & 
 \fr12 \biggl( \frac{\partial \phi^{ }_1}{\partial\tau}
 - i \mu\phi^{ }_2 \biggr)^2 +
 \fr12 \biggl( \frac{\partial \phi^{ }_2}{\partial\tau}
 + i \mu\phi^{ }_1 \biggr)^2 
 \nn 
 & = &
 \fr12 \biggl[ 
 \biggl( \frac{\partial \phi^{ }_1}{\partial\tau} \biggr)^2 + 
 \biggl( \frac{\partial \phi^{ }_2}{\partial\tau} \biggr)^2 \biggr]
 -  \mu \times \underbrace{
 i \biggl[ 
  \phi^{ }_2 \frac{\partial \phi^{ }_1}{\partial\tau} - 
  \phi^{ }_1 \frac{\partial \phi^{ }_2}{\partial\tau}
 \biggr]}_{\phi\, \partial^{ }_\tau \phi^* - \phi^* \partial^{ }_\tau \phi}
 - \fr12 \mu^2 (\phi_1^2 + \phi_2^2)
 \nn  
 & = &
 [(\partial^{ }_\tau + \mu)\phi^*]
 [(\partial^{ }_\tau - \mu)\phi]
 \;. 
\ea 
In total, then, the path integral representation for the grand canonical
partition function of a complex scalar field reads
\be
 \mathcal{Z}(T,\mu) 
 = C \int_\rmi{periodic} \!\!\!\!\!\!\!\! \mathcal{D} \phi \, 
 \exp \biggl\{  
- \int_0^\beta \! {\rm d}\tau \int_\vec{x} \, 
 \biggl[ 
 (\partial^{ }_\tau + \mu)\phi^*
 (\partial^{ }_\tau - \mu)\phi + 
 \partial^{ }_i \phi^* \partial^{ }_i\phi
 + m^2 \phi^*\phi + 
 \lambda(\phi^*\phi)^2
 \biggr] 
 \biggr\}
 \;.  \la{Z_sft_1}
\ee
As anticipated, $\mu$ appears in a way reminiscent
of an imaginary gauge field $ A_0^{ }$.

Let us work out the properties of the free theory in the 
presence of $\mu$. Going to momentum space with $P = (\omega^{ }_n,\vec{p})$, 
the quadratic part of the Euclidean action becomes
\ba
 S^{(0)}_\iE & = & 
 \Tint{ P}
 \tilde \phi^*( P) 
 \Bigl[ 
 (-i \omega^{ }_n + \mu)(i \omega^{ }_n - \mu) + {p}^2 + m^2 
 \Bigr]
 \tilde \phi( P)
 \nn & = & 
 \Tint{ P}
 \tilde \phi^*( P) 
 \tilde \phi( P)
 \Bigl[ 
  (\omega^{ }_n + i \mu)^2 + {p}^2 + m^2
 \Bigr] 
 \;. \la{SE_sft_mu_0}
\ea
We observe that the chemical potential induces a
{\em shift of the Matsubara frequencies by a constant imaginary term}
(this was the reason for considering 
the corresponding sum in \eq\nr{iEc_def}). 
In particular, the propagator reads
\be
 \langle \tilde \phi( P) \tilde \phi^*( Q) \rangle_0^{ }
 = 
 \; \deltabar( P -  Q) 
 \, \frac{1}{(\omega^{ }_n + i \mu)^2 + {p}^2 + m^2}
 \;, 
\ee
whereas the grand canonical free energy density 
(sometimes referred to as the grand potential in the literature) 
is obtained from \eqs\nr{jEc} and \nr{ZHO_mu}. 
We just need to  replace $c\to i \mu$ 
and note that for a {\em complex} scalar field, 
all Fourier modes are independent, 
whereby the structures in 
\eqs\nr{jEc} and \nr{ZHO_mu} are to be multiplied by a factor 2:  
\be
 f(T,\mu) =  
 \left. 
 \int_\vec{p} \,
 \Bigl\{
 \E^{ }_p + T  
 \Bigl[
   \ln \Bigl( 1 - e^{-\beta({\E^{ }_p-\mu})}\Bigr) 
   + \ln \Bigl( 1 - e^{-\beta({\E^{ }_p+\mu})}\Bigr) 
 \Bigr]
 \Bigr\}
 \right|^{ }_{\E^{ }_p = \sqrt{{p}^2 + m^2}}
 \;. \la{f_SFT_mu}
\ee

We may wonder how the existence of $\mu\neq 0$ affects the infrared
problem of finite-temperature field theory, discussed in 
\se\ref{se:Linde}. 
In \se\ref{se:highT} we
found that the high-temperature expansion 
($T \gg m$) of \eq\nr{f_SFT_mu} at $\mu=0$
has a peculiar structure, because of a branch cut
starting at $m^2 = 0$. {}From the second term in \eq\nr{f_SFT_mu}, we
note that this problem has become {\em worse} in the presence of $\mu > 0$: 
the integrand is complex-valued if $\mu > m$, 
because then $\exp(-\beta (\E^{ }_p-\mu))>1$ at small ${p}$.  
In an interacting theory, thermal corrections generate an 
effective mass $m_\rmi{eff}^2 \sim \lambda T^2$ (cf.\ \eq\nr{mmeff}), 
which postpones
the problem to a larger $\mu$. Nevertheless, for large enough $\mu$
it still exists. 

It turns out that there is a {\em physics consequence} from
this infrared problem: the existence of {\em Bose-Einstein condensation}, 
to which we now turn. 


\subsection*{Bose-Einstein condensation}
\la{se:Veff}

\index{Bose-Einstein condensation}
\index{Condensate}

In order to properly treat complex scalar field theory with a chemical
potential, two things need to be realized:
\bi
\item[(i)]
In contrast to gauge field theory, the infrared problem exists 
even in the {\em non-interacting limit}. Therefore it cannot 
be cured by a perturbatively or 
non-perturbatively generated effective mass. Rather, it 
corresponds to a strong dependence of the properties of the system
on the volume, so we should keep the volume finite to start with.

\item[(ii)]
The chemical potential $\mu$ is a most useful quantity in theoretical
computations, but it is somewhat ``abstract'' from a practical point 
of view; the physical properties of the system are typically best 
characterized not by $\mu$ but by the intensive 
variable conjugate to $\mu$, 
i.e.\ the number density of the conserved
charge. Therefore, rather than trying to give $\mu$ some specific value, 
we should fix the number density. 

\ei

Motivated by point (i), 
let us put the system in a periodic box, $V= L^{ }_1 L^{ }_2 L^{ }_3$. 
The spatial momenta get discretized like in \eq\nr{finV},
\be
 \vec{p} = 2\pi 
 \Bigl( \frac{n^{ }_1}{L^{ }_1}, \frac{n^{ }_2}{L^{ }_2},
 \frac{n^{ }_3}{L^{ }_3} \Bigr)
 \;, 
\ee 
with $n^{ }_i \in \ZZ$. The mode with $\omega^{ }_n =0, \vec{p}=\vec{0}$
will be called {\em the condensate}, and denoted by $\bar\phi$. 
Note that the condensate is a Matsubara zero mode but 
{\em in addition} a spatial zero mode.

We now rewrite the partition function of \eq\nr{Z_sft_1} as
\ba
 \mathcal{Z}(T,\mu) & = &
 \int_{-\infty}^{\infty} \! {\rm d}\bar\phi \, \biggl\{ 
 \int_{\rmi{periodic},\, P\neq 0} \hspace*{-1.5cm}
 \mathcal{D}\phi' \, 
 e^{-S^{ }_\iiE[\phi = \bar\phi + \phi']} \biggr\}
 \nn 
 & \equiv & 
 \int_{-\infty}^{\infty} \! {\rm d}\bar\phi \,
 \exp\biggl[
 - \frac{V}{T} V^{ }_\rmi{eff}(\bar\phi^* \bar\phi) 
 \biggr]
 \;. \la{Veff_def}
\ea
Here $\phi'$ contains all modes with $ P\neq 0$, 
and $V^{ }_\rmi{eff}$ is called the {\em (constrained) effective potential}. 
The factor $V/T$ is the trivial spacetime integral, 
$\int_0^\beta\! {\rm d}\tau \int_V \! {\rm d}^d\vec{x}$.

\index{Effective potential}
\index{Constrained effective potential}

Let us write down the effective potential explicitly for very weak 
interactions, $\lambda \approx 0$. It turns out that the limit
$\lambda\to 0$ is subtle, so for the moment 
we keep $\lambda$ non-zero 
in the zero-mode part. {}From \eqs\nr{Z_sft_1} and \nr{SE_sft_mu_0} we get
\ba
 S^{ }_\iE[\phi = \bar\phi + \phi']
 & = &  \frac{V}{T} 
 \Bigl[ (m^2 - \mu^2)\, \bar\phi^* \bar\phi 
 + \lambda\,  (\bar\phi^* \bar\phi)^2 \Bigr]
 \nn & & \quad
 + \,
 \Tint{ P \neq 0} \biggl\{ 
 \tilde \phi'^*( P) 
 \tilde \phi'( P)
 \Bigl[ 
  (\omega^{ }_n + i \mu)^2 + {p}^2 + m^2 
 \Bigr] + \rmO(\lambda) 
 \biggr\}
 \;,  \la{SE_sft_mu_1}
\ea
where we made use of the fact that the crossterm between 
$\bar\phi$ and $\phi'$ vanishes, given that by definition
$\phi'$ has no zero-momentum mode: 
\be
 \int_0^\beta\! {\rm d}\tau \int_V \! {\rm d}^d\vec{x} \, 
 \phi' = 0
 \;. 
\ee
The path integral over the latter term in \eq\nr{SE_sft_mu_1} 
yields then \eq\nr{f_SFT_mu}; in the limit of a large volume, 
the omission of a single mode does not matter
(its effect is $\propto (T/V) \ln(m^2 - \mu^2)$).  
Thereby the effective potential reads
\be
 V^{ }_\rmi{eff}(\bar\phi^* \bar\phi) = (m^2 - \mu^2)\,\bar\phi^* \bar\phi 
 + \lambda\, (\bar\phi^* \bar\phi)^2 
 + f(T,\mu) + \rmO\Bigl( \frac{1}{V},\lambda \Bigr)
 \;. \la{full_F_sft_mu}
\ee
Physically, the first two terms correspond to the contribution
of the particles that have formed a condensate, 
whereas the third term represents 
propagating particle modes in the plasma. 

Now, if we go to the limit of very small temperatures, $T \ll m$, 
and assume furthermore that $|\mu|\le m$, which is required
in order for \eq\nr{f_SFT_mu} to be defined, then 
the thermal part of \eq\nr{f_SFT_mu} vanishes.
(It has a ``non-relativistic'' limiting value 
for $\mu \to m^-$, which scales as $-T^4 (\tfr{m}{2\pi T})^{\tfr32}$.) 
The vacuum contribution to  \eq\nr{f_SFT_mu}
is on the other hand independent of $T$ and $\mu$,
and can be omitted. Therefore, 
\be
 V^{ }_\rmi{eff}(\bar\phi^* \bar\phi)
 \approx (m^2 - \mu^2)\,\bar\phi^* \bar\phi 
 + \lambda\, (\bar\phi^* \bar\phi)^2
 \;. \la{full_F_sft_mu_2}
\ee

The remaining task is to carry out the integral over 
$\bar\phi$ in \eq\nr{Veff_def}. At this point we need to make contact
with the particle number density. From 
$
 \mathcal{Z} = \tr[\exp(-\beta\hat H + \beta\mu\hat Q)]
$ and 
the definition of $V^{ }_\rmi{eff}$
in \eq\nr{Veff_def}, we obtain
\ba
 \rho & \equiv & 
 \frac{\langle \hat Q \rangle}{V}
 = \frac{T}{V} \frac{\partial\ln \mathcal{Z}}{\partial\mu}
 \\ \la{rho1} & = & 
 \frac{
 \int  {\rm d}\bar\phi \,
 2 \mu\,\bar\phi^*\bar\phi
 \exp\biggl[
 - \frac{V}{T} V^{ }_\rmi{eff}(\bar\phi^* \bar\phi) 
 \biggr]
  }{
 \int  {\rm d}\bar\phi \,
 \exp\biggl[
 - \frac{V}{T} V^{ }_\rmi{eff}(\bar\phi^* \bar\phi) 
 \biggr]
  }\;\equiv\; 2\mu\,\langle \bar\phi^* \bar\phi \rangle
 \;, \la{rho2}
\ea 
where $\langle \bar\phi^* \bar\phi \rangle$ is the expectation value
of $\bar\phi^* \bar\phi$.

Let us consider a situation where we decrease the temperature, 
$T \ll m$, and attempt simultaneously to keep the 
particle number density, the left-hand side of \eq\nr{rho1}, fixed. 
How should we choose $\mu$ in this situation? There are three 
possibilities: 
\bi
\item[(i)]
If $|\mu| < m$, the integrals
in \eq\nr{rho2} can be carried out even for $\lambda\to 0^+_{ }$. 
In fact the result corresponds to the ``propagator'' of $\bar\phi$: 
\be
 \lim_{\lambda\to 0^+_{ }} \rho = \frac{2 \mu  T}{V(m^2-\mu^2)}
 \;.  \la{rho3}
\ee
We note that if $T\to 0$, then $\rho \to 0$. This conflicts with our
assumption that the number density stays constant; therefore
this range of $\mu$ is not physically relevant for our situation.

\item[(ii)]
If $|\mu| > m$, the integrals 
in \eq\nr{rho2} are defined only for $\lambda > 0$.
For $V \to \infty$ they can be determined by the 
{\em saddle point approximation}: 
\be \index{Saddle point approximation}
 V_\rmi{eff}'(\bar\phi^* \bar\phi) = 0 
 \quad \Rightarrow \quad 
 \bar\phi^* \bar\phi = \frac{\mu^2 - m^2}{2\lambda}
 \quad \Rightarrow \quad 
 \rho = \frac{\mu(\mu^2 - m^2)}{\lambda}
 \;. \la{saddle_BEC}
\ee
We see that for $\lambda\to 0^+_{ }$, we need to send 
$\mu \to m^+$, in order to keep $\rho$ finite.  

\item[(iii)]
According to the preceding points, 
the only possible choice at $\lambda = 0^+_{ }$ is $|\mu| = m$. 
For $\rho > 0$ we need to choose $\mu = m$.
In this limit \eq\nr{rho2} can be expressed as  
\be
 \rho = 2 m \langle \bar\phi^* \bar\phi \rangle 
 \;, \la{rho4}
\ee
which should be thought of as a condition for the field $\bar\phi$.

\ei
\Eq\nr{rho4} manifests the phenomenon of {\em Bose-Einstein condensation}
(at zero temperature in the free limit): the conserved particle number
is converted into a non-zero scalar condensate. 

It is straightforward to include the effects of a finite 
temperature in these considerations, 
by starting from \eq\nr{full_F_sft_mu} so that $-\partial^{ }_\mu f(T,\mu)$
gives another contribution to the charge density, and the effects
of interactions, by keeping $\lambda > 0$. These very interesting 
developments go beyond the scope of the present lectures
(cf.\ e.g.\ refs.~\cite{mu1}--\hspace*{-1.1mm}\cite{mu3}). 
On the other hand, the concepts of a condensate and an effective potential 
will be met again in later chapters. 

\newpage

\subsection{Dirac fermion with a finite chemical potential}

\index{Chemical potential: Dirac field}

The Lagrangian of a Dirac fermion, 
\be
 \mathcal{L}^{ }_\iM 
 = \bar\psi^{ }_A (i \bsl{D}^{ }_{\!\!AB} - m \,\delta^{ }_{AB} )\psi^{ }_B
 \;, 
 \quad
 \bsl{D}^{ }_{\!\!AB}
 = \gamma_{ }^\mu (\delta^{ }_{AB}\, \partial^{ }_\mu - i g A^a_\mu T^a_{AB})
 \;, \la{LM_Dirac}
\ee
possesses a {\em global symmetry},
\be
 \psi^{ }_A \to e^{- i \alpha} \psi^{ }_A
 \;, \quad
 \bar\psi^{ }_A \to e^{i \alpha} \bar\psi^{ }_A
 \;, \la{DF_alpha}
\ee 
in addition to the usual non-Abelian (local) gauge symmetry. Therefore
there is a conserved quantity, and we can consider
the behaviour of the system in the presence of a chemical potential. 

\index{Path integral: Dirac field}
\index{Euclidean Lagrangian: Dirac field}

The conserved Noether current reads
\ba
 \mathcal{J}^{ }_\mu & = &
  \frac{\partial\mathcal{L}^{ }_\iM}{\partial(\partial^\mu \psi^{ }_A)} 
 \frac{\delta \psi^{ }_A}{\delta\alpha}
 \nn & = & 
 - \bar\psi^{ }_A \, i \gamma^{ }_\mu \, i \psi^{ }_A 
 = \bar\psi^{ }_A \gamma^{ }_\mu \psi^{ }_A
 \;. 
\ea
The corresponding charge is $Q = \int_\vec{x} \, \mathcal{J}_0^{ }$,
and as an operator
it commutes with the Hamiltonian, $[\hat H, \hat Q]=0$. Therefore, 
like with scalar field theory, we can treat 
the combination $\hat H - \mu \hat Q$ as an ``effective'' Hamiltonian, 
and directly write down the corresponding path integral, by adding 
\be
 -\mu\, Q = - \mu \int_\vec{x} \, \bar\psi^{ }_A \gamma_0^{ } \psi^{ }_A 
 \la{Dirac_Q}
\ee
to the Euclidean action. The path integral thereby reads
\be
 \mathcal{Z}(T,\mu) = 
 \int_\rmi{antiperiodic} \hspace*{-1cm}
 \mathcal{D} \bar\psi \, 
 \mathcal{D} \psi 
 \, 
 \exp\biggl\{
  -\int_0^{\beta}
 \! {\rm d}\tau \int_\vec{x} \,  
 \bar\psi\, 
  [\gamma^{ }_\mu  D^{ }_\mu -  \gamma_0^{ }\, \mu + m ] \,\psi
 \biggr\} 
 \;. \la{Z_fer_mu}
\ee

For perturbation theory, let us consider the quadratic part of the 
Euclidean action. Going to momentum space with $P = (\omega^{ }_n,\vec{p})$, 
we get 
\be
 S^{(0)}_\iE = \Tint{ \{ P \} }
 \,\tilde{\!\bar\psi}( P)
 [i  \gamma_0^{ }\, \omega^{ }_n + i  \gamma^{ }_i\, p^{ }_i  
 -  \gamma_0^{ }\, \mu + m ] \tilde \psi( P)
 \;. \la{SE_fer_mu_2} 
\ee
Therefore, just like in \se\ref{ss:scalar_mu}, the existence 
of a chemical potential corresponds to a shift 
$\omega^{ }_n\to \omega^{ }_n + i \mu$ of the Matsubara frequencies. 

Let us write down the free energy density of a single 
free Dirac fermion. Compared with a complex scalar field, there is 
an overall factor $-2$ (rather than $-4$ like in \eq\nr{f_f_final}, where we 
compared with a real scalar field). Otherwise, the chemical potential
appears in identical ways in \eqs\nr{SE_sft_mu_0} and
\eq\nr{SE_fer_mu_2}, so \eq\nr{SfSb_rel}, 
$
 \sigma^{ }_\fe(T) =
  2 \sigma^{ }_\bo\bigl( \tfr{T}{2} \bigr) - \sigma^{ }_\bo(T)
$, continues to apply. Employing it with \eq\nr{f_SFT_mu} we get 
\ba
  f(T,\mu) & = &   
 -2 
 \int_\vec{p} \,
 \biggl\{
 \E^{ }_p + T  
 \Bigl[
   \ln \Bigl( 1 - e^{-2\beta({\E^{ }_p-\mu})}\Bigr) 
   + \ln \Bigl( 1 - e^{-2\beta({\E^{ }_p+\mu})}\Bigr) 
 \nn & & \hspace*{2.2cm} 
   - \, \ln \Bigl( 1 - e^{-\beta({\E^{ }_p-\mu})}\Bigr) 
   - \ln \Bigl( 1 - e^{-\beta({\E^{ }_p+\mu})}\Bigr) 
 \Bigr]
 \Bigr\}
 \nn & = & 
 -2 
 \int_\vec{p} \,
 \Bigl\{
 \E^{ }_p + T  
 \Bigl[
   \ln \Bigl( 1 + e^{-\beta({\E^{ }_p-\mu}) }\Bigr) 
   + \ln \Bigl( 1 + e^{-\beta({\E^{ }_p+\mu}) }\Bigr) 
 \Bigr]
 \Bigr\}
 \;. \hspace*{1cm} \la{f_DF_mu}
\ea
The thermal part of 
this integral is well-defined for any $\mu$; thus
fermions do not suffer from 
infrared problems with $\mu\neq 0$, 
and do not undergo condensation
(in the absence of interactions). 


\subsection*{How about chemical potentials for gauge symmetries?}

\index{Chemical potential: gauge field?}
\index{QED}

It was mentioned after \eq\nr{delta_Q} that a chemical potential has 
some relation to a gauge field $ A_0^{ }$. However, in cases like
QCD, a chemical potential has no colour structure (i.e.\ it is an identity
matrix in colour space), whereas $ A_0^{ }$ is a traceless matrix in 
colour space (cf.\ \eq\nr{LM_Dirac}). 
On the other hand, in QED, $ A_0^{ }$ is not traceless. 
In fact, in QED, the gauge symmetry is nothing but a local version of
that in \eq\nr{DF_alpha}. We may therefore ask whether we can associate
a chemical potential to the electric charge of QED, and what 
the precise relation of $ A_0^{ }$ and $\mu$ is in this case. 

Let us first recall what happens in such a situation physically. 
A non-zero chemical potential in QED corresponds to a system which 
is {\em charged}. Moreover, if we want to describe it perturbatively 
with the QED Lagrangian, we had better choose a system where the charge 
carriers (particles) are essentially free; such a system could be a metal
or a plasma. In this situation, the free charge carriers 
interact repulsively with a long-range force, and hence all the 
net charge resides {\em on the surface}. In other words, the 
homogeneous ``bulk'' of the medium is {\em neutral} (i.e.\ has
no free charge). The charged body as 
a whole has a non-zero electric potential, $V_0^{ }$, with respect 
to the ground. 

Let us try to understand how to reproduce this behaviour
directly from the partition function, \eq\nr{Z_fer_mu}, adapted to QED:
\be 
 \mathcal{Z}(T,\mu) = 
 \int_\rmi{b.c.}
 \mathcal{D} A^{ }_\mu 
 \mathcal{D} \bar\psi
 \mathcal{D} \psi 
 \, 
 \exp\biggl\{
  -\int_0^{\beta}
 \!\! {\rm d}\tau \int_\vec{x} \,  
 \biggl[ \fr14 F_{\mu\nu}^2 
 + 
 \bar\psi\Bigl( 
    \gamma_0^{ } (\partial^{ }_\tau - i e  A_0^{ } - \mu)
  +  \gamma^{ }_i D^{ }_i + m \Bigr) \psi
 \biggr]
 \biggr\} 
 \;. \la{Z_qed_mu}
\ee
The usual boundary conditions (``b.c.'')\
over the time direction are assumed. The basic claim is that, 
according to the physical picture above, if we
assume the system to be {\em homogeneous}, i.e. consider the ``bulk'' 
situation, then the partition function {\em should not depend on $\mu$}. 
Indeed this would ensure the neutrality that we expect: 
\be  
 \rho = - \frac{\partial f}{\partial\mu} = 0 
 \;. \la{neutrality}
\ee
How does this arise?

\index{Effective potential}

The key observation is that we should again think of the system in terms
of an {\em effective potential}, like in \eq\nr{Veff_def}. The role of the 
condensate is now given to the field $ A_0^{ }$; let us denote it
by $\,\bar{ A}_0^{ }$. The last integral to be carried out is 
\be
 \mathcal{Z}(T,\mu) = 
 \int_{-\infty}^{\infty} \! {\rm d}\,\bar{ A}_0^{ } \, 
 \exp\biggl\{ - \frac{V}{T} V^{ }_\rmi{eff}( \,\bar{ A}_0^{ } )\biggr\}
 \;.  \la{MF_A0}
\ee

Now, we can deduce from \eq\nr{Z_qed_mu} that $\mu$ can only
appear in the combination $-i e \bar{ A}_0^{ } - \mu$, 
so that 
$
 V^{ }_\rmi{eff}( \bar{ A}_0^{ } ) = 
 f( \,\bar{ A}_0^{ } - i \mu/e)
$.
Moreover, we know from \eq\nr{Leff_EQCD} that in a large volume 
and high temperature, 
\be \index{Saddle point approximation}
 V^{ }_\rmi{eff}( \bar{ A}_0^{ } ) \approx
 \fr12 \mE^2\, ( \bar{ A}_0^{ } - i \mu/e)^2
 + \rmO ( \bar{ A}_0^{ } - i \mu/e)^4
 \;, \la{Veff_QED}
\ee
where $\mE^2 \sim e^2 T^2$. 
(The complete 1-loop $V^{ }_\rmi{eff}$ 
could be deduced from \eq\nr{f_ferm_expl} below, simply 
by substituting $\mu\to \mu + i e \bar{ A}_0^{ }$ there.)
In the infinite-volume limit, the integral in \eq\nr{MF_A0} 
can be carried out by making use of 
the {\em saddle point approximation}, like with Bose-Einstein
condensation in \eq\nr{saddle_BEC}.  
The saddle point is located
in the complex plane at the position 
where $V'_\rmi{eff}(\,\bar{ A}_0^{ }) = 0$, i.e.\ at
$
 \,\bar{ A}_0^{ } =  i \mu/e
$. 
The value of the potential at the saddle point, 
as well as the second derivative and so also the  
Gaussian integral around it, are independent of $\mu$. 
This leads to \eq\nr{neutrality}.

It is interesting to note that
the saddle point lies at a purely imaginary $\,\bar{ A}_0^{ }$. 
Recalling the relation of Minkowskian and Euclidean $A_0^{ }$ from
page~\pageref{AMAE_rel}, this corresponds to a real Minkowskian
$A_0^{ }$. Thus there indeed is a real electric potential $V_0^{ }\propto\mu$, 
just as we anticipated on physical grounds.

Finally, we note that in more complicated systems, like the Standard Model
of particle physics, the proper procedure involves introducing chemical
potentials for all global charges, and background values 
$\,\bar{ A}_0^a$ for all gauge fields. Subsequently, we need to
search for the saddle point, by minimizing the effective potential
as a function of the background fields~\cite{bg1,bg2}. 


\subsection*{Appendix A: Exact results in the free massless limit}

\index{Thermal sums: chemical potential}

The free energy density of a single Dirac fermion, 
\be
 f(T,\mu) = -2\, \Tint{ \{ P \} } 
 \Bigl\{ \ln[(\omega^{ }_n + i \mu)^2 + \E_p^2] - \mbox{const.} \Bigr\} 
 \;, 
\ee
can be computed explicitly for the case $m=0$ (i.e.\ $\E^{ }_p = {p}$).
We show that, subtracting the vacuum part, the result is 
\be
 f(T,\mu) = - 
 \biggl(
  \frac{7\pi^2 T^4}{180} + \frac{\mu^2 T^2}{6} + \frac{\mu^4}{12\pi^2} 
 \biggr)
 \;. \la{f_ferm_expl}
\ee

We start from \eq\nr{f_DF_mu}, subtracting
the vacuum term and setting $m=0,d=3$: 
\ba
 f(T,\mu) & = &  
  -2 T
 \int \! \frac{{\rm d}^3\vec{p}}{(2\pi)^3} \,
 \biggl\{
   \ln \biggl[ 1 + \exp\biggl({-\frac{{p}-\mu}{T} }\biggr)\biggr] 
   + \ln \biggl[ 1 + \exp\biggl({-\frac{{p}+\mu}{T} }\biggr)\biggr] 
 \biggr\}
 \nn 
 & = & 
 - \frac{T^4}{\pi^2}
 \int_0^\infty \! {\rm d}x \, x^2 
 \biggl\{
 \ln\Bigl( 1 + e^{-x+y} \Bigr) + 
 \ln\Bigl( 1 + e^{-x-y} \Bigr) 
 \biggr\}
 \;, \la{exe10_a}
\ea
where we set $x\equiv {p}/T$ and $y\equiv \mu/T$, 
and carried out the angular integration.

A possible trick now is to expand the logarithms in Taylor series, 
\be
 \ln(1+z) = \sum_{n=1}^{\infty} (-1)^{n+1} \frac{z^n}{n}
 \;, \quad |z| < 1
 \;. 
\ee
Assuming $y>0$, this is indeed possible with the second term of 
\eq\nr{exe10_a}, whereas in the first term a direct application 
is not possible, because the series does not converge for all $x$. 
However, if $e^{-x+y} > 1$, we can write
$1 + e^{-x+y} = e^{-x+y}(1 + e^{x-y})$, where $e^{x-y} < 1$. 
Thereby the Taylor expansion can be written as 
\be
 \ln\Bigl( 1 + e^{-x+y} \Bigr)
 = \theta(x-y) \sum_{n=1}^{\infty} \frac{(-1)^{n+1}}{n} e^{-xn}e^{yn}
 + \theta(y-x) \Bigl[ 
 y - x + \sum_{n=1}^{\infty} \frac{(-1)^{n+1}}{n} e^{xn}e^{-yn}
 \Bigr]
 \;. 
\ee
Inserting this into \eq\nr{exe10_a}, we get 
\ba
 f(T,\mu) & = & 
 -\frac{T^4}{\pi^2} \biggl\{ 
  \int_0^y \! {\rm d}x 
  \biggl[
   y x^2 - x^3 + 
  \sum_{n=1}^{\infty} \frac{(-1)^{n+1}}{n}  x^2 
  \Bigl(e^{xn}e^{-yn} + e^{-xn}e^{-yn} \Bigr)  \biggr] 
  \nn & & + 
  \int_y^\infty \! {\rm d}x 
  \biggl[
  \sum_{n=1}^{\infty} \frac{(-1)^{n+1}}{n}  x^2 
  \Bigl(e^{-xn}e^{yn} + e^{-xn}e^{-yn} \Bigr) \biggr] \biggr\} 
 \nn & = &  
 -\frac{T^4}{\pi^2} \biggl\{ 
  \int_0^y \! {\rm d}x 
  \biggl[
   y x^2 - x^3 + 
  \sum_{n=1}^{\infty} \frac{(-1)^{n+1}}{n}  x^2 
  \Bigl(e^{xn}e^{-yn} - e^{-xn}e^{yn} \Bigr)  \biggr] 
  \nn & & + 
  \int_0^\infty \! {\rm d}x 
  \biggl[
  \sum_{n=1}^{\infty} \frac{(-1)^{n+1}}{n}  x^2 
  \Bigl(e^{-xn}e^{yn} + e^{-xn}e^{-yn} \Bigr) \biggr] \biggr\}
 \;. \la{exe10_b}
\ea
All the $x$-integrals can be carried out: 
\ba
 \int_0^y  \! {\rm d}x \, (yx^2 - x^3) & = &
 \biggl( \fr13 - \fr14 \biggr) y^4 = \fr1{12} y^4 
 \;, \\ 
 \int_0^y  \! {\rm d}x \, x^2 e^{\alpha x} & = &
 -\frac{2}{\alpha^3} + e^{\alpha y}
 \biggl( \frac{2}{\alpha^3} - \frac{2 y}{\alpha^2} + \frac{y^2}{\alpha}\biggr)
 \;, \\ 
 \int_0^\infty \! {\rm d}x \, x^2 e^{-x n} & = & \frac{2}{n^3}
 \;.
\ea
Inserting these into \eq\nr{exe10_b} we get 
\ba
 f(T,\mu) & = & 
 -\frac{T^4}{\pi^2} \biggl\{ 
  \frac{y^4}{12} + \sum_{n=1}^{\infty} \frac{(-1)^{n+1}}{n}
 \biggl[ e^{-yn}
 \biggl( 
 -\frac{2}{n^3} + e^{y n}
 \biggl( \frac{2}{n^3} - \frac{2 y}{n^2} + \frac{y^2}{n}\biggr)
 \biggr)
 \nn & &  - e^{yn}
 \biggl( 
  \frac{2}{n^3} + e^{-y n}
 \biggl( - \frac{2}{n^3} - \frac{2 y}{n^2} - \frac{y^2}{n}\biggr)
 \biggr) + e^{yn} \frac{2}{n^3} + e^{-yn} \frac{2}{n^3} 
 \biggr] \biggr\} 
 \nn & = &
 -\frac{T^4}{\pi^2} \biggl\{ 
  \frac{y^4}{12} + \sum_{n=1}^{\infty} \frac{(-1)^{n+1}}{n}
 \biggl[
   \frac{4}{n^3} + \frac{2 y^2}{n}
 \biggr] \biggr\}
 \;, \la{exe10_c}
\ea
where a remarkable cancellation took place.
The sums can be carried out: 
\ba
 \eta(2) & \equiv & \sum_{n=1}^{\infty} \frac{(-1)^{n+1}}{n^2}
 = \frac{1}{1^2} - \frac{1}{2^2} + \frac{1}{3^2} - \frac{1}{4^2} + \cdots
 = 
 \zeta(2) - \frac{2}{2^2}\zeta(2) = \fr12 \zeta(2) = \frac{\pi^2}{12}
 \;, \hspace*{5mm} \\ 
 \eta(4) & \equiv & \sum_{n=1}^{\infty} \frac{(-1)^{n+1}}{n^4}
 = \frac{1}{1^4} - \frac{1}{2^4} + \frac{1}{3^4} - \frac{1}{4^4} + \cdots
 = 
 \zeta(4) - \frac{2}{2^4}\zeta(4) = \fr78 \zeta(4) = 
 \fr78 \frac{\pi^4}{90}
 \;. \hspace*{10mm}
\ea
Inserting into \eq\nr{exe10_c}, we end up with 
\be
 f(T,\mu) =
  -\frac{T^4}{\pi^2} \biggl\{ 
  \frac{y^4}{12} + 
  \frac{\pi^2 y^2}{6}  +
  \frac{7\pi^4}{180}
 \biggr\} 
 \;, 
\ee
which after the substitution $y=\mu/T$ reproduces \eq\nr{f_ferm_expl}.

\subsection*{Appendix B: Free susceptibilities}

\index{Susceptibility}

Important characteristics of dense systems are offered 
by {\em susceptibilities}, which define fluctuations of 
the particle number in a grand canonical ensemble. 
For a Dirac fermion,  
\ba
 \chi^{ }_\fe & \equiv &
 \lim_{V\to\infty} 
 \frac{\langle \hat N^2 \rangle - \langle \hat N \rangle^2}{V}
 = 
 \lim_{V\to\infty}  T \partial^{ }_\mu
 \biggl( \frac{ \langle \hat N \rangle }{V}  \biggr)
 = 
 \lim_{V\to\infty}  \frac{T^2 \partial_\mu^2 \ln\mathcal{Z} }{V}  
 = - T \partial_\mu^2 f(T,\mu)
 \hspace*{9mm}
 \la{chi_fe} \\
 & \stackrel{\rmi{\nr{f_DF_mu}}}{=} & 
 2 T \int_\vec{p} \partial_\mu^2 
 \Bigl\{ \E^{ }_p + T \Bigl[ 
    \ln \Bigl(1+ e^{-\beta(\E^{ }_p - \mu)}\Bigr)
 +  \ln \Bigl(1+ e^{-\beta(\E^{ }_p + \mu)}\Bigr) 
 \Bigr] \Bigr\} 
 \nn & = & 
 2 T \int_\vec{p} \partial^{ }_\mu 
 \biggl\{
 \frac{1}{ e^{\beta(\E^{ }_p - \mu)} + 1 }
 - \frac{1}{ e^{\beta(\E^{ }_p + \mu)} + 1 }   
 \biggr\} 
 \nn & = & 
 2  \int_\vec{p} 
 \biggl\{  \frac{e^{\beta(\E^{ }_p - \mu)}}
 { [ e^{\beta(\E^{ }_p - \mu)} + 1 ]^2 } + 
   \frac{e^{\beta(\E^{ }_p + \mu)}}{[ e^{\beta(\E^{ }_p + \mu)} + 1 ]^2 }
 \biggr\} 
 \nn \!\! & = & \!\! 
 \frac{1}{\pi^2} \int_0^\infty \!\!\! {\rm d}p \, p^2 
 \Bigl\{ 
  \nF{}(\E^{ }_p - \mu) \bigl[ 1 - \nF{}(\E^{ }_p - \mu) \bigr]
  + 
  \nF{}(\E^{ }_p + \mu) \bigl[ 1 - \nF{}(\E^{ }_p + \mu) \bigr]
 \Bigr\} 
 \;. 
\ea
In the massless limit, \eq\nr{f_ferm_expl} 
directly gives 
$
 \chi^{ }_\fe = \frac{T^3}{3} + \frac{\mu^2 T}{\pi^2}
$.
On the other hand, for $m\neq 0$, $\mu = 0$, 
one gets 
$
 \chi^{ }_\fe = \frac{2 m^2 T}{\pi^2}
 \sum_{n=1}^{\infty} (-1)^{n+1} K^{ }_2\bigl( \frac{n m }{T} \bigr)
$, 
where $K^{ }_2$ is a modified Bessel function. 
In the bosonic case of a complex scalar field, 
\eq\nr{f_SFT_mu} similarly leads to 
\be
 \chi^{ }_\bo = 
 \frac{1}{2 \pi^2} \int_0^\infty \! {\rm d}p \, p^2 
 \Bigl\{ 
  \nB{}(\E^{ }_p - \mu) \bigl[ 1 + \nB{}(\E^{ }_p - \mu) \bigr]
  + 
  \nB{}(\E^{ }_p + \mu) \bigl[ 1 + \nB{}(\E^{ }_p + \mu) \bigr]
 \Bigr\} 
 \;. \la{chi_bo}
\ee
In this case the massless limit is only relevant at $\mu=0$ 
(otherwise the integrand is singular at $p = |\mu|$), 
where we obtain $ \chi^{ }_\bo = T^3/3$. For $m\neq 0$, 
the susceptibility at $\mu=0$ can again be expressed in terms
of modified Bessel functions,  
$
 \chi^{ }_\bo = \frac{m^2 T}{\pi^2}
 \sum_{n=1}^{\infty}  K^{ }_2\bigl( \frac{n m }{T} \bigr)
$.

\subsection*{Appendix C: Finite density QCD at next-to-leading order}

\index{Susceptibility: next-to-leading order}

Extending the description of finite density systems
to higher perturbative orders has become an actively 
studied topic, cf.\ e.g.\ refs.~\cite{bg2,av}. 
An example is the susceptibility defined in \eq\nr{chi_fe}, 
evaluated at $\mu=0$ within QCD, 
and the generalization thereof to the case of 
several quark flavours.  
These quantities probe finite density, 
but can nevertheless be compared with lattice
QCD simulations that are well under control
only at vanishing chemical potentials. The
basic strategy in their evaluation follows the above leading-order computation
in the sense that it is technically easier to first compute the
entire free energy density 
at finite $\mu$, and only afterwards to take 
derivatives with respect to $\mu$~\cite{av}. 
As a by-product of evaluating 
susceptibilities at $\mu=0$, we therefore obtain the behaviour 
of the pressure at finite density.

At 2-loop order, the $\mu$-dependent part of the QCD free energy density
gets contributions from one single diagram, 
namely the same as in \eq\nr{ggpsipsi}. 
Like in \eq\nr{ggpsipsi} it is easy to see  
that in the limit of massless quarks (an approximation that 
significantly simplifies higher-order computations) this diagram 
can be written in the form
\ba
  \ToptVS(\Aqu,\Aqu,\Lgl) 
  &=&\dA g^2\, \frac{d-1}{2}\, \Tint{ \{ P \}Q  } 
 \bigg[ \frac{1}{\tilde{P}^2 (\tilde{P}-Q)^2}
 -\frac{2}{\tilde{P}^2Q^2}\bigg]
 \;, \la{deltafmu}
\ea
where $\dA\equiv \Nc^2-1$, we set $\Nf =1$ and, 
in accordance with \eq\nr{SE_fer_mu_2}, 
the fermionic Matsubara frequencies have been shifted by 
$\omega^{ }_n\to \omega^{ }_n + i \mu \equiv \tilde{p}^{ }_n$. 
Both terms in this result clearly factorize into products 
of 1-loop sum-integrals that (up to the shift of the 
fermionic Matsubara frequencies) can be identified as the functions 
$I(0,T)= I^{ }_T(0)$ and $\tilde{I}(0,T) = \tilde I^{ }_T(0)$ 
studied in \ses\ref{se:highT} and \ref{se:Dirac}, respectively.

For completeness, let us next inspect the more general fermionic sum-integral 
\ba
 \tilde{I}(m=0,T,\mu,\alpha)&\equiv& 
 \Tint{ \{ P \}} \frac{1}{(\tilde{P}^2)^\alpha_{ }}
 \;, 
\ea
following a strategy similar to that in \eq\nr{IpmT}.
In other words we first perform the $3-2\epsilon$ -dimensional integral 
over the spatial momentum $\mathbf{p}$, and afterwards take care 
of the Matsubara sum. Applying the familiar result of \eq\nr{fmdA}, we obtain
\ba
 \tilde{I}(m=0,T,\mu,\alpha)&=& \fr{1}{\(4\pi\)^{3/2-\epsilon}}
 \fr{\Gamma(\alpha-3/2+\epsilon)}{\Gamma(\alpha)}
 \, T \sum_{k=-\infty}^{\infty}
\fr{1}{\big[\bigl(\(2k+1\)\pi T+i\mu\bigr)^2\big]^{\alpha-3/2+\epsilon}}  \nn
&=&  2^{-2\alpha}\pi^{-2\alpha+3/2-\epsilon}
 T^{-2\alpha+4 - 2\epsilon} 
 \; \fr{\Gamma(\alpha-3/2+\epsilon)}{\Gamma(\alpha)} \nn
&& \times 
 \Big[\zeta\Bigl(2\alpha-3+2\epsilon,\fr12-i \bar{\mu}\Bigr)
 +\zeta\Bigl(2\alpha-3+2\epsilon,\fr12+i \bar{\mu}\Bigr)\Big]\; ,
\ea
where $\bar{\mu}\equiv \mu/(2\pi T)$ 
and we have expressed the infinite sums in terms of 
the generalized (Hurwitz) zeta-function
\ba
 \zeta(z,q) \; \equiv \; \sum_{n=0}^{\infty} \fr{1}{\(q+n\)^z}\; .
\ea
Specializing now to $\alpha=1$ and 
dropping terms of ${\mathcal O}(\epsilon)$, we easily get 
\ba
 \tilde{I}(m=0,T,\mu,1)&=&-\frac{T^2}{24}-\frac{\mu^2}{8\pi^2}
 + \rmO(\epsilon)
 \;.
\ea
Plugging this and $I^{ }_T(0) = T^2/12$ into \eq\nr{deltafmu} produces
\ba
  \ToptVS(\Aqu,\Aqu,\Lgl) 
  &=&\frac{\dA g^2T^4}{576} 
 \bigg[ 5+\frac{18\mu^2}{(\pi T)^2}+\frac{9\mu^4}{(\pi T)^4}\bigg]
  + \rmO(\epsilon)
 \; . \la{f_fe_nlo}
\ea
From here the next-to-leading order  
contribution to the quark number susceptibility
can be extracted according to \eq\nr{chi_fe}, 
\ba
 \chi^{ }_\fe|^{ }_{\mu=0} = T^3\,\biggl(
  \frac{\Nc}{3} - \frac{\dA g^2}{16\pi^2}
 \biggr)
 \;, \la{chi_fe_nlo}
\ea
where we have added the appropriate colour factor to the leading-order term.

\subsection*{Appendix D: Cold and dense limit}

\index{Cold and dense limit}

As discussed in \ses\ref{se:IR}, \ref{se:fTg3} and \ref{se:DR_QCD}, 
at higher orders of perturbation theory one encounters
uncancelled IR divergences that necessitate the use of either 
diagrammatic resummations, or an effective theory framework for
obtaining well-defined weak-coupling results.
At high temperature, these IR problems can typically be
traced back to the static Matsubara zero mode sector of bosonic
fields, but the situation is qualitatively different in the limit of
zero (or very small) temperature, where the discrete Matsubara modes
$k^{ }_n$ merge into a continuous (Euclidean) frequency
$k^{ }_0$. This is 
an active topic of research for physical systems
characterized by sizable chemical potentials for fermionic
fields. An example is dense quark matter
possibly found inside neutron stars, where densities greatly
exceed the so-called saturation density of nuclear matter, $n^{ }_s\approx
0.16$ baryons per fm$^3$. In this appendix, 
we review tools used in higher-order computations
of thermodynamic quantities in dense $T=0$ quark matter, giving
special emphasis to how IR problems are treated in the evaluation of
its pressure. Our treatment largely follows that of ref.~\cite{tyler}.

The strict zero-temperature limit can be approached either by starting
from $T\neq 0$ and then gradually lowering the temperature, following
steps similar to those taken in sec.~\ref{se:lowT}, or by setting
$T=0$ from the outset. Choosing the latter option, we end up
considering the same vacuum, or bubble, diagrams as at non-zero temperature,
but with two crucial differences. First, in order to yield a
non-zero contribution, each Feynman graph needs to contain at least one
closed fermion loop, as purely bosonic integrals vanish in the $T=0$
limit. And second, bosonic and fermionic 
sum-integrals get replaced by continuous integrals in
$D\equiv d+1 = 4-2\epsilon$ dimensions, according to
\ba
\Tint{K} &\rightarrow& 
 \int_{-\infty}^\infty\frac{{\rm d}k_0}{2\pi}
 \int  \frac{{\rm d}^d\vec{k}}{(2\pi)^d} \; =\; 
 \int_{K}\;=\; \Lambda^{-2\epsilon}\bigg[\Lambda^{2\epsilon}\int_{K}\bigg]
  \,  \quad \mbox{(bosons)}\;, \la{T0measure1} \\
  \Tint{\{K\}} &\rightarrow& 
  \int_{-\infty+i\mu}^{\infty+i\mu}\frac{{\rm d}k_0}{2\pi}
 \int  \frac{{\rm d}^d\vec{k}}{(2\pi)^d} \; =\; 
 \int_{\widetilde{K}}\;=\; \Lambda^{-2\epsilon}
 \bigg[\Lambda^{2\epsilon}\int_{\widetilde{K}}\bigg]
  \,  \quad \mbox{(fermions)} \;, \la{T0measure2}
\ea
where $\Lambda$ is a scale parameter. 
For QCD, the chemical potentials $\mu$ corresponding to
different quark flavors are all independent if one altogether neglects
the weak interactions. In a neutron-star setting, weak interactions,
however, play an important role in enforcing the so-called
$\beta$ equilibrium, which together with local charge neutrality only
leaves one independent chemical potential. This is normally taken as
that of the down quark and denoted by 
$\mu^{ }_q\equiv \mu^{ }_d =\mu_\iB^{ }/3$,
where $\mu^{ }_\iB$ is the baryon number chemical potential.
The set of quark chemical potentials is denoted by $\{ \mu^{ }_f \}$.

For the first two orders of perturbation theory, few subtleties arise
in the determination of the pressure. In practice, one first takes
care of the temporal momentum integrations using the residue theorem,
which for the case of massless QCD yields the non-interacting-limit
result (cf.\ \eq\nr{f_ferm_expl})
\ba
 p_\rmii{QCD}^\rmii{LO}(\{\mu_f\})&=&2\Nc \sum_f 
 \int_\vec{k} (\mu^{ }_f-k)\theta(\mu^{ }_f - k)
 \;=\; \frac{\Nc}{12\pi^2} \sum_f \mu_f^4 \; .
\ea
Here the $\theta$-function
originates from taking the limit $\nF{}^{ }(\E-\mu) \stackrel{T\to
0}{\longrightarrow}\theta(\mu-\E)$. The next order of
the weak-coupling expansion (NLO) is similarly straightforward and
automatically leads to an UV and IR finite result
(cf.\ \eq\nr{f_fe_nlo} for $f=-p$).

At Next-to-Next-to-Leading Order (NNLO), or three loops in naive
perturbation theory, one finally encounters uncancelled IR
divergences, closely analogous to those discussed in
sec.~\ref{se:IR}. This time they come from one diagram, namely 
the 3-loop vacuum graph containing two closed fermion loops,
which is built by joining together two 1-loop self-energy diagrams
from \eq\nr{fe_loop}. Technically, the divergence occurs due to
the non-vanishing low-momentum limit of this self-energy, which will
be thoroughly discussed in section \ref{se:htl}. That discussion
culminates in the derivation of eqs.~(\ref{Pi_T}) and (\ref{Pi_E}),
the Euclidean versions of the so-called Hard Thermal Loop (HTL)
self-energies. These results can immediately be adopted to our $T=0$
discussion, with the only modifications being the replacement 
$k^{ }_n\to k^{ }_0$ and the use of the zero-temperature Debye mass from
\eq\nr{mmE_2}, $\mE^2=g^2/(2\pi^2)\sum_f
\mu_f^2$. Importantly, the broader HTL effective theory, also to be
discussed in section \ref{se:htl}, represents the correct IR effective
theory of QCD at zero temperature, replacing the dimensionally reduced
EQCD valid at high temperatures.

Studying the HTL self-energies, we can make an important
observation: while both components reduce to mere numbers
in the static $k^{ }_0=0$ limit, relevant for the IR physics of high
temperatures, they remain non-trivial functions of the dimensionless
variable $k^{ }_0/k$ if $k^{ }_0\sim k\sim \mE^{ }$. Given that the latter
represents the relevant IR limit at zero temperature, we realize that
the IR physics of cold and dense QCD is qualitatively different 
from hot quark-gluon
plasma. As a result, the leading
non-analytic behavior of the weak-coupling expansion of the pressure
in $\alphas$ is this time not proportional to $T^4 \alphas^{3/2}$, like at
high temperatures, but to $\mu^4 \alphas^2\ln\alphas $. In what follows, we
derive this contribution and the corresponding prefactor using
relatively simple effective theory arguments.

To obtain the leading pressure contribution from the soft momentum
scale $\mE^{ }$, one can follow two different but in principle equivalent
routes. The first, closely analogous to the corresponding discussion
in section \ref{se:IR}, involves an
explicit resummation of the IR sensitive ring diagrams of the full
theory and analyzing it in the limit of soft gluonic momentum. The
second approach involves the use of effective
theories for soft physics. In
both cases, one ends up inspecting the leading-order pressure of the
HTL theory, which takes the form 
\ba
 p_\rmii{HTL}^\rmii{LO}&=&-\frac{\dA}{2}
 \int_K\Bigl\{
  (d-1)\ln\bigl[ K^2+ \Pi^{ }_\rmii{T}(K)\bigr]
 +\ln\bigl[ K^2+\Pi^{ }_\rmii{E}(K) \bigr]
  \Bigr\} \la{eq:pHTL} \\
&=&-\frac{\dA \mE^{D}}{2}
 \int_{\hat K}  
 \Bigl\{ (
 d-1)\ln\bigl[ \hat{K}^2 + \hat{\Pi}^{ }_\rmii{T}(\phi)\bigr]
 +\ln\bigl[ \hat{K}^2+\hat{\Pi}^{ }_\rmii{E}(\phi)\bigr]
 \Bigr\} \, , \la{eq:pHTL2}
\ea
where 
the coefficients $d-1$ and $1$ correspond to the traces of the 
projectors in \eqs\nr{PT} and \nr{PE}, respectively. Otherwise
the structure is like in \eq\nr{f_gluons}, 
omitting the longitudinal part which has no HTL self-energy.   
The two self-energies can be taken from \eqs\nr{Pi_T} and \nr{Pi_E}. 
In \eq\nr{eq:pHTL2} we have scaled out the parameter
$\mE^{ }$ from the integral, 
by defining $K\equiv \mE^{ } {\hat K},\;
\Pi^{ }_i(K)\equiv \mE^2\,{\hat \Pi}_{i}(\phi)$, with $\tan\phi\equiv
k/k^{ }_0$.\footnote{To obtain ${\hat \Pi}^{ }_{i}(\phi)$
  from $\Pi^{ }_i(K)$, we first remove the overall factor
  $\mE^2$ and then write $|K|=\sqrt{k_0^2+k^2}$ everywhere. This allows us
  to write ${\hat \Pi}^{ }_{i}$ as a function of the dimensionless variable
  $k/k^{ }_0$ that can be replaced by $\tan\phi$. The resulting
  functions ${\hat \Pi}^{ }_{i}(\phi)$ can be read off from eq.~(B.28) 
  of ref.~\cite{tyler}.} 
From the latter form of the result, 
recalling $\mE^2 \propto \alphas \mu^2$, 
it becomes clear that the integral yields a contribution 
proportional to $\alpha_s^2$.

To evaluate the integral in eq.~(\ref{eq:pHTL2}), we may take
advantage of eq.~(\ref{fmdA}), specifically
\ba
 \int\! \frac{{\rm d}^D K}{(2\pi)^D}
 \ln(K^2+m^2) &=&
  -  \lim_{A\to 0} \frac{\Phi(m,D,A)}{A} \; = \; 
  - \frac{m^D_{ }\Gamma(-D/2)  }{(4\pi)^{\frac{D}{2}}}
 \;, 
\ea
which allows us to perform the radial integration in
eq.~(\ref{eq:pHTL2}), leaving over the angular one. 
Utilizing the integration measure given in
eq.~(\ref{meas2}), with $d\to D$ and $z\to\cos\phi$, 
and recalling $\Gamma(1/2) = \sqrt{\pi}$, this leads to
\ba
 p_\rmii{HTL}^\rmii{LO}&=&
 \frac{\dA\mE^{D} \,\Gamma(D/2)\Gamma(-D/2)}
      {(4\pi)^{\frac{D+1}{2}}\Gamma((D-1)/2)}
 \int_0^\pi{\rm d}\phi\,\sin^{2-2\epsilon}\!\phi 
 \Big[(d-1)\,\hat{\Pi}^{D/2}_\rmii{T}(\phi)
 +\hat{\Pi}^{D/2}_\rmii{E}(\phi)\Big]\, . \;\;\;\;
\ea
The prefactor here evaluates to 
\ba
  \frac{\dA\mE^{D} \,\Gamma(D/2)\Gamma(-D/2)}
  {(4\pi)^{\frac{D+1}{2}}\Gamma((D-1)/2)}
  &=&\frac{\dA}{32\pi^3}
 \frac{\mE^{4-2\epsilon}}{\epsilon}+\rmO(\epsilon^0)\, ,
\ea
which produces the desired logarithm of $\mE$ (and thus
of $\alphas$). This allows us to set
$\epsilon=0$ in the angular integral, which thereby becomes 
analytically calculable, giving
\ba
 \int_0^\pi{\rm d}\phi\,\sin^{2-2\epsilon}\!\phi 
 \Big[(d-1)\,\hat{\Pi}^{D/2}_\rmii{T}(\phi)+\hat{\Pi}^{D/2}_\rmii{E}(\phi)
 \Big]&=&\frac{\pi}{4}+\rmO(\epsilon)\, .
\ea

Collecting all the results and explicitly reinstating the
scale parameter $\Lambda$ in the integration measure, we obtain
\ba
  p_\rmii{HTL}^\rmii{LO}&=&
  \frac{\dA\mE^4\, \Lambda^{-2\epsilon}}{8(4\pi)^2}
  \biggl(\frac{\Lambda^2}{\mE^2}\biggr)^\epsilon_{ }
 \frac{1}{\epsilon}+\rmO(\epsilon^0)\, ,
\ea
which upon an expansion in powers of $\epsilon$ and the insertion
of $\mE^2 \propto \alphas^{ } \mu^2$
leads us 
to the final result
for the non-analytic part, 
\ba
  p_\rmii{HTL}^\rmii{LO}
 &\supset & -\frac{\dA \mE^4}{8(4\pi)^2}\ln\alphas
 + \rmO(\epsilon) \, .
\ea
The coefficient of the $\alphas^2\ln\alphas$ term 
was originally derived back in 1977 through a tedious
brute-force calculation in full QCD \cite{fmcl}. More than forty years
later, a series of systematic calculations in the HTL theory have
generalized this result to include all soft contributions up to and
including the full order $\alphas^3$ \cite{tyler}, leaving only the
so-called hard and mixed contributions missing from the 4-loop
pressure.

%

\newpage


\newpage 

\section{Real-time observables} 
\la{se:real-time}

\paragraph{Abstract:}

Various real-time correlation functions are defined
(Wightman, retarded, advanced, time-ordered, spectral). Their analytic
properties are discussed, and general relations between them are worked out 
for the case of a system in thermal equilibrium. Examples are given for 
free scalar and fermion fields. A physically 
relevant spectral function related to a composite operator
is analyzed in detail. 
The so-called real-time formalism is introduced, and it is shown how
it can be used to compute the same spectral function that was previously
determined with the imaginary-time formalism.
The need for resummations in order to systematically determine 
spectral functions in weakly coupled systems is stated. 
The concept of Hard Thermal Loops (HTLs), which implement a particular
resummation, is introduced. HTL-resummed gauge field
and fermion propagators are derived. The main plasma physics 
phenomena that the HTL resummation captures are pointed out. 
A warning is issued that although necessary,  
HTL resummation is in general 
not sufficient for obtaining a systematic weak-coupling expansion. 

\paragraph{Keywords:} 

Wick rotation, time ordering, Heisenberg operator, Wightman function,  
retarded and advanced correlators, 
Kubo-Martin-Schwinger relation, spectral representation, sum rule, 
analytic continuation, density matrix, 
Schwinger-Keldysh formalism, Hard Thermal Loops, 
Landau damping, plasmon, plasmino, dispersion relation.

\index{Green's functions: time orderings}
\index{Real-time observables}
\index{Wick rotation}
\index{Heisenberg-operator: bosonic}


\subsection{Different Green's functions}
\la{se:diff_G}

We now move to a new class of observables including both 
a Minkowskian time $t$ and a temperature $T$. Examples are production
rates of weakly interacting 
particles from a thermal plasma; 
oscillation and damping rates of long-wavelength fields in 
a plasma; as well as transport coefficients of a plasma such as
its electric
and thermal conductivities and bulk and shear viscosities. 
We start by developing some aspects of the general formalism, and return 
to specific applications later on. Let us stress 
that we do remain in thermal 
equilibrium in the following,  
even though some of the results also 
apply to an off-equilibrium ensemble.

Many observables of interest 
can be reduced to {\em 2-point correlation functions} of elementary
or composite operators. Let us therefore list 
some common definitions and relations 
that apply to such correlation 
functions~\cite{old4}--\hspace*{-1.1mm}\cite{old3}.

We denote Minkowskian spacetime coordinates by 
$\mathcal{X}=(t,x^i)$ and
momenta by $\mathcal{K} = (k^0_{ },k^i)$, 
whereas their Euclidean counterparts
are denoted by $X = (\tau,x^i)$, 
$ K = (k^{ }_n,k^{ }_i)$. Wick rotation is carried 
out by $\tau \leftrightarrow i t$, 
$k^{ }_n \leftrightarrow - i k^0_{ }$. Scalar products are defined as
$
 \mathcal{K}\cdot \mathcal{X} = k^{ }_0 t + k^{ }_i x^i =
 k^0_{ } t - \vec{k}\cdot \vec{x}
$, 
$
  K \cdot  X = k_n^{ } \tau + k^{ }_i x^i = 
 k_n^{ } \tau  - \vec{k}\cdot \vec{x}
$.
Arguments of operators
denote implicitly whether we are in Minkowskian or Euclidean spacetime.
In particular, Heisenberg-operators are defined as
\be
 \hat O(t,\vec{x}) \,\equiv\, 
 e^{i \hat H t} \, \hat O(0,\vec{x}) \, e^{- i \hat H t}
 \;, \quad
 \hat O(\tau,\vec{x}) \,\equiv\, 
 e^{\hat H \tau} \, \hat O(0,\vec{x}) \, e^{- \hat H \tau}
 \;. \la{op_defs}
\ee
The thermal ensemble is normally defined by the density matrix
$\hat\rho = \mathcal{Z}^{-1} \exp(-\beta \hat H)$, even though it is
also possible to include a chemical potential, as will be done 
in \eq\nr{rho_mu_T}. Expectation values of (products of) 
operators are defined through 
$\langle \cdots \rangle \equiv \tr \big[\hat\rho\,(\cdots)\big]$. 

\newcommand{\A}{\hat\phi^{ }_\alpha}
\renewcommand{\B}{\hat{\phi}^\dagger_\beta}
\newcommand{\I}{\int_\mathcal{X} \,e^{i \mathcal{K}\cdot\mathcal{X}}}

\subsection*{Bosonic case}

We start by considering operators that are {\em bosonic} in nature, 
i.e.\ commuting (modulo possible contact terms).
We denote the operators 
by $\A$, $\B$. These may be 
either elementary fields or composite operators built from them.
In order to simplify the notation, functions and their Fourier transforms
are to be recognized through the argument, $\mathcal{X}$ vs.\
$\mathcal{K}$.

\index{Spectral function: bosonic} \index{Retarded correlator: bosonic}

We can define various classes of 
correlation functions. ``Physical'' 
correlators are defined as
\ba \index{Wightman function}
  \Pi^{>}_{\alpha\beta}(\mathcal{K}) & \equiv & 
 \I \Bigl\langle \A(\mathcal{X})\, \B(0) \Bigl\rangle
 \;,   
 \la{bL}
 \\
  \Pi^{<}_{\alpha\beta}(\mathcal{K}) & \equiv & 
 \I \Bigl\langle \B(0)\,  \A(\mathcal{X}) \Bigl\rangle
 \;,   
 \la{bS}
 \\
  \rho^{ }_{\alpha\beta}(\mathcal{K}) & \equiv & 
 \I \Bigl\langle \fr12 \Bigl[ \A(\mathcal{X}) , \B(0) \Bigr] \Bigl\rangle
 \;,   
 \la{brho}
 \\
  \Delta^{ }_{\alpha\beta}(\mathcal{K}) & \equiv & 
 \I \Bigl\langle \fr12 \Bigl\{ \A(\mathcal{X}) , \B(0) \Bigr\} \Bigl\rangle
 \;,   
 \la{bdelta}
\ea 
where $ \Pi^{>}_{ }$ and $ \Pi^{<}_{ }$ are called Wightman
functions and $ \rho_{ }$ the {\em spectral function}, 
whereas $\Delta_{ }$ is sometimes referred 
to as the statistical correlator.
We are implicitly assuming the presence of an UV regulator so that there
are no short-distance singularities in the Fourier transforms.

The ``retarded''/``advanced'' correlators can be defined as 
\ba
  \Pi^\iR_{\alpha\beta}(\mathcal{K}) & \equiv & 
 i \I \Bigl\langle 
 \Bigl[ \A(\mathcal{X}) , \B(0) \Bigr] \theta(t) \Bigl\rangle
 \;, 
 \la{bR}
 \\
  \Pi^\iA_{\alpha\beta}(\mathcal{K}) & \equiv & 
 i \I \Bigl\langle  - \Bigl[ \A(\mathcal{X}) , \B(0) \Bigr] \theta(-t)
 \Bigl\rangle
 \;. 
 \la{bA} 
\ea
Note that since $ \Pi^\iR$ involves positive times only, 
$
 e^{i k^0_{ } t} = e^{i [\re k^0_{ } + i \im k^0_{ } ] t} 
 = e^{i \re k^0_{ } t} e^{- \im k^0_{ } t}
$  
is exponentially suppressed for $\im k^0_{ } > 0$. 
Therefore $ \Pi^\iR$ can be considered an analytic function 
of $k^0_{ }$ in the upper half of the complex $k^0_{ }$-plane (it can 
develop distribution-like singularities at the physical boundary 
$\im k^0_{ } \to 0^+_{ }$). Similarly, $ \Pi^\iA$ is an analytic function 
in the lower half of the complex $k^0_{ }$-plane. 
These turn out to be strong and useful properties, 
and do not apply to general correlation functions. 

On the other hand, from the computational point of view one 
is often faced with ``time-ordered'' correlators, 
\ba \index{Time-ordered correlator: bosonic}
  \Pi^\iT_{\alpha\beta}(\mathcal{K}) & \equiv & 
  \I \Bigl\langle \A(\mathcal{X})\,  \B(0) \, \theta(t)
                 +\B(0)\,  \A(\mathcal{X}) \, \theta(-t) \Bigl\rangle
 \;, \la{bT}
\ea\index{Euclidean correlator: bosonic}%
which appear in time-dependent perturbation theory at zero temperature, 
or with the ``Euclidean'' correlator
\ba
  \Pi^\iE_{\alpha\beta}( K) & \equiv & 
 \int_X
  e^{i  K \cdot  X}
  \Bigl\langle \A( X)\,  \B(0) \Bigl\rangle
 \;,
 \la{bE}
\ea
which appears in non-perturbative formulations.
Restricting to $0 \le \tau \le \beta$, 
the Euclidean correlator is also time-ordered, 
and can be computed with standard imaginary-time functional integrals. 
If the correlator is periodic (cf.\ text below \eq\nr{b_kms}), 
then $k^{ }_n$ is a {\em bosonic} Matsubara frequency. 

It follows from \eq\nr{op_defs}, by using the cyclicity 
of the trace, that
\be
  \bigl\langle \A(t-i\beta,\vec{x})\,\B(0,\vec{0}) \bigr\rangle 
 = \frac{1}{\mathcal{Z}} \tr
 \Bigl[ 
  e^{-\beta \hat H} e^{\beta\hat H}
 \A(t,\vec{x}) e^{-\beta\hat H}\B(0,\vec{0})
 \Bigr]
 = 
  \bigl\langle \B(0,\vec{0})\, \A(t,\vec{x}) \bigr\rangle 
 \;. \la{b_kms}
\ee 
This is a configuration-space version of the so-called 
Kubo-Martin-Schwinger (KMS) relation, 
which relates $\Pi^{>}_{\alpha\beta}$ and $\Pi^{<}_{\alpha\beta}$
to each other, provided that we are in thermal equilibrium. If we set
$t\to 0$ and keep $\vec{x}\neq \vec{0}$, then  
$\A(0,\vec{x})$ and $\B(0,\vec{0})$ commute with each other. In this case, 
the KMS relation implies that the integrand in \eq\nr{bE} is 
a periodic function of $\tau$, with periodicity defined in the 
same sense as around \eq\nr{Fousum}.

\index{Kubo-Martin-Schwinger: bosonic}

It turns out that
{\em all} of the correlation functions defined  
can be related to each other in thermal equilibrium.
In particular, all correlators can be expressed in terms of the spectral
function, which in turn can be determined as a certain analytic 
continuation of the Euclidean correlator. In order to show this, 
we may first insert sets of energy eigenstates into the definitions
of $ \Pi^{>}_{\alpha\beta}$ and $ \Pi^{<}_{\alpha\beta}$: 
\ba
  \Pi^{>}_{\alpha\beta}(\mathcal{K}) & = & 
 \frac{1}{\mathcal{Z}}
 \I \tr \Bigl[ e^{-\beta\hat H + i \hat H t} 
  \!\!\underbrace{\unit}_{\sum_m |m\rangle\,\langle m|}\!\! \A(0,\vec{x}) \, 
  e^{-i\hat H t} 
  \!\!\underbrace{\unit}_{\sum_n |n\rangle\,\langle n|}\!\! 
 \B(0,\vec{0}) \Bigl]
 \nn & = & 
 \frac{1}{\mathcal{Z}}
 \sum_{m,n} \I e^{(-\beta+it)\E^{ }_m}e^{-it\E^{ }_n}
 \langle m | \A(0,\vec{x}) | n \rangle \, \langle n | \B(0,\vec{0}) | m \rangle
 \nn & = & 
 \frac{1}{\mathcal{Z}}
 \int_\vec{x} e^{-i\vec{k}\cdot\vec{x}}
 \sum_{m,n} e^{-\beta \E^{ }_m} \, 2\pi
 \, \delta(k^0_{ } + \E^{ }_m - \E^{ }_n)
 \langle m | \A(0,\vec{x}) | n \rangle \, \langle n | \B(0,\vec{0}) | m \rangle
 \;, \hspace*{0.5cm} \la{der_kms_1} \\  
  \Pi^{<}_{\alpha\beta}(\mathcal{K}) & = & 
 \frac{1}{\mathcal{Z}}
 \I \tr \Bigl[ e^{-\beta\hat H} 
 \!\! \underbrace{\unit}_{\sum_n |n\rangle\,\langle n|}\!\! \B(0,\vec{0}) \,
  e^{i \hat H t} 
  \!\! \underbrace{\unit}_{\sum_m |m\rangle\,\langle m|}\!\! \A(0,\vec{x}) \,
  e^{-i\hat H t} 
   \Bigl]
 \nn & = & 
 \frac{1}{\mathcal{Z}}
 \sum_{m,n} \I e^{(-\beta-it)\E^{ }_n}e^{it\E^{ }_m}
 \langle n | \B(0,\vec{0}) | m \rangle \, 
 \langle m | \A(0,\vec{x}) | n \rangle  
 \nn 
 & = & 
 \frac{1}{\mathcal{Z}}
 \int_\vec{x} e^{-i\vec{k}\cdot\vec{x}}
 \sum_{m,n} e^{-\beta \E^{ }_n} \, 2\pi \, 
 \underbrace{\delta(k^0_{ } + \E^{ }_m - \E^{ }_n)}_
            {\E^{ }_n = \E^{ }_m + k^0_{ }}
 \langle m | \A(0,\vec{x}) | n \rangle \, \langle n | \B(0,\vec{0}) | m \rangle
 \nn & = &
 e^{-\beta k^0_{ }} \, \Pi^{>}_{\alpha\beta}(\mathcal{K})
 \;. \hspace*{0.5cm}   \la{der_kms_2}
\ea
This is a Fourier-space version of the KMS relation. 
Consequently 
\be
  \rho^{ }_{\alpha\beta}(\mathcal{K})
 \; = \; \fr12 [ \Pi^{>}_{\alpha\beta}(\mathcal{K})
  -  \Pi^{<}_{\alpha\beta}(\mathcal{K})]
 \; = \; \fr12 (e^{\beta k^0_{ }} - 1) 
  \Pi^{<}_{\alpha\beta}(\mathcal{K}) \la{rhorel}
\ee
and, conversely, 
\ba
  \Pi^{<}_{\alpha\beta}(\mathcal{K}) 
 & = & 2 \nB{}(k^0_{ })  \rho^{ }_{\alpha\beta}(\mathcal{K})
 \;, \la{bLSrel0} \\
  \Pi^{>}_{\alpha\beta}(\mathcal{K}) & = & 
 2 \frac{e^{\beta k^0_{ }}}{e^{\beta k^0_{ }} - 1}
  \rho^{ }_{\alpha\beta}(\mathcal{K})
 \; = \; 2[1 + \nB{}(k^0_{ })] \,  \rho^{ }_{\alpha\beta}(\mathcal{K})
 \;, \la{bLSrel}
\ea
where $\nB{}(k^0_{ })\equiv 1/[\exp(\beta k^0_{ }) - 1]$ is the Bose
distribution. Moreover, 
\be
  \Delta^{ }_{\alpha\beta}(\mathcal{K}) = 
 \fr12 \bigl[ \Pi^{>}_{\alpha\beta}(\mathcal{K}) +
  \Pi^{<}_{\alpha\beta}(\mathcal{K}) \bigr] 
 \; = \; 
 \bigl[1 + 2 \nB{}(k^0_{ })\bigr]\, \rho^{ }_{\alpha\beta}(\mathcal{K})
 \;. \la{bDelta}
\ee
Note that $1+ 2 \nB{}(-k^0_{ }) = - [1 + 2 \nB{}(k^0_{ })]$, so that 
if $\rho$ is odd in $\mathcal{K}\to -\mathcal{K}$, 
then $\Delta$ is even. 

Inserting the representation
\be
 \theta(t) = i \int_{-\infty}^{\infty} \! \frac{{\rm d}\omega}{2\pi}
 \frac{e^{-i\omega t}}{\omega + i 0^+_{ }}
 \la{theta}
\ee
into the definitions of $ \Pi^\iR$, $ \Pi^\iA$, 
in which the commutator is represented as an inverse transformation
of \eq\nr{brho}, we obtain
\ba
 \Pi^\iR_{\alpha\beta}(\mathcal{K}) 
 & = &
 i \I \; 2 \theta(t) \int_\mathcal{P} \, 
 e^{-i \mathcal{P}\cdot \mathcal{X}} \rho^{ }_{\alpha\beta}(\mathcal{P})
 \nn 
 & = &
 - 2  \int \! {\rm d}t \int \! \frac{{\rm d}\omega}{2\pi} 
 \int \! \frac{{\rm d} p^0_{ }}{2\pi} 
 \frac{e^{i (k^0_{ } - p^0_{ } -\omega ) t}}{\omega + i 0^+_{ }} 
 \,  \rho^{ }_{\alpha\beta}(p^0_{ },\vec{k})
 \nn 
 & = &
 - 2 \int \! \frac{{\rm d}\omega}{2\pi} 
 \int \! \frac{{\rm d} p^0_{ }}{2\pi} 
 \frac{2\pi \delta(k^0_{ } - p^0_{ } -\omega )}{\omega + i 0^+_{ }}
 \,  \rho^{ }_{\alpha\beta}(p^0_{ },\vec{k})
 \nn 
 & = &  \int_{-\infty}^{\infty} \! \frac{{\rm d} p^0_{ }}{\pi} 
 \frac{ \rho^{ }_{\alpha\beta}(p^0_{ },\vec{k})}{p^0_{ } -k^0_{ }- i 0^+_{ }}
 \;, \la{PiR_rho}
\ea
and similarly
\be
  \Pi^\iA_{\alpha\beta}(\mathcal{K})
 =  \int_{-\infty}^{\infty} \! \frac{{\rm d} p^0_{ } }{\pi} 
 \frac{ \rho^{ }_{\alpha\beta}(p^0_{ },\vec{k})}{p^0_{ } -k^0_{ }+ i 0^+_{ }}
 \;. 
\ee
Note that these can be considered to be limiting values from the upper
half-plane for $ \Pi^\iR$ 
(since it is the combination $k^0_{ } + i 0^+_{ }$
that appears in the kernel) 
and from the lower
half-plane for $ \Pi^\iA$ 
(since it is the combination $k^0_{ } - i 0^+_{ }$
that appears). 

Making use of 
\be
 \frac{1}{\Delta \pm i0^+_{ }} = \mathbbm{P}\Bigl(\frac{1}{\Delta}\Bigr) 
 \mp i \pi \delta(\Delta)
 \;,  \la{delta}
\ee
and assuming that $ \rho^{ }_{\alpha\beta}$ is real, 
we find
\be
 \im  \Pi^\iR_{\alpha\beta} (\mathcal{K}) 
   = \rho^{ }_{\alpha\beta}(\mathcal{K}) 
 \;, \quad
 \im  \Pi^\iA_{\alpha\beta} (\mathcal{K}) = 
  -  \rho^{ }_{\alpha\beta}(\mathcal{K}) 
 \;.  \la{PiR_rho_rel}
\ee
Furthermore, the real parts of $\Pi^\iR$ and $\Pi^\iA$ agree, so that 
$
 -i [\Pi^\iR_{\alpha\beta}  - \Pi^\iA_{\alpha\beta} ]
 = 2 \rho^{ }_{\alpha\beta} 
$. 
 
\index{Fluctuation-dissipation theorem}

We note in passing that \eqs\nr{bDelta} and \nr{PiR_rho_rel} can be
combined into 
\be
 \fbox{$\displaystyle
  \Delta^{ }_{\alpha\beta}(\mathcal{K}) = 
  \bigl[1 + 2 \nB{}(k^0_{ })\bigr]
  \,
  \im  \Pi^\iR_{\alpha\beta} (\mathcal{K}) 
 $}
 \;. \la{f-t}
\ee
This important equality is sometimes referred to as the 
{\em fluctuation-dissipation theorem}; the physical 
reason for this nomenclature will be discussed 
in \se\ref{se:transport}.

Moving on to $\Pi^\iT_{\alpha\beta}$
and making use 
of \eqs\nr{bLSrel0} and \nr{bLSrel} as well as of \eq\nr{theta}, 
we find
\ba
 \Pi^\iT_{\alpha\beta}(\mathcal{K})
 & = &
 \I \; \int_\mathcal{P} \, 
 e^{-i \mathcal{P}\cdot \mathcal{X}}
 \Bigl[ \theta(t) 2 e^{\beta p^0_{ }}\! \nB{}(p^0_{ })
 + \theta(-t)2 \nB{}(p^0_{ }) \Bigr]
 \rho^{ }_{\alpha\beta}(\mathcal{P})
 \nn 
 & = &
 2 i  \int \! {\rm d}t \int \! \frac{{\rm d}\omega}{2\pi} 
 \int \! \frac{{\rm d} p^0_{ }}{2\pi} \biggl[ 
 \frac{e^{i (k^0_{ } - p^0_{ } -\omega ) t}}
 {\omega + i 0^+_{ }} e^{\beta p^0_{ }}
 +  
 \frac{e^{i (k^0_{ } - p^0_{ } + \omega ) t}}{\omega + i 0^+_{ }}
 \biggr] \nB{}(p^0_{ })
 \rho^{ }_{\alpha\beta}(p^0_{ },\vec{k})
 \nn 
 & = &
 2 i \int \! \frac{{\rm d}\omega}{2\pi} 
 \int \! \frac{{\rm d} p^0_{ }}{2\pi} \biggl[ 
 \frac{2\pi \delta(k^0_{ } - p^0_{ } -\omega )}
 {\omega + i 0^+_{ }} e^{\beta p^0_{ }}
 + 
 \frac{2\pi \delta(k^0_{ } - p^0_{ } +\omega )}{\omega + i 0^+_{ }}
 \biggr] \nB{}(p^0_{ }) 
 \rho^{ }_{\alpha\beta}(p^0_{ },\vec{k})
 \nn 
 & = &
 i 
 \int \! \frac{{\rm d} p^0_{ }}{\pi} \biggl[ 
 \frac{e^{\beta p^0_{ }} }{k^0_{ } - p^0_{ } + i 0^+_{ }} 
 - 
 \frac{1}{k^0_{ } - p^0_{ } - i 0^+_{ }}
 \biggr] \nB{}(p^0_{ }) 
 \rho^{ }_{\alpha\beta}(p^0_{ },\vec{k})
 \nn 
 & = &
 \int_{-\infty}^{\infty} \! \frac{{\rm d} p^0_{ }}{\pi} 
 \frac{i\rho^{ }_{\alpha\beta}(p^0_{ },\vec{k})}
 {k^0_{ } - p^0_{ } + i 0^+_{ }} + 
 2 \rho^{ }_{\alpha\beta}(k^0_{ },\vec{k}) \nB{}(k^0_{ })
 \nn & = & 
 - i \Pi^\iR_{\alpha\beta}(\mathcal{K}) + 
 \Pi^{<}_{\alpha\beta}(\mathcal{K})
 \;, \la{bTrho_rel}
\ea
where in the penultimate step we inserted 
the identity
$ 
 \nB{}(p^0_{ }) e^{\beta p^0_{ }} = 1 + \nB{}(p^0_{ })
$
as well as 
\eq\nr{delta}.
Note that \eq\nr{bTrho_rel} can be obtained also directly from 
the definitions in \eqs\nr{bS}, \nr{bR} and \nr{bT}, by inserting 
$1 = \theta(t) + \theta(-t)$ into \eq\nr{bS}. It can similarly
be seen that 
$
  \Pi^\iT_{\alpha\beta} = 
 - i \Pi^\iA_{\alpha\beta} + 
 \Pi^{>}_{\alpha\beta}
$. 

We note that both sums on the second row 
of \eq\nr{der_kms_1} are exponentially 
convergent for $0 < it < \beta$. Therefore we can formally 
relate the two functions 
\be
 \Bigl\langle \A(\mathcal{X})\, \B(0) \Bigl\rangle
 \quad
 \mbox{and}  
 \quad 
 \Bigl\langle \A( X)\,  \B(0) \Bigl\rangle
\ee
by a direct analytic continuation $t\to - i \tau$, or $it\to \tau$, 
with $0 < \tau < \beta$. Thereby
\ba
  \Pi^\iE_{\alpha\beta}( K) & = & 
 \int_X \, 
  e^{i  K \cdot  X}
 \biggl[ \int_\mathcal{P} 
  e^{- i \mathcal{P} \cdot \mathcal{X}} 
 \Pi^{>}_{\alpha\beta}(\mathcal{P}) \biggr]^{ }_{it\to\tau}
 \nn & = & 
 \int_0^\beta \! {\rm d}\tau\, e^{i k^{ }_n \tau}
 \int_{-\infty}^{\infty} \frac{{\rm d}p^0_{ } }{2\pi} e^{-p^0_{ } \tau} 
 \, \Pi^{>}_{\alpha\beta}(p^0_{ },\vec{k})
 \nn & = & 
 \int_0^\beta \! {\rm d}\tau\, e^{i k^{ }_n \tau}
 \int_{-\infty}^{\infty} \frac{{\rm d}p^0_{ } }{2\pi} e^{-p^0_{ } \tau} 
 \frac{2 e^{\beta p^0_{ }}}{e^{\beta p^0_{ }} -1} 
 \, \rho^{ }_{\alpha\beta}(p^0_{ },\vec{k})
 \nn & = & 
 \int_{-\infty}^{\infty} \frac{{\rm d}p^0_{ } }{\pi}
 \frac{\rho^{ }_{\alpha\beta}(p^0_{ },\vec{k})}{1 - e^{-\beta p^0_{ }}} 
 \biggl[ 
   \frac{e^{(i k^{ }_n - p^0_{ })\tau}}{i k^{ }_n - p^0_{ }}
 \biggr]^{\beta}_{0}
 \nn & = & 
 \int_{-\infty}^{\infty} \frac{{\rm d}p^0_{ } }{\pi}
 \frac{\rho^{ }_{\alpha\beta}(p^0_{ },\vec{k})}{1 - e^{-\beta p^0_{ }}} 
   \frac{e^{- \beta p^0_{ }} - 1}{i k^{ }_n - p^0_{ }}
 \nn & \stackrel{p^0_{ }\to k^0_{ }}{=} &
 \int_{-\infty}^{\infty} \! \frac{{\rm d} k^0_{ } }{\pi} 
 \frac{ \rho^{ }_{\alpha\beta}(k^0_{ },\vec{k})}{k^0_{ } - i k^{ }_n}
 \;, \la{bErhorel} \la{spectral}
\ea
where we inserted \eq\nr{bLSrel} for $\Pi^{>}(\mathcal{K})$,
and changed orders of integration.  
This relation is called the {\em spectral representation}
of the Euclidean correlator.\footnote{%
 It is more difficult but not impossible 
 to find spectral representations 
 for higher-point functions, 
 cf.\ appendix~A of ref.~\cite{3pt1} for the 3-point function,
 and ref.~\cite{3pt2} for a general discussion.
 }

\index{Spectral representation}
\index{Sum rule}

It is useful to note that \eq\nr{spectral} 
implies the existence of a simple ``sum rule'': 
\be
 \int_{-\infty}^{\infty} \! \frac{{\rm d} k^0_{ }}{\pi}
 \frac{ \rho^{ }_{\alpha\beta}(k^0_{ },\vec{k})}{k^0_{ }}
 = 
 \int_0^\beta \! {\rm d}\tau \, \Pi^\iE_{\alpha\beta}(\tau,\vec{k}) 
 \;. \la{sum_rule}
\ee
Here we set $k^{ }_n = 0$ and used the definition in \eq\nr{bE} on the 
left-hand side of \eq\nr{spectral}. The usefulness of the sum rule
is that it relates integrals over Minkowskian and Euclidean
correlators to each other. (Of course, we have implicitly assumed
that both sides are integrable which, as already alluded to,  
necessitates a suitable ultraviolet 
regularization in the spatial directions.)

Finally, the spectral representation in \eq\nr{spectral} can be 
inverted by making use of \eq\nr{delta}, 
\ba \index{Analytic continuation}
 \rho^{ }_{\alpha\beta}(\mathcal{K}) 
 & = &  \frac{1}{2i}
 \disc \Pi^\iE_{\alpha\beta}(k^{ }_n\to -i k^0_{ },\vec{k})
 \la{ancont} \\ 
 & \equiv &
 \frac{1}{2i} 
 \Bigl[ 
   \Pi^\iE_{\alpha\beta}(-i [k^0_{ } + i 0^+_{ }],\vec{k})
 -   \Pi^\iE_{\alpha\beta}(-i [k^0_{ } - i 0^+_{ }],\vec{k})
 \Bigr]  
 \;. \la{Discdef}
\ea
Furthermore, a comparison of \eqs\nr{PiR_rho} and \nr{spectral}
shows that 
\be
  \Pi^\iR_{\alpha\beta}(\mathcal{K}) = 
  \Pi^\iE_{\alpha\beta}(k^{ }_n\to -i [k^0_{ }+i0^+_{ }],\vec{k})
 \;. \la{PiR_PiE}
\ee
This last relation, which 
can be justified also through a more rigorous mathematical 
analysis~\cite{cuniberti},
captures the essence of the analytic continuation from the 
imaginary-time (Matsubara) formalism 
to physical Minkowskian spacetime.\footnote{%
 The more general function 
 $
   \Pi^\iE_{\alpha\beta}(k^{ }_n\to -i z,\vec{k})
  = 
  \int_{-\infty}^{\infty} \! \frac{{\rm d} k^0_{ } }{\pi} 
  \frac{ \rho^{ }_{\alpha\beta}(k^0_{ },\vec{k})}{k^0_{ } - z}
 $, 
 $ z \in \mathbbm{C}$, 
 is often referred to as the ``resolvent''. \index{Resolvent}   
 } 

In the context of the spectral representation, \eq\nr{spectral}, 
it will often be useful to note from \eq\nr{Gtau}, {\it viz.}\ 
\be
  T \sum_{\omega^{ }_n} \frac{e^{i\omega^{ }_n \tau}}
 {\omega_n^2 + \omega^2}
 = \frac{\nB{}(\omega)}{2 \omega} 
 \Bigl[
   e^{(\beta-\tau)\omega} + e^{\tau\omega}
 \Bigr]
 \;, \la{Gebo_new}
\ee 
that, for $0 <  \tau < \beta$,  
\index{Thermal sums: non-relativistic boson}
\ba
 T \sum_{\omega^{ }_n} 
 \frac{1}
 {k^0_{ } - i \omega^{ }_n} e^{i \omega^{ }_n \tau}
 & = & 
 T \sum_{\omega^{ }_n} 
 \frac{i \omega^{ }_n + k^0_{ }}
 {\omega_n^2 + (k^0_{ })^2} e^{i \omega^{ }_n \tau}
 \nn & = & 
 (\partial^{ }_\tau + k^0_{ }) 
  T \sum_{\omega^{ }_n} 
 \frac{e^{i \omega^{ }_n \tau}}
 {\omega_n^2 + (k^0_{ })^2} 
 \nn
 & = & 
 \frac{\nB{}(k^0_{ })}{2\, k^0_{ }} \Bigl[ 
  (-k^0_{ } + k^0_{ }) e^{(\beta - \tau) k^0_{ }}
 + (k^0_{ } + k^0_{ }) e ^{\tau k^0_{ }}
 \Bigr] 
 \nn[3mm] & = & 
 \nB{}(k^0_{ }) e^{\tau k^0_{ }} \;.
 \la{bsum}
\ea
This relation turns out to be valid both for $k^0_{ } <  0$
and $k^0_{ } > 0$ (to show this, 
substitute $\omega^{ }_n \to -\omega^{ }_n $ and use \eq\nr{bsum2}). 
We also note that, again for $0 <  \tau < \beta$, 
\ba
 T \sum_{\omega^{ }_n } 
 \frac{1}
 {k^0_{ } - i \omega^{ }_n } e^{-i \omega^{ }_n \tau}
 & = & 
 T \sum_{\omega^{ }_n } 
 \frac{1}
 {k^0_{ } - i \omega^{ }_n } e^{i \omega^{ }_n (\beta- \tau)}
  =  
 \nB{}(k^0_{ }) e^{(\beta - \tau) k^0_{ }} \;.
 \la{bsum2}
\ea
In particular, 
taking the inverse Fourier transform 
($T\sum_{k^{ }_n} e^{-i k^{ }_n\tau}$)
from the left-hand side of \eq\nr{spectral}, 
and employing \eq\nr{bsum2}, we get the relation 
\ba
  & & \hspace*{-1.5cm}
 \int_\vec{x}\,
  e^{- i \vec{k}  \cdot \vec{x} }
 \Bigl\langle \A(\tau,\vec{x})\,  \B(0,\vec{0}) \Bigl\rangle
 \nn
 \!\! & = & \!\!\! 
 \int_{-\infty}^{\infty} \! \frac{{\rm d} k^0_{ } }{\pi} 
 \rho^{ }_{\alpha\beta}(\mathcal{K})
 \, \nB{}(k^0_{ }) e^{(\beta - \tau) k^0_{ }}
 \nn \!\! & = & \!\!\! 
 \int_{0}^{\infty} \! \frac{{\rm d} k^0_{ } }{\pi} 
  \left\{
     \frac{\rho^{ }_{\alpha\beta}(k^0_{ },\vec{k}) + 
           \rho^{ }_{\alpha\beta}(-k^0_{ },\vec{k})}{2}
     \frac{\sinh\Bigl[ \Bigl( \frac{\beta}{2} - \tau \Bigr) k^0_{ } \Bigr]}
          {\sinh\Bigl( \frac{\beta}{2} k^0_{ } \Bigr)} \right.
 \nn & & \hspace*{1.0cm}
    + \, \left.
     \frac{\rho^{ }_{\alpha\beta}(k^0_{ },\vec{k}) - 
           \rho^{ }_{\alpha\beta}(-k^0_{ },\vec{k})}{2}
     \frac{\cosh\Bigl[ \Bigl( \frac{\beta}{2} - \tau \Bigr) k^0_{ } \Bigr]}
          {\sinh\Bigl( \frac{\beta}{2} k^0_{ } \Bigr)}
  \right\}
 \;, \la{latt_relation}
\ea
where we symmetrized and anti-symmetrized the ``kernel''
$\nB{}(k^0_{ }) e^{(\beta - \tau) k^0_{ }}$ with respect to $k^0_{ }$.
Normally (when $\A$ and $\B$ are identical) the spectral function
is antisymmetric in $k^0_{ }\to -k^0_{ }$, and only the second term
on the last line of \eq\nr{latt_relation} contributes. Thereby 
we obtain a useful identity: 
if the left-hand side of \eq\nr{latt_relation} can be measured 
non-perturbatively on a Euclidean lattice with Monte Carlo simulations
as a function of $\tau$, then an ``inversion'' 
of \eq\nr{latt_relation} could lead to a non-perturbative 
estimate of the Minkowskian spectral function. Issues related
to this inversion are discussed in ref.~\cite{rev5_2}.

\subsection*{Example: free boson}

Let us illustrate the  relations obtained 
with the example of a free propagator in scalar field theory: 
\ba
  \Pi^\iE(K) & = & \frac{1}{k_n^2 + \E_{k}^2}
 \; = \; 
 \frac{1}{2 \E^{ }_{k}}
 \biggl( 
  \frac{1}{i k^{ }_n + \E^{ }_{k}} + 
  \frac{1}{- i k^{ }_n + \E^{ }_{k}} 
 \biggr)
 \;, 
\ea
where $\E^{ }_{k} = \sqrt{{k}^2 + m^2}$.
According to \eq\nr{PiR_PiE}, 
\ba
  \Pi^\iR(\mathcal{K}) & = &
  \frac{1}{-(k^0_{ }+ i 0^+_{ })^2 + \E_{k}^2 }
 \nn 
 & = &
 - \frac{1}{\mathcal{K}^2 - m^2 + i \sign(k^0_{ }) 0^+_{ }}
 \nn 
 & = &
 -   \mathbbm{P}\,\biggl(\frac{1}{(k^0_{ })^2 - \E_{k}^2}\biggr)
 + \frac{i \pi}{2 \E^{ }_{k}} 
 \Bigl[\delta(k^0_{ } - \E^{ }_{k}) - \delta(k^0_{ } + \E^{ }_{k}) \Bigr] 
 \;, 
\ea
and according to \eq\nr{PiR_rho_rel}, 
\ba
 \rho(\mathcal{K}) & = &
 \frac{\pi}{2 \E^{ }_{k}}
 \Bigl[ \delta(k^0_{ } - \E^{ }_{k}) - 
 \delta(k^0_{ } + \E^{ }_{k})
 \Bigr]
 \;. \la{free_S_rho}
\ea
Finally, according to \eqs\nr{bLSrel0} and \nr{bTrho_rel}, 
\index{Time-ordered propagator: free boson}
\ba
 \Pi^\iT(\mathcal{K}) & = &   
  \mathbbm{P}\,\biggl(\frac{i}{(k^0_{ })^2 - \E_{k}^2}\biggr) + 
 \frac{\pi}{2 \E^{ }_{k}}
 \Bigl\{ \delta(k^0_{ } - \E^{ }_{k}) \bigl[1 + 2 \nB{}(k^0_{ })\bigr]
  - 
 \delta(k^0_{ } + \E^{ }_{k}) \bigl[1 + 2 \nB{}(k^0_{ })\bigr]
 \Bigr\}
  \nn & = &
  \mathbbm{P}\,\biggl(\frac{i}{(k^0_{ })^2 - \E_{k}^2}\biggr) + 
 \frac{\pi}{2 \E^{ }_{k}}
 \Bigl[ \delta(k^0_{ } - \E^{ }_{k}) 
  +
 \delta(k^0_{ } + \E^{ }_{k}) \Bigr] \bigl[1 + 2 \nB{}(|k^0_{ }|)\bigr]
 \nn & = & 
  \mathbbm{P}\,\biggl(\frac{i}{(k^0_{ })^2 - \E_{k}^2}\biggr) + 
  {\pi}
  \delta\Bigl( (k^0_{ })^2 - \E_{k}^2 \Bigr) \bigl[1 + 2 \nB{}(|k^0_{ }|)\bigr]
 \nn & = & 
  \frac{i}{(k^0_{ })^2 - \E_{k}^2 + i 0^+_{ }} + 
  {2 \pi}
  \delta\Bigl( (k^0_{ })^2 - \E_{k}^2 \Bigr) \nB{}(|k^0_{ }|)
 \nn & = & 
 \frac{i}{\mathcal{K}^2-m^2 + i 0^+_{ }} + 2\pi\, \delta(\mathcal{K}^2 - m^2)
 \,\nB{}(|k^0_{ }|)
 \;, \la{PiT_free}
\ea
where in the second step we made use of the identity
$
 1 +2 \nB{}(-\E^{ }_{k}) = -[1 + 2 \nB{}(\E^{ }_{k})]
$.

It is useful to note that \eq\nr{PiT_free} is  
closely related to \eq\nr{Sfn_res}. However, 
\eq\nr{Sfn_res} is true in general, whereas \eq\nr{PiT_free} was derived
for the special case of a free propagator; 
thus it is not always true that  
thermal effects can be obtained by simply replacing the 
zero-temperature time-ordered propagator by \eq\nr{PiT_free}, 
even if surprisingly often such a simple recipe does function.
We return to a discussion of this point in \se\ref{se:realtime}.

\subsection*{Fermionic case} \index{Heisenberg-operator: fermionic}

Let us next consider 2-point correlation functions built out of fermionic 
operators~\cite{old4}--\hspace*{-1.1mm}\cite{old3}. 
In contrast to the bosonic case, 
we take for generality the density matrix to be of the form 
\be
 \hat\rho = \frac{1}{\mathcal{Z}} \exp[-\beta (\hat H - \mu \hat Q)]
 \;, \la{rho_mu_T} 
\ee
where $\hat Q$ is an operator commuting with $\hat H$ and $\mu$ is 
the associated chemical potential. 

\renewcommand{\A}{\hat j^{ }_\alpha}
\renewcommand{\B}{\,\hat{\bar{j}}^{ }_\beta}

We denote the operators appearing in the 2-point 
functions by $\A$, $\B$. They could
be elementary field operators, in which case 
the indices $\alpha,\beta$ label Dirac and/or flavour components, 
but they could also be composite operators consisting of
a product of elementary field operators. Nevertheless, 
we assume the validity of the relation
\ba
 {[} \A(t,\vec{x}), \hat Q {]} & = &   \A(t,\vec{x})
 \;. \la{prereq_2}
\ea
To motivate this, note that for $\A \equiv \hat \psi^{ }_\alpha$, 
$\B = \,\hat{\bar{\!\psi}}^{ }_\beta$, the canonical commutation relation
of \eq\nr{Dirac_can},
\be
 \{ \hat\psi^{ }_\alpha(x^0,\vec{x}) , \hat\psi_\beta^\dagger(x^0,\vec{y}) \}  
 = \delta^{(d)}(\vec{x-y}) \delta^{ }_{\alpha\beta}
 \;, 
\ee
and the expression for the conserved charge in \eq\nr{Dirac_Q},
\be
 \hat Q =  
 \int_\vec{x} \,\,\hat{\!\bar\psi} \gamma_0^{ } \hat\psi =   
 \int_\vec{x} \,\hat{\psi}^\dagger_\alpha \hat\psi^{ }_\alpha
 \;, 
\ee 
as well as the identity
$
 [\hat A, \hat B\hat C] = \hat A \hat B \hat C - \hat B \hat C \hat A
 = \hat A \hat B \hat C  + \hat B \hat A \hat C - 
 \hat B \hat A \hat C - \hat B \hat C \hat A
 = \{ \hat A, \hat B\} \hat C - \hat B \{ \hat A, \hat C\} 
$, 
indicate that \eq\nr{prereq_2} is indeed satisfied
for $\hat\psi^{ }_\alpha$. 
\Eq\nr{prereq_2} implies that 
\be
 e^{\beta\mu\hat Q} \A(t,\vec{x}) = \sum_{n=0}^{\infty} \frac{1}{n!}
 (\beta\mu)^n (\hat Q)^n \A(t,\vec{x}) = 
 \sum_{n=0}^{\infty} \frac{1}{n!}
 (\beta\mu)^n  \A(t,\vec{x}) (\hat Q - \hat\unit)^n
 = \A(t,\vec{x}) e^{\beta\mu\hat Q} e^{-\beta\mu}
 \;, 
\ee
and consequently that 
\ba
 \Bigl\langle \A(t-i\beta,\vec{x})\, \B(0,\vec{0}) \Bigr\rangle  
 & = &
 \frac{1}{\mathcal{Z}} \tr\Bigl[ e^{-\beta(\hat H - \mu \hat Q)}
 e^{\beta \hat H} \A(t,\vec{x}) e^{-\beta \hat H} \B(0,\vec{0}) \Bigr] 
 \nn & = & 
 \frac{1}{\mathcal{Z}} \tr\Bigl[ 
 \A(t,\vec{x}) e^{-\beta \mu }e^{-\beta (\hat H -\mu\hat Q) } 
 \B(0,\vec{0}) \Bigr] 
 \nn & = & 
 \frac{1}{\mathcal{Z}}\, e^{-\mu\beta}\, \tr\Bigl[ 
 \A(t,\vec{x}) e^{-\beta (\hat H -\mu\hat Q)} \B(0,\vec{0}) \Bigr] 
 \nn & = &
 e^{-\mu\beta}\Bigl\langle 
   \B(0,\vec{0})\, \A(t,\vec{x})
 \Bigr\rangle 
 \;. \la{fKMS}
\ea
This is a fermionic version of the KMS relation.

\index{Kubo-Martin-Schwinger: fermionic}

With this setting,  we can again define various classes of 
correlation functions.
The ``physical'' correlators are now set up as
\ba \index{Spectral function: fermionic} \index{Retarded correlator: fermionic}
 \Pi^{>}_{\alpha\beta}(\mathcal{K}) & \equiv & 
 \I \Bigl\langle \A(\mathcal{X})\, \B(0) \Bigl\rangle
 \;,   
 \la{fL}
 \\
 \Pi^{<}_{\alpha\beta}(\mathcal{K}) & \equiv & 
 \I \Bigl\langle - \B(0)\,  \A(\mathcal{X}) \Bigl\rangle
 \;,   
 \la{fS}
 \\
 \rho^{ }_{\alpha\beta}(\mathcal{K}) & \equiv & 
 \I \Bigl\langle \fr12 \Bigl\{ \A(\mathcal{X}) , \B(0) \Bigr\} \Bigl\rangle
 \;,   
 \la{frho}
 \\ 
 \Delta^{ }_{\alpha\beta}(\mathcal{K}) & \equiv & 
 \I \Bigl\langle \fr12 \Bigl[ \A(\mathcal{X}) , \B(0) \Bigr] \Bigl\rangle
 \;,   
 \la{fdelta}
\ea
where $\rho^{ }_{\alpha\beta}$ is the spectral function. 
The retarded and advanced correlators can be defined as 
\ba 
 \Pi^\iR_{\alpha\beta}(\mathcal{K}) & \equiv & 
 i \I \Bigl\langle \Bigl\{ \A(\mathcal{X}) , \B(0) \Bigr\}\,
 \theta(t) \Bigl\rangle
 \;, 
 \la{fR}
 \\
 \Pi^\iA_{\alpha\beta}(\mathcal{K}) & \equiv & 
 i \I \Bigl\langle  - \Bigl\{ \A(\mathcal{X}) , \B(0) \Bigr\}\,
 \theta(-t) \Bigl\rangle
 \;. 
 \la{fA} 
\ea 
On the other hand, the time-ordered correlation function reads 
\ba \index{Time-ordered correlator: fermionic} 
 \Pi^\iT_{\alpha\beta}(\mathcal{K}) & \equiv & 
  \I \Bigl\langle \A(\mathcal{X})\,  \B(0)\, \theta(t)
                 -\B(0)\,  \A(\mathcal{X})\, \theta(-t) \Bigl\rangle
 \;, \la{fT}
\ea
whereas the Euclidean correlator is
\ba \index{Euclidean correlator: fermionic}
 \Pi^\iE_{\alpha\beta}( K) & \equiv & 
 \int_0^\beta\!{\rm d}\tau \int_\vec{x}\,
  e^{(i k^{ }_n + \mu)\tau - i \vec{k} \cdot \vec{x} }
  \Bigl\langle \A( X)  \B(0) \Bigl\rangle
 \;.
 \la{fE}
\ea
Note that the Euclidean correlator is time-ordered by definition
($0 \le \tau \le \beta$), 
and can be computed with standard imaginary-time functional integrals. 

If the two operators in the integrand of \eq\nr{fE} anticommute 
with each other at $t=0$, then the KMS relation in \eq\nr{fKMS} asserts
that 
$
  \bigl\langle \A(-i\beta,\vec{x})\, \B(0,\vec{0}) \bigr\rangle  
 = 
  e^{-\mu\beta}\bigl\langle 
   \B(0,\vec{0})\, \A(0,\vec{x})
 \bigr\rangle 
 = 
 -  e^{-\mu\beta}\bigl\langle 
   \A(0,\vec{x})\, \B(0,\vec{0}) 
 \bigr\rangle 
$. 
The additional term in the Fourier transform with respect to $\tau$
in \eq\nr{fE} cancels the 
multiplicative factor $e^{-\mu\beta}$ at $\tau = \beta$, 
so that the $\tau$-integrand is antiperiodic. 
Therefore the Matsubara frequencies $k^{ }_n$ are fermionic.

We can establish relations between the different Green's functions
just like in the bosonic case: 
\ba
  \Pi^{>}_{\alpha\beta}(\mathcal{K}) \!\!\! & = &  \!\!\!
 \frac{1}{\mathcal{Z}}
 \I \tr \Bigl[ e^{-\beta\hat H + i \hat H t} 
  \!\!\underbrace{\unit}_{\sum_m |m\rangle\,\langle m|}\!\! e^{\beta\mu\hat Q} 
 \A(0,\vec{x})\, 
  e^{-i\hat H t} 
  \!\!\underbrace{\unit}_{\sum_n |n\rangle\,\langle n|}\!\!
 \B(0,\vec{0}) \Bigl]
 \nn & = & 
 \frac{1}{\mathcal{Z}}
 \sum_{m,n} \I e^{(-\beta+it)\E^{ }_m}e^{-it\E^{ }_n} e^{-\beta\mu}
 \langle m | \A(0,\vec{x}) e^{\beta\mu\hat Q} | n \rangle \, 
 \langle n | \B(0,\vec{0}) | m \rangle
 \nn & = & \!\!\!
 \frac{1}{\mathcal{Z}}
 \int_\vec{x} e^{-i\vec{k}\cdot\vec{x}}
 \sum_{m,n} e^{-\beta (\E^{ }_m+\mu)}
  \, 2\pi \, \delta(k^0_{ } + \E^{ }_m - \E^{ }_n)
 \langle m | \A(0,\vec{x}) e^{\beta\mu\hat Q} | n \rangle \, 
 \langle n | \B(0,\vec{0}) | m \rangle
 \;, \hspace*{0.4cm} \nn \la{f_der_kms_1} \\  
 \Pi^{<}_{\alpha\beta}(\mathcal{K}) & = & 
 - \frac{1}{\mathcal{Z}}
 \I \tr \Bigl[ e^{-\beta\hat H} e^{\beta\mu\hat Q} 
 \!\!\underbrace{\unit}_{\sum_n |n\rangle\,\langle n|}\!\! 
  \B(0,\vec{0})\,
  e^{i \hat H t} 
  \!\!\underbrace{\unit}_{\sum_m |m\rangle\,\langle m|}\!\! 
 \A(0,\vec{x})\, 
  e^{-i\hat H t} 
   \Bigl]
 \nn & = & 
 - \frac{1}{\mathcal{Z}}
 \sum_{m,n} \I e^{(-\beta-it)\E^{ }_n}e^{it\E^{ }_m}
 \langle n | \B(0,\vec{0}) | m \rangle \,  
 \langle m | \A(0,\vec{x}) e^{\beta\mu\hat Q} | n \rangle  
 \nn 
 & = & 
 - \frac{1}{\mathcal{Z}}
 \int_\vec{x} e^{-i\vec{k}\cdot\vec{x}}
 \sum_{m,n} e^{-\beta \E^{ }_n} \, 2\pi \, 
 \underbrace{\delta(k^0_{ } + \E^{ }_m - \E^{ }_n)}_{\E^{ }_n 
 = \E^{ }_m + k^0_{ }}
 \langle m | \A(0,\vec{x}) e^{\beta\mu\hat Q} | n \rangle \,
 \langle n | \B(0,\vec{0}) | m \rangle
 \nn & = &
 - e^{-\beta (k^0_{ } - \mu)} \,  \Pi^{>}_{\alpha\beta}(\mathcal{K})
 \;. \hspace*{0.5cm}   \la{f_der_kms_2}
\ea
Using the fact that
$
 \rho^{ }_{\alpha\beta}(\mathcal{K})
 = [\Pi^{>}_{\alpha\beta}(\mathcal{K}) -
  \Pi^{<}_{\alpha\beta}(\mathcal{K})]/2
$, we subsequently obtain 
\be
 \Pi^{>}_{\alpha\beta}(\mathcal{K}) = 
 2[1 - \nF{}(k^0_{ }-\mu)] \rho_{\alpha\beta}(\mathcal{K})
 \;, \quad
 \Pi^{<}_{\alpha\beta}(\mathcal{K}) = 
 - 2 \nF{}(k^0_{ }-\mu) \rho_{\alpha\beta}(\mathcal{K})
 \;, \la{fLSrel}
\ee
where $\nF{}(k^0_{ })\equiv 1/[\exp(\beta k^0) + 1]$ 
is the Fermi distribution. 
Moreover, the statistical correlator can be expressed as 
$
 \Delta^{ }_{\alpha\beta}(\mathcal{K}) = 
 [1 - 2 \nF{}(k^0_{ }-\mu)] \rho^{ }_{\alpha\beta}(\mathcal{K})
$.

The relation of 
$\Pi^\iR,\Pi^\iA$ and $\Pi^\iT$ to the spectral function can 
be derived in complete analogy with \eqs\nr{theta}--\nr{bTrho_rel}. 
For brevity we only cite the final results: 
\ba
 \Pi^\iR_{\alpha\beta}(\mathcal{K}) 
 \!\! & = & \!\! \int_{-\infty}^{\infty} \! \frac{{\rm d}\omega}{\pi} 
 \frac{\rho^{ }_{\alpha\beta}(\omega,\vec{k})}{\omega -k^0_{ }- i 0^+_{ }}
 \;, \quad 
 \Pi^\iA_{\alpha\beta}(\mathcal{K}) 
 =  \int_{-\infty}^{\infty} \! \frac{{\rm d}\omega}{\pi} 
 \frac{\rho^{ }_{\alpha\beta}(\omega,\vec{k})}{\omega -k^0_{ }+ i 0^+_{ }}
 \;,  \la{fRrhorel} \\ 
 \Pi^\iT_{\alpha\beta}(\mathcal{K}) & = &  
 \int_{-\infty}^{\infty} \! \frac{{\rm d}\omega}{\pi} 
 \frac{i\rho^{ }_{\alpha\beta}(\omega,\vec{k})}
 {k^0_{ } - \omega + i 0^+_{ }} - 
 2 \nF{}(k^0_{ }-\mu) \rho^{ }_{\alpha\beta}(k^0_{ },\vec{k}) 
 \nn 
 & = & 
 -i\, \Pi^\iR_{\alpha\beta}(\mathcal{K}) 
 + \Pi^{<}_{\alpha\beta}(\mathcal{K})
 \;. \la{fPiT_rho} 
\ea
Note that when written in a ``generic form'', where no distribution
functions are visible, the end results are identical to the bosonic ones. 
In addition, \eq\nr{fPiT_rho} can again 
be crosschecked using the right-hand sides
of \eqs\nr{fS}, \nr{fR} and \nr{fT}, and the alternative representation
$
  \Pi^\iT_{\alpha\beta} = 
 - i \Pi^\iA_{\alpha\beta} + 
 \Pi^{>}_{\alpha\beta}
$
also applies. The latter derivation 
implies that these ``operator relations'' apply even in 
a non-thermal situation, described by a generic density matrix
(cf.\ \se\ref{se:realtime}).

Finally, writing the argument inside the $\tau$-integration
in \eq\nr{fE} as a Wick rotation of
the inverse Fourier transform of \eq\nr{fL}, 
inserting \eq\nr{fLSrel}, and changing orders of integration, we get 
a spectral representation analogous to \eq\nr{spectral}, 
\ba \index{Spectral representation}
  \Pi^\iE_{\alpha\beta}( K) & = & 
 \int_0^\beta \! {\rm d}\tau\, e^{(i k^{ }_n +\mu ) \tau}
 \int_{-\infty}^{\infty} \frac{{\rm d}p^0_{ } }{2\pi} e^{-p^0_{ } \tau} 
 \, \Pi^{>}_{\alpha\beta}(p^0_{ },\vec{k})
 \nn & = & 
 \int_0^\beta \! {\rm d}\tau\, e^{(i k^{ }_n +\mu ) \tau}
 \int_{-\infty}^{\infty} \frac{{\rm d}p^0_{ } }{2\pi} e^{-p^0_{ } \tau} 
 \frac{2 e^{\beta(p^0_{ }-\mu)}}{e^{\beta(p^0_{ }-\mu)}+1}
 \,  \rho^{ }_{\alpha\beta}(p^0_{ },\vec{k})
 \nn & = & 
 \int_{-\infty}^{\infty} \frac{{\rm d}p^0_{ } }{\pi} 
 \frac{e^{\beta(p^0_{ }-\mu)}}{e^{\beta(p^0_{ }-\mu)}+1}
 \, \rho^{ }_{\alpha\beta}(p^0_{ },\vec{k})
 \int_0^\beta \! {\rm d}\tau\, e^{(i k^{ }_n +\mu -p^0_{ }) \tau}
 \nn  & = & 
 \int_{-\infty}^{\infty} \frac{{\rm d}p^0_{ } }{\pi} 
 \frac{e^{\beta(p^0_{ }-\mu)}}{e^{\beta(p^0_{ }-\mu)}+1}
 \, \rho^{ }_{\alpha\beta}(p^0_{ },\vec{k})
 \biggl[ \frac{e^{(i k^{ }_n +\mu -p^0_{ }) \tau}}
 {i k^{ }_n +\mu -p^0_{ }}\biggr]_0^{\beta}
 \nn  & = & 
 \int_{-\infty}^{\infty} \frac{{\rm d}p^0_{ } }{\pi} 
 \frac{e^{\beta(p^0_{ }-\mu)}}{e^{\beta(p^0_{ }-\mu)}+1}
 \, \rho^{ }_{\alpha\beta}(p^0_{ },\vec{k})
 \frac{-e^{-\beta(p^0_{ }-\mu)}-1}{i k^{ }_n +\mu -p^0_{ }}
 \nn  & \stackrel{p^0_{ }\to k^0_{ }}{=} & 
 \int_{-\infty}^{\infty} \! \frac{{\rm d}k^0_{ } }{\pi} 
 \frac{\rho^{ }_{\alpha\beta}(k^0_{ },\vec{k})}{k^0_{ } - i [k^{ }_n - i \mu]}
 \;. \la{fErhorel} \la{fspectral}
\ea
Like in the bosonic case, this relation can be 
inverted by making use of \eq\nr{delta}, 
\be \index{Analytic continuation}
  \rho^{ }_{\alpha\beta}(\mathcal{K}) = \frac{1}{2i} \disc 
  \Pi^\iE_{\alpha\beta}(k^{ }_n - i \mu \to -i k^0_{ },\vec{k})
 \;, \la{ffinal}
\ee
where the discontinuity is defined like in \eq\nr{Discdef}.

\index{Thermal sums: non-relativistic fermion}

The fermionic Matsubara sum over the structure 
in \eq\nr{fspectral} can be carried out explicitly. This could be 
verified by making use of \eq\nr{Gefe}, in analogy with the bosonic
analysis in \eqs\nr{bsum} and \nr{bsum2}, but let us 
proceed in another way for a change. 
We may recall (cf.\ footnote on p.~\pageref{delta_sum}) that 
\be
 T  \sum_{\omega^{ }_n}  e^{i \omega^{ }_n \tau} 
 = \delta(\tau \;\mbox{mod}\; \beta)
 \;. 
\ee 
According to \eq\nr{SfSb_rel}, viz.\ 
$
 \sigma^{ }_\fe(T)
 = 2 \sigma^{ }_\bo\bigl( \tfr{T}{2} \bigr) - \sigma^{ }_\bo(T)
$, 
we can thus write
\be
 T  \sum_{ \{ \omega^{ }_n \} }  e^{i \omega^{ }_n \tau} 
 = 2 \delta(\tau \;\mbox{mod}\; 2\beta) - \delta(\tau \;\mbox{mod}\; \beta)
 \;. \la{ferm_complete}
\ee
Let us assume for a moment that $k^0_{ } - \mu > 0$. 
Employing the representation
\be
 \frac{1}{\alpha + i \beta} = \int_0^\infty \! {\rm d}s \, 
 e^{-(\alpha + i \beta)s}
 \;, \quad \alpha > 0 
 \;, 
\ee
and inserting subsequently \eq\nr{ferm_complete}, we get
\ba
  T \sum_{ \{ \omega^{ }_n \} } 
 \frac{1}{k^0_{ } - \mu - i \omega^{ }_n}  
 e^{i \omega^{ }_n \tau} 
 & = & 
 \int_0^\infty \! {\rm d}s \, 
 T \sum_{ \{ \omega^{ }_n \} } e^{i \omega^{ }_n \tau - k^0_{ } s + \mu s
  + i \omega^{ }_n s}
 \nn & = & 
 \int_0^\infty \! {\rm d}s \, e^{-(k^0_{ } - \mu)s}
 \Bigl[
  2 \delta(\tau + s \;\mbox{mod}\; 2 \beta) - 
  \delta(\tau + s \;\mbox{mod}\; \beta)
 \Bigr]
 \nn & = & 
 2 \sum_{n=1}^{\infty} e^{-(k^0_{ } - \mu)(-\tau + 2 \beta n)}
 - \sum_{n=1}^{\infty} e^{-(k^0_{ } - \mu)(-\tau + \beta n)}
 \nn & = & 
 e^{(k^0_{ } - \mu)\tau} 
 \biggl[\underbrace{
  2 \underbrace{\sum_{n=1}^{\infty} e^{-2\beta (k^0_{ } - \mu) n}}_{
   \frac{e^{-2\beta(k^0_{ } - \mu)}}{1-e^{-2\beta(k^0_{ } - \mu)}}
  }
 -  \underbrace{\sum_{n=1}^{\infty} e^{-\beta (k^0_{ } - \mu) n}}_{
  \frac{e^{-\beta(k^0_{ } - \mu)}}{1-e^{-\beta(k^0_{ } - \mu)}}
 }}_{\frac{2}{(e^{\beta(k^0_{ }-\mu)}-1)(e^{\beta(k^0_{ }-\mu)}+1)} - 
 \frac{1}{e^{\beta(k^0_{ }-\mu)}-1} }
 \biggr]
 \nn & = & 
 - e^{(k^0_{ }-\mu)\tau} \nF{}(k^0_{ }-\mu)  \;, 
 \la{fsum}
\ea
where we assumed $0 < \tau < \beta$.
As an immediate consequence, 
\be
  T \sum_{ \{ \omega^{ }_n \} } 
 \frac{1}{k^0_{ } -\mu - i \omega^{ }_n}  
 e^{ - i \omega^{ }_n \tau} 
 = 
 - T \sum_{ \{ \omega^{ }_n \} } 
 \frac{1}{k^0_{ } -\mu - i \omega^{ }_n}  
 e^{ i \omega^{ }_n(\beta- \tau)} 
 = 
 e^{(\beta - \tau) (k^0_{ } - \mu)} \nF{}(k^0_{ }-\mu) 
 \;. 
 \la{fsum2}
\ee
Furthermore, it is not difficult to show 
(by substituting $\omega^{ }_n \to - \omega^{ }_n$)
that these relations continue to hold also for $k^0_{ } - \mu < 0$.

As a consequence of \eq\nr{fsum}, we note that 
\ba \index{Thermal sums: with chemical potential}
 T \sum_{ \{ \omega^{ }_n \} } \frac{e^{i (\omega^{ }_n + i \mu)\tau}}
 {(\omega^{ }_n + i \mu)^2 + \omega^2} & = & 
 e^{-\mu\tau} T \sum_{ \{ \omega^{ }_n \} } e^{i \omega^{ }_n \tau}
 \frac{1}{(\omega - i \omega^{ }_n + \mu)(\omega + i \omega^{ }_n - \mu)}
 \nn & = & 
 e^{-\mu\tau} T \sum_{ \{ \omega^{ }_n \} } e^{i \omega^{ }_n \tau}
 \frac{1}{2 \omega}
 \biggl[ 
 \frac{1}{\omega - \mu + i \omega^{ }_n }+  
 \frac{1}{\omega + \mu - i \omega^{ }_n }
 \biggr]
 \nn & = & 
 \frac{e^{-\mu\tau}}{2 \omega}
 \Bigl[
  e^{-(\omega-\mu)\tau} \nF{}(-\omega+\mu) - 
  e^{(\omega+\mu)\tau} \nF{}(\omega+\mu) 
 \Bigr]
 \nn & = & 
 \frac{e^{-\mu\tau}}{2 \omega}
 \Bigl[
  e^{(\beta-\tau)(\omega-\mu)}  \nF{}(\omega-\mu) - 
  e^{\tau(\omega+\mu)}  \nF{}(\omega+\mu)
 \Bigr]
 \nn & = &
 \frac{1}{2 \omega}
 \Bigl[
   \nF{}(\omega-\mu) e^{(\beta-\tau)\omega-\beta\mu} - 
   \nF{}(\omega+\mu) e^{\tau\omega} 
 \Bigr]
 \;. \la{Gefe_new}
\ea 
This constitutes a generalization of \eq\nr{Gefe} to the case of
a finite chemical potential. 

\subsection*{Example: free fermion}

We illustrate the relations obtained by 
considering the structure of the free fermion
propagator in the presence of a chemical potential. 
With fermions, one has 
to be quite careful with definitions. Suppressing spatial coordinates
and indices, \eq\nr{psiprop} and the presence of a chemical potential
{\em \`a la} \eq\nr{SE_fer_mu_2} imply that the free propagator can 
be written in the schematic form (here $A$ and $B$ 
carry dependence on the spatial momentum and the Dirac matrices)
\be
 \langle \hat\psi(\tau) \,\hat{\!\bar\psi}(0) \rangle
 = T \sum_{ \{ p^{ }_n \} } e^{i (p^{ }_n + i \mu )\tau} 
 \frac{- i A (p^{ }_n + i \mu) + B}{(p^{ }_n + i \mu)^2 + \E_k^2}
 \;, \la{F_prop_mu}
\ee
where an additional exponential has been inserted into the Fourier
transform, in order to respect the KMS property in \eq\nr{fKMS}.
The correlator in \eq\nr{fE} then becomes
\ba
  \Pi^\iE(k^{ }_n) & = & \int_0^\beta\!{\rm d}\tau \, 
 e^{(i k^{ }_n + \mu)\tau}
  T \sum_{ \{ p^{ }_n \} } e^{i (p^{ }_n + i \mu )\tau} 
 \frac{- i A (p^{ }_n + i \mu) + B}{(p^{ }_n + i \mu)^2 + \E_k^2} 
 \nn  & = & 
 \frac{ i A (k^{ }_n - i \mu) + B}{(k^{ }_n - i \mu)^2 + \E_k^2} 
 \;. \la{F_prop_mu_2}
\ea
The analytic continuation in \eq\nr{ffinal} yields the retarded correlator
\be
  \Pi^\iR (k^0_{ }) = 
   \frac{A (k^0_{ } + i 0^+_{ }) + B}{- (k^0_{ } + i 0^+_{ })^2 + \E_k^2}
 = - \frac{A k^0_{ }  + B}{(k^0_{ })^2 -  \E_k^2 + i \sign(k^0_{ }) 0^+_{ }}
 \;,
\ee
and its discontinuity gives 
\ba
 \rho(k^0_{ }) & = &
  \pi (A k^0_{ }  + B) \sign(k^0_{ })
 \, \delta\bigl( (k^0_{ } - \E^{ }_k)(k^0_{ } + \E^{ }_k) \bigr)
 \nn & = & 
 \pi (A k^0_{ }  + B) \frac{ \sign(k^0_{ }) }{2\E^{ }_k}
 \Bigl[ 
 \delta(k^0_{ } - \E^{ }_k) + 
 \delta(k^0_{ } + \E^{ }_k)
 \Bigr]
 \nn & = &  \frac{\pi}{2\E^{ }_k} 
 (A k^0_{ } + B)
 \Bigl[ \delta(k^0_{ } - \E^{ }_k) - \delta(k^0_{ } + \E^{ }_k) \Bigr]
 \;. \la{rho_F_free}
\ea
Any dependence on temperature and chemical potential has  
disappeared here.
Note that (if $B$ is odd in $\vec{k}$)
$\rho$ is even in $\mathcal{K}\to - \mathcal{K}$.
{}From \eqs\nr{fLSrel} and
\nr{fPiT_rho}, the time-ordered propagator can be determined
after a few steps: 
\ba \index{Time-ordered propagator: free fermion}
  \Pi^\iT (k^0_{ }) & = & 
 (A k^0_{ } + B)\biggl\{
 \frac{-i}{2\E^{ }_k} \biggl( \frac{1}{\E^{ }_k-k^0_{ }-i0^+_{ }} + 
 \frac{1}{\E^{ }_k+k^0_{ }+i0^+_{ }} \biggr)
 \nn & & 
  -\, \frac{2\pi}{2 \E^{ }_k} 
 \nF{}(k^0_{ } - \mu)
 \Bigl[ \delta(k^0_{ } - \E^{ }_k) - \delta(k^0_{ } + \E^{ }_k)\Bigr]
  \biggr\}
 \nn & = &  
 \frac{A k^0_{ } + B}{2\E^{ }_k} 
 \biggl\{
   -i \mathbbm{P}\,\biggl( \frac{1}{\E^{ }_k-k^0_{ }} \biggr)
   -i \mathbbm{P}\,\biggl( \frac{1}{\E^{ }_k+k^0_{ }} \biggr)
 \nn & & 
 +\, \pi \delta(k^0_{ } - \E^{ }_k ) \Bigl[1 - 2 \nF{}(k^0_{ } -\mu ) \Bigr]
 - \pi \delta(k^0_{ } + \E^{ }_k ) \Bigl[1 - 2 \nF{}(k^0_{ } -\mu ) \Bigr]
 \biggr\}
 \nn & = &  
 \frac{A k^0_{ } + B}{2\E^{ }_k} 
 \biggl\{
   - i  \mathbbm{P}\,\biggl( \frac{2 \E^{ }_k}{\E_k^2-(k^0_{ })^2} \biggr)
 \nn & & 
 +\, \pi \delta(k^0_{ } - \E^{ }_k ) \Bigl[1 - 2 \nF{}(k^0_{ } -\mu ) \Bigr]
 + \pi \delta(k^0_{ } + \E^{ }_k ) \Bigl[1 - 2 \nF{}(-k^0_{ } +\mu ) \Bigr]
 \biggr\}
 \nn & = &  
 \frac{A k^0_{ } + B}{2\E^{ }_k} 
 \biggl\{
   - i  \mathbbm{P}\,\biggl( \frac{2 \E^{ }_k}{\E_k^2-(k^0_{ })^2} \biggr) + 
 2 \E^{ }_k \, \pi \delta\Bigl((k^0_{ })^2-\E_k^2\Bigr)
 \nn & & 
 -\, 2 \pi \Bigl[ \delta(k^0_{ } - \E^{ }_k )\, \nF{}(k^0_{ } -\mu )
 + \delta(k^0_{ } + \E^{ }_k )\, \nF{}(-k^0_{ } +\mu ) \Bigr]
 \biggr\}
 \nn & = & 
 (A k^0_{ } + B)
 \biggl\{
  \frac{i}{\mathcal{K}^2 - m^2 + i 0^+_{ }} 
 - 2 \pi\, \delta\bigl( \mathcal{K}^2 - m^2 \bigr) 
 \, \nF{}\Bigl(|k^0_{ }| - \sign(k^0_{ }) \mu\Bigr)
 \biggr\}
 \;. \hspace*{7mm} \la{free_f_RTF}
\ea
Medium effects are seen to reside in 
the on-shell part and, to some extent, one could hope to account
for them simply by replacing free zero-temperature Feynman
propagators by \eq\nr{free_f_RTF}. The proper procedure, however, 
is to carry out the analytic continuation for the {\em complete observable}
considered, and this may not always amount to the simple replacement
of vacuum time-ordered propagators 
through \eq\nr{free_f_RTF}, cf.\ \se\ref{se:realtime}.

\def\Scata{\pic{%
 \Laqu(0,15)(15,30)%
 \Lgh(0,15)(15,15)%
 \Ldsc(0,15)(15,0)%
 \Textsr(17,21,1)%
 \Textsr(19,10.5,K)%
 \Textsr(17,0,2)%
}}
\def\Scatb{\picc{%
 \Ldsc(15,15)(30,15)%
 \Laqu(30,15)(45,25)%
 \Lgh(30,15)(45,5)%
 \Textsl(5,10.5,2)%
 \Textsr(36,19,1)%
 \Textsr(38,2,K)%
}}
\def\Scatc{\picc{%
 \Lqu(15,15)(30,15)%
 \Ldsc(30,15)(45,25)%
 \Lgh(30,15)(45,5)%
 \Textsl(5,10.5,1)%
 \Textsr(36,19,2)%
 \Textsr(38,2,K)%
}}
\def\Scatd{\picc{%
 \Lqu(15,25)(30,15)%
 \Ldsc(15,5)(30,15)%
 \Lgh(30,15)(45,15)%
 \Textsl(6,21,1)%
 \Textsl(6,2,2)%
 \Textsr(40,10.5,K)%
}}

\newpage

\subsection{From a Euclidean correlator to a spectral function}
\la{se:NI_rho}

As an application of the relations derived in \se\ref{se:diff_G}, 
let us carry out an explicit 1-loop computation illustrating 
the steps.\footnote{%
 A classic example of this kind of a  
 computation can be found in ref.~\cite{art}.
 It is straightforward to generalize the techniques to
 the 2-loop level, cf.\ e.g.\ ref.~\cite{master};
 at that order the novelty arises that 
 there are infrared divergences in ``real'' and ``virtual'' parts
 of the result, which only cancel in the sum. 
 } 
The computation performed here will turn out to 
be directly relevant in the context of particle production, 
discussed in more detail in \se\ref{se:ppr}.

\index{Yukawa interaction}

\def\Lwidth{1}

\def\TopoSBnew(#1,#2,#3){\piccc{#1(0,15)(15,15) #2(30,15)(15,0,180)%
 #3(30,15)(15,180,360) #1(45,15)(60,15)%
 \Textl(45,30,\tilde R)\Textl(45,0,\tilde K+\tilde R)\Textb(3,18,\tilde K)}}
\def\Bqu(#1,#2)(#3,#4,#5){\SetWidth{2.0}\ArrowArc(#1,#2)(#3,#4,#5)%
\SetWidth{1.0}}
\renewcommand{\A}{\hat\nu^{ }_\alpha}
\renewcommand{\B}{\,\hat{\!\bar{\nu}}^{ }_\beta}

Our goal is to work out the leading non-trivial 
contribution to the spectral function of a right-handed lepton ($N$) that 
originates from its Yukawa interaction with Standard Model particles, 
\be
 \delta \mathcal{L}^{ }_\iM \; \equiv \; 
 - h_{ } \bar L\, \tilde \phi\, \aR  N
 - h_{ }^* \bar{ N} \tilde \phi^\dagger \aL L
 \;.  \la{LM}  
\ee
Here $\tilde \phi \equiv i \tau^{ }_2 \phi^*$ 
is a conjugated Higgs doublet, $L$ is a lepton doublet,  
$\aL \equiv (1-\gamma^{ }_5)/2$ and $\aR \equiv (1+\gamma^{ }_5)/2$ are
chiral projectors, and $h$ is a Yukawa coupling constant. 
The Higgs and lepton doublets have the forms 
\be
 \tilde \phi = \frac{1}{\sqrt{2}}
 \biggl( \begin{array}{c} 
  \phi_0^{ } + i \phi^{ }_3 \\ -\phi^{ }_2 + i \phi^{ }_1 
 \end{array} \biggr) 
 \;, 
 \quad
 L =  
  \biggl( \begin{array}{c} 
  \nu \\ e
 \end{array} \biggr) 
 \equiv 
 \biggl( \begin{array}{c} 
  \ell^{ }_1 \\ \ell^{ }_2
 \end{array} \biggr) 
 \;, \la{doubs}
\ee
where $\phi^{ }_\mu$, $\mu \in \{0,1,2,3\}$, are real scalar fields. 
The neutral component $\phi^{ }_0$ is the physical Higgs field, whereas
the $\phi^{ }_i$ represent Goldstone modes after electroweak symmetry breaking.

Anticipating the results of \se\ref{se:ppr}, 
we consider the Euclidean correlator of the operators coupling to 
the right-handed lepton through the interaction in \eq\nr{LM},  
\be
  \Pi^\iE_{}( K)
 \, \equiv \,  
 \int_X e^{i K\cdot  X}
 \, \aL \, \bigl\langle \, (\tilde \phi^\dagger L)( X) \; 
 (\bar L\, \tilde \phi)(0) \, \bigr\rangle \, \aR
 \;. \la{ex_PiE}
\ee
This has the form of \eq\nr{fE}; 
the coupling constant $|h|^2$ has been omitted for simplicity.  
The four-momentum  $ K$ is {\em fermionic}.
The operators in \eq\nr{ex_PiE} are of a mixed ``boson-fermion'' type; 
similar computations will be carried out for ``fermion-fermion''
and ``boson-boson'' cases below, cf.\ \eqs\nr{G_2} and \nr{bG_2}, 
respectively. The ``boson-fermion'' analysis 
is furthermore generalized to include 
a chemical potential around \eq\nr{Seff_F_1}.

\index{Thermal sums: boson-fermion loop}

Inserting \eq\nr{doubs} and 
carrying out the contractions, we can rewrite \eq\nr{ex_PiE} in the form
\ba
  \Pi^\iE ( K)
 & = & \frac{1}{2}
 \int_X e^{i K\cdot  X}
 \, \aL \, \langle \ell( X) \bar\ell(0) \rangle_0^{ } 
 \, 
 \langle \phi( X) \phi(0) \rangle_0^{ } \, \aR 
 \nn & = & 
 \frac{1}{2}  
 \int_X
 \Tint{ \{  P \}  R }
 e^{i( K +  P +  R)\cdot  X}
 \, \aL \, 
 \frac{-i \bsl{ P} + m^{ }_{\ell}}{ P^2 + m_{\ell}^2}
 \frac{1}{ R^2 + m_{\phi}^2} 
 \, \aR 
 \nn & = & 
 \frac{1}{2} 
 \int_{\vec{p}} T \sum_{ \{ p^{ }_n \} }
 \frac{-i \bsl{ P}\, \aR}{ p_n^2 + \E_1^2}
 \frac{1}{(p^{ }_n+k^{ }_n)^2 + \E_2^2} 
 \;, \la{raw1}
\ea
where we inserted the free scalar and fermion propagators, and 
denoted
\be
 \E^{ }_1 \equiv \sqrt{\vec{p}^2 +  m_{\ell}^2  }
 \;, \quad
 \E^{ }_2 \equiv \sqrt{ (\vec{p+k})^2 + m_{\phi}^2 }
 \;. \la{E_1_E_2}
\ee  
Moreover the left and right projectors removed 
the mass term from the numerator. We have been 
implicit about the assignment of the masses $m^{ }_{\ell}$, $m^{ }_{\phi}$
to the corresponding fields, as well as about the summation over 
the different field components, 
as details of this kind are unnecessary for now.

The essential issue in handling \eq\nr{raw1}
is the treatment of the Matsubara sum. More generally, 
let us inspect the structure 
\be
 \mathcal{F} \equiv
 T \sum_{ \{ p^{ }_n \} } \frac{f(i p^{ }_n,ik^{ }_n,\vec{v})}
 {[p_n^2 + \E_1^2][(p^{ }_n+k^{ }_n)^2 + \E_2^2]}
 \;,  \la{F_sum}
\ee
where we assume that the book-keeping function $f$  
depends linearly on its arguments 
(this assumption will become crucial below), and $\vec{v}$ is a dummy 
variable representing spatial momenta. We can write
\ba
 \mathcal{F} & = &
 T \sum_{ \{ p^{ }_n \} } T \sum_{r^{ }_n} 
 \beta\, \delta(r^{ }_n -  p^{ }_n - k^{ }_n)
 \frac{f(ip^{ }_n,ik^{ }_n,\vec{v})}
  {[p_n^2 + \E_1^2][r_n^2 + \E_2^2]} 
 \nn & = & 
 \int_0^\beta \! {\rm d}\tau \, e^{-i k^{ }_n \tau} 
 \biggl\{ T \sum_{\{  p^{ }_n\} } e^{-i  p^{ }_n \tau} 
 \frac{f(ip^{ }_n,ik^{ }_n,\vec{v})}
  {p_n^2 + \E_1^2}
 \biggr\}
 \biggl\{ T \sum_{r^{ }_n} \frac{e^{i r^{ }_n \tau}}{r_n^2 + \E_2^2} \biggr\}
 \;, 
\ea
where we have used the relation
\be \index{Saclay method}
 \beta \, \delta(r^{ }_n -  p^{ }_n - k^{ }_n) = 
 \int_0^\beta \! {\rm d}\tau \, e^{i (r^{ }_n -  p^{ }_n - k^{ }_n) \tau}
 \;. \la{insert_delta}
\ee
This way of handling the Matsubara sums is sometimes called
the ``Saclay method'', cf.\ e.g.\ refs.~\cite{pisarski,parwani}. 
Now we can make use of \eqs\nr{Gebo_new} and \nr{Gefe_new} 
and time derivatives thereof: 
\ba 
 T \sum_{r^{ }_n} \frac{e^{i r^{ }_n \tau}}{r_n^2 + \E_2^2} 
 & = & 
 \frac{\nB{}(\E^{ }_2)}{2 \E^{ }_2}
 \Bigl[e^{(\beta-\tau)\E^{ }_2} + e^{\tau \E^{ }_2} \Bigr]
 \;, \\
 T \sum_{\{  p^{ }_n \}} \frac{e^{\pm i p^{ }_n \tau}}{p_n^2 + \E_1^2} 
 & = & 
 \frac{\nF{}(\E^{ }_1)}{2 \E^{ }_1}
 \Bigl[e^{(\beta-\tau)\E^{ }_1} - e^{\tau \E^{ }_1} \Bigr]
 \;, \la{mid_minus} \\
  T \sum_{ \{  p^{ }_n \} } 
 \frac{i p^{ }_n e^{- i p^{ }_n \tau}}{p_n^2 + \E_1^2} 
 & = & 
 \frac{\nF{}(\E^{ }_1)}{2 \E^{ }_1}\Bigl[\E^{ }_1 e^{(\beta-\tau)\E^{ }_1} 
 + \E^{ }_1 e^{\tau \E^{ }_1} \Bigr]
 \;.
\ea
Accounting for the minus sign in \eq\nr{mid_minus}
within the arguments of the linear function, we then get 
\ba
  & & \hspace*{-4cm} \mathcal{F} = 
   \int_0^\beta \! {\rm d}\tau \, e^{-i k^{ }_n \tau} \;  
   \frac{\nF{}(\E^{ }_1)\nB{}(\E^{ }_2)}{4 \E^{ }_1 \E^{ }_2} \nn
  \times \biggl\{ & & 
  e^{(\beta-\tau)(\E^{ }_1 + \E^{ }_2)} f(\E^{ }_1,ik^{ }_n,\vec{v}) 
  \nn & + & 
  e^{(\beta-\tau)\E^{ }_2 + \tau \E^{ }_1} f(\E^{ }_1,-ik^{ }_n,-\vec{v}) 
  \nn[3mm] & + & 
  e^{(\beta-\tau)\E^{ }_1 + \tau \E^{ }_2} f(\E^{ }_1,ik^{ }_n,\vec{v}) 
  \nn & + & 
  e^{ \tau(\E^{ }_1 + \E^{ }_2)} f(\E^{ }_1,-ik^{ }_n,-\vec{v}) 
  \quad \biggr\} 
 \;. \la{raw2}
\ea

As an example, 
let us focus on the third structure in \eq\nr{raw2}; the other three  
follow in an analogous way. The 
$\tau$-integral can be carried out, noting that $k^{ }_n$ is fermionic:  
\ba
  \int_0^\beta \! {\rm d}\tau \, e^{\beta \E^{ }_1} 
  e^{\tau (-i k^{ }_n - \E^{ }_1 + \E^{ }_2)}  
 & = &
 \frac{e^{\beta \E^{ }_1}}{-i k^{ }_n - \E^{ }_1 + \E^{ }_2}
 \Bigl[
 - e^{\beta(\E^{ }_2 - \E^{ }_1)} - 1 
 \Bigr]
 \nn & = & 
 \frac{e^{\beta \E^{ }_2} + e^{\beta \E^{ }_1}}
 {i k^{ }_n + \E^{ }_1 - \E^{ }_2} 
 \nn & = & 
 \frac{1}{i k^{ }_n + \E^{ }_1 - \E^{ }_2}
 \Bigl[
 \nB{}^{-1}(\E^{ }_2) + \nF{}^{-1}(\E^{ }_1) 
 \Bigr]
 \;. \la{raw3}
\ea
Thus
\be
 \left. \mathcal{F} \right|^{ }_\rmi{3rd} = 
 \frac{1}{4 \E^{ }_1 \E^{ }_2} 
 \Bigl[ \nF{}(\E^{ }_1) + \nB{}(\E^{ }_2) \Bigr]
 \frac{f(\E^{ }_1,ik^{ }_n,\vec{v})}{i k^{ }_n + \E^{ }_1 - \E^{ }_2}
 \;. \la{raw4}
\ee

Finally we set $k^{ }_n \to - i (k^0_{ } + i 0^+_{ })$
and take the imaginary part according to \eq\nr{ffinal}. 
Making use of \eq\nr{delta}, we note that   
\be
 \frac{1}{2i} \Bigl[ \frac{1}{k^0_{ } + \Delta + i 0^+_{ }} - 
 \frac{1}{k^0_{ } + \Delta - i 0^+_{ }} \Bigr]
 = - \pi \delta(k^0_{ } + \Delta)
 \;. \la{delta_disc}
\ee
Thereby $1/(i k^{ }_n + \E^{ }_1 - \E^{ }_2)$ 
in \eq\nr{raw4} gets replaced with
$-\pi\,\delta(k^0_{ } + \E^{ }_1 - \E^{ }_2)$.
Special attention needs to be paid to the possibility that 
$k^{ }_n$ could also appear in the numerator in \eq\nr{raw4}; however, 
we can then write
\be
 i k^{ }_n =  
 \underbrace{i k^{ }_n + \E^{ }_1 - \E^{ }_2}^{ }_{\rmi{no discontinuity}} 
 + \E^{ }_2 - \E^{ }_1
 \;, \la{no_disc}
\ee
so that in total 
\ba
 & & \hspace*{-1.2cm} \left. \im\Bigl\{  
 \mathcal{F}(i k^{ }_n \to k^0_{ } + i 0^+_{ }) \Bigr\} \right|^{ }_\rmi{3rd}
  =  
 -\frac{\pi}{4 \E^{ }_1 \E^{ }_2} 
 \Bigl[ \nF{}(\E^{ }_1) + \nB{}(\E^{ }_2) \Bigr]
 \, \delta(k^0_{ } + \E^{ }_1 - \E^{ }_2)
  \, f(\E^{ }_1,\underbrace{\E^{ }_2 - \E^{ }_1}_{k^0_{ }},\vec{v})
 \nn & = & 
 -\frac{ 2\pi \delta(k^0_{ } + \E^{ }_1 - \E^{ }_2) }{8 \E^{ }_1 \E^{ }_2}  
 f(\E^{ }_1,k^0_{ },\vec{v})\, \nF{}^{-1}(k^0_{ }) \,
 \frac{e^{\beta \E^{ }_1} + e^{\beta \E^{ }_2}}
 {[e^{\beta(\E^{ }_2 - \E^{ }_1)}+1]
 (e^{\beta \E^{ }_1}+1)(e^{\beta \E^{ }_2} - 1)}
 \nn & = & 
 -\frac{ 2\pi \delta(k^0_{ } + \E^{ }_1 - \E^{ }_2) }{8 \E^{ }_1 \E^{ }_2} 
 f(\E^{ }_1,k^0_{ },\vec{v})\, \nF{}^{-1}(k^0_{ }) \,
 \frac{e^{\beta \E^{ }_1} [ 1 + e^{\beta (\E^{ }_2-\E^{ }_1)}]}
 {[e^{\beta(\E^{ }_2 - \E^{ }_1)}+1]
 (e^{\beta \E^{ }_1}+1)(e^{\beta \E^{ }_2} - 1)}
 \nn &  = & 
 -\frac{ 2\pi \delta(k^0_{ } + \E^{ }_1 - \E^{ }_2) }{8 \E^{ }_1 \E^{ }_2}  
 f(\E^{ }_1,k^0_{ },\vec{v})\, \nF{}^{-1}(k^0_{ }) \, \nB{}(\E^{ }_2) 
 [1 - \nF{}(\E^{ }_1)]
 \;. \la{raw5}
\ea
We have chosen to factor out $\nF{}^{-1}(k^0_{ })$ because in typical 
applications it gets cancelled against $\nF{}(k^0_{ })$, cf.\ \eq\nr{master_2}.
Moreover, we remember that 
$\E^{ }_2 = \sqrt{m_{\phi}^2 + (\vec{p+k})^2}$, 
and can therefore use the trivial identity 
\be
 g(\vec{p+k}) = \int_{\vec{p}^{ }_2}
 (2\pi)^d \delta^{(d)}(\vec{k} + \vec{p} - \vec{p}^{ }_2) \,
 g(\vec{p}^{ }_2)
 \; \la{gp2}
\ee
to write the result in a somewhat more symmetric form (see below).

Let us now return to \eq\nr{raw1}. We had there the object 
$i \bsl{ P}$, which plays the role of the function~$f$, 
and according to \eq\nr{raw5} becomes
\be 
 i \bsl{ P} = i p^{ }_n  \gamma_0^{ } + i p^{ }_j  \gamma^{ }_j 
 \rightarrow \E^{ }_1 \gamma_{ }^0  + i p^{ }_j (-i \gamma_{ }^j) 
 \; \equiv \; \bsl{\mathcal{P}}
 \;, 
\ee
where we made use of the definition of the Euclidean 
Dirac-matrices in \eq\nr{Dirac_E} 
(\eq\nr{raw5} shows that any possible $i \bsl{ K}$
can also be replaced by $\bsl{\mathcal{K}}$).
Furthermore, two factors of 
$-1/2$ in \eq\nr{raw1} and \nr{raw5} combine into $1/4$. 
Renaming also $\mathcal{P}\to \mathcal{P}^{ }_1$ and inserting \eq\nr{gp2}, 
the spectral function finally becomes
\ba
 & &  \hspace*{-3.5cm}
  \rho(\mathcal{K}) = 
 \frac{ \nF{}^{-1}(k^0_{ }) }{4}   
 \int_{\vec{p^{ }_1},\vec{p^{ }_2}} 
  \frac{\bsl{\mathcal{P}}^{ }_{\!\! 1} \, \aR}{4\E^{ }_1 \E^{ }_2} 
 \nn 
 & \times \biggl\{ & \!\!\!
 (2\pi)^D \delta^{(D)}(\mathcal{P}^{ }_1+\mathcal{P}^{ }_2-\mathcal{K}) \, 
 \nF{1}\nB{2} 
 \; \Scatd \nn & + & \!\!\!
 (2\pi)^D \delta^{(D)}(\mathcal{P}^{ }_1-\mathcal{P}^{ }_2-\mathcal{K}) \, 
 \nF{1}(1+\nB{2})
 \; \Scatc \nn & + & \!\!\!
 (2\pi)^D \delta^{(D)}(\mathcal{P}^{ }_2-\mathcal{P}^{ }_1-\mathcal{K}) \,
 \nB{2}(1-\nF{1})
 \; \Scatb \nn & + & \!\!\!
 (2\pi)^D \delta^{(D)}(\mathcal{P}^{ }_1+\mathcal{P}^{ }_2+\mathcal{K} )\, 
 (1-\nF{1})(1+\nB{2})
 \biggr\}
 \;, \Scata  
 \la{pert1}
\ea
where the results of the other channels were added; 
$D \equiv d+1$; 
and we denoted
$\nF{i} \equiv \nF{}(\E^{ }_i)$, $\nB{i} \equiv \nB{}(\E^{ }_i)$.
The graphs in \eq\nr{pert1} illustrate the various processes that 
the energy-momentum constraints correspond to, 
with a dashed line for $\phi$, a solid 
for $L$, and a dotted for $N$. One immediate implication of 
these constraints is that for a positive $k^0_{ }$, the last of 
the four structures in \eq(\ref{pert1}) does not contribute at all. 
In general, depending on the particle masses, 
some of the other channels are also kinematically forbidden. 

The physics lesson to draw from \eq\nr{pert1} is that the spectral
function, as extracted here from an analytic continuation and cut of
a Euclidean correlator, represents real scatterings of on-shell
particles, whose distribution functions are given by the Bose and
Fermi distributions. The Bose and Fermi distributions appear in a form
reminiscent of a Boltzmann equation, save for the ``external'' line
carrying the momentum $\mathcal{K}$ which appears differently
(this is discussed in more detail in \se\ref{se:ppr}).
If we went to the 2-loop level, 
then there would also be virtual corrections, with the closed loops 
experiencing thermal modifications weighted by $\nB{}$ or $-\nF{}$.

As a final remark we note that 
the spectral function $\rho$ has the important property that, 
in a CP-symmetric situation, it is even in $\mathcal{K}$: 
\be
  \rho(-\mathcal{K}) = \rho(\mathcal{K}) 
 \;. \la{even}
\ee
(In contrast, bosonic spectral functions are odd in $\mathcal{K}$.)
Let us demonstrate this explicitly with the 2nd channel 
in \eq\nr{pert1}. Its energy-dependent part satisfies 
\ba
 \nF{}^{-1}(k^0_{ }) \delta(\E^{ }_1 - \E^{ }_2 - k^0_{ }) \nF{1}(1+\nB{2})
 \!\! & \stackrel{\mathcal{K}\to - \mathcal{K}}{\longrightarrow} & \!\!
 \nF{}^{-1}(-k^0_{ }) \delta(\E^{ }_1 - \E^{ }_2 + k^0_{ }) \nF{1}(1+\nB{2})
 \nn & = & \!\!\!
 \delta(\E^{ }_1 - \E^{ }_2 + k^0_{ })
 \frac{(e^{-\beta k^0_{ }} + 1)e^{\beta \E^{ }_2}}
 {(e^{\beta \E^{ }_1} + 1)(e^{\beta \E^{ }_2} - 1)}
 \nn & = & \!\!\!
  \delta(\E^{ }_1 - \E^{ }_2 + k^0_{ }) (e^{\beta k^0_{ } } + 1)
 \frac{e^{\beta (\E^{ }_2-k^0_{ })}}
 {(e^{\beta \E^{ }_1} + 1)(e^{\beta \E^{ }_2} - 1)}
 \nn & = & \!\!\!
 \delta(\E^{ }_1 - \E^{ }_2 + k^0_{ })\,\nF{}^{-1}(k^0_{ })\, 
 \frac{e^{\beta \E^{ }_1}}
 {(e^{\beta \E^{ }_1} + 1)(e^{\beta \E^{ }_2} - 1)}
 \nn & = & \!\!\!
 \nF{}^{-1}(k^0_{ })\, \delta(\E^{ }_2  - \E^{ }_1 - k^0_{ })\,
 \nB{2} (1-\nF{1})
 \;,
\ea
which is exactly the structure of the 3rd channel. 
The spatial change $\vec{k}\to -\vec{k}$ only has an effect 
on the three-dimensional $\delta$-function, turning it into that 
on the 3rd row of \eq(\ref{pert1}). Similarly, it can be checked 
that the 4th term goes over into the 1st term, and vice versa. 

There are a number of general remarks to make about the determination
of spectral functions of the type that we have considered here;  
these have been deferred to the end of appendix~A.


\subsection*{Appendix A: What if the internal lines are treated
non-perturbatively?}

\index{Thermal sums: non-perturbative case}
\index{Spectral representation}
\index{Propagator: HTL-resummed}

Above we made use of tree-level propagators, but in 
general the propagators need to be resummed
(cf.\ \se\ref{se:htl}), and have 
a more complicated appearance. It is then useful to 
express them as in the spectral representation of \eq\nr{spectral}. 
In particular, the scalar propagator can be written as
\be
 \langle \tilde \phi( K) \tilde \phi ( Q) \rangle_0^{ } =  
 \frac{\deltabar( K+ Q)}
 {k_n^2 + \vec{k}^2 + \Pi^{ }_\rmii{S}(k^{ }_n,\vec{k})}
 = 
 \;\deltabar( K+ Q) \, 
 \int_{-\infty}^{\infty} \! \frac{{\rm d} k^0_{ }}{\pi}
 \frac{\rho^{ }_\rmii{S}(k^0_{ },\vec{k})}{k^0_{ } - i k^{ }_n}
 \;,
\ee
whereas the fermion propagator contains two possible structures in 
the chirally symmetric case of a vanishing mass
(more general cases have 
been considered in ref.~\cite{hw}):
\ba
 \langle \tilde \psi( K) \,\bar{\!\tilde\psi} ( Q) \rangle_0^{ }
 & = &  
 \deltabar( K- Q) 
 \biggl[
  \frac{-i k^{ }_n \gamma^{ }_0}
  {k_n^2 + \vec{k}^2 + \Pi^{ }_\rmii{W}(k^{ }_n,\vec{k})} 
 + \frac{-i k^{ }_j \gamma^{ }_j}{k_n^2 + \vec{k}^2
 + \Pi^{ }_\rmii{P}(k^{ }_n,\vec{k})}
 \biggr]
 \nn 
 & = & 
 \deltabar( K- Q) 
 \biggl[
   i k^{ }_n \gamma^{ }_0  \int_{-\infty}^{\infty}
  \! \frac{{\rm d} k^0_{ }}{\pi}
   \frac{\rho^{ }_\rmii{W}(k^0_{ },\vec{k})}{k^0_{ } - i k^{ }_n}
   + i k^{ }_j \gamma^{ }_j \int_{-\infty}^{\infty}
 \! \frac{{\rm d} k^0_{ }}{\pi}
   \frac{\rho^{ }_\rmii{P}(k^0_{ },\vec{k})}{k^0_{ } - i k^{ }_n}
 \biggr]
 \;. \la{HTL_F_spec}
\ea
Here minus signs have been incorporated into the definitions of the 
spectral functions $\rho^{ }_\rmii{W}$ and $\rho^{ }_\rmii{P}$ for 
later convenience.
Let us carry out the steps from \eq\nr{raw1} to \nr{pert1} in this situation.  

The structure in \eq\nr{F_sum} now has the form
\be
 \mathcal{F} = 
 T \sum_{\{ p^{ }_n \}} \sum_{\rmii{F=W,P}}
 \int_{-\infty}^{\infty} \! \frac{{\rm d} \omega^{ }_1}{\pi} 
 \int_{-\infty}^{\infty} \! \frac{{\rm d} \omega^{ }_2}{\pi} 
 \frac{f^{ }_\rmii{F}(ip^{ }_n,i k^{ }_n,\vec{v})
 \, \rho^{ }_\rmii{F}(\omega^{ }_1,\vec{p}) 
 \, \rho^{ }_\rmii{S}(\omega^{ }_2,\vec{p+k})}
 {[\omega^{ }_1 - i p^{ }_n][\omega^{ }_2 - i (p^{ }_n+k^{ }_n)]}
 \;,  \la{eF_sum}
\ee
where the book-keeping function $f^{ }_\rmii{F}$ 
is again assumed to 
depend {linearly} on its arguments. We can write 
\ba
 \mathcal{F} & = &
 \sum_{\rmii{F=W,P}}
 \int_{-\infty}^{\infty} 
 \! \frac{{\rm d} \omega^{ }_1\,{\rm d} \omega^{ }_2}{\pi^2} 
 \,
 \rho^{ }_\rmii{F}(\omega^{ }_1,\vec{p})
 \rho^{ }_\rmii{S}(\omega^{ }_2,\vec{p+k})
 \nn & & \quad \times \, 
 T \sum_{ \{  p^{ }_n \} } T \sum_{r^{ }_n}
 \beta\, \delta(r^{ }_n -  p^{ }_n - k^{ }_n)
 \frac{f^{ }_\rmii{F}(ip^{ }_n,i k^{ }_n,\vec{v})}
 {[\omega^{ }_1 - i p^{ }_n][\omega^{ }_2 - i r^{ }_n]}
 \;. 
\ea
Employing \eqs\nr{insert_delta}, \nr{bsum} and \nr{fsum2}, 
as well as the time derivative of the last one,\footnote{
 We are somewhat sloppy here: a part of the sums leads
 to Dirac-$\delta$'s (cf.\ \eq\nr{ferm_complete}),
 which can give a contribution to $\mathcal{F}$. 
 That term is, however, independent of $k^{ }_n$ and thus 
 drops out when taking the discontinuity. 
 }  
\be
  T \sum_{ \{ p^{ }_n \} } 
 \frac{i p^{ }_n}{\omega^{ }_1 - i p^{ }_n}  
 e^{ - i p^{ }_n \tau} 
 = - \frac{{\rm d}}{{\rm d}\tau}
 \Bigl[ 
 \nF{}(\omega^{ }_1 ) e^{(\beta - \tau) \omega^{ }_1 }
 \Bigr]
 = \nF{}(\omega^{ }_1 )\, \omega^{ }_1 \, e^{(\beta - \tau) \omega^{ }_1 }
 \;, \quad 0 < \tau < \beta
 \;, 
 \la{fsum3}
\ee
we get
\ba
 \mathcal{F} & = & 
 \sum_{\rmii{F=W,P}}
 \int_{-\infty}^{\infty} 
 \! \frac{{\rm d} \omega^{ }_1\,{\rm d} \omega^{ }_2}{\pi^2} 
 \,
 \rho^{ }_\rmii{F}(\omega^{ }_1,\vec{p})
 \rho^{ }_\rmii{S}(\omega^{ }_2,\vec{p+k})
 \nn & & \quad \times \, 
   \int_0^\beta \! {\rm d}\tau \, e^{-i k^{ }_n \tau}  
   \,
   \nF{}(\omega^{ }_1)
   \, \nB{}(\omega^{ }_2)
   \, f^{ }_\rmii{F}(\omega^{ }_1,ik^{ }_n, \vec{v}) 
   \, e^{(\beta-\tau)\omega^{ }_1 + \tau \omega^{ }_2} 
 \;. \la{eraw2}
\ea
The $\tau$-integral can now be carried out, 
noting that $k^{ }_n$ is fermionic:  
\ba
  \int_0^\beta \! {\rm d}\tau \, 
 \nF{}(\omega^{ }_1)\, \nB{}(\omega^{ }_2) \, e^{\beta \omega^{ }_1} 
  e^{\tau (-i k^{ }_n - \omega^{ }_1 + \omega^{ }_2)}  
 & = &
 \frac{\nF{}(\omega^{ }_1)\, \nB{}(\omega^{ }_2) e^{\beta \omega^{ }_1}}
 {-i k^{ }_n - \omega^{ }_1 + \omega^{ }_2}
 \Bigl[
 - e^{\beta(\omega^{ }_2 - \omega^{ }_1)} - 1 
 \Bigr]
 \nn & = & 
 \frac{\nF{}(\omega^{ }_1)\, \nB{}(\omega^{ }_2)}
 {i k^{ }_n + \omega^{ }_1 - \omega^{ }_2} \Bigl[
 e^{\beta \omega^{ }_2} + e^{\beta \omega^{ }_1}
 \Bigr]
 \nn & = & 
 \frac{\nF{}(\omega^{ }_1)\, \nB{}(\omega^{ }_2)}
 {i k^{ }_n + \omega^{ }_1 - \omega^{ }_2}
 \Bigl[
 \nF{}^{-1}(\omega^{ }_1) + \nB{}^{-1}(\omega^{ }_2) 
 \Bigr]
 \nn & = & 
 \frac{1}{i k^{ }_n + \omega^{ }_1 - \omega^{ }_2}
 \Bigl[
 \nF{}(\omega^{ }_1) + \nB{}(\omega^{ }_2) 
 \Bigr]
 \;. \la{eraw3}
\ea

Finally we set $k^{ }_n \to - i (k^0_{ } + i 0^+_{ })$
and take the discontinuity. 
The appearance of $k^{ }_n$ inside $f^{ }_\rmii{F}$ 
can be handled like in \eq\nr{no_disc}.
Making use of \eq\nr{delta_disc}, 
the denominator in \eq\nr{eraw3} simply gets replaced with
$(-\pi)$ times a Dirac $\delta$-function, so that in total 
\ba
 & & \hspace*{-1.2cm}
  \im\Bigl\{  
 \mathcal{F}(i k^{ }_n \to k^0_{ } + i 0^+_{ }) \Bigr\} 
 \nn 
 & = &  
 -{\pi}
 \sum_{\rmii{F=W,P}}
 \int_{-\infty}^{\infty} 
 \! \frac{{\rm d} \omega^{ }_1\,{\rm d} \omega^{ }_2}{\pi^2} 
 \,
 \rho^{ }_\rmii{F}(\omega^{ }_1,\vec{p}) 
 \rho^{ }_\rmii{S}(\omega^{ }_2,\vec{p+k})
 \nn & & \quad \times \, 
 \Bigl[ \nF{}(\omega^{ }_1) + \nB{}(\omega^{ }_2) \Bigr]
 \delta(k^0_{ } + \omega^{ }_1 - \omega^{ }_2) 
 f^{ }_\rmii{F}(\omega^{ }_1,\omega^{ }_2-\omega^{ }_1,\vec{v})
 \nn & = & 
 -\frac{1}{2}
 \sum_{\rmii{F=W,P}}
 \int_{-\infty}^{\infty} 
 \! \frac{{\rm d} \omega^{ }_1\,{\rm d} \omega^{ }_2}{\pi^2} 
 \,
 \rho^{ }_\rmii{F}(\omega^{ }_1,\vec{p})
 \rho^{ }_\rmii{S}(\omega^{ }_2,\vec{p+k})
 \nn & & \quad \times \, 
 2\pi\, \delta(k^0_{ } + \omega^{ }_1 - \omega^{ }_2) 
 \, f^{ }_\rmii{F}(\omega^{ }_1,k^0_{ },\vec{v})
 \,  \nF{}^{-1}(k^0_{ })
 \,  \nB{}(\omega^{ }_2) 
 [1 - \nF{}(\omega^{ }_1)]
 \;, \hspace*{1cm} \la{eraw5}
\ea
where we parallelled the steps in \eq\nr{raw5}.
Finally, making use of \eq\nr{gp2}
and defining 
$\mathcal{P}^{ }_1 \equiv (\omega^{ }_1,\vec{p}) 
 \equiv (\omega^{ }_1,\vec{p}^{ }_1)$, 
$\mathcal{P}^{ }_2 \equiv (\omega^{ }_2,\vec{p}^{ }_2)$, the spectral function 
corresponding to \eq\nr{pert1} becomes
\ba
 & &  \hspace*{-1.5cm}
  \rho(\mathcal{K}) = 
 - \nF{}^{-1}(k^0_{ }) 
 \int_{\mathcal{P}^{ }_1}
 \int_{\mathcal{P}^{ }_2}
 \Bigl[ \omega^{ }_1 \gamma_{ }^0 \,\rho^{ }_\rmii{W}(\mathcal{P}^{ }_1)
   + \bsl{\vec{p}^{ }_1} \rho^{ }_\rmii{P}(\mathcal{P}^{ }_1) 
 \Bigr] \, \aR \, \rho^{ }_\rmii{S}(\mathcal{P}^{ }_2) 
 \nn 
 & \times \biggl\{ & \!\!\!
 (2\pi)^D \delta^{(D)}(\mathcal{P}^{ }_2-\mathcal{P}^{ }_1-\mathcal{K}) \,
 \nB{2}(1-\nF{1})
  \Scatb 
 \biggr\}
 \;, \la{pert2}
\ea
where $\bsl{\vec{p}^{ }_1} \equiv p^{ }_{1j} \gamma_{ }^j$, 
$\nF{i} \equiv \nF{}(\omega^{ }_i)$ and $\nB{i} \equiv \nB{}(\omega^{ }_i)$.
If we insert here the free spectral shape from 
\eq\nr{free_S_rho}, recalling the extra minus
sign that was incorporated into $\rho^{ }_\rmii{W}$ and $\rho^{ }_\rmii{P}$
in \eq\nr{HTL_F_spec},  
then it can be shown that this result goes over into \eq\nr{pert1}, with
the four channels originating from the on-shell points
$\omega^{ }_i = \pm \E^{ }_i$, $i=1,2$.

A few concluding remarks are in order: 

\bi

\item
Expressions such as \eq\nr{pert2} are useful particularly if the scalar
and fermion propagators are Hard Thermal Loop (HTL) resummed, 
cf.\ \se\ref{se:htl}. In that case $\rho^{ }_\rmii{W}$ and $\rho^{ }_\rmii{P}$
are given by \eq\nr{rho0}. 

\item
HTL resummed spectral functions contain in general two types of 
contributions. First of all, there are ``pole contributions'', represented by 
Dirac $\delta$-functions. In these contributions the pole locations are 
shifted from the free vacuum spectral functions 
by thermal mass corrections. Consequently, 
kinematic channels which would be forbidden in vacuum (such as a 
$1\to 2$ decay between three massless particles) may open up. 

\item
The second type of HTL corrections originates from a ``cut contribution''. 
An HTL resummed fermion or gauge field spectral function $\rho(\omega,k)$
has a non-zero continuous part in the spacelike domain $k > |\omega|$. 
Physically, this originates from real $2\leftrightarrow 1$
scatterings experienced by
such off-shell fields. Inserted into  \eq\nr{pert2} this turns the
full process into a real $2\to 2$ scattering, which tends to play an 
important role for the physics of nearly massless particles, because
$2\to 2$ processes are not kinematically 
suppressed even in the massless limit. 

\ei

A classic example of an HTL computation in which both 
``pole'' and ``cut'' contributions
play a role can be found in ref.~\cite{htl1}. Further processes, contributing
at the same order even though not accounted for just by using HTL spectral
functions, have been discussed in ref.~\cite{lpm3}. A complete
leading-order computation of the observable considered in the present section,
related to right-handed fermions interacting with the Standard Model
particles through Yukawa interactions, 
is presented in refs.~\cite{bb2,sum3}, and a similar analysis for
the production rate of photons from a QCD plasma can be found
in refs.~\cite{photon1,photon2}.
We return to some of these issues in \se\ref{se:ppr}.

\index{Hard Thermal Loops (HTL)}

\newpage 

\subsection{Real-time formalism}
\la{se:realtime}

\index{Real-time formalism}

In the previous section, we considered a particular spectral function, 
obtained from the Euclidean correlator in \eq\nr{ex_PiE}
through the basic relation in \eq\nr{ffinal}. The question may be posed, 
however, whether it really is necessary to go through Euclidean
considerations at all. 
It turns out that, within perturbation theory, the
answer is negative: in the so-called real-time formalism, real-time
observables can be directly expressed as Feynman diagrams containing
real-time propagators. The price to pay for this simplification
is that the field content of the theory gets effectively ``doubled'' 
and, in a general situation, 
every propagator turns into a $2\times 2$ matrix, and every 
vertex splits into multiple vertices. 

A full-fledged formulation of the real-time formalism proceeds 
through the Schwinger-Keldysh or closed time-path framework; reviews 
can be found in refs.~\cite{sk1,sk2}. A frequently appearing concept
is that of Kadanoff-Baym equations, which are analogues of 
Schwinger-Dyson equations within this formalism.  
In the following, we only provide a short motivation for the field
doubling, and then demonstrate  how the result
of \eq(\ref{pert1}) can be obtained
directly within the real-time formalism. 

%
\subsection*{Basic definitions}

\index{Non-equilibrium ensemble}

One advantage of the real-time formalism is that it also applies
to systems out of equilibrium. In quantum statistical mechanics
a general out-of-equilibrium situation
is described by a {\em density matrix}, denoted by $\hat{\rho}(t)$.
The density matrix is assumed normalized such that 
$
 \tr( \hat{\rho} )= 1 
$, 
and statistical expectation values are defined as
\be \index{Density matrix}
 \bigl\langle
  \hat{O}(t^{ }_1,\vec{x}^{ }_1)
  \, \hat{O}(t^{ }_2,\vec{x}^{ }_2)\, ... 
 \bigr\rangle\; \equiv \;
 \tr
 \Bigl[  \hat\rho(t)
   \,\hat{O}(t^{ }_1,\vec{x}^{ }_1)
   \,\hat{O}(t^{ }_2,\vec{x}^{ }_2)\, ... 
 \Bigr] 
 \;, \la{noneq}
\ee
where $ \hat{O} $ is a Heisenberg operator defined like 
in \eq\nr{op_defs}. The same 2-point functions as in \se\ref{se:diff_G}
can be considered in this general ensemble, and some of the operator
relations also continue to hold, such as 
$ 
  \Pi^\iT_{ } = 
  - i \Pi^\iR_{ } + 
 \Pi^{<} = 
  - i \Pi^\iA_{ } + 
 \Pi^{>}_{ }
$.

An important difference between the out-of-equilibrium and equilibrium cases 
is that in the former situation the considerations leading to the KMS
relation, cf.\ \eqs\nr{der_kms_1} and \nr{der_kms_2} for the bosonic case, 
no longer go through. 
However, we can still work out the trace in \eq\nr{noneq}
in a given basis and 
learn something from the outcome. 

\index{Wightman function}

Consider the same Wightman function $\Pi^{>}_{ }$ 
as in \eq\nr{der_kms_1}. With a view of obtaining
a perturbative expansion, we now choose as the basis not energy
eigenstates, but rather eigenstates of elementary field operators; 
for the moment we denote these by $|\alpha_i\rangle$.
Simplifying also the operator notation
somewhat from that in \se\ref{se:diff_G}, we can write  
\ba
  \Pi^{>}_{ }(t) & \equiv & 
  \tr\Bigl[ \hat\rho(t) \, e^{i \hat{H}t}\, \hat{O}(0)
  \, e^{- i \hat{H}t}  \, \hat{O}(0) \Bigr]
 \nn 
 & = & 
 \int\! \Pi_{i=1}^{5} {\rm d}\alpha^{ }_i 
 \, 
 \langle \alpha^{ }_1 | 
 \hat\rho(t) 
 | \alpha^{ }_2 \rangle
 \, 
 \langle \alpha^{ }_2 |
 e^{i \hat{H}t}
 | \alpha^{ }_3 \rangle
 \, 
 \langle \alpha^{ }_3 |
 \hat{O}(0)
 | \alpha^{ }_4 \rangle
 \, 
 \langle \alpha^{ }_4 |
 e^{- i \hat{H}t}  
 | \alpha^{ }_5 \rangle
 \, 
 \langle \alpha^{ }_5 |
 \hat{O}(0)
 | \alpha^{ }_1 \rangle
 \;. \nn 
\ea
If the operators $\hat{O}$ contain only the field operators $\hat\alpha$ 
and no conjugate
momenta, then we can directly write 
$
 \langle \alpha^{ }_i | \hat{O}[\hat\alpha] | \alpha^{ }_j \rangle = 
 O[\alpha^{ }_j] 
 \delta^{ }_{ \alpha^{ }_i, \alpha^{ }_j}
$. 
For the time evolution, we insert the usual Feynman path 
integral, 
\be
 \langle \alpha^{ }_4 | e^{-i \hat{H}t} | \alpha^{ }_5 \rangle = 
 \int_{\alpha(0)=\alpha^{ }_5}^{\alpha(t) = \alpha^{ }_4} \! 
 \mathcal{D}\alpha\, e^{i \mathcal{S}^{ }_\iiM }
 \;, 
\ee
while the ``backward'' time evolution 
$
  \langle \alpha^{ }_2 | e^{i \hat{H}t} | \alpha^{ }_3 \rangle
$
is obtained from the Hermitian (complex) conjugate of this relation. 
Denoting the ``forward-propagating'' 
field interpolating between $\alpha^{ }_5$ and $\alpha^{ }_4$
now by $\phi^{ }_1$, and that 
interpolating between $\alpha^{ }_3$ and $\alpha^{ }_2$
by $\phi^{ }_2$, we thereby get
\be
 \Pi^{>}_{ }(t) = 
 \int\! \mathcal{D}\phi^{ }_{1} \mathcal{D}\phi^{ }_{2} \,
 O[\phi^{ }_{2}(t)]\, 
 O[\phi^{ }_1(0)]
 e^{\, i \mathcal{S}^{ }_\iiM[\phi^{ }_1] - i \mathcal{S}^{ }_\iiM[\phi^{ }_2]}
 \, \langle \phi^{ }_1(0) | \hat{\rho}(t) | \phi^{ }_2(0) \rangle 
 \;. \la{rtm1}
\ee
Note that $\phi^{ }_{1}(t) = \phi^{ }_{2}(t) = \alpha^{ }_3 = \alpha^{ }_4$ 
in this example because $t$ 
is the largest time value appearing; 
however $\phi^{ }_{2}(0) \neq \phi^{ }_{1}(0)$
and both are integrated over.
It is helpful to use $\phi^{ }_2(t)$ rather than $\phi^{ }_1(t)$
inside $O[\phi^{ }_{2}(t)]$
in \eq\nr{rtm1}, because this makes it explicit that $O[\phi^{ }_2(t)]$ stands
to the left of the operator 
$O[\phi^{ }_1(0)]$, as is indeed implied by the definition of 
the Wightman function $\Pi^{>}_{ }(t)$.
One should think of the field
$\phi^{ }_1$ as corresponding to the operators
positioned on the right and 
with time arguments increasing to the left, followed by $\phi^{ }_2$ for the 
operators positioned on the left.
 
A similar computation for the other Wightman function yields
\be
 \Pi^{<}_{ }(t) = 
 \int\! \mathcal{D}\phi^{ }_{1} \mathcal{D}\phi^{ }_{2} \,
 O[\phi^{ }_2(0)]\,
 O[\phi^{ }_{1}(t)] 
 e^{\, i \mathcal{S}^{ }_\iiM[\phi^{ }_1] - i \mathcal{S}^{ }_\iiM[\phi^{ }_2]}
 \, \langle \phi^{ }_1(0) | \hat{\rho}(t) | \phi^{ }_2(0) \rangle 
 \;. \la{rtm2}
\ee
This time we have indicated the field with the largest time
argument by $\phi^{ }_1(t)$ rather than $\phi^{ }_2(t)$, because the
corresponding operator stands to the utmost right, i.e.\ closest
to the origin of time flow. Note that within \eq\nr{rtm2}, 
$ O[\phi^{ }_2(0)] $ and 
$ O[\phi^{ }_{1}(t)] $ are just complex numbers and ordering
plays no role (in the bosonic case), so we could also write
$
 \Pi^{<}_{ }(t) = 
 \langle
 O[\phi^{ }_{1}(t)] \,
 O[\phi^{ }_2(0)]
 \rangle 
$.
Here 
$\langle ... \rangle$
refers to an expectation value in the sense 
of the Schwinger-Keldysh functional integral, 
\be \index{Schwinger-Keldysh formalism}
 \langle ... \rangle \; \equiv \;  
 \int\! \mathcal{D}\phi^{ }_{1} \mathcal{D}\phi^{ }_{2} \,
 (...) \,
 e^{\, i \mathcal{S}^{ }_\iiM[\phi^{ }_1] - i \mathcal{S}^{ }_\iiM[\phi^{ }_2]}
 \, \langle \phi^{ }_1(0) | \hat{\rho}(t) | \phi^{ }_2(0) \rangle 
 \;. \la{SK_exp}
\ee

If $\hat{\rho}$ happens to be a time-independent
thermal density matrix, $\hat{\rho} = e^{-\beta \hat{H}} / \mathcal{Z}$, 
then the remaining expectation value 
$
 \langle \phi^{ }_1(0) | \hat{\rho}(t) | \phi^{ }_2(0) \rangle
$
can be represented
as an imaginary-time path integral as was discussed for 
a scalar field in \se\ref{ss:pi_sft}. For many formal 
considerations it is however not necessary to write down this part
explicitly. 

The lesson to be drawn from \eqs\nr{rtm1} and \nr{rtm2} is 
that the two Wightman functions $\Pi^{>}$ and $\Pi^{<}$
are independent objects if 
$\hat{\rho}$ is non-thermal, and that representing them as
path integrals necessitates a doubling of the field
content of the theory ($\phi\to \{ \phi^{ }_1,\phi^{ }_2 \} $).  

If we specialize to the case in which the operators in \eqs\nr{rtm1}
and \nr{rtm2} are directly elementary fields, rather than composite
operators, then it is conventional to assemble these propagators into
a $2\times 2$ matrix. If we add a time-ordered structure, 
\ba
 && \hspace*{-2cm}
 \theta(t^{ }_2 - t^{ }_1)\, \hat{\phi}(t^{ }_2)\,\hat{\phi}(t^{ }_1) + 
 \theta(t^{ }_1 - t^{ }_2)\, \hat{\phi}(t^{ }_1)\,\hat{\phi}(t^{ }_2)  
 \nn 
 & = & 
 \theta(t^{ }_2 - t^{ }_1)\, e^{i \hat{H}t^{ }_2} 
 \, \hat{\phi}(0) \, e^{-i \hat{H}(t^{ }_2 - t^{ }_1)}
 \, \hat{\phi}(0) \, e^{- i \hat{H} t^{ }_1} 
 \nn
 & + &  
 \theta(t^{ }_1 - t^{ }_2)\, e^{i \hat{H}t^{ }_1} 
 \, \hat{\phi}(0) \, e^{-i \hat{H}(t^{ }_1 - t^{ }_2)}
 \, \hat{\phi}(0) \, e^{- i \hat{H} t^{ }_2} \;, 
\ea
it corresponds to time evolution  
along the forward-propagating branch, 
denoted above by the field~$\phi^{ }_1$. Similarly, an anti-time-ordered
propagator can be represented in terms of the $\phi^{ }_2$-field. 
The general propagator is then 
\be
 \Biggl( \begin{array}{cc}
   \displaystyle
   \langle \phi^{ }_1(t) \phi^{ }_1(0) \rangle & 
   \langle \phi^{ }_1(t) \phi^{ }_2(0) \rangle \\[1mm]
   \langle \phi^{ }_2(t) \phi^{ }_1(0) \rangle & 
   \langle \phi^{ }_2(t) \phi^{ }_2(0) \rangle \\
 \end{array}
 \Biggr)
 \; = \; 
 \Biggl( \begin{array}{cc}
   \displaystyle
    \Pi_\phi^\iT(t) & \Pi_\phi^<(t) \\[1mm] 
   \displaystyle
    \Pi_\phi^>(t) & \Pi_\phi^\ibT(t) 
 \end{array}
 \Biggr)
 \;, \la{matr_prop}
\ee
where $\overline{T}$ denotes anti-time-ordering. 
The action 
$
 \mathcal{S}^{ }_\iM[\phi^{ }_1] - \mathcal{S}^{ }_\iM[\phi^{ }_2]
$
contains vertices for both types of fields, and the non-diagonal matrix
structure of \eq\nr{matr_prop} implies that when interactions are included, 
both types of vertices contribute to a given observable. 

In the literature,
the field basis introduced above 
is referred to as the 1/2-basis. There is another
possible choice, referred to as the $r/a$-basis, which is beneficial
for many practical computations. It is obtained by the linear 
transformation
\be
 \phi^{ }_r \equiv \fr12 (\phi^{ }_1 + \phi^{ }_2) \;, \quad
 \phi^{ }_a \equiv \phi^{ }_1 - \phi^{ }_2
 \;. \la{ra_def}
\ee
Consequently, inserting the 1/2 propagators from \eq\nr{matr_prop}, we get 
\ba
 \langle \phi^{ }_r(t) \phi^{ }_r(0) \rangle & = & 
 \fr14 \Bigl( \Pi_\phi^\iT + \Pi_\phi^\ibT
  + \Pi_\phi^{>} + \Pi_\phi^{<}\Bigr)  
 \; = \;
 \fr12 \bigl(  \Pi_\phi^{>} + \Pi_\phi^{<} \bigr) \; = \; \Delta^{ }_\phi(t)
 \;, \la{ra_prop1} \\ 
 \langle \phi^{ }_r(t) \phi^{ }_a(0) \rangle & = & 
 \fr12 \Bigl( \Pi_\phi^\iT - \Pi_\phi^\ibT
  + \Pi_\phi^{>} - \Pi_\phi^{<}\Bigr)  
 \; = \; 
 \theta(t) \bigl(  \Pi_\phi^{>} - \Pi_\phi^{<} \bigr)
 \; = \; -i \Pi^\iR_\phi(t)
 \;, \la{ra_prop2}
\ea
and similarly 
$
  \langle \phi^{ }_a(t) \phi^{ }_r(0) \rangle = -i \Pi^\iA_\phi(t)
$
and
$
  \langle \phi^{ }_a(t) \phi^{ }_a(0) \rangle = 0.
$

Among the advantages of the $r/a$-basis are that the $aa$ 
element vanishes, and that closed loops containing only
the advanced
$
 \langle \phi^{ }_a(t) \phi^{ }_r(0) \rangle
$
or the retarded
$
 \langle \phi^{ }_r(t) \phi^{ }_a(0) \rangle
$
also vanish. In addition, the statistical function $\Delta^{ }_\phi$, 
containing the Bose distribution in the bosonic case
(cf.\ \eq\nr{bDelta}), is the only
element surviving in the classical limit (because it is not 
proportional to a commutator), and may thus dominate the 
dynamics if we consider a soft 
regime $\E\ll T$ such as in the situation described in \se\ref{se:Linde}
(cf.\ ref.~\cite{schz} for a detailed discussion).

Let us conclude by remarking that at higher orders of perturbation
theory, the real-time formalism quickly becomes technically rather
complicated, and for a long time only leading-order results
existed. The past few years have, however, witnessed significant
progress in the field, which is related in particular to the handling
of soft contributions in the computations, as alluded to above. 
Examples of next-to-leading order computations can be found
in refs.~\cite{sch_2}--\hspace*{-1.1mm}\cite{photon}.

\subsection*{Practical illustration}

In order 
to illustrate how the real-time formalism works, 
let us return to the 1-loop spectral function of the operator
coupling to a right-handed fermion in
the Standard Model, discussed in \se\ref{se:NI_rho}. 
Concretely, we wish to obtain the spectral function 
corresponding to \eq\nr{raw1}. In order to do this in the $r/a$-basis, 
let us first go back to the Lagrangian in \eq\nr{LM}, expressed
with the field components of \eq\nr{doubs}. We need to 
write the interaction in the form 
$
 \mathcal{S}^{ }_\iM[\phi^{ }_1] - \mathcal{S}^{ }_\iM[\phi^{ }_2]
$
that appears in
the exponential in \eq\nr{SK_exp}, so we insert
$
 \phi^{ }_1 = \phi^{ }_r + \phi^{ }_a / 2
$, 
$
 \phi^{ }_2 = \phi^{ }_r - \phi^{ }_a / 2
$
from \eq\nr{ra_def}, 
and similarly for the fermions. For the two structures in 
\eq\nr{LM} this yields
\ba
 \bar{\ell}^{ }_1\, \phi^{ }_1\, \aR N^{ }_1 - 
 \bar{\ell}^{ }_2\, \phi^{ }_2\, \aR N^{ }_2
 & = & 
 \bigl( 
    \bar{\ell}^{ }_r \phi^{ }_a + \bar{\ell}^{ }_a \phi^{ }_r
 \bigr)
 \aR N^{ }_r 
 + 
 \Bigl( 
    \bar{\ell}^{ }_r \phi^{ }_r + \frac{ \bar{\ell}^{ }_a \phi{ }_a }{4} 
 \Bigr) \aR N^{ }_a
 \;, \la{SM1} \\ 
 \bar{N}^{ }_1\, \phi^{ }_1\, \aL \ell^{ }_1 -  
 \bar{N}^{ }_2\, \phi^{ }_2\, \aL \ell^{ }_2
 & = & 
 \bar{N}^{ }_r \aL 
 \bigl( 
   \phi^{ }_a \ell^{ }_r + \phi^{ }_r \ell^{ }_a
 \bigr) 
 + 
 \bar{N}^{ }_a \aL 
 \Bigl(
   \phi^{ }_r \ell^{ }_r + \frac{ \phi^{ }_a \ell^{ }_a }{4}  
 \Bigr)
 \;. 
\ea 
The retarded correlator is obtained by considering the operators
coupling to $\bar{N}^{ }_a$ and $N^{ }_r$. We observe
that the part $\phi^{ }_a \ell^{ }_a  / 4 $ plays no role, because
the propagators 
$
 \langle \phi^{ }_a \phi^{ }_a \rangle
$
and 
$
 \langle \ell^{ }_a \bar{\ell}^{ }_a \rangle
$
vanish. The part $\phi^{ }_r \ell^{ }_r$ 
can be contracted with both operators coupling to 
$\aR N^{ }_r$ in \eq\nr{SM1}, so \eq\nr{raw1}
is replaced by 
\be
 -i \Pi^\iR_{ } (\mathcal{K})
  =  \frac{1}{2}
 \int_\mathcal{X} e^{i\mathcal{K}\cdot \mathcal{X} }
 \, \aL \, 
 \Bigl\{ 
 \langle \ell^{ }_r(\mathcal{X})\, \bar\ell^{ }_r(0) \rangle 
 \, 
 \langle \phi^{ }_r(\mathcal{X})\, \phi^{ }_a (0) \rangle
 + 
 \langle \ell^{ }_r(\mathcal{X})\, \bar\ell^{ }_a(0) \rangle 
 \, 
 \langle \phi^{ }_r(\mathcal{X})\, \phi^{ }_r (0) \rangle
 \Bigr\} \, \aR 
 \;. \la{raw1_new}
\ee 
Going to momentum space; inserting propagators from 
\eqs\nr{ra_prop1} and \nr{ra_prop2}; noting that $\Delta$ is real; 
and taking the imaginary part, we obtain
\ba
 \im \Pi^\iR_{ }(\mathcal{K})
 & = &  
 \frac{1}{2} \int_\mathcal{K} 
 \Bigl\{
   \Delta^{ }_{\ell}(\mathcal{P}) \im \Pi^\iR_{\phi} (\mathcal{K-P}) 
 + \im \Pi^\iR_{\ell}(\mathcal{P}) \Delta^{ }_{\phi} (\mathcal{K-P})
 \Bigr\} 
 \nn 
 & = & 
 \frac{1}{2} 
 \int_{\mathcal{P}^{ }_1,\mathcal{P}^{ }_2}
 (2\pi)^D_{ }\delta^{(D)}_{ }(\mathcal{P}^{ }_1 + \mathcal{P}^{ }_2 - K)
 \Bigl\{ 
   \Delta^{ }_{\ell}(\mathcal{P}^{ }_1)
   \rho^{ }_{\phi}(\mathcal{P}^{ }_2)
 + 
   \rho^{ }_{\ell}(\mathcal{P}^{ }_1)
   \Delta^{ }_{\phi}(\mathcal{P}^{ }_2)
 \Bigr\} 
 \;. \hspace*{5mm} \la{real-time-res}
\ea 
Here we have introduced a second momentum variable 
by inserting the relation
\ba
1 
 & = & 
 \int_{\mathcal{P}^{ }_2} \!\!\!
 (2\pi)^D \delta^{(D)}( \mathcal{P}^{ }_1 + \mathcal{P}^{ }_2 - \mathcal{K}) 
\ea
into the integral, and identified $\im \Pi^\iR_{ } = \rho$.
For simplicity we have also hid
the chiral projectors 
into the definition of the lepton propagator.  

According to  
\eq\nr{bLSrel} and the line below \eq\nr{fLSrel} we can write 
(with
$
 \mathcal{P}^{ }_i \equiv (\omega^{ }_i,\vec{p}^{ }_i)
$)
\be
 \Delta^{ }_{\phi}(\mathcal{P}^{ }_2) = 
 \bigl[ 1 + 2\nB{}(\omega^{ }_1)\bigr]\rho^{ }_{\phi}(\mathcal{P}^{ }_2)
 \;, \quad
 \Delta^{ }_{\ell}(\mathcal{P}^{ }_1) = 
 \bigl[ 1 - 2\nF{}(\omega^{ }_1)\bigr]\rho^{ }_{\ell}(\mathcal{P}^{ }_1)
 \;. 
\ee
Combining the two terms, \eq\nr{real-time-res} thereby becomes
\ba
 \rho(\mathcal{K})
 & = & 
 \int_{\mathcal{P}^{ }_1,\mathcal{P}^{ }_2} \!\!\!
 (2\pi)^D_{ } \delta^{(D)}_{ }
 ( \mathcal{P}^{ }_1 + \mathcal{P}^{ }_2 - \mathcal{K})
 \Bigl[ 1 - \nF{}(\omega^{ }_1) + \nB{}(\omega^{ }_2) \Bigr]
 \, \rho^{ }_\ell(\mathcal{P}^{ }_1) 
 \, \rho^{ }_\phi(\mathcal{P}^{ }_2)
 \;.  \hspace*{5mm} \label{rhoreal}
\ea

In order to make \eq\nr{rhoreal} more explicit, we insert
the free spectral functions (cf.~\eqs\nr{free_S_rho} and \nr{rho_F_free}), 
\ba
 \rho^{ }_\phi(\omega^{ }_2,\vec{p}^{ }_2)
 & \equiv &
  \frac{\pi}{2 \E^{ }_{2}}
 \Bigl[ \delta(\omega^{ }_2 - \E^{ }_{2}) - 
 \delta(\omega^{ }_2 + \E^{ }_{2})
 \Bigr] 
 \;, \la{rhos_free1} \\
 \rho^{ }_\ell(\omega^{ }_1,\vec{p}^{ }_1)
 & \equiv &
 \frac{\pi }{2\E^{ }_1} \, \aL \, \bsl{\mathcal{P}^{ }_1} \, \aR \, 
 \Bigl[ \delta(\omega^{ }_1 - \E^{ }_1)
  - \delta(\omega^{ }_1 + \E^{ }_1) \Bigr]
 \;, \la{rhos_free2}
\ea
where $\E^{ }_1$ and $\E^{ }_2$ are defined in accordance 
with \eq\nr{E_1_E_2}
(but with spatial momenta adjusted as appropriate).
Further re-organizing the phase space distributions 
in analogy with \eq\nr{raw5}, 
\be
 \delta(\omega^{ }_1 + \omega^{ }_2 - k_{ }^0 )
 \bigl[
   1  - \nF{}(\omega^{ }_1) + \nB{}(\omega^{ }_2)
 \bigr]
 \; = \; 
 \delta(\omega^{ }_1 + \omega^{ }_2 - k_{ }^0 )
 \, \nF{}^{-1}(k_{ }^0) \, \nF{}(\omega^{ }_1) \, \nB{}(\omega^{ }_2) 
 \;, 
\ee
we arrive at the result
\ba
 \rho(\mathcal{K}) & = & 
 \nF{}^{-1}(k^0_{ }) 
 \int_{-\infty}^{\infty} \! \frac{{\rm d}\omega^{ }_1}{2\pi}
 \int_{-\infty}^{\infty} \! \frac{{\rm d}\omega^{ }_2}{2\pi}
 \int_{\vec{p^{ }_1,p^{ }_2}} 
 (2\pi)^D \delta^{(D)}( \mathcal{P}^{ }_1 + \mathcal{P}^{ }_2 - \mathcal{K})
 \, \nF{}(\omega^{ }_1) \, \nB{}(\omega^{ }_2)
 \nn 
 & \times &
 \frac{\pi^2 }{4\E^{ }_1\E^{ }_2}
 \bsl{\mathcal{P}^{ }_1} \aR 
 \Bigl[ \delta(\omega^{ }_1 - \E^{ }_1)
  - \delta(\omega^{ }_1 + \E^{ }_1) \Bigr]
 \Bigl[ \delta(\omega^{ }_2 - \E^{ }_2)
  - \delta(\omega^{ }_2 + \E^{ }_2) \Bigr]
 \;.  
\ea
If we now integrate over $\omega^{ }_1$ and $\omega^{ }_2$, re-adjust the 
notation so that $\mathcal{P}^{ }_i \equiv (\E^{ }_i,\vec{p}^{ }_i)$,  
and in addition make the substitution $\vec{p}^{ }_i \to - \vec{p}^{ }_i$ 
where necessary, we obtain 
\ba
 & &  \hspace*{-3.5cm}
  \rho(\mathcal{K}) = 
 \frac{ \nF{}^{-1}(k^0_{ }) }{4}   
 \int_{\vec{p^{ }_1},\vec{p^{ }_2}} 
  \frac{\bsl{\mathcal{P}}^{ }_{\!\! 1} \, \aR}{4\E^{ }_1 \E^{ }_2} 
 \nn 
 & \times \biggl\{ & \!\!\!
 (2\pi)^D \delta^{(D)}(\mathcal{P}^{ }_1+\mathcal{P}^{ }_2-\mathcal{K}) \, 
 \nF{}(\E^{ }_1)\nB{}(\E^{ }_2) 
 \nn[-0.5mm] & - & \!\!\!
 (2\pi)^D \delta^{(D)}(\mathcal{P}^{ }_1-\mathcal{P}^{ }_2-\mathcal{K}) \, 
 \nF{}(\E^{ }_1)\nB{}(-\E^{ }_2)
 \nn[1.5mm] & + & \!\!\!
 (2\pi)^D \delta^{(D)}(\mathcal{P}^{ }_1-\mathcal{P}^{ }_2+\mathcal{K}) \,
 \nF{}(-\E^{ }_1)\nB{}(\E^{ }_2)
 \nn & - & \!\!\!
 (2\pi)^D \delta^{(D)}(\mathcal{P}^{ }_1+\mathcal{P}^{ }_2+\mathcal{K} )\, 
 \nF{}(-\E^{ }_1)\nB{}(-\E^{ }_2)
 \biggr\}
 \;.
 \la{pertX}
\ea
This 
becomes identical with \eq\nr{pert1} upon using the relations
\be
 \nF{}(-\E^{ }_1) = 1 - \nF{}(\E^{ }_1)
 \;, \quad
 \nB{}(-\E^{ }_2) = -1 - \nB{}(\E^{ }_2)
 \;.
\ee

The above example confirms our expectation that real-time quantities may
indeed be determined through the 
real-time formalism.  The imaginary-time formalism is, however,
equally valid for problems in thermal equilibrium, and applicable
on the non-perturbative level as well. Within perturbation theory, 
the main difference between the two formalisms is that in the
imaginary-time case Matsubara sums need to be carried out 
before taking the discontinuity, but there is only one expression 
under evaluation (cf.\ \eq\nr{raw1}), 
whereas in the real-time case
only integrations appear like in vacuum computations, 
with the price that there are more terms
(cf.\ \eq\nr{raw1_new}). 

\newpage 

\subsection{Hard Thermal Loops}
\la{se:htl}

\index{Hard Thermal Loops (HTL)}

For ``static'' observables, we realized in \se\ref{se:naive_lam2}
that the perturbative series suffers from infrared
divergences. However, as discussed in \se\ref{se:Linde},
in weakly coupled theories these divergences
can only be associated with bosonic Matsubara zero modes. They can therefore
be isolated by constructing an effective field theory for the 
bosonic Matsubara zero modes, as we did in \se\ref{se:DR_QCD}.

The situation is more complicated in the case of real-time
observables discussed in the present chapter. 
Indeed, as \eq\nr{ancont} shows, the dependence on {\em all} 
Matsubara modes is needed in order to carry out the analytic continuation 
leading to the spectral function, even if we were only interested in its 
behaviour at small frequencies $|k^0_{ }| \ll \pi T$. (The same holds also 
in the opposite direction: as the sum rule in \eq\nr{sum_rule} shows, 
the information contained in the Matsubara zero mode is spread out 
to {\em all} $k^0_{ }$'s in the Minkowskian formulation.) Therefore, it is 
non-trivial to isolate the soft/light degrees of freedom for 
which to write down the most general effective Lagrangian.\footnote{%
 The corresponding discussion in the real-time formalism, introduced 
 in \se\ref{se:realtime}, can be found in ref.~\cite{schz}.
 } 

Nevertheless, it turns out that the dimensionally reduced effective 
field theory of \se\ref{se:DR_QCD} {\em can} to some extent be 
generalized to real-time observables as well. In the case
of QCD, the generalization is known as the {\em Hard Thermal Loop} 
effective theory. The effective theory dictates what kind of 
resummed propagators should be used for instance in the computation
of \se\ref{se:NI_rho}, in order to alleviate infrared problems 
appearing in perturbative computations. 
An example of a computation showing that (logarithmic) infrared 
divergences get cancelled this way can be found in 
ref.~\cite{by}. 

More precisely, Hard Thermal Loops (HTL) can
operationally be defined via the following steps that refer 
to the computation of 2 or higher-point 
functions~\cite{ht1}--\hspace*{-1.1mm}\cite{ht4}:
\bi
 \item
 Consider ``soft'' external frequencies and momenta: 
 $|k^0_{ }|, |\vec{k}| \sim gT$. 

 \item
 Inside the loops, sum over all Matsubara frequencies $p^{ }_n$.

 \item
 Subsequently, integrate over ``hard'' spatial loop momenta, 
 $|\vec{p}|\gsim \pi T$, Taylor-expanding the result 
 to leading non-trivial order 
 in $|k^0_{ }|/|\vec{p}|$, $|\vec{k}|/|\vec{p}|$.
\ei 

\noindent
The soft momenta $|k^0_{ }|, |\vec{k}|$ 
are the analogues of the small mass $m$
considered in \se\ref{se:Linde}, and the scale $\sim \pi T$
plays
the role of the heavy mass $M$. According to \eq\nr{param_error}, 
the parametric error made through a given truncation
might be expected to be $\sim (g/\pi)^k$ with some $k > 0$, however
as will be discussed below this is unfortunately 
difficult to establish in general.

In order to illustrate the procedure, let us compute the gauge field 
self-energy in this situation. The computation is much like that 
in \se\ref{se:mmeff_gluon}, except that now we keep the
external momentum ($ K$) non-zero while carrying out the 
Matsubara sum, because the full dependence on $k^{ }_n$ is needed
for the analytic continuation. 
It is crucial to take $k^0_{ },\vec{k}$ soft only 
{\em after the analytic continuation}. 

As a starting point, we take the self-energy in Feynman
gauge, $\Pi^{ }_{\mu\nu}( K)$, as defined in \eq\nr{Fprop_Pi}. 
This will be interpreted as being a part of an ``effective action'', 
\be
 S^{ }_\rmi{eff} = 
 \Tint{K}
 \fr12 \tilde A^a_\mu (K)\, 
 \Bigl[
    K^2 \delta^{ }_{\mu\nu} - K^{ }_\mu K^{ }_\nu 
          + \frac{1}{\xi} K^{ }_\mu K^{ }_\nu 
 + \Pi^{ }_{\mu\nu}( K)
 \Bigr] \, \tilde A^a_\nu(-K) + \ldots
 \;. \la{htl_Seff}
\ee
Summing together 
results from \eqs\nr{Gtadpole}, \nr{Gloop}, \nr{DD2}, 
\nr{cloop2} and \nr{psiloop_2}, setting the fermion mass to zero
for simplicity, and expressing the spacetime dimensionality 
as $D \equiv d+1$, the 1-loop self-energy reads
\ba \index{Self-energy: gluon}
 \Pi^{ }_{\mu\nu}(K) & = & 
 \frac{g^2 \CA}{2} 
 \Tint{P} 
 \frac{ 
   \delta^{ }_{\mu\nu}\bigl[ -4 K^2 + 2 (D-2) P^2 \bigr]
   + (D+2) K^{ }_\mu K^{ }_\nu - 4 (D-2) P^{ }_\mu P^{ }_\nu
 }{P^2(K-P)^2}
   \nn &  - &  
 g^2 \Nf\, \Tint{\{ P \} }
 \frac{ 
   \delta^{ }_{\mu\nu} \bigl[ -K^2 + 2 P^2 \bigr]
   + 2 K^{ }_\mu K^{ }_\nu - 4 P^{ }_\mu P^{ }_\nu  
 }{P^2(K-P)^2}
 \;. \la{Pi_xi}
\ea 
The bosonic part is discussed in appendix A; here
we focus on the fermionic part.  

Consider first the spatial components, $\Pi_{ij}$. Shifting $P\to K-P$ 
in one term,
we can write 
\ba
 \Pi_{ij}^{(\fe)} ( K) & = &  
 -\,  g^2 \Nf\,
 \int_\vec{p} T \sum_{\{ p^{ }_n \}}
 \biggl[ 
 \frac{2 \delta^{ }_{ij}}{ P^2} 
 + \frac{
   - K^2 \delta^{ }_{ij} 
   + 2 k^{ }_i k^{ }_j - 4 p^{ }_i p^{ }_j  
 }{ P^2( K -  P)^2}
 \biggr]
 \;. \la{Pi_ij_1}
\ea
For generality we assume that, 
like in \eq\nr{F_prop_mu}, the Matsubara frequency 
is of the form\footnote{%
 As mentioned between \eqs\nr{psiloop} and \nr{psiloop_2}, 
 shifts like $P\to K-P$ are
 delicate in the presence of chemical potentials, given that
 we have to keep track of the relative signs of $\omega^{ }_n$ and 
 $i\mu$ in the Matsubara frequency. It turns out that there is 
 no problem with the present example, since the final result is an 
 even function of $\mu$. To be careful, 
 one could employ the unshifted form, 
 given in \eq\nr{Pimunuf}, 
 and verify that the results do remain the same.
 }  
\be
  p^{ }_n \to \tilde  p^{ }_n \equiv \omega^{ }_n + i \mu
 \;, \quad 
 \omega^{ }_n = 2\pi T \Bigl( n + \fr12 \Bigr)
 \;. \la{ppfe}
\ee

\index{Thermal sums: fermion-fermion loop}

The Matsubara sum can now be carried out, in analogy 
with the procedure described in \se\ref{se:NI_rho}.  Denoting
\be
 \E^{ }_1 \equiv |\vec{p}|
 \;, \quad
 \E^{ }_2 \equiv |\vec{p-k}|
 \;, \la{e1e2prime}
\ee
we can read from \eq\nr{Gefe_new} that 
\ba
 T \sum_{ \{ \omega^{ }_n \} } \frac{1}
 {(\omega^{ }_n + i \mu)^2 + \E_1^2} & = &
  \frac{1}{2 \E^{ }_1}
 \Bigl[
   \nF{}(\E^{ }_1-\mu) e^{\beta(\E^{ }_1-\mu)} - 
   \nF{}(\E^{ }_1+\mu)  
 \Bigr]
 \nn & = & 
  \frac{1}{2 \E^{ }_1}
 \Bigl[ 1 - 
   \nF{}(\E^{ }_1-\mu) - 
   \nF{}(\E^{ }_1+\mu)  
 \Bigr]
 \;. \la{G_0}
\ea
It is somewhat more tedious to carry out the other sum. 
Proceeding in analogy with the analysis following \eq\nr{F_sum}
and denoting the result by $\mathcal{G}$, we get 
\ba
 \mathcal{G} & = &  
 T \sum_{\{ p^{ }_n \} } \frac{1}
 {[\tilde p_n^2 + \E_1^2][(k^{ }_n -\tilde p^{ }_n)^2 + \E_2^2]}
 \la{G_def} \\ & = & 
 T \sum_{\{ p^{ }_n \}} T \sum_{\{ r^{ }_n \}}
  \beta\, \delta(\tilde r^{ }_n + k^{ }_n - \tilde p^{ }_n)
 \frac{1}
  {[\tilde p_n^2 + \E_1^2][\tilde r_n^2 + \E_2^2]} 
 \nn & = & 
 \int_0^\beta \! {\rm d}\tau \, e^{i k^{ }_n \tau} 
 \biggl\{ T \sum_{\{ p^{ }_n \}} \frac{e^{-i \tilde p^{ }_n \tau}} 
  {\tilde p_n^2 + \E_1^2}
 \biggr\}
 \biggl\{ T \sum_{\{  r^{ }_n \}} \frac{e^{i \tilde r^{ }_n \tau}} 
 {\tilde r_n^2 + \E_2^2} \biggr\}
 \;, \la{G_1}
\ea
where we used the trick in \eq\nr{insert_delta}.
The sums can be carried out by making use of \eq\nr{Gefe_new},  
\ba
 T \sum_{\{  r^{ }_n \}} \frac{e^{i \tilde r^{ }_n \tau}}
 {\tilde r_n^2 + \E_2^2} & = & 
 \frac{1}{2 \E^{ }_2}
 \Bigl[
   \nF{}(\E^{ }_2-\mu) e^{(\beta-\tau)\E^{ }_2-\beta\mu} - 
   \nF{}(\E^{ }_2+\mu) e^{\tau \E^{ }_2} 
 \Bigr]
 \;, \\
 T \sum_{\{ p^{ }_n \}} \frac{e^{-i \tilde p^{ }_n \tau}}
 {\tilde p_n^2 + \E_1^2} & = & 
 - e^{\mu\beta} T \sum_{\{ p^{ }_n \}}
 \frac{ e^{i \tilde p^{ }_n(\beta - \tau)}}
 {\tilde p_n^2 + \E_1^2}
 \nn 
 & = & 
 \frac{1}{2 \E^{ }_1}
 \Bigl[
   \nF{}(\E^{ }_1+\mu) e^{(\beta-\tau)\E^{ }_1+\beta\mu} - 
   \nF{}(\E^{ }_1-\mu) e^{\tau \E^{ }_1} 
 \Bigr]
 \;, 
\ea
where in the latter equation attention needed to be paid to the 
fact that \eq\nr{Gefe_new} only applies for $0 \le \tau \le \beta$
and that there is a shift due to the chemical potential in $\tilde p^{ }_n$.

Inserting these expressions
into \eq\nr{G_1} and carrying out the integral over $\tau$, 
we get
\ba
 \mathcal{G} & = & 
 \int_0^\beta \! {\rm d}\tau 
 \, e^{i k^{ }_n \tau} \frac{1}{4 \E^{ }_1 \E^{ }_2}
 \biggl\{
    \nF{}(\E^{ }_1 + \mu) \nF{}(\E^{ }_2 - \mu)
  e^{(\beta-\tau)(\E^{ }_1 + \E^{ }_2)}
 \nn & & \hspace*{2.55cm} 
  -\,  \nF{}(\E^{ }_1 + \mu) \nF{}(\E^{ }_2 + \mu)
  e^{\tau(\E^{ }_2 - \E^{ }_1) + \beta(\E^{ }_1+\mu)}
 \nn[2mm] & & \hspace*{2.55cm} 
  -\,  \nF{}(\E^{ }_1 - \mu) \nF{}(\E^{ }_2 - \mu)
  e^{\tau(\E^{ }_1 - \E^{ }_2) + \beta(\E^{ }_2-\mu)}
 \nn & & \hspace*{2.55cm} 
  +\,  \nF{}(\E^{ }_1 - \mu) \nF{}(\E^{ }_2 + \mu)
  e^{\tau(\E^{ }_1 + \E^{ }_2)}
 \biggr\}
 \nn & = &
 \frac{1}{4 \E^{ }_1 \E^{ }_2}
 \biggl\{
    \nF{}(\E^{ }_1 + \mu) \nF{}(\E^{ }_2 - \mu)
    \frac{1}{ik^{ }_n - \E^{ }_1 - \E^{ }_2}
    \Bigl[1 -  
        e^{\beta(\E^{ }_1 + \E^{ }_2)}
    \Bigr]
 \nn & &      \hspace*{0.75cm} 
  -\,  \nF{}(\E^{ }_1 + \mu) \nF{}(\E^{ }_2 + \mu)
  \frac{1}{ik^{ }_n + \E^{ }_2 - \E^{ }_1}
  \Bigl[ 
   e^{\beta(\E^{ }_2 +\mu)} - e^{\beta(\E^{ }_1 + \mu)}
  \Bigr]
 \nn[2mm] & & \hspace*{0.75cm} 
  -\,  \nF{}(\E^{ }_1 - \mu) \nF{}(\E^{ }_2 - \mu) 
  \frac{1}{ik^{ }_n + \E^{ }_1 - \E^{ }_2}
  \Bigl[ 
   e^{\beta(\E^{ }_1 - \mu)} - e^{\beta(\E^{ }_2 - \mu)}
  \Bigr]
 \nn & &      \hspace*{0.75cm} 
  +\,  \nF{}(\E^{ }_1 - \mu) \nF{}(\E^{ }_2 + \mu) 
    \frac{1}{ik^{ }_n + \E^{ }_1 + \E^{ }_2}
    \Bigl[ 
        e^{\beta(\E^{ }_1 + \E^{ }_2)} - 1
    \Bigr]
 \biggr\}
 \nn & = &
  \frac{1}{4 \E^{ }_1 \E^{ }_2}
 \biggl\{
   \frac{1}{ik^{ }_n - \E^{ }_1 - \E^{ }_2}
    \Bigl[ 
    \nF{}(\E^{ }_1 + \mu)+  \nF{}(\E^{ }_2 - \mu) - 1
    \Bigr]
 \nn & &      \hspace*{0.75cm} 
  +\, \frac{1}{ik^{ }_n + \E^{ }_2 - \E^{ }_1}
  \Bigl[ \nF{}(\E^{ }_2 + \mu)
   -  \nF{}(\E^{ }_1 + \mu) 
  \Bigr]
 \nn[1mm] & & \hspace*{0.75cm} 
  +\, \frac{1}{ik^{ }_n + \E^{ }_1 - \E^{ }_2}
   \Bigl[ \nF{}(\E^{ }_1 - \mu) - \nF{}(\E^{ }_2 - \mu) 
  \Bigr]
 \nn & &      \hspace*{0.75cm} 
  +\,  \frac{1}{ik^{ }_n + \E^{ }_1 + \E^{ }_2}
    \Bigl[ 
    1 -  \nF{}(\E^{ }_1 - \mu) - \nF{}(\E^{ }_2 + \mu) 
    \Bigr]
 \biggr\}
 \;. \la{G_2} 
\ea
At this point we could carry out the analytic continuation 
$i k^{ }_n \to k^0_{ } + i 0^+_{ }$, but it will be convenient to postpone 
it for a moment; we just need to keep in mind that after the 
analytic continuation, $i k^{ }_n$ becomes a {\em soft} quantity. 

The next step is to Taylor-expand to leading order in $k^0_{ },\vec{k}$.
To this end we can write 
\be
 \E^{ }_1 = p \equiv |\vec{p}|  \;, \quad
 \E^{ }_2 = |\vec{p-k}| \approx 
 p - k^{ }_i \frac{\partial}{\partial p^{ }_i} |\vec{p}| 
 = p - k^{ }_i v^{ }_i
 \;,  \la{E_exps}
\ee
where
\be
 v^{ }_i \equiv \frac{p^{ }_i}{p}
 \;, \quad i \in \{ 1,2,3 \}
 \;,
 \la{v_i}
\ee
are referred to as the {\em velocities of the hard particles}. 

It has to be realized that a Taylor expansion
is sensible only in terms  in which there is 
a thermal distribution function providing an external scale $T$ 
and thereby guaranteeing that the integral obtains its dominant 
contributions from hard momenta, $p\sim \pi T$. We cannot Taylor-expand in 
the vacuum part, which has no scale with respect to which to expand. 
It can, however, be separately verified that the vacuum part vanishes 
as a power of $k^0_{ },\vec{k}$, which is consistent with the fact 
that there is no gluon mass in vacuum. Here we simply 
omit the temperature-independent part. 

With these approximations, the function $\mathcal{G}$ reads
\ba
 \mathcal{G} & \approx & 
 \frac{1}{4 p^2}
 \biggl\{
   \frac{1}{2 p}
    \Bigl[ 
     - \nF{}(p + \mu) -   \nF{}(p - \mu) 
    \Bigr]
 \nn & & \hspace*{0.5cm} 
  +\, \frac{1}{ik^{ }_n - \vec{k}\cdot\vec{v}} (- \vec{k}\cdot\vec{v} )
   \nF{}'(p + \mu) 
 \nn[1mm] & & \hspace*{0.5cm} 
  +\, \frac{1}{ik^{ }_n + \vec{k}\cdot\vec{v}}
   (+\vec{k}\cdot\vec{v})
   \nF{}'(p - \mu)
 \nn & & \hspace*{0.5cm} 
  +\, \frac{1}{2 p}
    \Bigl[ 
     -  \nF{}(p - \mu) - \nF{}(p + \mu) 
    \Bigr]
 \biggr\} + \rmO(k^0_{ },\vec{k})
 \;. \la{G_3} 
\ea
Now we insert \eqs\nr{G_0} and \nr{G_3} into \eq\nr{Pi_ij_1}. 
Through the substitution $\vec{p}\to -\vec{p}$ 
(whereby $\vec{v}\to -\vec{v}$), the 3rd row in \eq\nr{G_3} 
can be put in the same form as the 2nd row. Furthermore, 
terms containing $k^{ }_n$ or $\vec{k}$ in the numerator in 
\eq\nr{Pi_ij_1} are seen to be of higher order. Thereby
\ba
 \Pi^{(\fe)}_{ij}( K) & \approx &  
 - {g^2 \Nf} 
 \int_\vec{p} 
 \biggl\{ 
 \frac{\delta^{ }_{ij}}{p}
    \Bigl[ 
     - \nF{}(p + \mu) -   \nF{}(p - \mu) 
    \Bigr]
  \nn & & \hspace*{1.4cm}
  -\, \frac{p^{ }_i p^{ }_j}{p^2} \frac{1}{p}
    \Bigl[ 
     - \nF{}(p + \mu) -   \nF{}(p - \mu) 
    \Bigr]
  \nn & & \hspace*{1.4cm}
  -\, \frac{p^{ }_i p^{ }_j}{p^2} 
   \frac{i k^{ }_n - \vec{k}\cdot\vec{v}  - i k^{ }_n}
   {ik^{ }_n - \vec{k}\cdot\vec{v}} 
    \Bigl[ 
      \nF{}'(p + \mu) +  \nF{}'(p - \mu) 
    \Bigr]
 \biggr\}
 \nn & = & 
  - {g^2 \Nf} 
 \int_\vec{p} 
 \biggl\{ 
 \frac{- \delta^{ }_{ij}}{p}
    \Bigl[ 
      \nF{}(p + \mu) +  \nF{}(p - \mu) 
    \Bigr]
  \nn & & \hspace*{1.5cm}
  +\, \frac{v^{ }_i v^{ }_j}{p}
    \Bigl[ 
     \nF{}(p + \mu) +   \nF{}(p - \mu) 
    \Bigr]
  \nn & & \hspace*{1.5cm}
  -\, {v^{ }_i v^{ }_j} 
    \Bigl[ 
      \nF{}'(p + \mu) +  \nF{}'(p - \mu) 
    \Bigr]
  \nn & & \hspace*{1.5cm}
  +\, \frac{v^{ }_i v^{ }_j\, i k^{ }_n }
   {i k^{ }_n - \vec{k}\cdot\vec{v}} 
    \Bigl[ 
      \nF{}'(p + \mu) +  \nF{}'(p - \mu) 
    \Bigr]
 \biggr\}
 \;. \la{Pi_ij_2}
\ea

The remaining integration can be factorized into a radial
and an angular part, 
\be 
 \int_\vec{p} 
 = \int_p \int \! {\rm d}\Omega^{ }_v
 \;, \la{htl_meas}
\ee
where the {\em angular integration} goes over the directions 
of $\vec{v} = \vec{p}/p$, and is normalized to unity: 
\be
 \int\! {\rm d}\Omega^{ }_v \equiv 1
 \;. 
\ee
Then, the following identities can be verified 
(for \eqs\nr{exe12_a} and \nr{exe12_b} details are given 
in appendix~C; 
\eq\nr{exe12_c} is a trivial consequence of rotational symmetry 
and $\vec{v}^2 = 1$): 
\ba
 \int_p \Bigl[ 
      \nF{}'(p + \mu) +  \nF{}'(p - \mu) 
    \Bigr]
 & = & -(d-1) \int_p \frac{1}{p}
    \Bigl[ 
      \nF{}(p + \mu) +  \nF{}(p - \mu) 
    \Bigr]
 \;, \la{exe12_a} \\
 \int\! {\rm d}\Omega^{ }_v v^{ }_i v^{ }_j 
 & = & 
 \frac{\delta^{ }_{ij}}{d}
 \;,  \la{exe12_c}
\ea
and, for $d=3$, 
\ba
 \int_p \frac{1}{p}
    \Bigl[ 
      \nF{}(p + \mu) +  \nF{}(p - \mu) 
    \Bigr] & \stackrel{d=3}{=} & 
    \frac{1}{4} \biggl( \frac{T^2}{3} + \frac{\mu^2}{\pi^2} \biggr)
 \;. \la{exe12_b} 
\ea
The integration 
\be
  \int\! {\rm d}\Omega^{ }_v 
 \frac{v^{ }_i v^{ }_j}{i k^{ }_n - \vec{k}\cdot \vec{v}}
  \la{exe12_d}
\ee
can also be carried out 
(cf.\ appendix~C) 
but we do not need its value for the moment. 

With these ingredients, \eq\nr{Pi_ij_2} becomes
\ba
  \Pi^{(\fe)}_{ij}(K) & = &  
 - {g^2 \Nf} \int_p \frac{1}{p}    
    \Bigl[ 
      \nF{}(p + \mu) +  \nF{}(p - \mu) 
    \Bigr]
 \nn & \times & \biggl\{ 
 \delta^{ }_{ij} \biggl( -1 + \fr1{d} + \fr{d-1}{d} \biggr)
 - (d-1) \int \! {\rm d}\Omega^{ }_v 
 \frac{v^{ }_i v^{ }_j\, i k^{ }_n }{i k^{ }_n - \vec{k}\cdot \vec{v}}
 \biggr\} 
 \nn & = & 
 g^2 \Nf (d-1) 
 \int_p \frac{1}{p}    
    \Bigl[ 
      \nF{}(p + \mu) +  \nF{}(p - \mu) 
    \Bigr]
 \int \! {\rm d}\Omega^{ }_v 
 \frac{v^{ }_i v^{ }_j\, i k^{ }_n}{i k^{ }_n - \vec{k}\cdot \vec{v}}
 \;. \la{Pi_ij_3}
\ea
Including also gauge fields and ghosts, 
the complete result reads
\be
   \Pi^{ }_{ij}( K) = \mE^2
 \int \! {\rm d}\Omega^{ }_v 
 \frac{v^{ }_i v^{ }_j\, i k^{ }_n}
 {i k^{ }_n - \vec{k}\cdot \vec{v}} + \rmO(i k^{ }_n,\vec{k})
 \;, \la{Pi_ij_4}
\ee
where $\mE$ is the generalization of the {\em Debye mass}
in \eq\nr{mmE} to the case 
of a fermionic chemical potential, 
\ba
 \mE^2 & \equiv & g^2 (d-1) \int_{p} \frac{1}{p} 
 \Bigl\{ \Nf \Bigl[ \nF{}(p+\mu) + \nF{}(p-\mu) 
 \Bigr] + (d-1) \Nc \nB{}(p) \Bigr\}
 \la{mmE_D} \\ 
 & \stackrel{d=3}{=} & 
 g^2
  \biggl[
    \Nf \biggl(\frac{T^2}{6} + \frac{\mu^2}{2\pi^2} \biggr)
       + 
    \frac{\Nc T^2}{3}  
  \biggr]
 \;. 
 \la{mmE_2}
\ea

\index{Debye mass}
\index{QED}
\index{Landau damping}
\index{Resummed self-energy: gluon}

\Eq\nr{Pi_ij_4}, known for QED since 
a long time~\cite{qed1}--\hspace*{-1.1mm}\cite{qed3},
is a remarkable expression.\footnote{%
 Recently the derivation of $\mE^2$ has been extended to NLO
 in QED~\cite{htl_nlo}, confirming that the result agrees with 
 that obtained in the dimensionally reduced
 effective theories.  
 } 
Even though it is of $\rmO(1)$ is we count $ik^{ }_n$ and $\vec{k}$
as quantities of the same order, it depends non-trivially 
on the ratio $i k^{ }_n/|\vec{k}|$. In particular, 
for $k^0_{ } = i k^{ }_n \to 0$, 
i.e.\ {\em in the static limit}, $\Pi^{ }_{ij}$ vanishes. This 
corresponds to the result in \eq\nr{GselfE}, i.e.\ that 
spatial gauge field components do not develop a thermal 
mass at 1-loop order.  
On the other hand, for $0 < |k^0_{ }| < |\vec{k}|$, it contains both a real 
and an imaginary part, cf.\ \eqs\nr{vivj_1} and \nr{Lq}. The imaginary
part is related to the physics of {\em Landau damping}: it means
that spacelike gauge fields can lose energy to hard 
particles in the plasma through real $2 \leftrightarrow 1$ scatterings. 

So far, we were only concerned with the spatial part
$\Pi^{ }_{ij}$. An interesting question is to generalize the computation
to the full self-energy $\Pi^{ }_{\mu\nu}$. 
Fortunately, it turns out that all the 
information needed can be extracted from \eq\nr{Pi_ij_4}, as we now show. 

\index{Slavnov-Taylor identities}

Indeed, the self-energy $\Pi^{ }_{\mu\nu}$, obtained by integrating out 
the hard modes, must produce a structure which is gauge-invariant in 
``soft'' gauge transformations, and 
therefore it must obey a Slavnov-Taylor identity and 
be {\em transverse} with respect to the external four-momentum. 
However, the meaning of transversality changes from the case of
zero temperature, because the heat bath introduces a preferred frame, 
and thus breaks Lorentz invariance. More precisely, we can now introduce
{\em two different} projection operators,  
\ba
 \mathbbm{P}^\rmii{T}_{\mu\nu}(K) & \equiv &
  \delta^{ }_{\mu i} \delta^{ }_{\nu j}
 \biggl( \delta^{ }_{ij} - \frac{k^{ }_i k^{ }_j}{k^2} \biggr)
 \;, \la{PT} \\ 
 \mathbbm{P}^\rmii{E}_{\mu\nu}(K) & \equiv & 
 \delta^{ }_{\mu\nu} - \frac{K^{ }_\mu K^{ }_\nu }{K^2}
 - \mathbbm{P}^\rmii{T}_{\mu\nu}(K)
 \;, \la{PE}
\ea
which both are four-dimensionally transverse,
\be
 \mathbbm{P}^\rmii{T}_{\mu\nu}(K)\,  K^{ }_\nu =
 \mathbbm{P}^\rmii{E}_{\mu\nu}(K)\,  K^{ }_\nu = 0 
 \;,
\ee
and of which $ \mathbbm{P}^\rmii{T}_{\mu\nu}(K)$ is 
in addition three-dimensionally transverse,
\be
 \mathbbm{P}^\rmii{T}_{\mu i}(K)\,  k^{ }_i = 0\; .
\ee
The two projectors are also orthogonal to each other, 
$
 \mathbbm{P}^\rmii{E}_{\mu\alpha} \mathbbm{P}^\rmii{T}_{\alpha\nu} = 0
$.

With the above projectors, we can write
\be \index{Self-energy: gluon}
 \Pi^{ }_{ij}( K) \; = \; 
 \mE^2
 \int \! {\rm d}\Omega^{ }_v 
 \frac{v^{ }_i v^{ }_j\, i k^{ }_n}{i k^{ }_n - \vec{k}\cdot \vec{v}}
 \;\; \equiv \;\;
 \mathbbm{P}^\rmii{T}_{ij}( K) \, \Pi^{ }_\rmii{T}( K) 
 +  \mathbbm{P}^\rmii{E}_{ij}( K) \, \Pi^{ }_\rmii{E}( K)
 \;. \la{equal}
\ee
Note that this decomposition applies for 
$(...)^{ }_{ij} \to (...)^{ }_{\mu\nu}$ as well. 
Contracting \eq\nr{equal} with $\delta^{ }_{ij}$ and with 
$k^{ }_i k^{ }_j$ leads to the equations
\ba
 \mE^2\, i k^{ }_n L & = & 
 (d-1)\, \Pi^{ }_\rmii{T} 
 + \biggl( 1  -\frac{k^2}{k_n^2 + k^2} \biggr) \Pi^{ }_\rmii{E}
 \;, \hspace*{1cm} \\ 
 \mE^2  
 \int \! {\rm d}\Omega^{ }_v \,
 \frac{(\vec{k}\cdot\vec{v})^2 i k^{ }_n}{i k^{ }_n - \vec{k}\cdot \vec{v}}
 & = & 0\,\Pi^{ }_\rmii{T} 
 + \biggl( {k}^2  -\frac{(k^2)^2}{k_n^2 + k^2} \biggr) \Pi^{ }_\rmii{E}
 \;, \la{eq2_p}
\ea
where 
\be
 L \equiv \int \! {\rm d}\Omega^{ }_v\, 
 \frac{1}{i k^{ }_n - \vec{k}\cdot \vec{v}}
 \;. \la{L_def}
\ee
The integral on the left-hand side of \eq\nr{eq2_p} 
can furthermore be written as
\ba
 \int \! {\rm d}\Omega^{ }_v \,
 \frac{(\vec{k}\cdot\vec{v})^2 i k^{ }_n}{i k^{ }_n - \vec{k}\cdot \vec{v}}
 & = & 
  \int \! {\rm d}\Omega^{ }_v 
 \frac{(-\vec{k}\cdot\vec{v}+ i k^{ }_n - i k^{ }_n)(-\vec{k}\cdot\vec{v}) 
 i k^{ }_n}{i k^{ }_n - \vec{k}\cdot \vec{v}}
 \nn & = & 
 (i k^{ }_n)^2 \int \! {\rm d}\Omega^{ }_v\, 
 \frac{\vec{k}\cdot\vec{v}}{i k^{ }_n - \vec{k}\cdot \vec{v}}
 \nn & = & 
 (ik^{ }_n)^2 \bigl( -1 + i k^{ }_n L \bigr)
 \;, 
\ea
where we have in the second step dropped a term that vanishes upon 
angular integration. Solving for $\Pi^{ }_\rmii{T}, \Pi^{ }_\rmii{E}$ 
and subsequently inserting the expression for $L$ from \eq\nr{exe12_h}, 
we thus get
\ba
 \Pi^{ }_\rmii{T}( K) & = &
 \frac{\mE^2}{d-1} 
 \biggl\{ 
   -\frac{k_n^2}{k^2} + \frac{K^2}{k^2} \; i k^{ }_n L
 \biggr\}
 \\ 
 & \stackrel{d=3}{=} & 
 \frac{\mE^2}{2} \biggl\{ 
  \frac{(i k^{ }_n)^2}{{k}^2} + 
  \frac{i k^{ }_n}{2 k}
  \biggl[
   1 - \frac{(i k^{ }_n)^2}{{k}^2} 
  \biggr]
  \ln \frac{ik^{ }_n + k}{ik^{ }_n - k} 
 \biggr\}
  \;, \la{Pi_T} \\
 \Pi^{ }_\rmii{E}( K) & = & \frac{\mE^2 K^2}{k^2}
  (1 - i k^{ }_n L)
 \\ 
 & \stackrel{d=3}{=} &
 \mE^2 
  \biggl[
   1 - \frac{(i k^{ }_n)^2}{{k}^2} 
  \biggr]
  \biggl[ 
    1 -  \frac{i k^{ }_n}{2 k} 
    \ln \frac{ik^{ }_n + k}{ik^{ }_n - k} 
  \biggr]
  \;. \la{Pi_E}
\ea

\Eqs\nr{Pi_T} and \nr{Pi_E} have a number of interesting limiting 
values. For $ik^{ }_n\to 0$ but with $k \neq 0$, 
$\Pi^{ }_\rmii{T} \to 0$, 
$\Pi^{ }_\rmii{E} \to \mE^2$. This corresponds to the physics
of {\em Debye screening}, familiar to us from \eq\nr{A0_effprop}. 
On the contrary, if we consider homogeneous but time-dependent 
waves, i.e.\ take ${k} \to 0$ with $i k^{ }_n \neq 0$, 
it can be seen that 
$\Pi^{ }_\rmii{T}$, $\Pi^{ }_\rmii{E} \to \mE^2/3$. This genuinely
Minkowskian structure in the resummed self-energy 
corresponds to {\em plasma oscillations}, or {\em plasmons}.  

\index{Debye screening}
\index{Plasma oscillations}
\index{Plasmon}
\index{HTL: gluon propagator}
\index{HTL: spectral representation}
\index{Propagator: HTL-resummed}
\index{Screening}

We can also write down a resummed 
propagator: in a general covariant gauge, where the tree-level propagator
has the form in \eq\nr{Aprop} and the static Feynman gauge propagator
the form in \eq\nr{A0_effprop}, we get
\be
 \langle A^a_{\mu} ( X) A^b_\nu ( Y) \rangle^{ }_0 = 
 \delta^{ab} \Tint{ K} e^{i K\cdot ( X -  Y)}
 \biggl[
   \frac{\mathbbm{P}^\rmii{T}_{\mu\nu}( K)}
     { K^2 + \Pi^{ }_\rmii{T}( K)} + 
   \frac{\mathbbm{P}^\rmii{E}_{\mu\nu}( K)}
     { K^2 + \Pi^{ }_\rmii{E}( K)} + 
   \frac{\xi \,  K^{ }_\mu  K^{ }_\nu}{( K^2)^2} 
 \biggr] 
 \;,  \la{prop}
\ee
where $\xi$ is the gauge parameter. 

\index{Spectral representation}

If the propagator of \eq\nr{prop} is used in practical 
applications, it is often useful to express it in terms of the
{\em spectral representation}, cf.\ \eq\nr{spectral}. The spectral
function appearing in the spectral representation can be obtained
from \eq\nr{Discdef}, 
where now $1/[ K^2 + \Pi^{ }_\rmii{T(E)}( K)]$
plays the role of $ \Pi^\iE_{\alpha\beta}$. 
After analytic continuation, $ik^{ }_n \to k^0_{ } + i 0^+_{ }$, 
\be
 \frac{1}{ K^2 + \Pi^{ }_\rmii{T(E)}(k^{ }_n,\vec{k})} \to 
 \frac{1}{-(k^0_{ } + i 0^+_{ })^2 + \vec{k}^2 + 
 \Pi^{ }_\rmii{T(E)}(-i(k^0_{ } + i 0^+_{ }),\vec{k})}
 \;, \la{PropR}
\ee
where
\ba \index{HTL: gluon self-energy}
 \Pi^{ }_\rmii{T}(-i(k^0_{ } + i 0^+_{ }),\vec{k}) & = & 
 \frac{\mE^2}{2} 
 \biggl\{ 
   \frac{(k^0_{ })^2}{{k}^2} + 
   \frac{k^0_{ }}{2k}
   \biggl[
     1 -  \frac{(k^0_{ })^2}{{k}^2}
   \biggr] 
   \ln\frac{k^0_{ }  + k + i 0^+_{ }}{k^0_{ } - k + i 0^+_{ }}
 \biggr\} 
 \;, \la{PiT_final} \\
 \Pi^{ }_\rmii{E}(-i(k^0_{ } + i 0^+_{ }),\vec{k}) & = & 
 \mE^2 
   \biggl[
     1 -  \frac{(k^0_{ })^2}{{k}^2}
   \biggr] 
   \biggl[
     1 -     
     \frac{k^0_{ }}{2k}
   \ln\frac{k^0_{ } + k + i 0^+_{ }}{k^0_{ } - k + i 0^+_{ }}
   \biggr] 
 \;. \la{PiE_final}
\ea
For $|k^0_{ }| > {k}$, $\Pi^{ }_\rmii{T}, \Pi^{ }_\rmii{E}$ are real, 
whereas for $|k^0_{ }| < {k}$, they have an imaginary part. 
Denoting 
$
 \eta \equiv \frac{k^0_{ }}{k}
$,
a straightforward computation 
(utilizing the fact that $\ln z$ has a branch cut 
on the negative real axis) leads to the spectral functions
$
 \rho^{ }_\rmii{T(E)} \equiv \im \bigl( \frac{1}{K^2 + \Pi^{ }_\rmii{T(E)}}
  \bigr)^{ }_{i k^{ }_n \to k^0_{ } + i 0^+_{ }}
$, where 
\ba
 \rho^{ }_\rmii{T}(\mathcal{K}) & = & 
 \left\{
   \begin{array}{ll} 
      \displaystyle\frac{\Gamma^{ }_\rmii{T}(\eta)}
      {\Sigma^2_\rmii{T}(\mathcal{K})+\Gamma^2_\rmii{T}(\eta)} \;,  &
      |\eta| < 1 \;, \\[3mm]
      \displaystyle
      \pi \mathop{\mbox{sign}} (\eta)
      \, \delta(\Sigma^{ }_\rmii{T}(\mathcal{K})) \;, & 
      |\eta| > 1 \;, 
   \end{array} 
 \right.  \la{rho_T} \\
 (\eta^2 - 1) {\rho}^{ }_\rmii{E}(\mathcal{K}) & = & 
 \left\{
   \begin{array}{ll} 
      \displaystyle\frac{ \Gamma^{ }_\rmii{E}(\eta)}
      {\Sigma^2_\rmii{E}(\mathcal{K})+\Gamma^2_\rmii{E}(\eta)} \;,  &
      |\eta| < 1 \;, \\[3mm]
      \displaystyle
      \pi \mathop{\mbox{sign}} (\eta)\, 
      \, \delta(\Sigma^{ }_\rmii{E}(\mathcal{K})) \;, & 
      |\eta| > 1 \;. 
   \end{array} 
 \right. \la{rho_E}
\ea
Here we have introduced the well-known 
functions~\cite{qed1}--\hspace*{-1.1mm}\cite{qed3}
\ba
 \Sigma^{ }_\rmii{T}(\mathcal{K}) & \equiv & 
 -\mathcal{K}^2 + \frac{\mE^2}{2}\biggl[
  \eta^2 + \frac{\eta(1-\eta^2)}{2}
  \ln\left| \frac{1+\eta}{1-\eta} \right| \biggr] 
 \;, \la{SigT} \\
 \Gamma^{ }_\rmii{T}(\eta) & \equiv & \frac{\pi \mE^2\, \eta (1-\eta^2)}{4}
 \;, \la{GamT} \\
 \Sigma^{ }_\rmii{E}(\mathcal{K}) & \equiv & 
  k^2 + \mE^2 \biggl[ 1 - \frac{\eta}{2}
  \ln\left| \frac{1+\eta}{1-\eta} \right| \biggr] 
 \;, \la{SigE} \\
 \Gamma^{ }_\rmii{E}(\eta) & \equiv & \frac{\pi \mE^2\, \eta}{2}
 \;. \la{GamE} 
\ea
The essential structure is that 
in each case there is 
a ``plasmon'' pole, i.e.\ a $\delta$-function analogous
to the $\delta$-functions in the free propagator of 
\eq\nr{free_S_rho} but displaced by an amount $\propto \mE^2$, 
as well as a cut at $|k^0_{ }|<k$, representing Landau damping. 

\index{Plasmon}
\index{Landau damping}

So far, we have only computed the resummed 
propagator. A very interesting question
is whether also an {\em effective action} can be written down, 
which would then 
not only contain the inverse propagator
like \eq\nr{htl_Seff}, but also new vertices,
in analogy with the dimensionally reduced effective theory 
of \eq\nr{Leff_EQCD}. Such effective vertices are needed for 
properly describing how the soft modes interact with each other.  
Note that since our observables 
are now non-static, the effective action should be gauge-invariant
also in time-dependent gauge transformations. 

\index{HTL: effective action}

Most remarkably, such an effective action can indeed be 
found~\cite{htl5,htl6}.
We simply cite here the result for the gluonic case. Expressing
everything in Minkowskian notation (i.e.\ after setting
$ik^{ }_n \to k^0_{ }$ and
using the Minkowskian $A^a_0$), the effective Lagrangian reads
\be
 \mathcal{L}^{ }_\iM = 
 - \fr12 \tr [F^{ }_{\mu\nu} F^{\mu\nu}]
 + \frac{\mE^2}{2} \int \! {\rm d} \Omega^{ }_v 
 \, \tr 
 \biggl[ 
 \biggl( \frac{1}{\mathcal{V}\cdot \mathcal{D}}\, \mathcal{V}^\alpha
    F^{ }_{\alpha\mu} \biggr)
 \biggl( \frac{1}{\mathcal{V}\cdot \mathcal{D}}\, \mathcal{V}^\beta
   {F^{ }_{\beta}}^\mu \biggr)
 \biggr]
 \;. \la{L_HTL}
\ee
Here $\mathcal{V}\equiv (1,\vec{v})$ is a light-like four-velocity, 
and $\mathcal{D}$ represents the covariant derivative in 
the adjoint representation. 

Several remarks on \eq\nr{L_HTL} are in order: 
\bi

\item
A somewhat tedious analysis, making use of the velocity integrals
listed in \eqs\nr{vel_first}--\nr{vel_last} below, shows that in the static
limit the second term in \eq\nr{L_HTL} reduces to the mass term
in \eq\nr{Leff_EQCD} (modulo Wick rotation and the
Minkowskian vs.\ Euclidean convention for $A^a_0$).

\item
In the static limit, we found quarks to always be infrared-safe, 
but this situation changes after the analytic continuation. 
Therefore a ``dynamical'' quark part should be added 
to \eq\nr{L_HTL}~\cite{htl5,htl6}; some details
are given in appendix B.

\item
In the presence of chemical potentials, additional operators, 
which break charge conjugation invariance, should be added to 
\eq\nr{L_HTL}~\cite{htl6x}.

\item
\Eq\nr{L_HTL} has the unpleasant feature that it is {\em non-local}: 
derivatives appear in the denominator. This we do not 
usually expect from effective theories. Indeed, if non-local structures
appear, it is difficult to analyze what kind of 
higher-order operators have been omitted and, 
hence, what the relative accuracy of the effective description is. 

In some sense, the appearance of non-local terms is a manifestation of the
fact that the proper infrared degrees of freedom have not been identified. 
It turns out that the HTL theory can be reformulated by introducing 
additional degrees of freedom, which gives the theory 
a local appearance~\cite{schz},\cite{cl1}--\hspace*{-1.1mm}\cite{cl3}
(for a pedagogic introduction, see ref.~\cite{cl5}). However
the reformulation contains classical on-shell particles rather
than quantum fields, whereby it continues to be difficult to 
analyze the accuracy of the effective description. 

\item
We arrived at \eq\nr{L_HTL} by integrating out the hard modes,
with momenta $p\sim \pi T$. However, like in the static limit, the theory
still has multiple dynamical momentum scales, ${k}\sim gT$
and $k \sim g^2T/\pi$.
It can be asked what happens if the momenta ${k}\sim gT$ are also
integrated out. This question has been analyzed in the 
literature, and leads indeed to a simplified (local) 
effective description~\cite{db1}--\hspace*{-1.1mm}\cite{db6},
which can be used for non-perturbatively studying 
observables only sensitive to ``ultrasoft''
momenta, ${k}\sim g^2 T/\pi$.

\item
Remarkably, for certain light-cone observables, ``sum rules'' can be 
established which allow to reduce gluonic HTL structures \index{Sum rule}
to the dimensionally reduced theory~\cite{sum3,sum1,sum2}.\footnote{%
 Picking out one spatial component and denoting it by $k^{ }_\parallel$, 
 so that $\vec{k}\equiv (k^{ }_\parallel,\vec{k}^{ }_\perp)$,   
 the sum rules can be expressed as 
 \ba
  \int_{-\infty}^{\infty} \! \frac{{\rm d} k^{ }_\parallel}{2 \pi}
  \biggl\{ 
    \frac{ \rho^{ }_\rmii{T}(k^{ }_\parallel,\vec{k}) }{k^{ }_\parallel} 
   - \frac{ \rho^{ }_\rmii{E}(k^{ }_\parallel,\vec{k}) }{k^{ }_\parallel} 
  \biggr\} 
  \frac{k_\perp^4}{ k_\perp^2 + k_\parallel^2}
  & = &   
  \frac{1}{2}  
  \frac{m_\rmii{E}^2}{k_\perp^2 + m_\rmii{E}^2}   
  \;, \\
  \int_{-\infty}^{\infty} \! \frac{{\rm d} k^{ }_\parallel}{2 \pi}
  \, k^{ }_\parallel  \Bigl\{ 
    \rho^{ }_\rmii{P}(k^{ }_\parallel,\vec{k})
   - \rho^{ }_\rmii{W}(k^{ }_\parallel,\vec{k}) 
  \Bigr\} 
  & = &   
    \frac{1}{4}
  \frac{m_\ell^2}{k_\perp^2 + m_\ell^2}    
  \;, 
 \ea
 where $\rho^{ }_\rmii{T}$, $\rho^{ }_\rmii{E}$, 
 $\rho^{ }_\rmii{W}$ and $\rho^{ }_\rmii{P}$
 are the spectral functions from \eqs\nr{rho_T}, \nr{rho_E}, and 
 \nr{rho0} for the last two. 
 } 
This is an important development, 
because the dimensionally reduced theory can be 
studied with standard non-perturbative techniques~\cite{sum4}.

\ei


\subsection*{Appendix A: Hard gauge boson loop}

Here a few details are given concerning the handling of the bosonic
part of \eq\nr{Pi_xi}. 
We follow the steps from \eq\nr{Pi_ij_1} onwards.
The spatial part of the self-energy can be written as 
\ba
 \Pi^{(\bo)}_{ij}(K) & = & 
 \frac{g^2 \CA}{2} 
 \Tint{P}
 \biggl\{ 
    (D-2) \biggl[
             \frac{2\delta^{ }_{ij}}{P^2} + 
             \frac{k^{ }_i k^{ }_j - 4 p^{ }_i p^{ }_j}{P^2(K-P)^2} 
          \biggr]
    - 4   \,   \frac{k^2\delta^{ }_{ij} - k^{ }_i k^{ }_j}{P^2(K-P)^2} 
 \biggr\} 
 \;, 
\ea
where all terms containing $k^{ }_i$ in the numerator are subleading. 
The bosonic counterpart of \eq\nr{G_0} (cf.~\eq\nr{Gebo_new}) reads
\be
 T\sum_{p^{ }_n} \frac{1}{p_n^2 + \E_1^2} = \frac{1}{2\E^{ }_1}
 \Bigl[ 1 + 2 \nB{}(\E^{ }_1) \Bigr]
 \;,
\ee 
whereas \eqs\nr{G_def}--\nr{G_2} get replaced with 
\ba \index{Thermal sums: boson-boson loop}
 \mathcal{G}' & \equiv & 
 T \sum_{p^{ }_n} \frac{1}{[p_n^2 + \E_1^2][(k^{ }_n-p^{ }_n)^2+\E_2^2]}
 \\ 
 & = & 
 \frac{1}{4 \E^{ }_1 \E^{ }_2}
 \biggl\{
   \frac{1}{i k^{ }_n - \E^{ }_1 - \E^{ }_2}
    \Bigl[ 
    - \nB{}(\E^{ }_1)-  \nB{}(\E^{ }_2) - 1
    \Bigr]
 \nn & & \hspace*{0.8cm} 
  +\, \frac{1}{i k^{ }_n + \E^{ }_2 - \E^{ }_1}
  \Bigl[ \nB{}(\E^{ }_1) -  \nB{}(\E^{ }_2) 
  \Bigr]
 \nn[1mm] & & \hspace*{0.8cm} 
  +\, \frac{1}{i k^{ }_n + \E^{ }_1 - \E^{ }_2}
   \Bigl[ \nB{}(\E^{ }_2) - \nB{}(\E^{ }_1) 
  \Bigr]
 \nn & & \hspace*{0.8cm} 
  +\,  \frac{1}{i k^{ }_n + \E^{ }_1 + \E^{ }_2}
    \Bigl[ 
    1 +  \nB{}(\E^{ }_1) + \nB{}(\E^{ }_2) 
    \Bigr]
 \biggr\}
 \;. \la{bG_2}
\ea
We observe that the bosonic results can be obtained from 
the fermionic ones simply by setting $\nF{}\to -\nB{}$.
The expansions of \eqs\nr{E_exps}--\nr{G_3} proceed as before, 
although one must be careful in making sure that the IR behaviour 
of the Bose distribution still permits for a Taylor expansion 
in powers of the external momentum. The partial integration 
identity in \eq\nr{exe12_a} can in addition be seen to 
retain its form, so that, effectively, 
\be
 \mathcal{G}'\to \frac{\nB{}(p)}{2 p^3}
 \biggl[ 1 - (D-2) 
 \frac{\vec{k}\cdot\vec{v}}{ik^{ }_n - \vec{k}\cdot\vec{v}} \biggr]
 = 
 \frac{\nB{}(p)}{2 p^3}
 \biggl[ D-1 - (D-2)
 \frac{ik^{ }_n}{ik^{ }_n - \vec{k}\cdot\vec{v}}  \biggr]
 \;. 
\ee
The final steps are like in \eq\nr{Pi_ij_3} and lead to \eq\nr{Pi_ij_4}, 
with $\mE^2$ as given in \eq\nr{mmE_D}.


\subsection*{Appendix B: Fermion self-energy}

\index{Resummed self-energy: fermion}
\index{QED}
\index{Self-energy: fermion}

Next, we consider a Dirac fermion at a finite temperature $T$ and 
a finite chemical potential $\mu$, interacting with an Abelian gauge 
field (this is no restriction at the current order: for a non-Abelian case 
simply replace $e^2 \to g^2 \CF$, where
$\CF \equiv (\Nc^2 - 1)/(2\Nc)$). 
The action is of the form in \eq\nr{Z_fer_mu} with 
$D^{ }_\mu = \partial^{ }_\mu - i e A^{ }_\mu$. To second order in $e$, 
the ``effective action'', or generating functional, takes the form
$
 S^{ }_\rmi{eff} = S_0^{ } 
 + \langle S^{ }_\iI - \fr12 S_\iI^2 + \rmO(e^3) \rangle_\rmi{1PI}
$, 
where $S_0^{ }$ is the quadratic part of the Euclidean action 
and $S^{ }_\iI$ contains the interactions. Carrying out the Wick
contractions, this yields
\be
 S^{ }_\rmi{eff} = 
 \Tint{ \{ K \} }
 \tilde{\!\bar\psi\,}( \tilde K) 
 \biggl[
   i \bsl{\tilde K} + m 
 + e^2\, \Tint{ \{  P \} } 
 \frac{ \gamma^{ }_\mu (- i \bsl{\tilde P} + m)  \gamma^{ }_\mu}
 {(\tilde P^2 + m^2)(\tilde P-\tilde K)^2} 
 \; + \; \rmO(e A^{ }_\mu) \; 
 \biggr]
 \tilde\psi(\tilde K)
 \;, \la{Seff_F_1}
\ee
where we have for simplicity 
employed the Feynman gauge,  
and $\tilde P, \tilde K$ are 
fermionic Matsubara momenta where the zero component
contains the chemical potential as indicated in
\eq\nr{ppfe}: $\tilde k^{ }_n \equiv k^{ }_n + i \mu$. 
In the momentum $\tilde P - \tilde K$, carried by $A^{ }_\mu$, 
the chemical potential drops out.

The Dirac structures appearing in \eq\nr{Seff_F_1} can be simplified: 
$
  \gamma^{ }_\mu  \gamma^{ }_\mu = D \, \unit^{ }_\rmi{$4\!\times\! 4$}
$, 
$
 \gamma^{ }_\mu \bsl{\tilde P} \gamma^{ }_\mu 
 = (2-D)\mbox{$\bsl{\tilde P}$} 
$. 
Denoting 
\be 
  f(i \tilde p^{ }_n,\vec{v}) \; \equiv \;
  i (D-2) \bsl{\tilde P} + D \, m \, \unit^{ }_\rmi{$4\!\times\! 4$} 
 \la{f_def}
\ee
where $\vec{v}$ is a dummy variable for both $\vec{p}$ and $m$; 
as well as 
\be
 \E^{ }_1 \equiv \sqrt{p^2 + m^2} 
 \;, \quad
 \E^{ }_2 \equiv \sqrt{(\vec{p}-\vec{k})^2}
 \;, \la{e1e2}
\ee
we are led to consider the sum 
(a generalization of \eq\nr{F_sum})
\be \index{Thermal sums: boson-fermion loop}
 \mathcal{F} \equiv 
 T \sum_{\{  p^{ }_n\} } \frac{f(i \tilde p^{ }_n,\vec{v})}
 {[\tilde p_n^2 + \E_1^2][(\tilde p^{ }_n-\tilde k^{ }_n)^2 + \E_2^2]}
 \;.  \la{raw1_F}
\ee
We can now write
\ba
 \mathcal{F} & = &
 T \sum_{\{ p^{ }_n \} } T \sum_{ r^{ }_n } 
 \beta \, \delta(\tilde p^{ }_n - \tilde k^{ }_n - r^{ }_n)
 \frac{f(i\tilde p^{ }_n,\vec{v})}
  {[\tilde p_n^2 + \E_1^2][r_n^2 + \E_2^2]} 
 \nn & = & 
 \int_0^\beta \! {\rm d}\tau \, e^{-i \tilde k^{ }_n \tau} 
 \biggl\{ T \sum_{\{ p^{ }_n\}} e^{i \tilde p^{ }_n \tau} 
 \frac{f(i\tilde p^{ }_n,\vec{v})}
  {\tilde p_n^2 + \E_1^2}
 \biggr\}
 \biggl\{ T \sum_{ r^{ }_n } 
 \frac{e^{- i r^{ }_n \tau}}{ r_n^2 + \E_2^2} \biggr\}
 \;, 
\ea
where we used a similar representation as before, 
\be
 \beta\, \delta( \tilde p^{ }_n - \tilde k^{ }_n - r^{ }_n) = 
 \int_0^\beta \! {\rm d}\tau 
 \, e^{i ( \tilde p^{ }_n - \tilde k^{ }_n - r^{ }_n ) \tau}
 \;. \la{insert_delta_F}
\ee
Subsequently \eqs\nr{Gebo_new} and \nr{Gefe_new} 
and their time derivatives can be inserted: 
\ba 
 T \sum_{ r^{ }_n } \frac{e^{- i r^{ }_n \tau}}{r_n^2 + \E_2^2} 
 & = & 
 \frac{\nB{}(\E^{ }_2)}{2 \E^{ }_2}
 \Bigl[e^{(\beta-\tau)\E^{ }_2} + e^{\tau \E^{ }_2} \Bigr]
 \;, \\
 T \sum_{ \{ p^{ }_n \} }
 \frac{e^{i \tilde p^{ }_n \tau}}{\tilde p_n^2 + \E_1^2} 
 & = & 
 \frac{1}{2 \E^{ }_1}
 \Bigl[\nF{}(\E^{ }_1-\mu)e^{(\beta-\tau)\E^{ }_1-\beta\mu} -
       \nF{}(\E^{ }_1+\mu) e^{\tau \E^{ }_1} \Bigr]
 \;, \\
  T \sum_{ \{ p^{ }_n \} } 
 \frac{i \tilde p^{ }_n e^{i \tilde p^{ }_n \tau}}{\tilde p_n^2 + \E_1^2} 
 & = & 
  - \fr12 \Bigl[\nF{}(\E^{ }_1-\mu) e^{(\beta-\tau)\E^{ }_1-\beta\mu} 
 + \nF{}(\E^{ }_1+\mu) e^{\tau \E^{ }_1} \Bigr]
 \;.
\ea
Thereby we obtain 
\ba
   \mathcal{F} = 
   \int_0^\beta \! {\rm d}\tau \, e^{-i \tilde k^{ }_n \tau}  
   \frac{\nB{}(\E^{ }_2)}{4 \E^{ }_1 \E^{ }_2}
  \; \Bigl\{ & & 
  \nF{}(\E^{ }_1-\mu) e^{(\beta-\tau)(\E^{ }_1 + \E^{ }_2)-\beta\mu} 
  f(-\E^{ }_1,\vec{v}) 
  \nn & + & 
  \nF{}(\E^{ }_1-\mu) e^{(\beta-\tau)\E^{ }_1 + \tau \E^{ }_2 - \beta\mu} 
  f(-\E^{ }_1,\vec{v}) 
  \nn[2mm] & + & 
  \nF{}(\E^{ }_1+\mu) e^{(\beta-\tau)\E^{ }_2 + \tau \E^{ }_1} 
  f(-\E^{ }_1,-\vec{v}) 
  \nn[2mm] & + & 
  \nF{}(\E^{ }_1+\mu) e^{ \tau(\E^{ }_1 + \E^{ }_2)} 
  f(-\E^{ }_1,-\vec{v}) 
  \; \Bigr\} 
 \;. \la{raw2_F}
\ea

As an example, 
let us focus on the second structure in \eq\nr{raw2_F}. The 
$\tau$-integral can be carried out, 
noting that $\tilde k^{ }_n$ is fermionic:  
\ba
  \int_0^\beta \! {\rm d}\tau \, e^{\beta (\E^{ }_1-\mu)} 
  e^{\tau (-i \tilde k^{ }_n - \E^{ }_1 + \E^{ }_2)}  
 & = &
 \frac{e^{\beta (\E^{ }_1-\mu) }}{-i \tilde k^{ }_n - \E^{ }_1 + \E^{ }_2}
 \Bigl[
 - e^{\beta(\E^{ }_2 - \E^{ }_1 + \mu)} - 1 
 \Bigr]
 \nn & = & 
 \frac{e^{\beta \E^{ }_2} + e^{\beta (\E^{ }_1-\mu)}}
  {i \tilde k^{ }_n + \E^{ }_1 - \E^{ }_2} 
 \nn & = & 
 \frac{1}{i \tilde k^{ }_n + \E^{ }_1 - \E^{ }_2}
 \Bigl[
 \nB{}^{-1}(\E^{ }_2) + \nF{}^{-1}(\E^{ }_1 - \mu) 
 \Bigr]
 \;. \la{raw3_F}
\ea
The inverse distribution functions nicely combine
with those appearing explicitly in \eq\nr{raw2_F}: 
\ba
 \mathcal{F}   =   
 \frac{1}{4 \E^{ }_1 \E^{ }_2} \biggl\{ & & 
  \frac{f(-\E^{ }_1,\vec{v})}{i \tilde k^{ }_n + \E^{ }_1 + \E^{ }_2}
   \Bigl[ 1  + \nB{}(\E^{ }_2) - \nF{}(\E^{ }_1-\mu) \Bigr] 
 \nn & + & \; 
 \frac{f(-\E^{ }_1,\vec{v})}{i \tilde k^{ }_n + \E^{ }_1 - \E^{ }_2}
 \Bigl[ \nF{}(\E^{ }_1-\mu) + \nB{}(\E^{ }_2) \Bigr]
 \nn & + & \; 
  \frac{f(\E^{ }_1,\vec{v})}{i \tilde k^{ }_n - \E^{ }_1 + \E^{ }_2}
   \Bigl[ - \nF{}(\E^{ }_1+\mu) - \nB{}(\E^{ }_2)  \Bigr] 
 \nn & + & \; 
  \frac{f(\E^{ }_1,\vec{v})}{i \tilde k^{ }_n - \E^{ }_1 - \E^{ }_2}
   \Bigl[ - 1  - \nB{}(\E^{ }_2) + \nF{}(\E^{ }_1+\mu) \Bigr] 
 \biggr\}
 \;. \la{raw4_F}
\ea

We now make the assumption, akin to that leading to 
\eq\nr{G_3}, that all four components of the (Minkowskian) 
external momentum $\mathcal{K}$ 
are small compared with the loop three-momentum $p=|\vec{p}|$, whose
scale is fixed by the temperature and the chemical potential
(this argument does not apply to the vacuum terms which are 
omitted; they amount e.g.\ to a radiative 
correction to the mass parameter $m$). Furthermore,  in order to 
simplify the discussion, we assume that the (renormalized) mass
parameter is small compared with $T$ and $\mu$. 
Thereby the ``energies'' of \eq\nr{e1e2} become
\be
 \E^{ }_1 \approx p + \frac{m^2}{2 p} + \rmO\Bigl( \frac{m^4}{p^3} \Bigr)
 \;, \quad
 \E^{ }_2 \approx p - \vec{k}\cdot\vec{v} + \rmO\Bigl(\frac{k^2}{p}\Bigr)
 \la{e1e2_appro}
\ee
where again 
\be
 \vec{v} \equiv \frac{\vec{p}}{p}
 \;.
\ee

Combining \eqs\nr{f_def} and \nr{raw4_F} with \eq\nr{e1e2_appro}, 
and noting that (for $m \ll  p$)
\be
 f(\pm \E^{ }_1,\vec{v}) \approx
 (D-2) (\pm \gamma_{ }^0 + v^{ }_i \gamma_{ }^i) p
 \;, 
\ee
where we returned to Minkowskian conventions for the Dirac
matrices (cf.\ \eq\nr{Dirac_E}), it is easy to see that the dominant
contribution, of order $1/\mathcal{K}$, 
arises from the 2nd and 3rd terms
in \eq\nr{raw4_F} which contain the difference $\E^{ }_1-\E^{ }_2$ in the 
denominator. Writing $-\vec{v}\cdot\bm{\gamma} \equiv v^{ }_i \gamma_{ }^i$
and substituting $\vec{v}\to -\vec{v}$ in the 3rd term, 
\eq\nr{Seff_F_1} becomes
$
 S_\rmi{eff}^{(0)} = 
 \Tinti{ \{ \tilde K \} }
 \tilde{\!\bar\psi\,}(\tilde K) 
 [
   i \bsl{\tilde K} + m +  \bsl{\Sigma}(\tilde K) 
 ]
 \tilde{\psi}(\tilde K)
$, 
where the superscript indicates that terms of $\rmO(e A^{ }_\mu)$ have
been omitted, and 
\be
 \bsl{\Sigma}(\tilde K) \approx - m_\rmii{F}^2 \int \! {\rm d}\Omega^{ }_v
 \, \frac{\gamma_0^{ } + \vec{v}\cdot\bm{\gamma}}
 {i \tilde k^{ }_n + \vec{k}\cdot\vec{v}}
 \;. \la{Sigma_F_res}
\ee
Here we have defined 
\ba \index{Thermal mass: fermion}
 m_\rmii{F}^2  & \equiv  & \frac{ (D-2) e^2}{4}
 \int_{\vec{p}} \frac{1}{p} 
 \Bigl[ 2 \nB{}(p) + \nF{}(p+\mu) + \nF{}(p-\mu)  \Bigr]
  \la{mmF} \\ 
 & \stackrel{D=4}{=} & 
 e^2 \biggl( \frac{T^2}{8} + \frac{\mu^2}{8\pi^2} \biggr)
 \;, \la{mmF2}
\ea
and carried out the integrals for $D=4$ 
(the bosonic part gives $2 \int_{\vec{p}} \nB{}(p)/p = T^2/6$; 
the fermionic part is worked out in appendix C). 
The angular integrations
can also be 
carried out, cf.\ \eqs\nr{htl_L_1} and \nr{htl_L_2} below. 

Next, we want to determine the corresponding spectral representation. 
As discussed in 
connection with the example following \eq\nr{F_prop_mu}, sign 
conventions are tricky with fermions. Our $S_\rmi{eff}^{(0)}$
defines the inverse propagator, representing therefore a 
generalization of the object in \eq\nr{F_prop_mu}, 
with the frequency variable appearing as $\tilde{k}^{ }_n = k^{ }_n + i \mu$.
Aiming for a spectral representation directly in terms of this variable,
needed in \eq\nr{HTL_F_spec},  
we define the analytic continuation as $i \tilde k^{ }_n \to \omega$
where $\omega$ has a small positive imaginary part.
Carrying out the angular integrals in \eq\nr{Sigma_F_res}
as explained in appendix~C, 
the analytically continued 
inverse propagator becomes (we set $m\to 0$)
\ba \index{HTL: fermion self-energy}
 \bsl{\mathcal{K}} + \bsl{\Sigma}(-i\omega,\vec{k}) = 
 \omega\gamma_0^{ } 
 \biggl[
    1 - \frac{m_\rmii{F}^2}{2 k \omega} \ln\frac{\omega+k}{\omega-k} 
 \biggr]
 - \vec{k}\cdot\bm{\gamma}
 \biggl[
    1 + \frac{m_\rmii{F}^2}{k^2} \biggl(
    1 - \frac{\omega}{2 k} \ln\frac{\omega+k}{\omega-k} 
 \biggr) \biggr]
 \;. \la{invS}
\ea
Introducing the concept of an ``asymptotic mass''
$m_\ell^2 \equiv 2 m_\rmii{F}^2$ and denoting
$L \equiv \frac{1}{2k} \ln \frac{\omega + k}{\omega - k}$,  
the corresponding spectral function reads
\be \index{Asymptotic mass} \index{HTL: fermion propagator}
 \im\Bigl\{ \big[\bsl{\mathcal{K}} 
 + \bsl{\Sigma}(-i\omega,\vec{k})\big]^{-1} \Bigr\} \; = \; 
 \bsl{\hspace*{0.3mm}\rho}(\omega,\vec{k})
 \;, \quad
 \rho \; \equiv \; (\omega {\rho}^{ }_\rmii{W}, \vec{k}\, {\rho}^{ }_\rmii{P})
 \; \equiv \; 
 \frac{1}{2} \bigl(  n^{ }_+ \rho^{ }_+ + n^{ }_- \rho^{ }_- \bigr)
 \;, 
\ee
where two separate basis choices have been introduced. 
Here $n^{ }_{\pm} \equiv (1,\pm \vec{k}/k)$, and 
\ba
 {\rho}^{ }_\rmii{W} & = & 
 \im \Biggl\{ \frac{1 - \frac{m_\ell^2 L}{2\omega}}
 {\bigl[\omega - \frac{ m_\ell^2 L}{2}\bigr]^2 -
  \bigl[k + \frac{m_\ell^2 (1-\omega L)}{2 k}\bigr]^2} \Biggr\}
 \;, \la{rho0} \quad
 {\rho}^{ }_\rmii{P} \; = \; 
 \im \Biggl\{ \frac{ 1 + \frac{m_\ell^2 (1-\omega L)}{2 k^2}  }
 {\bigl[\omega - \frac{ m_\ell^2 L}{2}\bigr]^2 -
  \bigl[k + \frac{m_\ell^2 (1-\omega L)}{2 k}\bigr]^2} \Biggr\}
 \;, \nn \la{rhos} \\
 \rho^{ }_+ & = & 
 \im \Biggl\{
   \frac{1}{ 
   (\omega - k ) 
   \bigl[ 1 + \frac{m_\ell^2 L}{2 k} \bigr]
            - \frac{m_\ell^2}{2 k}
   }
 \Biggr\}
 \;, \quad
 \rho^{ }_- \; = \; 
 \im \Biggl\{
   \frac{1}{ 
   (\omega + k ) 
   \bigl[ 1 - \frac{m_\ell^2 L}{2 k} \bigr]
            + \frac{m_\ell^2}{2 k}
   }
 \Biggr\}
 \;. 
\ea
These are well-known results~\cite{qed2,hw},
generalized to the presence
of a finite chemical potential~\cite{qed4};
note that the chemical 
potential only appears ``trivially'', in \eq\nr{mmF2}, 
without affecting the functional form of the momentum dependence. 
The corresponding ``dispersion relations'', relevant for computing
the ``pole contributions'' mentioned below \eq\nr{pert2}, 
have been discussed in the literature~\cite{pls1},
and can be shown to comprise two branches. There is 
a novel branch, dubbed a ``plasmino'' branch, 
with the peculiar property that 
\be \index{Plasmino} \index{Dispersion relation}
 \omega \approx m^{ }_\rmii{F} - \frac{k}{3} 
 + \frac{k^2}{3 m^{ }_\rmii{F}}  < m^{ }_\rmii{F}
 \;, \quad k \ll m^{ }_\rmii{F}
 \;. 
 \la{plasmino}
\ee
If the zero-temperature mass $m$
is larger than $m^{ }_\rmii{F}$, the plasmino branch 
decouples~\cite{pls2}.
For large momenta, the dispersion relation of the normal 
branch is of the form 
\be
 \omega \approx k + \frac{m_\ell^2}{2 k }
 \;, \quad
 k \gg m^{ }_\ell
 \;,
\ee
which explains why $m^{ }_\ell$ is called an asymptotic mass. 
A discussion of the dispersion relation in various
limits can be found in ref.~\cite{pls3}. 


\subsection*{Appendix C: Radial and angular momentum integrals}
\la{htl_app}

We compute here the radial and angular integrals defined in 
\eqs\nr{exe12_a}--\nr{exe12_d}.

For generality, and because this is necessary in loop computations, 
it is useful to keep the space dimensionality open for as long as 
possible. Let us recall that the dimensionally regularized 
integration measure can be written as
\be
 \int \! \frac{{\rm d}^d\vec{p}}{(2\pi)^d} 
 \to 
 \frac{4}{(4\pi)^{\frac{d+1}{2}} \Gamma(\frac{d-1}{2})}
 \int_0^{\infty} \! {\rm d}p \, p^{d-1} 
 \int_{-1}^{+1} \! {\rm d}z \, (1-z^2)^{\frac{d-3}{2}}
 \;, \la{meas2}
\ee 
where $d\equiv D-1$ and $z = \vec{k}\cdot\vec{p}/(kp)$ parametrizes
an angle with respect to some external vector. 
An important use of \eq\nr{meas2} is that it allows us to carry out
partial integrations with respect to both $p$ and $z$.  
If the integrand is independent of $z$, the $z$-integral yields
\be
 \int_{-1}^{+1} \! {\rm d}z \, (1-z^2)^{\frac{d-3}{2}}
 = \frac{\Gamma(\frac{1}{2})\Gamma(\frac{d-1}{2})}{\Gamma(\frac{d}{2})} 
 \;, \la{meas25}
\ee
and we then denote 
(cf.\ \eq\nr{k_measure}, now divided by $(2\pi)^d$)
\be
 c(d) \equiv
 \frac{2}{(4\pi)^{\frac{d}{2}} \Gamma(\frac{d}{2})}
 \;, 
\ee
so that 
$
 \int_\vec{p} = \int_p \,\equiv\, c(d) \int_0^\infty \! {\rm d}p\, p^{d-1}
$.

Now, \eq\nr{exe12_a} can be verified through
partial integration as follows: 
\ba \index{HTL: radial integrals}
 \int_{p} \frac{1}{p} 
 \Bigl[ 
      \nF{}(p + \mu) +  \nF{}(p - \mu) 
 \Bigr]
 & = & 
 c(d) \int_0^\infty \! {\rm d}p \, \frac{{\rm d}p}{{\rm d}p} \, p^{d-2} \, 
 \Bigl[ 
      \nF{}(p + \mu) +  \nF{}(p - \mu) 
 \Bigr]
 \nn 
 & = &  
 - (d-2)\, c(d) \int_0^\infty \! {\rm d}p \, p^{d-2} \, 
 \Bigl[ 
      \nF{}(p + \mu) +  \nF{}(p - \mu) 
 \Bigr]
 \nn & & 
 - c(d) \int_0^\infty \! {\rm d}p \, p^{d-1} \, 
 \Bigl[ 
      \nF{}'(p + \mu) +  \nF{}'(p - \mu) 
 \Bigr]
 \;. 
\ea
Moving the first term to the left-hand side
leads directly to \eq\nr{exe12_a}.

In order to derive the explicit expression in \eq\nr{exe12_b}, 
we set $d=3$; then a possible starting point is  
a combination of \eqs\nr{f_DF_mu} and \nr{f_ferm_expl}: 
\ba
 -f(T,\mu) & = & 
 2 \int_p  
 \biggl\{
 p + T  
 \biggl[
   \ln \Bigl( 1 + e^{-\frac{p-\mu}{T} }\Bigr) 
   + \ln \Bigl( 1 + e^{-\frac{p+\mu}{T} }\Bigr) 
 \biggr]
 \biggr\}
 \nn & \stackrel{d=3}{=} & 
 \frac{7\pi^2 T^4}{180} + \frac{\mu^2 T^2}{6} + \frac{\mu^4}{12\pi^2} 
 \;. \la{exe12_e}
\ea
Taking the second partial derivative with respect to $\mu$, we get
\ba
 -\frac{\partial^2 f(T,\mu)}{\partial\mu^2} & = & 
 2 T \int_p  
  \frac{\partial^2}{\partial\mu^2}  
 \biggl[
   \ln \biggl( 1 + e^{-\frac{p-\mu}{T} }\biggr) 
   + \ln \biggl( 1 + e^{-\frac{p+\mu}{T} }\biggr) 
 \biggr]
 \nn & = & 
 2 T \int_p  
  \frac{{\rm d}^2}{{\rm d} p^2}  
 \biggl[
   \ln \biggl( 1 + e^{-\frac{p-\mu}{T} }\biggr) 
   + \ln \biggl( 1 + e^{-\frac{p+\mu}{T} }\biggr) 
 \biggr]
 \nn & \stackrel{d=3}{=} & 
 - 4 T \int_p  
 \frac{1}{p} \frac{{\rm d}}{{\rm d} p} 
 \biggl[
   \ln \biggl( 1 + e^{-\frac{p-\mu}{T} }\biggr) 
   + \ln \biggl( 1 + e^{-\frac{p+\mu}{T} }\biggr) 
 \biggr]
 \la{exe12_f} \\ & = &
 \frac{T^2}{3} + \frac{\mu^2}{\pi^2}
 \;, \la{exe12_ff}
\ea
where in the penultimate step we carried out one partial integration. 
On the other hand, the integral in \eq\nr{exe12_f} can be rewritten as
\ba
 & & \hspace*{-1cm}
 -4 T \int_p  
 \frac{1}{p} \frac{{\rm d}}{{\rm d} p} 
 \biggl[
   \ln \biggl( 1 + e^{-\frac{p-\mu}{T} }\biggr) 
   + \ln \biggl( 1 + e^{-\frac{p+\mu}{T} }\biggr) 
 \biggr] 
 \nn 
 & = & 
 -4 T \int_p  
 \frac{1}{p}
 \biggl[ 
  \frac{e^{-\frac{p-\mu}{T} }}{1 + e^{-\frac{p-\mu}{T} } }
 + \frac{e^{-\frac{p+\mu}{T} }}{ 1 + e^{-\frac{p+\mu}{T} } } 
 \biggr] 
 \biggl( -\frac{1}{T} \biggr) 
 \nn & = & 
 4 \int_p \frac{1}{p} 
 \Bigl[ 
      \nF{}(p + \mu) +  \nF{}(p - \mu) 
 \Bigr]
 \;. \la{exe12_g}
\ea
\Eqs\nr{exe12_ff} and 
\eq\nr{exe12_g} combine into \eq\nr{exe12_b}.

\index{HTL: angular integrals}

As far as angular integrals go
(such as the one in \eq\nr{exe12_d}), 
we start with the simplest structure, 
defined in \eq\nr{L_def}: 
\ba
 L({K}) \;\equiv\; \int \! {\rm d}\Omega^{ }_v\, 
 \frac{1}{i k^{ }_n - \vec{k}\cdot \vec{v}} & \stackrel{d=3}{=} & 
 \frac{1}{4\pi}\, 2\pi \int_{-1}^{+1} \! {\rm d}z\,
  \frac{1}{i k^{ }_n - {k} z}
 \nn & = & 
 -\frac{1}{2 {k}}
 \int_{-1}^{+1} \! {\rm d}z \, \frac{{\rm d}}{{\rm d}z}
 \ln(i k^{ }_n - {k} z )
 \nn & = &
 \frac{1}{2 {k}} \ln \frac{i k^{ }_n + {k}}{i k^{ }_n - {k}}
 \;.  \la{exe12_h}
\ea
Further integrals can then be obtained by making use of rotational 
symmetry. For instance, 
\be
 \int \! {\rm d}\Omega^{ }_v\, 
 \frac{v^{ }_i}{i k^{ }_n - \vec{k}\cdot \vec{v}}
 = k^{ }_i \, f(i k^{ }_n, {k})
 \;, 
\ee
where, contracting both sides with $\vec{k}$, 
\ba
 f(i k^{ }_n, {k}) & = & 
 \frac{1}{{k}^2} \int \! {\rm d}\Omega^{ }_v \,
 \frac{\vec{k}\cdot\vec{v}}{i k^{ }_n - \vec{k}\cdot \vec{v}}
 = 
 \frac{1}{{k}^2} \biggl[ -1 + i k^{ }_n \int \! {\rm d}\Omega^{ }_v \,
 \frac{1}{i k^{ }_n - \vec{k}\cdot \vec{v}} 
 \biggr] 
 \;. 
\ea
Another trick, needed for having higher powers in the denominator, 
is to take derivatives of \eq\nr{exe12_h} with respect to $i k^{ }_n$.  

Without detailing further steps, 
we list the results for a number of velocity 
integrals that can be obtained this way.
Let us change the notation at this point: we 
replace $i k^{ }_n$ by $k^0_{ } + i 0^+_{ }$, as is relevant for retarded 
Green's functions ($i 0^+_{ }$ is not shown explicitly), and 
introduce the light-like four-velocity $\mathcal{V} \equiv (1,\vec{v})$.
Then the integrals read
($d=3$; $i,j=1,2,3$) 
\ba
 \int\! {\rm d}\Omega^{ }_v \,   & = & 1
 \;, \la{vel_first} \\
 \int\! {\rm d}\Omega^{ }_v \,  v^i & = & 0
 \;, \\
 \int\! {\rm d}\Omega^{ }_v \,  v^i v^j & = & \fr13 \delta^{ij}
 \;, \\  
 \int\! {\rm d}\Omega^{ }_v \,  \frac{1}{\mathcal{V}\cdot \mathcal{K}}
   & = & L(\mathcal{K})
 \;, \la{htl_L_1} \\
 \int\! {\rm d}\Omega^{ }_v \,  \frac{v^i}{\mathcal{V}\cdot \mathcal{K}} 
 & = & \frac{k^i}{{k}^2}
 \Bigl[ - 1 + k^0_{ } L(\mathcal{K}) \Bigr]
 \;, \la{htl_L_2} \\
 \int\! {\rm d}\Omega^{ }_v \,  \frac{v^i v^j}{\mathcal{V}\cdot \mathcal{K}}
  & = & 
 \frac{L(\mathcal{K})}{2}
 \biggl( 
  \delta^{ij} - \frac{k^i k^j}{{k}^2} 
 \biggr) + 
 \frac{k^0_{ }}{2 {k}^2}
 \Bigl[ 1 - k^0_{ } L(\mathcal{K}) \Bigr]
 \biggl( 
  \delta^{ij} -  \frac{3 k^i k^j}{{k}^2} 
 \biggr)  
 \;, \la{vivj_1} \\  
 \int\! {\rm d}\Omega^{ }_v \,  \frac{1}{(\mathcal{V}\cdot \mathcal{K})^2} 
 & = & \frac{1}{\mathcal{K}^2}
 \;, \\
 \int\! {\rm d}\Omega^{ }_v \,  \frac{v^i}{(\mathcal{V}\cdot \mathcal{K})^2} 
 & = & \frac{k^i}{{k}^2}
 \Bigl[ \frac{k^0_{ }}{\mathcal{K}^2} - L(\mathcal{K})  \Bigr]
 \;, \\
 \int\! {\rm d}\Omega^{ }_v \,
   \frac{v^i v^j}{(\mathcal{V}\cdot \mathcal{K})^2} & = & 
 \frac{1}{2 \mathcal{K}^2}
 \biggl( 
  \delta^{ij} - \frac{k^i k^j}{{k}^2} 
 \biggr) - 
 \frac{1}{2 {k}^2}
 \biggl[ 1 - 2 k^0_{ } L(\mathcal{K})
 + \frac{(k^0_{ })^2}{\mathcal{K}^2} \biggr]
 \biggl( 
  \delta^{ij} - \frac{3 k^i k^j}{{k}^2} 
 \biggr) 
 \;, \la{vel_last} \hspace*{0.8cm}
\ea
where 
$\mathcal{V}\cdot \mathcal{K} = k^0_{ } - \vec{v}\cdot\vec{k}$, 
and
\be
 L(\mathcal{K}) = \frac{1}{2{k}}
 \ln\frac{k^0_{ } + {k} + i0^+_{ }}{k^0_{ } - {k}+i 0^+_{ }}
 \quad \stackrel{|k^0_{ }| \ll k}{\approx} \quad
 - \frac{i \pi}{2{k}} + \frac{k^0_{ }}{{k}^2} + 
 \frac{(k^0_{ })^3}{3 {k}^4} + \ldots
 \;.  \la{Lq}
\ee


\subsection*{Appendix D: Photon polarization beyond the hard thermal limit}

\index{QED}
\index{Self-energy: photon}

The Hard Thermal Loop computations that we have reviewed in this section
relied on two separate approximations: the plasma particles were treated
as massless, and all components of 
the external momentum $\mathcal{K}$ were
assumed small compared with the loop momentum, $p\sim \pi T$. 
However, these approximations
are not always justified. 
In this appendix, we determine
the massive fermion contribution to the gauge field self-energy. 
The starting point is 
the sum-integral in \eq\nr{psiloop}, expressed 
in the same way as on 
the last line of \eq\nr{Pi_xi}, 
which we furthermore
consider in a system with 
a finite chemical potential (cf.\ \eq\nr{ppfe}), {\it viz.}\ 
\be
  \Pi^{ }_{\mu\nu}(K) \; \supset \; 
  - g^2 \Nf\, \Tint{\{ P \} }
 \frac{ 
   \delta^{ }_{\mu\nu} \bigl[ (K-\tilde P)^2 + \tilde{P}^2 + 2 m^2 - K^2 \bigr]
   + 2 (\tilde{P}^{ }_\mu K^{ }_\nu + \tilde{P}^{ }_\nu K^{ }_\mu)
   - 4 \tilde{P}^{ }_\mu \tilde{P}^{ }_\nu  
  }{  [\tilde{P}^2 + m^2 ]\, [ (K-\tilde P)^2 + m^2 ] }
  \;. \la{Pimunuf}
\ee

As a first step, consider 
the vacuum version of \eq\nr{Pimunuf}. This is obtained 
by setting $\Tinti{P}\to \int^{ }_P$ and $\tilde P \to P$. 
The integrals can be solved with standard techniques, 
and produce a wave function correction, i.e.\ a term 
$\propto K^2 \delta^{ }_{\mu\nu} - K^{ }_\mu K^{ }_\nu$.
In the following, this vacuum part is omitted. 

For the medium part, 
the Matsubara sums need to be carried out.
It is useful to focus on the spatial part, 
like in \eq\nr{Pi_ij_1}. 
Incorporating masses in the energies in \eq\nr{e1e2prime},
the results can be taken over from \eqs\nr{G_0} and \nr{G_2}.  
The terms without $\nF{}$ are vacuum ones, and omitted. 
In the terms containing $\nF{}(\E^{ }_2\pm\mu)$, we can substitute 
$\vec{p}\to \vec{k-p}$, whereby
$\E^{ }_2 \to \E^{ }_1$. 
The numerator is invariant in this substitution. 
Collecting the terms, the Fermi distributions can 
be factorized into 
$
 \nF{}(\E^{ }_p + \mu) + \nF{}(\E^{ }_p - \mu)
$, 
where we have denoted $\E^{ }_p \equiv \E^{ }_1 \equiv \sqrt{p^2 + m^2}$.

The next step is to carry out angular integration. 
First, tensor integrals can be reduced into
scalar ones, by making use of rotational invariance. This means
that if the numerator contains $p^{ }_i$, the result is proportional
to $k^{ }_i$, 
and if it contains $p^{ }_i \hspace*{0.2mm} p^{ }_j$, the result is 
a linear combination of terms proportional to 
$\delta^{ }_{ij}$ and $k^{ }_i k^{ }_j$. Subsequently, the 
angular integration can be carried out in the coefficient functions. 

With the result at hand, it may be projected into the basis
of \eqs\nr{PT}--\nr{equal}. Finally, 
analytic continuation,  
{\it viz.}\  $k^{ }_n \to -i [k^0_{ } + i 0^+_{ }]$, 
leads to a retarded correlator. Given
that the results are a bit lengthy, we employ an implicit
notation, suppressing $i0^+_{ }$ and using $k^{ }_0 = k^0_{ }$: 
\ba
 \Pi^{ }_\rmii{T}(\mathcal{K})
  \!\!& \supset &\!\! 
 \frac{g^2 \Nf^{ }}{2}
 \int_p \frac{\nF{}(\epsilon^{ }_p - \mu) + \nF{}(\epsilon^{ }_p + \mu)}
             {\epsilon^{ }_p}
  \nn 
 \!\!& \times &\!\!  
 \biggl[
  1 + \frac{k_0^2}{k^2} 
  - \, 
    \frac{k_0^4 - k^4 + 4 ( k_0^2 \epsilon_p^2 - k^2 p^2)}{8 k^3 p}
    \ln\bigl( \mathcal{R}^{ }_{+} \mathcal{R}^{ }_{-} \bigr)
  - \frac{k_0^{ }(k_0^2 - k^2)\epsilon^{ }_p}{2 k^3 p} 
    \ln\Bigl( \frac{ \mathcal{R}^{ }_{+} }{ \mathcal{R}^{ }_{-} } \Bigr)     
 \biggr]
 \;, \hspace*{8mm} \\ 
 \Pi^{ }_\rmii{E}(\mathcal{K}) 
 \!\!& \supset &\!\!  
 g^2 \Nf^{ }
 \int_p \frac{\nF{}(\epsilon^{ }_p - \mu) + \nF{}(\epsilon^{ }_p + \mu)}
             {\epsilon^{ }_p}
  \nn 
  \!\!& \times &\!\!  
 \biggl(   1 - \frac{k_0^2}{k^2} \biggr)
 \biggl[ 1 
  - \, 
    \frac{k_0^2 - k^2 + 4 \epsilon_p^2 }{8 k p}
    \ln\bigl( \mathcal{R}^{ }_{+} \mathcal{R}^{ }_{-} \bigr)
  - \frac{k_0^{ }\epsilon^{ }_p}{2 k p} 
    \ln\Bigl( \frac{ \mathcal{R}^{ }_{+} }{ \mathcal{R}^{ }_{-} } \Bigr)     
 \biggr]
 \;,
\ea
where
\be
 \mathcal{R}^{ }_+  \;\equiv\; 
 \frac{
    k_0^2 - k^2 
 + 2 ( k^{ }_0 \epsilon^{ }_p + k p  ) 
 }{
    k_0^2 - k^2 
 + 2 ( k^{ }_0 \epsilon^{ }_p - k p  ) 
 }
 \;, \quad
 \mathcal{R}^{ }_- \;\equiv\;
 \frac{
    k_0^2 - k^2 
 - 2 ( k^{ }_0 \epsilon^{ }_p - k p  ) 
 }{
    k_0^2 - k^2 
 - 2 ( k^{ }_0 \epsilon^{ }_p + k p  )
 }
 \;. 
\ee
In the Hard Thermal Loop limit, {\it viz.}\  
$
 \epsilon^{ }_p\to p 
$, 
$
 \mathcal{K}^2 \ll k p
$, 
the logarithms become
$
  \ln\bigl( \mathcal{R}^{ }_{+} \mathcal{R}^{ }_{-} \bigr)
 \to
 - {2k}/{p}
$
and 
$
 \ln\bigl( { \mathcal{R}^{ }_{+} } / { \mathcal{R}^{ }_{-} } \bigr)     
 \to 
 4 k L(\mathcal{K})
$, 
and \eqs\nr{PiT_final} and \nr{PiE_final} are reproduced. 

%

\newpage


\newpage 

\section{Applications}
\la{se:app}

\paragraph{Abstract:}

A number of physical applications of relativistic thermal field
theory are considered. First the basic formalism for addressing the existence
of a scalar field driven phase transition is developed 
(\se\ref{ss:transition}). 
Then the concept of
instantons is introduced with the example of a bubble nucleation rate 
related to a first order phase transition
(\se\ref{se:bubbles}). 
This is followed by a general
discussion concerning the formalism for particle production rate computations, 
relevant both for heavy ion collision experiments and cosmology
(\se\ref{se:ppr}). How a
particle production rate can be embedded in an expanding cosmological 
background is explained in detail
(\se\ref{se:dm}). 
Turning to so-called transport 
coefficients, we first consider the effective mass and 
friction coefficient that a scalar field evolving within a thermal 
environment feels
(\se\ref{se:field}). 
Then transport coefficients are discussed more
generally, culminating in the definition of shear and bulk viscosities, 
diffusion coefficients, and the electric conductivity of QCD matter
(\se\ref{se:transport}). 
Transport coefficients are closely related to the rate at which a 
slightly disturbed system equilibrates, and the corresponding formalism 
is introduced, stressing the idea of employing operator equations of
motion in order to simplify the correlation function to be computed
(\se\ref{se:decay}). 
Finally
a somewhat different but physically important topic, 
that of the behaviour of resonances made
of a heavy particle and antiparticle, is outlined, 
with emphasis on the roles that ``virtual'' and ``real'' corrections
play at finite temperature
(\se\ref{se:nrqcd}).

\paragraph{Keywords:} 

Effective potential, condensate, first order phase transition, 
semiclassical approximation, saddle point, instanton, fluctuation determinant, 
tunnelling, sphaleron, classical limit, critical bubble, latent heat, 
surface tension, particle production, on-shell field operator, 
Landau-Pomeranchuk-Migdal effect, decay rate, 
Friedmann equations, yield parameter, Boltzmann equation, 
friction coefficient, damping rate, 
thermal mass, dilaton, axion, Chern-Simons diffusion, 
equilibration, fluctuation-dissipation theorem, Kubo formula, transport peak, 
flavour diffusion, conductivity, viscosity, Brownian motion, 
Langevin equation, quarkonium, Debye screening, decoherence, thermal width,
real and virtual processes at finite temperature.


\subsection{Thermal phase transitions}
\la{ss:transition}

\index{Thermal phase transitions}

As a first application of the general formalism developed, we consider
the existence of thermal phase transitions in models of 
particle physics. Prime examples are the ``deconfinement'' transition 
in QCD, and the ``electroweak symmetry restoring'' 
transition in the electroweak theory,\footnote{%
 Here the standard terminology is used, even though 
 it is inappropriate in a strict sense, 
 given that both transitions are known to be of a crossover type, 
 i.e.~not genuine phase transitions.
 } 
both of which took place in the early universe.
For simplicity, though, the practical analysis will be carried out
within the scalar field theory discussed in \se\ref{se:int}.

In general, 
a phase transition can be defined as 
a line in the ($T,\mu$)-plane across which
the grand canonical free energy density $f(T,\mu)$ is non-analytic. 
In particular, if $\partial f/\partial T$
or $\partial f/\partial\mu$ is discontinuous, 
we speak of a {\em first order transition}. 
The energy density 
\be \index{Latent heat}
 e = \frac{1}{V \mathcal{Z}} 
 \tr \left[ \hat H e^{ -\beta ( \hat H - \mu \hat Q ) } \right]
 = \frac{T^2}{V} \frac{\partial}{\partial T} 
 \Bigl( \ln\mathcal{Z} \Bigr)^{ }_{\frac{\mu}{T}}
 = 
 f - T\left(\frac{\partial f}{\partial T}\right)^{ }_{\frac{\mu}{T}} 
\ee 
is then discontinuous, with the discontinuity known 
as the {\em latent heat}.  This means
that a closed system can proceed through the transition only 
if there is some mechanism for energy transfer and dissipation; 
thus, first order transitions possess non-trivial dynamics. 

\index{Order parameter}

It is often possible to associate an {\em order parameter} with 
a phase transition. In a strict sense, the order parameter should
be an elementary or composite field, the expectation value of which
vanishes in one phase and is non-zero in another. In a generalized sense, 
we may refer to an order parameter even if it does not vanish in either
phase, provided that (in a first order transition) it jumps across the 
phase boundary. In a particularly simple situation this role is taken
by some elementary field; in the following, we consider the case where
a real scalar field, $\phi$, plays the role of an order parameter. 
More realistically, $\phi$ could for instance be a neutral component
of the Higgs doublet (after gauge fixing).

Suppose now that the Euclidean Lagrangian of the $\phi$-field reads
\ba
  {L}^{ }_\iE & = & 
  \fr12 (\partial^{ }_\tau\phi)^2 + \fr12(\nabla\phi)^2 + V(\phi)
 \; .  \la{pot_phi}
\ea
We take the potential to be of the form 
$V(\phi) = -\fr12 m^2 \phi^2 + \fr14 \lambda\phi^4$,  
with positive real parameters $m$ and $\lambda$ and a discrete Z(2) symmetry. 
Then $\phi$ has a non-zero expectation value at zero temperature, 
as is suggested by a graphical illustration of the potential
in \fig\ref{fig:p5}. 

\begin{figure}[t]

\vspace*{0.2cm}

\centerline{\epsfysize=6.0cm\epsfbox{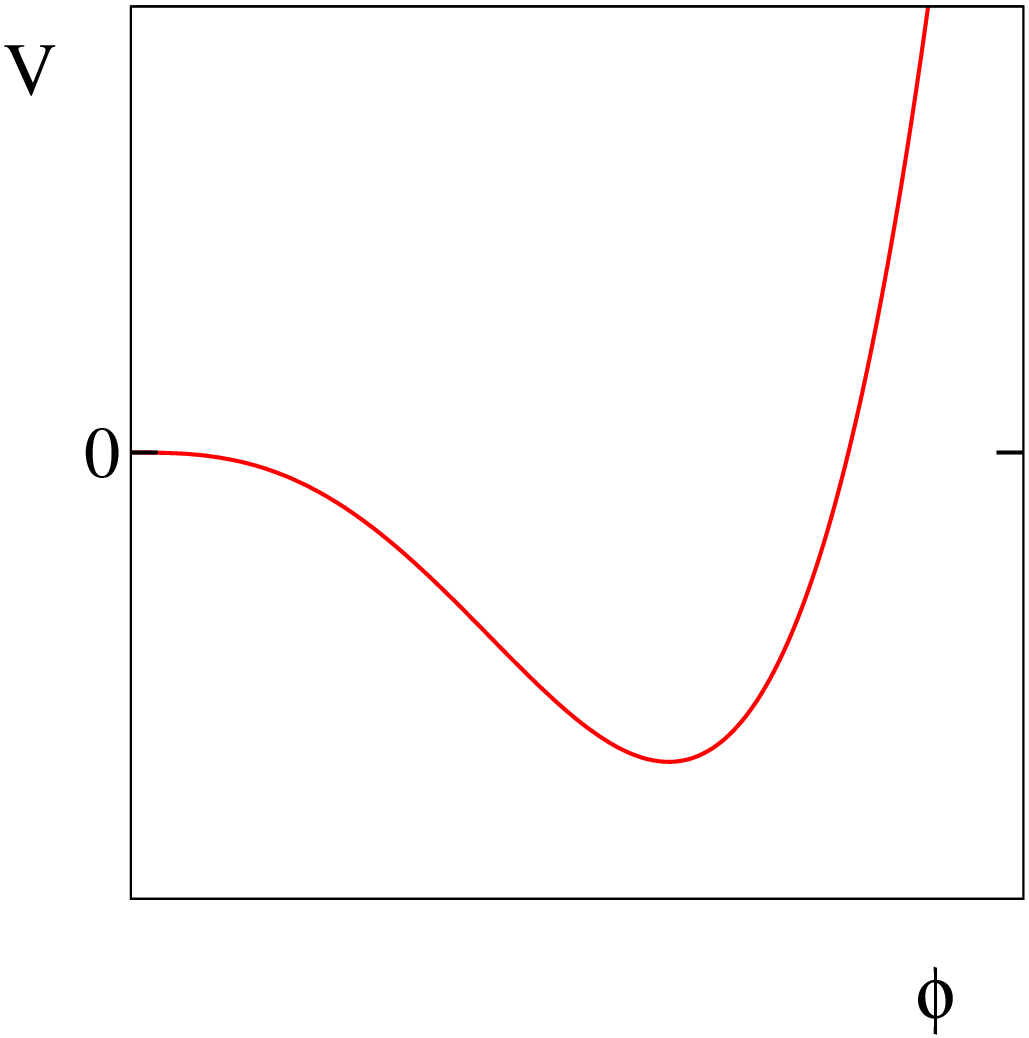}}

\caption[a]{\small
The potential from \eq\nr{pot_phi}
at zero temperature for $\phi > 0$.}

\la{fig:p5}
\end{figure}

\index{Condensate}
\index{Effective potential}

Let us now evaluate the partition function of the above system 
with the method of the {\em effective potential},
$V^{ }_\rmi{eff}(\bar\phi)$, introduced in \se\ref{se:Veff}.
In other words, we put the system in a finite volume $V$, 
and denote by $\bar\phi$ the {\em condensate}, i.e.\  
the mode with $p^{ }_n = 0, \vec{p} = 0$. 
As our system possesses no continuous symmetry, there are 
furthermore no conserved charges and thus we cannot introduce 
a chemical potential; only $T$ appears in the result after 
taking the $V\to\infty$ limit. We then write 
\ba
 \mathcal{Z}(V,T) 
 \; = \; 
 \exp\Bigl[ - \frac{V}{T}\, f(T)  \Bigr] 
 & = & 
 \int_{-\infty}^{\infty} \!\!\! {\rm d} \bar\phi 
 \int_\rmi{$ P \neq 0$} \!\!\!\!\!\!\!\! \mathcal{D} \phi'
 \;\; \exp\Bigl({ 
 -S^{ }_\iE[\phi = \bar\phi + \phi'] }\Bigr) \\
 & \equiv  &  
 \int_{-\infty}^{\infty} \!\!\! {\rm d} \bar\phi 
 \; \exp\left[ -\frac{V}{T} V^{ }_\rmi{eff}(\bar\phi) \right]
 \;. \la{final_int}
\ea 
We note that the thermodynamic limit 
$V \to\infty$ is to be taken only after 
the evaluation of $V^{ }_\rmi{eff}(\bar\phi)$, and that 
$\int_0^\beta\! {\rm d}\tau \int_\vec{x} \; \phi' = 0$, 
given that $\phi'$ by definition 
only has modes with $ P\neq 0$.

In order to carry out the integral in \eq\nr{final_int}, we expand 
$V^{ }_\rmi{eff}(\bar\phi)$ around 
its absolute minimum $\bar\phi^{ }_\rmi{min}$, 
and perform the corresponding Gaussian integral:\footnote{%
 To be precise an infinitesimal ``source'' should be added in order
 to pick a unique minimum. 
 } 
\ba
 V^{ }_\rmi{eff}(\bar\phi) & = &  
 V^{ }_\rmi{eff}(\bar\phi^{ }_\rmi{min}) + \fr12 
 V_\rmi{eff}''(\bar\phi^{ }_\rmi{min})
 (\bar\phi - \bar\phi^{ }_\rmi{min})^2 + \ldots
 \;, \\
 \int_{-\infty}^{\infty} \!\!\! {\rm d} \bar\phi \;
 \exp\Bigl[{ -\frac{V}{T} V^{ }_\rmi{eff}(\bar\phi) }\Bigr]  & \approx &
 \exp\Bigl[{ -\frac{V}{T} V^{ }_\rmi{eff}(\bar\phi^{ }_\rmi{min}) }\Bigr]\, 
 \sqrt{\frac{2\pi T}{ V_\rmi{eff}''(\bar\phi^{ }_\rmi{min}) V}} 
 \;.
\ea
Thereby the free energy density reads
\be
 f(T)  =   
 V^{ }_\rmi{eff}(\bar\phi^{ }_\rmi{min}) + 
 \mathcal{O}\left( \frac{\ln V}{V} \right)
 \;. \
\ee
In other words, in the thermodynamic limit $V\to \infty$, the problem
of computing $f(T)$
reduces to determining $V^{ }_\rmi{eff}$ and finding its minima. Note that 
$\bar\phi^{ }_\rmi{min}$ depends on the parameters
of the problem, particularly on~$T$.

Let us now ask under which conditions 
a first order transition could emerge. We can write
\ba
 \frac{{\rm d} f(T)}{{\rm d} T} \!\! & = & \!\!
 \left[ \frac{\partial V^{ }_\rmi{eff}(\bar\phi;T)}{\partial\bar\phi}
 \frac{{\rm d} \bar\phi^{ }_\rmi{min}}{{\rm d} T} + 
 \frac{\partial V^{ }_\rmi{eff}(\bar\phi;T)}{\partial T} 
 \right]^{ }_{\bar\phi = \bar\phi^{ }_\rmi{min}} 
 \\ 
 \!\! & = & \!\! 
 \left.  \frac{\partial V^{ }_\rmi{eff}(\bar\phi;T)}{\partial T} 
 \right|^{ }_{\bar\phi = \bar\phi^{ }_\rmi{min}}
 \;, \la{phi_disc}
\ea
where we have written out the explicit temperature dependence 
of the effective potential and made use of 
the fact that $\bar\phi^{ }_\rmi{min}$
minimizes $V^{ }_\rmi{eff}$. \Eq\nr{phi_disc} makes it clear 
(if $V^{ }_\rmi{eff}$ is an analytic function of its arguments)
that 
$
 \lim_{T\to T_c^+} \frac{{\rm d} f}{{\rm d} T} \neq 
 \lim_{T\to T_c^-} \frac{{\rm d} f}{{\rm d} T}
$
only if 
$
 \lim_{T\to T_c^+} \bar\phi^{ }_\rmi{min} \neq 
 \lim_{T\to T_c^-} \bar\phi^{ }_\rmi{min}
$.
In other words, a {\em first order transition necessitates a discontinuity
in $\bar\phi^{ }_\rmi{min}$}, such as is the case in the potential 
illustrated in \fig\ref{fig:p6}.

\begin{figure}[t]

\vspace*{0.2cm}

\centerline{~~~~~~~~\epsfysize=6.0cm\epsfbox{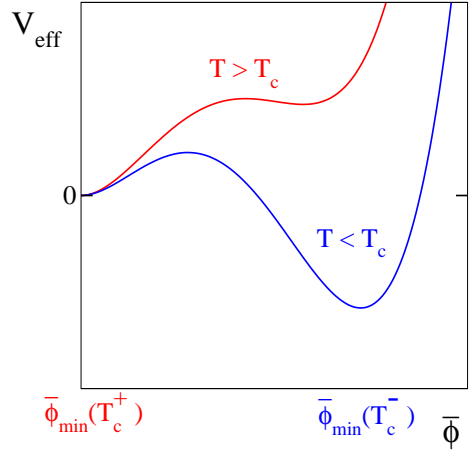}}

\caption[a]{\small
A thermal effective potential displaying
a first order phase transition.}

\la{fig:p6}
\end{figure}

Given the above considerations, our task 
becomes to evaluate $V^{ }_\rmi{eff}$. Before proceeding with 
the computation, let us formulate the generic rules
that follow from the analogy between the definition of the quantity,
\be
\exp\left[ -\frac{V}{T} V^{ }_\rmi{eff}(\bar\phi) \right] = 
 \int_\rmi{$ P \neq 0$} \!\!\!\!\!\!\!\! \mathcal{D} \phi'
 \;\; \exp\Bigl({ 
 -S^{ }_\iE[\phi = \bar\phi + \phi'] }\Bigr) \; ,
\ee 
and that of the free energy density, $f(T)$, discussed in \se\ref{se:wce}:
\bi
\item[(i)]
Write $\phi = \bar\phi + \phi'$ in ${L}^{ }_\iE$.

\item[(ii)]
The part only depending on $\bar\phi$ is the zeroth order, or tree-level, 
contribution to $V^{ }_\rmi{eff}$.

\item[(iii)]
Any terms linear in $\phi'$ should be omitted, because 
$\int_0^\beta\! {\rm d}\tau \int_\vec{x} \; \phi' = 0$.

\item[(iv)]
The remaining contributions to $V^{ }_\rmi{eff}$ are obtained like $f(T)$
before, cf.\ \eq\nr{compact_rule}, 
except that the masses and couplings of $\phi'$ now depend on 
the ``shift'' $\bar\phi$.

\item[(v)]
However, among all possible connected diagrams, 
one-particle-reducible graphs (i.e.\ graphs where the cutting of 
a single $\phi'$-propagator 
would split the graph into two disjoint parts) should be omitted, 
since such a $\phi'$-propagator would 
necessarily carry zero momentum, which is excluded by the above definition. 

\ei

\index{Effective potential}

Remarkably, as noted in ref.~\cite{fk}, these rules are {\em identical}
to the rules that follow~\cite{rj}
from a totally different (but ``standard'') 
definition of the effective potential, 
based on a Legendre transform of the generating functional:
\ba
 e^{-W[J]} & \equiv &
 \int \! \mathcal{D}\phi \, e^{-S^{ }_\iiE - \int_X \phi J}
 \;, \\
 \Gamma[\bar\phi] & \equiv & 
 W[J] - \int_X J \bar\phi 
 \;, \quad \bar\phi \; \equiv \; \frac{\delta W[J]}{\delta J}
 \;, \\
 V^{ }_\rmi{eff}(\bar\phi) & \equiv &  
 \frac{T}{V}\, \Gamma[\bar\phi]
 \quad \mbox{for} \quad \bar\phi = \; \mbox{constant}
 \;.  
\ea
However, our procedure is actually better than the Legendre transform one, 
because it is defined for any value of $\bar\phi$, 
whereas the existence of a Legendre transform requires
certain (invertibility) properties from the functions concerned, 
which has led to numerous discussions about whether
the effective potential necessarily needs to be a convex function. 

Let us now proceed to the practical computation. 
Implementing steps (i) and (ii), and indicating terms dropped 
in step (iii) by square brackets, we get
\ba
 \fr12 (\partial^{ }_\mu \phi)^2 & \to &  
 \fr12 (\partial^{ }_\mu \phi')^2 
 \;, \\
 -\fr12 m^2 \phi^2  & \to &   
 -\fr12 m^2 \bar\phi^2  
 -\left[ m^2 \bar\phi \phi' \right]  
 -\fr12 m^2 \phi'^2
 \;, \la{shift_2} \\
  \fr14\lambda \phi^4  & \to &   
 \fr14\lambda \bar\phi^4 + 
 \left[ \lambda \bar\phi^3 \phi' \right] + 
 \fr32 \lambda \bar\phi^2 \phi'^2 + 
 \lambda \bar\phi \phi'^3 + 
 \fr14 \lambda \phi'^4
 \;,  \la{shift_4} \\
 \int_0^\beta\! {\rm d}\tau \int_\vec{x}\; 
 & = & \frac{V}{T} 
 \;, \\
 V^{(0)}_\rmi{eff} (\bar\phi)  & = &  
 -\fr12 m^2\bar\phi^2 + \fr14 \lambda \bar\phi^4 
 \;, 
\ea
where one should in particular note that $V^{(0)}_\rmi{eff} (\bar\phi)$
is independent of the temperature $T$. 

The dominant ``thermal fluctuations'' or ``radiative corrections''
arise at the 1-loop order, and follow from the part quadratic 
in $\phi'$ as linear terms are dropped. 
Combining \eqs\nr{shift_2} and \nr{shift_4}, the ``effective'' mass 
of $\phi'$ reads $m_\rmi{eff}^2 \equiv - m^2 + 3 \lambda \bar\phi^2$, 
and the corresponding contribution to the effective potential becomes
\ba
 \exp\Bigl({-\frac{V}{T} V_\rmi{eff}^{(1)}}\Bigr)
 \!\! & = & \!\! 
 \int \! \mathcal{D} \phi' \;
 \exp\Bigl({- \int_0^\beta \! {\rm d}\tau \! \int_\vec{x} \;
 \fr12 \phi' \left[ -\partial_\mu^2 + m_\rmi{eff}^2 \right] \phi'}
 \Bigr) \\
 \!\! & = & \!\! 
 \int \! \mathcal{D} \tilde \phi' \;
 \exp\Bigl({- \frac{T}{V} \sum_{p^{ }_n,\vec{p}}
 \fr12 \tilde \phi' \left[ 
 p_n^2 + \vec{p}^2 + m_\rmi{eff}^2 
 \right] \tilde \phi' }\Bigr) \\
 \!\! & = & \!\! 
 \mathcal{C} 
 \biggl[ \prod_{ P \neq 0} (p_n^2 + \vec{p}^2 + m_\rmi{eff}^2)
 \biggr]^{-\fr12} 
 \; ; \\
 V_\rmi{eff}^{(1)}(\bar\phi) 
 \!\! & = & \!\! 
 \lim_{V\to\infty} \frac{T}{V} \sum_{ P \neq 0}
 \biggl[  \fr12 
 \ln(p_n^2 + \vec{p}^2 + m_\rmi{eff}^2) 
 - \mbox{const.} \biggr] 
 \;.
\ea
In the infinite-volume limit, 
where the omission of the zero mode is irrelevant,  
this goes over to the function
$J(m^{ }_\rmi{eff},T)$ defined in \eqs\nr{JmT_1} and \nr{JmT}. 
We return to the properties of this function presently, but 
let us first specify
how higher-order corrections to this result can be obtained. 

Higher-order corrections come from 
the remaining terms in \eq\nr{shift_4}, paying
attention to rules~(iv) and~(v): 
\ba
 \exp\Bigl[ { 
 -\frac{V}{T} V_\rmi{eff}^{(\ge 2)} (\bar\phi) 
 }\Bigr]
 & = & 
 \left\langle 
 \exp\Bigl( {- S^{ }_\rmi{$E$,$I$}[\bar\phi,\phi']} \Bigr) - 1  
 \right\rangle^{ }_\rmi{1PI}
 \;,  \\
 S^{ }_\rmi{$E$,$I$}[\bar\phi,\phi'] & = &  
 \int_0^\beta \! {\rm d}\tau \int_\vec{x}
 \left[ \lambda\, \bar\phi\, \phi'^3 + 
 \fr14 \lambda\, \phi'^4  
 \right]
 \;. 
\ea
Here the propagator to be used reads 
\be
 \left\langle
 \tilde\phi'( P) \tilde\phi'( Q) \right\rangle = 
 \frac{V}{T} \;\delta^{ }_{ P,-  Q}\; 
 \frac{1}{p_n^2 + \vec{p}^2 + m_\rmi{eff}^2 }
 \;.
\ee
The $V\to\infty$ limit leads to a scalar propagator, 
\eq\nr{prop_PQ}, with the mass $m^{ }_\rmi{eff}$.

\begin{figure}[t]
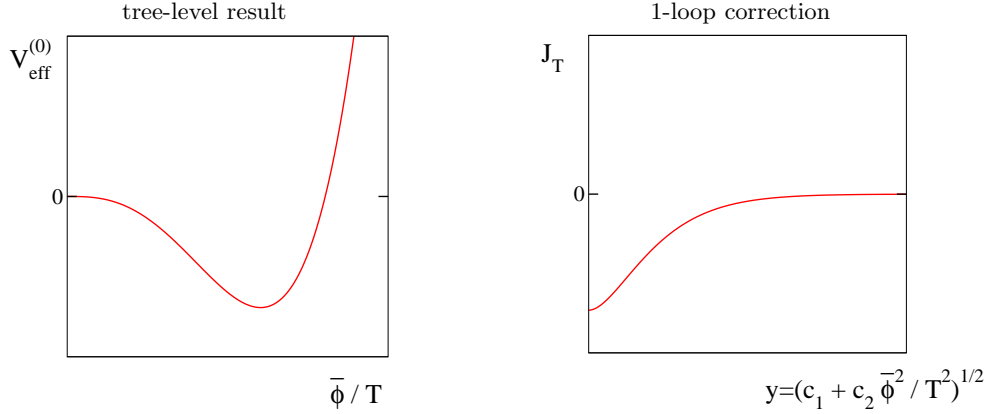


\hspace*{3.4cm}
{\small tree-level result} 
\hspace*{4.6cm}
{\small 1-loop correction} 

\vspace*{1mm}

\hspace*{2cm}\epsfysize=5.0cm\epsfbox{p8b.eps}%
\hspace*{2cm}\epsfysize=5.0cm\epsfbox{p8a.eps}

\caption[a]{\small
A comparison of the shapes of the tree-level zero-temperature
potential and the 1-loop thermal correction. The function $J^{ }_T$
is given in \eq\nr{JT_again}.}

\la{fig:p8}
\end{figure}

We now return to the evaluation of 
the 1-loop effective potential after taking $V \to \infty$. 
{}From \eq\nr{JmT_1} we have
\be
 V_\rmi{eff}^{(1)}(\bar\phi) = 
 \int_\vec{p}
 \left[ 
  \frac{\E^{ }_p}{2} + T \ln \left( 
  1 - e^{-\beta \E^{ }_p} \right)
 \right]^{ }_{\E^{ }_p = \sqrt{{p}^2 + m_\rmi{eff}^2 }}
 \;,
\ee
the temperature-dependent part of which is given by \eq\nr{JTm}:\footnote{%
  Note that even though $\bar\phi$-dependent, the $T=0$ part 
  ``only'' renormalizes the parameters $m^2$ and $\lambda$ that 
  appear in $V_\rmi{eff}^{(0)}$. These are important effects 
  in any quantitative
  study but can be omitted for a qualitative understanding. 
 }
\ba
 J^{ }_T(m^{ }_\rmi{eff}) \!\! & = & \!\! 
 \int_\vec{p}\;
 T \ln \left( 
  1 - e^{-\beta \E^{ }_p} \right)
 \; \stackrel{d=3}{=} \; 
 \frac{T^4}{2 \pi^2}
 \int_0^\infty \! {\rm d}x \, x^2 \ln 
 \left[ 
 1 - e^{-\sqrt{x^2 + y^2}}
 \right]^{ }_{y = \tfr{m^{ }_\rmi{eff}}{T} }
 \;. \la{JT_again}
\ea
This function was evaluated in \fig\ref{fig:exe3} 
on p.~\pageref{fig:exe3}; its shape in comparison with the 
zero-temperature potential is illustrated
in \fig\ref{fig:p8}.  
Clearly, the symmetric minimum becomes more favourable
(has a smaller free energy density) at higher temperatures.

In order to be more quantitative, let us study what happens
at $\pi T \gg m^{ }_\rmi{eff}$ where, from \eq\nr{JTm_res}, 
\be
  J^{ }_T(m^{ }_\rmi{eff})
 = -\frac{\pi^2 T^4}{90}
 + \frac{m_\rmi{eff}^2 T^2}{24}
 - \frac{m_\rmi{eff}^3 T}{12\pi}
 - \frac{m_\rmi{eff}^4}{2(4\pi)^2}
 \biggl[ 
  \ln\biggl( \frac{m^{ }_\rmi{eff} e^{\gammaE}}{4\pi T} \biggr) - \fr34
 \biggr]
 + \rmO\Bigl(\frac{m_\rmi{eff}^6}{\pi^4 T^2}\Bigr)
 \;. \la{JmT_res_2}
\ee
Keeping just the leading mass-dependent term leads to 
\be
 V_\rmi{eff}^{(0)} + 
 V_\rmi{eff}^{(1)} = 
 [\bar\phi\mbox{-indep.}] + 
 \fr12\left( -m^2 + \frac{\lambda  T^2}{4}
 \right)\bar\phi^2 + \fr14 \lambda \bar\phi^4 
 \;. \la{V_thermal_again}
\ee
We already knew that 
for $T = 0 $ 
the symmetry is broken; from here we observe that
for $T \gg 2 m/\sqrt{\lambda}$ 
it is restored. For the Standard Model Higgs field, 
this was realized in refs.~\cite{ki}--\hspace*{-1.1mm}\cite{swT}. 
Hence, there must 
be a phase transition of some kind; this is sketched in \fig\ref{fig:p8c}.

\begin{figure}[t]

\vspace*{3mm}

\centerline{\epsfysize=6.0cm\epsfbox{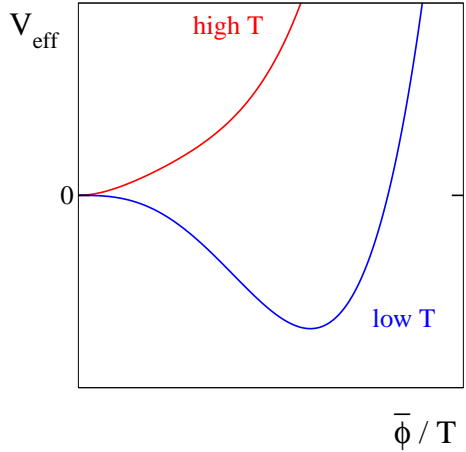}}

\vspace*{-1mm}

\caption[a]{\small
An illustration of the  
effective potential in \eq\nr{V_thermal_again}, 
possessing a phase transition.}

\la{fig:p8c}
\end{figure}

We may subsequently ask a refined question, namely, 
what is the order of the transition? In order to get 
a first impression, let us include the next term
from \eq\nr{JmT_res_2} in the effective potential. 
Proceeding for easier illustration 
to the $m^2 \to 0$ limit, we thereby obtain 
\be
 V_\rmi{eff}^{(0)} + 
 V_\rmi{eff}^{(1)} = 
 [\bar\phi\mbox{-indep.}] + 
 \frac{\lambda}{8} T^2 \bar\phi^2 
 - \frac{T}{12\pi}
 \left( 3\lambda \right)^{3/2} |\bar\phi|^3
 + \fr14 \lambda \bar\phi^4
 \;. \la{bumb}
\ee
This could describe a 
``fluctuation induced'' 
first order transition, 
as is illustrated in \fig\ref{fig:p8d}. \index{First order phase transition}

\begin{figure}[t]

\vspace*{3mm}

\centerline{\epsfysize=6.0cm\epsfbox{p8d.eps}}

\vspace*{-1mm}

\caption[a]{\small
A sketch of the structure described by \eq\nr{bumb}.}

\la{fig:p8d}
\end{figure}

We should not rush to conclusions, however. Indeed, it can 
be seen from \eq\nr{bumb} that the broken minimum appears where
the cubic and quartic terms are of similar magnitudes, i.e., 
\be
 \frac{ {T}\lambda^{\fr32} |\bar\phi|^3 }{\pi} 
 \;\sim\;
 \lambda |\bar\phi|^4
 \quad \Rightarrow \quad
 |\bar\phi| 
 \;\sim\;
 \frac{\lambda^{\fr12} T }{\pi}
 \;. 
\ee
However, the expansion parameter related to higher-order corrections, 
discussed schematically in \se\ref{se:Linde}, then becomes 
\be
 \frac{\lambda T}{\pi m^{ }_\rmi{eff}}
 \;\sim\; 
 \frac{\lambda T}{\pi \sqrt{3\lambda \bar\phi^2}}
 \;\sim\;
 \frac{\lambda^{\fr12} T}{\pi |\bar\phi|}
 \;\sim\;
 \mathcal{O}(1)
 \;. \la{exp_param}
\ee
In other words, the perturbative prediction is {\em not} 
reliable for the order of the transition.

On the other hand, a reliable analysis can again be carried out
with effective field theory techniques, as discussed in 
\se\ref{se:DR_QCD}. In the case of a scalar field theory, 
the dimensionally reduced action takes the form \index{Dimensional reduction}
\be
 S^{ }_\rmi{eff} = \frac{1}{T} \int_\vec{x}
 \left[ 
 \fr12 \left( \partial^{ }_i \phi^{ }_3 \right)^2 + 
 \fr12 m_3^2 \phi_3^2 + 
 \fr14 \lambda^{ }_3 \phi_3^4 + ...
 \right]
 \;, \la{Seff_phi}
\ee
with the effective couplings reading 
\ba
 m_3^2 & = & -m_\rmii{R}^2 
 \, [ 1 + \rmO(\lambda_\rmii{R}^{ })] 
 + \fr14 \lambda_\rmii{R}^{ } T^2\, [1+ \rmO(\lambda_\rmii{R}^{ })]
 \;, \\
 \lambda^{ }_3 & = & \lambda_\rmii{R}^{ }
 \, [ 1 + \rmO(\lambda_\rmii{R}^{ })]
 \;.
\ea
This system can be studied non-perturbatively 
(e.g.\ with lattice simulations) to show that there is a {\em second order 
transition} at 
$m_3^2 \approx 0$. The transition belongs to  
the 3d Ising universality class.\footnote{%
 To be precise we should note that scalar field theories suffer
 from the so-called ``triviality'' problem 
 (cf.\ e.g.\ refs.~\cite{tri1,tri2}): the only
 4-dimensional continuum 
 theory which is defined on a non-perturbative level is the one 
 with $\lambda^{ }_\rmii{R} = 0$.  Therefore, our discussion implicitly
 concerns a scalar field theory which has a finite ultraviolet cutoff. 
 }

\index{Triviality of scalar field theory}

Finally, we note that 
if the original theory is more complicated
(containing more fields and coupling constants), 
it is often possible to arrange the couplings so that the first
order signature seen in perturbation theory is physical.
Examples of systems where this happens include: 
\bi

\item
A theory with {\em two real scalar fields} can have a first order transition, 
if the couplings between the two fields are tuned 
appropriately~\cite{first1,twostep1}. An example
is given in appendix~A.

\item
A theory with {\em a complex scalar field and U(1) gauge symmetry}, 
which happens to form the Ginzburg-Landau theory of superconductivity, does
have a first order transition, if the quartic coupling
$\lambda_\rmii{R}^{ }$ is
small enough compared with the electric coupling squared, 
$e_\rmii{R}^2$~\cite{first2}.

\item
The standard electroweak theory, 
with a Higgs doublet and SU(2)$\times$U(1) gauge symmetry, 
can also have a first order transition if the scalar 
self-coupling $\lambda_\rmii{R}^{ }$ is small 
enough~\cite{first3,ae_2}. However, this possibility is not 
realized for the physical value of 
the Higgs mass $m^{ }_\iH \approx 125$~GeV
(for a review, see ref.~\cite{proc}). On the other hand, in many extensions
of the Standard Model, for instance in theories containing more than one scalar
field, first order phase transitions have been found
(for a review see, e.g.,\ ref.~\cite{rev4f_2}). 

\ei

\index{Spinodal decomposition}

If the transition is of first order, 
its real-time dynamics is non-trivial. 
Upon lowering the temperature, such a transition normally 
proceeds through supercooling and 
a subsequent nucleation of bubbles of the low-temperature phase, 
which then expand rapidly and fill the volume. (If bubble nucleation
does not have time to take place due to very fast cooling, it is  
possible to enter a regime of ``spinodal decomposition''
in which any ``barrier'' between the two phases disappears.)
We turn to this problem in the next section. 


\subsection*{Appendix A: Strong phase transition with two scalar fields}

Let us generalize the theory in \eq\nr{pot_phi} by including another
scalar field, denoted by $\chi$:
\ba
 {L}^{ }_{E} & = & 
  \fr12 \partial_{\mu}^{ }\phi\, \partial^{ }_{\mu} \phi
 + 
  \fr12 \partial_{\mu}^{ }\chi\, \partial^{ }_{\mu} \chi
 + 
  V(\phi,\chi)
 \;, 
 \\ 
  V(\phi,\chi)
 & = & 
 -\frac{m^2\phi^2}{2} + \frac{\lambda\phi^4}{4}
 -\frac{M^2\chi^2}{2} + \frac{\kappa\chi^4}{4}
 + \frac{\gamma\phi^2\chi^2}{2} 
 \;.
\ea
The form of the coupling between the fields has been constrained 
by imposing a Z(2) symmetry. 
The dimensionally reduced theory is like in \eq\nr{Seff_phi}, with the 
potential having the form 
\ba
  V^{ }_3(\phi^{ }_3,\chi^{ }_3)
 & = & 
  \frac{m_3^2\phi_3^2}{2} + \frac{\lambda^{ }_3\phi_3^4}{4}
 +\frac{M_3^2\chi_3^2}{2} + \frac{\kappa^{ }_3\chi_3^4}{4}
 + \frac{\gamma^{ }_3 \phi_3^2\chi_3^2}{2} 
 + ...
 \;. 
\ea
The most important parameters are the thermal masses, 
\be
 m_3^2 \;\approx\; -m^2 + \frac{(3\lambda + \gamma)T^2}{12} 
 \;, \quad 
 M_3^2 \;\approx\; -M^2 + \frac{(3\kappa + \gamma)T^2}{12} 
 \;. 
\ee 
Therefore, the respective symmetries tend to get restored at
(we assume $\gamma > 0$ here) 
\ba
 T_\phi^2 \;\approx\; \frac{12 m^2}{3\lambda + \gamma}
 \;, \quad
 T_\chi^2 \;\approx\; \frac{12 M^2}{3\kappa + \gamma}
 \;. \la{Tc_s}
\ea
Here the parameters are renormalized even if for simplicity we do not 
show the subscripts $\{\}_\rmii{R}^{ }$.

\index{Two-step phase transition}

The idea is now as follows. Suppose that as we cool down from very high
temperatures, the symmetry gets first broken
in the $\chi$-direction. According to \eq\nr{Tc_s}, this requires 
$T^{ }_\chi > T^{ }_\phi$, i.e.\ 
\be
 \frac{M^2}{m^2} > \frac{3\kappa + \gamma}{3\lambda + \gamma}
 \;. \la{constraint_1}
\ee
At a lower temperature, we assume that the minimum with 
a non-zero expectation value of $\phi$ becomes deeper, so that we 
go to the ``usual'' vacuum. 
Then, as we will see, the two minima may be connected
by a saddle point in between, and the transition
can become strong. 
This scenario is known as a ``two-stage'' or ``two-step'' 
transition~\cite{twostep1}.

In order to see if this can work, let us consider the extrema of the 
potential at low temperatures 
($T \ll T^{ }_{\phi}, T^{ }_{\chi}$). Extrema are 
found at 
\be
 \partial^{ }_\phi V = \partial^{ }_\chi V = 0 
 \; \Leftrightarrow \; 
 \left\{
 \begin{array}{ccl}  
 \displaystyle
 0 & = & \phi \, 
 ( - m^2 + \lambda \phi^2 + \gamma \chi^2 ) \;, \\[3mm] 
 0 & = & \chi \, 
 ( - M^2 + \kappa \chi^2 + \gamma \phi^2 ) \;.  
 \end{array}
 \right.
\ee
There are four solutions, which we label as 
\ba
 \mbox{(a)}: &&
 \phi = \chi = 0 \;, \la{case_a} \\ 
 \mbox{(b)}: && 
 \phi = 0\;, \quad \chi^2 = \frac{M^2}{\kappa} \;, \la{case_b} \\ 
 \mbox{(c)}: && 
 \chi = 0\;, \quad \phi^2 = \frac{m^2}{\lambda} \;, \la{case_c} \\ 
 \mbox{(d)}: && 
 \biggl( 
  \begin{array}{c} 
    \phi^2 \\ 
    \chi^2 
  \end{array} 
 \biggr) 
 = 
 \frac{1}{\lambda\kappa - \gamma_{ }^2}
 \biggl( 
  \begin{array}{c} 
    \kappa\, m^2 - \gamma\, M^2 \\ 
    \lambda\, M^2 - \gamma\, m^2  
  \end{array} 
 \biggr) 
 \;. \la{case_d}
\ea
Then we need to inspect which of these are minima, by considering
the ``mass matrix''  
\be
 \biggl( 
  \begin{array}{cc}
   \partial_\phi^2 V & \partial^{ }_{\phi}\partial^{ }_{\chi} V \\ 
   \partial^{ }_{\phi}\partial^{ }_{\chi} V & \partial_\chi^2 V
  \end{array}
 \biggr) 
 = 
 \biggl( 
  \begin{array}{cc}
    -m^2 + 3\lambda\phi^2 + \gamma\chi^2 \hspace*{-0.5cm}  & 
    2\gamma\phi\chi  \\ 
    2\gamma\phi\chi  & \hspace*{-0.5cm}
    -M^2 + 3 \kappa\chi^2 + \gamma\phi^2  
  \end{array}
 \biggr)
 \;.  \la{mass_matrix}
\ee
Inserting \eq\nr{case_a}, the extremum (a) is a local maximum. 
Inserting \eqs\nr{case_b} and \nr{case_c}, the extrema (b) and (c)
are local minima if 
\ba
 \mbox{(b) minimum} & \Leftrightarrow &
  \;\;  \frac{M^2}{m^2} > \frac{\kappa}{\gamma}  
 \;, \la{constraint_2}
 \\ 
 \mbox{(c) minimum} & \Leftrightarrow &
 \;\; \frac{M^2}{m^2} < \frac{\gamma}{\lambda} 
 \;. \la{constraint_3}
\ea
It follows from \eqs\nr{constraint_2} and \nr{constraint_3} that 
$  \gamma_{ }^2 > \lambda \kappa $ (recalling that we consider the 
case $\gamma > 0$ here). 
By inserting \eq\nr{case_d} into 
\eq\nr{mass_matrix} and taking the determinant, it can be verified
that under these conditions the determinant is negative, i.e.\ 
the extremum (d) is a saddle point. 

Finally, consider the value of the potential at the local minima. 
A substitution shows that 
\be
 V(0,\chi^{ }_\rmi{min}) = -\frac{M^4}{4\kappa}
 \;, \quad 
 V(\phi^{ }_\rmi{min},0) = -\frac{m^4}{4\lambda}
 \;. 
\ee 
The desired $\phi^{ }_\rmi{min}$-minimum is realized at
low temperatures if  
$
  | V(\phi^{ }_\rmi{min},0) | > | V(0,\chi^{ }_\rmi{min}) | 
$, 
i.e.\
\be
 \frac{M^2}{m^2} \; < \; \sqrt{\frac{\kappa}{\lambda}}
 \;. 
 \la{constraint_4}
\ee

In total, we thus have four constraints to satisfy, 
\eqs\nr{constraint_1}, \nr{constraint_2}, \nr{constraint_3} and 
\nr{constraint_4}. Denoting $\hat\kappa \equiv \kappa/\gamma$ 
and  $\hat\lambda \equiv \lambda/\gamma$, all of these can be 
respected simultaneously if 
\ba
 \hat\lambda \in \Bigl(0,\frac{1}{3}\Bigr)  \Rightarrow  
 \hat\kappa  \in \Bigl(
 \hat\lambda , \frac{1}{9\hat\lambda} \Bigr) \;, \quad  
 \hat\lambda \in \Bigl(\frac{1}{3},1\Bigr)  \Rightarrow  
 \hat\kappa \in \Bigl( 
 \frac{1}{9\hat\lambda} , \hat\lambda \Bigr) \;, \quad
 \hat\lambda > 1  \Rightarrow  
 \hat\kappa \in \Bigl( 
 \frac{1}{9\hat\lambda} , \frac{1}{\hat\lambda} 
 \Bigr)
 \;. 
\ea
Given that we did not rely on loop effects but only on tree-level
structures, and that the final transition can take place at a low 
temperature compared with vacuum mass scales, we expect that expansion
parameters such as \eq\nr{exp_param} can be kept small, and that 
there indeed {\em is} a first order transition in this system. 

\newpage 

\subsection{Bubble nucleation rate}
\la{se:bubbles}

\index{Bubble nucleation}
\index{Nucleation rate}
\index{Latent heat}

As was mentioned in the previous section, 
if a first order transition takes place
its dynamics is non-trivial, because the discontinuity 
in energy density (``latent heat'') released 
needs to be transported or 
dissipated away. The basic mechanism
for this is {\em bubble nucleation and growth}: the transition does
not take place exactly at the critical temperature, $T^{ }_\rmi{c}$, 
but upon lowering the temperature the system
first supercools to some nucleation temperature, $T^{ }_\rmi{n}$. Around this
point bubbles of the stable phase form, and start to grow; normally
(in the case of a ``deflagration'') the latent
heat is transported away in a hydrodynamic shock wave which precedes 
the expanding bubble. 

\index{Deflagration}

The purpose of this section is to determine the probability 
of bubble nucleation, per unit time and volume, at a given 
temperature $T < T^{ }_\rmi{c}$, having in mind the phase transitions 
taking place in the early universe. Combined with the cosmological 
evolution equation for the temperature~$T$, 
which determines the rate $\dot{T}(t)$ 
with which the system passes through the 
transition point (cf.\ \se\ref{se:dm}), 
this would in principle allow us to estimate $T^{ }_\rmi{n}$
(cf.\ appendix~B).
We will, however, not get into explicit estimates here, but rather 
illustrate aspects of the general formalism, given 
that it is analogous to several other ``rate'' computations in 
quantum field theory, such as the determination of the rate 
of baryon plus lepton number violation in the Standard Model.

\begin{figure}[t]

\vspace*{0.2cm}

\centerline{\epsfysize=6.0cm\epsfbox{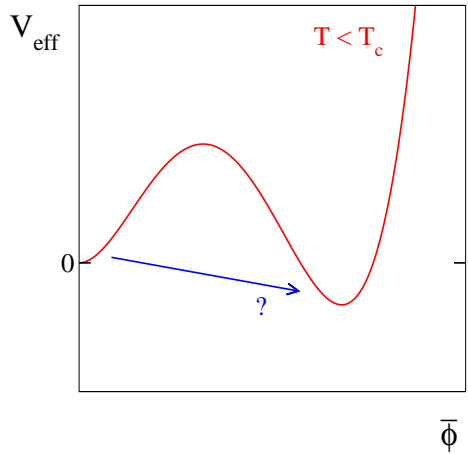}~~}

\vspace*{-0.3cm}

\caption[a]{\small
An illustration of the tunnelling process which a metastable
high-temperature state needs to undergo in a first order phase transition.} 

\la{fig:p10}
\end{figure}

In terms of the effective potential, 
the general setting can be illustrated as shown in 
\fig\ref{fig:p10}. 
For simplicity, we consider a situation in which a barrier 
between the minima
already exists in the tree-level potential $V(\bar\phi)$. 
For radiatively generated transitions, in which a barrier only appears
in $V^{ }_\rmi{eff}(\bar\phi)$, some degrees of freedom 
need to be integrated out for the discussion to apply.  

Our starting point now is an attempt at a {\em definition} of what is meant
with the nucleation rate. It turns out that this task is rather
non-trivial; in fact, it is not clear whether a completely general
definition can be given at all. Nevertheless,
for many practical purposes, the so-called Langer formalism~\cite{jl1,jl2}
appears sufficient. 

\index{Langer formalism}

The general idea is the following.  Consider first a system at zero 
temperature. Suppose we use {boundary conditions} at spatial 
infinity, $\lim_{|\vec{x}|\to\infty}\phi(\vec{x}) = 0$, in order 
to define metastable energy eigenstates. We could imagine that, 
as a result of the vacuum fluctuations taking place, 
the time evolution of these would-be states looks like
\ba
 & & |\phi(t)\rangle = 
     e^{-iEt} |\phi(0)\rangle = 
     e^{-i[\re (E) + i \im (E)]t} |\phi(0) \rangle \\
 & \Rightarrow &
   \langle \phi(t) | \phi(t) \rangle = e^{2 \im (E) \; t}
   \langle \phi(0) | \phi(0) \rangle
 \;. 
\ea
Thereby we could say that such a metastable state possesses
a decay rate, $\Gamma(E)$, given by
\be
   \Gamma(E) \simeq -2 \im (E)
 \;. \la{im_E}   
\ee
Moving to a thermal ensemble, we could analogously expect that  
\be
   \Gamma(T) \; \stackrel{?}{\simeq} \; -2 \im (F)
 \;, \la{im_Om}
\ee
where $F$ is the free energy of the system, defined in the usual way. 
It should be stressed, though, that this generalization 
is just a guess: it would 
be next to miraculous if a real-time observable, the nucleation rate, 
could be determined exactly from a Euclidean 
observable, the free energy. 

\index{Saddle point approximation}

To inspect the nature of 
our intuitive guess, we first pose the question whether 
$F$ could indeed develop an imaginary part. 
It turns out that the answer to this is positive, 
as can be seen via the following
argument~\cite{sc1}.
Consider the path integral expression for the partition function, 
\be
 F = -T \ln
 \biggl\{ \int_\rmi{b.c.}\! \mathcal{D}\phi\, \exp\Bigl( -S^{ }_\iE[\phi] 
 \Bigl) \biggr\} 
 \;, 
\ee
where ``b.c.'' refers to the usual periodic boundary conditions. 
Let us assume that we can find (at least) {\em two} different 
saddle points $\hat\phi$, 
each satisfying  
\be
 \left. \frac{\delta S^{ }_\iiE}{\delta\phi} 
   \right|^{ }_{\phi = \hat\phi} \!\!\!\! = 0
 \;, \quad
 \hat\phi(0,\vec{x}) = \hat\phi(\beta,\vec{x})
 \;, \quad
 \lim_{|\vec{x}|\to\infty} \hat\phi(\tau,\vec{x}) = 0
 \;. \la{bcs}
\ee
We assume that one of the solutions is the trivial one, 
$\hat\phi \equiv 0$, whereas the other is a non-trivial 
(i.e.\ $\vec{x}$-dependent) solution, which we henceforth 
denote by $\hat\phi(\tau,\vec{x})$.

Let us now consider fluctuations around the non-trivial saddle point, 
which we assume to have an unstable direction. Suppose for simplicity 
that the fluctuation operator around $\hat\phi$ has exactly one 
{\em negative eigenmode} 
\be
 \left. \frac{\delta^2 S^{ }_\iiE}{\delta\phi^2} \right|^{ }_{\phi = \hat\phi} 
 f^{ }_{-}(\tau,\vec{x}) = -\lambda_-^2 f^{ }_{-}(\tau,\vec{x})
 \;, \la{f_lam_def} 
\ee
whereas for the non-negative modes we define the eigenvalues through
\be
 \left. \frac{\delta^2 S^{ }_\iiE}{\delta\phi^2} \right|^{ }_{\phi = \hat\phi} 
 f^{ }_{n}(\tau,\vec{x}) = \lambda_n^2\, f^{ }_{n}(\tau,\vec{x})
 \;, \quad n \ge 0 
 \;.  \la{f_lam_pos_def} 
\ee
Writing now a generic deviation of the field $\phi$ from 
the saddle point solution in the form 
\be
 \delta \phi = \phi - \hat \phi = \sum_n \delta \phi^{ }_n 
 \equiv \sum_n c^{ }_n f^{ }_n
 \;, 
\ee
where $c^{ }_n$ are coefficients (which we assume, for simplicity, to be real),
and taking the eigenfunctions to be orthonormal
($\int_X f^{ }_m f^{ }_n = \delta^{ }_{mn}$), 
we can define the integration measure over the fluctuations as
\be \index{Fluctuation determinant}
 \int \! \mathcal{D}\phi \equiv 
 \prod_n \int \! \frac{{\rm d} c^{ }_n}{\sqrt{2\pi}}
 \;. \la{fluct_measure} 
\ee
In the vicinity of the saddle point, 
the action can be written in terms of the eigenvalues and coefficients as
\be
 S^{ }_\iE[\phi] 
 \; \approx \; 
 S^{ }_\iE[\hat\phi] + \int_X \fr12 
  \delta \phi\, \frac{\delta^2 S^{ }_\iiE[\hat\phi]}{\delta \phi^2}\,
  \delta \phi
 \; = \; 
 S^{ }_\iE[\hat\phi]  - \fr12 \lambda_-^2 c_-^2 + 
 \sum_{n\ge 0} \fr12 \lambda_n^2 c_n^2
 \;.
\ee
Then, denoting $\mathcal{Z}_0^{ } \equiv \mathcal{Z}[\hat\phi=0]$, 
we can use the {\em semiclassical approximation} to write the free energy 
in a form where  the contributions of both saddle points 
are separated,
\ba
 F \!\! & \sim & \!\! - T \ln \biggl\{ \mathcal{Z}_0^{ } \; + 
 e^{-S^{ }_\iiE[\hat\phi]}
 \int\! \frac{{\rm d}c_{-}}{\sqrt{2\pi}}\; e^{\fr12 \lambda_-^2 c_-^2}
 \int\! \prod_{n\ge 0} \frac{{\rm d} c^{ }_n}{\sqrt{2\pi}}\;  
 e^{-\fr12 \lambda_n^2 c_n^2} \biggr\} 
 \;. \la{Gamma_00}
\ea
Dealing with the negative eigenmode properly would require a careful
analysis, but in the end this 
leads (up to a factor 1/2) to the intuitive result 
\be
 \int\! \frac{{\rm d}c^{ }_{-}}{\sqrt{2\pi}}\; 
 e^{\fr12 \lambda_-^2 c_-^2}
 \sim \frac{1}{\sqrt{2\pi}} \sqrt{\frac{2\pi}{-\lambda_-^2}}
 \sim i \sqrt{\frac{1}{\lambda_-^2}} \;,  \la{neg_mode}
\ee
indicating that the partition function indeed obtains an imaginary part. 
Assuming furthermore that the contribution from the trivial
saddle point is much larger in absolute magnitude 
than that originating from the non-trivial one, 
the evaluation of \eq\nr{im_Om} leads to 
\be \index{Semiclassical approximation}
 \Gamma \sim \frac{T}{\mathcal{Z}_0^{ }} 
 \exp\left\{ -S^{ }_\iE[\hat\phi] \right\} 
 \left| 
 \det \left( 
  \delta^2 S^{ }_\iE[\hat\phi]/\delta \phi^2
 \right)
 \right|^{-\fr12}
 \;, \la{Gamma_0}
\ee 
where the determinant is simply the product of all eigenvalues. Somewhat more 
precise versions of this formula will be given
in \eqs\nr{Gamma_1} and \nr{Gamma_2} below.

\index{Instanton}
\index{Caloron}

The non-trivial saddle point contributing 
to the partition function is referred to as 
an {\em instanton}, or, if its shape is modified by a finite
temporal extent, a {\em caloron}. 
By definition, an instanton is a solution of 
the imaginary-time classical equations of motion, but it describes
the exponential factor in the rate of a real-time transition, as
suggested by the intuitive considerations above. 

\begin{figure}[t]

\vspace*{4mm}

\centerline{%
 \epsfysize=4.2cm\epsfbox{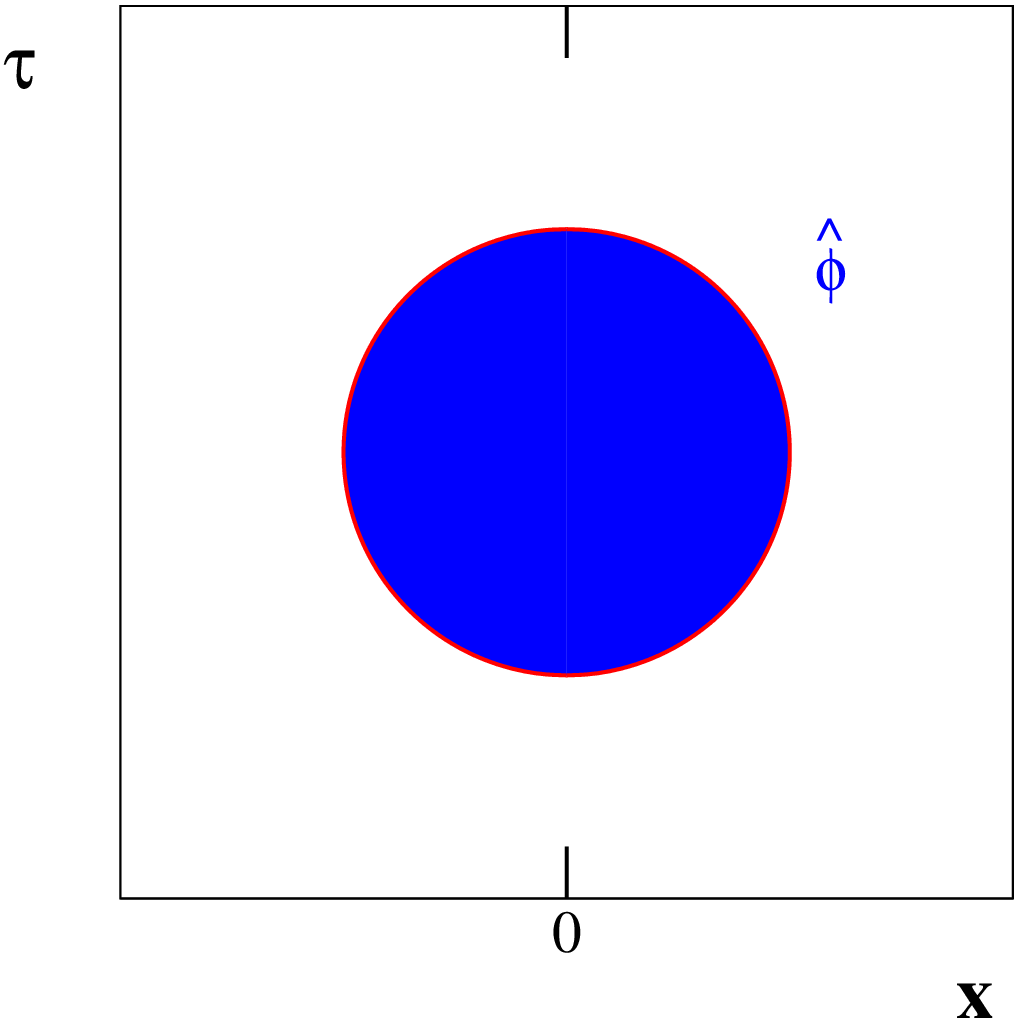}
 \hspace*{7mm}\epsfysize=4.2cm\epsfbox{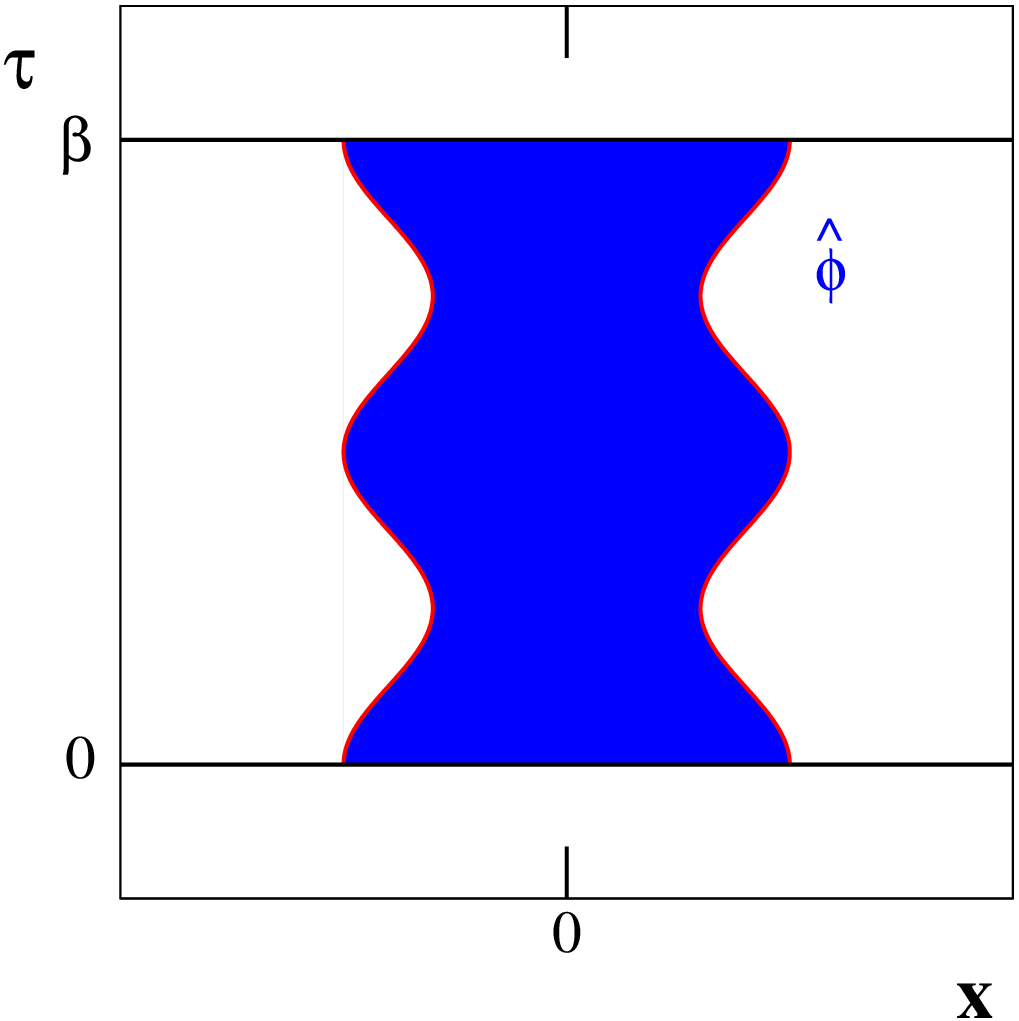}
 \hspace*{7mm}\epsfysize=4.2cm\epsfbox{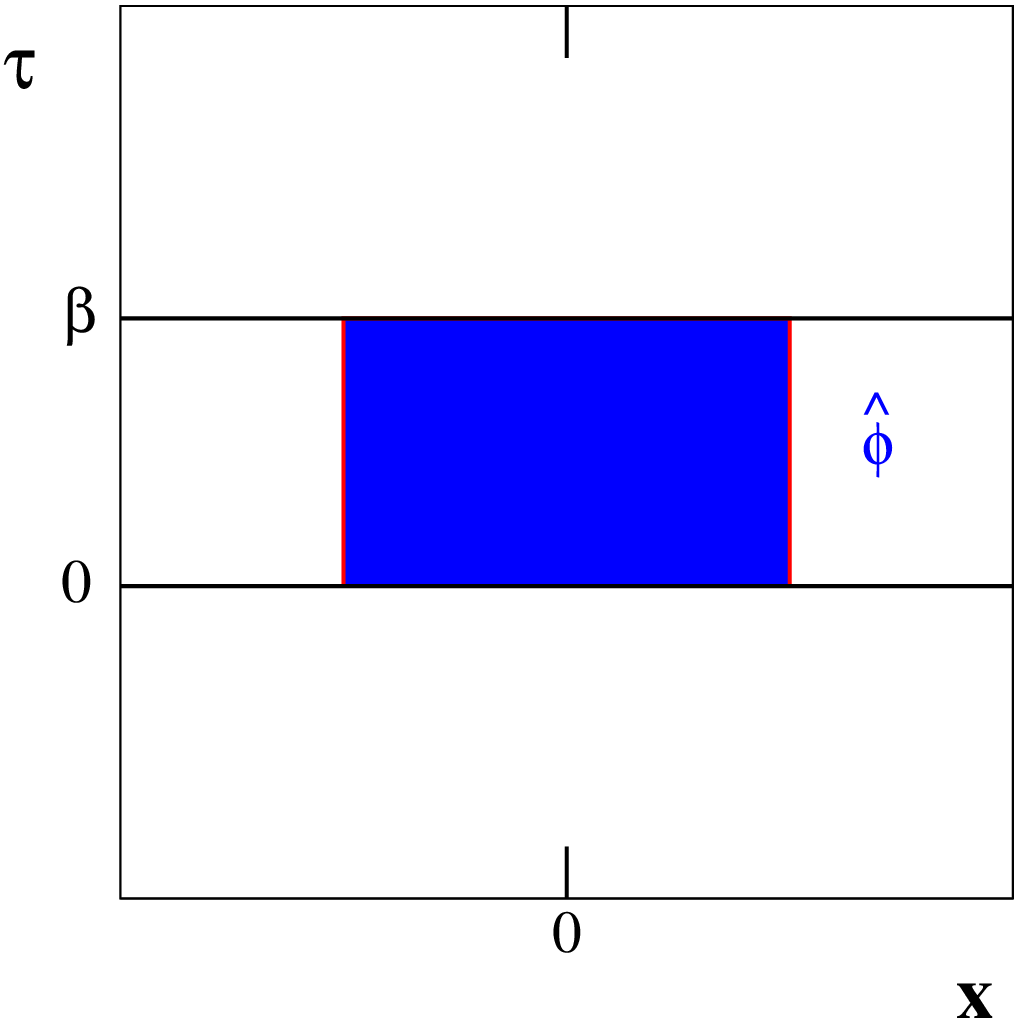}
}

\vspace*{-1mm}

\caption[a]{\small
The form of the non-trivial saddle point solution 
in various regimes: at zero temperature (left); at an intermediate
temperature (middle); and at a high temperature (right). }

\la{fig:p11}
\end{figure}

Of course, the instanton needs to respect the boundary conditions 
of \eq\nr{bcs}. Depending on the geometric shape of the instanton 
solution within these constraints, we can give different physical 
interpretations to the kind of ``tunnelling'' that the instanton 
describes. In the simplest case, when the temperature is very low
($\beta=1/T$ is very large), the Euclidean time direction is identical
to the space directions, and we can expect that the solution has
4d rotational symmetry, as illustrated in \fig\ref{fig:p11}(left). 
Such a solution is said to describe ``quantum tunnelling''. 
Indeed, had we kept $\hbar\neq 1$, \eq\nr{Gamma_0} would have had
the exponential $\exp\{ -S^{ }_\iE[\hat\phi]/\hbar \}$.

On the other hand, if the temperature increases and $\beta$ decreases, 
the four-volume becomes ``squeezed'', and this affects the form of
the solution~\cite{adl1}. 
This caloron  
is depicted in \fig\ref{fig:p11}(middle). 
Then we can say that ``quantum tunnelling'' and 
``thermal fluctuations'' both play a role. 

\index{Quantum tunnelling}
\index{Thermal fluctuations}

For very large $T$, 
the box becomes very squeezed, and we expect that the solution
only respects 3d rotational symmetry, as shown in 
\fig\ref{fig:p11}(right). 
In this situation, like in dimensional reduction, we can factorize
and perform
the integration over the $\tau$-coordinate, and the instanton action
becomes
\be
  \frac{1}{\hbar }S^{ }_\iE[\hat\phi] = \frac{1}{\hbar} 
  \beta\hbar \int_\vec{x}\; {L}^{ }_\iE
  \equiv \beta\; S^{ }_\rmi{3d}[\hat\phi]
 \;. \la{tun_clas}
\ee
We say that the transition takes place through 
``classical thermal fluctuations''.

In typical cases, the action appearing in the exponent is large, and 
thereby the exponential is very small. Just how small it is, 
is determined predominantly by the instanton action, rather than
the fluctuation determinant which does not have
any exponential factors, and is therefore ``of order unity''. 
Hence we can say that the instanton solution and its Euclidean
action $S^{ }_\iE[\hat\phi]$ play the dominant role in determining
the nucleation rate. \index{Zero mode: instanton}

At the same time, from a theoretical point of view,  
it can be said that the real ``art'' in solving  
the problem is the computation of the fluctuation 
determinant around the saddle point solution~\cite{callan}.
In fact, the eigenmodes of the fluctuation operator can be classified into: 
\bi
\item[(1)] one negative mode; 
\item[(2)] a number of zero modes; 
\item[(3)] infinitely many positive modes.  
\ei
We have already addressed the negative mode
(except for showing that there is only one), which is 
responsible for the imaginary part, so let us now look at the 
zero modes, whose normalization turns out to be somewhat non-trivial. 

The existence and multiplicity of the zero modes can be deduced from the 
classical equations of motion and from the expression of the 
fluctuation operator. Indeed, assuming the action to be of the form
\ba
 S^{ }_\iE & = & \int_0^\beta\! {\rm d}\tau \int_\vec{x}
                         \left[ \fr12 (\partial^{ }_\mu\phi)^2 + V(\phi) 
                         \right]
 \;, 
\ea
the classical equations of motion read
\ba
 \frac{\delta S^{ }_\iiE[\hat\phi]}{\delta\phi} = 0 & \Leftrightarrow &  
                         -\partial_\mu^2 \hat\phi + V'(\hat\phi) = 0
 \;. \la{cl_eom}
\ea
The fluctuation operator is thus given by 
\be
 \frac{\delta^2 S^{ }_\iiE[\hat\phi]}{\delta\phi^2} \; = \;   
                         -\partial_\mu^2 + V''(\hat\phi) 
 \;. 
\ee
Differentiating \eq\nr{cl_eom} by $\partial^{ }_\nu$
on the other hand yields the equation
\be
             \left[  -\partial_\mu^2 + V''(\hat\phi) \right] 
             \partial^{ }_\nu\hat\phi = 0
  \;,
\ee
implying that $\partial^{ }_\nu\hat\phi$ can be identified as 
one of the zero modes. Note that 
a zero mode exists (i.e.\ is non-trivial) only if the solution
$\hat\phi$ depends on the coordinate $x^\nu$; the trivial
saddle point $\hat\phi = 0$ does {\em not} lead to zero modes. 

Let us now turn to the normalization of the  zero modes.
It turns out that integrals over the zero modes are only 
defined in a finite volume, and are proportional to the volume, $V=L^d$, 
corresponding to 
translational freedom in where we place the instanton. 
A proper normalization amounts to
\be
 \int\! \frac{{\rm d}c^{ }_0}{\sqrt{2\pi}} = 
 \Bigl( \frac{\hat S^{ }_\iiE}{2\pi} \Bigr)^{\fr12} L 
 \la{prop_norm}
\ee
for $\partial^{ }_1\hat\phi$, $\partial^{ }_2\hat\phi$, 
$\partial^{ }_3\hat\phi$, and $L\to\beta$ for $\partial_0^{ } \hat\phi$.
This can be shown by considering (for simplicity) 
a one-dimensional case (of extent $L$), where 
the orthonormality condition  of the eigenmodes takes the form
\be
 \int_0^L \! {\rm d}x \, f^{ }_m f^{ }_n = \delta^{ }_{mn}
 \;. \la{normal}
\ee 
We note that the classical equation of motion, \eq\nr{cl_eom}, 
implies (upon multiplying with $\partial^{ }_x\hat\phi$ and fixing
the integration constant at infinity) a ``virial theorem'', 
$
 \fr12 (\partial^{ }_x \hat\phi)^2 =  V(\hat\phi)
$, 
where we assume $V(0) = 0$.
We then see that 
\be
 \int_0^L \! {\rm d} x \, (\partial^{ }_x\hat\phi)^2 
 = \int_0^L \! {\rm d} x \, 
 \Bigl[
   \fr12 (\partial^{ }_x \hat\phi)^2 +  V(\hat\phi)
 \Bigr]
 = S^{ }_\iE[\hat\phi] \; \equiv \; \hat S^{ }_\iE
 \;,
\ee
or in other words, that the properly normalized zero mode reads
\be
 f^{ }_0 = \tfr{1}{\sqrt{\hat S^{ }_\iiE}} \partial^{ }_x \hat\phi
 \;. \la{f0_norm} 
\ee
As the last step, we note that 
\ba
 c^{ }_0 \, f^{ }_0(x) = \tfr{c^{ }_0}{ \sqrt{\hat S^{ }_\iiE}} \, 
 \partial^{ }_x \hat\phi(x) \; \approx \;
 \hat\phi\Bigl( x + \tfr{c^{ }_0}{\sqrt{\hat S^{ }_\iiE}}  \Bigr)
 - \hat\phi(x)
 \;. 
\ea
This shows that the zero mode corresponds to translations of 
the saddle-point solution. 
Since the box is of size $L$ and assumed periodic, 
we should restrict the translations into the range
${c^{ }_0} / {\scriptstyle \sqrt{\hat S^{ }_\iiE}} \in (0,  L)$, i.e.\ 
$c^{ }_0 \in (0, L {\scriptstyle \sqrt{\hat S^{ }_\iiE}} )$. 
This directly leads to \eq\nr{prop_norm}.

We are now ready to put everything together. 
A more careful analysis~\cite{callan} shows
that the factor 2 in \eq\nr{im_E} cancels against a factor $1/2$ which we 
missed in \eq\nr{neg_mode}. Thereby \eq\nr{Gamma_0} can 
be seen to be accurate at low $T$, except for 
the treatment of the zero modes. Rectifying this point according to 
\eq\nr{prop_norm}, assuming that the number of zero modes is 4 
(according to the spacetime dimensionality), and 
expressing also $\mathcal{Z}_0^{ }$ in the Gaussian approximation, 
we arrive at  
\be
 \left. \frac{\Gamma}{V} \right|^{ }_\rmi{low $T$}
 \; \simeq \; \left( \frac{\hat S^{ }_\iiE}{2\pi} \right)^{\fr42}
 \left| \frac{\det'[-\partial^2 + V''(\hat\phi)]}
             {\det [-\partial^2 + V''(0)]} \right|^{-\fr12} 
 e^{-\hat S^{ }_\iiE} 
 \;, \la{Gamma_1} 
\ee
where $\det'$ means that zero modes have been omitted
(but the negative mode is kept). 

On the other hand, 
in the classical high-temperature limit, 
we can approximate $\partial^{ }_\tau\hat\phi = 0$, 
cf.\ \eq\nr{tun_clas}. Thereby there are 
only three zero modes, and 
\be
 - 2 \im F  \; \simeq \; 
 T V \left( \frac{\hat S^{ }_\rmi{3d}}{2\pi T} \right)^{\fr32}
 \left| \frac{\det'[-\nabla^2 + V''(\hat\phi)]}
             {\det [-\nabla^2 + V''(0)]} \right|^{-\fr12} 
 e^{-\beta \hat S^{ }_\rmi{3d}} 
 \;. 
\ee
Furthermore, it turns out that the  
guess $\Gamma \simeq -2 \im  F $ of \eq\nr{im_Om}
should in this case be corrected into~\cite{ia}
\be
 \Gamma  \simeq - \frac{\beta\lambda^{ }_-}{\pi} \im F 
 \;.  \la{im_Om_2}
\ee
A conjectured result for the nucleation rate is thus 
\be
 \left.\frac{\Gamma}{V}\right|^{ }_\rmi{high $T$} \; \simeq \; 
 \left( \frac{\lambda^{ }_-}{2\pi} \right)
 \left( \frac{\hat S^{ }_\rmi{3d}}{2\pi T} \right)^{\fr32}
 \left| \frac{\det'[-\nabla^2 + V''(\hat\phi)]}
             {\det [-\nabla^2 + V''(0)]} \right|^{-\fr12} 
 e^{- \beta{\hat S^{ }_\rmi{3d}}}
 \;. \la{Gamma_2}
\ee
Comparing \eqs\nr{im_Om} and \nr{im_Om_2}, we may expect
the high-temperature result of
\eq\nr{Gamma_2} to be more accurate than
the low-temperature result of \eq\nr{Gamma_1} above the regime
in which the prefactors cross each other, i.e.\ for 
\be
 T \gsim \frac{\lambda^{ }_-}{2\pi}
 \;. 
\ee
It should be stressed, however, that the 
simplistic approach based on the negative eigenmode $\lambda^{ }_-$
does not really give a theoretically consistent 
answer~\cite{nucl_pref1,nucl_pref2};
rather, we should understand the above analysis in the sense that 
a rate exists, and the formulae as giving its order of magnitude.

\index{Sphaleron}

Let us end by commenting on the analogous case 
of the baryon plus lepton number ($B+L$) 
violation rate~\cite{rev4a_2}. In that case, the vacua (there are 
infinitely many of them) are actually
degenerate, and the role of the field $\phi$ is played by the 
Chern-Simons number, which is a suitable coordinate for classifying
topologically distinct vacua. However, the formalism itself is 
identical: in particular, at low temperatures it may be assumed
that there is a saddle point 
solution with 4d symmetry, which is a usual instanton~\cite{instanton}, 
whereas at high temperatures (but still in the symmetry broken phase) 
the saddle point solution has 3d symmetry,  
i.e.\ is time-independent, 
and is referred to as a {\em sphaleron}~\cite{kli}. Again there are also 
zero modes, which have to be treated carefully~\cite{sphal}. 
At high temperatures a complete
analysis, even at leading order in couplings, requires non-perturbative
methods~\cite{clgt2,nucl_pref1,srate3}.


\subsection*{Appendix A: Nucleation action in the classical limit}

\index{Classical limit}
\index{Critical bubble}

In scalar field theory, the instanton 
solution (also known as the ``critical bubble'') and its Euclidean action
can be determined in a simple form, 
if we assume the classical limit of high temperatures 
and that the minima are almost degenerate. It can be shown
(cf.\ e.g.\ refs.~\cite{landau5,adl2}) that then 
\be
 \hat S^{ }_\rmi{3d} =  
 \frac{16\pi}{3} \frac{\sigma^3}{(\Delta p)^2}
 \;, \la{cl_S3d} 
\ee
where 
\be
 \sigma \equiv 
 \int_0^{\bar\phi^{ }_\rmii{broken}} \!\!\! {\rm d}\bar\phi 
 \sqrt{2 V(\bar\phi)}
\ee
is the {\em surface tension}, and 
\be
 \Delta p \equiv V(0) - V(\bar\phi^{ }_\rmi{broken}) 
 \la{Delta_p}
\ee
is the {\em pressure difference} in favour of the broken phase. 
In this limit the configuration $\hat\phi$ is called a 
{\em thin-wall bubble}.
It is important to note that 
\eq\nr{cl_S3d} implies that $\hat S^{ }_\rmi{3d} \to \infty$ for 
$\Delta p \to 0$; this is the reason why nucleation can only take 
place after some supercooling, 
when $\Delta p > 0$ and $\hat S^{ }_\rmi{3d}$ becomes finite.

\begin{figure}[t]

\vspace*{2mm}

\centerline{%
   \epsfysize=6.0cm\epsfbox{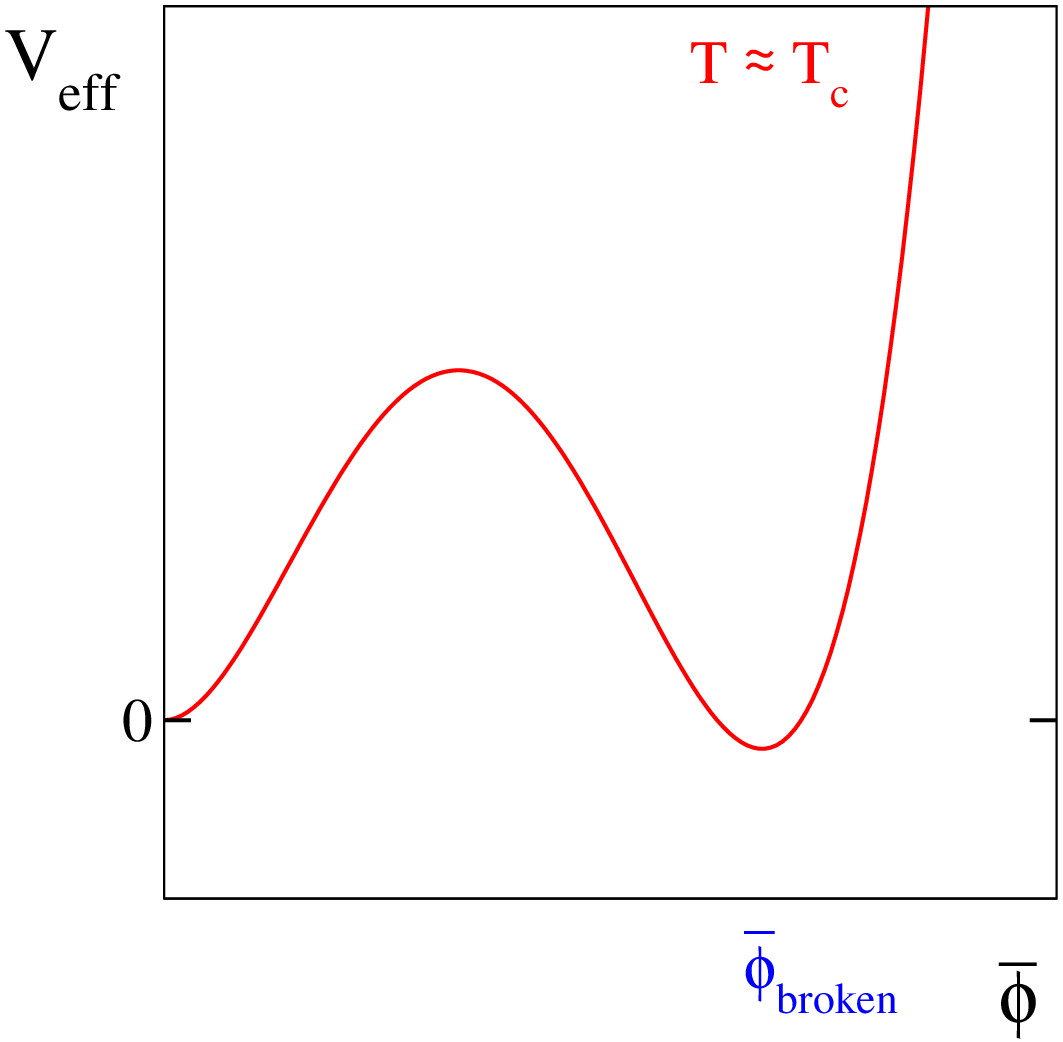}
   \hspace*{6mm}\epsfysize=6.0cm\epsfbox{p12b.eps}}

\caption[a]{\small 
Left: The effective potential describing a first order 
phase transition at the critical temperature.
Right: The profile of the critical bubble solution as a function
of the radial coordinate.} 

\la{fig:p12}
\end{figure}

The limit of almost degenerate minima is illustrated 
in \fig\ref{fig:p12}(left). 
In the classical limit, there is no dependence on $\tau$, 
and the equation of motion reads
\be
 -\nabla^2 \hat\phi + V'(\hat\phi) = 0 
 \;, \la{cl_eom_2}
\ee
which, assuming {\em spherical symmetry}, can be written as 
\be
 \frac{{\rm d}^2\hat\phi}{{\rm d}r^2} + 
 \frac{2}{r} \frac{{\rm d}\hat\phi}{{\rm d}r} = V'(\hat\phi)
 \;. \la{cl_eom_3} 
\ee
The boundary conditions in \eq\nr{bcs} can furthermore be rephrased as 
\be
 \left\{
 \begin{array}{l}
   \hat\phi(\infty) = 0 \\
   \frac{{\rm d}\hat\phi(r)}{{\rm d} r}|^{ }_{r=0} = 0 
 \end{array}
 \right. 
 \;, \la{bcs_2}
\ee
whereas the action reads
\be
 \hat S^{ }_\rmi{3d} = 4 \pi \int_0^\infty \! {\rm d}r \, r^2 \, 
 \biggl\{
  \fr12 \biggl( \frac{{\rm d}\hat\phi}{{\rm d} r} \biggr)^2
 + V(\hat\phi)
 \biggr\}
 \;. \la{S3d_2}
\ee

Before proceeding, it is useful to note that \eqs\nr{cl_eom_3} and \nr{bcs_2}
have a mechanical analogue. Indeed, rewriting $r\to t$, $V\to -U$, 
$\hat\phi\to x$, they correspond to a classical 
``particle in a valley'' problem with friction. The particle starts
at $t=0$ from $x > 0$, near the top of the hill in the potential $U$, 
and rolls then towards the other top of the hill at the origin. The 
starting point has to be slightly higher than the end point, because
the second term in \eq\nr{cl_eom_3} acts as friction. 
Therefore the broken minimum has to 
be {\em lower} than the symmetric one in order for a
non-trivial solution to exist. 

Proceeding now with the solution, we introduce the following ansatz. 
Suppose that at $r < R$, the field is constant and has a value close
to that in the broken minimum: 
\be
 \frac{{\rm d}\hat\phi}{{\rm d} r} \simeq 0
 \;, \quad
 V(\hat\phi) \simeq V(\bar\phi^{ }_\rmi{broken})
 \;. 
\ee
This is illustrated in \fig\ref{fig:p12}(right).  
The contribution to the action from this region is 
\be
 \delta \hat S^{ }_\rmi{3d} \simeq \fr43 \pi R^3 V(\bar\phi^{ }_\rmi{broken})
 \;. 
\ee
For $r > R$, we assume a similar situation, but now the field 
is close to the origin
\be
 \frac{{\rm d}\hat\phi}{{\rm d} r} \simeq 0
 \;, \quad
 V(\hat\phi) \simeq V(0) \equiv 0 
 \;, 
\ee
so that this region does not contribute to the action. 

Finally, let us inspect the region at $r\simeq R$. If $R$
is very large, the term $2\hat\phi'/R$ in \eq\nr{cl_eom_3}
is very small, and can be neglected. Thereby 
\be
 \frac{{\rm d}^2\hat\phi}{{\rm d}r^2} \simeq V'(\hat\phi)
 \;, 
\ee
which through multiplication with $\hat\phi'(r)$ can be 
integrated into 
\be
 \fr12 \biggl( \frac{{\rm d}\hat\phi}{{\rm d} r} \biggr)^2
 \simeq V(\hat\phi)
 \;. 
\ee
The contribution to the action thus becomes 
\ba
 \delta \hat S^{ }_\rmi{3d} & \simeq & 
 4 \pi R^2 \int_{R-\delta}^{R+\delta} \! {\rm d}r \, 
 \biggl( \frac{{\rm d}\hat\phi}{{\rm d} r} \biggr)^2 
 \nn & \simeq &
 4 \pi R^2 \int_0^{\bar\phi^{ }_\rmii{broken}}
 \! {\rm d}\hat\phi \, \frac{{\rm d}\hat\phi}{{\rm d} r} 
 \nn & \simeq &
 4 \pi R^2 \int_0^{\bar\phi^{ }_\rmii{broken}}
 \! {\rm d}\hat\phi \, \sqrt{2 V(\hat\phi)} 
 \;. 
\ea
The quantity 
\be
 \sigma \;\equiv\; \int_{R-\delta}^{R+\delta} \! {\rm d}r \,
 \biggl\{
  \fr12  \biggl( \frac{{\rm d}\hat\phi}{{\rm d} r} \biggr)^2 
  + V(\hat\phi)  
 \biggr\}
 \;\simeq\; \int_0^{\bar\phi^{ }_\rmii{broken}}
 \! {\rm d}\hat\phi \, \sqrt{2 V(\hat\phi)}
\ee
represents the energy density of a planar surface, 
i.e.\ a {\em surface tension}.

\index{Surface tension}

Summing up the contributions, we get
\ba
 \hat S^{ }_\rmi{3d}(R) \simeq  
 4\pi R^2 \sigma - \fr43 \pi R^3 \Delta p 
 \;, \la{S3d_R}
\ea
where $\Delta p > 0$ was defined according to \eq\nr{Delta_p}.
The so far undetermined parameter $R$ can be solved by extremizing the action, 
\be
 \delta_R \hat S^{ }_\rmi{3d}  =  0
 \;,
\ee
leading to the radius $R=2\sigma/\Delta p$.
Substituting this back to \eq\nr{S3d_R}  we get 
\be
 \hat S^{ }_\rmi{3d} =  
 4 \pi \sigma \frac{4\sigma^2}{(\Delta p)^2} -
 \fr43 \pi \frac{8\sigma^3}{(\Delta p)^2} = 
 \frac{16\pi}{3} \frac{\sigma^3}{(\Delta p)^2}
 \;. \la{hat_S3d}
\ee

Finally, 
if we are very close to $T^{ }_\rmi{c}$, $\Delta p$ can be related 
to basic characteristics of the first order transition. Indeed, 
the energy density is 
\be \index{Latent heat}
 e = T s - p \;, 
\ee 
where the entropy density reads $s = {\rm d} p/ {\rm d} T$.
Across the transition, the pressure is continuous but the energy 
density has a discontinuity called the {\em latent heat}
(here again $\Delta x \equiv x^{ }_\rmi{broken} - x^{ }_\rmi{symmetric}$): 
\be
 L \; \equiv \; - \Delta e = - T^{ }_\rmi{c}\, \Delta s 
 = - T^{ }_\rmi{c} \frac{{\rm d}  \Delta p}{{\rm d}T}
 \;. 
\ee
Therefore 
\be
 \Delta p (T) \approx \Delta p(T^{ }_\rmi{c}) + 
 \frac{{\rm d} \Delta p}{{\rm d}T}\, (T - T^{ }_\rmi{c})
 = L\, \biggl( 1 - \frac{T}{T^{ }_\rmi{c}} \biggr)
 \;.  \la{delta_p}
\ee
At the same time, the surface tension remains finite at the transition
point. Thereby the nucleation action in \eq\nr{hat_S3d} diverges
quadratically as $T \to T_\rmi{c}^-$.


\subsection*{Appendix B: Nucleation dynamics with many bubbles}

\index{Nucleation dynamics}

\begin{figure}[t]

\vspace*{2mm}

\centerline{%
   \epsfysize=5.0cm\epsfbox{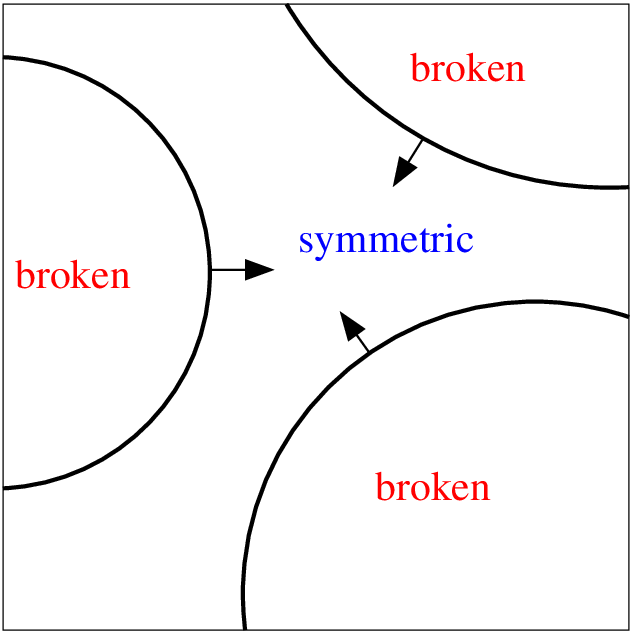}
}

\caption[a]{\small
A situation with many bubbles having 
nucleated recently and growing to fill the space.} 

\la{fig:p12d}
\end{figure}

Above we discussed the probability of nucleating a single 
bubble of the low-temperature phase. In an actual transition, many
bubbles nucleate with an increasing rate, and those that 
nucleated first have had time to grow. The transition ``completes'' 
when the bubbles have filled the space. 
This process is illustrated in \fig\ref{fig:p12d}. Let us 
sketch the practical approach that is often used for 
estimating the ``nucleation temperature'', $T^{ }_\rmi{n}$,  
around which this process effectively takes place.

The nucleation probability per time and volume is denoted 
by $p = \Gamma/V = p^{ }_0\, e^{-\hat{S}^{ }_{E}}$, where
in the classical limit $\hat{S}^{ }_{E} = \beta \hat{S}^{ }_\rmi{3d}$.
Let us write $\hat{S}^{ }_{E}$ as a function of time.  
Expanding 
$\hat{S}^{ }_{E}(t) \approx \hat{S}^{ }_{E}(t^{ }_\rmi{n}) + 
 \hat{S}'_{E}(t^{ }_\rmi{n})(t - t^{ }_\rmi{n})$ 
where $t^{ }_\rmi{n}$ is the effective 
nucleation time, and including the volume 
of a bubble growing with velocity~$v$, 
we integrate over all possible nucleation times prior to $t^{ }_\rmi{n}$, 
and assert that the transition has completed 
when the probability is of order 
unity~\cite{bubbles1}--\hspace*{-1.1mm}\cite{bubbles3}:  
\be
 1 \;\simeq\;  
 \int_{-\infty}^{t^{ }_\rmi{n}} \! {\rm d}t \,
 \frac{4\pi v^3(t^{ }_\rmi{n} - t)^3}{3}
 \, p^{ }_0\, e^{-\hat{S}^{ }_{E}(t)}
 \;\approx\; 
 \frac{8\pi v^3 \, 
  p^{ }_0}{|\hat{S}'_{E}(t^{ }_\rmi{n})|^4} 
 \, e^{-\hat{S}^{ }_{E}(t^{ }_\rmi{n})}
 \;. \la{nucl_1}
\ee
Here we noted that the nucleation action decreases with time, 
$\hat{S}'_{E}(t^{ }_\rmi{n}) < 0 $,
cf.\ the discussion below \eq\nr{delta_p}. 
Similarly, but this time not including the bubble growth factor, 
we can estimate the average inverse volume 
of the bubbles that did get nucleated: 
\be
 \frac{1}{V} 
 \;\simeq\; 
 \int_{-\infty}^{t_\rmi{n}} \! {\rm d}t \, 
  p^{ }_0\, e^{-\hat{S}^{ }_{E}(t)}
 \;\approx\;
 \frac{p^{ }_0}
  {|\hat{S}'_{E}(t^{ }_\rmi{n})|} 
 \, e^{-\hat{S}^{ }_{E}(t^{ }_\rmi{n})}
 \;\approx\; 
 \frac{|\hat{S}'_{E}(t^{ }_\rmi{n})|^3}{8\pi v^3}
 \;. 
\ee 
In the last step here, 
we inserted \eq\nr{nucl_1}. 
The average distance of the bubbles is therefore
\be
 \ell \;\equiv\; 
 \Bigl(\frac{1}{V}\Bigr)^{-1/3} \;\simeq\; 
 \frac{v}{|\hat{S}'_{E}(t^{ }_\rmi{n})|} 
 \;. \la{ell}
\ee

To summarize, the equilibrium properties of a first-order transition 
are characterized by the quantities $\Tc^{ }$, $L$, and $\sigma$, 
whereas the nucleation dynamics depends on $T^{ }_\rmi{n}$, $\ell$, 
and the velocity $v$. We can estimate $T^{ }_\rmi{n}$ from 
\eq\nr{nucl_1} after inserting 
the relation between $t^{ }_\rmi{n}$ and $T^{ }_\rmi{n}$
from \se\ref{se:dm}. For $\hat{S}^{ }_{E}$ this is trivial 
because we originally expressed $\hat{S}^{ }_{E}$ in terms of $T$
(cf.\ appendix~A), whereas 
for the time derivative in \eq\nr{nucl_1} 
we can make use of an explicit relation, 
to be derived in~\eq\nr{tT_rel_2}, in order to write 
\be
 |\hat{S}'_{E}(t^{ }_\rmi{n})| = 
 3 c_s^2 T^{ }_\rmi{n} H(T^{ }_\rmi{n}) \hat{S}'_{E}(T^{ }_\rmi{n})
 \;, 
\ee
where $H(T)$ is the Hubble rate.
According to the non-perturbative study in ref.~\cite{nucl_pref2}, 
the prefactor~$p^{ }_0$ in \eq\nr{nucl_1} can be approximated 
as $p^{ }_0 \simeq \alpha_w^5 (\frac{gT}{\mE^{ }})^2\ln(\frac{1}{g})T^4$, 
where $\alpha_w^{ } \equiv g^2/(4\pi)$ and $\mE^{ }$ is the Debye mass, 
provided that we make use of  
non-perturbative values of $L$ and $\sigma$ when 
evaluating \eq\nr{cl_S3d}. 
The velocity $v$ appearing in \eq\nr{ell} 
is a complicated function, because it depends not
only on microscopic processes~\cite{db_gdm} but also on global 
hydrodynamic aspects of the bubble dynamics which 
in turn depend on $T^{ }_\rmi{n}$~\cite{bubbles4}. 
All in all, it is fair to say that   
nucleation computations contain large uncertainties. 

\newpage 

\subsection{Particle production rate}
\la{se:ppr}

\index{Particle production rate: general}

Consider a system where some particles interact strongly
enough to be in thermal equilibrium, while others interact so weakly
that they are out of equilibrium. We can imagine that particles
of the latter type ``escape'' from the thermal system, either concretely
(if the system is of finite size) or in an abstract sense (being still
within the same volume but not interacting with 
the thermal particles). Familiar physical examples of such settings are 
the ``decoupling'' of weakly interacting dark matter 
particles in {cosmology};   
the production of electromagnetic ``hard probes'', such as photons
and lepton-antilepton pairs, in the QCD plasma generated in {heavy ion
collision experiments};  
as well as the neutrino ``emissivity'' of neutron stars, constituting the 
most important process by which neutron stars cool down. 

The purpose of this section is to develop a general formalism for 
addressing this phenomenon.\footnote{%
 Classic discussions of thermal particle production include
 refs.~\cite{dilepton1,dilepton2}, 
 which establish the dilepton and photon production
 rates from a QCD plasma as 
 \ba \index{Photon production rate} \index{Dilepton production rate}
 \frac{{\rm d} N^{ }_{\ell^-\ell^+}}
   {{\rm d}^4 \mathcal{X} {\rm d}^4 \mathcal{K}} & = &   
 \sum_{\ff\ff'}
 \frac{ -2 e^4 Q^{ }_{\ff} Q^{ }_{\ff'} 
 \, \theta(\mathcal{K}^2 - 4 m_\ell^2) } 
  {3 (2\pi)^5 \mathcal{K}^2} 
 \biggl( 1 + \frac{2 m_\ell^2}{\mathcal{K}^2}
 \biggr)
 \biggl(
 1 - \frac{4 m_\ell^2}{\mathcal{K}^2} 
 \biggr)^\fr12 n^{ }_\rmii{B}(k^0_{ }) \,\rho^{ }_{\ff\ff'}(\mathcal{K})
 \;, 
 \la{dilepton} \\
 \frac{{\rm d} N^{ }_{\gamma}}
   {{\rm d}^4 \mathcal{X} {\rm d}^3 \vec{k}} & = &   
 \sum_{\ff\ff'}
 \frac{ - e^2 Q^{ }_{\ff} Q^{ }_{\ff'}  } 
  {(2\pi)^3 k} 
 \left. n^{ }_\rmii{B}(k) \,\rho^{ }_{\ff\ff'}(\mathcal{K})
 \right|^{ }_{k^0_{ } = k}
 \;, 
 \la{photon} 
\ea 
 where $Q^{ }_{\ff}$ is the quark electric charge in units of $e$,
 and 
 $
   \rho^{ }_{\ff\ff'}(\mathcal{K}) = \int_\mathcal{X} 
   e^{i \mathcal{K}\cdot \mathcal{X}}
  \left\langle
    \fr12 [ 
    \hat \mathcal{J}_{\ff}^\mu(\mathcal{X}), 
    \hat \mathcal{J}^{ }_{\ff'\mu}(0)
    ]
  \right\rangle
 $ is a spectral function related to flavours $\ff$ and $\ff'$. 
 In the present section we follow the alternative
 formalism of ref.~\cite{als}.
  }
To keep the discussion concrete, we focus on
a simple model: the production rate 
of hypothetical scalar particles coupled to a 
gauge-invariant operator $\mathcal{J}$ composed
of Standard Model degrees of freedom. The classical Lagrangian
is assumed to take the form 
\be
 \mathcal{L}^{ }_\iM = \partial^{ }_\mu \phi^* \partial^\mu \phi
 - m^2 \phi^*\phi
 - h \, \phi^* \mathcal{J} - h^*   \mathcal{J}^* \phi + 
 \mathcal{L}^{ }_\rmi{bath}
 \la{L_c_sc}
 \;, 
\ee
where $\mathcal{L}^{ }_\rmi{bath}$ describes the thermalized
degrees of freedom. 
The first step is to derive a master equation relating 
the production rate of $\phi$'s to a certain Green's function
of ${\mathcal{J}}$'s.

\index{Density matrix}
\index{Liouville - von Neumann equation}

Let $\hat\rho$ be the density matrix of the full theory, 
incorporating all degrees of freedom, and $\hat H$ the corresponding
full Hamiltonian operator. Then the equation of motion for the density 
matrix is\footnote{%
 This is the {\em Liouville - von Neumann equation};
 its derivation proceeds roughly as 
 $$
  i \frac{{\rm d}}{{\rm d}t} | \psi \rangle = \hat H | \psi \rangle
  \;, \quad 
 -i \frac{{\rm d}}{{\rm d}t} \langle \psi | =  \langle \psi | \hat H
  \quad  
  \Rightarrow 
  \quad
  i \frac{{\rm d}}{{\rm d}t} | \psi \rangle \langle \psi | = 
  [\hat H, | \psi \rangle \langle \psi | ]
  \quad
  \Rightarrow
  \quad
  i \frac{{\rm d}}{{\rm d}t} \hat\rho(t) = 
  [\hat H, \hat\rho(t) ]
  \;. 
 $$
 } 
\be 
 i \frac{{\rm d} \hat\rho(t)}{{\rm d} t} =[\hat H,\hat\rho(t)]
 \;.
 \label{liuv}
\ee
We now split $\hat H$ up as
\be
 \hat H= \hat H^{ }_\rmi{bath} + \hat H^{ }_\rmi{$\phi$}
 + \hat H^{ }_\rmi{int}
 \;,
\ee
where $\hat H^{ }_\rmi{bath}$ is the Hamiltonian of the heat bath,  
$\hat H^{ }_\rmi{$\phi$}$ is the free Hamiltonian of the scalar fields, 
and $\hat H^{ }_\rmi{int}$, which is proportional to the
coupling constant $h$, contains the interactions between the two sets: 
\be
 \hat H^{ }_\rmi{int} = 
 \int_{\vec{x}} \bigl(h\, \hat \phi^\dagger \hat \mathcal{J} +
  h^* \hat \mathcal{J}^\dagger \hat \phi \bigr)
 \;. \la{H_c_sc} \la{Hint}
\ee
To find the density of the scalar particles, 
one has to solve \eq\nr{liuv}
with some initial conditions. We {\em
assume} that initially there were no $\phi$-particles, that is
\be
 \hat \rho (0) = \hat \rho^{ }_\rmi{bath}\otimes |0\rangle\langle 0|
 \;, 
 \label{in}
\ee
where  
$
 \hat \rho^{ }_\rmi{bath} 
 = \mathcal{Z}^{-1}_\rmi{bath} \exp(-\beta \hat H^{ }_\rmi{bath})
$, 
$
 \beta \equiv 1/T 
$,  
is the equilibrium density matrix of the heat bath 
at temperature $T$;  
and $|0\rangle$ is the vacuum 
state for the scalar particles.  

Denoting by $\hat H_0^{ } \equiv \hat H^{ }_\rmi{bath}
 + \hat H^{ }_\rmi{$\phi$}$  
a ``free'' Hamiltonian and by $\hat H^{ }_\rmi{int}$ an interaction
term, an equation of motion can be obtained for the density matrix in the
interaction picture, 
$ 
 \hat \rho_\iI^{ } \equiv 
 \exp(i \hat H_0^{ } t)\hat \rho \exp(-i \hat H_0^{ } t)
$, 
in the standard way:
\ba
 i \frac{{\rm d}}{{\rm d}t} \hat\rho_\iI^{ }(t) 
 \!\!\! & = & \!\!\!
 -\hat H_0^{ } \hat \rho_\iI^{ } + 
 e^{i \hat H_0^{ } t} [\hat H,\hat \rho(t)] e^{-i \hat H_0^{ } t}
 + \hat \rho_\iI^{ } \hat H_0^{ } 
 \nn
 \!\!\! & = & \!\!\! 
 -\hat H_0^{ } \hat \rho_\iI^{ } + 
 e^{i \hat H_0^{ } t} [\hat H_0^{ }\! +\! \hat H^{ }_\rmi{int},\hat \rho(t)] 
 e^{-i \hat H_0^{ } t}
 + \hat \rho_\iI^{ } \hat H_0^{ } 
 \nn[2mm] 
 \!\!\! & = & \!\!\! 
 e^{i \hat H_0^{ } t} [\hat H^{ }_\rmi{int},\hat \rho(t)] e^{-i \hat H_0^{ } t}
 \nn[2mm] 
 \!\!\! & = & \!\!\! 
 e^{i \hat H_0^{ } t} \hat H^{ }_\rmi{int} e^{-i \hat H_0^{ } t}  
 e^{i \hat H_0^{ } t} \hat \rho(t) e^{-i \hat H_0^{ } t} 
 -
 e^{i \hat H_0^{ } t} \hat \rho(t) e^{-i \hat H_0^{ } t}
 e^{i \hat H_0^{ } t} \hat H^{ }_\rmi{int} e^{-i \hat H_0^{ } t}  
 \nn[2mm]
 \!\!\! & = & \!\!\! 
 [\hat H_\iI^{ }(t),\hat \rho_\iI^{ }(t)]
 \;. 
\ea 
Here, as usual, 
$ \index{Interaction Hamiltonian}
 \hat H_\iI^{ }= \exp(i \hat H_0^{ } t)
 \hat H^{ }_\rmi{int} \exp(-i \hat H_0^{ } t)
$ is
the interaction Hamiltonian in the interaction picture. 

\begin{figure}[t]

\vspace*{3mm}

\centerline{\epsfysize=7.5cm\epsfbox{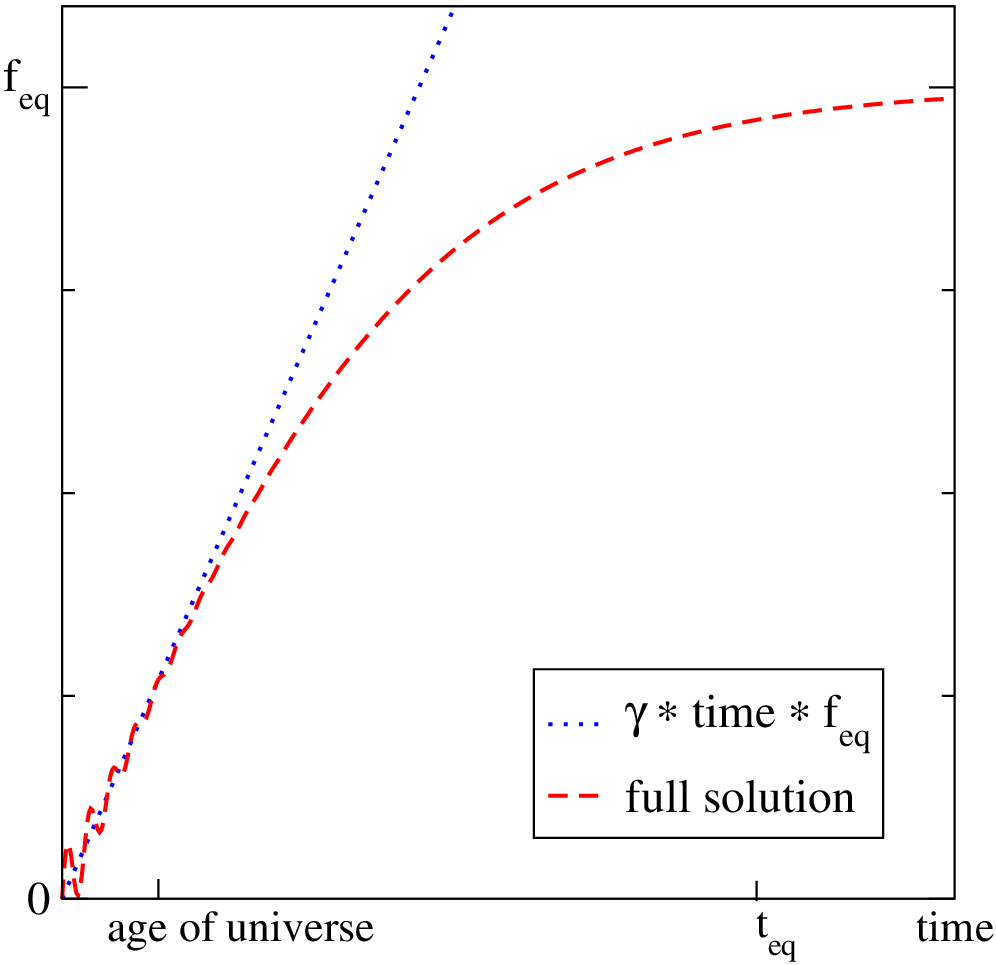}~~}

\caption[a]{\small
A sketch of how the phase space density of a weakly
interacting particle species evolves from zero towards its equilibrium
form. In many interesting cases, the equilibrium value is not reached
within the lifetime of the system (denoted here by ``age of universe''). 
Then it is important to know the rate $\gamma$, 
characterizing the linear
slope at intermediate times.}

\la{fig:equil}
\end{figure}

Now, perturbation theory with respect to $\hat H_\iI^{ }$ can be used 
to compute the time evolution of $\hat\rho_\iI^{ }$; 
the first two terms read
\ba
 \hat \rho_\iI^{ }(t) = \hat\rho_0^{ }
 - i \int_0^t \! {\rm d} t' \, 
 [\hat H_\iI^{ }(t'), \hat\rho_0^{ }]
 + (-i)^2 
 \int_0^t \! {\rm d} t' \,
 \int_0^{t'} \! {\rm d} t'' \,
 [\hat H_\iI^{ }(t'),[\hat H_\iI^{ }(t''), \hat\rho_0^{ }]]
 + \ldots \;,
 \label{pert}
\ea
where $\hat\rho_0^{ } \equiv \hat\rho(0) = \hat\rho_\iI^{ }(0)$.
We note that perturbation theory as an expansion
in $\hat H_\iI^{ }$ may break down at 
a certain time $t \simeq t^{ }_\rmi{eq}$ due to so-called secular terms. 
Physically, the reason is that 
for $t\gsim t^{ }_\rmi{eq}$ scalar particles enter thermal equilibrium 
and their concentration needs to be computed by other means (cf.\ below). 
Here we
assumed that $t \ll t^{ }_\rmi{eq}$ and thus perturbation theory should work.
At the same time, $t$ is also assumed to be much {\em larger} than the 
microscopic time scales characterizing the dynamics of the heat bath, 
say $t \gg 1/(\alpha^2 T)$, where $\alpha$ is a generic fine structure
constant. This guarantees that quantum-mechanical oscillations
get damped out, and the produced particles can be considered to constitute
a ``classical'' phase space distribution function. 
The situation is illustrated in \fig\ref{fig:equil}, with the slope $\gamma$
denoting the rate that we want to compute and initial
quantum-mechanical oscillations illustrated with small wiggles in the 
full solution.

More specifically, 
let us consider the distribution of scalar particles
``of type $a$'', generated by 
the creation operator $\hat a_\vec{k}^\dagger$.\footnote{%
 As our field $\phi$ is assumed to be complex-valued, 
 the expansion of the corresponding field operator, 
 cf.~\eq\nr{sc_wave}, contains two independent sets of 
 creation and annihilation operators, 
 denoted here by $\hat{a}^\dagger, \hat{a}$ and 
 $\hat{b}^\dagger, \hat{b}$.}
It is associated with the operator
\be
 \frac{{\rm d} \hat N^{ }_a}{{\rm d}^3\vec{x}\, {\rm d}^3\vec{k}}
 \; \equiv \; \frac{1}{V}\,   \hat a^\dagger_{\vec{k}} 
                      \hat a^{ }_{\vec{k}}
 \;, 
\ee
where $V$ is the volume of the system, and 
the normalization corresponds to  
\be
 [\,\hat a^{ }_{\vec{p}}, \hat a_{\vec{k}}^\dagger\,] = 
 [\,\hat b^{ }_{\vec{p}}, \hat b_{\vec{k}}^\dagger\,] = 
 \delta^{(3)}(\vec{p}-\vec{k})
 \;, \la{sc_a_q} \la{norm}
\ee
or in configuration space to
\be
 [\,\hat \phi(\mathcal{X}), \partial_0^{ } \hat\phi^\dagger(\mathcal{Y})\,]
 = 
 i \, \delta^{(3)}(\vec{x}-\vec{y})
 \quad 
 \mbox{for}
 \quad 
 x^0 = y^0
 \;.
 \la{sc_phi_q}
\ee 
Then the distribution function (in a translationally invariant system) 
is given by
\be
 {f}^{ }_a(t,\vec{k})
 \; \equiv \; (2\pi)^3
 \tr \biggl[ 
  \frac{{\rm d} \hat N^{ }_a }{{\rm d}^3\vec{x}\, {\rm d}^3\vec{k}}
  \,\hat\rho_\iI^{ }(t)
 \biggr]
 \;. \label{Np}
\ee
Inserting \eq\nr{pert}, 
the first term vanishes because 
$ 
 \langle 0 | \hat{a}^\dagger_{\vec{k}}\hat{a}_{\vec{k}}^{ }|0\rangle = 0
$, 
and the second term does not contribute since $\hat H_\iI^{ }$ 
is linear in $\hat a^\dagger_{\vec{k}}$ and  
$\hat a^{\mbox{ }}_{\vec{k}}$
(cf.\ \eqs\nr{Hint} and \nr{sc_wave}), 
so that the corresponding trace vanishes.
Thus, we get that 
the {\em rate} of particle production reads
\be
  \dot{f}^{ }_a(t,\vec{k})
 \; = \; 
   R^{ }_a(T,\vec{k})
 \; \equiv \;
 - \frac{(2\pi)^3}{V} 
 \tr \biggl\{ 
 \hat a^\dagger_{\vec{k}} \hat a^{\mbox{ }}_{\vec{k}}
 \int_0^t \! {\rm d} t' \,
 \bigl[\hat H_\iI^{ }(t),\bigl[\hat H_\iI^{ }(t'), \hat\rho_0^{ }
  \bigr]\bigr]
 \biggr\} + \rmO(|h|^4)
 \;. \label{rate}
\ee

The interaction Hamiltonian $\hat H_\iI^{ }$
appearing in \eq\nr{rate} has the form in \eq\nr{Hint}, 
except that we now interpret
the field operators as being in the interaction picture.
Since $\hat\phi$ evolves with the free Hamiltonian $\hat H^{ }_\rmi{$\phi$}$
in the interaction picture, it has the form of a 
free on-shell field operator, and can hence be written as 
\be \index{On-shell field operator}
 \hat\phi(\mathcal{X})
 = \int \! \frac{{\rm d}^3\vec{p}}{\sqrt{ (2\pi)^3 2 \E^{ }_p } }
 \Bigl(
   \hat a^{ }_{\vec{p}}\, e^{-i \mathcal{P}\cdot\mathcal{X}} + 
   \hat b_{\vec{p}}^\dagger\, e^{i \mathcal{P}\cdot\mathcal{X}}
 \Bigr) 
 \;, \la{sc_wave}
\ee
where we assumed the normalization in \eq\nr{norm}, 
and $p^0_{ }\equiv \E^{ }_{p} \equiv \sqrt{{p}^2 + M^2}$, 
$\mathcal{P} \equiv (p^0_{ },\vec{p})$. Inserting 
$\hat\phi(\mathcal{X})$ 
into (the interaction picture version of) \eq\nr{Hint}, 
we can rewrite $\hat H_\iI^{ }$ as 
\be
 \hat H_\iI^{ } = 
 \int_\vec{x}  
 \int\! \frac{{\rm d}^3 \vec{p}}{\sqrt{ (2\pi)^3 2 \E^{ }_p } }
 \biggl\{
   \Bigl[  
   h\, \hat a^\dagger_{\vec{p}} \, \hat{\mathcal{J}} +
   h^* \hat{\mathcal{J}}^\dagger \hat  b^\dagger_{\vec{p}} 
   \Bigr](\mathcal{X})\, e^{i \mathcal{P}\cdot \mathcal{X}} + 
   \Bigl[ 
   h^* \hat{\mathcal{J}}^\dagger \hat  a^{ }_{\vec{p}} +
   h\, \hat b^{ }_{\vec{p}} \, \hat{\mathcal{J}} 
   \Bigr](\mathcal{X})\, e^{-i \mathcal{P}\cdot \mathcal{X}}
 \biggr\} 
  \;. \la{HI2}
\ee

It remains to take the following steps:
\begin{itemize}
\item[(i)]
We insert \eq\nr{HI2} into \eq\nr{rate}. Denoting
\ba
 \hat A & \equiv & 
 \hat a^\dagger_{\vec{k}} \hat a^{\mbox{ }}_{\vec{k}}
 \;, \\[2mm]
 \hat B(t) & \equiv & 
 \int_\vec{x}  
 \int\! \frac{{\rm d}^3 \vec{p}}
 {\sqrt{\raise-0.1ex\hbox{$(2\pi)^3 2 \E^{ }_p$}}}
 \biggl\{ 
   \Bigl[  
   h\, \hat a^\dagger_{\vec{p}} \, \hat{\mathcal{J}} +
   h^* \hat{\mathcal{J}}^\dagger \hat  b^\dagger_{\vec{p}} 
   \Bigr](\mathcal{X})\, e^{i \mathcal{P}\cdot \mathcal{X}} + 
   \mbox{H.c.}
 \biggr\}
  \;, \\
 \hat C(t') & \equiv & 
 \int_\vec{y}  
 \int\! \frac{{\rm d}^3 \vec{r}}
 {\sqrt{\raise-0.1ex\hbox{$(2\pi)^3 2 \E^{ }_r$}}}
 \biggl\{ 
   \Bigl[  
   h\, \hat a^\dagger_{\vec{r}} \, \hat{\mathcal{J}} +
   h^* \hat{\mathcal{J}}^\dagger \hat  b^\dagger_{\vec{r}} 
   \Bigr](\mathcal{Y})\, e^{i \mathcal{R}\cdot \mathcal{Y}} + 
   \mbox{H.c.}
 \biggr\}
  \;, \hspace*{1.0cm}
\ea
with 
$
 \mathcal{X} \equiv (t,\vec{x})
$
and
$
 \mathcal{Y} \equiv (t',\vec{y})
$,
the trace can be re-organized as
\ba
 \tr\Bigl\{ \hat A\; [ \hat B, [\hat C, |0\rangle\,\langle 0 | ] ] \Bigr\}
 & = &
 \tr\Bigl\{ \hat A \Bigl( 
   \hat B \hat C |0\rangle\langle 0 |
 - \hat B |0\rangle\langle 0 | \hat C 
 - \hat C |0\rangle\langle 0 | \hat B
 + |0\rangle\langle 0 |  \hat C \hat B
 \Bigr) 
 \Bigr\} 
 \nn & = & 
 \langle 0 |
  \Bigl\{ 
    \hat A \hat B \hat C - \hat C \hat A \hat B 
  - \hat B \hat A \hat C + \hat C \hat B \hat A 
  \Bigr\} 
 | 0 \rangle
 \nn & = & 
 \langle 0 |
  \bigl[ \bigl[  
    \hat A , \hat B \bigr], \hat C
  \bigr] 
 | 0 \rangle
 \;. \la{trace}
\ea

\item[(ii)]
Since $\hat{A}$ commutes with $\hat b_\vec{p}^\dagger$
in $\hat{B}$, the part of $\hat{B}$
with $\hat b_\vec{p}^\dagger$ 
gives no contribution; this is also true for 
$\hat b_\vec{r}^\dagger$ 
in $\hat{C}$ since an odd
number of creation or annihilation operators yields nothing.  
A non-zero trace only arises from structures of the type 
$
 \langle 0 | \hat a \hat a^\dagger \hat a \hat a^\dagger | 0 \rangle
$, 
i.e.\ the second and third terms in the second line of \eq\nr{trace}, 
in which $\hat{A}$ is ``shielded'' from the vacuum state. 
Thus, \eq\nr{rate} becomes
\ba
 R^{ }_a(T,\vec{k}) & = & \frac{|h|^2(2\pi)^3}{V} \int_0^t \! {\rm d} t' \,
 \int_\vec{x}
 \int_\vec{y}
 \int\! \frac{{\rm d}^3 \vec{p}}
 {\sqrt{\raise-0.1ex\hbox{$(2\pi)^3 2 \E^{ }_p$}}}
 \int\! \frac{{\rm d}^3 \vec{r}}
 {\sqrt{\raise-0.1ex\hbox{$(2\pi)^3 2 \E^{ }_r$}}}
 \nn & \times & 
 \tr\Bigl\{ \hat \rho^{ }_\rmi{bath}
 \Bigl[ 
 \hat\mathcal{J}^\dagger (\mathcal{Y})
 \hat\mathcal{J}(\mathcal{X}) 
 e^{i \mathcal{P}\cdot \mathcal{X} -i \mathcal{R}\cdot \mathcal{Y}} 
 \langle 0 | 
 \hat a^{ }_{\vec{r}} 
 \hat a^\dagger_{\vec{k}} 
 \hat a^{\mbox{ }}_{\vec{k}} 
 \hat a^\dagger_{\vec{p}} 
 | 0 \rangle 
 \nn & & \hspace*{1.2cm} + \,     
 \hat \mathcal{J}^\dagger(\mathcal{X})
 \hat \mathcal{J}(\mathcal{Y})
 e^{-i \mathcal{P}\cdot \mathcal{X}+i \mathcal{R}\cdot \mathcal{Y}}
 \langle 0 | 
 \hat a^{ }_{\vec{p}}
 \hat a^\dagger_{\vec{k}} 
 \hat a^{\mbox{ }}_{\vec{k}} 
 \hat a^\dagger_{\vec{r}}
 | 0 \rangle 
 \Bigl] \Bigr\} 
 \;,
\ea
where $\hat \rho^{ }_\rmi{bath}$ has appeared from \eq\nr{in}.
Given \eq\nr{norm}, 
both expectation values evaluate to 
\be
  \langle 0 | 
 \hat a^{ }_{\vec{r}} 
 \hat a^\dagger_{\vec{k}} 
 \hat a^{\mbox{ }}_{\vec{k}} 
 \hat a^\dagger_{\vec{p}} 
 | 0 \rangle 
 = 
 \langle 0 | 
 \hat a^{ }_{\vec{p}}
 \hat a^\dagger_{\vec{k}} 
 \hat a^{\mbox{ }}_{\vec{k}} 
 \hat a^\dagger_{\vec{r}}
 | 0 \rangle 
 = 
 \delta^{(3)}(\vec{r-k})\, \delta^{(3)}(\vec{p-k})
 \;. 
\ee 
Thereby
\ba
  R^{ }_a(T,\vec{k}) & = & \frac{|h|^2}{V} \frac{1}{ 2 \E^{ }_{k}}
 \int_0^t \! {\rm d} t' \,
 \int_{\vec{x},\vec{y}} 
 \nn & \times & 
 \Bigl\langle
 \hat \mathcal{J}^\dagger(\mathcal{Y})
 \hat \mathcal{J}(\mathcal{X})
  e^{i \mathcal{K}\cdot (\mathcal{X} - \mathcal{Y})} + 
 \hat \mathcal{J}^\dagger(\mathcal{X})
 \hat \mathcal{J}(\mathcal{Y})
  e^{i \mathcal{K}\cdot (\mathcal{Y} - \mathcal{X})}
 \Bigr\rangle
 \;, \la{R_next}
\ea
where from now on the expectation value refers to that with 
respect to $\hat\rho^{ }_\rmi{bath}$.

\renewcommand{\A}{\hat\mathcal{J}}
\renewcommand{\B}{\hat\mathcal{J}^\dagger}
\renewcommand{\I}{\int_{\mathcal{X}} 
 e^{i \mathcal{K}\cdot (\mathcal{X}-\mathcal{Y})}}

\item[(iii)]
Recalling the notation in \eq\nr{bS}, 
\ba
 \Pi^{<}(\mathcal{K}) & \equiv & 
 \I \Bigl\langle \B(\mathcal{Y})  \A(\mathcal{X}) \Bigl\rangle
 \;,   
 \la{bbS}
\ea
where we made use of translational invariance, 
we can represent 
\ba
  \Bigl\langle \B(\mathcal{Y}) \A(\mathcal{X}) \Bigr\rangle
  & = &
  \int_\mathcal{P}
  e^{-i \mathcal{P}\cdot(\mathcal{X}-\mathcal{Y})} \Pi^{<}(\mathcal{P})
  \;, \la{iva} \\ 
  \Bigl\langle \B(\mathcal{X}) \A(\mathcal{Y}) \Bigr\rangle
  & = &
  \int_\mathcal{P}
  e^{-i \mathcal{P}\cdot(\mathcal{Y}-\mathcal{X})} \Pi^{<}(\mathcal{P})
 \;. \la{ivb}
\ea

\item[(iv)]
It remains to carry out the integrals over the space and time 
coordinates. At this point the result can be simplified by
taking the limit $t\to\infty$, which physically means that we
consider time scales large compared with the interaction rate 
within the heat bath
(cf.\ the figure on p.~\pageref{fig:equil}). 
Summing both terms 
in \eq\nr{R_next}  together and inserting \eqs\nr{iva} and \nr{ivb} yields 
\ba
 & & \lim_{t\to\infty}
 \int\! {\rm d}^3 \vec{x} \,
 \int\! {\rm d}^3 \vec{y} \,
 \int_0^t \! {\rm d}t'\,
 \Bigl[ 
  e^{i(\mathcal{K}-\mathcal{P})\cdot(\mathcal{X}-\mathcal{Y})} + 
  e^{i(\mathcal{P}-\mathcal{K})\cdot(\mathcal{X}-\mathcal{Y})}
 \Bigr] 
 \nn & = & 
 V (2\pi)^3 \delta^{(3)}(\vec{p-k}) 
 \lim_{t\to\infty} 
 \int_0^t \! {\rm d}t'\,
 \Bigl[ 
  e^{i(k^0_{ }-p^0_{ })(t-t')} + 
  e^{i(p^0_{ }-k^0_{ })(t-t')}
 \Bigr] 
 \nn & \stackrel{t'' = t' - t}{=} & 
 V (2\pi)^3 \delta^{(3)}(\vec{p-k}) 
 \lim_{t\to\infty} 
 \biggl\{ 
 \int_{-t}^{0} \! {\rm d}t''\,
 \Bigl[ 
  e^{i(p^0_{ }-k^0_{ })t''} + 
  e^{-i(p^0_{ }-k^0_{ })t''}
 \Bigr] 
 \biggr\} 
 \nn & \stackrel{t''' \equiv -t''}{=} & 
 V (2\pi)^3 \delta^{(3)}(\vec{p-k}) 
 \lim_{t\to\infty} 
 \biggl\{ 
 \int_{-t}^{0} \! {\rm d}t''\,
  e^{i(p^0_{ }-k^0_{ })t''} + 
 \int_{0}^{t} \! {\rm d}t'''\,
  e^{i(p^0_{ }-k^0_{ })t'''}
 \biggr\}
 \nn & = & 
 V (2\pi)^3 \delta^{(3)}(\vec{p-k}) 
 \int_{-\infty}^{\infty} \! {\rm d}\tilde t\,
  e^{i(p^0_{ }-k^0_{ })\tilde t}
 = V (2\pi)^4 \delta^{(4)}(\mathcal{P}-\mathcal{K})
 \;.
\ea
This allows us to
cancel $1/V$ in \eq\nr{R_next}
and remove $\int_\mathcal{P}$ from \eqs\nr{iva} and \nr{ivb}.

\end{itemize}

As a result of these steps we obtain (denoting $k^0_{ } \equiv \E^{ }_{k}$)
\be
 R^{ }_a(T,\vec{k})
 = \frac{|h|^2}{ 2 \E^{ }_k}\, 
  \Pi^{<}(\mathcal{K}) 
 +    \rmO(|h|^4)
 \;. \la{raw_a}
\ee
Using \eq\nr{bLSrel0}, {\em viz.}
$
 \Pi^{<}(\mathcal{K})   =  2 \nB{}(k^0_{ }) \rho(\mathcal{K})
$, 
we finally arrive at 
the master relation 
\be
  R^{ }_a(T,\vec{k})
 = \frac{\nB{}(\E^{ }_k)}{ \E^{ }_k}
   |h|^2
   \rho(\mathcal{K}) 
 +    \rmO(|h|^4)
 \;. \la{master}
\ee
We stress again that this relation is valid only provided that  the
number density of the particles created is much smaller than  
their equilibrium concentration.

For the production rate of $b$-particles, similar steps lead to 
\be
 R^{ }_b(T,\vec{k})
 = \frac{|h|^2}{ 2 \E^{ }_k} \, 
  \Pi^{>}(-\mathcal{K}) 
 +    \rmO(|h|^4)
 \;. \la{raw_b}
\ee
{}From \eq\nr{bLSrel} and 
the identity $\nB{}(-k^0_{ }) = -1 - \nB{}(k^0_{ })$, 
we get  
$
 \Pi^{>}(-\mathcal{K})   =  2[1 + \nB{}(-k^0_{ })] \rho(-\mathcal{K})
 = - 2 \nB{}(k^0_{ }) \rho(-\mathcal{K})
$, 
and subsequently
\be
  R^{ }_b(T,\vec{k})
 = \frac{\nB{}(\E^{ }_k)}{ \E^{ }_k}
   |h|^2
   \Bigl[ - \rho(-\mathcal{K}) \Bigr]
 +    \rmO(|h|^4)
 \;. \la{master_b}
\ee
In a CP-symmetric plasma (without chemical potentials), 
it can be shown that 
$\rho(-\mathcal{K}) = - \rho(\mathcal{K})$
in the bosonic case, 
so that in fact the two production rates coincide. 

In summary, we have obtained a relation connecting the particle production
rate, \eq\nr{rate}, to a finite-temperature spectral function, concerning
the operator to which the produced particle couples. We return to
a specific example in \se\ref{se:dm}.

Three concluding remarks are in order: 
\begin{itemize}

\item
In terms of the figure on p.~\pageref{fig:equil}, the rate $\gamma$
equals
$
 \gamma = \frac{|h|^2 \rho(\mathcal{K})}{ \E^{ }_k} + \rmO(|h|^4)
$
and
$
 f^{ }_\rmi{eq} = \nB{}(\E^{ }_k)
$.

\item
Once sufficiently many particles have been produced, they tend to equilibrate,
and the results above are no longer valid. 
We can expect that in this
situation \eq\nr{master} is modified into
\be
  \dot{f}^{ }_a(t,\vec{k})
 = \frac{|h|^2 \rho(\mathcal{K})}{ \E^{ }_k}
   \Bigl[
     \nB{}(\E^{ }_k) - f^{ }_a(t,\vec{k})
   \Bigr]
 +    \rmO(|h|^4)
 \;. \la{master_nonlin}
\ee
This equation is valid both for large 
and small deviations from equilibrium.\footnote{%
 A way to show this from the above formalism 
 has been presented in ref.~\cite{dmpheno}, and a general 
 analysis can be found in ref.~\cite{dbx}.
 } 
It is seen how the production stops when $f^{ }_a \to \nB{}$, 
as must be the case. 
The equilibration rate is the 
same $\gamma = |h|^2 \rho(\mathcal{K}) / \E^{ }_k$ as before. 

\item \index{Landau-Pomeranchuk-Migdal (LPM)}
In this section we have related a particle production rate to 
a general 
spectral function, $\rho(\mathcal{K})$. The {\em computation} of this 
spectral function represents a challenge of its own. 
In sections~\ref{se:NI_rho} and \ref{se:realtime} simple examples of 
such computations were given; 
however as alluded to below \eq\nr{pert2}, a proper computation
normally requires HTL resummation, the inclusion of $2\leftrightarrow 2$
scatterings, as well as a so-called Landau-Pomeranchuk-Migdal (LPM)
resummation of almost coherent $1+n\leftrightarrow 2+n$ scatterings. 
Computations including these processes for the example of 
sections~\ref{se:NI_rho} and \ref{se:realtime} have been 
presented in refs.~\cite{bb2_2,sum3_2}, and a similar analysis for
the production of photons from a QCD plasma can be found
in refs.~\cite{photon1_2,photon2_2}.

\end{itemize}


\subsection*{Appendix A: Streamlined derivation of 
the particle production rate}
\la{se:simple}

We outline here another derivation of the particle production rate, 
similar to the one employed in refs.~\cite{dilepton1,dilepton2}, 
which is technically simpler than the one presented above
but comes with the price of being somewhat heuristic and thus
implicit about the assumptions made.  

Let $|\vec{k}\rangle \equiv \hat a^\dagger_{\vec{k}} |0 \rangle$
be a state with one ``$a$-particle'' of momentum $\vec{k}$. Consider
an initial state $|I \rangle$ and a final state $| F \rangle$, with
\be
 |I\rangle \equiv |\g\rangle \otimes | 0 \rangle 
 \;, \quad
 |F\rangle \equiv |\f\rangle \otimes | \vec{k} \rangle
 \;, 
\ee
where $|\g\rangle$ and $|\f\rangle$ are the initial and final states, 
respectively, in the Hilbert space of the degrees of freedom constituting
the heat bath. The transition matrix element reads
\be
 T^{ }_\rmi{$F$$I$} =
 \langle F | \int_0^{t} \! {\rm d}t' \, \hat H_\iI^{ }(t') \, 
 | I \rangle
 \;, 
\ee
where $\hat H_\iI^{ }$ is the interaction Hamiltonian in the interaction
picture. The particle production rate can now be defined as 
\be
  \frac{\dot{f}^{ }_a(t,\vec{k})}{(2\pi)^3}
  \equiv 
  \lim_{t,V\to \infty}
  \sum_\rmii{$\f,\g$}
  \frac{e^{-\beta \E^{ }_\rmii{$\g$}}}{\mathcal{Z}^{ }_\rmii{bath}}
  \frac{|T^{ }_\rmi{$F$$I$}|^2}{t\, V}
 \;, \la{simp_rate} 
\ee
where a thermal average is taken over all initial states, whereas 
for final states no constraint other than that built into the 
transition matrix elements is imposed. Furthermore,  
$\mathcal{Z}^{ }_\rmi{bath} \equiv
 \sum^{ }_\rmi{$\g$} e^{-\beta \E^{ }_\rmii{$\g$}}$ 
is the partition function of the heat bath. 

By making use of 
$
 \langle \vec{k} | \hat H^{ }_\iI \, | 0 \rangle 
 = \langle 0 |\, \hat a^{ }_{\vec{k}} \, \hat H^{ }_\iI \,| 0 \rangle 
 = \langle 0 |\, [\hat a^{ }_{\vec{k}}, \hat H^{ }_\iI ] \, | 0 \rangle 
$
and \eq\nr{HI2}, 
we immediately obtain
\be
 \langle F | \int_0^{t} \! {\rm d}t' \, \hat H_\iI^{ }(t') \, 
 | I \rangle
 = h \int_{\mathcal{X}'} \frac{e^{i \mathcal{K}\cdot\mathcal{X}'}}
 {\sqrt{(2\pi)^3 2 \E^{ }_k}}
 \, 
 \langle \f | \, \hat \mathcal{J}(\mathcal{X}') \, | \g \rangle 
 \;, \la{stream_amp}
\ee
and subsequently
\be
 |T^{ }_\rmi{$F$$I$}|^2 = \frac{|h|^2}{(2\pi)^3 2 \E^{ }_k}
 \int_{\mathcal{X}',\mathcal{Y}'} 
 e^{i \mathcal{K}\cdot (\mathcal{X}' - \mathcal{Y}')}
 \; 
 \langle \f | \, \hat \mathcal{J}(\mathcal{X}') \, | \g \rangle 
 \langle \g | \, \hat \mathcal{J}^\dagger(\mathcal{Y}') \, | \f \rangle 
 \;. \la{TFI_sqr}
\ee
This can be inserted into \eq\nr{simp_rate}. Taking the thermodynamic
limit; making use of translational invariance in order to cancel 
the time and volume factors from the denominator; 
and paying attention to the ordering of the operators
($e^{-\beta \hat H}$ and therefore the states $|\g \rangle$
should appear at the ``outer edge''), we arrive at 
\be
 \dot{f}^{ }_a(t,\vec{k})
 =  
 \frac{|h|^2}{ 2 \E^{ }_k}
 \int_{\mathcal{X}} e^{i\mathcal{K}\cdot\mathcal{X}}
 \Bigl\langle 
   \hat \mathcal{J}^\dagger (0) \, 
   \hat \mathcal{J} (\mathcal{X}) \Bigr\rangle
 \;.
\ee
If we now compare the result with \eqs\nr{bS} and \nr{bLSrel0}, we can 
write the final expression as 
\be
 \dot{f}^{ }_a(t,\vec{k})
 =  
 \frac{|h|^2}{ 2 \E^{ }_k}\,
 \Pi^{<}(\mathcal{K}) 
 =  
 \frac{\nB{}(k^0_{ })}{ \E^{ }_k}\, 
  |h|^2 \, \rho (\mathcal{K}) 
 \;, \la{dNa}
\ee
where $k^0_{ } = \E^{ }_k$ and
it is understood that the spectral function corresponds
to the operator $\hat{\mathcal{J}}$.

A similar computation, making use of translational 
invariance and \eqs\nr{bL} and 
\nr{bLSrel}, yields the production rate of the ``$b$-particles'':
\ba
 \dot{f}^{ }_b(t,\vec{k})
 & = &   
 \frac{|h|^2}{ 2 \E^{ }_k}
 \int_{\mathcal{X}} e^{i\mathcal{K}\cdot\mathcal{X}}
 \Bigl\langle 
   \hat \mathcal{J} (0) \, 
   \hat \mathcal{J}^\dagger (\mathcal{X}) \Bigr\rangle
 \nn 
 & = & 
 \frac{|h|^2}{ 2 \E^{ }_k}
 \int_{\mathcal{X}} e^{i\mathcal{K}\cdot(\mathcal{X}-\mathcal{Y})}
 \Bigl\langle 
   \hat \mathcal{J} (\mathcal{Y}) \, 
   \hat \mathcal{J}^\dagger (\mathcal{X}) \Bigr\rangle
 \nn 
 & = & 
  \frac{|h|^2}{ 2 \E^{ }_k}
 \int_{\mathcal{Y}} e^{-i\mathcal{K}\cdot\mathcal{Y}}
 \Bigl\langle 
   \hat \mathcal{J} (\mathcal{Y}) \, 
   \hat \mathcal{J}^\dagger (0) \Bigr\rangle 
 \nn
 & = & 
 \frac{|h|^2}{ 2 \E^{ }_k}\,
 \Pi^{>}(-\mathcal{K}) 
 =  
 - \frac{\nB{}(k^0_{ })}{ \E^{ }_k}
 \, |h|^2 \, \rho (-\mathcal{K}) 
 \;,   \la{dNb} 
\ea
like in \eq\nr{master_b}.
Finally we note that the total
number density increases as 
\be
 \frac{{\rm d}(N^{ }_a+N^{ }_b)}{{\rm d}^4 \mathcal{X} }
 = 
 |h|^2 \int_\vec{k} \frac{\nB{}(\E^{ }_k)}{\E^{ }_k}
 \Bigl[
  \rho (\mathcal{K})
   -  \rho (-\mathcal{K})
 \Bigr]
 \;. \la{tot_dN}
\ee


\subsection*{Appendix B: Particle decay rate}

\index{Decay rate}

The discussion above concerned the production rate of 
particles whose total density remains below the equilibrium value. 
As we now outline, one can similarly consider an ``opposite'' limit, 
in which the Hilbert space corresponding
to weakly interacting particles is ``full'' in the initial state, 
and the particles are forced to decay.
 
Using the same notation as in appendix A, we consider
an initial state $|I \rangle$ and a final state $| F \rangle$, with
\be
 |I\rangle \equiv |\g\rangle \otimes | \vec{k} \rangle
 \;, \quad
 |F\rangle \equiv |\f\rangle \otimes | 0 \rangle 
 \;, 
\ee
where $|\g\rangle$ and $|\f\rangle$ are the initial and final states, 
respectively, in the Hilbert space of the degrees of freedom constituting
the heat bath. The transition matrix element can be defined and computed
as before, and in the end \eq\nr{TFI_sqr} gets replaced with
\be
 |T^{ }_\rmi{$F$$I$}|^2 = \frac{|h|^2}{(2\pi)^3 2 \E^{ }_k}
 \int_{\mathcal{X}',\mathcal{Y}'} 
 e^{i \mathcal{K}\cdot (\mathcal{Y}' - \mathcal{X}')}
 \; 
 \langle \f | \, \hat \mathcal{J}^\dagger(\mathcal{X}') \, | \g \rangle 
 \langle \g | \, \hat \mathcal{J}(\mathcal{Y}') \, | \f \rangle 
 \;. \la{x_TFI_sqr}
\ee
It follows that 
\be
 \dot{f}^{ }_a(t,\vec{k})
 =  
 - \frac{|h|^2}{ 2 \E^{ }_k} \, 
 \Pi^{>}(\mathcal{K}) 
 =  
 - \frac{|h|^2}{ \E^{ }_k}
 \, [1 +  \nB{}(k^0_{ })]\, \rho (\mathcal{K}) 
 \;. \la{x_dNa}
\ee
If we take
the zero-temperature limit by setting $\nB{}(k^0_{ }) \to 0$, 
the result remains non-zero, 
and equals the text-book decay rate in vacuum 
(which was denoted by $\gamma$ in the figure on p.~\pageref{fig:equil}).


\subsection*{Appendix C: Scattering on dense media}

\index{Scattering on dense media}

We have shown that if there is an 
interaction  of the form
$
  \hat H^{ }_\rmi{int} \sim \int_{\vec{x}} 
  ( h\, \hat\phi^\dagger \hat \mathcal{J} + \mbox{H.c.})
$, 
then an initial density matrix, 
\be
 \hat \rho (0) = \hat \rho^{ }_\rmi{bath} \otimes |0\rangle\langle 0|
 \;, 
\ee
evolves into a form whereby the number density operator 
of the $\phi$-particles takes a non-zero expectation value. 
The formal reason for this is that the amplitude 
$\langle \vec{k} |\, \hat H^{ }_\rmi{int} \, | 0 \rangle$ 
is non-zero because of $\hat\phi$
in $\hat H^{ }_\rmi{int}$, cf.\ \eq\nr{stream_amp}.

Now, a variant of this situation can be 
envisaged, namely that of {\em scattering} of a weakly 
interacting particle on a thermal medium. In this case, 
the initial state would be something like 
\be
 \hat \rho (0) = \hat \rho^{ }_\rmi{bath} \otimes 
 |\vec{k}^{ }_\rmi{i}\rangle \langle \vec{k}^{ }_\rmi{i} |
 \;, 
\ee
and the question is how fast the system evolves towards a state
\be
 \hat \rho (t) = \hat \rho^{ }_\rmi{bath} \otimes 
 |\vec{k}^{ }_\rmi{f}\rangle \langle \vec{k}^{ }_\rmi{f} |
 \;, 
\ee
with $\vec{k}^{ }_\rmi{i} \neq \vec{k}^{ }_\rmi{f}$. 
This is an interesting problem
because the weakly interacting probe could effectively scatter 
from a {\em collective excitation}, and the relation between 
$\vec{k}^{ }_\rmi{i}$ and  $\vec{k}^{ }_\rmi{f}$ 
could be used to determine the 
dispersion relation of the collective excitation. 
Cohen and Feynman originally proposed this method 
in order to determine the dispersion relation of collective
excitations in liquid helium through inelastic neutron 
scattering~\cite{cf},
and the proposal has been successfully realized~\cite{hewo}.

In the context of QCD, it is difficult to envisage how
a corresponding experiment could be realized, because QCD matter
cannot be confined to a container on which a scattering experiment
could be carried out. Nevertheless, on an adventurous note, we might
speculate that a monochromatic X-ray beam from a quasar scattering
on a very compact neutron star could experience similar phenomenology. 

In any case, 
the basic idea is the following. Suppose that the temperature 
is so low that the medium is in its ground state, without any 
kind of motion taking place ($T \ll m, \mu$), and suppose that
the $\phi$-particles are massless (e.g.\ photons). In an inelastic 
scattering, the momentum $\Delta p$ and the energy $\Delta E$
are transferred to the medium, with
\ba
 \Delta p & = & |\vec{k}^{ }_\rmi{i}-\vec{k}^{ }_\rmi{f}|
 = \sqrt{ k_\rmi{i}^2 + k_\rmi{f}^2
 - 2  k^{ }_\rmi{i} k^{ }_\rmi{f} \cos\theta }
 \;, \\ 
 \Delta E & = & k^{ }_\rmi{i} - k^{ }_\rmi{f}
 \;.
\ea
The scattering is most efficient, i.e.\ resonant, if the given 
$\Delta p$ and $\Delta E$ kick an on-shell collective excitation into 
motion. If the latter has the dispersion relation $\omega(k)$, 
resonant scattering takes place for 
\be
 \Delta E = \omega(\Delta p)
 \;. 
\ee
For a given $k^{ }_\rmi{i}$ and $\theta$, this can be viewed
as an equation for $k^{ }_\rmi{f}$. 
Consequently, if the peak wave number $k^{ }_\rmi{f}$
is measured as a function of the scattering angle~$\theta$, 
one can experimentally determine the function $\omega(k)$.  
If $\omega(k)$ contains a scale, such as $m^{ }_\rmii{F}$
(cf.\ \eq\nr{plasmino}),  
then non-trivial solutions are to be expected 
in the range $k^{ }_\rmi{i}, k^{ }_\rmi{f} \sim m^{ }_\rmii{F}$.

\index{Plasmino}

The phenomenon just discussed might be particularly
remarkable if $\Delta E < 0$, i.e.\ more energy comes
out than goes in.  This can happen with 
the dispersion relation of \eq\nr{plasmino}, and could be 
referred to as ``Compton scattering on a plasmino''. 
Of course, in any realistic situation, there is a background
to this process from thermal free electrons with
a non-trivial velocity distribution.

Formally, the amplitude for the scattering  
contains two appearances of $\hat H^{ }_\rmi{int}$, one
for absorption and the other for emission. The rate will 
therefore be proportional to a certain 4-point function 
of the currents, yielding a theoretical description of 
scattering more complicated than for particle production, 
where we only encountered 2-point functions.

\newpage 

\subsection{Embedding rates in cosmology}
\la{se:dm}

\index{Cosmological background}

In \se\ref{se:ppr} we considered the production rate of 
weakly interacting particles at a fixed temperature, $T$. 
In a cosmological setting, however, 
account needs to be taken of the expansion of the universe, 
which leads to an evolving temperature as well as 
red-shifting particle momenta. 
This has implications for practical computations 
of e.g.\ dark matter spectra, as will be illustrated in the current section. 

Let $f(t,\vec{k})$ denote the phase space density of a species 
of particles being produced, so that their number density reads 
\be
 \frac{N}{V} = 
 \int\!\frac{ {\rm d}^3 \vec{k} }{(2\pi)^3}
 \, f(t,\vec{k})
 \;, \la{nI_def}
\ee
and the corresponding production rate (cf.\ \eq\nr{rate}) equals
\be
  \dot{f} (t,\vec{k}) = R(T,\vec{k})
  \;. \la{rate_1} 
\ee
For the particular model considered in \se\ref{se:ppr} and 
particles of ``type $a$'', the production rate $R^{ }_a$ 
is given by \eq\nr{master}; in the following we use a slightly
more realistic example, where $f$ counts right-handed neutrinos 
in either polarization state. Then a computation similar to 
that in \se\ref{se:ppr} leads to 
\be
  R(T,\vec{k}) \;\equiv\; \sum_{a=\pm}^{ } \dot{f}^{ }_a (t,\vec{k})
 = \frac{\nF{}(k^0_{ })}{k^0_{ }}
  \, |h|^2 \left.
  \tr\Bigl\{ 
  \bsl{\mathcal{K}}
  \aL 
   \Bigl[
    \rho(-\mathcal{K})
  + \rho(\mathcal{K}) 
   \Bigr]
  \aR
 \Bigr\} \right|^{ }_{ k^0_{ } = \sqrt{k^2 + M^2 } }  
 \;, \la{master_2}
\ee
where the notation and the spectral function are as discussed
in \se\ref{se:NI_rho}, 
and the sum goes over the two polarization states of 
a massive Majorana fermion.

%
\subsection*{Basic cosmology}

Let us begin by recalling cosmological relations
between the time $t$ and the temperature $T$ and by setting up our notation. 
As usual, we assume that even if there were 
net number densities present, they are very small compared with the 
temperature, $\mu \ll \pi T$, so that thermodynamic quantities
are determined by the temperature alone
(this assumption will be relaxed in appendix~B). 
Assuming furthermore a homogeneous and isotropic metric, 
\be
 {\rm d}s^2 = {\rm d}t^2 - a^2(t) \, {\rm d} \vec{x}^2_\kappa
 \;, 
\ee
where $\kappa = 0, \pm 1$ characterizes the spatial geometry,
as well as the energy-momentum tensor of an ideal fluid, 
\be 
 {T^{ }_\mu}^{\nu} = \mbox{diag}(e,-p,-p,-p)
 \;,  
\ee
where $e$ denotes the energy density and $p$ the pressure, the Einstein
equations, \index{Einstein equations}
$
 {G^{ }_\mu}^\nu = 8 \pi\, G\, {T^{ }_\mu}^{\nu}
$,
reduce to the Friedmann equations \index{Friedmann equations}
\ba
 \biggl( \frac{\dot{a}}{a}\biggr)^2 + \frac{\kappa}{a^2} & = & 
 \frac{8\pi G\, e}{3} 
 \;, \la{AE1} \\
 {\rm d} (e a^3 ) & = & -p\, {\rm d}(a^3) 
 \;.  \la{AE2}
\ea
We assume a flat universe, $\kappa=0$, and denote
\be
 \frac{1}{m_\rmi{Pl}^2} \equiv G
 \;, \la{mPl}
\ee
where $m^{ }_\rmi{Pl} \approx 1.2\times 10^{19}$~GeV is the Planck mass. 
We may then introduce the ``Hubble parameter'' $H$ via 
\be
 H 
 \equiv \frac{ \dot{a}(t) } { a(t) }  
 = \sqrt{\frac{8\pi}{3}} \frac{\sqrt{e}}{m^{ }_\rmi{Pl}}
 \;. \la{Hubble}
\ee 

We now combine the Friedmann equations with basic thermodynamic relations. 
In a system with small chemical potentials, 
the energy and entropy densities are related by 
\be
 e = T s - p 
 \;, \la{stat_1}
\ee
where $s = {\rm d}p/{\rm d}T$ is the entropy density. From here, 
it follows that ${\rm d}e = T {\rm d}s$, which together 
with \eq\nr{AE2} leads to the relation
\ba
 0 & = &  {\rm d} (e a^3 ) + p\, {\rm d}(a^3) 
 \nn & = &  a^3 {\rm d} e + (p+e)\, {\rm d}(a^3) 
 \nn & = &  a^3 T {\rm d} s  + T s\, {\rm d}(a^3)
 \nn & = & T {\rm d} ( s a^3 ) 
 \;. \la{entr_conv}
\ea
This relation is known as the entropy conservation law, 
and can be re-expressed as 
\be
 \frac{a(t)}{a(t^{ }_0)} = \biggl[ \frac{s(T^{ }_0)}{s(T)} \biggr]^\fr13
 \;. \la{aT_rel}
\ee

We can also derive an evolution equation for the temperature. 
The entropy conservation law implies that 
\be
 \frac{{\rm d}s}{s} = - 3 \frac{{\rm d}a}{a}
 \;, \la{dsos}
\ee
whereas defining the ``heat capacity'' $c$ through 
\be
 \frac{{\rm d}e}{{\rm d}T} \; = \; T \frac{{\rm d}s}{{\rm d}T}
 \; \equiv \; T c
 \;,  \la{stat_2}
\ee
we get ${\rm d}s = c\, {\rm d}T$. Inserting this into \eq\nr{dsos} 
and dividing by ${\rm d}t$ leads to 
\ba
  & & \frac{c}{s} \frac{{\rm d}T}{{\rm d}t} = - \frac{3 \dot{a}}{a}
 \la{stat_3} \\ 
  & \stackrel{\rmi{\nr{Hubble}}}{\Rightarrow} &
  \frac{{\rm d}T}{{\rm d}t} = - 
 \frac{\sqrt{24\pi}}{m^{ }_\rmi{Pl}} 
 \frac{s(T) \sqrt{e(T)}}{c(T)}
 \;. \la{Tt_rel}
\ea

In cosmological literature, 
it is conventional to introduce two different ways to count 
the effective numbers of
massless bosonic degrees of freedom, $g^{ }_\rmi{eff}(T)$ and
$h^{ }_\rmi{eff}(T)$, defined via the relations
\be
 e(T) \;\equiv\; \frac{\pi^2 T^4}{30} g^{ }_\rmi{eff}(T) \;, \;\;\;
 s(T) \;\equiv\; \frac{2 \pi^2 T^3}{45} h^{ }_\rmi{eff}(T) \;, \la{sT_heff}
\ee
where the prefactors follow by applying \eq\nr{stat_1}
and the line below it to the free result $p(T) = \pi^2 T^4/90$
from \eq\nr{JT0}. Furthermore, for later reference, 
we note that the sound speed squared can be written in the forms
\be
  c_s^2(T) \equiv
 \frac{\partial p}{\partial e} = 
 \frac{p'(T)}{e'(T)} = 
 \frac{p'(T)}{Ts'(T)} = 
 \frac{s(T)}{T c(T)}  
 \;. \la{ccs}
\ee

%
\subsection*{Production equation and its solution}

In order to generalize \eq\nr{rate_1}
to an expanding background, we have to properly 
define our variables, the time $t$ and the momentum $\vec{k}$.
In the following we mean by these 
the {\em physical} time and momentum, 
i.e.\ quantities defined in a local Minkowskian frame. However, as is 
well known, local Minkowskian frames at different times are inequivalent 
in an expanding background; in particular, the physical momenta redshift.
Carrying out the derivation of the rate equation in this situation is 
a topic of general relativity, and we 
only quote the result here: the main effect of expansion is that  the time
derivative gets replaced as 
$
 \partial/\partial t \to \partial/\partial t - H k^i \partial/\partial k^i
$~\cite{be,kolb},
and \eq\nr{rate_1} becomes
\be \index{Hubble parameter}
  \biggl( \frac{\partial}{\partial t} - 
  H k^i \frac{\partial}{\partial k^i}\biggr) 
  f (t,\vec{k}) 
  = R(T,\vec{k})
 \;, 
 \la{kinetic}
\ee
where $H$ is the Hubble parameter from \eq\nr{Hubble}
and $k^i$ are the components of $\vec{k}$. 

It is important to stress that the production rate $R(T,\vec{k})$ 
in \eq\nr{kinetic} can 
be directly taken over from the flat spacetime result in \eq\nr{master_2}. 
The reason is that 
the time scale of the equilibration of the plasma and 
of the scattering reactions taking place within the plasma is
$\tau \lsim 1/(\alpha^2 T)$, where $\alpha$ is a 
generic fine-structure constant. Unless $T$ is exceedingly high, 
this is much smaller than 
the time scale associated with the expansion of the universe,  
$H^{-1}\sim m^{ }_\rmi{Pl}/T^2$. Therefore,
local Minkowskian coordinates can be used
for the duration of the plasma scatterings. 
Note however that the rate $R$ itself can be small; 
as has been discussed in \se\ref{se:ppr}, the coupling $|h|^2$,
connecting the non-equilibrium degrees of freedom to the plasma
particles, is by assumption small, $|h|^2 \ll \alpha$. 
In other words, the rate $R$ is determined by the physics of almost 
instantaneous scatterings taking place with a rate
$1/\tau \gg H $, 
but its numerical value could nevertheless be tiny, 
$R \ll H$. 

Now, because of rotational symmetry, $R(T,\vec{k})$
and consequently also $f(t,\vec{k})$ are typically only functions 
of $k \equiv |\vec{k}|$. Changing the notation correspondingly, 
and noting that $\partial k / \partial k^i = k^i / k$,
\eq\nr{kinetic} becomes
\be
  \biggl( \frac{\partial}{\partial t} - 
  H k \frac{\partial}{\partial k}\biggr) 
  f (t,{k}) 
  = R(T,{k}) 
 \;. \la{kinetic2}
\ee
Furthermore, if we are only interested in the total number density, 
$\int_\vec{k}  f(t,k)$, rather than the shape of the spectrum, 
we can integrate \eq\nr{kinetic2} on both sides. Partially integrating
$
 \int \! {\rm d}^3\vec{k} \, k \partial^{ }_k f(t,k) =
 -3 \int \! {\rm d}^3\vec{k} \, f(t,k)
$
then leads to an equation for the number density, 
\be
 (\partial^{ }_t + 3 H) \int_\vec{k}  f(t,k) = 
 \int_\vec{k}  R(T,k)
 \;. \la{kin_total}
\ee

\index{Method of characteristics}

\Eq\nr{kinetic2} can be integrated through a suitable 
change of variables, known as the method of characteristics. 
Introducing an ansatz $f(t,k) = f(t,k(t^{ }_0) \frac{a(t^{ }_0)}{a(t)})$,
and noting that 
\be
 \frac{{\rm d}}{{\rm d}t} \biggl[ k(t^{ }_0) \frac{a(t^{ }_0)}{a(t)} \biggr] = 
 - k(t^{ }_0) \frac{a(t^{ }_0)\dot{a}(t)}{a^2(t)} = 
 - H k 
 \;, 
\ee 
\eq\nr{kinetic2} can be re-expressed as 
\be
 \frac{{\rm d}f}{{\rm d}t} \biggl(t,k(t^{ }_0) \frac{a(t^{ }_0)}{a(t)}\biggr)
 = R\biggl(T,k(t^{ }_0) \frac{a(t^{ }_0)}{a(t)}\biggr)
 \;.  
\ee
This can immediately be solved as 
\be
 f(t^{ }_0,k(t^{ }_0)) = \int_0^{t^{ }_0} \! {\rm d}t \,  
 R\biggl(T(t),k(t^{ }_0) \frac{a(t^{ }_0)}{a(t)}\biggr)
 \;, \la{kinetic3}
\ee
where we assumed the initial condition 
$f(0,k) = 0$, i.e.\ that there were no
particles at $t=0$. Let us also note that the entropy conservation 
law of \eq\nr{entr_conv} implies that $(\partial^{ }_t + 3 H)s = 0$, 
permitting us to re-express \eq\nr{kin_total} as
\be
 \frac{{\rm d}}{{\rm d}t}
 \biggl[ 
   \frac{\int_\vec{k}  f(t,k)}{s(t)} 
 \biggr]
 = 
 \frac{\int_\vec{k}  R(T,k)}{s(t)}
 \;. \la{kin_tot_2}
\ee

In cosmology, it is convenient to measure time directly
in terms of the temperature. The corresponding change of variables, 
\eq\nr{Tt_rel}, is often implemented in some approximate form; 
in its exact form, we need information concerning the 
pressure $p(T)$ (appearing in $e(T) = T p'(T) - p(T)$), 
its first derivative $p'(T)$ 
(appearing in $e(T)$ as well as in $s(T) = p'(T)$), 
and its second derivative $p''(T)$ (appearing in $c(T) = s'(T)$). 
The particular combination defining the sound speed
squared, \eq\nr{ccs}, is close to 
$\tfr13$, so it is useful to factor it out. 
Inserting also \eq\nr{Hubble}, \eq\nr{Tt_rel} then becomes
\be
 \frac{{\rm d}T}{{\rm d}t}
 = - \sqrt{\frac{8\pi}{3}} \frac{T}{m^{ }_\rmi{Pl}} 
   \sqrt{e(T)}\, \bigl[3 c_s^2(T) \bigr]
 = - T H(T) \bigl[3 c_s^2(T) \bigr]
 \;. \la{tT_rel_2}
\ee
Further defining the so-called {\em yield parameter},
\be \index{Yield parameter}
 Y(t^{ }_0) \equiv  \frac{\int_\vec{k}  f(t^{ }_0,k)}{s(t^{ }_0)} 
 \;, \la{Y_def}
\ee   
\eq\nr{kin_tot_2} becomes
\be
 T\, \frac{{\rm d}Y}{{\rm d} T}
 = \frac{-1}{3 c_s^2(T) s(T) H(T) }
 \int_\vec{k}  R(T,k)
 \;. \la{kin_tot_3}
\ee
This equation implies, amongst other things, 
that close to a first order phase transition, where 
$c_s^2$ typically has a dip, the yield of produced 
particles is enhanced. The reason is that the system 
spends a long time at these temperatures, diluting the specific heat being 
released into the expansion of the universe, and that therefore there is 
a long period available for particle production. 

%
\subsection*{Example}

Let us write the main results derived above in an explicit form, 
by inserting into them the parametrizations of \eq\nr{sT_heff}.
Denoting $k\equiv k(t^{ }_0)$, inserting the red-shift 
factor from \eq\nr{aT_rel}, and changing the integration
variable from $t$ to $T$ according to \eq\nr{tT_rel_2}, 
the result of \eq\nr{kinetic3} can be expressed as~\cite{als2}
\be \index{Particle production rate: spectrum}
 f(t^{ }_0,k)=
 \sqrt{\frac{5}{4\pi^3}} \, 
 \int_{T^{ }_0}^{T^{ }_\rmii{max}} \! \frac{{\rm d}T}{T^3} 
 \, 
 \frac{m^{ }_\rmi{Pl}}{c_s^2(T)\sqrt{g^{ }_\rmi{eff}(T)}}
 \, R\biggl( {T},
 k\, \frac{T}{T^{ }_0}
 \left[\frac{h^{ }_\rmi{eff}(T)}{h^{ }_\rmi{eff}(T^{ }_0)}\right]^{\frac{1}{3}}
 \biggr)
 \;,
 \label{distribution}
\ee 
where $T^{ }_\rmi{max}$ corresponds to the highest temperature of
the universe. 
This gives the spectrum of particles produced as an integral
over the history of their production. The integral 
over \eq\nr{distribution}, after the substitution 
$
 k = z T^{ }_0 [{h^{ }_\rmi{eff}(T^{ }_0)}/{h^{ }_\rmi{eff}(T)}]^{1/3}
$
and followed by a division by $s(t^{ }_0)$, 
or a direct integration of \eq\nr{kin_tot_3}, gives their total yield: 
\ba
 Y(t^{ }_0) \!\! & = &  \!\!
 \frac{45\sqrt{5}}{(2\pi)^3 \pi^{5/2}}
 \int_{T^{ }_0}^{T^{ }_\rmii{max}}
 \! \frac{{\rm d} T}{T^3} 
 \frac{m^{ }_\rmi{Pl}}{c_s^2(T) h^{ }_\rmi{eff}(T) 
 \sqrt{g^{ }_\rmi{eff}(T)} }
 \int_0^{\infty}
 \! {\rm d}z \, z^2 \, 
 R\left({T}, T z\right)
 \;.
 \la{yield}
\ea
We note that if 
$
 \int_0^{\infty}
 \! {\rm d}z \, z^2 \,
 R\left({T}, T z\right)
$
vanishes sufficiently fast at low temperatures
(typically it contains a Boltzmann factor and becomes exponentially 
suppressed when $T$ falls below some mass scale), 
then the result is independent of $T^{ }_0$.

To be more explicit, we need to 
specify the function $R(T,k)$, which for our example 
can be obtained from \eqs\nr{pert1}, \nr{even} and \nr{master_2}.
The Dirac algebra in \eq\nr{master_2} can be trivially 
carried out, resulting in
\be
  \tr\Bigl\{ 
  \bsl{\mathcal{K}}
  \aL 
   \Bigl[
     \bsl{\mathcal{P}}^{ }_1
   \Bigr]
  \aR
 \Bigr\} 
 = 2 \mathcal{K}\cdot \mathcal{P}^{ }_1 
 \;.
\ee
Furthermore, the $\delta$-functions appearing in \eq\nr{pert1}
can be written in various ways depending on the channel
(setting $D\to 4$), 
\ba
 \delta^{(4)}(\mathcal{P}^{ }_1+\mathcal{P}^{ }_2-\mathcal{K})\;
  2 \mathcal{K} \cdot \mathcal{P}^{ }_1 & = & 
 \delta^{(4)}(\mathcal{P}^{ }_1+\mathcal{P}^{ }_2-\mathcal{K}) 
 [\mathcal{P}_1^2 + \mathcal{K}^2 -(\mathcal{K}-\mathcal{P}^{ }_1)^2]
 \nn & = &  
 \delta^{(4)}(\mathcal{P}^{ }_1+\mathcal{P}^{ }_2-\mathcal{K}) 
 [\mathcal{P}_1^2 + \mathcal{K}^2 - \mathcal{P}_2^2]
 \;, \nn  
 \delta^{(4)}(\mathcal{P}^{ }_1 - \mathcal{P}^{ }_2-\mathcal{K})\; 
 2 \mathcal{K} \cdot \mathcal{P}^{ }_1 & = & 
 \delta^{(4)}(\mathcal{P}^{ }_1 - \mathcal{P}^{ }_2-\mathcal{K}) 
 [\mathcal{P}_1^2 + \mathcal{K}^2 -(\mathcal{K}-\mathcal{P}^{ }_1)^2 ]
 \nn & = &  
 \delta^{(4)}(\mathcal{P}^{ }_1 - \mathcal{P}^{ }_2-\mathcal{K}) 
 [\mathcal{P}_1^2 + \mathcal{K}^2 - \mathcal{P}_2^2]
 \;, \nn  
 \delta^{(4)}(\mathcal{P}^{ }_2 - \mathcal{P}^{ }_1-\mathcal{K})\;
  2 \mathcal{K} \cdot \mathcal{P}^{ }_1 & = & 
 \delta^{(4)}(\mathcal{P}^{ }_2 - \mathcal{P}^{ }_1-\mathcal{K})
  [(\mathcal{K}+\mathcal{P}^{ }_1)^2 - \mathcal{P}_1^2 - \mathcal{K}^2 ]
 \nn & = &  
 \delta^{(4)}(\mathcal{P}^{ }_2 - \mathcal{P}^{ }_1-\mathcal{K})
  [\mathcal{P}_2^2 - \mathcal{P}_1^2 - \mathcal{K}^2]
 \;, \nn  
 \delta^{(4)}(\mathcal{P}^{ }_1 + \mathcal{P}^{ }_2+\mathcal{K})\; 
 2 \mathcal{K} \cdot \mathcal{P}^{ }_1 & = & 
 \delta^{(4)}(\mathcal{P}^{ }_1 + \mathcal{P}^{ }_2+\mathcal{K}) 
 [(\mathcal{K}+\mathcal{P}^{ }_1)^2 - \mathcal{P}_1^2 - \mathcal{K}^2 ]
 \nn & = &  
 \delta^{(4)}(\mathcal{P}^{ }_1 + \mathcal{P}^{ }_2+\mathcal{K}) 
 [\mathcal{P}_2^2 - \mathcal{P}_1^2 - \mathcal{K}^2]
 \;,
\ea
where the factors are all constants, independent of
$\vec{p}^{ }_1, \vec{p}^{ }_2$. Thereby we arrive at
\ba
 & &  \hspace*{-3.5cm}
 R(T,{k}) =  
 \frac{|h|^2}{ 2 \sqrt{k^2 + M^2}}
 \, (m_{\phi}^2 - m_{\ell}^2 - M^2)
 \int \! \frac{{\rm d}^3 \vec{p}^{ }_1}{(2\pi)^3 2 \E^{ }_1} \,
 \int \! \frac{{\rm d}^3 \vec{p}^{ }_2}{(2\pi)^3 2 \E^{ }_2} \,
 \times  
 \nn 
 \times \biggl\{ & - & \!\!\!
 (2\pi)^4 \delta^{(4)}(\mathcal{P}^{ }_1+\mathcal{P}^{ }_2-\mathcal{K}) \, 
 \nF{1}\nB{2} 
 \Scatd \nn 
 & - & \!\!\!
 (2\pi)^4 \delta^{(4)}(\mathcal{P}^{ }_1-\mathcal{P}^{ }_2-\mathcal{K}) \, 
 \nF{1}(1+\nB{2})
 \Scatc \nn 
 & + & \!\!\!
 (2\pi)^4 \delta^{(4)}(\mathcal{P}^{ }_2-\mathcal{P}^{ }_1-\mathcal{K}) \,
 \nB{2}(1-\nF{1})
 \Scatb \nn 
 & + & \!\!\!
 (2\pi)^4 \delta^{(4)}(\mathcal{P}^{ }_1+\mathcal{P}^{ }_2+\mathcal{K} )\, 
 (1-\nF{1})(1+\nB{2})
 \biggr\}
 \;, \Scata   
 \la{pert3}
\ea
where 
$
 \E^{ }_1 \equiv \sqrt{ p_1^2 + m_{\ell}^2 }
$
and
$
 \E^{ }_2 \equiv \sqrt{ p_2^2 + m_{\phi}^2 }
$.
In passing, we note that \eq\nr{pert3} 
is equivalent to a {\em collision term of a Boltzmann equation}; 
the structure of the latter is recalled in appendix A. 

\index{Boltzmann equation}

Let us analyze \eq\nr{pert3} in more detail, recalling that $k^0_{ } =
\sqrt{k^2 + M^2 } > 0$, with $M$ the mass of the produced particle. 
The first question is, 
when do the different channels get realized. 
Since all the particles are massive, we can go to the rest
frame of the decaying one; it is then clear that 
the first channel gets realized for $M > m^{ }_{\ell} + m^{ }_{\phi}$; 
the second for $m^{ }_{\ell} > M + m^{ }_{\phi}$;
the third  for $m^{ }_{\phi} > M + m^{ }_{\ell}$; 
and the last one never. 
As an example, assuming that the scalar mass (the Higgs mass)
is larger than those of the produced particles, 
$m^{ }_{\phi} \gg M, m^{ }_{\ell}$, we can focus on the third channel, 
where the integral to be considered reads 
\be
 I(k) \equiv 
 \int \! \frac{{\rm d}^3 \vec{p}^{ }_1}{(2\pi)^3 2 \E^{ }_1} \,
 \int \! \frac{{\rm d}^3 \vec{p}^{ }_2}{(2\pi)^3 2 \E^{ }_2} \,
 (2\pi)^4 \delta^{(4)}(\mathcal{P}^{ }_2-\mathcal{P}^{ }_1-\mathcal{K}) \,
 \nB{}(\E^{ }_2)[1-\nF{}(\E^{ }_1)] 
 \;.  \la{MS_int}
\ee 

The integral in \eq\nr{MS_int} can be 
simplified, if we go to the high-temperature limit 
where the masses $M^2 = \mathcal{K}^2$ 
and $m_{\ell}^2 = \mathcal{P}_1^2$ of the produced
particles can be neglected.\footnote{%
 It must be noted that, as discussed in 
 sections~\ref{se:NI_rho} and \ref{se:ppr},
 unresummed computations
 typically lose their validity in the ultra-relativistic
 limit when the temperature is much higher than particle 
 masses, cf.\ e.g.\ refs.~\cite{photon1_2,bb2_2}. 
 We assume here 
 that $M,m^{ }_{\ell} \ll \pi T \ll m^{ }_{\phi}$. 
 } 
Denoting
\be
 p \equiv |\vec{p}^{ }_1|
 \;, \quad
 k \equiv |\vec{k}|
 \;, 
\ee
we get 
\ba
 I(k) & = &
 \int \! \frac{{\rm d}^3 \vec{p}^{ }_1}{(2\pi)^3 2 p} \,
 \int \! \frac{{\rm d}^3 \vec{p}^{ }_2}{(2\pi)^3 2 \E^{ }_2} \,
 (2\pi)^3 \delta^{(3)}(\vec{p}^{ }_1 + \vec{k} - \vec{p}^{ }_2) \,
 (2\pi) \delta(p + k - \E^{ }_2)
 \nB{}(\E^{ }_2)[1-\nF{}(p)] 
 \nn & = & 
 \frac{1}{(4\pi)^2}
 \int \! \frac{{\rm d}^3 \vec{p}^{ }_1}{p(p+k)}
 \delta\,\Bigl(p + k - \sqrt{m_{\phi}^2 + (\vec{p}^{ }_1+ \vec{k})^2}\Bigr)
 \nB{}(p+k) [1 - \nF{}(p)]
 \nn & = & 
 \frac{1}{8\pi} \int_0^\infty \! \frac{{\rm d}p\, p}{p+k}
 \int_{-1}^{+1} \! {\rm d}z \, 
 \delta\Bigl( p + k - \sqrt{m_{\phi}^2 + p^2 + k^2 + 2 p k z} \Bigr)
 \nB{}(p+k) [1 - \nF{}(p)] 
 \;, \nn
\ea
where spherical coordinates were introduced in the last step.  
The Dirac-$\delta$ gets realized when
\be
 p^2 + k^2 + 2 p k = m_{\phi}^2 + p^2 + k^2 + 2 p k z
 \;,  
\ee
i.e.\ 
$
 z = 1 - m_{\phi}^2 / (2 p k)
$.
This belongs to the interval $(-1,1)$ if $p > m_{\phi}^2/4 k$, so that 
\be
 I(k) = 
 \frac{1}{8\pi} \int_{\frac{m_{\phi}^2}{4 k}}^\infty \! 
 \frac{{\rm d}p\, p}{p+k}
 \biggl| \frac{{\rm d}}{{\rm d}z}
 \sqrt{m_{\phi}^2 + p^2 + k^2 + 2 p k z} 
 \biggr|^{-1}%
_{ \sqrt{m_{\phi}^2 + p^2 + k^2 + 2 p k z} = p + k } 
 \nB{}(p+k) [1 - \nF{}(p)]
 \;. 
\ee
The derivative appearing in the above expression is taken trivially, 
\be
 \left. \frac{{\rm d}}{{\rm d}z} \sqrt{\cdots} \right|^{ }_{\cdots} 
 = \frac{pk}{p+k}
 \;, 
\ee
whereby we arrive at
\ba
  I(k) & = & \frac{1}{8\pi k} 
 \int_{\frac{m_{\phi}^2}{4 k}}^\infty \! {\rm d}p \, 
 \nB{}(p+k) [1 - \nF{}(p)]
 \;. \la{Ik_pref}
\ea
This describes how a fermion of momentum $k$ is produced
from a decay of a Higgs particle of energy $p+k$, with the part $p$ of the 
energy being carried away by the other fermionic decay product, which
experiences Pauli blocking in the final state.

\index{Pauli blocking}

The integration in \eq\nr{Ik_pref} can be performed by decoupling 
the $p$-dependence via the identity 
\be
 \nB{}(p+k) \bigl[ 1 - \nF{}(p) \bigr] = 
 \bigl[ \nB{}(p+k) + \nF{}(p) \bigr] \nF{}(k)
 \;,
\ee
leading to 
\ba
 I(k) & = & 
 \frac{T \nF{}(k)}{8\pi k}
 \Bigl[ \ln\Bigl( 1 - e^{-\beta(p+k)}\Bigr) - 
 \ln\Bigl( 1 + e^{-\beta p}\Bigr) \Bigr]^{\infty}_{\frac{m_\phi^2}{4k}}
 \nn & = & 
 \frac{T \nF{}(k)}{8\pi k}
 \ln \Biggl\{
 \frac{1 + \exp\bigl[{- \beta \bigl(\frac{m_{\phi}^2}{4 k} \bigr)}\bigr]}
 { 1 - \exp\bigl[{- \beta \bigl( k + \frac{m_{\phi}^2}{4 k} \bigr)}\bigr]}
 \Biggr\}
 \;. \la{I_estim}
\ea
Inserting \eq\nr{I_estim} into \eq\nr{pert3} 
(with $M = m^{ }_\ell = 0$) then yields
\be
 R(T,k) =  
 \frac{|h|^2 m_{\phi}^2 }{ 2 k}
 \, I(k)
 \;, 
\ee
which in combination with \eq\nr{yield} produces
\ba
 Y(t^{ }_0) & = &  \frac{45\sqrt{5}}{\pi^{5/2}}
 \frac{|h|^2 m_{\phi}^2}{16\pi^3}
  \int_{T^{ }_0}^{T^{ }_\rmii{max}}
  \! \frac{{\rm d} T}{T^4} 
  \frac{m^{ }_\rmi{Pl}}{c_s^2(T) h^{ }_\rmi{eff}(T) 
  \sqrt{g^{ }_\rmi{eff}(T)} }
  \int_0^{\infty}
  \! 
  {\rm d}z \, z \, I(T z)
  \;. \la{Y_final}
\ea
The remaining integrals can be carried out numerically.
They display the variables on which 
the ``dark matter'' abundance
depends on in this model: the coupling constant ($|h|^2$), 
the mass of the decaying particle ($m^{ }_{\phi}$), as well as
the thermal history of the universe 
(through the functions $c_s^2$, $h^{ }_\rmi{eff}$ 
and $g^{ }_\rmi{eff}$).\footnote{%
 A phenomenologically viable dark matter scenario analogous to the one 
 discussed here, albeit with a scalar field decaying into 
 two right-handed neutrinos, 
 has been suggested in refs.~\cite{fbez1,fbez2}. 
 }


\subsection*{Appendix A: Relativistic Boltzmann equation}
\la{se:Boltz}

\index{Boltzmann equation}

We recall here the structure of the collision term
in the relativistic Boltzmann equation, and compare the result 
with the quantum field theoretic formula in \eq\nr{pert3}.\footnote{%
  A concise discussion of the Boltzmann equation 
  can be found in the appendix of
  ref.~\cite{msm}. 
 }

\index{Fermi's Golden Rule}

To understand the logic of the Boltzmann equation, 
a possible starting point is Fermi's Golden Rule for a decay rate, 
\be
 \Gamma^{ }_{1 \to n}(\mathcal{K})  
 = \frac{1}{2 \E^{ }_{k}}  
 \int \! {\rm d}\Phi^{ }_{1 \to n} \, c \sum |\mathcal{M}^{ }_{1 \to n}|^2   
 \;, \la{Fermi}
\ee
where the phase space integration measure is defined as
\ba
 \int \! {\rm d}\Phi^{ }_{1+m \to n} & \equiv & 
 \int 
 \left\{ \prod_{i=1}^m \frac{{\rm d}^3 
 \vec{k}^{ }_{a_i}}{(2\pi)^3 2 \E^{ }_{a_i}}
 \right\}
 \left\{ \prod_{j=1}^n \frac{{\rm d}^3 
 \vec{p}^{ }_{b_j}}{(2\pi)^3 2 \E^{ }_{b_j}}
 \right\}
 (2\pi)^4 \delta^{(4)}
  \Bigl( 
 \mathcal{K}  
 + \sum_{i=1}^m \mathcal{K}^{ }_{a_i}
 - \sum_{j=1}^n \mathcal{P}^{ }_{b_j}
  \Bigr)
 \;, \nn
\ea
and $\{ a^{ }_i\}$ and $\{ b^{ }_j \}$ label 
initial- and final-state particles, respectively, with four-momenta
$\mathcal{K}^{ }_{a_i}\equiv (\E^{ }_{a_i},\vec{k}^{ }_{a_i})$ and  
$\mathcal{P}^{ }_{b_j} \equiv (\E^{ }_{b_j},\vec{p}^{ }_{b_j})$. 
Moreover, 
$c = \frac{1}{i^{ }_a! i^{ }_b!}$,  
where $i^{ }_a$, $i^{ }_b$ are the numbers of 
identical particles in the initial and final states, whose momenta
are integrated over;
$\mathcal{M}^{ }_{1\to n}$ is an invariant amplitude; 
and the sum in \eq\nr{Fermi} goes over unresolved polarization states.  

Let now $f(\mathcal{X},\vec{k})$ be a particle distribution function; we assume
its normalization to be so chosen that the total number density of 
particles at $\mathcal{X}$ is given by (cf.\ \eq\nr{nI_def})
\be
 n(\mathcal{X}) = \int \! \frac{{\rm d}^3\vec{k}}{(2\pi)^3} \, 
 f(\mathcal{X},\vec{k})
 \;. \la{nx_def}
\ee
In thermal equilibrium, $f(\mathcal{X},\vec{k})$ 
is uniquely determined
by the temperature and by possible chemical potentials,  
$f(\mathcal{X},\vec{k}) \equiv \nF{}(\E^{ }_{k}\pm\mu)$
(or $\nB{}(\E^{ }_{k}\pm\mu)$ for bosons).
At the same time, for a single plane wave in vacuum,
regularized by a finite volume $V$, we would have
\be
 f(\mathcal{X},\vec{k}) = \frac{(2\pi)^3}{V}
 \,\delta^{(3)}(\vec{k}-\vec{k}^{ }_0)
 \;,
\ee 
which would lead to $n(\mathcal{X})$ in \eq\nr{nx_def}
evaluating to $1/V$. 

To convert \eq\nr{Fermi} into a Boltzmann equation, we identify 
the decay rate $\Gamma$ by $-\partial^{ }_t f/ f$, and 
multiply that by $\E^{ }_k$ in order to identify
a Lorentz-covariant structure: 
\be
 - \E^{ }_{k} \frac{\partial f^{ }_{1}}{\partial t} \frac{1}{f^{ }_{1}}
 \Rightarrow 
 - \mathcal{K}^\alpha  
 \frac{\partial f^{ }_{1}}{\partial \mathcal{X}^\alpha} \frac{1}{f^{ }_{1}}
 \;. 
\ee
We also modify the right-hand side of \eq\nr{Fermi} by 
allowing for $1+m$ particles in the initial state, and by adding
Bose enhancement and Pauli blocking factors. Thereby we obtain
\ba \index{Bose enhancement} \index{Pauli blocking}
 & & \hspace*{-1cm} \mathcal{K}^\alpha 
 \frac{\partial f^{ }_{1}}{\partial \mathcal{X}^\alpha}
 = - \frac{1}{2} \sum_{m,n} 
 \int \! {\rm d}\Phi^{ }_{1+m \to n} 
 \, c \sum |\mathcal{M}|^2_{{1}+m\to n} \,
 \nn & \times & \!\!\! 
 \bigl\{  
 f^{ }_{1} f^{ }_{a_1} \cdots f^{ }_{a_m}
 (1\pm f^{ }_{b_1}) \cdots (1\pm f^{ }_{b_n}) 
 \; - \;  
 f^{ }_{b_1} \cdots f^{ }_{b_n}
 (1\pm f^{ }_{1})(1\pm f^{ }_{a_1}) \cdots (1\pm f^{ }_{a_m})
 \bigr\}
 \;, \hspace*{6mm} \la{collisions}
\ea 
where $+$ applies to bosons and $-$ to fermions. On the last row 
of \eq\nr{collisions}, inverse reactions (``gain terms'') have been
introduced, in order to guarantee detailed balance in the case that 
all distribution functions have their equilibrium forms. 

Let us finally compare \eq\nr{collisions}
with \eq\nr{pert3}. 
We observe that \eq\nr{pert3} corresponds to the 
{gain terms} of \eq\nr{collisions}; 
the reason is that in the quantum field theoretic formula  
the produced particles were (by assumption) non-thermal, 
$f^{ }_{1} \equiv 0$. (This can be corrected for as discussed
around \eq\nr{master_nonlin}.)
At the same time, to obtain a complete
match, we should work out the scattering matrix elements, $|\mathcal{M}|^2$,
and the factors $c$. One ``strength'' of the quantum field
theoretic computation leading to \eq\nr{pert3} is that these automatically
come with their correct values. Another strength is that 
at higher orders, there are also virtual effects in the quantum field
theoretic computation which lead to thermal masses, modified dispersion
relations, additional quasiparticle states (cf.\ e.g.\ the discussion 
concerning the plasmino branch in \se\ref{se:htl}), and, 
last but not least, which cancel IR divergences (mass singularities) 
from the real processes. It is not obvious
whether these can be 
accounted for by a simple modification of the Boltzmann equation. 

%
\subsection*{Appendix B: 
Evolution equations in the presence of a conserved charge}

Above we assumed that there were no chemical potentials 
affecting thermodynamic functions determining 
the evolution of the system; 
this is most likely a good
assumption in cosmology, as shown e.g.\ by the great success of the 
Big Bang Nucleosynthesis computation based on this ansatz. The assumption
is often quantified by the statement that the observed baryon asymmetry
of the universe corresponds to a chemical potential $\mu\sim 10^{-10}T$.
On the 
other hand, in heavy ion collision experiments and particularly 
in astrophysics, conserved charges and the associated chemical 
potentials do play an important role. Even in cosmology lepton
asymmetries could in principle be much larger than the baryon asymmetry, 
since they cannot be directly observed, hidden as they are in a 
neutrino background. 
Let us see how the presence of a chemical potential
would change the cosmological considerations presented above. 

We consider a system with one chemical potential, $\mu$, and the
corresponding total particle number, $N$. The energy, entropy, and number
densities are defined through 
$e\equiv E/V$, $s\equiv S/V$, $n\equiv N/V$, respectively, 
where $V$ is the volume. The total energy of the system is then 
\be
 E = TS - pV + \mu N
 \;, \la{stat_E}
\ee
while the corresponding differential reads 
\be
 {\rm d}E = T\, {\rm d}S - p \,{\rm d}V + \mu\, {\rm d} N
 \;. \la{stat_dE}
\ee
Dividing both equations by the volume, we get 
\be
  e + p = T s + \mu \, n  \;, 
  \la{e_stat}
\ee
as well as 
\ba
  {\rm d}e & = & {\rm d} \Bigl( \frac{E}{V} \Bigr)
  \; = \; \frac{{\rm d}E}{V} - E \frac{{\rm d}V }{V^2}
  \nn & = & 
  T\frac{{\rm d}S}{V} - p \frac{{\rm d}V}{V} + \mu\frac{{\rm d} N}{V} 
  -T S \frac{{\rm d}V }{V^2} + p \frac{{\rm d}V}{V} - \mu N
 \frac{{\rm d}V }{V^2} 
  \nn & = & T {\rm d} \Bigl( \frac{S}{V} \Bigr)
  + \mu \, {\rm d} \Bigl( \frac{N}{V} \Bigr)
 = T \, {\rm d}s + \mu \, {\rm d}n
 \;. \la{de_stat}
\ea
Taking a differential from \eq\nr{e_stat} and subtracting the 
result of \eq\nr{de_stat} yields the {\em Gibbs-Duhem} equation, 
\be \index{Gibbs-Duhem equation}
 {\rm d}p = s\, {\rm d}T + n \, {\rm d}\mu
 \;. \la{GD}
\ee
As indicated by this equation, the natural variables of $p$ 
are $T$ and $\mu$.

The system of equations
we now consider is composed of \nr{AE1} and \nr{AE2}, 
complemented by the comoving conservation law for the number density, 
\be
 {\rm d}(n a^3) = 0 
 \;, \la{dn_stat} 
\ee
as well as the thermodynamic relations just derived. 

As a first step let us show that the entropy conservation law, 
\eq\nr{entr_conv}, continues to hold in the presence of the new terms. 
Repeating the argument leading to it with the new thermodynamic relations
of \eqs\nr{e_stat} and \nr{de_stat}, we obtain 
\ba
 0 & = &  {\rm d} (e a^3 ) + p\, {\rm d}(a^3) 
 \nn & = &  a^3 {\rm d} e + (p+e)\, {\rm d}(a^3) 
 \nn & = &  a^3 \, \bigl[ T {\rm d} s + \mu\, {\rm d}n \bigr]  
 + \bigl[ T s + \mu n \bigr]\, {\rm d}(a^3)
 \nn & = & T {\rm d} ( s a^3 ) + \mu \, {\rm d} (n a^3) 
 \;. \la{new_entr_conv}
\ea
The relation in \eq\nr{dn_stat} then directly leads to \eq\nr{entr_conv}.

It is considerably more difficult to find a generalization 
of \eq\nr{Tt_rel}. In fact, we must {\em simultaneously} follow the 
time evolution of $T$ and $\mu$, solving a coupled set of non-linear
differential equations. 
\Eqs\nr{entr_conv} and \nr{dn_stat} can be written as 
$ 
 {\rm d} s / s = -3 {\rm d} a / a
$ 
and 
$ 
 {\rm d} n / n = -3 {\rm d} a / a
$,  
i.e.\ 
\ba
 \frac{\partial^{ }_\rmii{$T$} s}{s} \dot{T}
  + \frac{\partial^{ }_\mu s}{s} \dot{\mu} 
 & = & -\frac{3\dot{a}}{a}
 \;, \\ 
 \frac{\partial^{ }_\rmii{$T$} n}{n} \dot{T}
  + \frac{\partial^{ }_\mu n}{n} \dot{\mu} 
 & = & -\frac{3\dot{a}}{a}
 \;.
\ea
Denoting (from Gibbs-Duhem, \eq\nr{GD}) 
\be
 s = \partial^{ }_\rmii{$T$} p \equiv p^{ }_\rmii{$T$} \;, \quad
 n = \partial^{ }_\mu p \equiv p^{ }_{\mu} \;, \quad 
 \partial^{ }_\rmii{$T$} s \equiv p^{ }_{\rmii{$T$}\rmii{$T$}} \;, \quad
 \partial^{ }_\rmii{$T$} n = \partial^{ }_\mu s
  \equiv p^{ }_{\rmii{$T$}\mu} \;, \quad
 \partial^{ }_\mu n \equiv p^{ }_{\mu\mu} 
 \;, 
\ee
and inserting the right-hand side from \eq\nr{Hubble}, we obtain 
\ba
 \frac{{\rm d}T}{{\rm d}t} & = & 
 \frac{p^{ }_{\mu} p^{ }_{\rmii{$T$}\mu}
     - p^{ }_{\rmii{$T$}} p^{ }_{\mu\mu}}
      {p^{ }_{\rmii{$T$}\rmii{$T$}} p^{ }_{\mu\mu}
    - (p^{ }_{\rmii{$T$}\mu})^2}
 \frac{\sqrt{24\pi e (T,\mu) }}{m^{ }_\rmi{Pl}}
 \;, \la{pTu1} \\
 \frac{{\rm d}\mu}{{\rm d}t} & = & 
 \frac{p^{ }_{\rmii{$T$}} p^{ }_{\rmii{$T$}\mu}
            - p^{ }_{\mu} p^{ }_{\rmii{$T$}\rmii{$T$}}}
      {p^{ }_{\rmii{$T$}\rmii{$T$}} p^{ }_{\mu\mu}
    - (p^{ }_{\rmii{$T$}\mu})^2}
 \frac{\sqrt{24\pi e (T,\mu) }}{m^{ }_\rmi{Pl}}
 \;, 
\ea
where, according to \eq\nr{e_stat},
\be
 e(T,\mu) = -p + T \, p^{ }_\rmii{$T$} + \mu \, p^{ }_\mu
 \;.
\ee
Therefore, in a general case, the pressure and all its
first and second derivatives are needed for determining
the cosmological evolution. 

Finally we remark that in typical relativistic systems,
mixed derivatives are small,
$ p^{ }_{\rmii{$T$}\mu} \sim \mu T \ll 
  p^{ }_{\rmii{$T$}\rmii{$T$}}, p^{ }_{\mu\mu} \sim T^2
$.
Setting 
$
 p^{ }_{\rmii{$T$}\mu} \to 0
$, 
\eq\nr{pTu1} reduces to 
\be
 \frac{{\rm d}T}{{\rm d}t}  =  - 
 \frac{
       p^{ }_{\rmii{$T$}}  }
      {p^{ }_{\rmii{$T$}\rmii{$T$}} 
      }
 \frac{\sqrt{24\pi e (T,\mu) }}{m^{ }_\rmi{Pl}}
 \;,  
\ee
which agrees with \eq\nr{Tt_rel}. 

\newpage 

\subsection{Evolution of a long-wavelength field in a thermal environment}
\la{se:field}

We now move to a different class of observables: from single particles 
to collective ``fields'' that evolve within a thermal environment. 
In the present section we consider a field~$\varphi$ that is out
of equilibrium; 
the entire field has a non-zero expectation value, or 
``condensate'', consisting roughly speaking
of very many almost zero-momentum quanta ($k \ll g T$
rather than $k \sim \pi T$ as is the case for typical 
particle states). 
In \se\ref{se:transport}, we then move on to cases where there is no separate
field forming a condensate, but rather the degrees of freedom of 
a strongly interacting system contain almost conserved 
quantities which evolve analogously to separate weakly coupled fields.

Consider a system containing two sets
of elementary fields: a scalar field $\varphi$ as well as other fields
which we do not need to specify but which are 
contained in $\mathcal{J}^{ }_\rmi{int}$
and $\mathcal{L}^{ }_\rmi{bath}$. 
The setup is essentially the same as in \eq\nr{L_c_sc} but for
simplicity now with a real scalar field, 
being described by the Lagrangian density
\be
 \mathcal{L}^{ }_\iM =  \fr12 \varphi (-\square - m^2 ) \varphi
 - \varphi \mathcal{J}^{ }_\rmi{int} + \mathcal{L}^{ }_\rmi{bath}
 \;. \la{simp_1}
\ee
We assume that the scalar field is initially displaced from
its equilibrium value $\varphi^{ }_\rmi{eq} \equiv 0$, 
and then evolves towards it. 
We also assume that
this evolution is a ``slow'' and essentially ``classical'' process: 
the coupling between~$\varphi$ 
and the heat bath, described by $\mathcal{J}^{ }_\rmi{int}$, 
is taken to be weak, 
implying that  $\varphi$ 
evolves on time scales $\Delta t$ much longer than 
those associated with the  plasma interactions
($\Delta t \gg 1/(\alpha^2 T)$, where $\alpha$ is a generic 
fine-structure constant for plasma processes). 
Therefore multiple plasma collisions take place during the time interval 
in which $\varphi$ changes only a little, implying  a smooth and
decoherent evolution.  
Then, we may postulate a classical equation of motion for 
how $\varphi$ evolves towards equilibrium, which can be
expanded in gradients (since the field consists of 
small-$k$ quanta and evolves slowly) 
and powers of $\varphi$ (since we assume that the initial 
state is already close to equilibrium): 
\be \index{Friction coefficient}
 \square \varphi + V_\rmi{eff}'(\varphi) = - \Gamma \dot{\varphi}
 + \rmO(\raise-0.15em\hbox{$\stackrel{\dots}{\varphi}$}, 
 \nabla^2\dot\varphi,
 \dot{\varphi}^2,(\nabla\varphi)^2)
 \;. \la{eom_2}
\ee 
The coefficients appearing in this equation, such as $\Gamma$, are functions
of the properties of the heat bath, such as its temperature $T$ and 
the couplings $\alpha$.\footnote{%
 For simplicity 
 we consider a system in 
 flat spacetime. In cosmology, the expansion of the universe causes
 another type of ``dissipation'', with the Hubble rate $H$ playing
 a role similar to $\Gamma$ (more precisely the Hubble friction 
 amounts to $-3 H \dot{\varphi}$ in analogy with 
 \eq\nr{kin_total}). The total friction is the sum of these two contributions. 
 }
We note in passing that once $\varphi$ is already close to equilibrium, 
the right-hand side of \eq\nr{eom_2} should be completed with a noise
term representing thermal fluctuations; the general ideology for 
this is discussed in more detail in \se\ref{se:decay}. 

\index{Kramers-Kronig relations}

There are two separate effects that the interactions of the 
$\varphi$-field with the heat bath
lead to. The first one is that $\varphi$ obtains an ``effective mass'', 
appearing as a part of the effective potential $V^{ }_\rmi{eff}$. The 
second is that the interactions generate a friction coefficient
$\Gamma$ as defined by \eq\nr{eom_2}. 
The role of friction is to transmit energy
from the classical field to the heat bath or, 
equivalently, to increase the entropy of the system. 
Despite different physical manifestations, 
the effective mass and the friction 
are intricately related to each other; as we will
see, they are on the formal level 
related to the real and imaginary parts of a single
analytic function (the retarded correlator of 
$\mathcal{J}^{ }_\rmi{int}$), and as such have 
a relation to each other, analogous to 
Kramers-Kronig relations.  In practice, 
there are circumstances
in which the effective mass plays a more substantial role, 
leading to so-called underdamped oscillations, as well as ones 
where the friction dominates the dynamics, referred 
to as overdamped oscillations. Technically,  
the effective mass turns out to be related to a ``Euclidean susceptibility''
of the operator $\mathcal{J}^{ }_\rmi{int}$, whereas $\Gamma$ is related to a
``Minkowskian susceptibility'', which is a genuine real-time quantity. 


\subsubsection{Effective mass}

\index{Effective mass} \index{Effective potential} 

In general, an effective potential can be defined and computed as 
discussed around \eq\nr{Veff_def}. Note that for this computation 
$\varphi$ is treated as constant both in temporal and spatial coordinates. 
We integrate over all fields appearing in $\mathcal{J}^{ }_\rmi{int}$
and $\mathcal{L}^{ }_\rmi{bath}$. Denoting these by $\chi$, we obtain
\be
 \exp\Bigl( - \frac{V}{T} V^{ }_\rmi{eff} \Bigr)
 = 
 \exp\Bigl( - \frac{V}{T} V^{ }_0 \Bigr) \, 
 \int \! \mathcal{D}\chi \, \exp\Bigl( 
 - \int_X L^{ }_\rmi{bath} - \varphi \int_X \mathcal{J}^{ }_\rmi{int}
 \Bigr)
 \;, \la{Veff_varphi}
\ee
where $V^{ }_0 \equiv \fr12 m^2\varphi^2$ is the tree-level
potential appearing in $\mathcal{L}^{ }_\iM$;  
we have gone over to Euclidean spacetime as usual for static
observables; and $L^{ }_\rmi{bath}$ is the Euclidean Lagrangian of the 
$\chi$-fields. Assuming 
that $\langle \mathcal{J}^{ }_\rmi{int} \rangle = 0$, 
and defining the effective mass as
\ba
 V^{ }_\rmi{eff}(\varphi) &=& V^{ }_\rmi{eff}(0) + 
  \fr12 m_\rmi{eff}^2\, \varphi^2 + \rmO(\varphi^3)
 \;, 
\ea
the matching of the left and right sides of \eq\nr{Veff_varphi} leads to
\be
 \delta m^2 \;\equiv\; m_\rmi{eff}^2 -m^2
 = -\frac{T}{V} \int_{X,Y}
 \Bigl\langle \mathcal{J}^{ }_\rmi{int}(X) \,
 \mathcal{J}^{ }_\rmi{int}(Y) \Bigr\rangle^{ }_\rmi{c}
 = - \int_X  \Bigl\langle \mathcal{J}^{ }_\rmi{int}(X) 
 \, \mathcal{J}^{ }_\rmi{int}(0) \Bigr\rangle^{ }_\rmi{c}
 \;. \la{meff}
\ee
Here we made use of translational invariance, and 
$\langle ... \rangle^{ }_\rmi{c}$ indicates that only the 
connected contraction contributes 
as long as $\langle \mathcal{J}^{ }_\rmi{int} \rangle = 0$. 
The correlator in \eq\nr{meff} has the form of a susceptibility, 
cf.\ \eq\nr{chi_fe}, but with an additional integral over
$\tau$ which gives it the dimension of GeV$^2$.

\index{Susceptibility}

In general, the 2-point correlator in \eq\nr{meff}
has a temperature-independent divergent part, because the correlator
$ \bigl\langle \mathcal{J}^{ }_\rmi{int}(X) 
 \, \mathcal{J}^{ }_\rmi{int}(0) \bigr\rangle^{ }_\rmi{c} $
diverges at short distances.  
This amounts to a renormalization 
of the (bare) mass parameter $m^2$. In addition, there can be 
a finite $T$-dependent correction, which can be interpreted as 
a thermal mass. A simple example was previously
seen around \eq\nr{mmeff}, and we return to a couple of further
examples below. 

Before proceeding let us recall that, in terms of Minkowskian quantities, 
the Euclidean susceptibility corresponds to a particular integral over
the corresponding spectral function, cf.\ \eq\nr{sum_rule}: 
\be
 - \delta m^2 = 
 \int_X  \Bigl\langle \mathcal{J}^{ }_\rmi{int}(X) 
 \, \mathcal{J}^{ }_\rmi{int}(0) \Bigr\rangle^{ }_\rmi{c}
 = \int_0^\beta \! {\rm d}\tau \, \Pi^{ }_\iE(\tau,\vec{k=0})
 = \int_{-\infty}^\infty \! \frac{{\rm d}\omega}{\pi}
 \frac{\rho(\omega,\vec{0})}{\omega}
 \;. \la{sum_rule_2}
\ee
For simplicity we put $E$ in a subscript from now on
(rather than in a superscript like in \se\ref{se:diff_G}).


\subsubsection{Friction coefficient}

\index{Friction coefficient}  \index{Damping rate}

Turning next to the friction coefficient, let us transform \eq\nr{eom_2} 
to Fourier space, 
writing $\varphi \propto e^{-i\omega t + i \vec{k}\cdot\vec{x}}
\;\tilde\varphi$:
\be
 [-\omega^2 + \vec{k}^2 + m_\rmi{eff}^2 - i \omega \Gamma
 + \rmO(\omega^3, \omega \vec{k}^2)]\, \tilde\varphi 
 = \rmO(\tilde\varphi^2)
 \;. \la{w_eq}
\ee 
We compare this with the position of the ``pole'' appearing
in the (retarded) propagator obtained after setting 
$\omega^{ }_n \to -i (\omega + i 0^+_{ })$ in the Euclidean propagator, 
cf.\ \eq\nr{PiR_PiE}:
\ba
 && \frac{1}{\omega_n^2 + \vec{k}^2 + m^2
 - \Pi^{ }_\iE 
 }
 \to
 \frac{1}{-\omega^2 -i\omega 0^+_{ } + \vec{k}^2 + m^2 - 
 \re\Pi^{ }_\iE 
 - i \im \Pi^{ }_\iE 
 }
 \;. 
\ea
The minus sign in front of $\Pi^{ }_\iE$ can be associated 
with that in \eq\nr{meff}, 
so that $- \re\Pi^{ }_\iE$ corresponds to $\delta m^2$ (this is an alternative
interpretation for the susceptibility discussed 
in \eq\nr{sum_rule_2}; the general 
Kramers-Kronig relation
of $\re\Pi^{ }_\iE (-i[\omega+i0^+_{ }],\vec{k})$ and 
the spectral function can be
deduced from \eq\nr{PiR_rho}). 
Setting $\vec{k}\to \vec{0}$ we obtain
\be
 \Gamma =  \lim_{\omega\to m^{ }_\rmii{eff}}
 \frac{\im\Pi^{ }_\iE(-i[\omega+i0^+_{ }],\vec{0})}{\omega} 
 \;,
\ee
where 
\be
 \Pi^{ }_\iE(\omega^{ }_n,\vec{k}) = 
 \int_{X} e^{i \omega^{ }_n \tau - i \vec{k}\cdot\vec{x}}
 \left\langle 
 \mathcal{J}^{ }_\rmi{int}(\tau,\vec{x}) \, 
 \mathcal{J}^{ }_\rmi{int}(0,\vec{0})
 \right\rangle
 \;,  \la{PiE}
\ee
whose imaginary part is given by (cf.\ \eq\nr{Discdef})
\be 
 \im\Pi^{ }_\iE(-i[\omega+i0^+_{ }],\vec{0}) =   \rho(\omega,\vec{0}) 
 \;. \la{rhodef} 
\ee 
To summarize, we have obtained
\be
 \Gamma =  \frac{\rho(m^{ }_\rmi{eff},\vec{0})}{m^{ }_\rmi{eff}} 
 \;, \la{Gamma_meff}
\ee
where $m_\rmi{eff}^2$ is the mass parameter 
appearing in $V^{ }_\rmi{eff}(\varphi)$.\footnote{%
 To be precise, 
 we have here included also purely $\omega$-dependent 
 terms such as $\omega^3$ into 
 $\Gamma$; the 
 remaining corrections are of $\rmO(\omega \vec{k}^2)$.
 }
The expectation value in \eq\nr{PiE}
is taken with respect to the density matrix 
of the heat bath degrees of freedom. \Eq\nr{Gamma_meff}
should be contrasted with \eq\nr{sum_rule_2}: the information 
concerning both the mass and the friction coefficient is 
encoded in the same spectral function, however in different ways.

Now, if $m^{ }_\rmi{eff}$ 
is much smaller than the thermal scales characterizing 
the structure of $\rho(\omega,\vec{0})$, in particular
the {\em width of its transport peak} (this 
concept will be defined around \eq\nr{loren}), 
which is normally $\sim \alpha^2 T$ 
(cf.\ \se\ref{se:transport}), 
then we can to a good accuracy 
set $m^{ }_\rmi{eff}\to 0$ in the evaluation
of~$\Gamma$.
Then $\Gamma$ amounts to a ``transport coefficient'', 
cf.\ \se\ref{se:transport}.


\subsubsection{Examples}

As a first example, 
we let $\mathcal{J}^{ }_\rmi{int}$ be a scalar operator~\cite{bulk1,bulk2}, 
in which case $\varphi$
could be called a ``dilaton'' field. For instance, if the medium 
is composed of non-Abelian gauge fields, we could have 
\be  
 \mathcal{J}^{(\rmi{s})}_\rmi{int} = \frac{1}{M}\, F^{a\mu\nu}F^a_{\mu\nu}
 \;. \la{Hint2}
\ee 
As a second example, 
we consider a pseudoscalar operator~\cite{mms}, whereby
$\varphi$ could be an ``axion'' field. 
Then the operator appearing in the interaction term reads
\be  
 \mathcal{J}^{(\rmi{p})}_\rmi{int} = \frac{q^{ }_\iM}{M}
 \;, \quad
 q^{ }_\iM \equiv  
 \epsilon^{ }_{\mu\nu\rho\sigma} 
   \frac{g^2 F_{ }^{a\mu\nu} \, F_{ }^{a\rho\sigma}}{64\pi^2}
 \;, \la{Hint3}
\ee 
where $q^{ }_\iM$ is a (Minkowskian) 
``topological charge density''.\footnote{%
 In the axion literature $M$ is often denoted by $f^{ }_a$.
 } 

\index{Dilaton mass}

Now, in the case of $ \mathcal{J}^{(\rmi{s})}_\rmi{int} $, the effective mass
originating from \eq\nr{meff} is ultraviolet divergent. Therefore
a bare mass parameter needs to exist, and the susceptibility
simply corrects this. There is also a finite thermal mass correction which, 
on dimensional grounds, is of the form $\delta m^2(T) \sim T^4/M^2$.
If $M$ is large, this thermal correction is small. 

\index{Axion mass}

In the case of $ \mathcal{J}^{(\rmi{p})}_\rmi{int} $, 
in contrast, the Euclidean
susceptibility is finite~\cite{mlx}. Once we go to Euclidean spacetime, 
the ``Wick rotation'' $D^{ }_t\to i D^{ }_\tau$ (cf.\ \se\ref{ss:gauge_path})
implies that $q^{ }_\iM$ becomes purely imaginary and 
thus the susceptibility in \eq\nr{meff} is positive. 
Therefore we can consider $m_\rmi{eff}^2$ to be generated purely from 
the interaction. This is the usual scenario for axion mass
generation, and corresponding measurements of the Euclidean 
topological susceptibility as a function of the temperature 
have been carried out on the lattice
(cf.\ ref.~\cite{axion1} for a review). 
Consistent with the fact that the Euclidean 
topological susceptibility vanishes to all orders in perturbation theory
and that perturbation theory works at least qualitatively at high
temperatures, the measurements show a rapid decrease 
as the temperature increases above the confinement scale. 

\index{Dilaton damping coefficient}

As far as the friction coefficients go, 
the operator $ \mathcal{J}^{(\rmi{s})}_\rmi{int} $ is related to 
the ``trace anomaly'' of pure Yang-Mills theory, 
\be
 {T^{\mu}}^{ }_\mu \approx 
 -\frac{b^{ }_0}{2}\, 
 F^{a\mu\nu}F^a_{\mu\nu} 
 \;, \la{Theta}
\ee
where $b^{ }_0$ defines the 1-loop $\beta$-function related to the running 
coupling, $b^{ }_0 \equiv 11\Nc/[3(4\pi)^2]$. The trace anomaly 
determines a particular transport coefficient, namely the
bulk viscosity $\zeta$, cf.\ \eq\nr{res_zeta} below. 
This parameter has been determined in perturbation theory, 
with the result~\cite{adm}  
\be
 \zeta\sim \frac{b_0^2\, g^4 T^3}{4 \ln(1/\alpha) } 
 \;, 
 \quad \alpha \equiv \frac{g^2}{4\pi}
 \;. \la{zeta_res}
\ee
For $m^{ }_\rmi{eff} \ll \alpha^2 T$, 
\eqs\nr{Gamma_meff} 
and \nr{Hint2} now imply $\Gamma\sim 4\zeta/(b_0^2 M^2)$, 
which after the insertion of \eq\nr{zeta_res} shows that  
$\Gamma \sim g^4 T^3 / M^2$
in the weak-coupling limit, 
up to logarithms.

\index{Axion damping coefficient} \index{Sphaleron rate} 

Finally, for $ \mathcal{J}^{(\rmi{p})}_\rmi{int} $, we need the transport 
coefficient associated with $q^{ }_\iM$. This quantity has 
been studied in great detail, 
given its important relation to fermion number
non-conservation through the axial anomaly. The quantity normally
considered is the so-called ``Chern-Simons diffusion rate'', 
or (twice) the ``sphaleron rate''
(cf.\ e.g.\ ref.~\cite{gdm_x} for a discussion of these two rates). 
This can be defined from the time and volume
average of the operator $\hat{q}^{ }_\iM$ as~\cite{cs3}
\ba
 \Gamma^{ }_\rmi{diff} 
 & \equiv &  \lim_{\Omega\to\infty}
 \frac{\langle \int_\Omega {\rm d}^4 \mathcal{X}
 \, \hat q^{ }_\iM(\mathcal{X}) 
 \int_\Omega {\rm d}^4 \mathcal{Y} \, \hat q^{ }_\iM(\mathcal{Y})
 \rangle}{\Omega} 
 = \int \! {\rm d}^4 \mathcal{X} \, 
 \Bigl\langle
 \fr12 \bigl\{ \hat q^{ }_\iM(\mathcal{X}), \hat q^{ }_\iM(0) \bigr\} 
 \Bigr\rangle 
 \la{Gamma_diff_1} \\
 & \stackrel{\rmi{\nr{bdelta}}}{=} &
   \lim_{\omega\to 0 }\Delta(\omega,\vec{0})
 \nn 
 & \stackrel{\rmi{\nr{bDelta}}}{=} & 
 \lim_{\omega\to 0 } \frac{2 T  \rho(\omega,\vec{0})}{\omega}
 \;, \la{Gamma_diff_2}
\ea
where $\Omega = V t$ is the spacetime volume and we made use 
of translational invariance. In the last equation of \eq\nr{Gamma_diff_1} 
the integration goes over
all the spacetime (positive and negative $t$).\footnote{%
 To show the step in \eq\nr{Gamma_diff_1} more precisely, 
 let us denote $\int_V {\rm d}^3\vec{x} \, \hat{q}^{ }_\iiM(\mathcal{X}) 
  \equiv \hat{o}^{ }_\iiM(t)$. If a non-zero limit exists for the correlator
 after division by $t$, it means that the correlator grows linearly 
 with $t$ at large times,
 so we can replace $1/t$ through ${\rm d}/{\rm d}t$ in the prefactor.
 Then
 $
 \lim_{t\to\infty} \frac{{\rm d}}{{\rm d}t}
 \int_0^t \! {\rm d}t' \int_0^t\! {\rm d}t''\, 
 \langle \hat{o}^{ }_\iiM(t')\, \hat{o}^{ }_\iiM(t'') \rangle
 = 
 \lim_{t\to\infty}
 \int_0^t \! {\rm d}t' \, 
 \langle \{ \hat{o}^{ }_\iiM(t') , \hat{o}^{ }_\iiM(t)  \} \rangle
 =
 \lim_{t\to\infty}
 \int_{-t}^t \! {\rm d}t' \, 
 \langle \frac{1}{2} 
 \{ \hat{o}^{ }_\iiM(t') , \hat{o}^{ }_\iiM(0)  \} \rangle
 $.
 In the last step we made use of the facts that the correlator
 only depends on the time difference and is symmetric in the sign
 of the time difference. 
 }
In the step leading to \eq\nr{Gamma_diff_2}, 
we furthermore exploited 
the fact that for $|\omega| \ll T$, 
$\nB{}(\omega) \approx T/\omega$. 
The first equality in \eq\nr{Gamma_diff_1} suggests
that we call $\Gamma^{ }_\rmi{diff}$ a 
``Minkowskian topological susceptibility''.

\index{Classical field theory}

In order to estimate $\Gamma^{ }_\rmi{diff}$, 
it has been argued that at high temperatures 
the dominant contribution  
comes from the dynamics of ``soft modes'', which are 
Bose enhanced and can thus 
be described by {classical} field theory~\cite{clgt1,clgt2,clgt3}.
In the classical limit we can write
\be
 Q(t) \equiv  \int_0^t \! {\rm d}t' \int_V 
  \! {\rm d}^3 \vec{x}' \, q^{ }_\iM (\mathcal{X}') 
 \equiv N^{ }_{\mbox{\tiny\rm{CS}}}(t) - N^{ }_{\mbox{\tiny\rm{CS}}}(0)
 \;, \la{Qt}
\ee
where $N^{ }_{\mbox{\tiny\rm{CS}}}(t)$ is the Chern-Simons number.
Therefore \eq\nr{Gamma_diff_1} becomes 
\be
 \Gamma^{ }_\rmi{diff} = \lim_{V,t\to\infty}
 \frac{\langle\!\langle Q^2(t) \rangle\!\rangle}{V t}
 \la{Gammadef}
 \;,
\ee
where the expectation value refers to a classical thermal average.  
It is the resemblance of \eqs\nr{Qt} and \nr{Gammadef} to the usual process 
of particle diffusion in non-relativistic statistical mechanics
(with $N^{ }_{\mbox{\tiny\rm{CS}}}(t)\to x(t)$) that
gives rise to the above-mentioned concept of ``Chern-Simons diffusion''.

\index{Chern-Simons diffusion}

Practical measurements of $\Gamma^{ }_\rmi{diff}$ within classical
lattice gauge theory have been carried out 
for pure SU(2)~\cite{cs1}
and SU(3) gauge theory~\cite{cs2}, and indicate that 
$\Gamma^{ }_\rmi{diff}\sim \alpha^5 T^4$ in these cases, 
up to logarithms. Therefore, the axion 
friction coefficient scales as $\Gamma \sim \alpha^5 T^3 / M^2$.
Even though smaller than the friction coefficient
for $ \mathcal{J}^{(\rmi{s})}_\rmi{int} $ by $\rmO(\alpha^3)$, 
this is still parametrically larger than $m_\rmi{eff}^2$ which vanishes
to all orders in perturbation theory in the pseudoscalar case
(these arguments are relevant if $T \gg \Lambda$, where $\Lambda$
is the confinement scale). 
On the non-perturbative level we may write
$m_\rmi{eff}^2 \sim (\Lambda^4 / M^2) (\Lambda/T)^n$, $n > 0$, 
for $T \gg \Lambda$, 
leading to $\Gamma / m^{ }_\rmi{eff} 
\sim \alpha^5 (T/M) (T / \Lambda)^{2 + n/2}$.
Thus axion oscillations become overdamped in the regime 
$T \, \gsim \, (M \Lambda^{2+n/2} / \alpha^5)^{\frac{1}{3+n/2}} $.\footnote{%
 It has been proposed that such dynamics could play a role for
 warm inflation~\cite{warm1}--\hspace*{-1.1mm}\cite{warm4}. 
 } 

To conclude this section, let us 
stress again that ``Hubble friction'' $H \sim T^2/m^{ }_\rmi{Pl}$ 
has been omitted from the above estimates. Roughly speaking, if $M$
and $m^{ }_\rmi{Pl}$ are similarly large scales, then Hubble damping 
dominates over $\Gamma$ below a certain temperature, because it decreases
less rapidly with $T$. In particular, $H$ is generally assumed to 
dominate at $T \lsim \Lambda$, in which regime $m_\rmi{eff}^2$
also becomes ``large'', 
$
 m_\rmi{eff}^2 \sim \Lambda^4 / M^2
$.

\newpage 

\subsection{Linear response theory and transport coefficients}
\la{se:transport}

\index{Transport coefficients}

Transport coefficients parametrize the 
{\em small-frequency behaviour of long-wavelength excitations} 
of a multiparticle system, in close analogy to the friction 
coefficient $\Gamma$ of the field $\varphi$ in \se\ref{se:field}. 
If there is no separate field to consider, it is meaningful to speak of 
long-wavelength excitations only for quantities 
for which long-distance correlations exist. 
This prompts us to consider conserved
(or almost conserved) currents, such as the energy-momentum tensor
and various particle number currents. When the amplitude of 
a perturbation is so large that it requires many scatterings to  
change it, the dynamics of the system
should be classical in nature, governed for instance by the known 
differential equations of hydrodynamics. The transport coefficients 
are then the ``low-energy constants'' of this 
infrared theory, and encode the effects of the 
short-wavelength modes that have been ``integrated out''
in order to arrive at the effective description
(cf.\ e.g.\ refs.~\cite{rev5_3,ga}).

Apart from a similar physical origin, transport coefficients 
also possess the formal property that they can be extracted from 
the small-frequency limit of a spectral 
function (cf.\ \eq\nr{brho}) as
\be
 \lim_{\omega\to 0^+_{ }} \frac{\rho(\omega,\vec{0})}{\omega}
 \;. \la{sketch_intercept}
\ee
Note that the spatial momentum has been set to zero before
the frequency here. Even though partly just a convention, 
this may be thought of as guaranteeing that 
the system considered is ``large'' and consists of very many
small-$k$ quanta. 

One example of a transport coefficient has already been 
discussed around \eq\nr{Gamma_meff}. That case was 
particularly simple because there were 
explicitly two sets of fields, one exhibiting ``slow'' or ``soft'' 
dynamics and another corresponding 
to ``fast'' or ``hard'' thermal modes. In most
cases, we have just one set of fields and the task is to 
consistently split that set into two parts, with the 
transport coefficients characterizing the dynamics of the soft modes. 


\subsubsection{Generic case}

\index{Equilibration rate} 

We wish to illustrate generic aspects of the formalism related 
to transport coefficients with the example of an {\em equilibration rate}. 
To this end, let us assume that some external perturbation 
has displaced the system 
from equilibrium by giving it a net ``charge'' of some type.  
We assume, however, that the charge under consideration is not 
conserved (in the case of QCD, 
this is the case for instance for the {\em spatial} components
of the baryon number or energy current). 
In this case, the system will {\em relax back to 
equilibrium}, i.e.\ the net charge will disappear, and the equilibration
rate describes how fast this process takes place. 

\index{Heisenberg-operator: bosonic}

Let $\hat N(t)$ now be the Heisenberg operator 
of some almost, but not exactly conserved physical quantity.
Any possible dependence on spatial coordinates has been 
suppressed for simplicity. 
According to the discussion above, 
we assume the equilibrium expectation value to be zero, 
\be
 \langle \hat N(t) \rangle^{ }_\rmi{eq} = 0 
 \;. \la{eq_ass}
\ee
The non-vanishing non-equilibrium expectation value, 
$
 \langle \hat N(t) \rangle^{ }_\rmi{non-eq}
$, 
is assumed to evolve so slowly that all other quantities are in equilibrium. 
If 
$
 \langle \hat N(t) \rangle^{ }_\rmi{non-eq}
$
is small in some sense (even though it should 
still be larger than typical equilibrium thermal fluctuations), 
we can expect the evolution to be described
by an equation linear in 
$
 \langle \hat N(t) \rangle^{ }_\rmi{non-eq}
$, and can therefore write 
\be
  \frac{{\rm d}}{{\rm d} t} \langle \hat N(t) \rangle^{ }_\rmi{non-eq}
  =  
 -\Gamma \, \langle \hat N(t) \rangle^{ }_\rmi{non-eq}
 + \rmO\Bigl(\langle \hat N(t) \rangle_\rmi{non-eq}^2\Bigr)
 \;,  \la{Gamma}
\ee
where ${\rm d}t > 0$ is also implicitly assumed. 
The coefficient $\Gamma$ introduced here may be called 
the {\em equilibration rate}.
Our goal is to obtain an expression for $\Gamma$, describing 
``dissipation'', in terms of various {\em equilibrium} expectation 
values, of the type
$
 \langle ... \rangle^{ }_\rmi{eq}
$,
describing ``fluctuations''. 

\index{Linear response}
\index{Fluctuation-dissipation theorem}
\index{Matching: equilibration rate} 

In what follows, we derive an expression for $\Gamma$ in two different ways. 
The first one is called {\em matching}: we consider a Green's 
function which is well-defined both in the classical limit as well 
as in the full quantum theory, compute it on both sides, and equate 
the results. The second method is on the other hand 
called {\em linear response theory}: 
we stay in the quantum theory all the time and try to obtain 
an equation of the form of \eq\nr{Gamma}, from which $\Gamma$
can be identified. 

\index{Classical limit}

As far as the matching method goes, an appropriate Green's function is a 
{\em symmetric} 2-point function, since it has a classical limit. 
Let us thus define 
\be
 \Delta(t) \equiv 
 \Bigl\langle \fr12 
 \Bigl\{ \hat N(t), \hat N(0) \Bigr\} \Bigr\rangle^{ }_\rmi{eq} 
 \;, \la{Deltat}
\ee
as well as the corresponding Fourier transform, 
\be
 \tilde \Delta(\omega) \equiv 
 \int_{-\infty}^{\infty} \! {\rm d}t \, 
 e^{i \omega t} \, \Delta(t)
 \;. \la{Deltaw}
\ee
The value $\Delta(0)$ amounts to a ``susceptibility'', 
as defined in \eq\nr{chi_fe} or in \eq\nr{sum_rule_2},
\ba \index{Susceptibility}
 \Delta(0) = \langle \hat N^2 \rangle^{ }_\rmi{eq}
 & = &
 T \partial^{ }_\mu \langle \hat N \rangle^{ }_\rmi{eq} 
 \equiv \frac{T}{\mathcal{Z}}  \partial^{ }_\mu 
 \tr \Bigl\{ 
 \hat N \Bigl[ e^{-\beta(\hat H - \mu \hat N)}\Bigr]\Bigr\}^{ }_{\mu = 0}
 \la{susc_1} \\  & = & 
 \frac{T^2}{\mathcal{Z}} \partial^2_\mu 
 \tr \Bigl[ e^{-\beta(\hat H - \mu \hat N)}\Bigr]^{ }_{\mu = 0}
 \nn & = & T^2 \left. 
 \partial^2_\mu \ln \mathcal{Z}(T,\mu) \right|^{ }_{\mu = 0}
 \;,  \la{Delta0}
\ea
where in the last stage we 
used the fact that $\langle \hat N \rangle^{ }_\rmi{eq}=0$.
Furthermore, by time-translational invariance
it is clear that $\Delta(-t) = \Delta(t)$. 

Now, on the classical side, we replace 
$
 \langle \hat N(t) \rangle^{ }_\rmi{non-eq}
$
by $N(t)$, and instead of \eq\nr{Gamma} have 
\be
 \dot{N}(t) \approx -\Gamma N(t)
 \;, \la{diff_eq}
\ee
with the trivial solution $N(t) = N(0) \exp(-\Gamma t)$. 
Enforcing the correct symmetry by replacing $t \to |t|$
and taking a thermal average with respect to initial
conditions, denoted by $\langle\!\langle ... \rangle\!\rangle$,  
leads straightforwardly to 
\be
 \Delta^{ }_\rmi{cl}(t) 
 \; \equiv \;
   \langle\!\langle N(t) N(0) \rangle\!\rangle 
 = \langle\!\langle [N(0)]^2\, \rangle\!\rangle \,  e^{-\Gamma |t|}
 \;, \la{Gamma_cl}
\ee
the time integral of which yields 
\be
 \tilde \Delta^{ }_\rmi{cl}(0) = 
 \int_{-\infty}^{+\infty} \! {\rm d} t \, \Delta^{ }_\rmi{cl}(t) 
 = \frac{2 \Delta^{ }_\rmi{cl}(0)}{\Gamma}
 \;. \la{Delta_cl}
\ee
Thus, the ratio of the susceptibility 
$\Delta^{ }_\rmi{cl}(0)$ and
the equilibration rate $\Gamma$ can be determined from the 
zero-frequency limit of the Fourier transform of the symmetric correlator,
\be
 \frac{\Delta^{ }_\rmi{cl}(0)}{\Gamma}
 = \frac{ \tilde \Delta^{ }_\rmi{cl}(0) }{2}
 = \fr12 \lim_{\omega\to 0}
 \int_{-\infty}^{\infty} \! {\rm d}t \, 
 e^{i \omega t} \, \Delta^{ }_\rmi{cl}(t) 
 \;. \la{chi_Gamma_ratio}
\ee

Let us match this equation to the quantum side.
Identifying $\Delta^{ }_\rmi{cl}(0) \leftrightarrow \Delta(0)$, 
which makes sense if the
susceptibility is ultraviolet finite; 
inserting \eq\nr{susc_1} for the latter; 
and rewriting the result as in \eq\nr{Gamma_diff_2}, we obtain
\be
 \frac{\partial^{ }_\mu \langle \hat N \rangle^{ }_\rmi{eq}}{\Gamma}
 = \lim_{\omega\to 0} \frac{ \rho(\omega)}{\omega}
 \;,  \la{generic_transport}
\ee 
where $ \rho (\omega)$ is the spectral function corresponding
to $\hat N$. This is an example of a {\em Kubo formula}. \index{Kubo formula}
However, the right-hand side does not directly give the rate of interest 
(as was the case in \eq\nr{Gamma_diff_2}), but a susceptibility
needs to be known before $\Gamma$ can be extracted.
There is also the peculiar fact that $\Gamma$ appears in the denominator 
in \eq\nr{generic_transport}, to which we return in 
\se\ref{se:decay}. 

Let us now rederive \eq\nr{generic_transport} in another way, namely
through a linear response analysis of 
$
 \langle \hat N(t) \rangle^{ }_\rmi{non-eq}
$.
We can assume that at time $t=-\infty$ the system was
in full equilibrium, but then a source term was added to 
the Hamiltonian, 
\be
 \hat H \to \hat H(t) = \hat H - \mu(t) \hat N(t)
 \;, \la{Ht}
\ee
which slowly displaced 
$
 \langle \hat N(t) \rangle^{ }_\rmi{non-eq}
$
from zero. 
Solving the equation of motion for the density matrix $\hat \rho(t)$
(cf.\ \eq\nr{liuv}), 
\be \index{Density matrix}
 i \frac{{\rm d} \hat \rho(t)}{{\rm d} t}
 = 
 \Bigl[ 
  \hat H(t), \hat \rho(t)
 \Bigr]
 \;,
\ee
to first order in the perturbation yields 
\be
 \hat\rho(t) \approx
 \hat\rho(-\infty) - i 
 \int_{-\infty}^{t} \! {\rm d} t' \, 
 \Bigl[
  \hat H(t'), \hat\rho(-\infty) 
 \Bigr] + \ldots
 \;, \la{rho_soln}
\ee
where we can furthermore replace 
$\hat H(t')$ by $- \mu(t') \hat N(t')$, 
as $\hat H$ and $\hat\rho(-\infty)\equiv \tfr{1}{\mathcal{Z}} 
e^{-\beta \hat H}$ commute.
Using this result in the definition of 
$
 \langle \hat N(t) \rangle^{ }_\rmi{non-eq}
$ 
gives 
\ba
 \langle \hat N(t) \rangle^{ }_\rmi{non-eq}
 & = & 
 \tr [\hat \rho(t) \hat N(t) ]
 \nn & \approx & 
 i  \int_{-\infty}^{t} \! {\rm d} t' \, \mu(t')
 \tr\Bigl\{
   \Bigl[
  \hat N(t'), \hat\rho(-\infty) 
 \Bigr] \, \hat N(t)
 \Bigr\}
 \nn & = & 
 i \int_{-\infty}^{\infty} \! {\rm d} t' \,
 \Bigl\langle [
  \hat N(t), \hat N(t') ] 
 \Bigr\rangle^{ }_\rmi{eq} \theta(t-t')\, \mu(t')
 \;,\la{LR}
\ea
where we have denoted
\be
 \langle ... \rangle^{ }_\rmi{non-eq} 
 \equiv \tr [\hat \rho(t) (...) ]
 \;, \qquad
 \langle ... \rangle^{ }_\rmi{eq} 
 \equiv \tr [\hat \rho(-\infty) (...) ]
 \;.
 \la{vevs}
\ee
The leading term from \eq\nr{rho_soln}
disappeared because of the assumption
in \eq\nr{eq_ass}.\footnote{%
 Note that the time dependence of $\hat{N}(t) = e^{i \hat{H} t}
 \,\hat{N}(0)\, e^{-i \hat{H} t}$ commutes with 
 $\tfr{1}{\mathcal{Z}} 
 e^{-\beta \hat H}$, so 
 $\langle \hat{N}(t) \rangle^{ }_\rmi{eq} =
  \langle \hat{N}(0) \rangle^{ }_\rmi{eq} $.
 }
In addition, we have inserted $\theta(t-t')$
and extended the upper end of the integration to infinity  
to stress the retarded nature of the correlator.  
The equilibrium expectation value appearing in \eq\nr{LR} 
is called a {\em linear response function}. 

We define next the retarded correlator (cf.\ \se\ref{se:diff_G}), 
\be
 C^{ }_\iR(t) \equiv 
 \Bigl\langle i  
 \Bigl[ \hat N(t), \hat N(0) \Bigr] \, \theta(t) \Bigr\rangle^{ }_\rmi{eq} 
 \; , \la{CRt}
\ee
and its Fourier transform, 
\be
 \widetilde C^{ }_\iR (\omega) \equiv 
 \int_{-\infty}^{\infty} \! {\rm d}t \, 
 e^{i \omega t} \, C^{ }_\iR (t)
 \;. \la{CRw}
\ee
We recall that the
imaginary part of $\widetilde C^{ }_\iR$ yields
the spectral function (cf.\ \eqs\nr{Discdef}, \nr{PiR_PiE}): 
\ba
 \rho(\omega)
  & = &   \im \widetilde C^{ }_\iR(\omega+ i 0^+_{ })
 \;.  \la{relation} 
\ea
\Eq\nr{LR} can now be written as 
\be
 \langle \hat N(t) \rangle^{ }_\rmi{non-eq}
  \approx  \int_{-\infty}^{\infty} \! {\rm d} t' \,
 C^{ }_\iR(t-t')\, \mu(t')
 \;. \la{LR_2}
\ee

To proceed from here, we assume that $\mu$ {\it varies slowly} around 
$t' \approx t$ and subsequently expand it in a Taylor series around 
this point: 
\be
 \mu(t') \approx \mu(t) + \dot{\mu}(t) (t'-t) + \rmO(t'-t)^2
 \;. 
\ee
Plugging this into \eq\nr{LR_2}, the first term yields
\ba
\int_{-\infty}^{\infty} \! {\rm d} t' \,
 C^{ }_\iR(t-t') 
 &  \stackrel{\rmi{\nr{CRw}}}{=}  & 
 \widetilde C^{ }_\iR(0) 
 \; \stackrel{\rmi{\nr{spectral}}}{=} \; 
 \widetilde C^{ }_\iE(0)
 \nn  
 & \stackrel{\rmi{\nr{bE}}}{=}  &
 \int_0^\beta \! {\rm d}\tau \langle \hat N(\tau) \hat N(0) \rangle  
 \quad {\approx} \quad 
 \beta \langle \hat N^2 \rangle 
 \; \stackrel{\rmi{\nr{susc_1}}}{=} \;
 \partial^{ }_\mu \langle \hat N \rangle^{ }_\rmi{eq} 
 \;, \hspace*{1cm} \la{moment_0} 
\ea
whereas the second term produces the integral 
\be
 \int_{-\infty}^{\infty} \! {\rm d} t' \,
 (t'-t) C^{ }_\iR(t-t')  
 \; \stackrel{\rmi{\nr{CRw}}}{=}  \; 
 i \partial^{ }_\omega \widetilde C^{ }_\iR(0) 
 \; \stackrel{\rmi{\nr{relation}}}{=} \; 
 - \partial^{ }_\omega  \rho(0)
 \;. \la{moment_1}
\ee
In \eq\nr{moment_0} we assumed that the equilibration rate  
is small (slow) compared with the temperature, $\Gamma\beta \ll 1$, 
so that
the correlator can be assumed constant on the Euclidean time interval; 
in \eq\nr{moment_1} we on the other hand
used the fact that the real part of $\widetilde C^{ }_\iR$
is even in $\omega$ (cf.\ \eq\nr{PiR_rho} together 
with \eq\nr{delta} and the antisymmetry of $\rho(\omega)$), 
such that its derivative vanishes at $\omega=0$, whereas
its imaginary part gives the spectral function.

Assembling everything together, we obtain up to first order in gradients, 
\be
  \langle \hat N(t) \rangle^{ }_\rmi{non-eq}
  \; \approx \;
   \partial^{ }_\mu \langle \hat N \rangle^{ }_\rmi{eq} \; \mu(t)
  -  \partial^{ }_\omega  \rho(0) \; \partial^{ }_t \mu(t) 
 + \rmO(\partial_t^2 \mu )
 \;. \la{simple_LR}
\ee
The first term amounts to an equilibrium fluctuation 
and exists even in the absence of any time dependence,
if a non-zero $\mu$ is inserted in order to displace
$\langle \hat N \rangle$ from zero. Moreover, 
it implies that 
$
 \partial^{ }_t \langle \hat N(t) \rangle^{ }_\rmi{non-eq}
  \approx
   \partial^{ }_\mu \langle \hat N \rangle^{ }_\rmi{eq}
 \; \partial^{ }_t \mu(t)
 + \rmO(\partial_t^2 \mu )
$.
If this information is inserted into the second term, we obtain
\be
  \langle \hat N(t) \rangle^{ }_\rmi{non-eq}
  \approx
   \partial^{ }_\mu \langle \hat N \rangle^{ }_\rmi{eq} \; \mu(t)
  -  \frac{\partial^{ }_\omega  \rho(0)\;}
          {\partial^{ }_\mu \langle \hat N \rangle^{ }_\rmi{eq}} 
     \partial^{ }_t \langle \hat N(t) \rangle^{ }_\rmi{non-eq}
 + \rmO(\partial_t^2 \langle \hat N(t) \rangle^{ }_\rmi{non-eq} )
 \;. 
\ee
Setting $\mu(t)\to 0$, so that the first term
on the right-hand side can be omitted; omitting higher derivatives; 
and comparing with \eq\nr{Gamma}, we finally identify 
\be
 \frac{1}{\Gamma} = \frac{\partial^{ }_\omega  \rho(0)\;}
          {\partial^{ }_\mu \langle \hat N \rangle^{ }_\rmi{eq}}
 \;, \la{Gamma_LR}
\ee
which is indeed 
in agreement with \eq\nr{generic_transport}. 

\subsubsection{Transport peak}

\index{Transport peak}
\index{Classical limit}

Returning to the first method, i.e.\ matching with the classical limit, 
we note that one can  extract more than just $\Gamma$ from 
the calculation. Indeed,
the Fourier transform of \eq\nr{Gamma_cl} implies that, 
for small frequencies, 
\ba
  \Delta^{ }_\rmi{cl}(\omega) & \simeq  & 
 \Delta^{ }_\rmi{cl}(0) 
 \biggl[ \int_{-\infty}^0 \! {\rm d}t \, e^{(i\omega + \Gamma) t}
 + \int_0^\infty \! {\rm d}t \, e^{(i\omega - \Gamma)t}
 \biggr]
 \nn  &  =  & \Delta^{ }_\rmi{cl}(0) 
 \frac{2 \Gamma}{\omega^2 + \Gamma^2}
 \;. \la{asympt2}
\ea
According to \eq\nr{bDelta}, the corresponding spectral function
reads (for $\omega\ll T$)
\ba
  \rho (\omega) & = & \Delta^{ }_\rmi{cl}(0)\, 
 \frac{\beta\omega \Gamma}{\omega^2 + \Gamma^2}
 \la{peak_full} \\ 
 & = &  \Delta^{ }_\rmi{cl}(0)\, \beta
 \im \frac{i\Gamma}{\omega+i\Gamma} 
 \;. \la{loren}
\ea 
This ``Lorentzian'' structure is referred to \index{Lorentzian shape}
as a {\em transport peak}.\footnote{%
 Classical physics can also yield corrections to the Lorentzian 
 shape, cf.\ e.g.\ ref.~\cite{corr_tr}. 
 } 

These equations contain kind of a ``paradox'', 
which becomes manifest in the free limit. In the free limit, 
the equilibration rate $\Gamma$ should 
vanish and therefore, according to \eq\nr{Gamma_LR}, 
the first derivative of the spectral function
at zero frequency appears to diverge. In contrast,  
according to what we saw for the free spectral functions of single 
particle states (cf.\ \eqs\nr{free_S_rho} and \nr{rho_F_free}), 
we would normally expect the spectral function to be {\em zero} 
at small frequencies ($|\omega| < m$).

The resolution to the paradox is to consider the free case
as a limit, $\Gamma\to 0^+_{ }$, whereby
\ba
 \frac{T  \rho(\omega)}{\omega} & = & 
 \Delta^{ }_\rmi{cl}(0) 
 \, \im \biggl( \frac{1}{\omega} - \frac{1}{\omega+i0^+_{ }} \biggr) 
 \; = \; 
 \Delta^{ }_\rmi{cl}(0) \,
 \pi \, \delta(\omega)
 \la{peak_free}
 \;.
\ea
This indeed vanishes at $\omega > 0$ but nevertheless possesses a 
certain structure. 
The distribution encountered is typical of 
a spectral function related to a {\em conserved charge}.
Namely, plugging this into \eq\nr{latt_relation}, we get 
\be
 \Bigl\langle \hat N(\tau)  \hat N (0) \Bigl\rangle
 \; = \;
 \int_{-\infty}^{\infty} \! \frac{{\rm d} \omega }{\pi} 
 {\rho(\omega)} \nB{}(\omega) e^{(\beta - \tau) \omega}
 \; = \; \Delta^{ }_\rmi{cl}(0)
 \;, \la{tau_indep}
\ee 
where we made use of $\nB{}(\omega) \approx T / \omega$ for 
$|\omega| \ll T$. 
This shows that \eq\nr{peak_free} corresponds to a Euclidean correlator
which is independent of~$\tau$. If the quantity is not conserved in 
the presence of interactions, then the transport peak gets a finite width.
If, however, the quantity is exactly
conserved even in the presence of interactions, the spectral function
retains an infinitely narrow transport peak like in \eq\nr{peak_free}, 
and no transport coefficient can be defined. 

Let us conclude the discussion concerning the transport peak 
with two observations: 
\bi

\item

As shown by \eq\nr{peak_full}, the height ($\Delta^{ }_\rmi{cl}(0)/\Gamma$)
and width ($\Gamma$) of the transport peak in $T \rho(\omega)/\omega$
are two independent quantities: $\Gamma$ can be extracted from the height
only if $\Delta^{ }_\rmi{cl}(0)$ is known from other considerations. This, 
of course, was also the content of \eq\nr{generic_transport}.

\item

In order to identify $\Gamma$
from the transport peak, we need to 
compute the spectral function in the regime $\omega \lsim \Gamma$.
This is in general challenging 
because $\Gamma$ is generated by interactions, 
and is therefore of the type $\Gamma \sim \alpha^2 T$,  
where $\alpha$ is a fine-structure constant, typically assumed 
small in perturbative calculations. 
To compute $\rho(\omega)$ correctly 
for soft energies $\omega\lsim \alpha^2 T$ requires 
extensive resummations~\cite{amy1}  
(cf.\ \se\ref{se:htl} or the paragraph
below \eq\nr{master_nonlin}). 

\ei


\subsubsection{Appendix A: Transport coefficients in QCD}

Here we briefly discuss
the transport coefficients most often encountered in QCD, 
namely flavour diffusion coefficients, the electric 
conductivity, as well as the shear and bulk viscosities. 

The transport coefficients of QCD are all related to conserved 
currents. In the absence of weak interactions and non-diagonal 
entries in the quark mass matrix, there is a separate conserved
current related to each flavour: 
$\partial^{ }_\mu \mathcal{J}_{\ff}^\mu = 0$, 
$\f = 1, ..., \Nf$. The sum of all 
flavour currents (divided by $\Nc$) defines the baryon current, 
whereas a particular linear combination, weighted by the electric 
charges of each flavour, defines the electromagnetic current
(denoted by $\mathcal{J}^\mu_\rmi{em}$). In addition, 
a conserved energy-momentum tensor can be defined:
$\partial^{ }_\mu T^{\mu\nu} = 0$. Physically, 
diffusion coefficients describe how 
inhomogeneities in flavour distributions flatten out, and shear and 
bulk viscosities how excesses in energy or momentum flow disappear.

Now, to {\em define} the transport coefficients requires 
a specification of the classical description
onto which to match. Let us first \index{Flavour diffusion coefficient}
consider the case of a diffusion coefficient, 
denoted by $D^{ }_{\ff}$.\footnote{%
 To be precise, in the case of several conserved charges the diffusion
 coefficients constitute a matrix, cf.\ e.g.\ ref.~\cite{amy1}. For 
 simplicity we consider a case here where the fluctuations of the 
 different flavours are decoupled from each other. Physically 
 this amounts to the omission of electromagnetic effects and so-called
 disconnected quark contractions. In the deconfined phase of QCD both
 are assumed to be small effects.  
 } 
The way it is defined is that, akin to the 
discussion following \eq\nr{eq_ass}, we assume that the 
system is slightly perturbed around the equilibrium state, and 
express 
$
 \mathcal{J}_{\ff}^\mu  = 
 \langle \hat \mathcal{J}_{\ff}^\mu \rangle^{ }_\rmi{non-eq}
$ 
in terms of a gradient expansion. The equilibrium state itself is
characterized by the temperature ($T$) and chemical potentials
of the conserved charges ($\mu^{ }_{\ff}$), as well as a four-velocity
defining the fluid rest frame ($u^\mu$; $u^\mu u^{ }_\mu = 1$). 
Given $T$ and $\mu^{ }_{\ff}$, 
variables such as pressure ($p$), energy density ($e$), and 
average particle number densities ($n^{ }_{\ff}$) can be defined
via standard relations. The gradient expansion acts on these
variables, whereas the free coefficients allowed by Lorentz 
symmetry are the transport coefficients 
(their definitions through subsequent orders of 
the gradient expansion are sometimes called 
``constitutive relations''). For instance, we can expand 
\be
 \mathcal{J}_{\ff}^\mu = n^{ }_{\ff} u^\mu + 
 D^{ }_{\ff} \partial^\mu_\perp n^{ }_{\ff} + \rmO(\partial^2) 
 \qquad \mbox{($\f$ fixed)}
 \;, \la{const_J}
\ee
where the transverse derivative has been defined as 
\be 
 \partial^\mu_\perp \equiv (\eta^{\mu\nu} - u^\mu u^\nu) \partial^{ }_\nu
 \;, \quad
 \eta^{\mu\nu} \equiv  \mathop{\mbox{diag}}\mbox{($+$$-$$-$$-$)}
 \;.  
\ee
A particular convention (called the Landau-Lifshitz convention) 
has been chosen whereby
in the rest frame of the fluid the zeroth component of 
$\mathcal{J}_{\ff}^\mu$ is  
the number density, $u^{ }_\mu \mathcal{J}_{\ff}^\mu \equiv n^{ }_{\ff}$,
to all orders in the gradient expansion. The coefficient $D^{ }_{\ff}$
is called the flavour diffusion coefficient.

Now, in analogy with the procedure leading 
to \eq\nr{generic_transport}, one way to 
determine $D^{ }_{\ff}$ is via matching: we need to find
suitable 2-point functions on the classical side that we 
can equate with the corresponding quantum objects. 
To achieve this, we may go to the fluid rest frame 
and impose current conservation on \eq\nr{const_J}, producing
\be
 \partial^{ }_t\, n^{ }_{\ff} =
 D^{ }_{\ff} \nabla^2 n^{ }_{\ff} + \rmO(\nabla^3)
 \;. \la{Di_eq}
\ee 
It is important to stress that even though \eq\nr{Di_eq}
evidently takes a non-relativistic form, 
the ``low-energy constant'' $D^{ }_{\ff}$
itself is defined also for relativistic flow; the corresponding
covariant form of the diffusion equation follows from \eq\nr{const_J}
together with $\partial^{ }_\mu \mathcal{J}_{\ff}^\mu = 0$.
  
In order to solve \eq\nr{Di_eq} on the classical side, 
we Fourier transform in space coordinates, 
$
 \tilde n^{ }_{\ff}(t,\vec{k}) \equiv 
 \int_{\vec{x}} e^{-i \vec{k}\cdot\vec{x}} n^{ }_{\ff}(t,\vec{x})
$, 
to trivially obtain 
$
 \tilde n^{ }_{\ff}(t,\vec{k}) =
 \tilde n^{ }_{\ff}(0,\vec{k}) \exp(- D^{ }_{\ff} \vec{k}^2 t) 
$.
If we then define a 2-point function 
(replacing $t\to |t|$), average over the initial conditions, 
and integrate over time like in \eq\nr{asympt2}, we obtain
\be
 \int_{-\infty}^{+\infty} \! {\rm d}t \, e^{i\omega t}
 \, \langle\!\langle \tilde n^{ }_{\ff} (t,\vec{k})
  \tilde n^{ }_{\ff}(0, - \vec{k}) 
 \rangle\!\rangle
 = \frac{2 D^{ }_{\ff} \vec{k}^2}{\omega^2 + D_{\ff}^2\vec{k}^4} 
 \, 
 \langle\!\langle \tilde n^{ }_{\ff}(0,\vec{k})
  \tilde n^{ }_{\ff}(0,-\vec{k})
 \rangle\!\rangle
 \;. \la{n_ff_corr}
\ee
Let us now choose $\vec{k}$ along one of the coordinate
axes, $\vec{k} = (0,0,k)$; then current
conservation, $\partial^{ }_\mu \mathcal{J}_{\ff}^\mu = 0$, allows us 
to re-express \eq\nr{n_ff_corr} as
\be
 \int_{-\infty}^{+\infty} \! {\rm d}t \, e^{i\omega t}
 \, \langle\!\langle 
  \tilde \mathcal{J}^3_{\ff} (t,\vec{k})
  \tilde \mathcal{J}^3_{\ff}(0, - \vec{k}) 
 \rangle\!\rangle
 = \frac{2 D^{ }_{\ff}\, \omega^2}{\omega^2 + D_{\ff}^2\vec{k}^4} 
 \, \langle\!\langle \tilde n^{ }_{\ff}(0,\vec{k})
  \tilde n^{ }_{\ff}(0,-\vec{k})
 \rangle\!\rangle
 \;. \la{J3_ff_corr}
\ee
Taking subsequently 
$\vec{k}\to 0$ and $\omega\to 0$, and
making use of translational invariance, we obtain from here
\be
 \frac{1}{3} \sum_i \int_{\mathcal{X}} 
 \langle\!\langle 
    \mathcal{J}^i_{\ff}(t,\vec{x}) 
    \mathcal{J}^i_{\ff}(0,\vec{0})
 \rangle\!\rangle
 = 2 D^{ }_{\ff} \int_\vec{x} \langle\!\langle 
 n^{ }_{\ff}(0,\vec{x}) n^{ }_{\ff}(0,\vec{0}) \rangle\!\rangle
 \;. 
\ee 
The left-hand side of this expression can be matched onto 
the zero-frequency limit of 
a spectral function like in \eq\nr{Gamma_diff_2}, whereas 
on the right-hand side we identify the classical limit
of a Euclidean susceptibility, to be denoted by $\chi^{ }_{\ff}$,\footnote{%
 Different conventions are frequently used with regard to the trivial 
 factor $\beta$ appearing in the second equality in \eq\nr{chi00}.
 If it is included in the definition of $\chi^{ }_{\ff}$
 like here (this is natural within the imaginary-time formalism; 
 cf.\ also \eq\nr{meff}),
 then $\chi^{ }_{\ff}$ has the dimensionality
 $T^2$. If rather the conventions of standard ``canonical'' 
 statistical physics 
 are followed, like in \eq\nr{chi_fe} or \nr{chi_fe_nlo}, then 
 $\chi^{ }_{\ff}$ has the dimensionality $T^3$.
 }
\be \index{Susceptibility}
 \chi^{ }_{\ff} \equiv \int_0^\beta \! {\rm d}\tau \int_{\vec{x}}
 \langle 
  \hat \mathcal{J}^0_{\ff}(\tau,\vec{x}) 
  \hat \mathcal{J}^0_{\ff}(0,\vec{0}) \rangle
 = \beta \int_{\vec{x}}
 \langle 
  \hat \mathcal{J}^0_{\ff}(0,\vec{x}) 
  \hat \mathcal{J}^0_{\ff}(0,\vec{0}) \rangle
 \;, \la{chi00}
\ee  
where we again made use of current conservation. Factors of 2 
as well as of $T$ nicely cancel out at this point, and we finally
obtain a Kubo relation for the flavour diffusion coefficient: 
\be \index{Matching: transport coefficients} \index{Diffusion coefficient} 
 D^{ }_{\ff} = 
 \fr1{3 \chi^{ }_{\ff}} 
 \lim_{\omega\to 0^+_{ } } 
 \sum_{i=1}^{3}
 \frac{\rho^{ii}_{\ff}(\omega,\vec{0})}{\omega}
 \;. \la{Kubo_D}
\ee
We note that, in analogy with \eq\nr{generic_transport}, 
two independent pieces of information are needed for determining
$D^{ }_{\ff}$: the Minkowskian spectral function 
of the spatial components of the current, and the Euclidean 
susceptibility related to the temporal component. 

Let us briefly elaborate on how the structure of \eq\nr{Kubo_D} 
relates to the considerations following \eq\nr{peak_free}. 
If we were to compute the spectral function related to the 
zeroth component of the current for $\vec{k}\to\vec{0}$, 
then we would get precisely
the behaviour in \eq\nr{peak_free}, because the charge 
$\int_{\vec{x}} \hat \mathcal{J}^0$ is exactly conserved even in 
an interacting theory. In contrast, there is no conservation
law related to $\int_{\vec{x}} \hat \mathcal{J}^i$, and 
an actual transport peak exists once interactions are present. 
Its width, let us call it $\eta^{ }_{D^{ }_\rmii{\sl f}}$, 
scales like $\Gamma$ before, 
i.e.\ $\eta^{ }_{D^{ }_\rmii{\sl f} }\sim \alpha^2 T$,
in the massless and weakly coupled limit; 
at the same time $D^{ }_{\ff}$, which plays 
a role  similar to $1/\Gamma$ in \eq\nr{generic_transport}, 
diverges like $1/(\alpha^2 T)$.\footnote{%
 For a concise review, see ref.~\cite{lgy}. 
 Actual expressions for $D^{ }_{\ff}$ in the massless limit are given in
 refs.~\cite{amy1,amy2}, whereas the case of a heavy flavour
 (with a mass $M \gg T$) has been discussed in ref.~\cite{mt}.
 }
Physically, this is because inhomogeneities even out extremely fast in 
a free theory, given that there are no collisions to stop the process.


\index{Electric conductivity}

\index{Conductivity}

We end by summarizing the Kubo formulae 
for some other physically relevant transport coefficients. Let us first 
discuss the {\em electric conductivity}, $\sigma$,
which is closely related to the 
flavour diffusion coefficients. 
It can be defined through
\be
 \langle\hat\mathcal{\vec{J}}^{ }_{em}\rangle = \sigma \vec{E}
 \;,
\ee
where $\vec{E}$ is an external electric field. 
Recalling the (classical) Maxwell equation 
$\nabla\times\vec{B} - \partial \vec{E}/\partial t = 
\langle\hat\vec{J}^{ }_{em}\rangle$, 
and assuming that the external $\vec{B}$ 
has been set to zero, we obtain in analogy with \eq\nr{diff_eq}
\be
 \frac{\partial \vec{E}}{\partial t} 
 =- \sigma \vec{E}
 \;.   \la{conductivity}
\ee

Now, a Kubo formula for $\sigma$ can be derived almost
trivially if we choose a convenient gauge (note that $\sigma$
as defined by \eq\nr{conductivity} is manifestly gauge
independent). 
In particular, let us choose a gauge in which $\partial^{ }_i A^0 = 0$;
then \eq\nr{conductivity} takes the form 
\be
 \partial_t^2 A^i = -\sigma \partial^{ }_t A^i
 \;, 
\ee 
reproducing the form of \eq\nr{eom_2} in the homogeneous and 
massless limit. Recalling also that $A^i$ couples to  
vector currents like $\varphi$ to $\mathcal{J}^{ }_\rmi{int}$
in \eq\nr{simp_1}, we can immediately write 
down an expression for $\sigma$ from 
\eqs\nr{Gamma_meff} and \nr{Kubo_D},
\be
  \sigma =  \sigma^{ }_d + e^2 
  \sum_{\ff=1}^{\Nf} Q_{\ff}^2 
  \chi^{ }_{\ff} 
  D^{ }_{\ff,\rmi{c}}
 \;, \la{sigma}
\ee
where $Q^{ }_{\ff}$ denotes the electric charge of flavour $\f$ in units of
the elementary charge $e$, and 
the subscript $(...)^{ }_\rmi{c}$
refers to 
``connected'' (or ``non-singlet'')  
quark contractions. The term $\sigma^{ }_d$
corresponds to a 
 ``disconnected'' 
or ``singlet'' 
contraction, in which quark lines are contracted back to the same
position $X$ from which the propagation started.

It is worth noting that, compared with
\eq\nr{Kubo_D}, no
susceptibility is needed for determining~$\sigma$. 
The formal reason for this difference is 
that, like with the example of \se\ref{se:field}, 
there are really two sets of fields, and the electric
conductivity encodes the influence of the ``hard modes'' (charged particles)
on the dynamics of the ``soft ones'' (electromagnetic fields). 
With the diffusion 
coefficient, in contrast, there is only one set of 
degrees of freedom, but ``soft modes'' can be 
generated through fluctuations as described by the 
susceptibility. 


\index{Viscosities}

The last quantities to be considered are 
the {\it shear} and {\it bulk viscosities}. 
They are defined through constitutive relations 
concerning the leading gradient corrections 
to the energy-momentum tensor:
the shear viscosity coefficient $\eta$ is defined
to be a function that multiplies its traceless part, while 
the bulk viscosity coefficient $\zeta$ multiplies 
the trace part. The explicit forms of the corresponding 
structures are most simply displayed in 
a non-relativistic frame, where $|u^i| \ll 1$; then 
\be
 T^{ }_{ij} \approx \Bigl( p - \zeta \nabla\cdot\vec{v} \Bigr)\delta^{ }_{ij} 
 - \eta \Bigl( \partial^{ }_i v^j + \partial^{ }_j v^i 
       - \fr23 \delta^{ }_{ij} \nabla\cdot\vec{v} \Bigr)
 + \rmO(\vec{v}^2,\nabla^2)
 \;, 
\ee
where $\nabla\cdot\vec{v} = \partial^{ }_i v^i$ and $\delta^{ }_{ij}$
is the usual Kronecker symbol.

Once again, Kubo relations for the transport coefficients 
can be derived in (at least) two different ways: through
a matching between quantum and classical 2-point functions, 
and through a linear response type computation. The former
approach amounts to solving (linearized) Navier-Stokes 
equations for various independent hydrodynamic modes.\footnote{%
  A review can be found in appendix C of ref.~\cite{dt}.
}
The latter approach on the other hand proceeds by
coupling the energy-momentum tensor to a source field which in 
this case is taken to be a metric perturbation, i.e.\
the latter part of $g^{\mu\nu} = \eta^{\mu\nu} + h^{\mu\nu}$.\footnote{%
 A concise discussion can be found in ref.~\cite{kas}, 
 while the general approach dates back to ref.~\cite{kubo}. 
 }
In the following we leave out all details and simply state
final expressions for the two viscosities: 
\ba
 \eta & = & 
 \lim_{\omega\to 0^+_{ }} 
 \biggl\{ 
 \frac{1}{\omega}
 \int_{\mathcal{X}} e^{i \omega t}
 \left\langle 
 \fr12 \left[ \hat T^{12}(\mathcal{X}), 
 \hat T^{12}(0) \right] 
 \right\rangle
 \biggr\} 
 \;, \la{res_eta} \\ 
 \zeta & = & \fr19 \sum_{i,j=1}^{3}
 \lim_{\omega\to 0^+_{ }} 
 \biggl\{ 
 \frac{1}{\omega}
 \int_{\mathcal{X}} e^{i \omega t}
 \left\langle 
 \fr12 \left[ \hat T^{ii}(\mathcal{X}), 
 \hat T^{jj}(0) \right] 
 \right\rangle
 \biggr\}
 \;. \la{res_zeta}
\ea
Note that in the case of $\zeta$, the operator could also be
replaced by the full trace $\hat T^{ii} - \hat T^{00}$, given that
the $\hat T^{00}$-part does not contribute because of energy 
conservation (it leads to an infinitely narrow transport peak
like in \eq\nr{peak_free}). 

\newpage 

\subsection{Equilibration rates / damping coefficients}
\la{se:decay}

\index{Equilibration rate}
\index{Damping coefficient}

In the previous section, we already discussed
an equilibration rate which we denoted by $\Gamma$, cf.\ \eq\nr{Gamma}. 
However, it appears that the formalism for its determination merits
further development, and this is the purpose of the present section. 
In short, we show that the use of operator equations of motion 
may simplify the structure of the 2-point correlator from 
which $\Gamma$ is to be extracted, 
thus streamlining its determination.\footnote{%
  A classic example of the use of this logic 
  comes from cosmology where, through the anomaly 
  equation, the rate of baryon number violation can be related to 
  the rate of Chern-Simons number diffusion~\cite{KhSh}
  (the latter is defined around \eq\nr{Gamma_diff_1}).
 }

\subsubsection*{General analysis} 

Like in the previous section, the idea is to start with 
an ``effective'' classical picture, whose free parameters are 
subsequently matched to reproduce quantum-mechanical correlators. 
Large deviations of physical quantities from their respective 
equilibrium values tend to decrease with time, with rates that we want 
to determine; however, small deviations can also be  generated 
by the occasional inverse reactions. This is formally the same physics 
as that of Brownian motion, just with the momentum of the test particle 
replaced with the deviation of our
generic ``charge density'' from its equilibrium value.\footnote{%
 The discussion here follows
 the description of heavy quark kinetic~\cite{ct,eucl} or
 chemical~\cite{chem} equilibration, 
 and more generally the theory of statistical fluctuations~\cite{landau9}.
 } 
In this context, it is good to note that a {\em density}, 
averaged over a large volume, is a continuous observable whose 
changes may be given a classical interpretation.

Mathematically, Brownian
motion can be described via a Langevin equation, 
\ba
 \delta \dot{N}_{}(t) 
 &  = &   
 -\Gamma \, \delta N_{}(t) + \xi (t) 
 \;, \la{Lan1} \\ 
 \langle\!\langle \, \xi (t) \, \xi (t') \, \rangle\!\rangle & = & 
  \Omega \, \delta(t-t')
 \;, \qquad
 \langle\!\langle \xi(t) \rangle\!\rangle = 0 
 \;, \la{Lan2}
\ea
where $\delta N_{}$ is the non-equilibrium excess
in our ``density'' observable; 
$\xi$ is a Gaussian stochastic noise, 
whose autocorrelation function is parametrized by the coefficient $\Omega$; 
and $\langle\!\langle ... \rangle\!\rangle$ denotes 
an average over the noise. This description can only be 
valid if the rate $\Gamma$ is much slower than that of typical reactions
in the plasma, implying $\Gamma \ll \alpha^2 T$.
Then $\Gamma$ originates as a sum of very many incoherent plasma
scatterings, guaranteeing the classical nature of the evolution. 

\index{Brownian motion}
\index{Langevin equation}
\index{Classical limit}

Given an initial value $\delta N(t^{ }_0)$, \eq\nr{Lan1} admits 
a straightforward explicit solution,
\be
 \delta N_{}(t) = \delta N_{}(t^{ }_0)\, e^{-\Gamma (t-t^{ }_0)}
 + 
 \int_{t^{ }_0}^t \! {\rm d}t' \, e^{\Gamma(t'-t)} \xi(t')
 \;. \la{Lan_soln} 
\ee
Making use of this expression and taking an average 
over the noise, we can determine the 2-point 
unequal time correlation function of the $\delta N_{}$ fluctuations:
\ba
 \Delta^{ }_\rmi{cl}(t,t')
 & \equiv & 
 \lim_{t_0 \to -\infty} 
 \langle\!\langle 
 \, \delta N_{} ( t) 
 \, \delta N_{} ( t') \, \rangle\!\rangle
 \nn & = & 
 \lim_{t^{ }_0 \to -\infty} 
 \int_{t^{ }_0}^t \! {\rm d}t^{ }_1 \, e^{\Gamma(t^{ }_1-t)} 
 \int_{t^{ }_0}^{t'} \! {\rm d}t^{ }_2 \, e^{\Gamma(t^{ }_2-t')} 
 \langle\!\langle \, \xi(t^{ }_1)
 \, \xi (t^{ }_2) \, \rangle\!\rangle
 \nn & = & 
 \Omega \lim_{t^{ }_0 \to -\infty} 
 \int_{t^{ }_0}^t \! {\rm d}t^{ }_1 \, e^{\Gamma(t^{ }_1-t)} 
 \int_{t^{ }_0}^{t'} \! {\rm d}t^{ }_2 \, e^{\Gamma(t^{ }_2-t')} 
 \delta(t^{ }_1-t^{ }_2)
 \nn & = & 
 \Omega \lim_{t^{ }_0 \to -\infty} 
 \int_{t^{ }_0}^t \! {\rm d}t^{ }_1 \,
 e^{\Gamma(2 t^{ }_1-t-t')}\, \theta(t'-t^{ }_1)
 \nn & = & 
 \frac{\Omega}{2\Gamma} \lim_{t^{ }_0 \to -\infty} 
 \Bigl[ \theta(t'-t)
 \Bigl( e^{\Gamma(t-t')} - e^{\Gamma(2t^{ }_0-t-t')} \Bigr) 
       + \theta(t-t')
 \Bigl( e^{\Gamma(t'-t)} - e^{\Gamma(2t^{ }_0-t-t')} \Bigr)
 \Bigr] 
 \nn & = & 
  \frac{\Omega}{2\Gamma} \,  e^{- \Gamma | t-t'| }
  \;. \la{Delta_c}
\ea
The limit $t^{ }_0\to -\infty$ guarantees that 
any initial transients have died out, making $\Delta^{ }_\rmi{cl}$
an {\em equilibrium} correlation function.\footnote{%
 In Brownian motion, when $\delta N$ corresponds to momentum, 
 we are also interested in the correlator of particle positions, 
 which are obtained as time integrals of the momentum. 
 This yields 
 $
  \int_0^\tau \! {\rm d}t \int_0^\tau \! {\rm d}t' \, 
  \Delta^{ }_\rmi{cl}(t,t') 
  = \frac{\Omega}{\Gamma^2}
  [
    \tau +  ( e^{-\Gamma \tau} - 1 )/\Gamma 
  ]
 $. 
 The coefficient of the linear growth is proportional to $2D$, 
 where $D$ is the diffusion coefficient.  \index{Diffusion coefficient} 
 } 
Subsequently, making use of 
$
 \partial^{ }_t |t-t'| = \theta(t-t') - \theta(t'-t)
$, 
$
 \partial^{ }_{t'} |t-t'| = \theta(t'-t) - \theta(t-t')
$, and
$
 \partial^{ }_t \partial^{ }_{t'} |t-t'| = -2 \delta(t-t') 
$, 
we obtain
\be
 \partial^{ }_t \partial^{ }_{t'} \Delta^{ }_\rmi{cl}(t,t')
 =  - \frac{\Omega \Gamma}{2}\,  e^{- \Gamma | t-t'| }
 + \Omega\, \delta(t-t')
 \;. \la{ff_c} 
\ee
Fourier transforming
\eqs\nr{Delta_c} and \nr{ff_c} leads to\footnote{%
 In the latter case one can literally Fourier-transform \eq\nr{ff_c}, or 
 carry out partial integrations, whereby the result can be extracted 
 from \eq\nr{F_Delta_c}.
 }  
\ba
 \widetilde \Delta^{ }_\rmi{cl}(\omega) 
 & \equiv &  \int_{-\infty}^{\infty} \! {\rm d}t \, e^{i \omega (t-t')} 
 \Delta^{ }_\rmi{cl}(t,t')
  =  \frac{\Omega}{2\Gamma} \biggl[ 
  \int_0^\infty \! {\rm d}t \, e^{(i\omega - \Gamma)t} + 
  \int_{-\infty}^{0} \! {\rm d}t \, e^{(i\omega + \Gamma)t}
 \biggr]
 \nn & = & 
 \frac{\Omega}{ \omega^2 + \Gamma^2}
 \;, \la{F_Delta_c} \\ 
 \omega^2 \widetilde \Delta^{ }_\rmi{cl}(\omega) 
 & = &  \int_{-\infty}^{\infty} \! {\rm d}t \, e^{i \omega (t-t')} 
 \partial^{ }_t \partial^{ }_{t'} \Delta^{ }_\rmi{cl}(t,t')
 \nn & = &  
 \frac{\Omega\, \omega^2}{ \omega^2 + \Gamma^2}
 \;. \la{F_ff_c} 
\ea
It is also useful to note that, setting the time arguments equal, 
we can define a susceptibility as
\be \index{Susceptibility}
 \langle (\delta N_{})^2 \rangle^{ }_\rmi{cl}
 \equiv 
 \lim_{t^{ }_0 \to -\infty} 
 \langle\!\langle \, \delta N_{} ( t) 
 \, \delta N_{} ( t) \, \rangle\!\rangle 
 = \frac{\Omega}{2\Gamma}
 \;, \la{susc_c_def}
\ee
where we made use of \eq\nr{Delta_c}. 

Combining \eqs\nr{F_Delta_c}--\nr{susc_c_def}, various strategies
can be envisaged for determining the quantity that we are interested in, 
namely the equilibration rate $\Gamma$. 
One formally correct track would be 
to note from \eq\nr{F_Delta_c} that 
$
 \widetilde \Delta^{ }_\rmi{cl}(0) = \Omega / \Gamma^2
$, 
and to combine this with \eq\nr{susc_c_def}, in order to obtain 
\be
 \Gamma = \frac{2 \langle (\delta N_{})^2 \rangle^{ }_\rmi{cl}}
 {\widetilde \Delta^{ }_\rmi{cl}(0)}
 \;. 
\ee
This is equivalent to our previous approach, \eq\nr{Delta_cl}.
However, as discussed in connection with the transport peak 
(cf.~paragraphs around \eq\nr{tau_indep}), in practice it is difficult to 
determine $\Gamma$ from this relation, because the relevant 
information resides 
in the denominator of $\widetilde \Delta^{ }_\rmi{cl}(\omega)$ 
and we would need to 
evaluate this function at extremely ``soft'' values of $\omega$. 

Taking, in contrast, 
\eqs\nr{F_ff_c} and \nr{susc_c_def} 
as starting points, we obtain the alternative expressions
\ba
 \Omega & = & 
 \lim_{\Gamma \ll \omega \ll \omega^{ }_\rmii{UV}} 
 \omega^2 \widetilde \Delta^{ }_\rmi{cl}(\omega) 
 \;, \la{get_Lambda_c} \\ 
 \Gamma  & = & 
 \frac{\Omega}{2 \langle (\delta N_{})^2 \rangle^{ }_\rmi{cl}}
 \;. \la{get_Gamma_c} \la{Gam_Om}
\ea
Here $\omega^{ }_\rmii{UV}$ is a frequency scale around which 
physics beyond the classical picture 
sets in, $\omega^{ }_\rmii{UV} \gsim \alpha^2 T$. 
At the same time, it has been tacitly assumed that 
{\em $\Gamma$ is parametrically small
compared with $\omega^{ }_\rmii{UV}$}. This is the case if, for instance, 
$\Gamma$ is inversely proportional to a heavy 
mass scale $M \gg T$, or proportional to a very weak coupling constant 
which plays
no role in the dynamics of the heat bath. With these reservations, 
the information needed is now in the numerator
of $\widetilde \Delta^{ }_\rmi{cl}(\omega)$. Thus, 
if a hierarchy between $\Gamma$ and $\omega^{ }_\rmii{UV}$
can be identified and an 
error suppressed by $\Gamma / \omega^{ }_\rmii{UV}$ is tolerable, 
transport coefficients are most easily determined from 
``force--force'' correlation functions 
(i.e.\ those of $\delta \dot{N}_{}$, cf.\ 
\eqs\nr{ff_c} and \nr{F_ff_c}), rather 
than from ``momentum--momentum'' 
correlation functions (i.e.\ those of $\delta N_{}$, 
cf.\ \eqs\nr{Delta_c} and \nr{F_Delta_c}).

After these preparatory steps, we can promote the determination 
of $\Gamma$ to the quantum level. It just remains
to note that since 
observables commute in the classical limit, a suitable 
quantum version of the ``momentum--momentum'' correlator considered is 
\be
 \Delta^{ }_\rmi{qm}(t,t') \equiv  \Bigl\langle \fr12 
 \bigl\{ \delta \hat{N}_{}(t), \delta \hat{N}_{}(t') \bigr\} \Bigr\rangle
 \;. \la{Delta_q} 
\ee
With this convention, \eqs\nr{get_Lambda_c} and \nr{get_Gamma_c} 
can be rephrased as
\ba \index{Matching: equilibration rate}
 \Omega & = & 
 \lim_{\Gamma \ll \omega \ll \omega^{ }_\rmii{UV}} 
  \int_{-\infty}^{\infty} \! {\rm d}t \, e^{i \omega (t-t')} 
  \biggl\langle \fr12 
  \Bigl\{ \frac{{\rm d}  \hat{N}_{}(t)}{{\rm d}t}, 
  \frac{{\rm d} \hat{N}_{}(t')}{{\rm d}t'} 
  \Bigr\} 
  \biggr\rangle^{ }_\rmi{qm}
  \;, \la{get_Lambda_q} \\ 
 \Gamma  & = & 
 \frac{\Omega}{2 \langle (\delta \hat{N}_{})^2 \rangle^{ }_\rmi{qm}}
 \;, \la{get_Gamma_q}
\ea
where the susceptibility in the denominator of the latter equation is nothing 
but the variance of the number density operator, 
$
 \langle (\delta \hat{N}_{})^2 \rangle 
 = \langle \hat{N}_{}^2 \rangle - \langle \hat{N}_{} \rangle^2 
$.

The formulae introduced above can be applied on a non-perturbative
level as well, if we re-express them in the 
imaginary-time formalism. This means that we first define
a Euclidean correlator, $\Omega(\tau)$, like in \eq\nr{bE}; 
Fourier-transform it, 
$
   \Omega (\omega^{ }_n) = 
  \int_0^\beta \! {\rm d}\tau \, e^{i \omega^{ }_n\tau } \Omega (\tau)
$, 
where $\omega^{ }_n = 2\pi n T$, $n \in \mathbbm{Z}$; and 
obtain the spectral function from its imaginary part, 
$
  \rho (\omega) = 
  \im  \Omega (\omega^{ }_n \to -i [\omega + i 0^+_{ }])
$, 
cf.\ \eq\nr{Discdef}.
The symmetric combination needed in \eq\nr{get_Lambda_q} 
is subsequently given by 
$
 {2 T \rho(\omega)} / {\omega}
$, 
where we assumed $\omega \ll T$, 
cf.\ \eq\nr{bDelta}.

%
\subsubsection*{Example}

As an example of the use of \eqs\nr{get_Lambda_q}
and \nr{get_Gamma_q}, 
let us consider the Lagrangian 
of \eq\nr{L_c_sc},
\be
 \mathcal{L}^{ }_\iM = \partial^{ }_\mu \phi^* \partial^\mu \phi
 - m^2 \phi^*\phi
 - h \, \phi^* \mathcal{J} - h^*   \mathcal{J}^* \phi + 
 \mathcal{L}^{ }_\rmi{bath}
 \;. \la{L_c_sc_2}
\ee
For $h=0$, the system has a conserved 
current (cf.\ \eq\nr{sft_charge}), 
\be
 \mathcal{J}^{ }_\mu = 
 - i \bigl(\partial^{ }_\mu\phi^* \, \phi
  - \phi^* \, \partial^{ }_\mu \phi \bigr)
 \;.
\ee
If the composite object $\mathcal{J}$ of \eq\nr{L_c_sc_2} does 
{\em not} transform under the associated symmetry, 
\eq\nr{phi_sym}, 
then the coupling $h$ 
mediates transitions through which current conservation is violated. 

\index{Chemical equilibration rate} 

In \eq\nr{tot_dN} of \se\ref{se:ppr}, 
the rate at which particles and antiparticles are {\em produced}
from a plasma was obtained for this model; both rates are furthermore 
equal if the plasma is CP-symmetric. In the present section, 
an initial state is considered in which the number densities of 
particles and antiparticles are almost in thermal equilibrium but
not quite, being slightly different. Given that 
the reactions mediated by $h$ 
violate the conservation of the net 
particle {\em minus} antiparticle density, there is
a rate by which the difference 
of the two number densities evens out, returning the system
to full thermal equilibrium. This can be called a
{\em chemical equilibration rate}. 

Before proceeding with the calculation of 
the chemical equilibration rate, 
let us write down the canonically quantized
number density operator. At this point operator ordering plays 
a role: if we choose the explicitly Hermitean ordering, 
\be
 \hat N \equiv 
 - i \int_{\vec{x}} \bigl( 
 \partial_0^{ } \hat \phi^\dagger \, \hat \phi - 
  \hat\phi^\dagger \partial_0^{ } \hat\phi
 \bigr)
 \;, \la{hatN_sc_def}
\ee
and represent the fields in terms of the canonically normalized 
creation and annihilation operators like in \eq\nr{sc_wave}, then 
\be
 \hat N = \int \! {\rm d}^3\vec{p} \, 
 \bigl( 
   \hat a_{\vec{p}}^\dagger
   \hat a^{ }_{\vec{p}}
  - 
   \hat b^{ }_{\vec{p}}
   \hat b_{\vec{p}}^\dagger
 \bigr)
 \;. 
\ee
We note that this expression is not automatically ``normal-ordered''; 
rather, it contains an infinite constant from the latter term,
amounting to 
$\int {\rm d}^3\vec{p} \; \delta^{(3)}(\vec{0}) = \int_{\vec{p}} V$, 
where the Dirac-$\delta$ at vanishing momentum was regulated
through finite volume,  
$
 (2\pi)^3\delta^{(3)}(\vec{p} = \vec{0}) = \int_V \! {\rm d}^3\vec{x} = V
$. 
This infinity  could be hidden by tuning
the orderings in \eq\nr{hatN_sc_def}. However, 
the infinite constant drops out in the time derivative of $\hat N$, 
so we stick to \eq\nr{hatN_sc_def} in the following. 

Making next use of the operator equations of motion corresponding
to \eq\nr{L_c_sc_2}, 
\ba
 \partial_0^2 \hat\phi & = & 
 \nabla^2 \hat\phi - m^2 \hat\phi - h \, \hat{\mathcal{J}}
 \;, \\
 \partial_0^2 \hat\phi^\dagger & = & 
 \nabla^2 \hat\phi^\dagger - m^2 \hat\phi^\dagger 
  - h^*  \hat{\mathcal{J}}^\dagger
 \;,
\ea
and omitting boundary terms, the time derivative of \eq\nr{hatN_sc_def} yields
\be
 \partial_0^{ } \hat N 
 = - i \int_\vec{x} \bigl( 
   h \hat\phi^\dagger \hat{\mathcal{J}} - 
   h^* \hat{\mathcal{J}}^\dagger \hat\phi
 \bigr) 
 \;.
\ee
Then, according to the discussion above, we obtain to first order in $|h|^2$
\ba
 \Omega & = & \lim_{\Gamma \ll \omega \ll \omega^{ }_\rmii{UV}} 
 \frac{2 T \rho(\omega)}{\omega}
 \;, \la{Om_intercept}
\ea
where $\rho$ is the spectral function corresponding to the 
operator $\partial_0^{ } \hat N$.

Let us now define the 
Euclidean correlator corresponding 
to $\partial_0^{ } \hat N$:
\ba
 \Omega^{ }_\iE(\tau) & \equiv & 
 - \int_{\vec{x},\vec{y}}
 \Bigl\langle
   \bigl[ h\phi^* \mathcal{J} - h^* \mathcal{J}^* \phi \bigr]
   (\tau,\vec{x}) \, 
   \bigl[ h\phi^* \mathcal{J} - h^* \mathcal{J}^* \phi \bigr]
   (0,\vec{y}) 
 \Bigr\rangle 
 \nn & = & 
 |h|^2 \int_{\vec{x},\vec{y}}
 \Bigl[
   \langle \phi^*(X) \phi(Y) \rangle \;
   \langle \mathcal{J}(X) \mathcal{J}^* (Y) \rangle
 +  
   \langle \phi(X) \phi^*(Y) \rangle \;
   \langle \mathcal{J}^*(X) \mathcal{J} (Y) \rangle
 \Bigr]
 \;, \hspace*{5mm} \la{Omega_tau}
\ea
where we defined
$
 {X\equiv (\tau,\vec{x}), Y \equiv (0,\vec{y})}
$
and only considered contractions allowed by the U(1) invariance
of free $\phi$-particles.
Hats have been left out from these expressions 
because Euclidean correlators can 
be evaluated with path integral techniques. 

We wish to make as few assumptions about 
the operator $\mathcal{J}$ as possible, 
and simply express its 2-point correlation function in a general 
spectral representation. Inverting \eq\nr{bE} we can write 
\be
 \bigl\langle \mathcal{J}(X) \, \mathcal{J}^*(Y) \bigr\rangle
  = 
 \Tint{K} e^{-i K\cdot (X-Y)} \; \Pi^{ }_\iE(K)
 \;.
\ee
Inserting here \eq\nr{spectral}, i.e.\ 
\be
 \Pi^{ }_\iE(K) = \int_{-\infty}^{\infty} \! \frac{{\rm d}\omega}{\pi}
 \frac{\rho(\omega,\vec{k})}{\omega - i k^{ }_n}
 \;, 
\ee
as well as the sum in \eq\nr{bsum2}, i.e.\ 
\be
 T\sum_{k^{ }_n} \frac{e^{-i k^{ }_n\tau}}{\omega - i k^{ }_n} 
 = \nB{}(\omega) e^{(\beta-\tau)\omega}
 \;, \quad \mbox{for} \; 0 < \tau  < \beta
 \;, \la{sum_x}
\ee
and substituting $\omega\to k^0_{ }$, we obtain 
\be
 \bigl\langle \mathcal{J}(X) \, \mathcal{J}^*(Y) \bigr\rangle
 = 
 2 \int_{\mathcal{K}} 
 \, \rho(\mathcal{K})
 \, \nB{}(k^0_{ }) 
 \, e^{(\beta-\tau) k^0_{ }
 + i \vec{k}\cdot(\vec{x}-\vec{y})}
 \;.  \la{K_prop_1}
\ee
The other case is handled similarly: 
making use of \eq\nr{bsum} 
and renaming subsequently
$
 \omega\to -k^0_{ }
$
and
$
 \vec{k}\to -\vec{k}
$,
we obtain 
\ba
 \bigl\langle \mathcal{J}(Y) \, \mathcal{J}^*(X) \bigr\rangle
 & = & 
 \int_\vec{k} \int\! \frac{{\rm d}\omega}{\pi}
 \rho(\omega,\vec{k}) \, \nB{}(\omega)
 e^{\tau\omega + i \vec{k}\cdot(\vec{y}-\vec{x})}
 \nn & = &  
 - 2 \int_{\mathcal{K}}
 \, \rho(-\mathcal{K}) 
 \, \nB{}(k^0_{ })
 \, e^{(\beta-\tau) k^0_{ }
 +i \vec{k}\cdot(\vec{x}-\vec{y})}
 \;,  \la{K_prop_2}
\ea
where we also made use of
$
 \nB{}(-k^0_{ }) = - e^{\beta k^0_{ }} \nB{}(k^0_{ })
$. 

As far as the scalar propagators are concerned, they can be replaced
with their tree-level forms, given that the fields have been 
assumed to be weakly interacting and we work to leading order 
in $|h|^2$. If we furthermore assume that the scalar particles are
close to equilibrium, 
which is consistent with the linear response nature of our computation, then
\ba
 \langle \phi(X) \phi^*(Y) \rangle 
 = 
 \langle \phi(Y) \phi^*(X) \rangle 
 & = & 
 \Tint{P} \frac{e^{i P\cdot(X-Y)}}{p_n^2 + \E_p^2}
 \nn 
 & = & 
 \int_\vec{p} \frac{\nB{}(\E^{ }_p)}{2\E^{ }_p}
 \Bigl[ e^{(\beta-\tau)\E^{ }_p} + e^{\tau \E^{ }_p} \Bigr] 
 e^{ - i \vec{p}\cdot(\vec{x}-\vec{y})}
 \;, \la{phi_prop_0}
\ea
where we made use of \eq\nr{Gebo_new}.

Inserting \eqs\nr{K_prop_1}--\nr{phi_prop_0} into \eq\nr{Omega_tau}, 
and carrying out the integrals over $\vec{x},\vec{y}$, and $\vec{p}$, 
yields
\be
 \Omega^{ }_\iE(\tau) = 
 |h|^2 V \int_{\mathcal{K}} \frac{\nB{}(\E^{ }_k) \nB{}(k^0_{ })}{\E^{ }_k}
 \Bigl[ e^{(\beta-\tau)\E^{ }_k} + e^{\tau \E^{ }_k}  \Bigr] 
 e^{(\beta-\tau)k^0_{ }}
 \Bigl[ \rho(\mathcal{K}) - \rho(-\mathcal{K}) \Bigr]
 \;. 
\ee
The Euclidean Fourier transform (cf.\ \eq\nr{bE}) turns this into
\ba
  \Omega^{ }_\iE(\omega^{ }_n) & = & 
 \int_0^\beta \! {\rm d}\tau
 \, e^{i \omega^{ }_n \tau} \, \Omega^{ }_\iE(\tau) 
 \nn & = & 
 |h|^2 V \int_{\mathcal{K}} \frac{\nB{}(\E^{ }_k) \nB{}(k^0_{ })}{\E^{ }_k}
 \Bigl[ \rho(\mathcal{K}) - \rho(-\mathcal{K}) \Bigr]
 \biggl[ 
   \frac{1-e^{\beta(\E^{ }_k + k^0_{ })}}{i\omega^{ }_n - \E^{ }_k - k^0_{ }}
  + \frac{e^{\beta \E^{ }_k} - e^{\beta k^0_{ }}}
  {i\omega^{ }_n + \E^{ }_k - k^0_{ }}
 \biggr]
 \;, \nn 
\ea
and the corresponding spectral function reads (cf.\ \eq\nr{Discdef})
\ba
 \rho(\omega) & = &
 \im   \Omega^{ }_\iE(\omega^{ }_n\to -i[\omega + i 0^+_{ }])
 \nn 
 & = & 
 \pi |h|^2 V \int_{\mathcal{K}} 
 \frac{ \rho(\mathcal{K}) - \rho(-\mathcal{K}) }{\E^{ }_k}
 \biggl\{ 
   \delta(\omega - \E^{ }_k - k^0_{ }) 
     \Bigl[ 1 + \nB{}(\E^{ }_k) + \nB{}(k^0_{ }) \Bigr]
 \nn & & \hspace*{3.7cm} 
  + \;
    \delta(\omega + \E^{ }_k - k^0_{ }) 
     \Bigl[ \nB{}(\E^{ }_k) - \nB{}(k^0_{ }) \Bigr]
 \biggr\}
 \;,
\ea
where we applied \eq\nr{delta} as well as relations satisfied 
by the Bose distribution.

To extract the limit needed in \eq\nr{Om_intercept}, we note that
\ba
   \delta(\omega - \E^{ }_k - k^0_{ }) 
     \Bigl[ 1 + \nB{}(\E^{ }_k) + \nB{}(k^0_{ }) \Bigr]
   & = & 
   \delta(\omega - \E^{ }_k - k^0_{ }) 
     \Bigl[ \nB{}(\E^{ }_k) - \nB{}(\E^{ }_k - \omega) \Bigr]
   \nn & = & 
   \delta(\omega - \E^{ }_k - k^0_{ }) 
     \Bigl[ \omega \, \nB{}'(\E^{ }_k) + \rmO(\omega^2) \Bigr]
   \nn & = & 
   - \delta(\E^{ }_k + k^0_{ }) \, \beta\, \omega \, 
   \nB{}(\E^{ }_k) [1 + \nB{}(\E^{ }_k)] + \rmO(\omega^2)
   \;, \hspace*{5mm} \nn \\ 
    \delta(\omega + \E^{ }_k - k^0_{ }) 
     \Bigl[ \nB{}(\E^{ }_k) - \nB{}(k^0_{ }) \Bigr]
   & = & 
    \delta(\omega + \E^{ }_k - k^0_{ }) 
     \Bigl[ \nB{}(\E^{ }_k) - \nB{}(\E^{ }_k + \omega) \Bigr]
   \nn & = & 
   \delta(\omega + \E^{ }_k - k^0_{ }) 
     \Bigl[ - \omega \, \nB{}'(\E^{ }_k) + \rmO(\omega^2) \Bigr]
   \nn & = & 
   \delta(\E^{ }_k - k^0_{ }) \, \beta\, \omega \, 
   \nB{}(\E^{ }_k) [1 + \nB{}(\E^{ }_k)] + \rmO(\omega^2)
    \;, \nn
\ea
where we made use of 
$
 \nB{}(-E) = -1 - \nB{}(E)
$ as well as 
$
 \nB{}'(E) = - \beta \nB{}(E) [1 + \nB{}(E)]
$.
Therefore, for small $\omega$,
\ba
 \frac{2T\rho(\omega)}{\omega} 
 & \approx & 
 2 \pi |h|^2 V \int_{\mathcal{K}} 
 \frac{ \rho(\mathcal{K}) - \rho(-\mathcal{K}) }{\E^{ }_k}
 \Bigl[ 
   \delta(\E^{ }_k - k^0_{ }) 
   - \delta(\E^{ }_k + k^0_{ })
 \Bigr] 
 \, \nB{}(\E^{ }_k) [1 + \nB{}(\E^{ }_k)] 
 \nn 
 & = & 
 2 |h|^2 V \int_\vec{k} 
 \frac{ \rho(\mathcal{K}) - \rho(-\mathcal{K}) }{\E^{ }_k}
 \, \nB{}(\E^{ }_k) [1 + \nB{}(\E^{ }_k)] 
 \;,
\ea
where in the second step we substituted $\vec{k}\to-\vec{k}$
in some of the terms, and also implicitly changed the notation: 
from now on $\mathcal{K}$ denotes an on-shell four-vector, 
$\mathcal{K} \equiv (\E^{ }_k, \vec{k})$. 

In order to apply \eq\nr{get_Gamma_q}, we also need the susceptibility. 
This can most easily be extracted from \eq\nr{f_SFT_mu}, which gives 
the grand canonical free energy density
for a complex scalar field. 
In the thermodynamic limit, 
the susceptibility
is obtained from the second partial derivative of this quantity 
with respect to $\mu$, 
evaluated at $\mu = 0$ (cf.\ \eq\nr{chi_bo}):
\ba
  \langle (\delta \hat N)^2 \rangle & = & 
  T^2 \partial_\mu^2 \left.  \ln\mathcal{Z} \right|^{ }_{\mu = 0}
  = -V T \partial_\mu^2 \left.  f(T,\mu) \right|^{ }_{\mu = 0}
 \nn  & = &
 -V T \partial_\mu^2 
 \int_\vec{p} \Bigl\{ 
  \E^{ }_p + T \Bigl[       
     \ln \Bigl( 1 - e^{-\beta({\E^{ }_p+\mu})}\Bigr) 
   + \ln \Bigl( 1 - e^{-\beta({\E^{ }_p-\mu})}\Bigr) 
  \Bigr]  
  \Bigr\}^{ }_{\mu = 0}
 \nn  & = &
 -V T \partial^{ }_\mu
 \int_\vec{p} \biggl\{ 
   \frac{1}{e^{\beta({\E^{ }_p+\mu})} - 1} - 
   \frac{1}{e^{\beta({\E^{ }_p-\mu})} - 1} 
  \biggr\}^{ }_{\mu = 0}
 \nn & = & 
 2 V \int_{\vec{p}} \nB{}(\E^{ }_p) \bigl[  1 + \nB{}(\E^{ }_p) \bigr]
 \;. \hspace*{5mm}
\ea
Putting everything together, and making use of the relation
$
 \rho(-\mathcal{K}) = - \rho(\mathcal{K})
$,
valid for a CP-symmetric plasma, \eq\nr{get_Gamma_q} yields
\be
 \Gamma
 \; = \;  
 \frac{ |h|^2 \int_\vec{k} \displaystyle 
 \frac{\rho(\mathcal{K})}{\E^{ }_k}
 \, \nB{}(\E^{ }_k) [1 + \nB{}(\E^{ }_k)] } 
 { \mbox{\raise-1ex\hbox{$\int_\vec{k} \displaystyle 
 \nB{}(\E^{ }_k) [1 + \nB{}(\E^{ }_k)]$} } }  
 \;. \la{Gamma_final}
\ee

We conclude with a discussion concerning 
the physical interpretation of \eq\nr{Gamma_final}. 
According to \eqs\nr{master} and \nr{master_b}, 
$
 |h|^2 \frac{\rho(\mathcal{K})}{\E^{ }_k}
 \nB{}(\E^{ }_k)
$
gives the production rate of particles or antiparticles
of momentum $\vec{k}$; the appearance of $\nB{}(\E^{ }_k)$ indicates
that the production necessitates the presence of a plasma in which
collisions take place, because the energy $\E^{ }_k$ 
needs to be extracted from thermal fluctuations. 
In the chemical equilibration rate, in contrast, we could think of 
$
 |h|^2 \frac{\rho(\mathcal{K})}{\E^{ }_k} [1 + \nB{}(\E^{ }_k)]
$
as the rate at which particles {\em or} antiparticles decay
if they were initially in excess (cf.\ \eq\nr{x_dNa}); only 
one of these processes needs to take place if there is
an imbalance.
Because we posed a question about 
the ``chemical'' rather than the ``kinetic'' equilibration of the 
system, \eq\nr{Gamma_final} contains
an average over $\vec{k}$ of the decay rate, weighted by 
the kinetically equilibrated 
momentum distribution given by $\nB{}(\E^{ }_k)$. 


\subsubsection{Appendix A: Relation of Langevin and Fokker-Planck equations}

\index{Fokker-Planck equation}

We have seen that the dynamics following from the Langevin equation, 
\eqs\nr{Lan1} and \nr{Lan2}, is simple enough to be exactly solvable. 
Nevertheless,
there are circumstances where it may be advantageous to reformulate
the (stochastic) Langevin equation as a deterministic (non-stochastic) 
equation, for the corresponding probability density. The latter 
is known as the Fokker-Planck equation. 

Let us start by generalizing $\delta N^{ }_{ }$ to a multi-component
variable, and \eq\nr{Lan_soln} to a corresponding solution, 
$
 \delta N^{ }_{i}(t) = \delta N^{ }_{i}(t^{ }_0)\, e^{-\Gamma (t-t^{ }_0)}
 + 
 \int_{t^{ }_0}^t \! {\rm d}t'_i \, e^{\Gamma(t'_i-t)} \xi^{ }_i(t'_i)
$,
where 
$
 \langle\!\langle \, \xi^{ }_i (t'_i) \,
                     \xi^{ }_j (t'_j) \, \rangle\!\rangle  =  
 \Omega\, \delta^{ }_{ij} \delta(t'_i - t'_j)
$.
Consider now the $n$-point function
\be
 G^{ }_{i^{ }_1 \cdots i^{ }_n}(t^{ }_1,...,t^{ }_n) 
 \; \equiv \;  
 \langle\!\langle \,
   \delta N^{ }_{i^{ }_1}(t^{ }_1) \cdots
   \delta N^{ }_{i^{ }_n}(t^{ }_n)
 \, \rangle\!\rangle
 \;. \la{G_n}
\ee 
A time derivative yields
\be
 \partial^{ }_{\tilde t} \,
 G^{ }_{i^{ }_1 i^{ }_2 \cdots i^{ }_n}(\tilde{t},t,...,t) 
 \; =  \;  
 \langle\!\langle \,
   [-\Gamma \delta N^{ }_{i^{ }_1}(\tilde{t})
  + \xi^{ }_{i^{ }_1}(\tilde{t})]\,
   \delta N^{ }_{i^{ }_2}(t^{ }) \cdots
   \delta N^{ }_{i^{ }_n}(t^{ })
 \, \rangle\!\rangle
 \;. \la{dG_n_1}
\ee
The first term is proportional to 
$ G^{ }_{i^{ }_1 i^{ }_2 \cdots i^{ }_n} $, whereas
the second term requires knowledge about $n$-point correlators of the noise.
Let us assume that the noise is ``white'', i.e.\ Gaussian. Then the noise
correlator factorizes, and we get
\ba
 \langle\!\langle \,
   \xi^{ }_{i^{ }_1}(\tilde{t})\,
   \delta N^{ }_{i^{ }_2}(t^{ }) \cdots
   \delta N^{ }_{i^{ }_n}(t^{ })
 \, \rangle\!\rangle
 & = & 
 \sum_{j=2}^n
 \langle\!\langle \,
  \xi^{ }_{i^{ }_1}(\tilde{t})\, 
  \delta N^{ }_{i^{ }_j}(t)
 \, \rangle\!\rangle \, 
 \frac{\partial
 \langle\!\langle \,
   \delta N^{ }_{i^{ }_2}(t^{ }) \cdots
   \delta N^{ }_{i^{ }_n}(t^{ })
 \, \rangle\!\rangle
 }{\partial \delta N^{ }_{i^{ }_j}(t)}
 \nn 
 & = & 
 \sum_{j=2}^n
 \langle\!\langle \,
  \xi^{ }_{i^{ }_1}(\tilde{t})\, \int_{t^{ }_0}^t \! {\rm d}t'_j \, 
  e^{\Gamma(t'_j - t)}\, \xi^{ }_{i^{ }_j}(t'_j)
 \, \rangle\!\rangle \, 
 \frac{\partial
 \langle\!\langle \,
   \delta N^{ }_{i^{ }_2}(t^{ }) \cdots
   \delta N^{ }_{i^{ }_n}(t^{ })
 \, \rangle\!\rangle
 }{\partial \delta N^{ }_{i^{ }_j}(t)}
 \nn 
 & \stackrel{ \tilde{t}\to t }{=} & 
 \theta(t - \tilde{t}) \, \Omega \, 
 \frac{\partial
 \langle\!\langle \,
   \delta N^{ }_{i^{ }_2}(t^{ }) \cdots
   \delta N^{ }_{i^{ }_n}(t^{ })
 \, \rangle\!\rangle
 }{\partial \delta N^{ }_{i^{ }_1}(t)} 
 \;. \la{dG_n_2}
\ea
In the third step, we inserted the noise autocorrelator. 
In terms of the original correlator from \eq\nr{G_n}, 
we see that {\em two insertions} 
of the fluctuation $\delta N^{ }_{i^{ }_1}$ have been eliminated. 

Now, we return to \eq\nr{dG_n_1}, but set $\tilde{t}\to t^{-}_{  }$, 
whereby the time derivative can act on any of the fluctuations. 
Implementing \eq\nr{dG_n_2}
as a second derivative, then yields
\be
 \partial^{ }_t G^{ }_{i^{ }_1 \cdots i^{ }_n}(t,...,t) = 
 - n \, \Gamma \, G^{ }_{i^{ }_1 \cdots i^{ }_n}(t,...,t)
 + \frac{\Omega}{2}
   \sum_i \frac{\partial^2 G^{ }_{i^{ }_1 \cdots i^{ }_n}(t,...,t)}
               {\partial \delta N^{2}_{\!i}(t) }
 \;. \la{dG_n_3}
\ee

On the other hand, 
let us define a probability distribution, $\mathcal{W}$, as 
\be
 \mathcal{W}(\{ \delta N^{ }_{i1} \} ,t^{ }_1 ; 
 \{ \delta {N}^{ }_{i0} \} ,t^{ }_0)
 \; \equiv \; 
 \bigl\langle\hspace*{-0.8mm}\bigl\langle \, 
   \prod_i 
   \delta \bigl[ \delta {N}^{ }_{i1} - \delta N^{ }_{i}(t^{ }_1) \bigr]
 \bigr\rangle\hspace*{-0.8mm}\bigr\rangle
 \;, \quad
 \delta N^{ }_i(t^{ }_0) \; \equiv \, \delta N^{ }_{i0}
 \;. \la{def_W}
\ee
The initial condition at time $t^{ }_1 = t^{ }_0$ reads 
$
 \mathcal{W}(\{ \delta N^{ }_{i1} \} ,t^{ }_0 ; 
 \{ \delta {N}^{ }_{i0} \} ,t^{ }_0)
 = 
   \prod_i 
   \delta \bigl[ \delta {N}^{ }_{i1} - \delta N^{ }_{i0} \bigr]
$, 
and the normalization amounts to 
$
 \int \prod_i {\rm d} \delta N^{ }_{i1} 
 \mathcal{W}(\{ \delta N^{ }_{i1} \} ,t ; 
 \{ \delta {N}^{ }_{i0} \} ,t^{ }_0)
 = 
 1
$.

With the probability distribution, the moments appearing
in \eq\nr{dG_n_3} are expressed as 
\be
 G^{ }_{i^{ }_1 \cdots i^{ }_n}(t,...,t)
 = 
 \int \bigl\{ \prod_i {\rm d} \delta N^{ }_{i} \bigr\} 
 \, 
  \delta N^{ }_{i^{ }_1} \cdots
  \delta N^{ }_{i^{ }_n} 
 \, 
 \mathcal{W}(\{ \delta N^{ }_{i} \} ,t ; 
 \{ \delta {N}^{ }_{i0} \} ,t^{ }_0)
 \;. \la{W_n}
\ee
The time derivative then yields 
\be
 \partial^{ }_t
 G^{ }_{i^{ }_1 \cdots i^{ }_n}(t,...,t)
 = 
 \int \bigl\{ \prod_i {\rm d} \delta N^{ }_{i} \bigr\} 
 \, 
  \delta N^{ }_{i^{ }_1} \cdots
  \delta N^{ }_{i^{ }_n} 
 \, 
 \partial^{ }_t
 \mathcal{W}(\{ \delta N^{ }_{i} \} ,t ; 
 \{ \delta {N}^{ }_{i0} \} ,t^{ }_0)
 \;. \la{dW_n}
\ee
We envisage that the time derivative of $\mathcal{W}$ 
equals a differential operator, 
acting with respect to~$\delta N^{ }_i$, and carry out partial integrations. 
In order to reproduce the structure in \eq\nr{dG_n_3}, with the 
correct signs and the ``counting'' of insertions in the first term
(i.e.\ the factor $n$), the differential operator must read
\be
 \partial^{ }_t\mathcal{W} = 
 \sum_i \frac{\partial}{\partial \delta N^{ }_i}
 \biggl[
   \biggl( \Gamma \delta N^{ }_i
          + \frac{\Omega}{2}
          \frac{\partial}{\partial \delta N^{ }_i}
   \biggr) \mathcal{W} 
 \biggr]
 \;. \la{fp}
\ee
This is the Fokker-Planck equation (which bears some 
similarity with the Schr\"odinger equation). 

It is easy to search for a {\em stationary} solution of the 
Fokker-Planck equation. Physically, this corresponds to the 
form that $\mathcal{W}$ obtains at late times. 
Requiring the square brackets in \eq\nr{fp} to vanish, 
we see that 
\be
 \lim_{t\to\infty}
 \mathcal{W}(\{ \delta N^{ }_{i} \} ,t ; 
 \{ \delta {N}^{ }_{i0} \} ,t^{ }_0)
 = \mathcal{C}
 \, \exp\biggl( - \frac{\Gamma}{\Omega} \sum_i \delta N_i^2 \biggr)
 \;. \la{fp_static_soln}
\ee
The prefactor $\mathcal{C}$ can be fixed from the normalization 
condition as 
$
 \mathcal{C} = \prod^{ }_i ( \frac{\Gamma}{\pi\Omega} )^{1/2}_{ }
$.
The existence of a ``fixed point''
shows that the system equilibrates. As a crosscheck, 
the width of the Gaussian distribution, as visible in 
\eq\nr{fp_static_soln},  
agrees with what we had obtained in \eq\nr{susc_c_def}.

\newpage 

\subsection{Resonances in medium}
\la{se:nrqcd}

\index{Quarkonium states}

As a final observable, we consider the behaviour of a pair of 
heavy particles
within a thermal medium. In the QCD context, this would be relevant
for quarkonium physics~\cite{satz}.
Physically, heavy quarkonium refers to a bound
state of a charm and anti-charm quark ($c\bar{c}$) or a bottom and 
anti-bottom quark ($b\hspace*{0.2mm}\bar{b}$). 
Formally, quarkonium physics refers 
to observables that can be extracted from 2-point correlation functions 
of the conserved vector current, 
$
 \hat{\mathcal{J}}^\mu \equiv \,\hat{\!\bar{\psi}}\gamma_{ }^\mu \hat{\psi}
$,
around the 2-particle thresfold, i.e.\ for energies $\omega \sim 2 M$
(more precisely, $|\omega-2M| \ll M$), where $M$ denotes a heavy 
quark ``pole mass''.\footnote{\la{p_rate}%
 These considerations are related to the
 production rate of $e^-e^+$ or $\mu^-\mu^+$ pairs with a total 
 energy close to the mass of a quarkonium resonance, 
 cf.\ \eq\nr{dilepton}. Analogous physics may also play
 a role in cosmology, in connection with the thermal annihilation
 process of non-relativistic dark matter particles into Standard Model
 particles, if
 the former interact attractively through gauge boson 
 exchange~\cite{threshold}.
} 
In the following, we are 
interested in the limit $ M \gg 1$~GeV, so that the 
situation should be at least partly perturbative.
The goal is to illustrate
with yet another example (cf.\ \se\ref{se:field}) how in a thermal medium
two types of effects operate in parallel: ``virtual corrections'', which 
modify masses or effective parameters (in the present case, a potential);
and ``real corrections'', which represent real scatterings not taking
place in vacuum.

A useful starting point for our analysis is the observation 
that in the heavy-mass limit, the QCD Lagrangian can be simplified. 
Considering the extreme case in which the quarks 
do not move in the spatial directions at all, because the kicks they
receive from medium or vacuum fluctuations are insufficient to 
excite them, we may keep only 
the temporal part of the theory, resulting  in the Lagrangian
\be
 \mathcal{L}^{ }_\iM \; \equiv \; \bar\psi (i \gamma_{ }^0 D^{ }_0 - M) \psi
 \;. \la{LM_dt} 
\ee
This expression can be further split up into 
a form that contains explicitly a ``quark'' and an 
``antiquark'', e.g.\ by adopting a representation for the Dirac matrices 
with $\gamma_{ }^0 = \mathop{\mbox{diag}}
 (\unit^{ }_\rmii{$2\times 2$},-\unit^{ }_\rmii{$2\times 2$})$
and by writing 
\be
 \psi \; \equiv \;   
  \left( 
  \begin{array}{c} 
    \theta \\ \chi
  \end{array}
 \right)
 \;, \quad
 \bar\psi \; \equiv \; 
 ( \theta^\dagger \;, \; - \chi^\dagger ) 
 \;, \la{nr}
\ee
or more abstractly by defining 
$\theta \equiv \tfr12 (\unit + \gamma_{ }^0) \psi$, 
$\chi \equiv \tfr12(\unit - \gamma_{ }^0) \psi.$ Since fermions are
Grassmann fields, we must recall a minus sign  
when fields are commuted; in the following this concerns in particular  
the ordering of $\chi^*_\alpha$, $\chi^{ }_\beta$. 
Noting furthermore that 
(the Minkowskian four-vector $\mathcal{X}$  here
is not to be confused with the Grassmann field $\chi$)
\be
 \int_\mathcal{X} 
 - f(\mathcal{X}) \overleftarrow{\,{D}}_{\!\mu}^\dagger\, g(\mathcal{X}) 
 =
 \int_\mathcal{X} 
 f(\mathcal{X}) \overrightarrow{\!{D}}^{ }_{\!\mu}\, g(\mathcal{X}) 
 \;, \la{ibp_2}
\ee
where the arrow indicates the side on which the derivative operates, 
we obtain
\be
 \int_\mathcal{X} \chi^*_\alpha 
 {[\overrightarrow{\!{D}}^{ }_{\!\mu}]}^{ }_{\alpha\beta}
 \chi^{ }_\beta
 = 
 - \int_\mathcal{X} \chi^{ }_\beta 
 {[\overleftarrow{\,{D}}^{ }_{\!\mu}]}^{ }_{\alpha\beta}
 \chi^*_\alpha 
 = 
 \int_\mathcal{X} \chi^{ }_\beta 
 {[\overrightarrow{\!{D}}_{\!\mu}^*]}^{ }_{\beta\alpha}
 \chi^*_\alpha 
 \;.
\ee
Therefore the action corresponding to \eq\nr{LM_dt} can be written as
\ba
 \mathcal{S}^{ }_\iM & = &  
 \int_\mathcal{X}
 \Bigl\{
   \theta^\dagger (i {D}^{ }_0 - M) \theta 
 + \chi^\dagger (i {D}^{ }_0 + M) \chi
  \Bigr\} \nn 
 & = & 
 \int_\mathcal{X}
 \Bigl\{
   \theta^\dagger (i {D}^{ }_0 - M) \theta 
 + \chi^{*\dagger} (i {D}_0^* - M) \chi^*
  \Bigr\}
 \;.
\ea
This shows that the (charge-conjugated)
field $\chi^*$ represents an antiparticle to $\theta$, 
having the same mass but an opposite gauge charge. 

Next, we need a representation for the current 
${\mathcal{J}}^\mu$ in terms of the spinors $\theta,\chi$. 
If we consider the ``zero-momentum'' projection 
$\int_\vec{x} \mathcal{J}^\mu$, then current conservation 
implies that the component $\mu = 0$ must be constant in time, and hence that 
all interesting dynamics resides in the spatial components. 
It follows directly from the field redefinition in \eq\nr{nr} that 
in the standard representation\footnote{ 
 This refers to  $ \displaystyle
 \gamma_{ }^0 \equiv
 \left( 
  \begin{array}{cc}
   \mathbbm{1} & 0 \\   
   0 & -\mathbbm{1} 
  \end{array}
 \right)
 \,, \;\;
 \gamma_{ }^k \equiv
 \left( 
  \begin{array}{cc}
   0 & \sigma^{ }_k \\   
   -\sigma^{ }_k & 0 
  \end{array}
 \right)
 \,, \;\; k \in\{ 1,2,3 \}
 $, 
 where $\sigma^{ }_k$ are the Pauli matrices.
 \index{Dirac matrices}
 }
the spatial components can be expressed as
\be
  \mathcal{J}^k = \theta^\dagger \sigma^{ }_k \chi 
                + \chi^\dagger \sigma^{ }_k \theta
  \; \equiv \; \mathcal{J}^k_\rmii{NRQCD}
 \;. \la{Jk_nrqcd}
\ee
Let us note in passing that this relation does experience
corrections of $\rmO(M^0)$ through loop effects;
in fact at the next-to-leading order the relation reads~\cite{nrqcd1,nrqcd2} 
\be
 \mathcal{J}_\rmii{QCD}^k = \mathcal{J}_\rmii{NRQCD}^k 
 \left(1 - \frac{g^2 \CF}{2\pi^2} + \ldots \right)
 \;. \la{V_norm}
\ee

In the following, we omit the spin structure 
from \eq\nr{Jk_nrqcd}, but simultaneously also separate the quark
and antiquark from each other, 
connecting them with a Wilson line defined at the time slice~$t$ 
and denoted by $W^{ }_t$, which is needed to keep the 
quantity gauge invariant. This produces the structure
\be
 \mathcal{J}^k_\rmii{NRQCD} \to 
 \mathcal{J}^{ }_\vec{r}(t,\vec{x}) \;\equiv\; 
 \theta^\dagger\Bigl(t,\vec{x} + \frac{\vec{r}}{2} \Bigr)
 \, W^{ }_t \,  
 \chi\Bigl(t,\vec{x} - \frac{\vec{r}}{2} \Bigr)
 +
 \chi^\dagger\Bigl(t,\vec{x} - \frac{\vec{r}}{2} \Bigr)
 \, W_t^\dagger \,  
 \theta\Bigl(t,\vec{x} + \frac{\vec{r}}{2} \Bigr)
 \;. \la{J_nrqcd}
\ee
A typical Green's function could be of the form
$ 
 \int_\vec{x} \langle \hat{\mathcal{J}}^{ }_\vec{r}(t,\vec{x}) 
 \hat{\mathcal{J}}^{ }_\vec{r'}(0,\vec{0}) \rangle
$.
Given that we have assumed the heavy quarks not to move, 
we can however even set $\vec{x}=\vec{0}$ and 
$\vec{r'} = \vec{r}$, leaving us with 
\be
 C_r^>(t) 
 \; \equiv \; 
 \langle \hat{\mathcal{J}}^{ }_\vec{r}(t,\vec{0})
 \hat{\mathcal{J}}^{ }_\vec{r}(0,\vec{0}) \rangle
 \la{Gr_nrqcd}
 \;, \quad r \equiv |\vec{r}|
 \;, 
\ee
where we have used the notation of \se\ref{se:diff_G}
for the particular time-ordering chosen. 

Now, time-translation invariance guarantees that 
$C_r^>(t) = C_r^<(-t)$ and 
$\rho^{ }_r(t) = \fr12[C_r^>(t)-C_r^<(t)] = - \rho^{ }_r(-t)$, which
in frequency-space imply that 
$ C_r^>(\omega) =  C_r^<(-\omega)$ and 
$\rho^{ }_r(\omega) = - \rho^{ }_r(-\omega)$.
All of these functions contain 
the same information but, as indicated 
by \eqs\nr{bLSrel0} and \nr{bLSrel}, for 
$\omega \gg T$ it is $ C_r^>(\omega)$ that 
approximates $\rho^{ }_r(\omega)$ well, whereas
for $\omega \ll - T$ the spectral information
is dominantly contained in $ C_r^<(\omega)$.
We also note that for $|\omega| \ll T$, corresponding to the 
classical limit, the two orderings agree.

If we allow the Wilson lines $W^{ }_t$ to have an arbitrary shape, the
operators $\hat{\mathcal{J}}^{ }_\vec{r}$ constitute a whole set of
possible choices. 
Upon operating on the vacuum state, they generate a basis 
of gauge-invariant states in the sense 
of refs.~\cite{cwb_2}--\hspace*{-1.1mm}\cite{ml_2}.
These basis states are in general not eigenstates 
of the Hamiltonian, but the latter can be expressed
as linear combinations of the basis states.  

\index{Fock space}

Let now $|n\rangle$ denote the 
gauge-invariant eigenstates of the QCD Hamiltonian in the 
sector of the Fock space that contains no heavy quarks or antiquarks
(but does contain glueballs and their scattering states, 
as well as light hadrons), 
and $|n';r \rangle$ those in the sector with 
one heavy quark and one antiquark, 
separated by a distance $r$. 
If $\hat{\mathcal{J}}^{ }_\vec{r}$
operates on a state of the type $|n\rangle$, the result should have
a non-zero overlap with some of the $|n';r \rangle$.
If on the other hand the Hamiltonian operates
on the states $|n\rangle$, the state does not change but 
gets multiplied by an $r$-independent eigenvalue $\E^{ }_n$. 
And finally, if the Hamiltonian operates on the $|n';r \rangle$, 
the result is a multiplication of the state by 
an $r$-dependent eigenvalue $\E^{ }_{n'}(r)$, 
conventionally referred to as the (singlet) static potential. 
Numerical results for several 
of the lowest-lying values of $\E^{ }_{n'}(r)$  
in pure SU(3) gauge theory can be found in ref.~\cite{string}
(to be precise, these measurements concern $\E^{ }_{n'}(r) - \E^{ }_0$, 
cf.\ below).

We now expand the equilibrium correlator of \eq\nr{Gr_nrqcd} in the energy
eigenbasis. A key observation~\cite{akr} is that in the 
limit $T \ll 2M$, no states of the type $|n';r\rangle$ need to be 
put next to the density matrix $\mathcal{Z}^{-1}\exp(-\beta\hat{H})$, 
because such contributions are suppressed by $\sim e^{-2 M / T}$.
Thus, we obtain
\ba
 C_r^{>}(t) & \approx & 
 \frac{1}{\mathcal{Z}} \sum_{n,n'}
 \langle n | e^{-\beta\hat{H}}
             e^{i\hat{H} t} 
             \,\hat{\mathcal{J}}^{ }_\vec{r}\,
             e^{- i\hat{H} t} 
             | n';r\rangle 
             \langle n';r | 
             \hat{\mathcal{J}}^{ }_\vec{r}
             | n \rangle 
 \nn 
 & = & 
 \frac{1}{\mathcal{Z}} \sum_{n,n'}
 e^{-\beta \E^{ }_n} 
 e^{i [\E^{ }_n - \E^{ }_{n'}(r)] t}
  \; \bigl| \,  \langle n | 
   \hat{\mathcal{J}}^{ }_\vec{r}
  | n';r\rangle \, \bigr|^2
 \;. \la{Cr_spectral}
\ea
This function contains all relevant information about the dynamics 
of gauge-invariant quark-antiquark states as long as $T \ll 2M$. Next, 
we discuss its basic physical features. 

To start with, let us consider the case of zero temperature 
($\beta\to\infty$) and furthermore carry out a Wick rotation
to Euclidean spacetime like in \eq\nr{op_defs}, i.e.\ $it\to \tau$, 
$0 < \tau < \beta$. Then the sum over $n$, with  
the weight $e^{(\tau-\beta)\E^{ }_n}$, is dominated by the ground 
state $n=0$, whereas the sum over $n'$, with the weight 
$e^{-\tau \E^{ }_{n'}(r)}$, is dominated by $n' \equiv 0'$. 
Noting that $\mathcal{Z} \approx e^{-\beta \E^{ }_0}$ in this limit,
we observe that  $\E^{ }_{0'}(r) - \E^{ }_0$ can be extracted from 
the asymptotic $\tau$-dependence of $ C_r^{>} $
in the range $0 \ll \tau \ll \beta/2$, independent of 
the details of the operator $\hat{\mathcal{J}}^{ }_\vec{r}$ considered. 
This is how the results of ref.~\cite{string}
for $\E^{ }_{n'}(r) - \E^{ }_0$ have been obtained. 

Suppose then that we modify the setup by returning
to Minkowskian signature but keeping still $T=0$. 
The sum over $n$ is clearly still saturated by the ground state, but 
in the sum over $n'$ we now get a contribution from several
states. The excited $n'\neq 0'$ states, however, lead to more rapid
oscillations than the ground state, so that the ground state energy 
could be identified as the smallest oscillation frequency of the correlator. 
In pure SU($\Nc$) gauge theory, $\E^{ }_{n'}(r)$ is believed to display
a ``string spectrum'', representing vibrations of a colour ``flux tube''
between the quark and antiquark.  
The string spectrum is expected to be discrete, 
with level spacings $\Delta \E^{ }_{n'} \approx \pi/r$ at large~$r$.
Therefore $C^{>}_r$ remains periodic, or ``coherent''; 
this means that no information gets lost but after a certain period
the time evolution of $C^{>}_r$ repeats itself.\footnote{%
 Strictly speaking this is true only if the Hilbert space
 is finite-dimensional. 
 } 

Next, we switch on a finite temperature, which implies that 
the sum over $n$ becomes non-trivial as well. 
An immediate consequence of this is that, 
for any given $n'$, there are contributions to \eq\nr{Cr_spectral}  
which make the oscillations {\em slower}, decreasing
$\E^{ }_{n'}(r)$ to $\E^{ }_{n'}(r) - \E^{ }_n$, $n \ge 1$. 
Of course, the overlaps
$
  |   \langle n | 
   \hat{\mathcal{J}}^{ }_\vec{r}
  | n';r\rangle  |^2
$
also depend on $n$, so that the correlator may now be 
dominated by another value of $n'$, and the ``effective'' 
magnitude of $\E^{ }_{n'}(r) - \E^{ }_n$ is not easily deduced. 
Nevertheless we could refer 
to this phenomenon as 
{\em Debye screening}: \index{Debye screening} \index{Screening}
in the presence of a medium the energy associated with 
a quark-antiquark pair separated by a distance $r$ changes from that
in vacuum, because of the presence of states other than 
the vacuum one in the thermal average.  
In the language introduced at the beginning of this section, such a 
change in the energetics could be considered a ``virtual correction''. 

The temperature may also lead to a more dramatic 
effect. Indeed, in an infinite volume the spectrum $\E^{ }_n$ contains 
a {\em continuous part}, consisting e.g.\  
of pionic states, with the pions moving with respect to 
the rest frame that we have chosen to represent the heat bath.
If the temperature is high enough to excite
this part, we may expect to 
find a ``resonance''-type feature: the density 
of states grows with $\E^{ }_n$, 
whereas $e^{-\beta \E^{ }_n}$ decreases with $\E^{ }_n$.
To be explicit, let us model the resulting energy dependence by 
a Breit-Wigner shape, which implies writing
\ba
 \sum_n e^{-\beta \E^{ }_n + i \E^{ }_n t}  
  \, \bigl|   \langle n | 
   \hat{\mathcal{J}}^{ }_\vec{r}
  | n';r\rangle  \bigr|^2
 & \rightarrow & 
 \int \! {\rm d}\E^{ }_n \, \rho(\E^{ }_n) 
  e^{-\beta \E^{ }_n + i \E^{ }_n t}  
  \, \bigl|  \langle n |  \hat{\mathcal{J}}^{ }_\vec{r} | n';r\rangle  \bigr|^2
 \nn[-1mm] 
 & \simeq & 
  \int \! {\rm d}\E^{ }_n \,   e^{ i \E^{ }_n t}  
  \frac{\mathcal{F}(n';r)}{[\E^{ }_n-\mathcal{E}(n';r)]^2 +  \Gamma^2(n';r) }
 \nn 
 & \simeq & 
  \frac{\pi \mathcal{F}(n';r)}{\Gamma(n';r)}
  \exp\Bigl\{ i \mathcal{E}(n';r)\, t  - \Gamma(n';r)\, t \Bigr\}
 \;. \la{decoher}
\ea
We observe that, 
apart from the energy shift that was referred to as Debye screening 
above and is now represented by $\mathcal{E}(n';r)$, the absolute value
of the correlator also decreases with time. This phenomenon may
be referred to as {\em decoherence}: the coherent 
quantum-mechanical state $|n';r\rangle$ loses ``information'' through
a continuum of random scattering processes with the heat bath. 
According to 
\eq\nr{decoher}, we can also talk about a ``thermal width''
affecting the time evolution, 
or of an ``imaginary part'' in the effective energy shift.

\index{Decoherence} \index{Thermal width}
 
Physically, 
imaginary parts or widths correspond to ``real scatterings''.\footnote{%
 A classic example of this is the optical theorem of scattering theory. 
 } 
We expect that the farther apart the quark and the antiquark
are from each other, the larger should the width $\Gamma(n';r)$ be.
In the partonic language of quarks and gluons, 
this is because at a large separation the 
quark-antiquark pair carries a large ``colour dipole'', which scatters 
efficiently on the medium gluons. These frequent scatterings
lead to a loss of coherence of the initially quantum-mechanical
quark-antiquark state. (Computations of these reactions
have been reviewed, e.g.,  in ref.~\cite{jgx}.)

\index{Quarkonium dissociation}

We can also illustrate the physical meaning of the thermal width
in a gauge-invariant hadronic language. In this picture real  
scatterings are possible because
thermal fluctuations can excite colour-neutral states, like pions, 
from the medium, with which the heavy quarks can interact. 
This mechanism can for instance dissociate 
the quarkonium bound state into so-called
``open charm'' or ``open bottom'' hadrons, i.e.\ ones in which the 
heavy quarks form mesons 
with light antiquarks, or  vice versa (cf.\ e.g.\ ref.~\cite{mott}):
\be
 \lxi
 \hspace*{2cm}
 \vspace*{0.5cm}
 \;.
\ee
This is a purely thermal effect which would not be kinematically
allowed in vacuum, because the energy of the two open states is
higher than that of the single bound state. 
In full equilibrium, i.e.\ at 
time scales much larger than how often the process
shown occurs, the opposite
reaction would also take place at an equivalent rate, and 
the equilibrium ensemble would contain both open and
bound states. The entropy of this state is maximal, i.e.\ all 
information about the coherent quantum-mechanical initial 
state in which the quark-antiquark pair was generated, has been lost. 

%

\newpage


%
\appendix
\renewcommand{\thesection}{Appendix~\Alph{section}}

\newpage 

\addcontentsline{toc}{section}{Appendix: Extended Standard Model 
in Euclidean spacetime}

\section*{Appendix: Extended Standard Model in Euclidean spacetime}

\setcounter{section}{10}
\setcounter{equation}{0}

\index{Euclidean Lagrangian: Standard Model}
\index{Yukawa interaction}

In \eq\nr{Z_full}, the gauge-fixed imaginary-time Lagrangian of QCD was
given. Here we display the corresponding structure for the Standard
Model. For simplicity, terms related to gauge fixing are omitted. On
the other hand, we include right-handed neutrinos as degrees of freedom,
even though they are not considered to be part of the
``classic'' Standard Model, in which neutrinos
are postulated to be massless. 
We include them because this 
does not change any of the construction principles and 
yet offers for a simple way to solve a number of 
short-comings of the classic Standard Model (for a review, 
see ref.~\cite{drewes}). 
A reader preferring not to include 
right-handed neutrinos may decouple them by 
setting the corresponding Yukawa coupling matrix $h^{ }_\nu$ to zero.   

In QCD literature, it is common to keep the number of colours, $\Nc$, 
as a free parameter. When considering the Standard Model, 
however, its proper inclusion requires care~\cite{nc1,nc2}. 
For simplicity we restrict to $\Nc = 3$ in the following. 
The gauge group is then 
U$^{ }_\rmii{Y}$(1)$\times$SU$^{ }_\rmii{L}$(2)$\times$SU$^{ }_\rmi{c}$(3), 
where Y refers to the hypercharge
degree of freedom and L to left-handed fermions. 

The fermionic matter fields of the Standard Model carry a specific
chirality. Denoting chiral projectors by 
$
 \aL \equiv (1- \gamma^{ }_5)/2
$
and
$
 \aR \equiv (1+ \gamma^{ }_5)/2 
$,
left-handed doublets are defined as
\be
 Q^{ }_a \equiv \aL Q^{ }_a \equiv 
 \left( \begin{array}{c} \aL u^{ }_a \\ \aL d^{ }_a \end{array} \right)
 \;, \quad
 L^{ }_a \equiv \aL L^{ }_a \equiv 
 \left( \begin{array}{c} \aL \nu^{ }_a \\ \aL e^{ }_a \end{array} \right)
 \;,
\ee
where $a \in\{1,2,3\}$ is a ``family'' or ``generation'' index. 
Here the fermion fields are 4-component Dirac spinors. 
The quark fields carry an additional colour index 
that has been suppressed in the notation, i.e.\
they are really 12-component spinors. Subsequently we redefine the notation
in order to denote the right-handed components by 
\be
 u^{ }_a \equiv \aR u^{ }_a \;, \quad 
 d^{ }_a \equiv \aR d^{ }_a \;, \quad
 \nu^{ }_a \equiv \aR  \nu^{ }_a \;, \quad
 e^{ }_a \equiv \aR e^{ }_a \;. 
\ee
The scalar (Higgs) doublet is denoted by $\phi$, 
and $\tilde \phi \equiv i \sigma^{ }_2\phi^*$, 
where $\sigma^{ }_2$ is a Pauli matrix, 
is a conjugated version thereof,
which transforms in the same way under  
SU$^{ }_\rmii{L}$(2) but has an opposite ``charge''
under U$^{ }_\rmii{Y}$(1). Note that the right-handed neutrino 
field $\nu^{ }_a$ was denoted
by $N$ in \eq\nr{LM}.

With this field content, 
the Euclidean Lagrangian can be written 
as ($Q = (Q^{ }_1\,Q^{ }_2\,Q^{ }_3)^T$, etc.)
\ba
 L^{ }_\iE & \equiv & 
 \frac{1}{4} 
 F_{\mu\nu}^{a_i} F_{\mu\nu}^{a_i} 
 \; + \; 
 (D^{ }_\mu \phi)^\dagger D^{ }_\mu\phi 
 \; - \; m^2 \phi^\dagger\phi + \lambda (\phi^\dagger\phi)^2 
 \nn[0.0mm] 
 & + & 
   \bar{Q}^{ }_{ } \bsl{D}\! Q^{ }_{ }
 + \bar{u}^{ }_{ } \bsl{D}\! u^{ }_{ } 
 + \bar{d}^{ }_{ } \bsl{D}\! d^{ }_{ } 
 + \bar{L}^{ }_{ } \bsl{D}\! L^{ }_{ }
 + \bar{\nu}^{ }_{ } \bsl{D}\! \nu^{ }_{ } 
 + \bar{e}^{ }_{ } \bsl{D}\! e^{ }_{ } 
 +  \frac{1}{2}
 \bigl(
 \bar{\nu}_{ }^c M^{ }_{ } \nu^{ }_{ } 
 + 
 \bar{\nu}_{ }\, M^{\dagger}_{ } \nu^c_{ } 
 \bigr)
 \nn[0.5mm] 
 & + & 
 \bar{Q}^{ }_{ }\, h^{ }_{u} \, u^{ }_{ } \, \tilde\phi 
 + 
  \bar{Q}^{ }_{ }\, h_{d}^{ } \, d^{ }_{ } \, \phi 
 + 
 \bar{L}^{ }_{ }\, h_{\nu}^{ } \, \nu^{ }_{ } \, \tilde\phi
 + 
  \bar{L}^{ }_{ }\, h_{e}^{ } \, e^{ }_{ } \, \phi
 \nn[2mm] 
 & + & 
  \tilde\phi^\dagger \bar{u}^{ }_{ }\,  h_{u}^\dagger Q^{ }_{ }
 + 
  \phi^\dagger \bar{d}^{ }_{ }\,  h_{d}^\dagger Q^{ }_{ }
 + 
 \tilde\phi^\dagger \bar{\nu}^{ }_{ }\,  h_{\nu}^\dagger L^{ }_{ }
 + 
  \phi^\dagger \bar{e}^{ }_{ }\,  h_{e}^\dagger L^{ }_{ }
 \;. \la{SM}
\ea
A number of undefined symbols will be explained below.  
Starting from the end of the second row, 
$
 \nu^c \equiv C \bar{\nu}^T 
$
denotes a charge-conjugated spinor, with the 
charge conjugation matrix defined for instance as 
$
 C \equiv i \gamma_{ }^2 \gamma_{ }^0 
$, 
where $\gamma_{ }^\mu$ are Dirac matrices. 
It is possible to verify that the ``Majorana'' mass 
matrix $M$ is symmetric, $M^T = M$; 
through the so-called Takagi factorization 
(a special case of singular value decomposition),
it can consequently be written as $M = V \Delta\, V^T$, where $V$ is
unitary and $\Delta$ is a diagonal matrix with real non-negative entries, 
referred to as the Majorana masses of the right-handed neutrinos.

\index{Takagi factorization}

The theory defined by \eq\nr{SM} contains a number of parameters: 
the real gauge couplings $g^{ }_1,g^{ }_2,g^{ }_3$; 
the Higgs mass parameter $m^2$ and self-coupling $\lambda$; 
the complex $3\times 3$ Yukawa matrices 
$h^{ }_u$, $h^{ }_d$, $h^{ }_\nu$ and $h^{ }_e$; 
and the complex $3\times 3$ Majorana mass matrix $M$. 
In the quantized theory, all of 
these are to be understood as bare parameters. Note that there is a fairly 
large redundancy in this parameter set, which could be reduced by 
various field redefinitions, 
but for simplicity we display the general expression. 

Let us now define the 
gauge interactions. In the fermionic case, gauge interactions reside
in $\bsl{D} = \gamma^{ }_\mu D^{ }_\mu$, where $D^{ }_\mu$ is 
a covariant derivative. When acting on the Higgs doublet
the covariant derivative takes the form 
\be
 D^{ }_\mu\phi  \; \equiv \;  
 \Bigl(
   \partial^{ }_\mu + \frac{i g^{ }_1}{2}  A^{ }_\mu
  - i g^{ }_2 
   T^{ a_2 }_{ } B^{ a_2}_\mu 
 \Bigr)\, \phi 
 \;, 
\ee
where $g^{ }_1$ and $g^{ }_2$ are gauge couplings related to 
the U$_\rmii{Y}$(1) and SU$_\rmii{L}$(2) gauge fields
$A^{ }_\mu$ and $B_\mu^{a_2}$, 
respectively, and $T^{a_2}$ are Hermitean generators of SU$^{ }_\rmii{L}$(2), 
normalized as $\tr [T^{a_2} T^{b_2}] = \fr12 \delta^{a_2 b_2}$.
We employ a notation whereby the index 
$a^{ }_2 \in \{1,...,d^{ }_2\equiv 3 \}$ 
implies the use of SU(2) generators, 
and repeated indices are summed over (in addition, $d^{ }_1 \equiv 1$, 
and sums over $a^{ }_1$ are omitted whenever possible).  
When acting on leptons, 
the covariant derivative reads
\ba
 D^{ }_\mu L^{ }_a  & \equiv &  
 \Bigl(
   \partial^{ }_\mu - \frac{i g^{ }_1}{2}  A^{ }_\mu
  - i g^{ }_2 
    T^{ a_2 } B^{ a_2}_\mu 
 \Bigr)\, L^{ }_a 
 \;, \\
 D^{ }_\mu \nu^{ }_a  & \equiv & 
 \Bigl( \,\partial^{ }_\mu\, \Bigr)\, \nu^{ }_a
 \;, \\[0mm]
 D^{ }_\mu e^{ }_a  & \equiv &
 \Bigl(
   \partial^{ }_\mu -  i g^{ }_1 A^{ }_\mu
 \Bigr)\, e^{ }_a 
 \;. 
\ea
In the case of quarks, the SU$^{ }_\rmi{c}$(3) gauge coupling $g^{ }_3$ and 
the generators $T^{a^{ }_3}$ and 
the gauge fields $C_\mu^{a^{ }_3}$ appear as well: 
\ba
 D^{ }_\mu Q^{ }_a  & \equiv &  
 \Bigl(
   \partial^{ }_\mu + \frac{i g^{ }_1}{6}  A^{ }_\mu
  - i g^{ }_2 
   T^{ a_2 } B^{a_2}_\mu 
  - i g^{ }_3 
   T^{ a_3 } C^{ a_3}_\mu 
 \Bigr)\, Q^{ }_a 
 \;, \\
 D^{ }_\mu u^{ }_a  & \equiv & 
 \Bigl(
   \partial^{ }_\mu + \frac{2 i g^{ }_1}{3}  A^{ }_\mu
  - i g^{ }_3 
   T^{ a_3 } C^{ a_3}_\mu 
 \Bigr)\,  u^{ }_a
 \;, \\[0mm]
 D^{ }_\mu d^{ }_a  & \equiv &
 \Bigl(
   \partial^{ }_\mu - \frac{i g^{ }_1}{3}  A^{ }_\mu
  - i g^{ }_3 
   T^{ a_3 } C^{ a_3}_\mu 
 \Bigr)\,  d^{ }_a
 \;. 
\ea
The colour index $a^{ }_3$ is summed over the 
set $a^{ }_3 \in \{1,...,d^{ }_3 \equiv 8\}$.
Finally, field strength tensors are defined in accordance with 
\eq\nr{LM_gauge}, 
\ba
 F^{a_1}_{\mu\nu} & \equiv & 
 \partial^{ }_\mu A^{ }_\nu - \partial^{ }_\nu A^{ }_\mu
 \;, \\
 F^{a_2}_{\mu\nu} & \equiv & 
 \partial^{ }_\mu B^{a_2}_\nu - \partial^{ }_\nu B^{a_2}_\mu
 + g^{ }_2\, \epsilon^{a_2 b_2 c_2}
 B^{b_2}_\mu B^{c_2}_\nu
 \;, \\
 F^{a_3}_{\mu\nu} & \equiv & 
 \partial^{ }_\mu C^{a_3}_\nu - \partial^{ }_\nu C^{a_3}_\mu
 + g^{ }_3\, f^{a_3 b_3 c_3} C^{b_3}_\mu C^{c_3}_\nu
 \;,
\ea
where $\epsilon^{a_2 b_2 c_2}$ is the Levi-Civita symbol and 
$f^{a_3 b_3 c_3}$ are
the structure constants of SU(3). 

We remark that we have not included so-called $\theta$-terms in \eq\nr{SM}, 
which would have the form
$
 \delta L^{ }_\iE = i \sum_{n=1}^3 \theta^{ }_n 
 \frac{ \epsilon^{ }_{\mu\nu\rho\sigma}
  g_n^2 F^{a_n}_{\mu\nu} F^{a_n}_{\rho\sigma} }{ 64 \pi^2 } 
$. 
This is because our Yukawa couplings are complex. The complex Yukawa couplings
corresponding to quark masses can be tuned to be real by a chiral rotation,
but this induces a QCD $\theta$-term, leading to the so-called strong 
CP problem, i.e.\ the unnatural-looking 
fact that phenomenologically $\theta^{ }_3\approx 0$. 
(It is an interesting exercise to contemplate why the 
U$^{ }_\rmii{Y}$(1) and the SU$^{ }_\rmii{L}$(2) $\theta$-angles, 
$\theta^{ }_1$ and $\theta^{ }_2$, do not pose similar problems.)

Finally we recall that the quantization of 
chiral gauge theories is highly non-trivial. 
Even staying within perturbation theory, 
$\gamma^{ }_5$ has to be carefully defined
in the context of dimensional regularization~\cite{g5a,g5b}, 
but this tends to break spacetime and/or
gauge symmetries, leading e.g.\ to a complicated pattern of operator 
mixings~\cite{JK,BW}. 


%

%
\clearpage
\addcontentsline{toc}{section}{Index}
\printindex


\begin{thebibliography}{9.99}

\bibitem{jk}
  J.I.~Kapusta, {\it Finite-temperature Field Theory} 
  (Cambridge University Press, Cambridge, 1989).

\bibitem{rev2}
  M. Le Bellac, {\it Thermal Field Theory} 
  (Cambridge University Press, Cambridge, 2000).

\bibitem{rev3}
  J.I.~Kapusta and C.~Gale, 
  {\it Finite-Temperature Field Theory: Principles and Applications} 
  (Cambridge University Press, Cambridge, 2006).

\bibitem{rev4}
  P.~Arnold, 
  {\it Quark-Gluon Plasma and Thermalization,} 
  Int.\ J.\ Mod.\ Phys.\  E {16} (2007) 2555 
  {[0708.0812]}.

\bibitem{rev4a}
  V.A.~Rubakov and M.E.~Shaposhnikov,
  {\it Electroweak baryon number non-conservation in the early Universe 
  and in high-energy collisions,}
  Usp.\ Fiz.\ Nauk {166} (1996) 493
  [Phys.\ Usp.\  {39} (1996) 461]
  [hep-ph/9603208].

\bibitem{rev4aa}
  L.S.~Brown and R.F.~Sawyer,
  {\it Nuclear reaction rates in a plasma,}
  Rev.\ Mod.\ Phys.\  {69} (1997) 411
  [astro-ph/9610256].

\bibitem{rev4b}
  J.P.~Blaizot and E.~Iancu,
  {\it The quark-gluon plasma: collective dynamics and hard thermal loops,}
  Phys.\ Rept.\  {359} (2002) 355
  [hep-ph/0101103].

\bibitem{rev4c}
  D.H.~Rischke,
  {\it The quark-gluon plasma in equilibrium,}
  Prog.\ Part.\ Nucl.\ Phys.\  {52} (2004) 197
  [nucl-th/0305030].

\bibitem{rev4d}
  U.~Kraemmer and A.~Rebhan,
  {\it Advances in perturbative thermal field theory,}
  Rept.\ Prog.\ Phys.\  {67} (2004) 351
  [hep-ph/0310337].

\bibitem{rev4e}
  S.~Davidson, E.~Nardi and Y.~Nir,
  {\it Leptogenesis,}
  Phys.\ Rept.\  {466} (2008) 105
  [0802.2962].

\bibitem{rev4ef}
  P.~Kovtun,
  {\it Lectures on hydrodynamic fluctuations in relativistic theories,}
  J.\ Phys.\ A {45} (2012) 473001
  [1205.5040].

\bibitem{rev4f}
  D.E.~Morrissey and M.J.~Ramsey-Musolf,
  {\it Electroweak baryogenesis,}
  New J.\ Phys.\ {14} (2012) 125003
  [1206.2942].

\bibitem{rev4g}
  J.~Ghiglieri and D.~Teaney,
  {\it Parton energy loss and momentum broadening at NLO
  in high temperature QCD plasmas,}
  Int.\ J.\ Mod.\ Phys.\ E {24} (2015) 1530013
  [1502.03730].

\bibitem{rev4h}
  J.~Ghiglieri, A.~Kurkela, M.~Strickland and A.~Vuorinen,
  {\it Perturbative thermal QCD: Formalism and applications,}
  Phys.\ Rept.\ {880} (2020) 1
  [2002.10188].

\bibitem{rev4hh}
  M.B.~Hindmarsh, M.~L\"uben, J.~Lumma and M.~Pauly,
  {\it Phase transitions in the early universe,}
  SciPost Phys.\ Lect.\ Notes {24} (2021) 1
  [2008.09136].

\bibitem{rev4i}
  D.~B\"odeker and W.~Buchm\"uller,
  {\it Baryogenesis from the weak scale to the grand unification scale,}
  Rev.\ Mod.\ Phys.\ {93} (2021) 035004
  [2009.07294].

\bibitem{rev5}
  H.B.~Meyer, 
  {\it Transport properties of the quark-gluon plasma: A lattice QCD
  perspective,} 
  Eur.\ Phys.\ J.\ A {47} (2011) 86
  [1104.3708].

\end{thebibliography}

\begin{thebibliography}{9.99}

\setcounter{enumiv}{0} 

\bibitem{fh}
  R.P.~Feynman and A.R.~Hibbs,
  {\it Quantum Mechanics and Path Integrals} 
  (McGraw-Hill, New York, 1965).
  
\end{thebibliography}

\begin{thebibliography}{9.99}

\setcounter{enumiv}{0} 

\bibitem{gm}
  L.~Giusti and H.B.~Meyer,
  {\it Thermodynamic potentials from shifted boundary conditions: 
  the scalar-field theory case,}
  JHEP {11} (2011) 087
  [1110.3136].

\bibitem{dj}
  L.~Dolan and R.~Jackiw,
  {\it Symmetry behavior at finite temperature,}
  Phys.\ Rev.\  D {9} (1974) 3320.

\bibitem{az}
  P.~Arnold and C.~Zhai,
  {\it Three-loop free energy for pure gauge QCD,}
  Phys.\ Rev.\  {D 50} (1994) 7603
  [hep-ph/9408276].

\end{thebibliography}

\begin{thebibliography}{9.99}

\setcounter{enumiv}{0} 

\bibitem{phi1}
  A.~Gynther, M.~Laine, Y.~Schr\"oder, C.~Torrero and A.~Vuorinen,
  {\it Four-loop pressure of massless O(N) scalar field theory,}
  JHEP {04} (2007) 094
  [hep-ph/0703307].

\bibitem{phi2}
  J.O.~Andersen, L.~Kyllingstad and L.E.~Leganger,
  {\it Pressure to order $g^8 \log g$ of massless $\phi^4$ theory 
  at weak coupling,}
  JHEP {08} (2009) 066
  [0903.4596].

\bibitem{ae}
  P.B.~Arnold and O.~Espinosa,
  {\it Effective potential and first-order phase transitions: 
  Beyond leading order,}
  Phys.\ Rev.\ D {47} (1993) 3546; 
  {\it ibid.} {50} (1994) 6662 (E)
  [hep-ph/9212235].

\end{thebibliography}

\begin{thebibliography}{9.99}

\setcounter{enumiv}{0} 

\bibitem{fb}
 F.A.~Berezin, 
 {\it The method of second quantization,}
 Pure Appl.\ Phys.\ {24} (1966) 1--228.

\end{thebibliography}

\begin{thebibliography}{9.99}

\setcounter{enumiv}{0} 

\bibitem{cwb}
  C.W.~Bernard, 
  {\it Feynman rules for gauge theories at finite temperature,}
  Phys.\ Rev.\  D {9} (1974) 3312.

\bibitem{kogut}
  J.B.~Kogut and L.~Susskind,
  {\it Hamiltonian formulation of Wilson's lattice gauge theories,}
  Phys.\ Rev.\  D {11} (1975) 395.

\bibitem{ml}
  M.~L\"uscher,
  {\it Construction of a selfadjoint, strictly positive transfer matrix for
  Euclidean lattice gauge theories,}
  Commun.\ Math.\ Phys.\  {54} (1977) 283.

\bibitem{fp}
  L.D.~Faddeev and V.N.~Popov,
  {\it Feynman diagrams for the Yang-Mills field,}
  Phys.\ Lett.\ B {25} (1967) 29.

\bibitem{tv}
  T.~Vachaspati,
  {\it Progress on cosmological magnetic fields,}
  Rept.\ Prog.\ Phys.\ {84} (2021) 074901
  [2010.10525].

\bibitem{form}
  J.~Kuipers, T.~Ueda, J.A.M.~Vermaseren and J.~Vollinga,
  {\it FORM version 4.0,}
  Comput.\ Phys.\ Commun.\  {184} (2013) 1453
  [1203.6543].

\bibitem{meg}
  M.E.~Carrington,
  {\it Effective potential at finite temperature in the Standard Model,}
  Phys.\ Rev.\ D {45} (1992) 2933.

\bibitem{Ghi-Sch}
  I.~Ghisoiu, J.~M\"oller and Y.~Schr\"oder,
  {\it Debye screening mass of hot Yang-Mills theory to three-loop order,}
  JHEP {11} (2015) 121
  [1509.08727].

\bibitem{rebhan}
  A.K.~Rebhan,
  {\it Non-Abelian Debye mass at next-to-leading order,}
  Phys.\ Rev.\ D {48} (1993) 3967
  [hep-ph/9308232].

\bibitem{ay}
  P.B.~Arnold and L.G.~Yaffe,
  {\it Non-Abelian Debye screening length beyond leading order,}
  Phys.\ Rev.\ D {52} (1995) 7208
  [hep-ph/9508280].

\bibitem{jk2}
  J.I.~Kapusta,
  {\it Quantum Chromodynamics at high temperature,}
  Nucl.\ Phys.\ B {148} (1979) 461.

\bibitem{es}
  E.V.~Shuryak,
  {\it Theory of hadronic plasma,}
  Sov.\ Phys.\ JETP {47} (1978) 212.

\bibitem{ch}
  S.A.~Chin,
  {\it Transition to hot quark matter in relativistic heavy-ion collision,}
  Phys.\ Lett.\ B {78} (1978) 552.

\bibitem{tt}
  T.~Toimela,
  {\it The next term in the thermodynamic potential of QCD,}
  Phys.\ Lett.\ B {124} (1983) 407.

\bibitem{az_2}
  P.~Arnold and C.~Zhai,
  {\it Three-loop free energy for pure gauge QCD,}
  Phys.\ Rev.\  {D 50} (1994) 7603
  [hep-ph/9408276].

\bibitem{zk}
  C.~Zhai and B.~Kastening,
  {\it Free energy of hot gauge theories with fermions through $g^5$,}
  Phys.\ Rev.\  {D 52} (1995) 7232
  [hep-ph/9507380].

\bibitem{bn}
  E.~Braaten and A.~Nieto,
  {\it Free energy of QCD at high temperature,}
  Phys.\ Rev.\ D 53 (1996) 3421 \la{bn}
  [hep-ph/9510408].

\bibitem{klrs}
  K.~Kajantie, M.~Laine, K.~Rummukainen and Y.~Schr\"oder,
  {\it Pressure of hot QCD up to $g^6 \ln (1/g)$},
  Phys.\ Rev.\ D 67 (2003) 105008
  [hep-ph/0211321]. 

\end{thebibliography}

\begin{thebibliography}{9.99}

\setcounter{enumiv}{0} 

\bibitem{linde}
  A.D.~Linde,
  {\it Infrared problem in thermodynamics of the Yang-Mills gas,}
  Phys.\ Lett.\ {B 96} (1980) 289.

\bibitem{jc}
  J.C.~Collins, 
  {\it Renormalization} 
  (Cambridge University Press, 1984).

\bibitem{kj}
  K.~Jansen {\it et al.},
  {\it Non-perturbative renormalization of lattice QCD at all scales,}
  Phys.\ Lett.\  B {372} (1996) 275 
  [hep-lat/9512009].
 
\bibitem{pw}
  P.~Weisz,
  {\it Renormalization and lattice artifacts,}
  arXiv:1004.3462.

\bibitem{dr1}
  P.~Ginsparg, 
  {\it First and second order phase transitions 
  in gauge theories at finite temperature,}
  Nucl.\ Phys.\ B 170 (1980) 388.
  
\bibitem{dr2}
  T.~Appelquist and R.D.~Pisarski,
  {\it High-temperature Yang-Mills theories and three-dimensional 
  Quantum Chromodynamics,}
  Phys.\ Rev.\ D 23 (1981) 2305.

\bibitem{generic}
  K.~Kajantie, M.~Laine, K.~Rummukainen and M.E.~Shaposhnikov,
  {\it Generic rules for high temperature dimensional reduction and 
  their application to the Standard Model,}
  Nucl.\ Phys.\ B {458} (1996) 90
  [hep-ph/9508379].

\bibitem{sc}
  S.~Chapman,
  {\it New dimensionally reduced effective action 
  for QCD at high temperature,}
  Phys.\ Rev.\ D {50} (1994) 5308 
  [hep-ph/9407313].

\bibitem{mu}
  A.~Hart, M.~Laine and O.~Philipsen,
  {\it Static correlation lengths in QCD at high temperatures 
  and finite densities,}
  Nucl.\ Phys.\ B {586} (2000) 443
  [hep-ph/0004060].

\bibitem{nadkarni}
  S.~Nadkarni,
  {\it Dimensional reduction in finite-temperature Quantum Chromodynamics. II,}
  Phys.\ Rev.\ D {38} (1988) 3287.
 
\bibitem{npl}
  N.P.~Landsman,
  {\it Limitations to dimensional reduction at high temperature,}
  Nucl.\ Phys.\ B {322} (1989) 498.

\bibitem{hl}
  S.~Huang and M.~Lissia,
  {\it The relevant scale parameter in the high temperature phase of QCD,}
  Nucl.\ Phys.\ B {438} (1995) 54
  [hep-ph/9411293].

\bibitem{gE2}
  M.~Laine and Y.~Schr\"oder,
  {\it Two-loop QCD gauge coupling at high temperatures,}
  JHEP {03} (2005) 067
  [hep-ph/0503061].

\bibitem{parity}
  K.~Kajantie, M.~Laine, K.~Rummukainen and M.E.~Shaposhnikov,
  {\it High temperature dimensional reduction and parity violation,}
  Phys.\ Lett.\  B {423} (1998) 137
  [hep-ph/9710538].

\bibitem{gpy}
  D.J.~Gross, R.D.~Pisarski and L.G.~Yaffe,
  {\it QCD and instantons at finite temperature,}
  Rev.\ Mod.\ Phys.\  {53} (1981) 43.

\bibitem{dn}
  A.~Boccaletti and D.~Nogradi,
  {\it The semi-classical approximation at high temperature revisited,}
  JHEP {03} (2020) 045
  [2001.03383].

\bibitem{ua1}
  T.~Kanazawa and N.~Yamamoto,
  {\em U(1) axial symmetry and Dirac spectra in QCD at high temperature,}
  JHEP {01} (2016) 141
  [1508.02416].

\bibitem{sch}
  S.~Caron-Huot and G.D.~Moore,
  {\it Heavy quark diffusion in perturbative QCD at next-to-leading order,}
  Phys.\ Rev.\ Lett.\  {100} (2008) 052301
  [0708.4232].

\bibitem{nspt}
  F.~Di Renzo, M.~Laine, V.~Miccio, Y.~Schr\"oder and C.~Torrero,
  {\it The leading non-perturbative coefficient in the weak-coupling 
  expansion of hot QCD pressure,}
  JHEP {07} (2006) 026
  [hep-ph/0605042].

\bibitem{sw}
  S.~Weinberg,
  {\it Baryon- and Lepton-Nonconserving Processes},
  Phys.\ Rev.\ Lett.\  {43} (1979) 1566.

\bibitem{wz}
  F.~Wilczek and A.~Zee,
  {\it Operator Analysis of Nucleon Decay},
  Phys.\ Rev.\ Lett.\  {43} (1979) 1571.

\bibitem{ay_2}
  P.B.~Arnold and L.G.~Yaffe,
  {\it Non-Abelian Debye screening length beyond leading order,}
  Phys.\ Rev.\ D {52} (1995) 7208
  [hep-ph/9508280].

\bibitem{srate}
  M.~D'Onofrio, K.~Rummukainen and A.~Tranberg,
  {\it Sphaleron Rate in the Minimal Standard Model,}
  Phys.\ Rev.\ Lett.\  {113} (2014) 141602
  [1404.3565].

\bibitem{abbott}
  L.F.~Abbott,
  {\it The background field method beyond one loop,}
  Nucl.\ Phys.\ B {185} (1981) 189.

\bibitem{Ghi-Sch-2}
  I.~Ghisoiu, J.~M\"oller and Y.~Schr\"oder,
  {\it Debye screening mass of hot Yang-Mills theory to three-loop order,}
  JHEP {11} (2015) 121
  [1509.08727].

\end{thebibliography}

\begin{thebibliography}{9.99}

\setcounter{enumiv}{0} 

\bibitem{anomaly}
  G.~'t Hooft,
  {\it Symmetry Breaking through Bell-Jackiw Anomalies,}
  Phys.\ Rev.\ Lett.\  {37} (1976) 8.

\bibitem{srate2}
  M.~D'Onofrio, K.~Rummukainen and A.~Tranberg,
  {\it Sphaleron Rate in the Minimal Standard Model,}
  Phys.\ Rev.\ Lett.\  {113} (2014) 141602
  [1404.3565].

\bibitem{mu1}
  J.I.~Kapusta,
  {\it Bose-Einstein condensation, spontaneous symmetry breaking, 
  and gauge theories,}
  Phys.\ Rev.\ D {24} (1981) 426.

\bibitem{mu2}
  H.E.~Haber and H.A.~Weldon,
  {\it Finite-temperature symmetry breaking as Bose-Einstein condensation,}
  Phys.\ Rev.\ D {25} (1982) 502.

\bibitem{mu3}
  K.M.~Benson, J.~Bernstein and S.~Dodelson,
  {\it Phase structure and the effective potential at fixed charge,}
  Phys.\ Rev.\ D {44} (1991) 2480.

\bibitem{bg1}
  S.Y.~Khlebnikov and M.E.~Shaposhnikov,
  {\it Melting of the Higgs vacuum: Conserved numbers at high temperature,}
  Phys.\ Lett.\ B {387} (1996) 817
  [hep-ph/9607386].
  
\bibitem{bg2}
  D.~B\"odeker and M.~Sangel,
  {\it Order $g^2$ susceptibilities
  in the symmetric phase of the Standard Model,}
  JCAP {04} (2015) 040
  [1501.03151].
  
\bibitem{av}
  A.~Vuorinen,
  {\it Pressure of QCD at finite temperatures and chemical potentials,}
  Phys.\ Rev.\ D {68} (2003) 054017
  [hep-ph/0305183].

\bibitem{tyler}
  T.~Gorda, A.~Kurkela, R.~Paatelainen, S.~S\"appi and A.~Vuorinen,
  {\it Cold quark matter at N3LO: Soft contributions,}
  Phys.\ Rev.\ D {104} (2021)  074015
  [2103.07427].

\bibitem{fmcl}
  B.A.~Freedman and L.D.~McLerran,
  {\it Fermions and gauge vector mesons at finite temperature and density. 
  III. The ground-state energy of a relativistic quark gas,}
  Phys.\ Rev.\ D {16} (1977) 1169.

\end{thebibliography}

\begin{thebibliography}{9.99}

\setcounter{enumiv}{0} 

\bibitem{old4}
  L.P.~Kadanoff and G.A.~Baym, 
  {\it Quantum Statistical Mechanics} 
  (Benjamin, Menlo Park, 1962). 

\bibitem{old1}
   A.L.~Fetter and J.D.~Walecka, 
  {\it Quantum Theory of Many-Particle Systems}
  (McGraw-Hill, New York, 1971). 
  
\bibitem{old2}
  S.~Doniach and E.H.~Sondheimer, 
  {\it Green's Functions for Solid State Physicists}
  (Benjamin, Reading, 1974).
  
\bibitem{old3}
  J.W.~Negele and H.~Orland,
  {\it Quantum Many Particle Systems}
  (Addison-Wesley, Redwood City, 1988). 

\bibitem{3pt1}
  D.~B\"odeker and M.~Sangel,
  {\it Lepton asymmetry rate from quantum field theory: 
  NLO in the hierarchical limit,}
  JCAP {06} (2017) 052
  [1702.02155].

\bibitem{3pt2}
  S.~Chaudhuri, C.~Chowdhury and R.~Loganayagam,
  {\it Spectral representation of thermal OTO correlators,}
  JHEP {02} (2019) 018
  [1810.03118].

\bibitem{cuniberti}
  G.~Cuniberti, E.~De Micheli and G.A.~Viano, \la{cuniberti}
  {\it Reconstructing the thermal Green functions at real times from those at
  imaginary times,}
  Commun.\ Math.\ Phys.\  {216} (2001) 59
  [cond-mat/0109175]. 

\bibitem{rev5_2}
  H.B.~Meyer, 
  {\it Transport properties of the quark-gluon plasma: A lattice QCD
  perspective,} 
  Eur.\ Phys.\ J.\ A {47} (2011) 86
  [1104.3708].

\bibitem{art}
  H.A.~Weldon,
  {\it Simple rules for discontinuities in finite-temperature field theory,}
  Phys.\ Rev.\ D {28} (1983) 2007.

\bibitem{master}
  M.~Laine,
  {\it Thermal 2-loop master spectral function at finite momentum,}
  JHEP {05} (2013) 083
  [1304.0202].

\bibitem{pisarski}
  R.D.~Pisarski,
  {\it Computing finite temperature loops with ease,}
  Nucl.\ Phys.\ B {309} (1988) 476.

\bibitem{parwani}
  R.R.~Parwani,
  {\it Resummation in a hot scalar field theory,}
  Phys.\ Rev.\ D {45} (1992) 4695; 
  {\it ibid.}\ {48} (1993) 5965 (E)
  [hep-ph/9204216].

\bibitem{hw}
  H.A.~Weldon,
  {\it Effective fermion masses of order $gT$ 
  in high-temperature gauge theories
  with exact chiral invariance,}
  Phys.\ Rev.\ D {26} (1982) 2789.

\bibitem{htl1}
  E.~Braaten, R.D.~Pisarski and T.-C.~Yuan,
  {\it Production of soft dileptons in the quark--gluon plasma,}
  Phys.\ Rev.\ Lett.\  {64} (1990) 2242.

\bibitem{lpm3}
  G.D.~Moore and J.-M.~Robert,
  {\it Dileptons, spectral weights, and conductivity 
  in the quark-gluon plasma,}
  hep-ph/0607172.

\bibitem{bb2}
 A.~Anisimov, D.~Besak and D.~B\"odeker,
 {\it Thermal production of relativistic Majorana neutrinos:
 strong enhancement by multiple soft scattering,}
 JCAP {03} (2011) 042
 [1012.3784].

\bibitem{sum3}
  D.~Besak and D.~B\"odeker,
  {\it Thermal production of ultrarelativistic right-handed neutrinos:
  complete leading-order results,}
  JCAP {03} (2012) 029
  [1202.1288].

\bibitem{photon1}
  P.B.~Arnold, G.D.~Moore and L.G.~Yaffe,
  {\it Photon emission from ultrarelativistic plasmas,}
  JHEP {11} (2001) 057
  [hep-ph/0109064].

\bibitem{photon2}
  P.B.~Arnold, G.D.~Moore and L.G.~Yaffe,
  {\it Photon emission from quark-gluon plasma:
  complete leading order results,}
  JHEP {12} (2001) 009
  [hep-ph/0111107].

\bibitem{sk1}
  K.~Chou, Z.~Su, B.~Hao and L.~Yu,
  {\it Equilibrium and nonequilibrium formalisms made unified},
  Phys.\ Rept.\ 118 (1985) 1.

\bibitem{sk2}
  N.P.~Landsman and C.G.~van Weert,
  {\it Real- and imaginary-time field theory 
  at finite temperature and density,}
  Phys.\ Rept.\  {145} (1987) 141.

\bibitem{schz}
  S.~Caron-Huot,
  {\it Hard thermal loops in the real-time formalism,}
  JHEP {04} (2009) 004
  [0710.5726].

\bibitem{sch_2}
  S.~Caron-Huot and G.D.~Moore,
  {\it Heavy quark diffusion in perturbative QCD at next-to-leading order,}
  Phys.\ Rev.\ Lett.\  {100} (2008) 052301
  [0708.4232].

\bibitem{sum2}
  S.~Caron-Huot,
  {\it $O(g)$ plasma effects in jet quenching,}
  Phys.\ Rev.\ D {79} (2009) 065039
  [0811.1603].

\bibitem{photon}
  J.~Ghiglieri, J.~Hong, A.~Kurkela, E.~Lu, G.D.~Moore and D.~Teaney,
  {\it Next-to-leading order thermal photon production 
  in a weakly coupled quark-gluon plasma,}
  JHEP {05} (2013) 010
  [1302.5970].

\bibitem{by}
  E.~Braaten and T.C.~Yuan,
  {\it Calculation of screening in a hot plasma,}
  Phys.\ Rev.\ Lett.\  {66} (1991) 2183.

\bibitem{ht1}
  R.D.~Pisarski,
  {\it Scattering amplitudes in hot gauge theories,}
  Phys.\ Rev.\ Lett.\  {63} (1989) 1129.

\bibitem{ht2}
  J.~Frenkel and J.C.~Taylor,
  {\it High-temperature limit of thermal QCD,}
  Nucl.\ Phys.\ B {334} (1990) 199.

\bibitem{ht3}
  E.~Braaten and R.D.~Pisarski,
  {\it Soft amplitudes in hot gauge theories: A general analysis,}
  Nucl.\ Phys.\ B {337} (1990) 569.
  
\bibitem{ht4}
  J.C.~Taylor and S.M.H.~Wong,
  {\it The effective action of hard thermal loops in QCD,}
  Nucl.\ Phys.\ B {346} (1990) 115.
  
\bibitem{qed1}
  V.P.~Silin, 
  {\it On the electromagnetic properties of a relativistic plasma,}
  Sov.\ Phys.\ JETP {11} (1960) 1136
  [Zh.\ Eksp.\ Teor.\ Fiz.\ {38} (1960) 1577].
  
\bibitem{qed2}
  V.V.~Klimov,
  {\it Collective Excitations in a Hot Quark Gluon Plasma,}
  Sov.\ Phys.\ JETP {55} (1982) 199
  [Zh.\ Eksp.\ Teor.\ Fiz.\  {82} (1982) 336].
  
\bibitem{qed3}
  H.A.~Weldon,
  {\it Covariant calculations at finite temperature: The relativistic plasma,}
  Phys.\ Rev.\ D {26} (1982) 1394.

\bibitem{htl_nlo}
  S.~Carignano, M.E.~Carrington and J.~Soto,
  {\it The HTL Lagrangian at NLO: the photon case,}
  Phys.\ Lett.\ B {801} (2020) 135193
  [1909.10545].

\bibitem{htl5}
  J.~Frenkel and J.C.~Taylor,
  {\it Hard thermal QCD, forward scattering and effective actions,}
  Nucl.\ Phys.\ B {374} (1992) 156.
 
\bibitem{htl6}
  E.~Braaten and R.D.~Pisarski,
  {\it Simple effective Lagrangian for hard thermal loops,}
  Phys.\ Rev.\ D {45} (1992) 1827. 

\bibitem{htl6x}
  D.~B\"odeker and M.~Laine,
  {\it Finite baryon density effects on gauge field dynamics,}
  JHEP {09} (2001) 029
  [hep-ph/0108034].

\bibitem{cl1}
  J.P.~Blaizot and E.~Iancu,
  {\it Kinetic equations for long-wavelength excitations of the 
  quark-gluon plasma,}
  Phys.\ Rev.\ Lett.\  {70} (1993) 3376
  [hep-ph/9301236].

\bibitem{cl2}
  P.F.~Kelly, Q.~Liu, C.~Lucchesi and C.~Manuel,
  {\it Deriving the hard thermal loops of QCD from 
  classical transport theory,}
  Phys.\ Rev.\ Lett.\  {72} (1994) 3461
  [hep-ph/9403403].

\bibitem{cl3}
  F.T.~Brandt, J.~Frenkel and J.C.~Taylor,
  {\it High temperature QCD and the classical Boltzmann equation in curved
  space-time,}
  Nucl.\ Phys.\ B {437} (1995) 433
  [hep-th/9411130].
 
\bibitem{cl5}
  R.D.~Pisarski,
  {\it Kinetic Theory of Hot Gauge Theories: Overview, Details \& Extensions,}
  NATO Sci.\ Ser.\ C {511} (1998) 195
  [hep-ph/9710370].

\bibitem{db1}
  D.~B\"odeker, 
  {\it Effective dynamics of soft non-Abelian gauge fields at finite
  temperature,}
  Phys.\ Lett.\ B {426} (1998) 351
  [hep-ph/9801430].
 
\bibitem{db2}
  P.~Arnold, D.T.~Son and L.G.~Yaffe,
  {\it Effective dynamics of hot, soft non-Abelian gauge fields:
  Color conductivity and $\log(1/\alpha)$ effects,}
  Phys.\ Rev.\ D {59} (1999) 105020
  [hep-ph/9810216].

\bibitem{db3}
  D.~B\"odeker,
  {\it Diagrammatic approach to soft non-Abelian dynamics
  at high temperature,}
  Nucl.\ Phys.\ B {566} (2000) 402
  [hep-ph/9903478].

\bibitem{db5}
  P.B.~Arnold and L.G.~Yaffe,
  {\it High temperature color conductivity at next-to-leading log order,}
  Phys.\ Rev.\ D {62} (2000) 125014
  [hep-ph/9912306].

\bibitem{db6}
  D.~B\"odeker,
  {\it Perturbative and non-perturbative aspects of the non-Abelian
  Boltzmann-Langevin equation,}
  Nucl.\ Phys.\ B {647} (2002) 512
  [hep-ph/0205202].
 
\bibitem{sum1}
  P.~Aurenche, F.~Gelis and H.~Zaraket,
  {\it A simple sum rule for the thermal gluon spectral 
  function and applications,}
  JHEP {05} (2002) 043
  [hep-ph/0204146].

\bibitem{sum4}
  M.~Panero, K.~Rummukainen and A.~Sch\"afer,
  {\it Lattice Study of the Jet Quenching Parameter,}
  Phys.\ Rev.\ Lett.\  {112} (2014)  162001
  [1307.5850].

\bibitem{qed4}
  J.-P.~Blaizot and J.-Y.~Ollitrault,
  {\it Collective fermionic excitations in systems with a large chemical
  potential,}
  Phys.\ Rev.\  D {48} (1993) 1390
  [hep-th/9303070].

\bibitem{pls1}
  H.A.~Weldon,
  {\it Dynamical holes in the quark-gluon plasma}, 
  Phys.\ Rev.\  D {40} (1989) 2410.

\bibitem{pls2}
  E.~Petitgirard,
  {\it Massive fermion dispersion relation at finite temperature,}
  Z.\ Phys.\  C {54} (1992) 673.

\bibitem{pls3}
  P.M.~Chesler, A.~Gynther and A.~Vuorinen,
  {\it On the dispersion of fundamental particles in QCD and 
  $\mathcal{N} = 4$ Super Yang-Mills theory,}
  JHEP {09} (2009) 003
  [0906.3052].

\end{thebibliography}

\begin{thebibliography}{9.99}

\setcounter{enumiv}{0} 

\bibitem{fk}
  R.~Fukuda and E.~Kyriakopoulos,
  {\it Derivation of the effective potential,}
  Nucl.\ Phys.\ B {85} (1975) 354.

\bibitem{rj}
  R.~Jackiw,
 {\it Functional evaluation of the effective potential},  
  Phys.~Rev.~D 9 (1974) 1686.

\bibitem{ki}
  D.A.~Kirzhnits,
  {\it Weinberg model in the hot universe,}
  JETP Lett.\  {15} (1972) 529
  [Pisma Zh.\ Eksp.\ Teor.\ Fiz.\  {15} (1972) 745].

\bibitem{kil}
  D.A.~Kirzhnits and A.D.~Linde,
  {\it Macroscopic consequences of the Weinberg model,}
  Phys.\ Lett.\ B {42} (1972) 471.

\bibitem{dj_2}
  L.~Dolan and R.~Jackiw,
  {\it Symmetry behavior at finite temperature,}
  Phys.\ Rev.\  D {9} (1974) 3320.

\bibitem{swT}
  S.~Weinberg,
  {\it Gauge and global symmetries at high temperature,}
  Phys.\ Rev.\ D {9} (1974) 3357.
 
\bibitem{tri1}
  M.~L\"uscher and P.~Weisz,
  {\it Scaling laws and triviality bounds in the lattice 
  $\varphi^4$ theory: (I). One-component model in the symmetric phase,}
  Nucl.\ Phys.\ B {290} (1987) 25.

\bibitem{tri2}
  M.~L\"uscher and P.~Weisz,
  {\it Scaling laws and triviality bounds in the lattice 
  $\varphi^4$ theory: (II). One-component model in the phase 
  with spontaneous symmetry breaking,}
  Nucl.\ Phys.\ B {295} (1988) 65.

\bibitem{first1}
 J.~Rudnick, 
 {\it First-order transition induced by cubic anisotropy}, 
 Phys.\ Rev.\ B 18 (1978) 1406.

\bibitem{twostep1}
  D.~Land and E.D.~Carlson,
  {\it Two stage phase transition in two Higgs models,}
  Phys.\ Lett.\ B {292} (1992) 107
  [hep-ph/9208227].

\bibitem{first2}
  B.I.~Halperin, T.C.~Lubensky and S.-K.~Ma,
  {\it First-Order Phase Transitions in Superconductors and 
  Smectic-A Liquid Crystals},  
  Phys.\ Rev.\ Lett.\ 32 (1974) 292.

\bibitem{first3}
  D.A.~Kirzhnits and A.D.~Linde, 
  {\it Symmetry behavior in gauge theories,}
  Annals Phys.\  {101} (1976) 195.

\bibitem{ae_2}
  P.B.~Arnold and O.~Espinosa,
  {\it Effective potential and first-order phase transitions: 
  Beyond leading order,}
  Phys.\ Rev.\ D {47} (1993) 3546; 
  {\it ibid.} {50} (1994) 6662 (E)
  [hep-ph/9212235].

\bibitem{proc}
  M.~Laine and K.~Rummukainen,
  {\it What's new with the electroweak phase transition?,}
  Nucl.\ Phys.\ Proc.\ Suppl.\  {73} (1999) 180
  [hep-lat/9809045].

\bibitem{rev4f_2}
  D.E.~Morrissey and M.J.~Ramsey-Musolf,
  {\it Electroweak baryogenesis,}
  New J.\ Phys.\ {14} (2012) 125003
  [1206.2942].

\bibitem{jl1}
  J.S.~Langer, 
  {\it Theory of the condensation point}, 
  Ann.\ Phys.\ 41 (1967) 108.

\bibitem{jl2}
  J.S.~Langer, 
  {\it Statistical theory of the decay of metastable states}, 
  Ann.\ Phys.\ 54 (1969) 258.

\bibitem{sc1}
  S.R.~Coleman,
  {\it Fate of the false vacuum: Semiclassical theory,}
  Phys.\ Rev.\  D {15} (1977) 2929; 
  {\it ibid.}  {16} (1977) 1248 (E).
 
\bibitem{adl1}
  A.D.~Linde,
  {\it Fate of the false vacuum at finite temperature: 
  Theory and applications,}
  Phys.\ Lett.\  B {100} (1981) 37.

\bibitem{callan}
  C.G.~Callan and S.R.~Coleman, 
  {\it Fate of the false vacuum. II. First quantum corrections,}
  Phys.\ Rev.\  D {16} (1977) 1762.

\bibitem{ia}
  I.\ Affleck, 
  {\it Quantum-Statistical Metastability,}
  Phys.\ Rev.\ Lett.\  {46} (1981) 388.

\bibitem{nucl_pref1}
  P.~Arnold, D.~Son and L.G.~Yaffe,
  {\it The hot baryon violation rate is $\rmO(\alpha_w^5 T^4)$,}
  Phys.\ Rev.\  D {55} (1997) 6264
  [hep-ph/9609481].

\bibitem{nucl_pref2}
  G.D.~Moore and K.~Rummukainen,
  {\it Electroweak bubble nucleation, nonperturbatively,}
  Phys.\ Rev.\ D {63} (2001) 045002
  [hep-ph/0009132].
 
\bibitem{rev4a_2}
  V.A.~Rubakov and M.E.~Shaposhnikov,
  {\it Electroweak baryon number non-conservation in the early Universe 
  and in high-energy collisions,}
  Usp.\ Fiz.\ Nauk {166} (1996) 493
  [Phys.\ Usp.\  {39} (1996) 461]
  [hep-ph/9603208].

\bibitem{instanton}
  A.A.~Belavin, A.M.~Polyakov, A.S.~Schwartz and Y.S.~Tyupkin,
  {\it Pseudoparticle solutions of the Yang-Mills equations,}
  Phys.\ Lett.\ B {59} (1975) 85.

\bibitem{kli}
  F.R.~Klinkhamer and N.S.~Manton,
  {\it A saddle-point solution in the Weinberg-Salam theory,}
  Phys.\ Rev.\ D {30} (1984) 2212.

\bibitem{sphal}
  P.~Arnold and L.D.~McLerran,
  {\it Sphalerons, small fluctuations, and baryon-number violation in 
  electroweak theory,}
  Phys.\ Rev.\  D {36} (1987) 581.

\bibitem{clgt2}
  J.~Ambj{\o}rn, T.~Askgaard, H.~Porter and M.E.~Shaposhnikov,
  {\it Sphaleron transitions and baryon asymmetry:
  A numerical, real-time analysis,}
  Nucl.\ Phys.\ B {353} (1991) 346.

\bibitem{srate3}
  M.~D'Onofrio, K.~Rummukainen and A.~Tranberg,
  {\it Sphaleron Rate in the Minimal Standard Model,}
  Phys.\ Rev.\ Lett.\ {113} (2014) 141602
  [1404.3565].

\bibitem{landau5}
  L.D.~Landau and E.M.~Lifshitz,
  {\it Statistical Physics, Part 1}, \S162
  (Butterworth-Heinemann, Oxford).

\bibitem{adl2}
  A.D.~Linde,
  {\it Decay of the false vacuum at finite temperature,}
  Nucl.\ Phys.\ B {216} (1983) 421; 
  {\it ibid.} {223} (1983) 544 (E).

\bibitem{bubbles1}
  A.H.~Guth and E.J.~Weinberg,
  {\it Cosmological consequences of a first-order phase transition
  in the SU(5) Grand Unified Model,}
  Phys.\ Rev.\ D {23} (1981) 876.

\bibitem{bubbles2}
  G.M.~Fuller, G.J.~Mathews and C.R.~Alcock,
  {\it Quark-hadron phase transition in the early Universe:
  Isothermal baryon number fluctuations and primordial nucleosynthesis,}
  Phys.\ Rev.\ D {37} (1988) 1380.

\bibitem{bubbles3}
  K.~Enqvist, J.~Ignatius, K.~Kajantie and K.~Rummukainen,
  {\it Nucleation and bubble growth in a first order cosmological
  electroweak phase transition,}
  Phys.\ Rev.\ D {45} (1992) 3415.

\bibitem{db_gdm}
  D.~B\"odeker and G.D.~Moore,
  {\it Electroweak bubble wall speed limit,}
  JCAP {05} (2017) 025
  [1703.08215].

\bibitem{bubbles4}
  H.~Kurki-Suonio and M.~Laine,
  {\it On bubble growth and droplet decay in cosmological phase transitions,}
  Phys.\ Rev.\ D {54} (1996) 7163
  [hep-ph/9512202].

\bibitem{dilepton1}
  L.D.~McLerran and T.~Toimela,
  {\it Photon and dilepton emission from the 
  quark-gluon plasma: Some general considerations,}
  Phys.\ Rev.\ D {31} (1985) 545.

\bibitem{dilepton2}
  H.A.~Weldon,
  {\it Reformulation of finite-temperature dilepton production,}
  Phys.\ Rev.\ D {42} (1990) 2384.

\bibitem{als}
  T.~Asaka, M.~Laine and M.~Shaposhnikov,
  {\it On the hadronic contribution to sterile neutrino production,}
  JHEP {06} (2006) 053
  [hep-ph/0605209].

\bibitem{dmpheno}
  J.~Ghiglieri and M.~Laine,
  {\it Improved determination of sterile neutrino dark matter spectrum,}
  JHEP {11} (2015) 171
  [1506.06752].

\bibitem{dbx}
  D.~B\"odeker, M.~Sangel and M.~W\"ormann,
  {\it Equilibration, particle production, and self-energy,}
  Phys.\ Rev.\ D {93} (2016) 045028
  [1510.06742].

\bibitem{bb2_2}
  A.~Anisimov, D.~Besak and D.~B\"odeker,
  {\it Thermal production of relativistic Majorana neutrinos:
  strong enhancement by multiple soft scattering,}
  JCAP {03} (2011) 042
  [1012.3784].

\bibitem{sum3_2}
  D.~Besak and D.~B\"odeker,
  {\it Thermal production of ultrarelativistic right-handed neutrinos:
  complete leading-order results,}
  JCAP {03} (2012) 029
  [1202.1288].

\bibitem{photon1_2}
  P.B.~Arnold, G.D.~Moore and L.G.~Yaffe,
  {\it Photon emission from ultrarelativistic plasmas,}
  JHEP {11} (2001) 057
  [hep-ph/0109064].

\bibitem{photon2_2}
  P.B.~Arnold, G.D.~Moore and L.G.~Yaffe,
  {\it Photon emission from quark-gluon plasma:
  complete leading order results,}
  JHEP {12} (2001) 009
  [hep-ph/0111107].

\bibitem{cf}
  M.~Cohen and R.P.~Feynman,
  {\it Theory of Inelastic Scattering of Cold Neutrons from Liquid Helium,}
  Phys.\ Rev.\  {107} (1957) 13.

\bibitem{hewo}
  D.G.~Henshaw and A.D.B.~Woods, 
  {\it Modes of Atomic Motions in Liquid Helium by Inelastic Scattering
  of Neutrons,}
  Phys.\ Rev.\  {121} (1961) 1266.

\bibitem{be}
  J. Bernstein, {\it Kinetic Theory in the Expanding Universe}
  (Cambridge University Press, Cambridge, 1988).

\bibitem{kolb}
  E.W.~Kolb and M.S.~Turner,
  {\it The Early Universe,}
  Front.\ Phys.\ {69} (1990) 1--547.

\bibitem{als2}
  T.~Asaka, M.~Laine and M.~Shaposhnikov,
  {\it Lightest sterile neutrino abundance within the $\nu$MSM,}
  JHEP {01} (2007) 091; 
  {\it ibid.} {02} (2015) 028 (E)
  [hep-ph/0612182].

\bibitem{fbez1}
  M.~Shaposhnikov and I.~Tkachev,
  {\it The $\nu$MSM, inflation, and dark matter,}
  Phys.\ Lett.\ B {639} (2006) 414
  [hep-ph/0604236].

\bibitem{fbez2}
  K.~Petraki and A.~Kusenko,
  {\it Dark-matter sterile neutrinos in models 
  with a gauge singlet in the Higgs sector,}
  Phys.\ Rev.\ D {77} (2008) 065014
  [0711.4646].

\bibitem{msm}
  T.~Matsui, B.~Svetitsky and L.D.~McLerran,
  {\it Strangeness production in ultrarelativistic heavy-ion 
  collisions.\ I.\ Chemical kinetics in the quark-gluon plasma,}
  Phys.\ Rev.\ D {34} (1986) 783; 
  {\it ibid.}  {37} (1988) 844 (E).

\bibitem{bulk1}
  D.~B\"odeker, 
  {\it Moduli decay in the hot early Universe,}
  JCAP {06} (2006) 027
  [hep-ph/0605030].

\bibitem{bulk2}
  M.~Laine,
  {\it On bulk viscosity and moduli decay,}
  Prog.\ Theor.\ Phys.\ Suppl.\  {186} (2010) 404
  [1007.2590].

\bibitem{mms}
  L.D.~McLerran, E.~Mottola and M.E.~Shaposhnikov,
  {\it Sphalerons and axion dynamics in high-temperature QCD,}
  Phys.\ Rev.\ D {43} (1991) 2027.

\bibitem{mlx}
  M.~L\"uscher,
  {\it Topological effects in QCD and the problem of 
  short distance singularities,}
  Phys.\ Lett.\ B {593} (2004) 296
  [hep-th/0404034].

\bibitem{axion1}
  M.P.~Lombardo and A.~Trunin,
  {\it Topology and axions in QCD,}
  Int.\ J.\ Mod.\ Phys.\ A {35} (2020) 2030010
  [2005.06547].

\bibitem{adm}
  P.B.~Arnold, C.~Dogan and G.D.~Moore,
  {\it Bulk viscosity of high-temperature QCD,}
  Phys.\ Rev.\ D {74} (2006) 085021
  [hep-ph/0608012].

\bibitem{gdm_x}
  G.D.~Moore,
  {\it Motion of Chern-Simons number at high temperatures 
  under a chemical potential,}
  Nucl.\ Phys.\ B {480} (1996) 657
  [hep-ph/9603384].

\bibitem{cs3}
  D.T.~Son and A.O.~Starinets,
  {\it Minkowski-space correlators in AdS/CFT correspondence: 
  recipe and applications,}
  JHEP {09} (2002) 042
  [hep-th/0205051].

\bibitem{clgt1}
  D.Y.~Grigoriev and V.A.~Rubakov,
  {\it Soliton pair creation at finite temperatures. Numerical study
  in (1+1)-dimensions,}
  Nucl.\ Phys.\ B {299} (1988) 67.

\bibitem{clgt3}
  D.~B\"odeker,
  {\it Classical real time correlation functions and 
  quantum corrections at finite temperature,}
  Nucl.\ Phys.\ B {486} (1997) 500
  [hep-th/9609170].

\bibitem{cs1}
  D.~B\"odeker, G.D.~Moore and K.~Rummukainen,
  {\it Chern-Simons number diffusion and hard thermal loops on the lattice,}
  Phys.\ Rev.\ D {61} (2000) 056003
  [hep-ph/9907545].

\bibitem{cs2}
  G.D.~Moore and M.~Tassler,
  {\it The sphaleron rate in SU(N) gauge theory,}
  JHEP {02} (2011) 105
  [1011.1167].

\bibitem{warm1}
  K.V.~Berghaus, P.W.~Graham and D.E.~Kaplan,
  {\it Minimal warm inflation,}
  JCAP {03} (2020) 034
  [1910.07525].

\bibitem{warm2}
  S.~Das, G.~Goswami and C.~Krishnan, 
  {\it Swampland, axions and minimal warm inflation}, 
  Phys.\ Rev.\ D {101} (2020) 10
  [1911.00323]. 

\bibitem{warm3}
  Y.~Reyimuaji and X.~Zhang,
  {\it Warm-assisted natural inflation,}
  JCAP {04} (2021) 077
  [2012.07329].

\bibitem{warm4}
  M.~Laine and S.~Procacci,
  {\it Minimal warm inflation with complete medium response,}
  JCAP {06} (2021) 031
  [2102.09913].

\bibitem{rev5_3}
  H.B.~Meyer, 
  {\it Transport properties of the quark-gluon plasma: A lattice QCD
  perspective,} 
  Eur.\ Phys.\ J.\ A {47} (2011) 86
  [1104.3708].

\bibitem{ga}
  G.~Aarts,
  {\it Transport and spectral functions in high-temperature QCD,}
  PoS {LAT2007} (2007) 001
  [0710.0739].

\bibitem{corr_tr}
  P.~Kovtun, G.D.~Moore and P.~Romatschke,
  {\it Stickiness of sound: An absolute lower limit on viscosity and
  the breakdown of second-order relativistic hydrodynamics,}
  Phys.\ Rev.\ D {84} (2011) 025006
  [1104.1586].

\bibitem{amy1}
  P.B.~Arnold, G.D.~Moore and L.G.~Yaffe,
  {\it Transport coefficients in high temperature gauge theories (I): 
  leading-log results,}
  JHEP {11} (2000) 001
  [hep-ph/0010177].
 
\bibitem{lgy}
  L.G.~Yaffe,
  {\it Dynamics of hot gauge theories,}
  Nucl.\ Phys.\ B (Proc.\ Suppl.)\  {106} (2002) 117
  [hep-th/0111058].

\bibitem{amy2}
  P.B.~Arnold, G.D.~Moore and L.G.~Yaffe,
  {\it Transport coefficients in high temperature gauge theories (II):
  Beyond leading log,}
  JHEP {05} (2003) 051
  [hep-ph/0302165].
 
\bibitem{mt}
  G.D.~Moore and D.~Teaney,
  {\it How much do heavy quarks thermalize in a heavy ion collision?,}
  Phys.\ Rev.\  C {71} (2005) 064904
  [hep-ph/0412346].

\bibitem{dt}
  D.~Teaney,
  {\it Finite temperature spectral densities of momentum 
  and $R$-charge correlators in ${\cal N} = 4$ Yang-Mills theory,}
  Phys.\ Rev.\  D {74} (2006) 045025
  [hep-ph/0602044].

\bibitem{kas}
  G.D.~Moore and K.A.~Sohrabi,
  {\it Kubo Formulas for Second-Order Hydrodynamic Coefficients,}
  Phys.\ Rev.\ Lett.\  {106} (2011) 122302
  [1007.5333].

\bibitem{kubo}
  R.~Kubo,
  {\it Statistical-Mechanical Theory of Irreversible Processes. 
  I. General Theory and Simple Applications 
  in Magnetic and Conduction Problems,}
  J.\ Phys.\ Soc.\ Jap.\  {12} (1957) 570.

\bibitem{KhSh}
  S.Y.~Khlebnikov and M.E.~Shaposhnikov, 
  {\it The statistical theory of anomalous fermion number non-conservation,}
  Nucl.\ Phys.\ B {308} (1988) 885. 

\bibitem{ct}
  J.~Casalderrey-Solana and D.~Teaney,
  {\it Heavy quark diffusion in strongly coupled ${\cal N} = 4$ Yang Mills,}
  Phys.\ Rev.\  D {74} (2006) 085012
  [hep-ph/0605199].

\bibitem{eucl}
  S.~Caron-Huot, M.~Laine and G.D.~Moore,
  {\it A way to estimate the heavy quark
  thermalization rate from the lattice,}
  JHEP 04 (2009) 053
  [0901.1195].

\bibitem{chem}
  D.~B\"odeker and M.~Laine,
  {\it Heavy quark chemical equilibration rate as a transport coefficient,}
  JHEP {07} (2012) 130
  [1205.4987].

\bibitem{landau9} 
  E.M.~Lifshitz and L.P.~Pitaevskii, 
  {\it Statistical Physics, Part 2}, \S88-89
  (Butterworth-Heinemann, Oxford).

\bibitem{satz}
  T.~Matsui and H.~Satz,
  {\it $J/\psi$ suppression by quark-gluon plasma formation,}
  Phys.\ Lett.\ B {178} (1986) 416.

\bibitem{threshold}
  S.~Kim and M.~Laine,
  {\it On thermal corrections to near-threshold annihilation,}
  JCAP {01} (2017) 013
  [1609.00474].

\bibitem{nrqcd1}
  A.~Czarnecki and K.~Melnikov,
  {\it Two-Loop QCD Corrections to the Heavy Quark Pair Production
  Cross Section in $e^+ e^-$ Annihilation near Threshold,}
  Phys.\ Rev.\ Lett.\  {80} (1998) 2531
  [hep-ph/9712222].

\bibitem{nrqcd2}
  M.~Beneke, A.~Signer and V.A.~Smirnov,
  {\it Two-Loop corrections to the Leptonic Decays of Quarkonium,}
  Phys.\ Rev.\ Lett.\  {80} (1998) 2535
  [hep-ph/9712302].

\bibitem{cwb_2}
  C.W.~Bernard, 
  {\it Feynman rules for gauge theories at finite temperature,}
  Phys.\ Rev.\  D {9} (1974) 3312.

\bibitem{kogut_2}
  J.B.~Kogut and L.~Susskind,
  {\it Hamiltonian formulation of Wilson's lattice gauge theories,}
  Phys.\ Rev.\  D {11} (1975) 395.

\bibitem{ml_2}
  M.~L\"uscher,
  {\it Construction of a selfadjoint, strictly positive transfer matrix for
  Euclidean lattice gauge theories,}
  Commun.\ Math.\ Phys.\  {54} (1977) 283.

\bibitem{string}
  K.J.~Juge, J.~Kuti and C.~Morningstar,
  {\it Fine Structure of the QCD String Spectrum,}
  Phys.\ Rev.\ Lett.\  {90} (2003) 161601
  [hep-lat/0207004].

\bibitem{akr}
  A.~Rothkopf, T.~Hatsuda and S.~Sasaki, 
  {\it Proper heavy-quark potential from a spectral decomposition 
  of the thermal Wilson loop,}
  PoS {LAT2009} (2009) 162
  [0910.2321].

\bibitem{jgx}
  J.~Ghiglieri,
  {\it Review of the EFT treatment of quarkonium at finite temperature,}
  PoS ConfinementX (2012) 004
  [1303.6438].

\bibitem{mott}
 D.~Blaschke, G.~Burau, Y.~Kalinovsky and T.~Barnes,
  {\it Mott effect and $J / \psi$ dissociation at the 
  quark hadron phase transition,}
  Eur.\ Phys.\ J.\ A {18} (2003) 547
  [nucl-th/0211058].
 
\end{thebibliography}

\newpage

\begin{thebibliography}{99.9}

\setcounter{enumiv}{0} 

\bibitem{drewes}
  L.~Canetti, M.~Drewes, T.~Frossard and M.~Shaposhnikov,
  {\it Dark matter, baryogenesis and neutrino oscillations 
  from right-handed neutrinos,}
  Phys.\ Rev.\ D {87} (2013) 093006
  [1208.4607].

\bibitem{nc1}
  A.~Abbas,
  {\em Anomalies and charge quantization in the Standard Model 
  with arbitrary number of colours,}
  Phys.\ Lett.\ B {238} (1990) 344.

\bibitem{nc2}
  O.~B\"ar and U.-J.~Wiese,
  {\it Can one see the number of colors?,}
  Nucl.\ Phys.\ B {609} (2001) 225
  [hep-ph/0105258].

\bibitem{g5a}
  G.~'t Hooft and M.J.G.~Veltman,
  {\em Regularization and renormalization of gauge fields,}
  Nucl.\ Phys.\  {B 44} (1972)  189.

\bibitem{g5b}
  P.~Breitenlohner and D.~Maison,
  {\em Dimensional renormalization and the action principle,}
  Commun.\ Math.\ Phys.\ \ {52} (1977) 11.

\bibitem{JK}
  J.G.~K\"orner, N.~Nasrallah and K.~Schilcher,
  {\em Evaluation of the flavor-changing vertex 
  $b \to s H$ using the Breitenlohner-Maison-'t Hooft-Veltman 
  $\gamma^{ }_5$ scheme,}
  Phys.\ Rev.\  {D 41} (1990)  888.

\bibitem{BW}
  A.J.~Buras and P.H.~Weisz,
  {\em QCD nonleading corrections to weak decays in 
  dimensional regularization and 't Hooft-Veltman schemes,}
  Nucl.\ Phys.\  {B 333} (1990)  66.

\end{thebibliography}
\end{document}